\documentclass[twocolumn]{aastex631}

\usepackage{amsmath}
\usepackage[caption=false]{subfig}
\usepackage[shortlabels]{enumitem}
\usepackage{array}
\newcolumntype{P}[1]{>{\centering\arraybackslash}p{#1}}
\usepackage{booktabs}
\usepackage{multirow}
\usepackage{bm}
\usepackage{graphicx}

\definecolor{burntorange}{rgb}{0.8, 0.33, 0.0}
\definecolor{black}{rgb}{0.0, 0.0, 0.0}
\newcommand{\reviewTwo}[1]{\textcolor{black}{#1}}
\newcommand{\reviewThree}[1]{\textcolor{black}{#1}}
\newcommand{\reviewFour}[1]{\textcolor{black}{#1}}
\newcommand{\reviewFive}[1]{\textcolor{black}{#1}}

\begin{document}

\title{Deconvolution for Large Astronomical Surveys: A Study of the Scaled Gradient Projection Method on Zwicky Transient Facility Data}
\correspondingauthor{Ashish~A.~Mahabal}

\author[0000-0002-6646-4225]{Yash Gondhalekar}
\affiliation{Department of CSIS, BITS Pilani K.K Birla Goa Campus, Goa, 403726, Goa, India}

\author[0000-0002-9330-8738]{Richard M. Feder}
\affiliation{Berkeley Center for Cosmological Physics, University of California, Berkeley, CA 94720, USA}
\affiliation{California Institute of Technology, 1200 E California Boulevard, CA 91125, USA}

\author[0000-0002-3168-0139]{Matthew J. Graham}
\affiliation{Division of Physics, Mathematics and Astronomy, California Institute of Technology, Pasadena, CA 91125, USA}

\author[0000-0003-0598-4152]{Ajit K. Kembhavi}
\affiliation{Inter-University Centre for Astronomy and Astrophysics, Pune, India}

\author[0000-0003-4893-6150]{Margarita Safonova}
\affiliation{M. P. Birla Institute of Fundamental Research, Bangalore, India}

\author[0000-0002-8458-604X]{Snehanshu Saha}
\affiliation{Department of CSIS and APPCAIR, BITS Pilani K.K Birla Goa Campus, Goa, 403726, Goa, India}

\author[0000-0003-2242-0244]{Ashish~A.~Mahabal}
\affiliation{Division of Physics, Mathematics and Astronomy, California Institute of Technology, Pasadena, CA 91125, USA}
\affiliation{Center for Data Driven Discovery, California Institute of Technology, Pasadena, CA 91125, USA}

\begin{abstract}

Ground-based astronomical observations will continue to produce resolution-limited images due to atmospheric seeing. Deconvolution reverses such effects and thus can benefit extracted science in multifaceted ways. We apply the Scaled Gradient Projection (SGP) algorithm for the single-band deconvolution of several observed images from the Zwicky Transient Facility and mainly discuss the performance on stellar sources. The method shows good photometric flux preservation, which deteriorates for fainter sources but significantly reduces flux uncertainties even for the faintest sources. Deconvolved sources have a well-defined Full-Width-at-Half-Maximum (FWHM) of roughly one pixel (one arcsecond for ZTF) regardless of the observed seeing. Detection after deconvolution results in catalogs with $\gtrsim$99.6\% completeness relative to detections in the observed images. A few observed sources that could not be detected in the deconvolved image are found near saturated sources, whereas for others, the deconvolved counterparts are detected when slightly different detection parameters are used. \reviewTwo{The deconvolution reveals new faint sources previously undetectable, which are confirmed by crossmatching with the deeper DESI Legacy DR10 and with Pan-STARRS1 through forced photometry}. The method could identify examples of serendipitous potential deblends that exceeded SExtractor's deblending capabilities, with as extreme as $\Delta m \approx 3$ and separations as small as one arcsecond between the deblended components. \reviewThree{Our survey-agnostic approach} is better and \reviewFive{eight} times faster than Richardson-Lucy deconvolution and could be a reliable method for incorporation into survey pipelines.
\end{abstract}

\keywords{Sky surveys(1464), Astronomy image processing (2306), Deconvolution (1910), Maximum likelihood estimation (1901)}

\section{Introduction}\label{sec:intro}
\subsection{Background}

Deconvolution of astronomical images observed from ground-based telescopes aims to deduce the true yet unknown emission from the observed images degraded by atmospheric seeing, instrumental aberrations, diffraction, and other sources of noise. The idea of deconvolution is crucial in astronomy because our current theoretical models for characterizing astronomical sources are not very accurate, so they cannot be directly used to forward model observed data \citep{Molina2001}. Despite the increasing quality of telescopes, the reversal of blurring and other distortions due to the atmosphere and instrument is critical for ground-based observations, where the Point Spread Functions (PSFs) have large wings or have elliptical or otherwise distorted shapes. If left untreated, the scientific value of these images is reduced, since (i) unwanted systematic errors in measurements of detected source properties may be introduced that preclude robust statistical analyses, and (ii) they may miss true astrophysical phenomena and thus inhibit their exploration. Deconvolution can help increase the authenticity of source detections by eliminating ambiguous ones and better elucidate the astrophysical properties of detections \citep[see, e.g.,][for applications in galaxy studies]{Bock2000, Burud2002, Faure2002, Chung2021, Leist2024}, thus demonstrating the scientific importance of deconvolution. Some of these works have demonstrated how deconvolution helped cross-correlate information from different datasets and gain further insights from available data.


Since deconvolution tackles the essence of the problem (i.e., degradation by PSF), it may benefit a wide range of downstream tasks, such as deblending closely separated sources, identifying faint sources submerged in the background, revealing substructure inside extended sources such as galaxies, and improving transient detections in time-series photometry. It may also help alleviate the common practice of tuning several detection parameters used to optimize certain types of detection in source extraction packages such as `SExtractor' \citep{Bertin1996}. The merits of deconvolution could become even more pertinent to make full use of the increasingly larger volume and deeper datasets from surveys such as ZTF \citep{Bellm2019,Masci2019}, and the upcoming Vera Rubin Observatory \citep{Ivezic2019}, {\it Nancy Grace Roman Space Telescope} \citep{Spergel2015}, and the {\it Euclid} mission \citep{Laureijs2011}. For example, the Wide survey of the Hyper Suprime-Cam (HSC) finds 58\% sources to be blended \citep{Bosch2018}; \citet{Dawson2016} found $\sim$14\% galaxies to be unrecognizable blends in a survey similar to the Vera Rubin Observatory, 

However, there are a number of challenges that impede deconvolution. This inverse problem is ill-posed, implying that many feasible solutions are possible given observed data, but not all solutions are practical. Any solution that does exist is generally not robust to noise, which is particularly challenging for deconvolving low signal-to-noise ratio data. In addition, a sufficiently accurate model of the PSF is required since small imperfections in the PSF modeling may produce drastically different solutions. However, it is challenging to yield precise PSF estimations, especially if the model is derived using observed data, because of the complex simultaneous mathematical description of atmospheric and instrumental imperfections. To reduce the ill-posedness of the problem, one makes simple yet effective assumptions about the noise and any possible prior information about the astronomical source: the Poisson noise model to describe photons reaching the CCD detector, non-negativity of pixel values, and conservation of total flux. In view of these inherent challenges of the deconvolution problem, it has been found that iterative methods with suitable regularization schemes to control noise amplification are generally desired since they are more flexible and robust to the above issues than, say, one-step inversion schemes.


\citealt{Starck2002} and \citet{Bertero2021}, for example, provide a good review of statistical deconvolution theory and the deconvolution methods used in astronomy. Some examples of traditional deconvolution algorithms include the minimization of the Tikhonov functional \citep{Tikhonov1977}; Wiener filtering, a Fourier-based technique well known in the sciences that is fast and easy to implement but generates ringing artifacts or over-smooths intricate structures within astronomical sources; the {\sc CLEAN} algorithm developed for radio astronomy \citep{Hogbom1974}, which can deconvolve point-like sources but fails to handle extended emissions. The Maximum Entropy Method was developed by \citet{Skilling1984, Gull1984} to perform a constrained maximization of the entropy of the restored image, but may be inefficient in simultaneously handling point-like and extended sources, and its multiscale extension was proposed by \citet{Pantin1996}. Deconvolution using shapelets was applied in \citet{Refregier2003} in the context of weak lensing, and an approach based on blind deconvolution was proposed in \citet{Jefferies1993}. The Richardson-Lucy (RL; also called the Expectation-Maximization method) \citep{Richardson1972, Lucy1974}, is an iterative method to obtain a maximum likelihood solution for data corrupted with Poisson noise and is widely used in astronomy. RL, however, possesses slow convergence, especially for images with low Poisson noise, and tends to amplify noise during its iteration.  

Now we mention some more recent attempts to perform deconvolution. \citet{Hirsch2011} developed an online multi-frame blind deconvolution algorithm using multiple frames to achieve super-resolution and handle saturated pixels. \citet{Farrens2017} developed a deconvolution method with regularization methods such as low-rank approximation and object-based deconvolution and applied it to galaxy images. \citet{Sureau2020} combined deep learning and classical deconvolution methods for galaxy deconvolution. \citep{Millon2024} developed the {\sc STARRED} method for optimized joint deconvolution of photometric time series images along with an empirical PSF reconstruction approach. \citet{Long2021} presented a general structure learning framework based on deep learning to tackle some topics related to the restoration of astronomical images. \citet{Akhaury2024} used Vision Transformers to identify substructures within galaxies. \citet{Donath2024} developed a multi-frame deconvolution method that optimizes the posterior distribution using the joint Poisson likelihood and a specialized prior distribution, which was found to be better than the RL algorithm.

Some of the recent works on astronomical image deconvolution have relied on deep learning, which aims to build predictive models by typically learning the inverse convolution kernel in a data-driven manner. This means that a lot of good-resolution training data is required for training, but this may be challenging to acquire, particularly in some wavelength ranges (e.g., X-ray). In addition to this, current deep learning models typically lack interpretability, and it is unclear whether they can generalize to out-of-distribution data by seamlessly handling differing observational systematics across datasets \citep[although see, e.g., ][for a recent attempt based on physics-informed neural networks]{Ni2024}.

In this paper, we focus on a prescriptive rather than a predictive approach that is not based on machine learning, but instead based on a purely mathematical optimization routine. We thus do not require any training or confronting the challenges with neural network deployment mentioned above. Specifically, we employ the Scaled Gradient Projection (SGP) deconvolution algorithm \citep{Bonettini2009}, described in more detail in Sect.~\ref{sec:deconv}, which is a more generalized (more parameters) and efficient version of the RL algorithm and which has shown better convergence than RL in simulation studies. We apply SGP on twelve $\sim$3k $\times$ 3k arcsec quadrants from field images from ZTF at varying sky locations and at different seeing conditions. Although the SGP framework can be extended to multi-band deconvolution, we focus on single-band deconvolution and leverage the $r$-band images as the default. A discussion of the deconvolution performance in the $g$ and $i$ bands is also presented separately. We select two of these quadrants that contain a dwarf galaxy (extended source) and a globular cluster (crowded field), and briefly demonstrate the performance of the deconvolution. 




\subsection{Scope of this work}

While deconvolution is valuable for space-based observations, as demonstrated by past applications to Hubble Space Telescope images for resolving spherical aberration \citep{White1991,Hanisch1994}, this paper focuses on the deconvolution of images acquired from ground-based telescopes affected by atmospheric seeing. The overarching aim of this study is to present a deconvolution pipeline that can potentially be incorporated into ZTF to improve the scientific outcome extracted from the observed images. In this regard, one of our primary envisioned use cases of deconvolution in ZTF is deblending merged sources from non-optimal seeing images so that they can be told apart. We aim to successfully deblend sources whose component sources have similar magnitudes and, in addition, component sources that have a non-trivial magnitude difference (e.g., a bright source merged with a faint one). The retention of original sources in the deconvolved images is also vital, and ideally, we require that the deconvolution identify faint sources that have not been detected in the original image.

As mentioned before, the scope of this study is in single-band deconvolution of astronomical images where the PSF is assumed to be known (non-blind). We do not consider the case of a spatially varying PSF, but its incorporation may be a straightforward extension of the current implementation. The specific deconvolution algorithm used here, the Scaled Gradient Projection method, has been extensively tested for reconstruction accuracy and convergence using simulated data in previous studies, while its application to observational data has been less examined. According to the authors, this work is the first study to extensively test the capabilities of SGP for deconvolution on different types of observed data. Although deconvolution of extended sources such as galaxies is relevant for ongoing and future astronomical surveys, in this study, we mainly evaluate the deconvolution performance on star-like sources (small full width at half maximum (FWHM), low ellipticity) and those lying within a plausibly restricted magnitude range. This is primarily for simplicity, as it is easier to analyze star-like sources, unlike galaxies, which contain substructures. On a technical note, \citet{Magain1998} suggested that, for sampled data (which is true for images acquired through a modern CCD), attempting to deconvolve an observed image in the hope of entirely reversing the blurring and distortion by PSF would violate the sampling theorem and thus cause unwanted artifacts (e.g., speckling, ringing). Although a solution to this challenge was detailed in the reference cited above, which involves deconvolution with a PSF narrower than the observed PSF, such treatments are not considered in our implementation. 

It is well known that the practical viability of a deconvolution algorithm generally depends on the parameter space of application (e.g., type of source, signal-to-noise ratio of observation, sampling) and specific scientific objectives \citep{Schade1993}. Thus, it becomes essential to test a deconvolution algorithm under different conditions and to highlight the merits and possible pitfalls of a deconvolution method. Thus, given the above-mentioned constraints of this study, our aim is to extensively test the abilities of SGP on different field images and discuss its strengths and weaknesses.

The paper is organized as follows. Sect.~\ref{sec:deconv} describes the basic deconvolution theory and details of the SGP deconvolution algorithm. Sect.~\ref{sec:data} describes the data used in this study, and Sect.~\ref{subsec:sgp} describes the implementation details of SGP. Sect.~\ref{subsec:experimental-details} describes the experimental details regarding source selection criteria and catalog crossmatching, and Sects.~\ref{sec:exec-time}--\ref{subsec:rl-sgp-compare-sec} discuss the results. A summary of the results is presented in Sect.~\ref{sec:discussion}, followed by concluding remarks in Sect.~\ref{sec:conclusion}. 

\section{Deconvolution and the Scaled Gradient Projection Algorithm}\label{sec:deconv}


The imaging equation is given by: $\mathbf{g} = A\mathbf{f} + \mathbf{b}$, where $\mathbf{g}$ is the observed image, modeled as the convolution of the PSF matrix, $A$, with the unknown undegraded image, $\mathbf{f}$, (where $A\mathbf{f} = \mathbf{K} \ast \mathbf{f}$, where $\mathbf{K}$ is the PSF), and $\mathbf{b}$ is the background emission. Deconvolution is an inverse modeling problem that aims to estimate $\mathbf{f}$ from $\mathbf{g}$; however, this problem is ill-posed: a solution may not exist or may not be unique, and small changes in the observations can lead to drastic changes in the solution. Thus, iterative approaches are required to estimate $\mathbf{f}$ with suitable regularization, and constraints must be applied at each iteration to reduce the number of possible solutions and yield stable solutions\footnote{Constraints in such optimization problems can be viewed as a way to impose \emph{a priori} information about the `unknown' object $\mathbf{f}$. Simple constraints are often preferred to keep computational costs under limit}.

The Poisson noise model is typically assumed for deconvolution algorithms in astronomy, which is a suitable assumption since astronomical images are described by count statistics of the true object and background emission. Other noise sources, such as (additive Gaussian) readout noise, are expected to exist in the data but are generally excluded from the imaging model for simplicity or are sometimes even modeled as Poisson noise \citep{Snyder1994} so that they can be included with the background term, $\mathbf{b}$. We follow the first approach by excluding the readout noise component. Under this assumption, the data are modeled as only Poisson noise, so the objective function can be chosen to be the Kullback-Leibler (KL) divergence of $A\mathbf{f} + \mathbf{b}$ from $\mathbf{g}$, since in this case, minimizing the KL divergence is equivalent to maximizing the likelihood.

We first start by briefly introducing the RL algorithm, followed by a description of SGP. The traditional RL algorithm, assuming readout noise to be negligible or included in the background, iteratively yields the deconvolved estimate at the $n^{th}$ iteration as follows (see \citealt{Snyder1991}):
\begin{equation}
    \mathbf{f}^n = \mathbf{f}^{n-1} \odot A^\intercal \frac{\mathbf{g}}{A\mathbf{f}^{n-1} + \mathbf{b}}
\end{equation}\label{eqn:traditional-rl}
where $A^\intercal$ is the transposed PSF matrix and $\odot$ denotes element-wise matrix multiplication. An equivalent version of this iteration can be written as:
\begin{equation}
    \mathbf{f}^n = \mathbf{f}^{n-1} - \mathbf{f}^{n-1} \nabla J(\mathbf{f}; \mathbf{g})
\end{equation}\label{eqn:rl-scaled-gradient-method}
This simple scheme ensures a non-negative solution at each iteration and, if $\mathbf{b} = \mathbf{0}$, total and local flux are preserved. Moreover, if these iterations converge, they would lead to the maximum likelihood estimate for the case of Poisson noise data and thus one that minimizes the KL divergence (note that convergence to a solution when $\mathbf{b}$ is non-zero has not been proved yet). However, the primary disadvantages of RL are that it may require several iterations for bright, star-like sources and that it has no robustness to noise: the algorithm is forced to fit noise and generate artifacts (e.g., non-physical ``speckled'' structures) in the deconvolution if suitable regularization is not used. Thus, regularized versions of RL that tackle these challenges have been the focus of modern RL applications.

\citet{Bonettini2009} proposed the scaled gradient projection method (SGP), an iterative first-order gradient projection-type optimization method to minimize any continuously differentiable objective function, such as the KL divergence ($J(\mathbf{f}; \mathbf{g})$) for the setup mentioned above. The estimate is updated in SGP using the following scheme:
\begin{equation}
    \mathbf{f}^n = \mathbf{f}^{n-1} - \alpha_{n-1} D_{n-1} \nabla J(\mathbf{f}; \mathbf{g})
\end{equation}
where $D_{n-1}$ is called the scaling matrix and $\alpha_{n-1}$ as the steplength. Compared to the previous equation, it can be observed that RL iterations can be reproduced by setting $D_{n-1} = \mathbf{f}^{n-1}$ and $\alpha_{n-1} = 1$. In this sense, SGP can be seen as a generalization of RL. In SGP, the gradient directions are diagonally scaled using a diagonal scaling matrix and effective steplength selection rules designed for these directions. The highlight is that the scaling matrix and the steplength parameter are updated at each iteration (instead of keeping them fixed as in RL, if we view it as a type of scaled gradient method; see Eqn.~\ref{eqn:rl-scaled-gradient-method}). These considerations cumulatively help achieve a better convergence rate in SGP than in RL.

Another difference in SGP compared to RL is the projection step in SGP (since it is a gradient `projection' method), in which a special operation is used to project the deconvolved estimate onto a feasible region determined by constraints such as non-negativity and flux conservation. Non-negativity is necessary since photon counts cannot be negative, and flux conservation is also crucial since distortions by the PSF do not affect the photon counts except for cases such as pixels near the edges. However, in the RL algorithm, flux conservation is not guaranteed unless the background level is null, which generally does not hold. RL is robust to small errors in the PSF \citep{Bertero2009}, and is expected to be similar for SGP due to some similarities between the RL and SGP iteration schemes discussed above.

The time complexity of SGP is the same as that of RL ($\mathcal{O}(N\log{}N)$ per iteration), where $N$ is the number of pixels in the image and for images with equal dimensions along the axes of size $n \times n, N = n^2$ \citep{Bonettini2009}. This is because even though SGP contains more computations to find the solution, these additional calculations are efficient: for example, the projection step in SGP is only of linear complexity ($\mathcal{O}(N)$). Although SGP requires around 70\% more time per iteration than RL, SGP's better convergence ensures that SGP requires, in general, $\sim$20 times less time than RL \citep{Bonettini2009}. SGP contains more parameters than RL; however, extensive tests have found an optimal parameter set, thus mitigating the requirement of parameter fine-tuning \citep{Prato2012,Bertero2013}. The reader is referred to the works of \citet{Bonettini2009,Prato2012} for more specific details on the SGP algorithm and to \citet{Bertero2009} for a broader discussion.

SGP has been studied in a few applications in the literature: \citet{Zanella2009} used SGP for denoising with a modified constraint; \citet{Benvenuto2010} used SGP to accelerate the convergence of a few algorithms for least-squares image deblurring; \citet{Prato2013}, \citet{Wei2015}, and \citet{Jia2017} performed blind deconvolution in the SGP framework. These studies mainly used simulations, making it essential to test SGP on observed images where the source of the ground truth is not necessarily known, and observational artifacts could be present unknowingly. \citet{Gondhalekar2023} modified SGP by using a flexible objective function and applied it to globular cluster fields.

\section{Methods}
\subsection{Data}\label{sec:data}
ZTF is a 48-inch Schmidt telescope with a 47 $\mathrm{deg}^2$ field of view, containing 16 CCD detectors and a limiting depth of $m \approx 20.5$ \citep{Bellm2019,Masci2019}. The pixel size of the ZTF camera is 1$\arcsec$.012, and each CCD detector contains four 3k $\times$ 3k pixels readout-amplifier channels (or CCD quadrants). This work uses instrumentally calibrated science image files and estimated PSF models in the center of science images, as obtained through the NASA/IPAC Infrared Science Archive (IRSA) service \citep{Masci2019}. ZTF data products are CCD-quadrant-based, resulting in science images of 3k $\times$ 3k pixels dimensions. Although these CCD quadrants are used throughout the paper, which are not technically entire field images, we generally refer to them as `fields' ahead. Other data products, such as source catalogs, are not used. Instead, source extraction and cataloging are performed downstream (details in Sect.~\ref{subsec:experimental-details}).

\begin{deluxetable*}{cllcccll}[ht]\label{tab:data-description}
\tabletypesize{\footnotesize}
\centering
\caption{Details of the images used for deconvolution. All images are CCD quadrants of size $\sim$3k $\times$ 3k pixels. The RA and DEC values are of the image center; the location of NGC 1569 in field ID 811 is (RA, DEC) = (67.7 , 64.85) deg, and the location of NGC 7006 in field ID 594 is (RA, DEC) = (315.37,  16.19) deg. All images are of 30-second exposure. $m_{\mathrm{lim}}$ denotes the 5$\sigma$ magnitude limit obtained from ZTF's PSF-fit catalog, as noted in the image header. $b$ denotes the galactic latitude. The two-digit CCD identifier and the quadrant identifier within the CCD are also specified.} 
\tablehead{
\colhead{ZTF Field ID} & \colhead{(RA, DEC) deg} & \colhead{Field Description and Galactic Latitude} & \colhead{Filter} & \colhead{CCD} & \colhead{Quadrant} & \colhead{Seeing ($\arcsec$)} & \colhead{$m_{\mathrm{lim}}$}
}
\startdata
626 & \reviewTwo{(192.394, +27.501)} & High ($b \approx 90$ deg) & $r$ & 12 & 2& 3.12 & 20.27\\
& \reviewTwo{(192.401, +27.51)} & & $g$ & 12 & 2 & 3.13 & 20.8\\
& \reviewTwo{(192.414, +27.512)} & & $i$ & 12 & 2 & 3.32 & 20.04\\
251 & \reviewTwo{(42.974, -21.812)} & High ($b \approx -60$ deg) & $r$ & 16 & 4 & 2.9 & 20.11\\
619 & \reviewTwo{(142.124, +27.549)} & Intermediate ($b \approx 45$ deg) & $r$ & 11 & 2 &2.31 & 21.32\\
635 & \reviewTwo{(260.739, +26.665)} & Intermediate ($b \approx 30$ deg) & $r$ & 12 & 4 & 1.69 & 20.94\\
829 & \reviewTwo{(316.37, +62.645)} & Low ($b \approx 10$ deg) & $r$ & 10 & 4 & 1.85 & 21.39\\
& \reviewTwo{(316.354, +62.645)} & & $g$ & 10 & 4 & 1.91 & 21.43\\
& \reviewTwo{(316.376, +62.643)} & & $i$ & 10 & 4 & 1.83 & 20.39\\
233 & \reviewTwo{(266.132, -28.982)} & Low ($b \approx 0$ deg) & $r$ & 15 & 3 & 2.65 & 19.5\\
811 & \reviewTwo{(67.86, +64.477)} & Has dwarf galaxy NGC 1569 ($b \approx 11.2$ deg) & $r$ & 16 & 4 & 2.33 & 20.72\\
594 & \reviewTwo{(315.446, +16.44)} & Has globular cluster NGC 7006 ($b \approx -19.4$ deg) & $r$ & 04 & 1 & 1.84 & 21.17\\
\enddata
\end{deluxetable*}

Table~\ref{tab:data-description} details the images used for deconvolution in this study, selected across different galactic latitudes ($-60 \lesssim b \lesssim 90$ deg), filters ($g$, $r$, and $i$), and seeing conditions ($\sim$$1\arcsec.7 - 3\arcsec.3$). Apart from these, no specific selection criteria were applied, such as avoidance of bright sources or artifacts, to ensure minimal selection bias.

\subsection{SGP implementation}\label{subsec:sgp}
The Python version of SGP released by \citet{Gondhalekar2023} is an adapted reimplementation of the \textsc{MATLAB} SGP code for single image deconvolution of \cite{Prato2012}, which uses a two-dimensional background estimate to handle background gradients across the image. This work closely follows this Python implementation of SGP\footnote{Code can be found at 
\url{https://github.com/Yash-10/deconv_ztf}
}. Following the common choice, the initialization of the scaling matrix used in SGP is from the RL method since, as discussed in Sect.~\ref{sec:deconv}, SGP is a generalization of RL \citep{Bonettini2009,Prato2012}. Although it must be noted that SGP does not reduce to the standard RL algorithm with such a choice of initialization, since the projection step in SGP, which provides a direct way for flux preservation, is absent in RL.

The predefined parameters of SGP are taken from \citet{Prato2012}. The observed image is used as the initialization of the deconvolved image. Regularization is obtained by early stopping the iterations to handle the ill-posedness of the deconvolution problem. The convergence of the KL divergence is used as the criterion for stopping the iterations, i.e., when the following holds:
\begin{equation}
    |J(\bm{f}^{k+1}; \bm{g}) - J(\bm{f}^{k}; \bm{g})| \leq tol \enspace J(\bm{f}^{k}; \bm{g})\,
\end{equation}
In practice, iterations are stopped when either the above condition is met or if a certain sufficient number of iterations (500 iterations used here) are reached to prevent the deconvolution run indefinitely. High $tol$ values may prioritize deconvolution for brighter sources in the images and may even remove fainter sources \citep{Prato2012}. Such suppression of relatively fainter sources is typical of iterative deconvolution methods \citep[see also, e.g.,][]{Wu1998, Starck2002}. For extended sources, too low $tol$ values may produce undesired discontinuities since non-negative minimizers of KL divergence are sparse objects \citep{Barrett2003}. $tol = 10^{-4}$ is used in this work as a rough compromise.

As discussed in Sect.~\ref{sec:deconv}, non-negativity and total flux conservation constraints are imposed at each iteration of SGP for physical and photometric plausibility. Since our implementation of SGP preserves only the total flux, a separate validation check is generally required to determine the level of agreement of individual source fluxes. It was decided to apply SGP to $512 \times 512$ subdivisions extracted from the entire field images with an overlap of 10 pixels instead of the entire field at once -- we speculate that this choice may better impose flux conservation of individual sources since the algorithm needs to handle fewer sources at a time.  

We assume a constant PSF across the $\mathrm{3k} \times \mathrm{3k}$ ZTF fields, so the PSF model estimated at the center of the entire field image is used to restore all subdivisions. This choice may be sub-optimal for fields with significant spatial variation of the PSF. Extensions of our procedure using a space-variant PSF can be incorporated using different PSFs for different subdivisions during the deconvolution. Deconvolution can be affected by saturated pixels or by cosmic rays. Any pixel that is saturated or contains a non-finite value is treated as a bad pixel; such pixels are replaced using interpolation from surrounding finite-valued pixels before deconvolution. The interpolation is performed using the PSF as the kernel. No boundary effect correction is applied since we are concerned with sources that are completely contained in the image, and those that are not are not considered in our study \citep[see, e.g.,][for a discussion on this issue]{Bertero2005}.

\section{Results}\label{sec:results}

\subsection{Experimental details}\label{subsec:experimental-details}
\textsc{Sextractor} \citep{Bertin1996} is used to detect sources in the original and deconvolved images. Table~\ref{tab:sextractorParams} describes a few essential detection parameters. The \texttt{MAG\_ZEROPOINT}, \texttt{SEEING\_FWHM}, and \texttt{GAIN} parameters are catered for each image using the metadata information from the FITS header. A $3 \times 3$ pyramidal function with $\mathrm{FWHM} = 2$ pixels is used to filter images prior to detection. Weighting is used to handle variable spatial noise through variance maps calculated using the background level in the image, although the specific weighting mechanism is different from that used in the ZTF pipeline. For deconvolved images, we use \texttt{DETECT\_MINAREA} = 1 to detect sources that (expectedly) may be too compact (since deconvolution reverses the effect of PSF; see Fig.~\ref{fig:visual}), turn off filtering and cleaning for similar reasons\footnote{While filtering can help smooth out and optimize certain detections, it may not be strictly required for deconvolved images since the background emission is removed. Turning off filtering and cleaning also means that some spurious sources will be detected, but given our selection criteria (see below in the text), we have found that a good fraction of the spurious ones were eliminated.}, and set the background level required to zero for detection by SExtractor (since the deconvolution would have removed the background; see Sect.~\ref{sec:deconv})\footnote{This implies that the detected deconvolved sources may have higher contamination from true detections; however, the selection criteria for the deconvolved sources, discussed further in the text, helps reduce the contamination.}. We note that these choices for the deconvolved images are motivated by our internal experiments rather than being arbitrary choices, which suggested that these settings help detect deconvolved sources that would otherwise have been undetected by the SExtractor configurations used for the original images. The isophotal mode is used to measure the properties of the detected sources, which differs from the aperture-based photometry outputs used in the ZTF image processing pipeline. The sources in the original images with $< 5$ connected pixels above the detection threshold are removed to exclude extremely compact original sources.

\begin{deluxetable}{lclc}[ht] \label{tab:sextractorParams}
\centering
\caption{A few critical \textsc{SExtractor} detection parameters used for detecting sources in the original and deconvolved images. We use \texttt{DETECT\_MINAREA} = 3 for the original images, whereas \texttt{DETECT\_MINAREA} = 1 for the deconvolved images; see main text for explanation.}
\tablehead{
\colhead{Parameter} & \colhead{Value} & \colhead{Parameter} & \colhead{Value}
}
\startdata
\texttt{DETECT\_MINAREA} & 3/1  & \texttt{WEIGHT\_GAIN} & {\sc N} \\
\texttt{DETECT\_THRESH} & 2.5 & \texttt{WEIGHT\_TYPE} & {\sc BACKGROUND} \\
\texttt{FILTER}, \texttt{CLEAN} & {\sc Y}, {\sc Y} & \texttt{BACK\_SIZE} & 64\\
\texttt{DEBLEND\_NTHRESH} & 4 & \texttt{BACK\_FILTERSIZE} & 3\\
\enddata
\end{deluxetable}

For each considered field, we crossmatch sources from the observed and the deconvolved images and conduct detailed analyses of different types of matched and unmatched sources. The original and deconvolved source catalogs are crossmatched using the \textsc{STILTS} package \citep[][; version 3.4-7]{Taylor2006}, a command-line implementation of \textsc{Topcat} \citep[][; version 4.9-1 is used]{Taylor2005}. A simple proximity-based crossmatching is used with a 1.383 pix ($\sim$1.4$\arcsec$) error, chosen to allow a maximum of $\sim$1 arcsec shifts in x and y coordinates. A match is declared if the sources in the two catalogs are separated by less than this error. We do not consider the errors in the centroid positions for crossmatching for simplicity, which mainly has the effect that sources with large positional errors, which otherwise had the potential to crossmatch with one or more sources in the other catalog, will not be crossmatched.

We apply the following selection criteria to filter the catalogs. Only original sources matching the following three criteria are considered: (i) $15 \leq m \leq 20.5$ ($m$ being the apparent magnitude), (ii) $0 < \mathrm{FWHM} \leq 4$ pix, and (iii) $0 \leq \mathrm{ellipticity} < 0.5$ (the ellipticity here is defined as $1 - \dfrac{b}{a}$, where $a$ and $b$ are the lengths of the semimajor and semiminor axis, respectively). Original sources with internal SExtractor extraction flags greater than seven, $\mathrm{FLAGS} > 7$, corresponding to truncated sources and memory limits/overflows, are also excluded. We choose this threshold as sources with a flag value lower than that (i.e., those affected by nearby sources or bad pixels affecting at least 10\% of the area, are deblended, or have saturated pixels) are not necessarily forbidden for our analysis. This choice has also been used in previous works \citep[e.g.,][]{Wolf2004}. These cuts help us to select mostly astrophysical and star-like sources, which is the focus of this study. \reviewFour{For similar reasons,} only deconvolved sources with (i) $m \leq 21.5$\reviewFour{--one magnitude fainter than the limiting magnitude of observed images to account for large scatter in recovered magnitudes for faint observed sources (Fig.~\ref{fig:mag-one-to-one-comparison}) and identify newly found faint deconvolved sources--}and (ii) $0 < \mathrm{FWHM} \leq 4$ pix are considered. For one-to-one, one-to-many, and many-to-one crossmatches, we only consider results that meet the above criteria for original and deconvolved sources. We also aim to reduce false-positive one-to-many and many-to-one matches, so we remove associations that are $<1$ pix apart, which we consider unreliable, in these two types of matched catalogs. These criteria serve to reduce some obvious spurious detections, but it should be noted that this is not rigorous\footnote{Other sophisticated treatments to determine contamination from spurious sources in the detection catalogs exist, such as testing detections on negative images or leveraging fields from other filters/epochs.}. However, Table~\ref{tab:crossmatching-results} shows crossmatching statistics with and without these criteria for transparency.

\subsection{Execution time of SGP}\label{sec:exec-time}

Here, we highlight some general computational details of SGP, namely the execution time, before discussing the scientific results in the upcoming subsections. Recall from Sect.~\ref{subsec:sgp} that the deconvolution of the entire $\sim$3k $\times$ 3k pixel fields is run on subdivisions extracted from the field of size $512 \times 512$ pixels with a 10-pixel overlap and that the iterations of SGP are stopped when the KL divergence criterion converges. For all images considered, typically $\sim$10-40 iterations were required for this convergence for a subdivision. Our implementation of SGP is based on Python with substantial use of NumPy, an efficient numerical library, and FFT-based convolutions based on FFTW using the pyFFTW library\footnote{\url{https://github.com/pyFFTW/pyFFTW}}. FFTW has the advantage of being multi-threaded, but we only use a single thread in our current implementation. We also use the \texttt{interfaces} package of pyFFTW, which is expected to be slower than the core FFTW routines.

The time required to deconvolve the entire $\sim$3k $\times$ 3k pixel fields is shown in Table~\ref{tab:execution-time}. SGP requires $\sim$4-6 minutes to process the entire field, and no specific correlation could be found between execution time and the type of field. The time required for crossmatching catalogs and creating subdivisions is negligible compared to the deconvolution time. A time profiling of our SGP code suggested that the most time-consuming steps are the convolution operations, followed by the flux conservation projection step, both of which are run in each iteration of SGP.

\begin{table}

\centering
\caption{Elapsed wall-clock time on CPU to run the SGP deconvolution on the entire field image of size $\sim$3k $\times$ 3k pixels. This time is the sum of the execution time of each subdivision of size $512 \times 512$ pixels, and accounting for the overlap, this leads to 49 subdivisions. The execution times were calculated using Python 3.10.13 and the processor Intel(R) Xeon(R) CPU @ 2.20GHz.}
\label{tab:execution-time}
\begin{tabular}{ccc}
\hline\hline
ZTF Field ID & Filter & Execution time (min) \\ \midrule
626               & $r$ & 4.95\\
\multirow{2}{*}{} & $g$ & 5.08\\
                  & $i$ & 5.27\\
251               & $r$ & 4.45\\
619               & $r$ & 4.36\\
635               & $r$ & 3.64\\
829               & $r$ & 3.80\\
\multirow{2}{*}{} & $g$ & 4.15\\
                  & $i$ & 4.13\\
233               & $r$ & 6.23\\
811               & $r$ & 4.45\\
594               & $r$ & 4.38\\ \bottomrule
\end{tabular}
\end{table}

\subsection{Visual inspection of deconvolution}\label{sec:visualization}

\begin{figure*}
    \centering
      \includegraphics[keepaspectratio,width=0.8\linewidth]{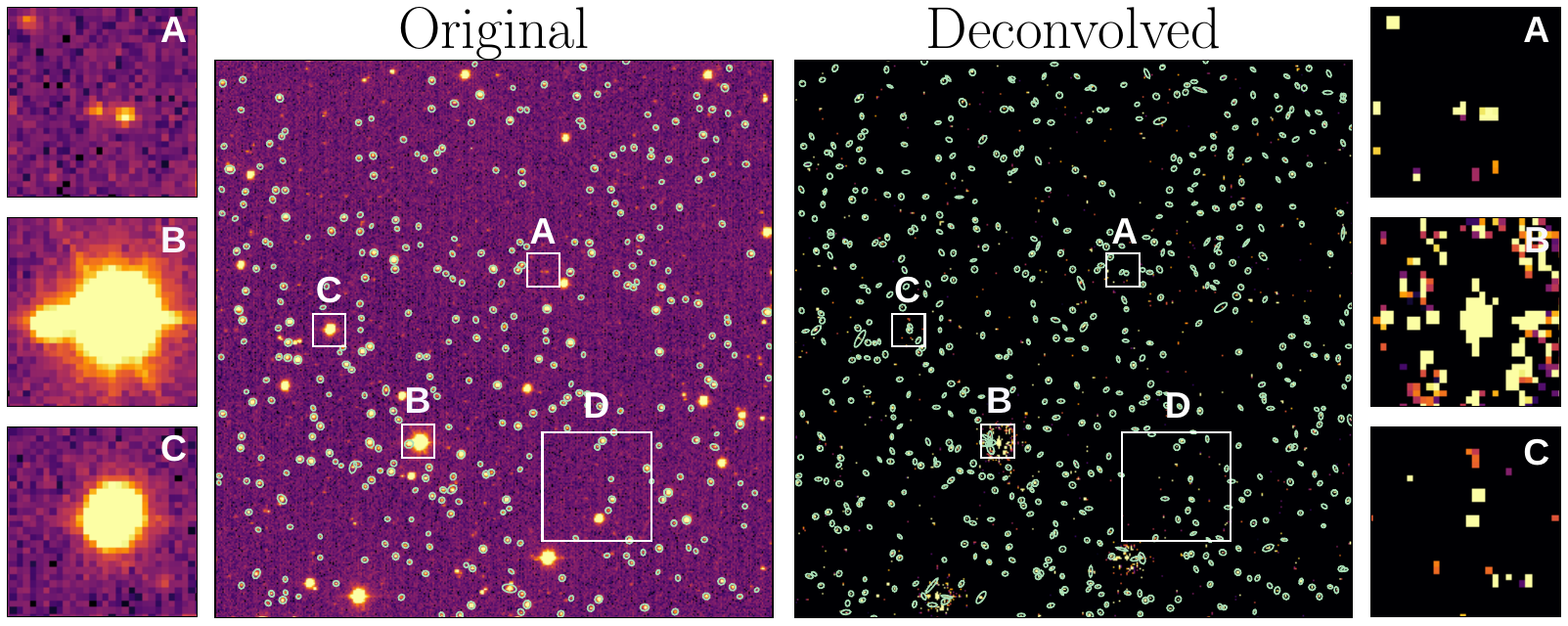}
      \includegraphics[keepaspectratio,width=0.6\linewidth]{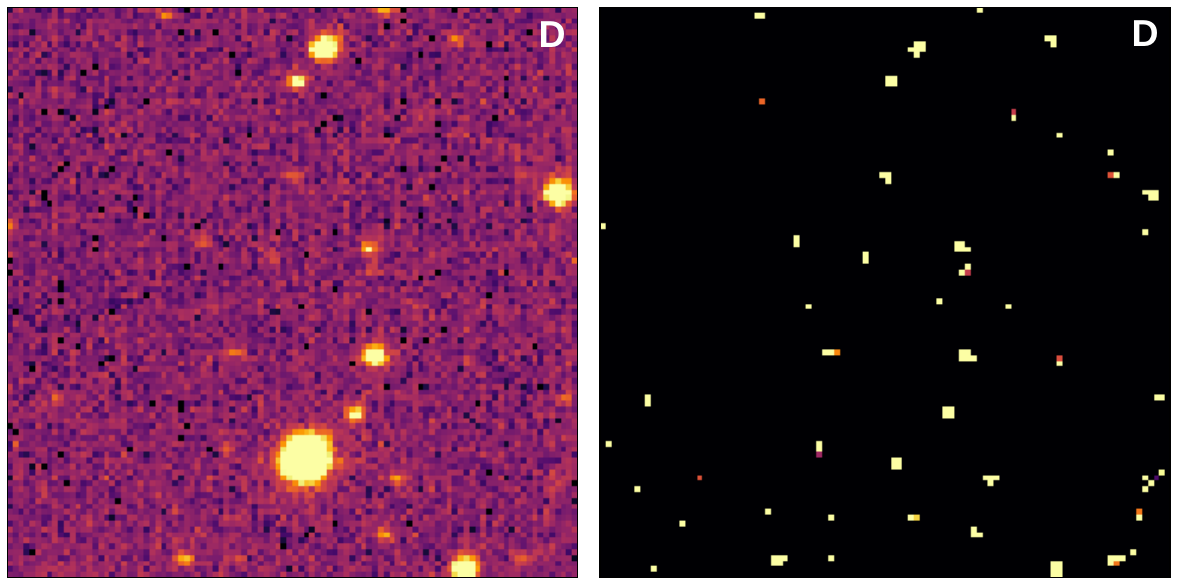}
    \caption{Example visualization of a $512 \times 512$ pixels patch extracted from the original and deconvolved images of the $r$-band low galactic latitude field (ZTF Field ID = 829; see Table~\ref{tab:data-description}). Deconvolved sources are more compact than the original sources and visually reveal some fainter sources that are challenging to identify in the original image (see `D'). In particular, apart from isolated faint original sources, even those near brighter sources are revealed, suggesting that the deconvolution does {\it not} suppress faint sources near bright ones. Detected sources in the original and deconvolved images that match the respective selection criteria (see Sect.~\ref{subsec:experimental-details}) are marked by whitish ellipses. The sizes of the ellipses are conveniently expanded by a convenient, constant factor that is larger for the sources in the deconvolved image than those in the original image to aid visualization. Some regions are marked by squares, discussed further in the text, and denote the same region in the sky for the original and deconvolved images. Images are shown using a combination of square root stretching and clipping pixel values beyond the central 98\% percentile.} \label{fig:visual}
\end{figure*}

We show a sample visualization of the deconvolution of a patch extracted from the 829 ID field in the $r$ band in Fig.~\ref{fig:visual}. As noted in Sect.~\ref{subsec:sgp}, SGP deconvolution produces white spots embedded on a black background, supporting the observation in Fig.~\ref{fig:visual}. In addition, virtually all sources seen in the original image are also present in the deconvolved image, and the deconvolved sources are more compact, which is expected because deconvolution reverses the effect of the PSF. There are also no visible ring artifacts, which is a common phenomenon observed with methods such as RL \citep[see, e.g.,][for discussion]{Magain1998}.

Astronomical images acquired using a CCD are sampled rather than continuous. As a result, a perfect deconvolution, even under ideal noiseless conditions, would not produce a point source (represented by the Dirac delta function) \citep{Magain1998}. Therefore, we should expect the ideal deconvolution of stellar original sources to produce a profile with a finite width instead of a point source. However, in realistic scenarios, noise is present, and factors such as errors in PSF estimation, inaccuracies in background estimation (relevant for the flux constraint in SGP), and the regularized nature of the SGP algorithm can lead to deviations from the ideal deconvolved profile. These factors could explain why deconvolved sources are generally spread across a few pixels (see, for example, the annotated box marked `A' and `D' in the figure). Furthermore, the ellipses marked on the deconvolved image appear more elliptical than those on the original image because the deconvolved sources do not look spherically symmetrical. This lack of near-circular shapes may arise due to the reasons mentioned above.

In `A', we observe a situation where the original source appears extremely dim and is perceived as two nearby sources. SExtractor detected this as a single source, which is not marked by an ellipse in the figure since it did not meet our selection criteria. However, the deconvolution made this faint source more visually apparent and resulted in two separate detections instead of one, thus deblending the original source. These two detections passed our selection criteria and are marked by two close ellipses.

`B' highlights a case where our deconvolution produces a high concentration of detected sources near bright original sources. Some of these detections may be dubious, and this pattern is also observed in the deconvolved image around other corresponding bright original sources.

In `C', we observe a single bright original source that SExtractor detected, although it did not meet our selection criteria. The deconvolution visually deblended this source, separating it into two nearby sources that were detected by SExtractor and met our selection criteria.

Overall, we conclude that the deconvolution approach is effective in recovering faint sources that are otherwise difficult to detect in the original images. We also saw that it has the potential to deblend overlapping sources. Both aspects are discussed in more detail in the following sections.

\subsection{Crossmatching results}\label{subsec:crossmatch-results}

\begin{table*}
\renewcommand{\arraystretch}{1.25}
\centering
\caption{Crossmatching results for the images considered in this study. The columns denote (from left to right): (a) the ZTF Field ID, (b) the filter in which the image was observed, (c) no. of detected original sources, (d) no. of detected deconvolved sources, (e) no. of one-one matches (original sources having exactly one match in the deconvolved and vice-versa), (f) no. of many-to-one matches (original sources having more than one match in the deconvolved), (g) no. of one-to-many matches (deconvolved sources having more than one match in the original), (h) original sources having no match in the deconvolved, and (i) deconvolved sources having no match in the original. In `one-to-many', for example, `one' corresponds to the deconvolved, and `many' corresponds to the original. Values outside the parentheses denote the crossmatching statistics without imposing any selection criteria on the magnitude, FWHM, and ellipticity of the original sources, whereas values inside the parentheses denote the crossmatching statistics after imposing the selection criteria; see the main text for details on the selection criteria. For one-to-many and many-to-one, the numbers shown denote the total no. of rows in the entire table and not the no. of deconvolved and original sources, respectively; we have found that all one-to-many and many-to-one matches are two-matches, i.e., two deconvolved (original) sources are matched to the same original (deconvolved) source, except one case in the 233 ID image where a single original source was matched to three deconvolved sources. This means that the values for the one-to-many and many-to-one matches can be divided by two to get the no. of sources. Also, for one-to-many and many-to-one, it is possible that the selection criteria may remove one or more crossmatches for a given source; in such cases, we also exclude all other crossmatches for that source. For the rest of the paper, only crossmatches imposing the selection criteria on original sources (i.e., inside the parentheses) are discussed unless otherwise specified.}
\label{tab:crossmatching-results}
\begin{tabular*}{\textwidth}{p{0.3in} p{0.3in} p{0.8in} p{0.73in} p{0.8in} p{0.73in} p{0.73in} p{0.73in} p{0.73in}}
\hline\hline
ZTF Field ID & Filter & Original sources & Deconvolved sources & One-to-one matches & Many-to-one matches & One-to-many matches & Unmatched (original) & Unmatched (deconvolved) \\ 
(a) & (b) & (c) & (d) & (e) & (f) & (g) & (h) & (i)\\ \midrule
626               & $r$ & 1255 (746)& 6352 (1831)& 1227 (722)& 20 (0)& 2 (0)& 27 (3)& 5115 (744) \\
\multirow{2}{*}{} & $g$ & 983 (527)& 9106 (1319)& 966 (506)& 14 (0)& 966 (0)& 17 (1)& 8133 (501) \\
                  & $i$ & 1610 (856)& 11802 (2957)& 1537 (835)& 14 (2)& 4 (0)& 71 (0)& 10258 (1566) \\
251               & $r$ & 1131 (732)& 3406 (1375)& 1104 (711)& 16 (0)& 4 (0)& 25 (2)& 2294 (362)  \\
619               & $r$ & 2315 (1523)& 35791 (3276)& 2281 (1423)& 14 (2)& 12 (0)& 28 (0)& 33503 (1277) \\
635               & $r$ & 6824 (5824)& 12997 (7333)& 6771 (5170)& 20 (2)& 10 (0)& 48 (7)& 6216 (1508)  \\
829               & $r$ & 16122 (12363)& 56506 (19864)& 15982 (11357)& 96 (14)& 18 (0)& 131 (16)& 40476 (5466)  \\
\multirow{2}{*}{} & $g$ & 8163 (6044)& 54029 (10342)& 8099 (5573)& 42 (14)& 10 (0)& 59 (9)& 45909 (3012)  \\
                  & $i$ & 15115 (13009)& 31682 (19589)& 14987 (11769)& 90 (14)& 26 (0)& 115 (3)& 16650 (6277)  \\
233               & $r$ & 18633 (14501)& 25107 (20048)& 18260 (13473)& 161 (6)& 32 (0)& 357 (6)& 6766 (3605) \\
811               & $r$ & 9229 (7815)& 20523 (11014)& 9135 (7254)& 62 (14)& 8 (0)& 91 (8)& 11357 (2755) \\
594               & $r$ & 13275 (10596)& 54542 (17305)& 13153 (9739)& 76 (16)& 14 (0)& 115 (18)& 41351 (5503)  \\ \bottomrule
\end{tabular*}
\end{table*}

Table~\ref{tab:crossmatching-results} shows the crossmatching results for all images. Here, the notation used is that the deconvolved and the original source catalogs are the first and second tables, respectively, and should be noted for the definitions below.

Two types of statistics are shown: normal crossmatching results (outside the parentheses) and `filtered' crossmatching results (inside the parentheses). The filtered results apply the selection criteria to the original sources, as outlined in Sect.~\ref{subsec:experimental-details}. For clarity of the discussion in the remainder of the paper, we will focus only on the `filtered' crossmatching results. Match statistics are divided into three categories: (i) one-to-one: the original source has exactly one match in the deconvolved, and the corresponding deconvolved source has the same original source as the only match; (ii) many-to-one: the original source has more than one match with distinct deconvolved sources\footnote{In colloquial terms, many-to-one matches include one-to-one matches; however, here we explicitly exclude one-to-one matches from the many-to-one matches. This distinction also applies to one-to-many matches.}, and (iii) one-to-many: the deconvolved source has more than one match with distinct original sources. The unmatched statistics are divided into two categories: (i) unmatched (original): the original source has no match with any deconvolved source, and (ii) unmatched (deconvolved): the deconvolved source has no match with any original source. Many-to-one matches may include cases where an original source was deblended into more than one deconvolved source, although this may not be necessary. The ``unmatched original sources'' refers to original sources that do not appear in the deconvolved catalog. The ``unmatched deconvolved sources'' indicate `new' sources that were not detectable in the original image but were detectable in the deconvolved image. A detailed discussion of matched and unmatched sources across different types of fields is presented in the next section.

While the table shows the statistics of unmatched deconvolved sources after applying the basic selection criteria detailed in Sect.~\ref{subsec:experimental-details}, for our analysis ahead, unless otherwise noted, we further discard unmatched deconvolved sources with (a) $\lvert \frac{(\mathrm{FWHM} - \mathrm{FWHM}_{\mathrm{median},\,\mathrm{1\,to\,1}})}{1.4826 * \mathrm{FWHM}_{\mathrm{M.A.D},\,\mathrm{1\,to\,1}}} \rvert > 3.5$ (where $\mathrm{FWHM}_{\mathrm{median}, \mathrm{1 to 1}}$ and $\mathrm{FWHM}_{\mathrm{M.A.D}, \mathrm{1 to 1}}$ are the median and median absolute deviation of FWHM of deconvolved sources from one-to-one matches for that image, and the 1.4826 scaling factor arises because statistically, the expectation of the median absolute deviation is $\frac{1}{1.4826}\sigma$, $\sigma$ being the standard deviation; see \citealt{Iglewicz1993}), (b) $\mathrm{FLAGS} > 7$, and (c) ellipticity $>$ 0.8. (a) is a modified z-score measure and discards sources with outlier FWHM, since, as we will see later, the FWHM of deconvolved sources generally does not show huge variability and is tightly distributed around roughly 1 pixel\footnote{This argument assumes that the FWHM of deconvolved sources from one-to-one matches is representative of the FWHM of unmatched deconvolved sources}. These criteria were chosen empirically based on our visualizations of such sources, which suggested that these sources are likely spurious and can be removed to obtain a more astrophysically representative sample. \reviewFour{For all images, we have internally verified that the modified z-score criterion (a) accounts for the vast majority ($\sim$98-99\%) of reductions in deconvolved sources, followed by the FLAGS and ellipticity criteria, in that order.} The discussion of the number of sources remaining after applying these criteria is detailed in the following sections.

In general, there are more deconvolved sources than the corresponding original sources by a factor $\sim$1.25 - 4 (after filtering based on the selection criteria). There is also a non-trivial reduction in the number of deconvolved sources before and after applying the filtering, with, on average, a reduction of $\sim$3-4 times\footnote{We have found that the dominant selection criterion contributing to this drastic decrease is the $m \leq 21.5$ selection criteria mentioned in Sect.~\ref{subsec:experimental-details}, which suggests that the deconvolution reveals many sources fainter than 21.5 mag; some of these might be true, \reviewFour{and thus we may have excluded some very faint but plausible deconvolved sources}. However, the faintest ones are likely artifacts introduced due to low signal-to-noise regions in the original image.}. A large fraction of successful matches between original and deconvolved sources are one-to-one matches with relatively fewer asymmetric matches (i.e., many-to-one and one-to-many), which may be the result of using a strict crossmatching distance threshold, which is $\sim$1.2-2.4 times smaller than the typical seeing in the considered images. The number of many-to-one crossmatches ranges from none for the high-galactic latitude fields (IDs 626 and 251) to a few tens in the low galactic latitude fields (IDs 829, 811) and the globular cluster field (ID 594). There are no instances of one-to-many crossmatches in all cases, which is inconsequential or desirable depending on the crowding in the field and suggests that there is no deconvolved source matched to more than one original source within the crossmatching threshold\footnote{Note that this does not suggest the deconvolution does not artificially group nearby original sources into a single source since if this does happen, the centroid might be shifted more than the crossmatching threshold and thus unaccounted in the one-to-many cross matches. Although extremely rare, such cases have been found; see the discussion on unmatched original sources in Sect.~\ref{subsec:intermediate-gal-lat} and Sect.~\ref{subsubsec:low-gal}, for example.}. Also, it is desired that the number of unmatched original sources be extremely low since these denote original sources that could not be detected in the deconvolved image -- the number of such unmatched sources is $\lesssim$0.4\% (considering the `filtered' results), which is negligible. The table also shows that there are generally $\sim$$1/4^{\mathrm{th}} - 1/2$ deconvolved sources that are not matched by any original source. When comparing the number of original sources detected in the $g$, $r$, and $i$ bands for field IDs 626 and 829, it can be observed that fewer sources are detected in the $g$-band than in the other two, which could be due to the stronger response of the $g$-band to extinction than in the other two bands. In the following sections, a detailed investigation of each type of source is conducted.

It must be cautioned that for crowded fields such as globular cluster and low galactic latitude fields (Sects.~\ref{subsubsec:gc} and \ref{subsubsec:low-gal}, respectively), crossmatching may overestimate the number of matches (i.e., produce false positives) due to closely separated sources, and in particular may produce dubious matches.

\subsection{Field-specific results}\label{sec:field-specific-results}

\begin{figure*}[hbt!]
    \centering
      \includegraphics[keepaspectratio,width=0.32\linewidth]{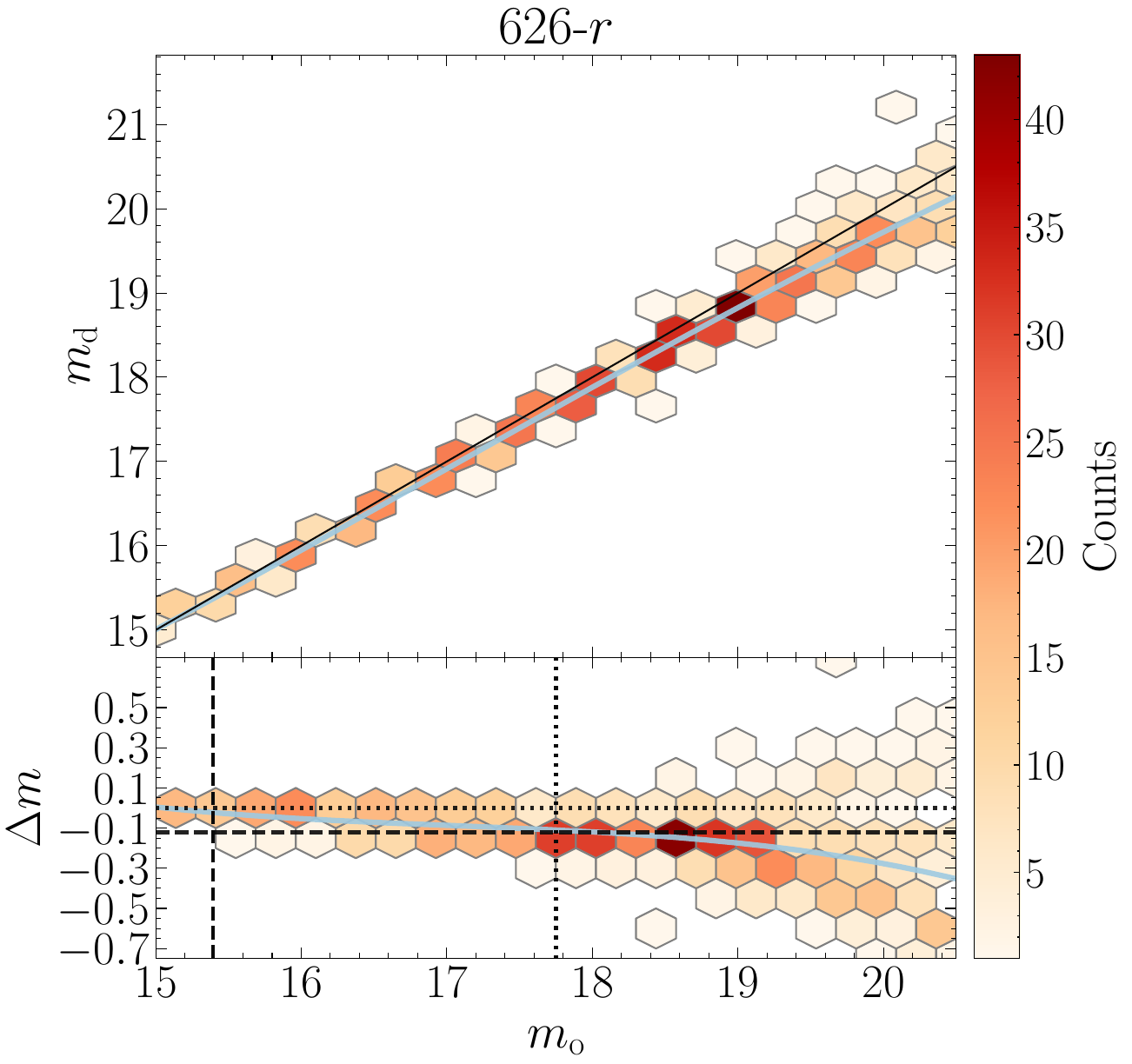}
      \includegraphics[keepaspectratio,width=0.32\linewidth]{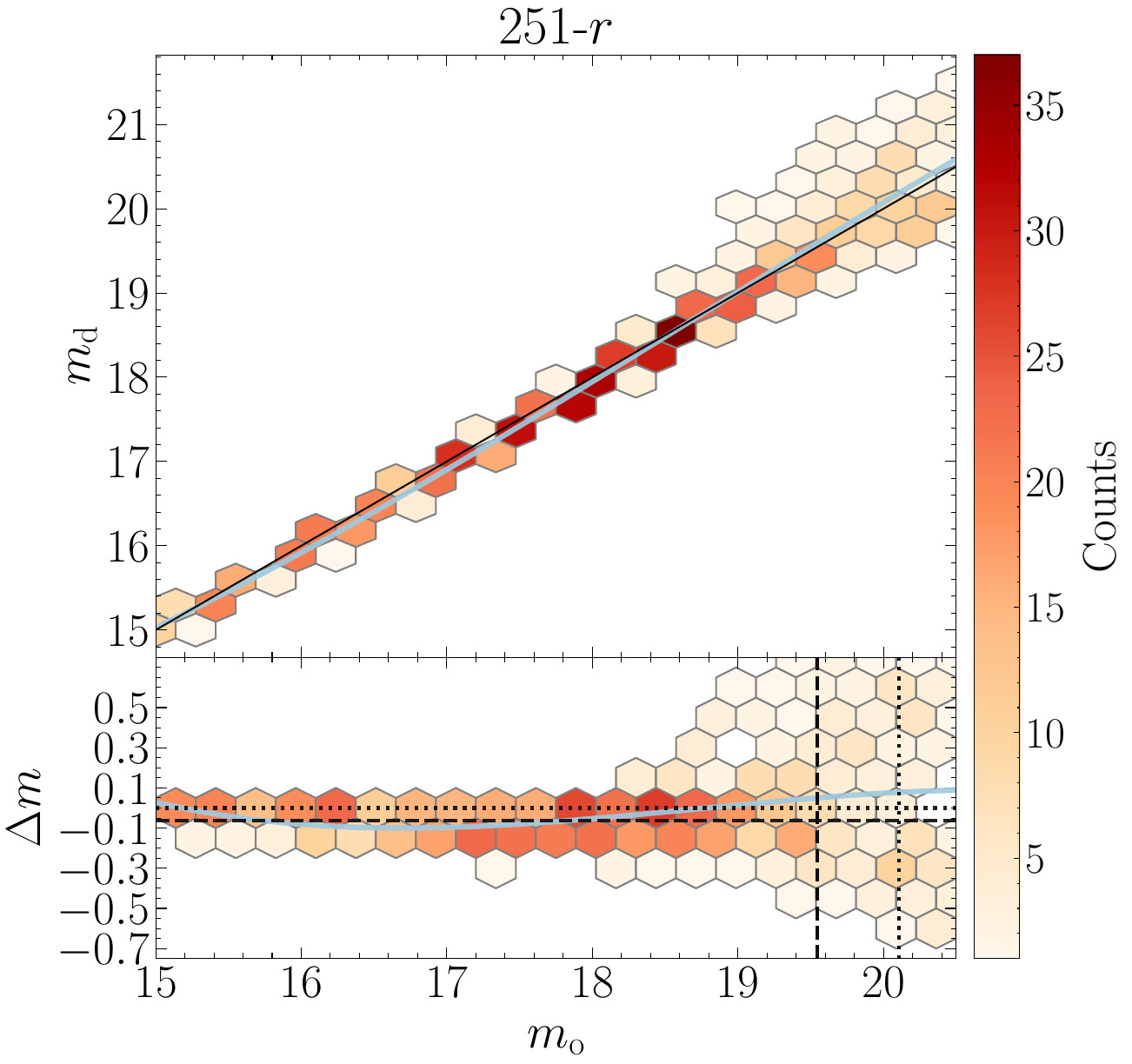}
      \includegraphics[keepaspectratio,width=0.32\linewidth]{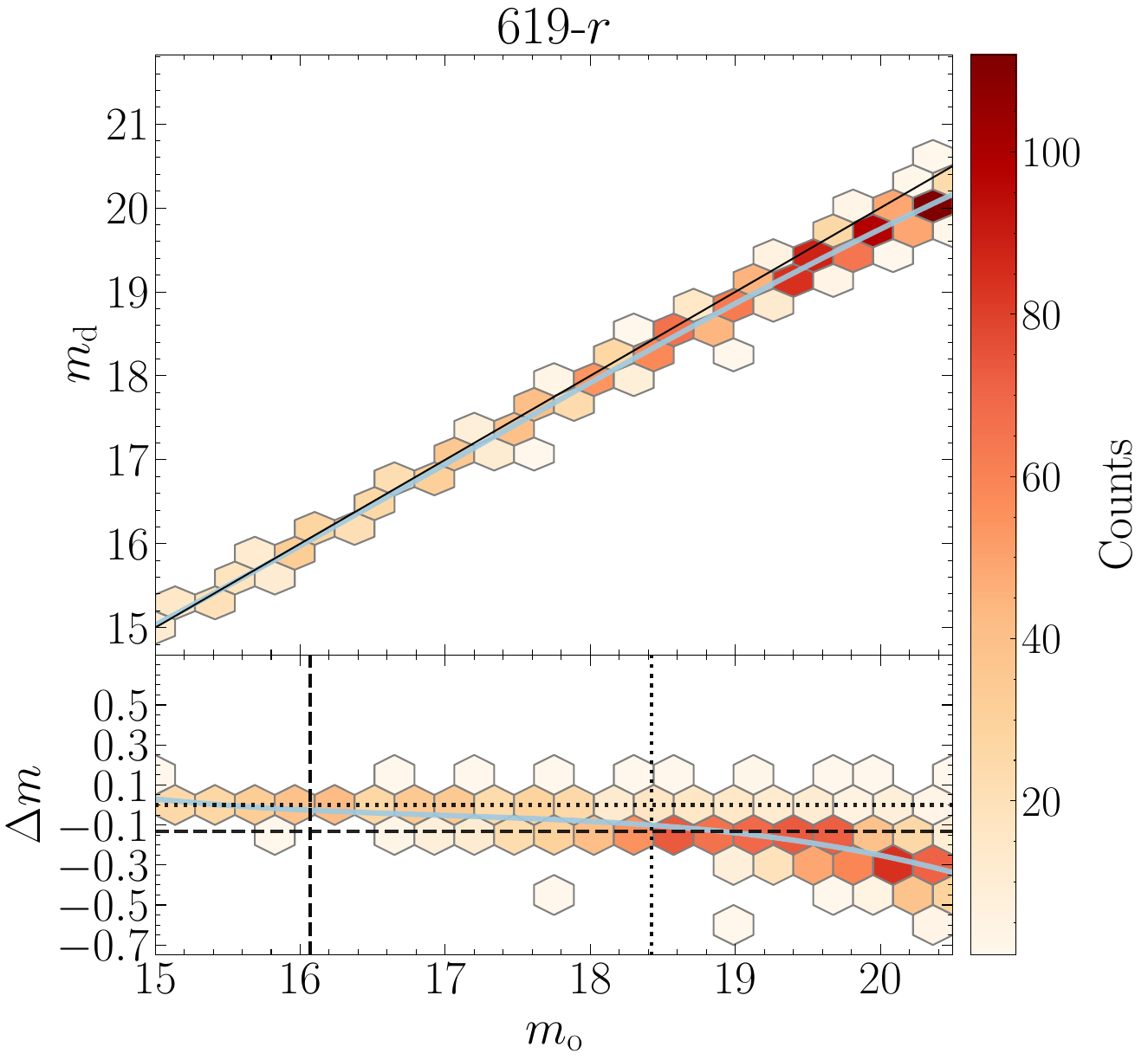}
      \includegraphics[keepaspectratio,width=0.32\linewidth]{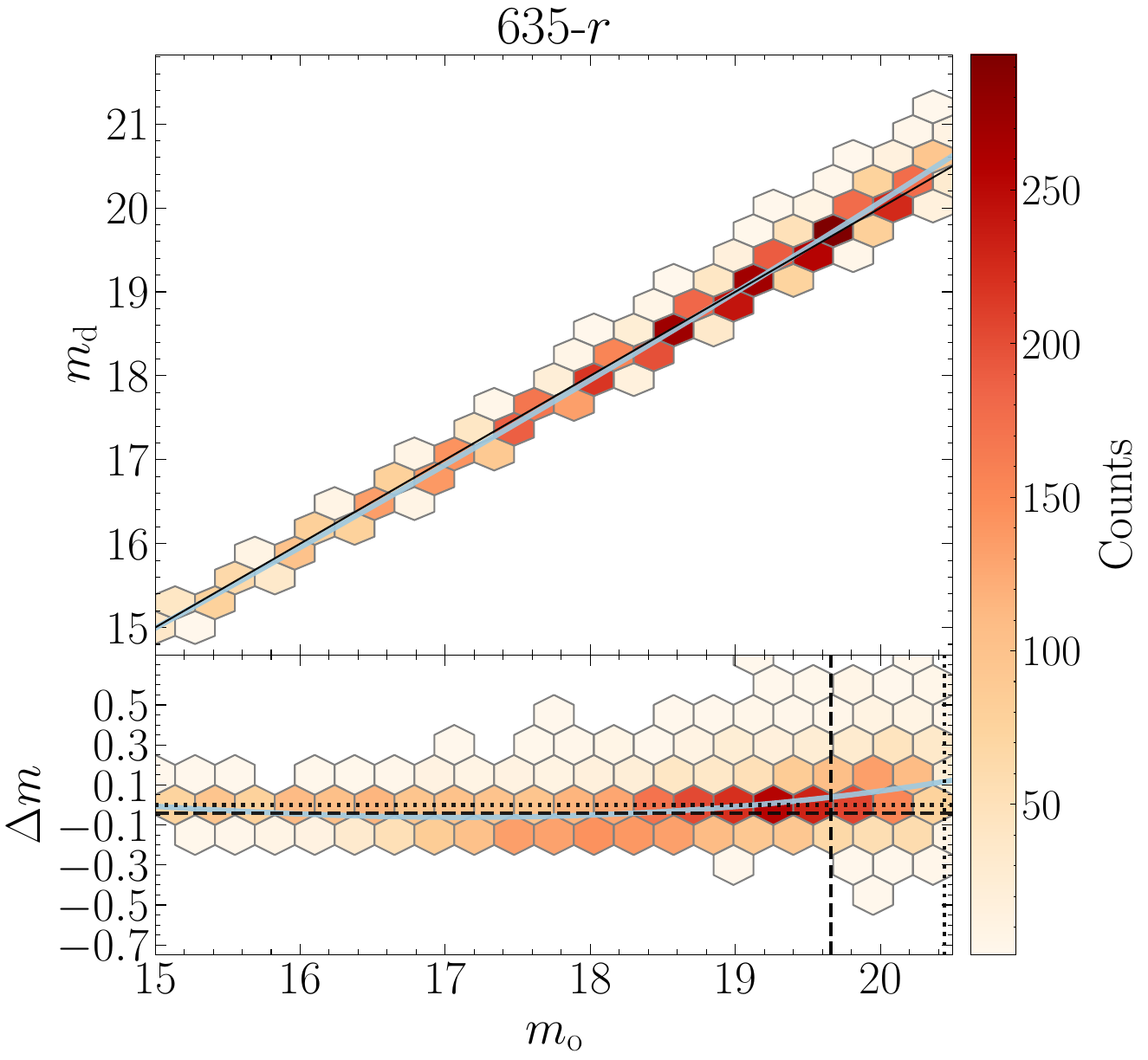}
      \includegraphics[keepaspectratio,width=0.32\linewidth]{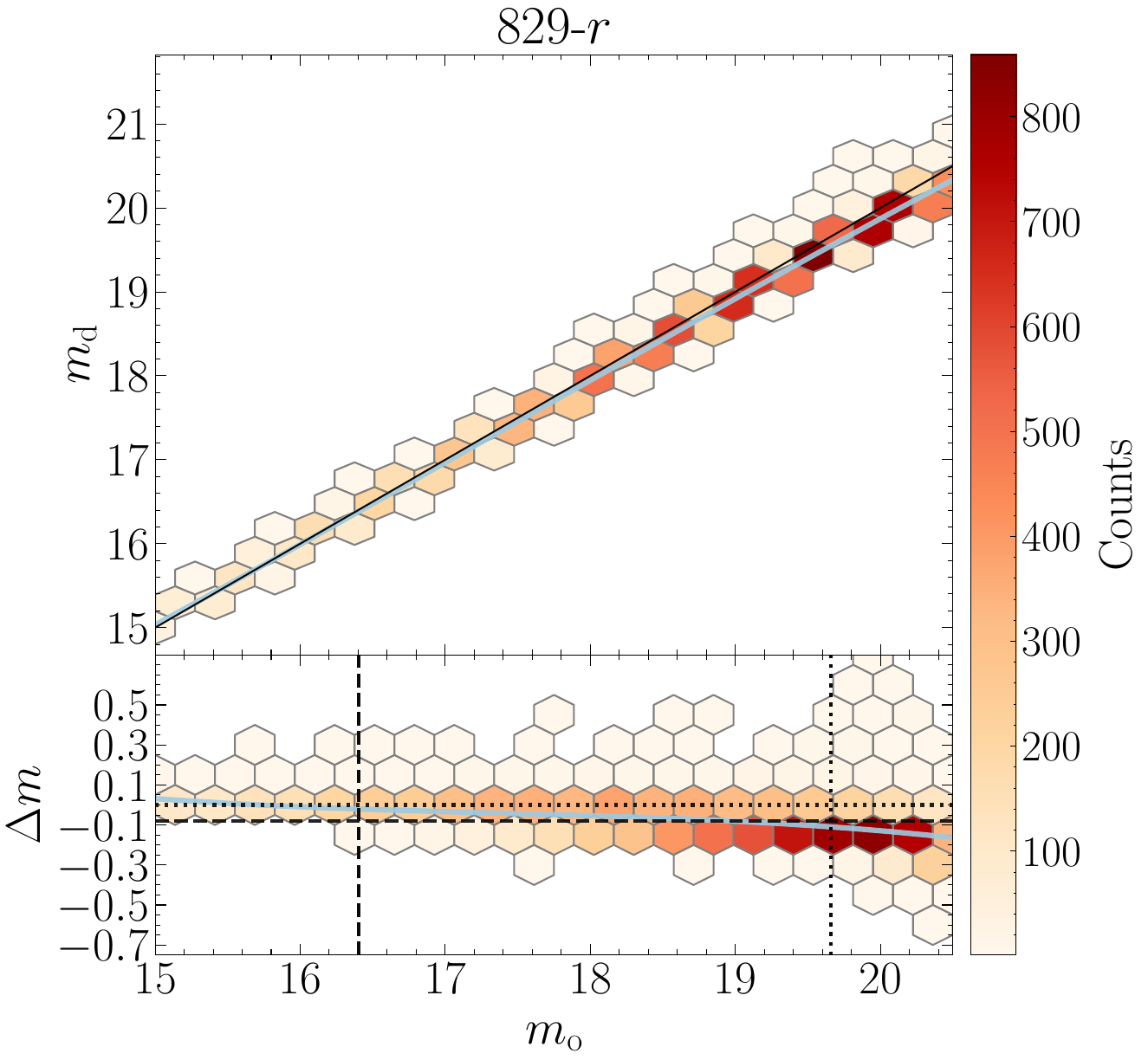}
      \includegraphics[keepaspectratio,width=0.32\linewidth]{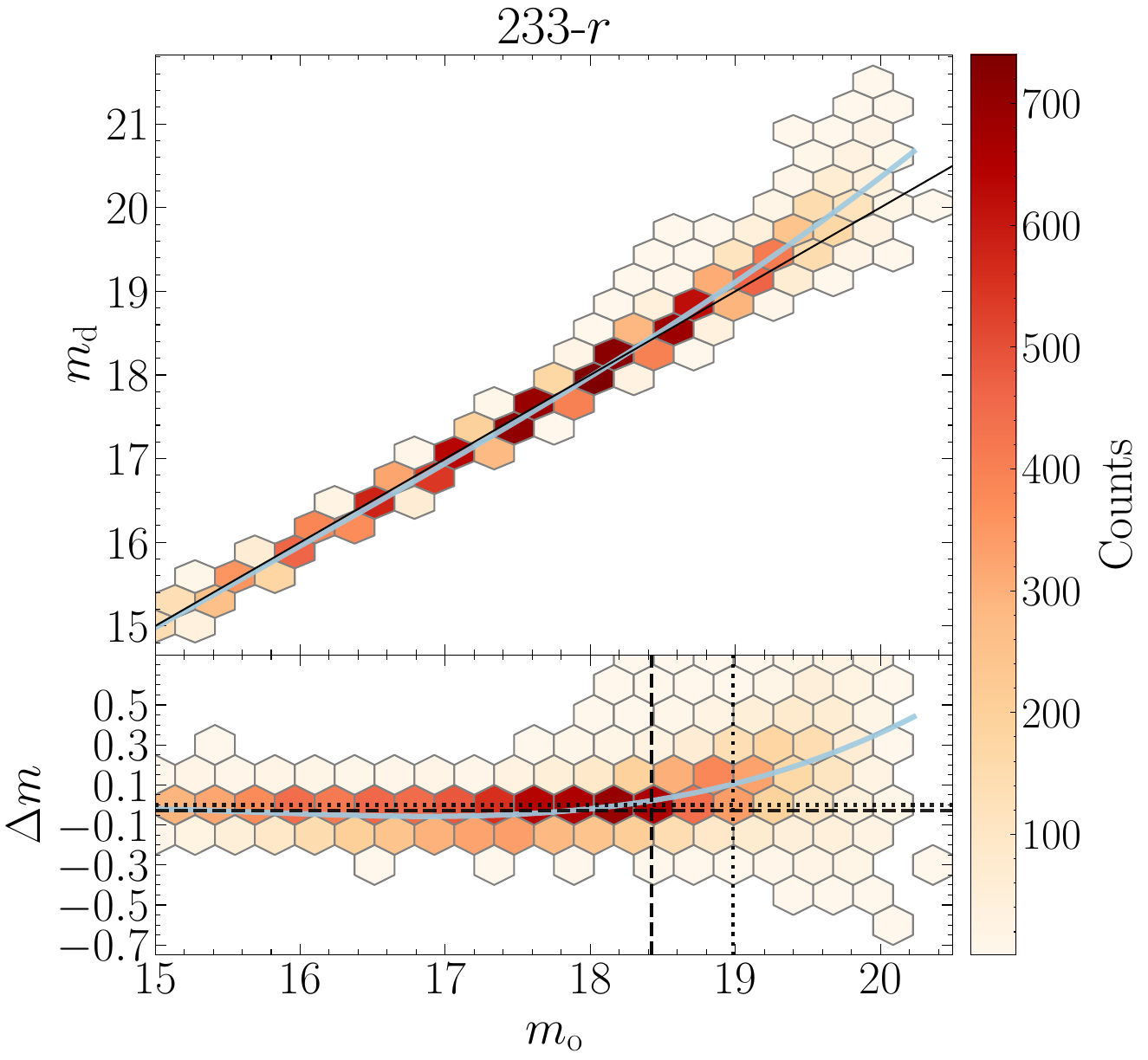}
      \includegraphics[keepaspectratio,width=0.32\linewidth]{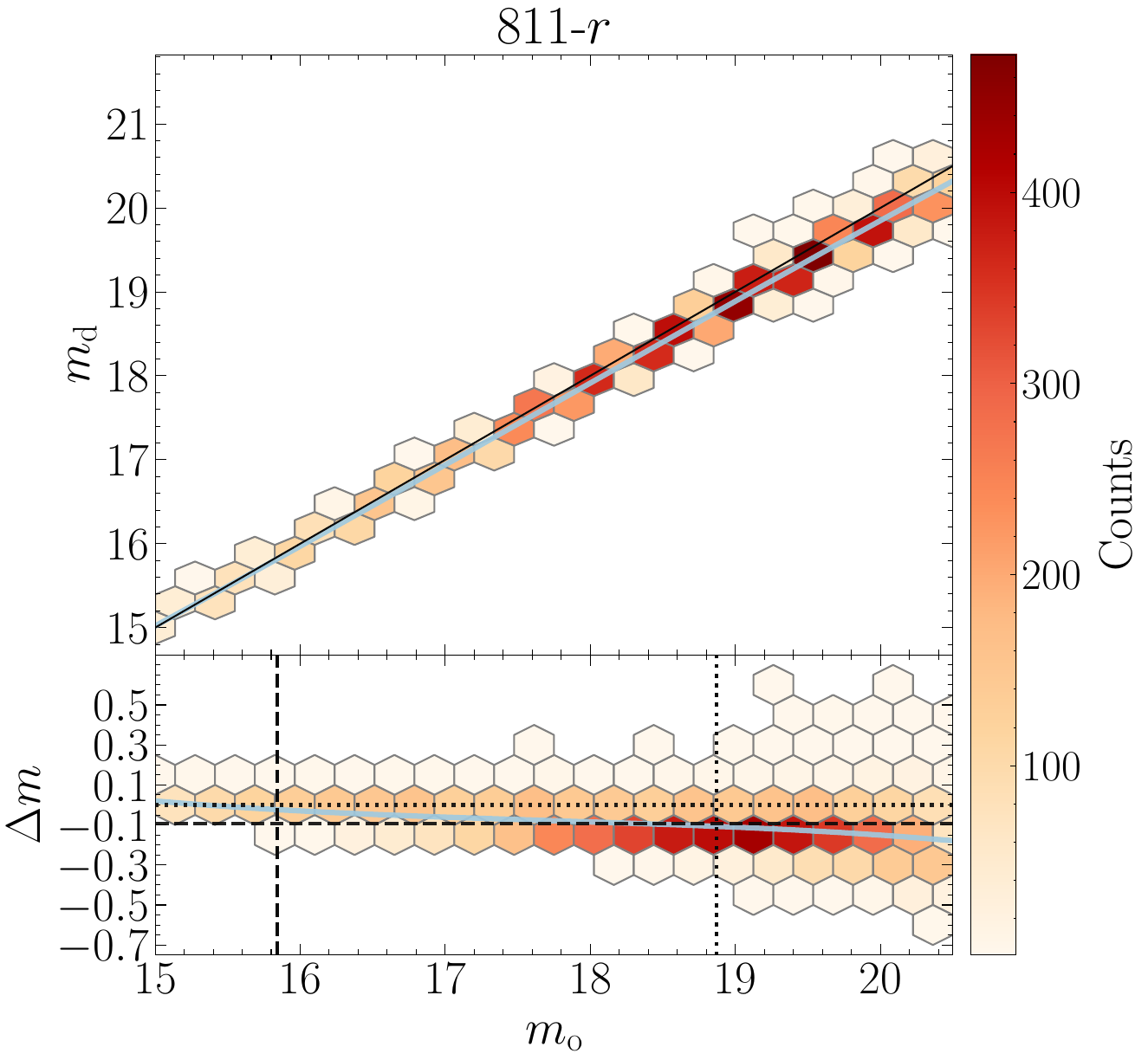}
      \includegraphics[keepaspectratio,width=0.32\linewidth]{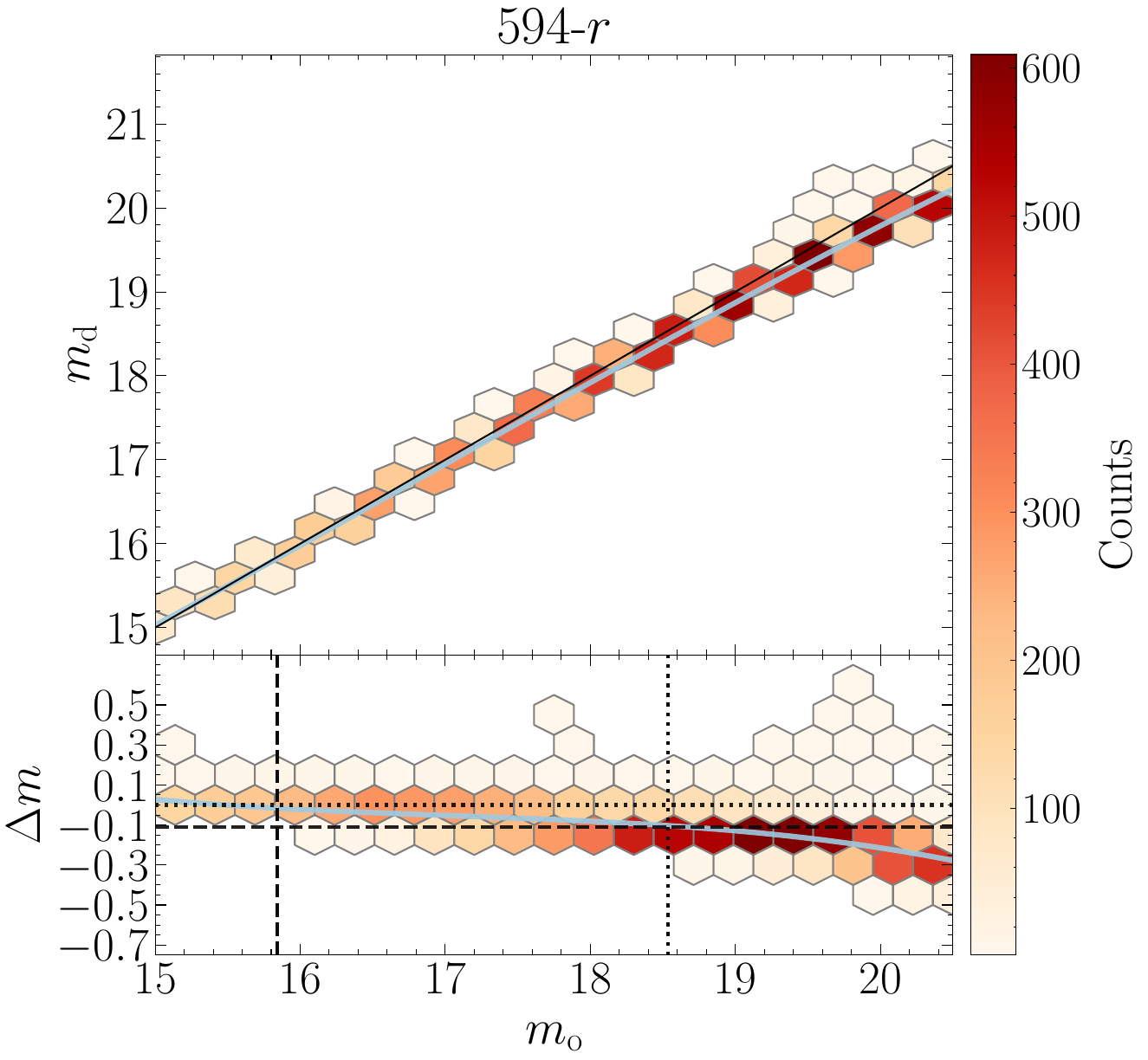}
        \caption{Comparison of magnitudes of the original ($m_{\mathrm{o}}$) and the one-to-one matched deconvolved ($m_{\mathrm{d}}$) sources for all the $r$ filter images considered in this study. The bottom subpanels of each panel show the difference in the magnitudes (residuals), $\Delta m = m_{\mathrm{d}} - m_{\mathrm{o}}$. The vertical dashed and dotted lines denote the faintest original source, resulting in 98\% and 90\% flux conservation, respectively, as obtained by binning the x-axis into 50 equally spaced bins, calculating the median $\Delta m$ in each bin, and identifying the `faintest' bin center corresponding to 98\% and 90\% flux conservation (or $\Delta m \approx 0.02$ and 0.1). The solid line in the upper panel and the dotted horizontal line in the bottom subpanels denote perfect agreement $\Delta m = 0$. The horizontal dashed line in the lower panel denotes the median $\Delta m$. The blue line in the upper and lower panels denotes the trend line. The title of each panel denotes the ZTF field ID. Magnitude errors are excluded in this comparison and are discussed separately in Appendix~\ref{appn:photo-error-compare}, where we found that the photometric magnitude uncertainties are much smaller and also scale sublinearly as the original source gets fainter, unlike the uncertainties on the original source magnitudes.} \label{fig:mag-one-to-one-comparison}
\end{figure*}

\begin{figure*}
    \centering
      \includegraphics[keepaspectratio,width=0.32\linewidth]{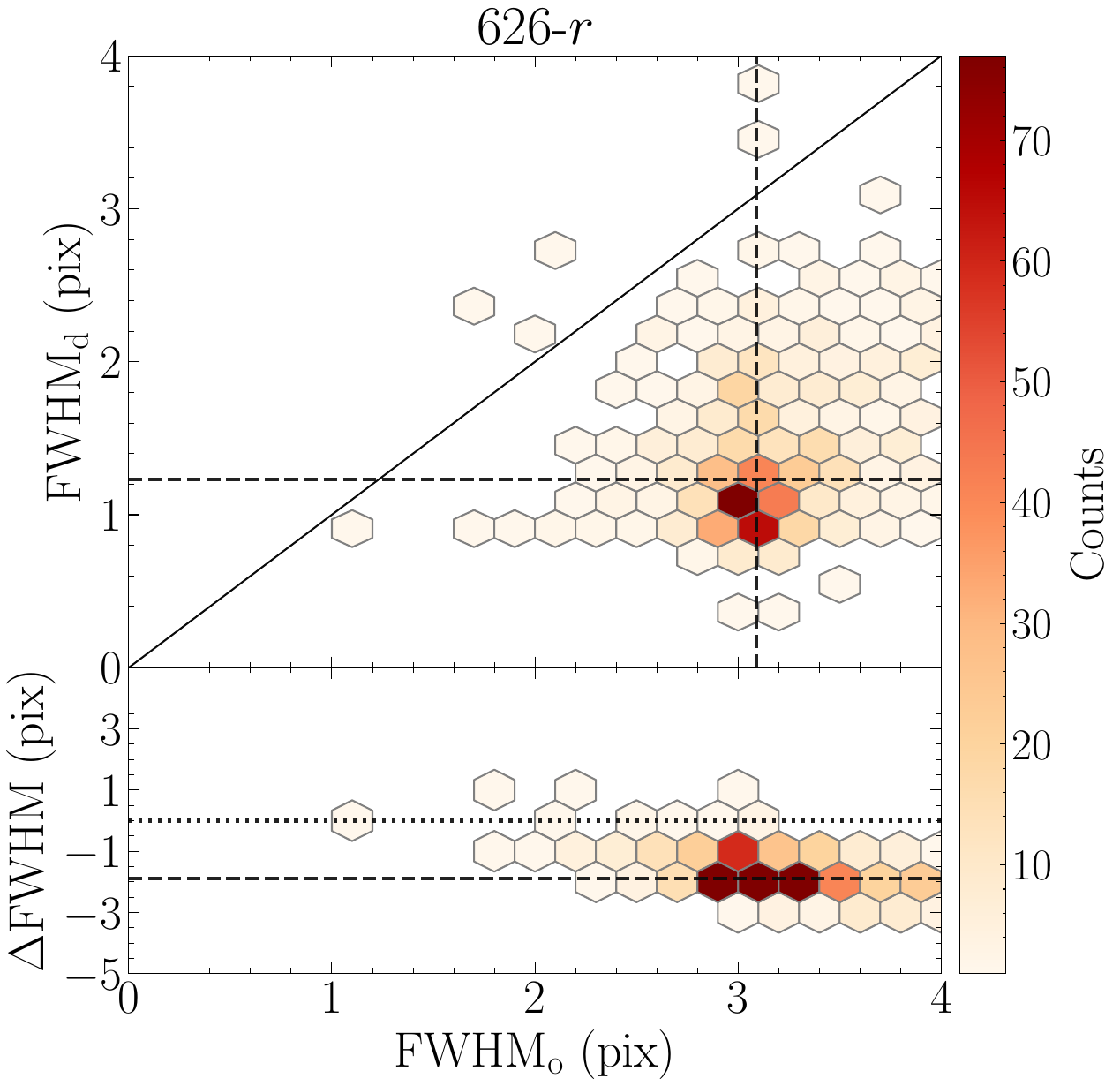}
      \includegraphics[keepaspectratio,width=0.32\linewidth]{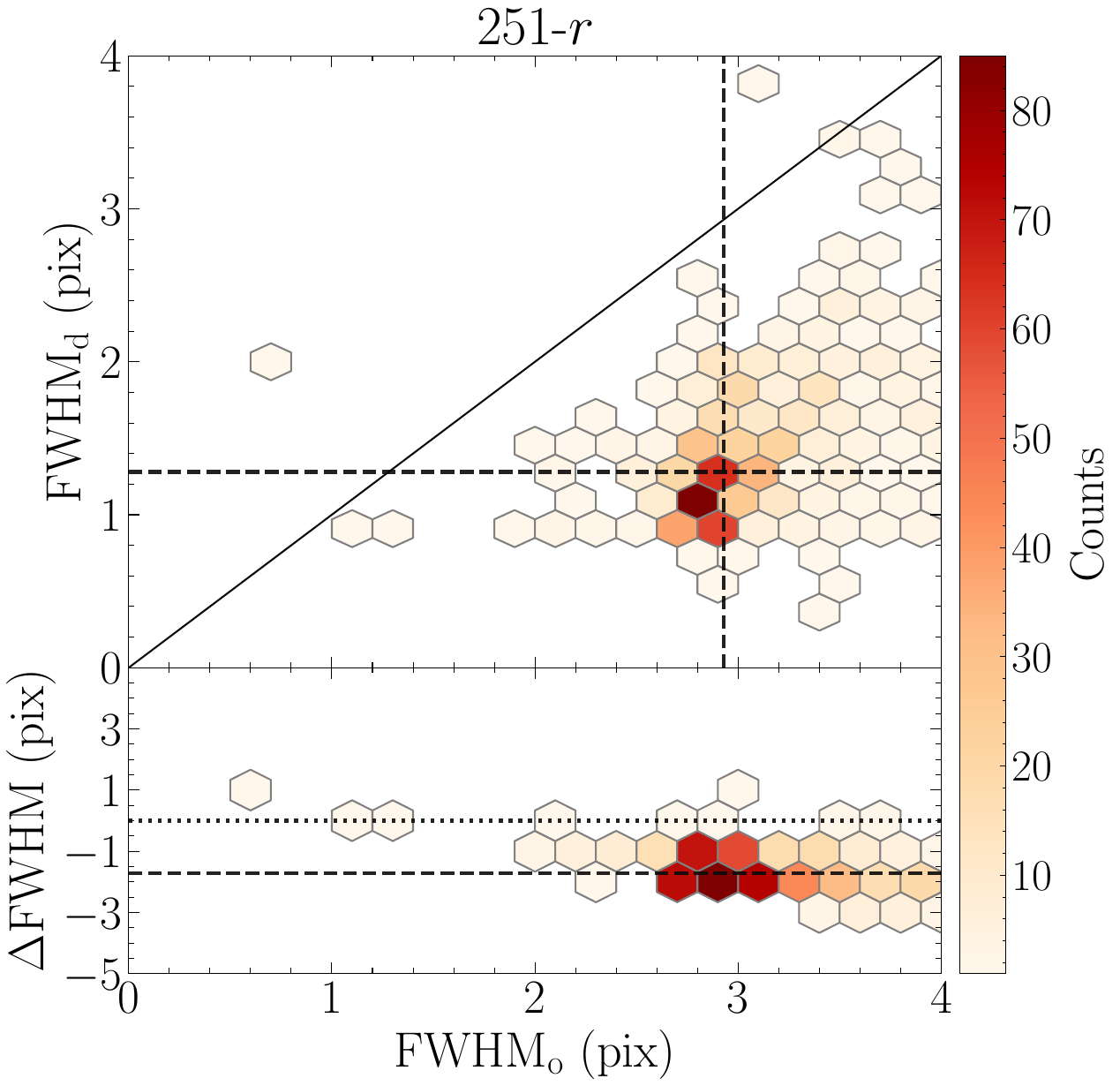}
      \includegraphics[keepaspectratio,width=0.32\linewidth]{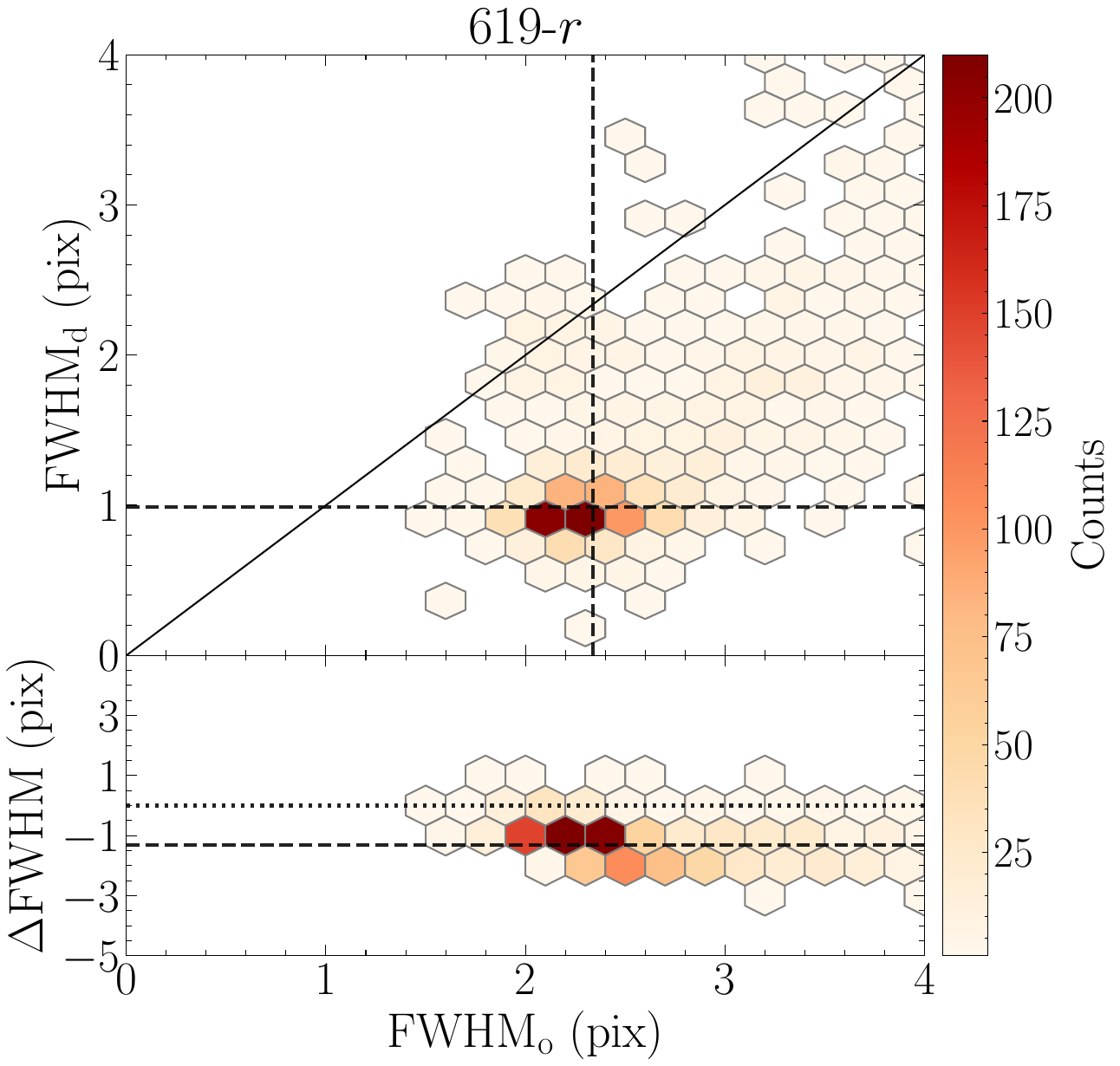}
      \includegraphics[keepaspectratio,width=0.32\linewidth]{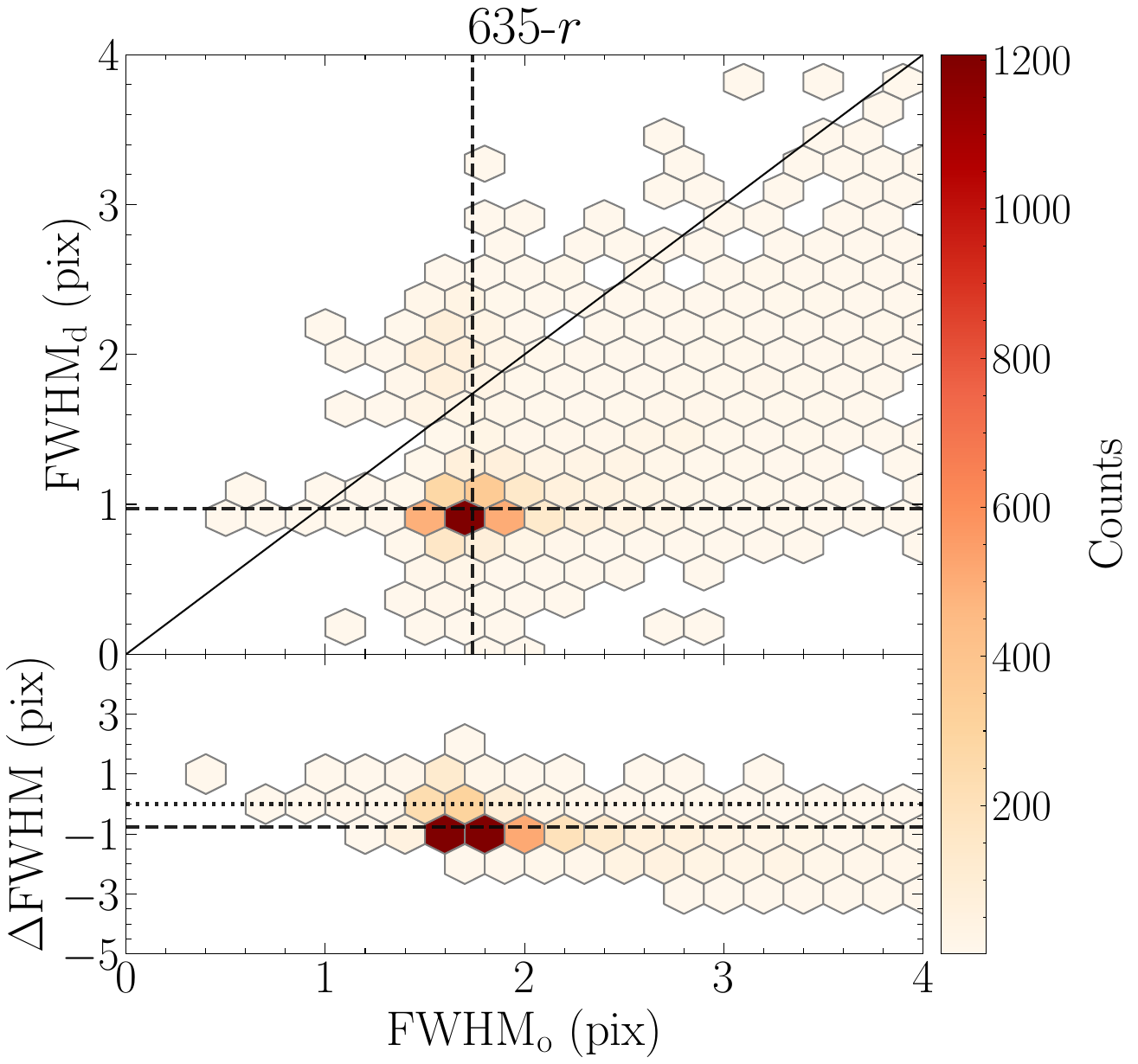}
      \includegraphics[keepaspectratio,width=0.32\linewidth]{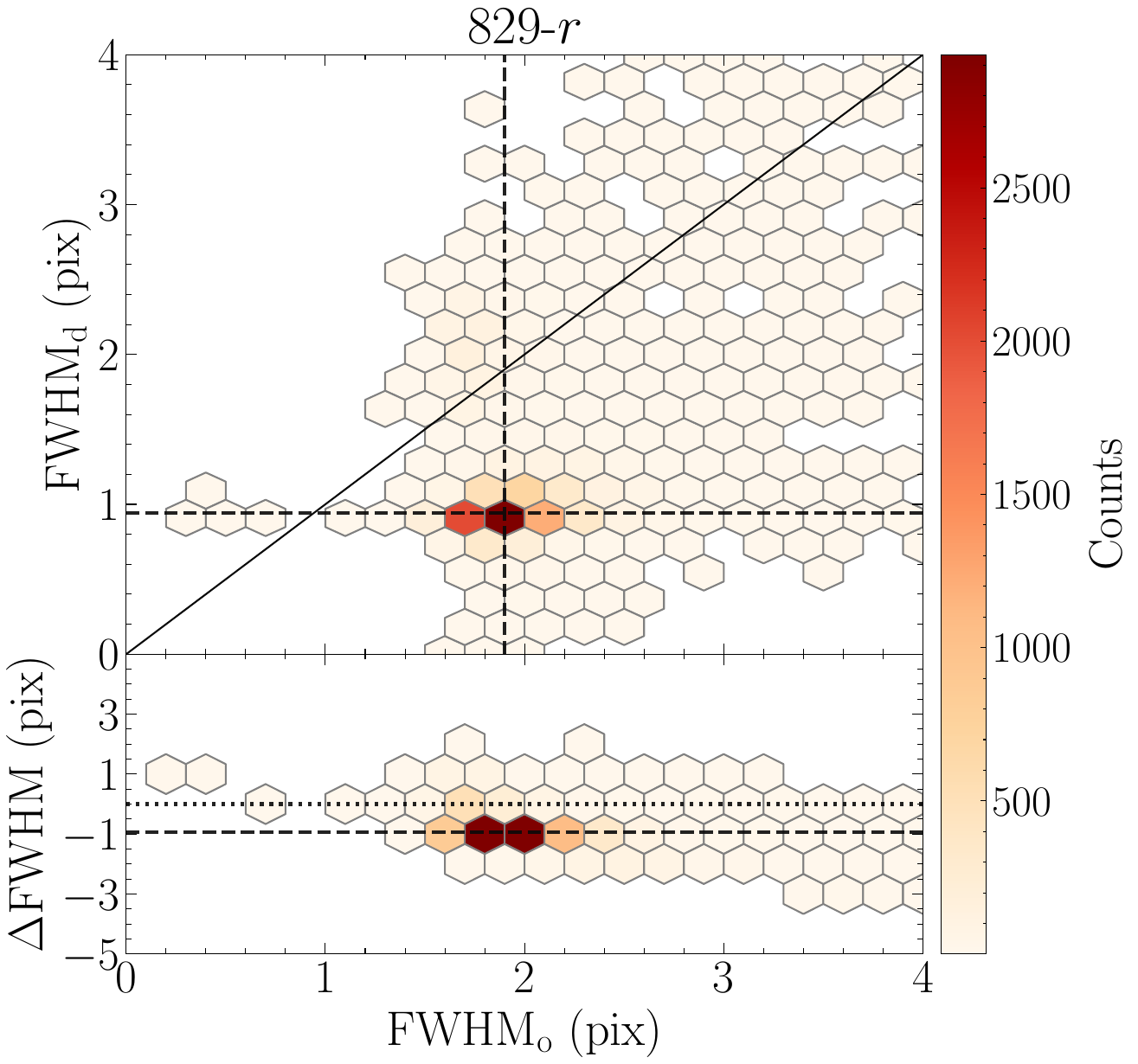}
      \includegraphics[keepaspectratio,width=0.32\linewidth]{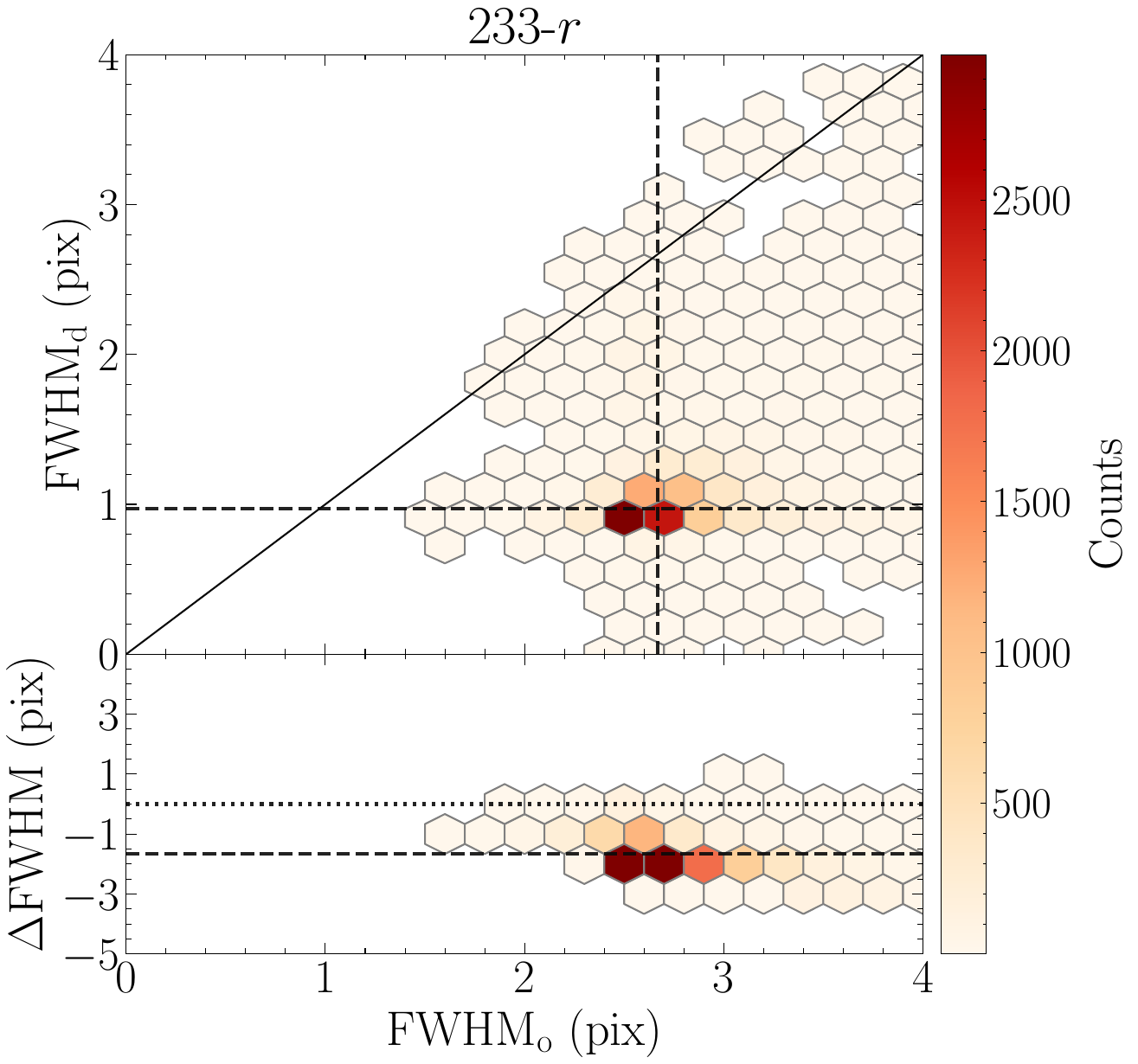}
      \includegraphics[keepaspectratio,width=0.32\linewidth]{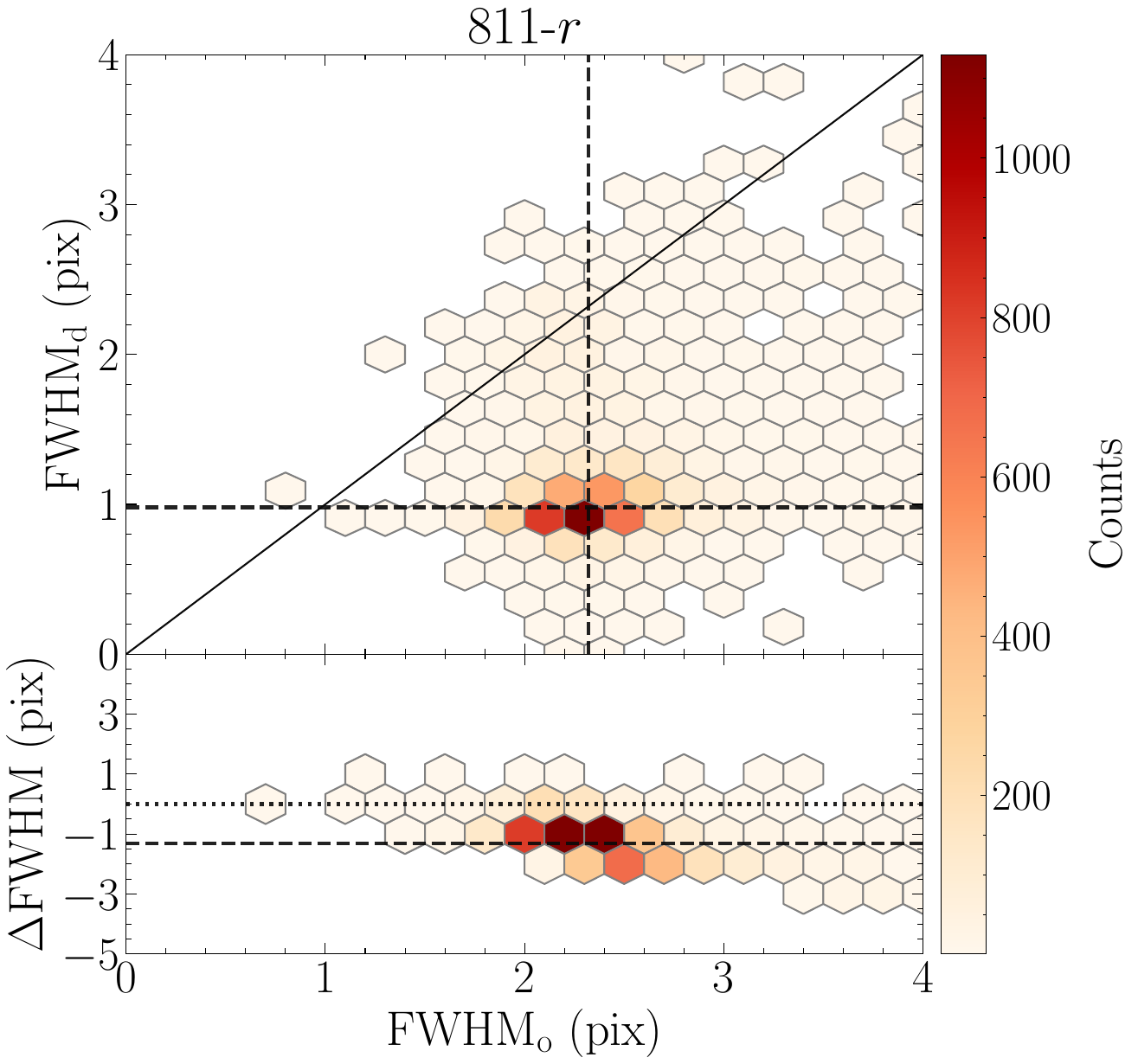}
      \includegraphics[keepaspectratio,width=0.32\linewidth]{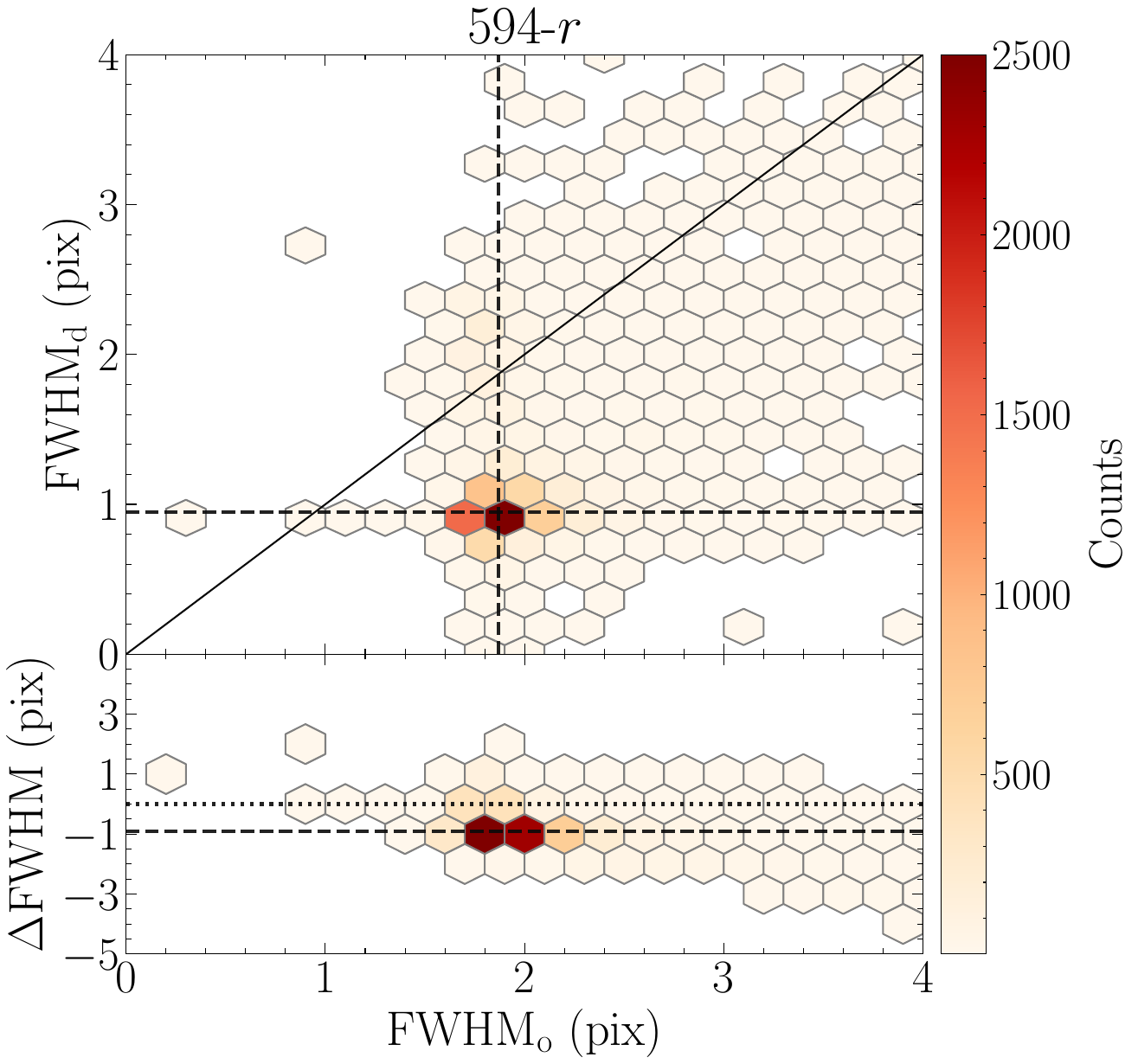}
        \caption{Comparison of FWHM of the original ($\mathrm{FWHM}_{\mathrm{o}}$) and the one-to-one matched deconvolved ($\mathrm{FWHM}_{\mathrm{d}}$) sources for all the $r$ filter images considered in this study. The bottom subpanels of each panel show the difference in the FWHM, $\Delta \mathrm{FWHM} = \mathrm{FWHM}_{\mathrm{d}} - \mathrm{FWHM}_{\mathrm{o}}$. The solid line in the upper panel and the dotted horizontal line in the bottom subpanels denote perfect agreement $\Delta \mathrm{FWHM} = 0$. The vertical and horizontal dashed lines in the upper panel denote the median FWHM. The horizontal dashed line in the lower panel denotes the median $\Delta \mathrm{FWHM}$. The title of each panel denotes the ZTF field ID. The distribution of the data points in the upper panel indicates that there is a weak correlation between the FWHM of the original and deconvolved sources. Instead, the deconvolved FWHM are generally roughly $\sim$1 pix irrespective of the original FWHM. The typical scatter, quantified in Table~\ref{tab:summary-deconv-results}, is also small. While the deconvolved FWHM are higher than the FWHM of the central maximum of the Airy disk (the theoretical limit due to diffraction) in the $r$-band of the 1.2 m telescope used by ZTF ($\approx$0$\arcsec$.11), Table~\ref{tab:summary-deconv-results} quantifies that deconvolution still offers 2-2.5 times improvement in resolution.} \label{fig:fwhm-one-to-one-comparison}
\end{figure*}

\begin{figure*}
    \centering
      \includegraphics[keepaspectratio,width=0.32\linewidth]{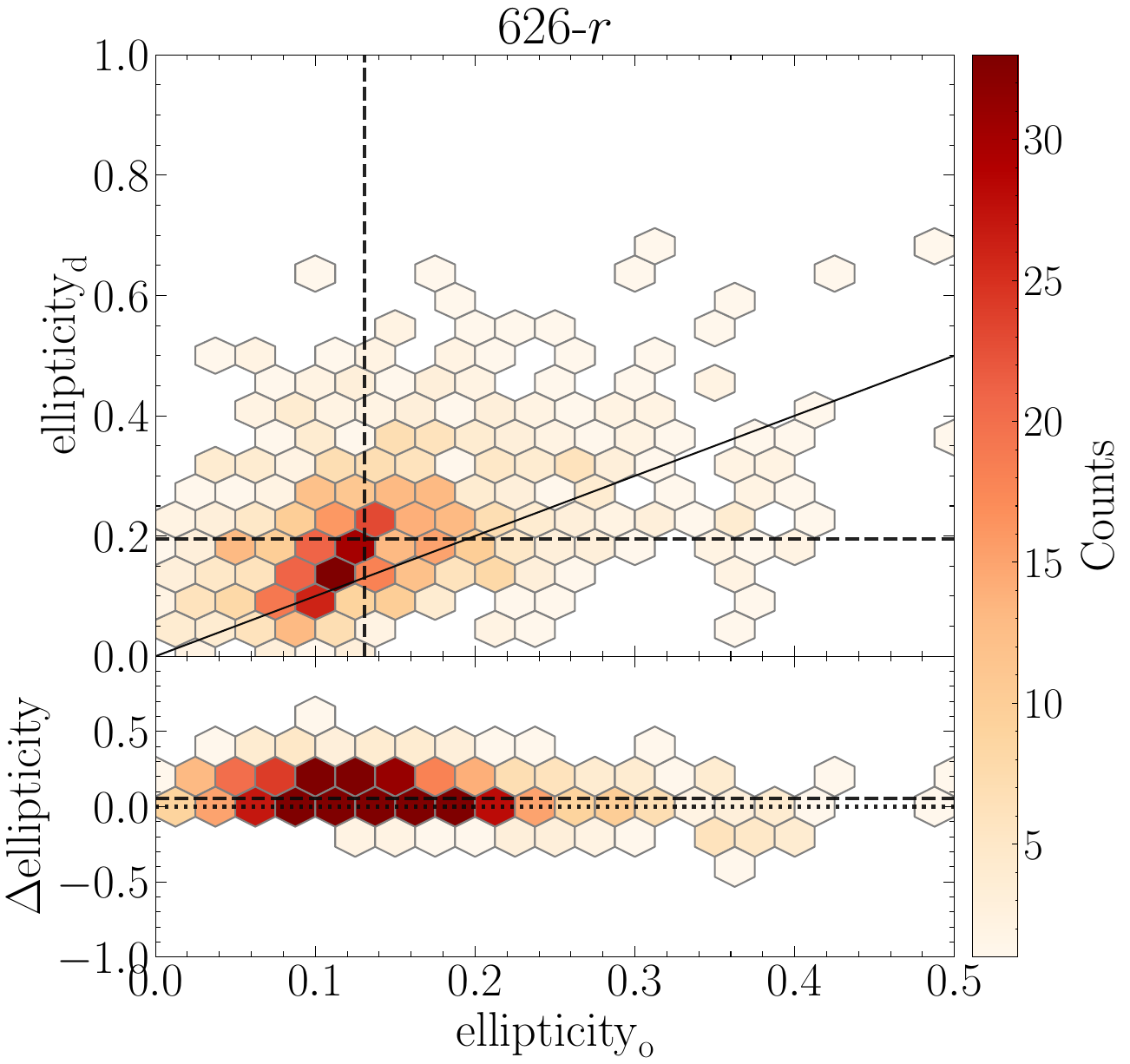}
      \includegraphics[keepaspectratio,width=0.32\linewidth]{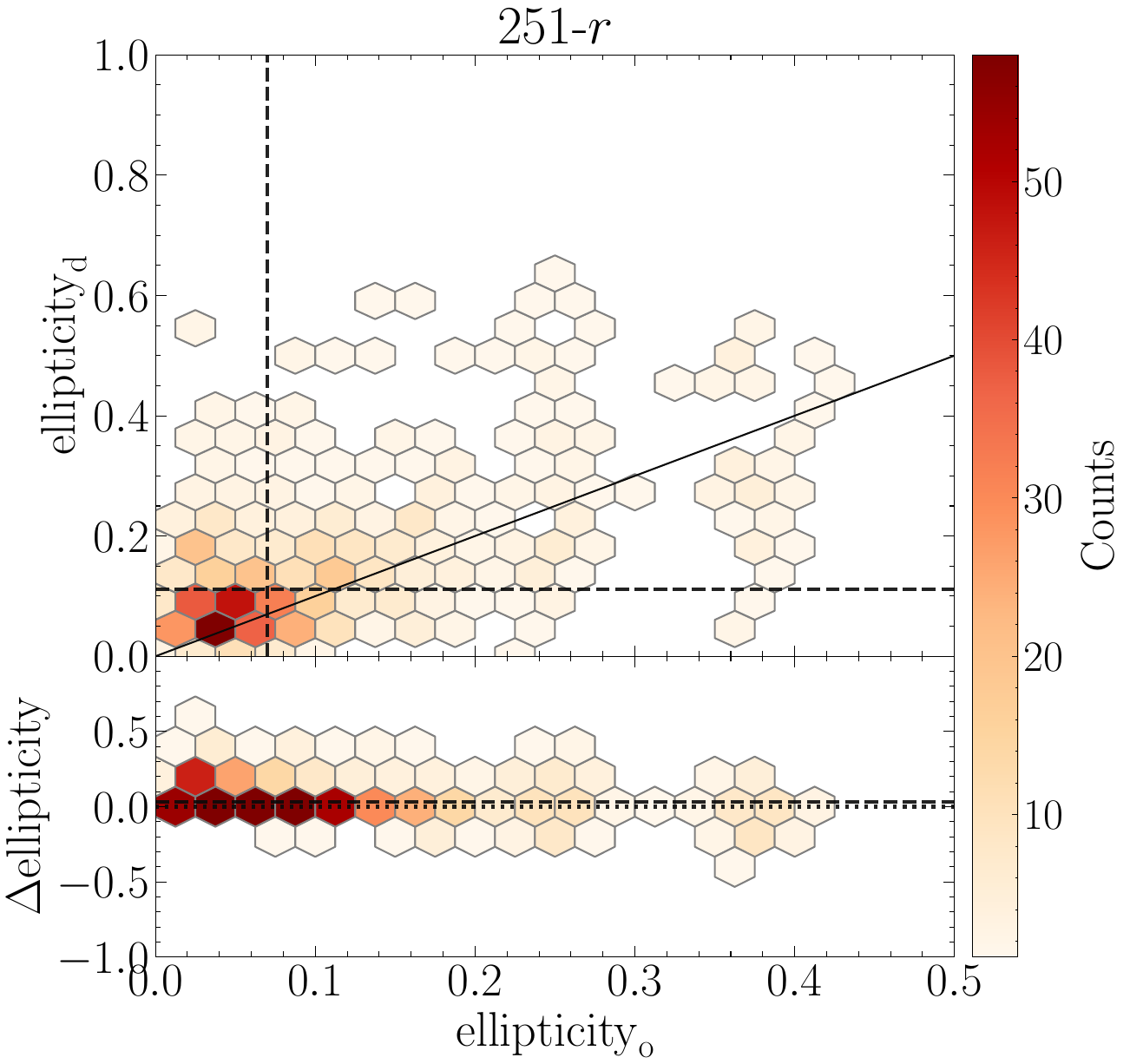}
      \includegraphics[keepaspectratio,width=0.32\linewidth]{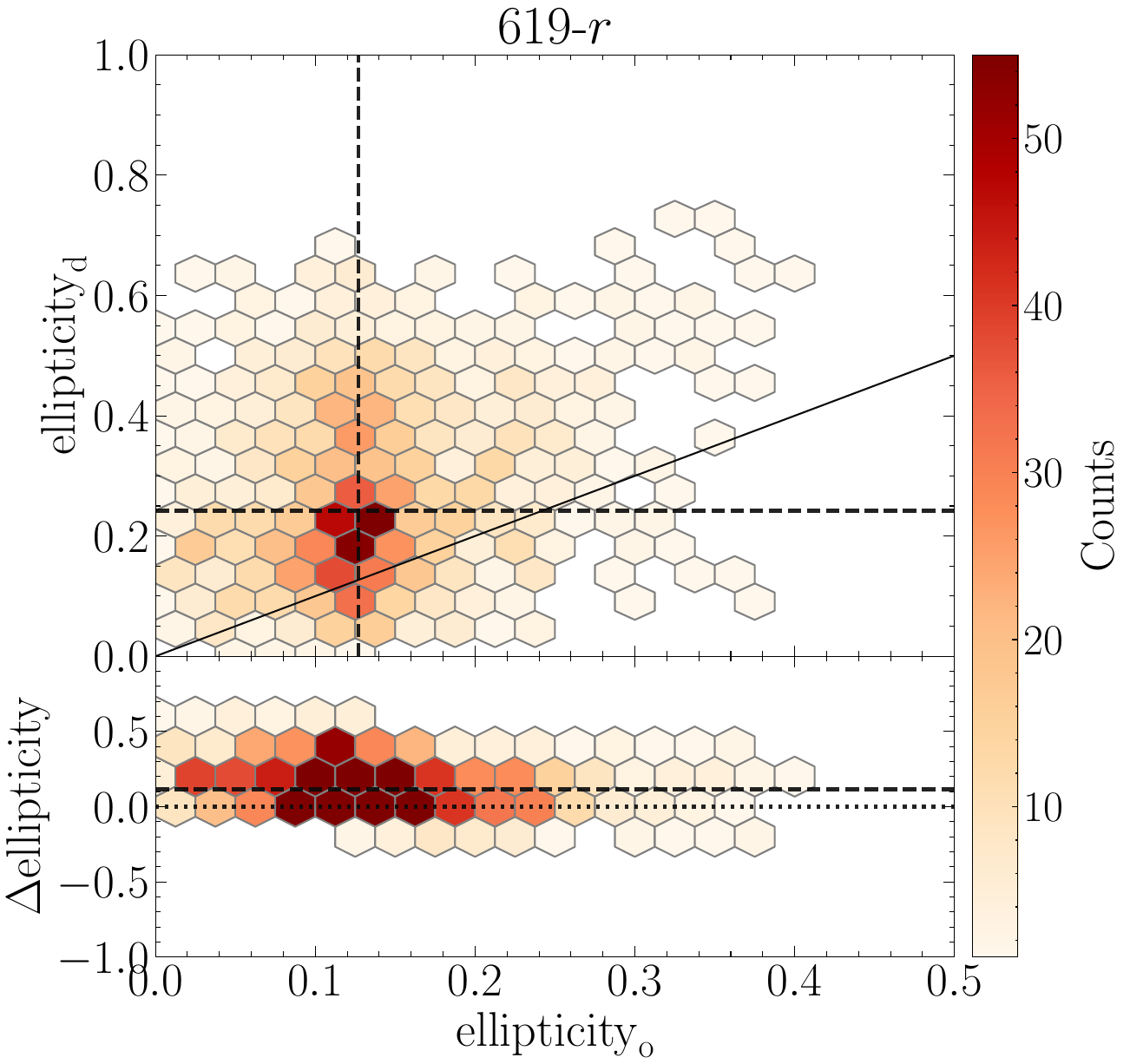}
      \includegraphics[keepaspectratio,width=0.32\linewidth]{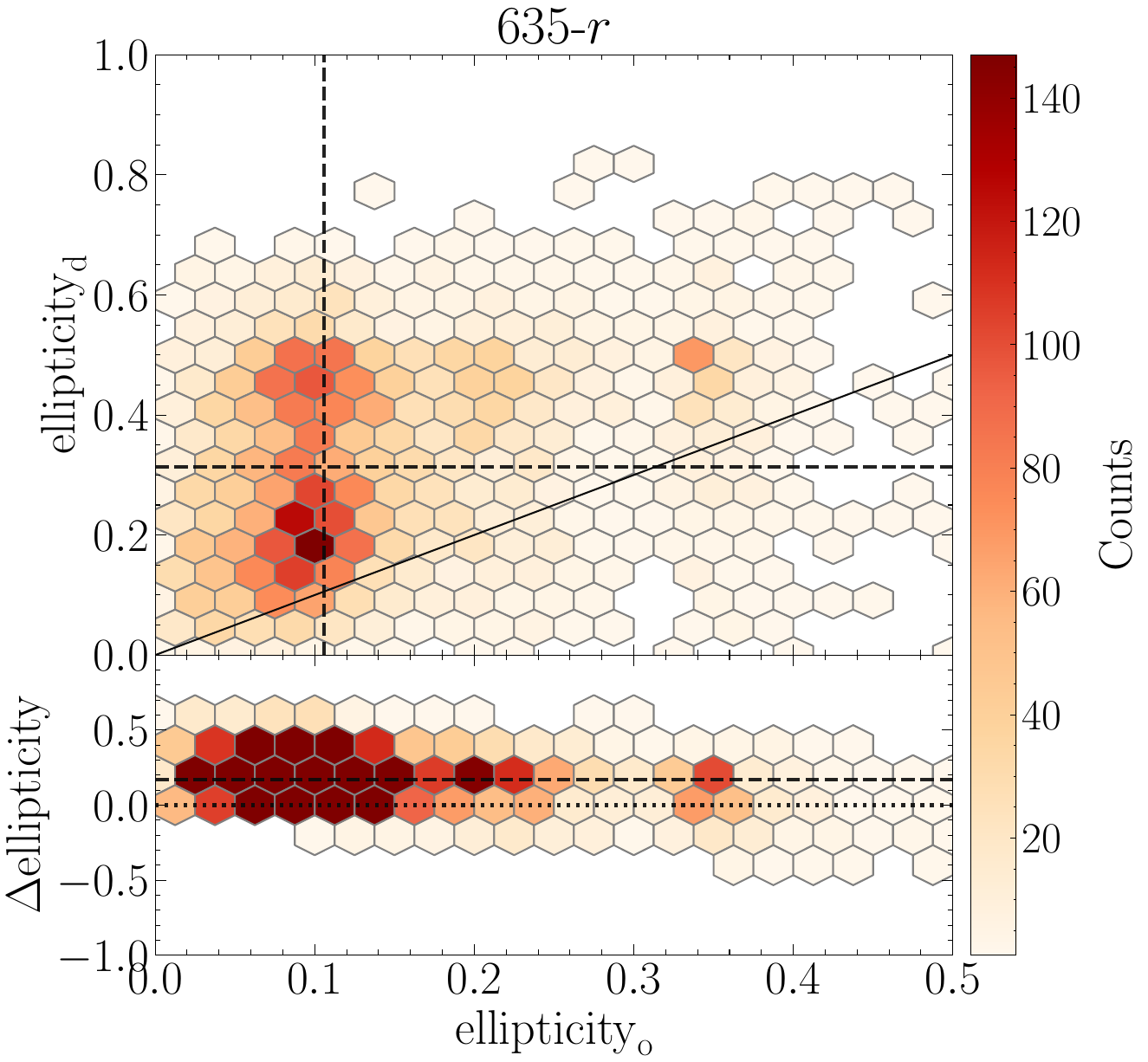}
      \includegraphics[keepaspectratio,width=0.32\linewidth]{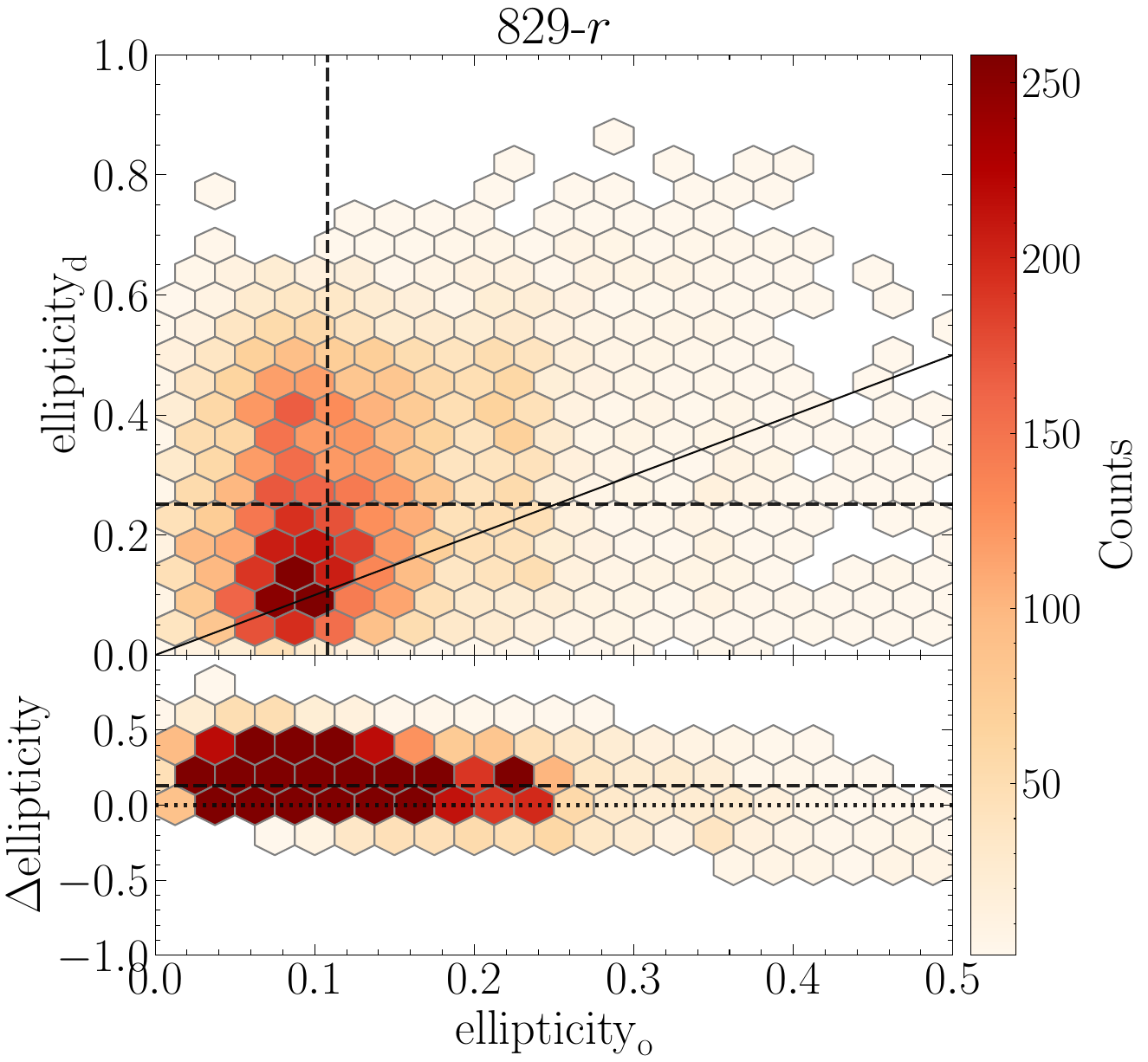}
      \includegraphics[keepaspectratio,width=0.32\linewidth]{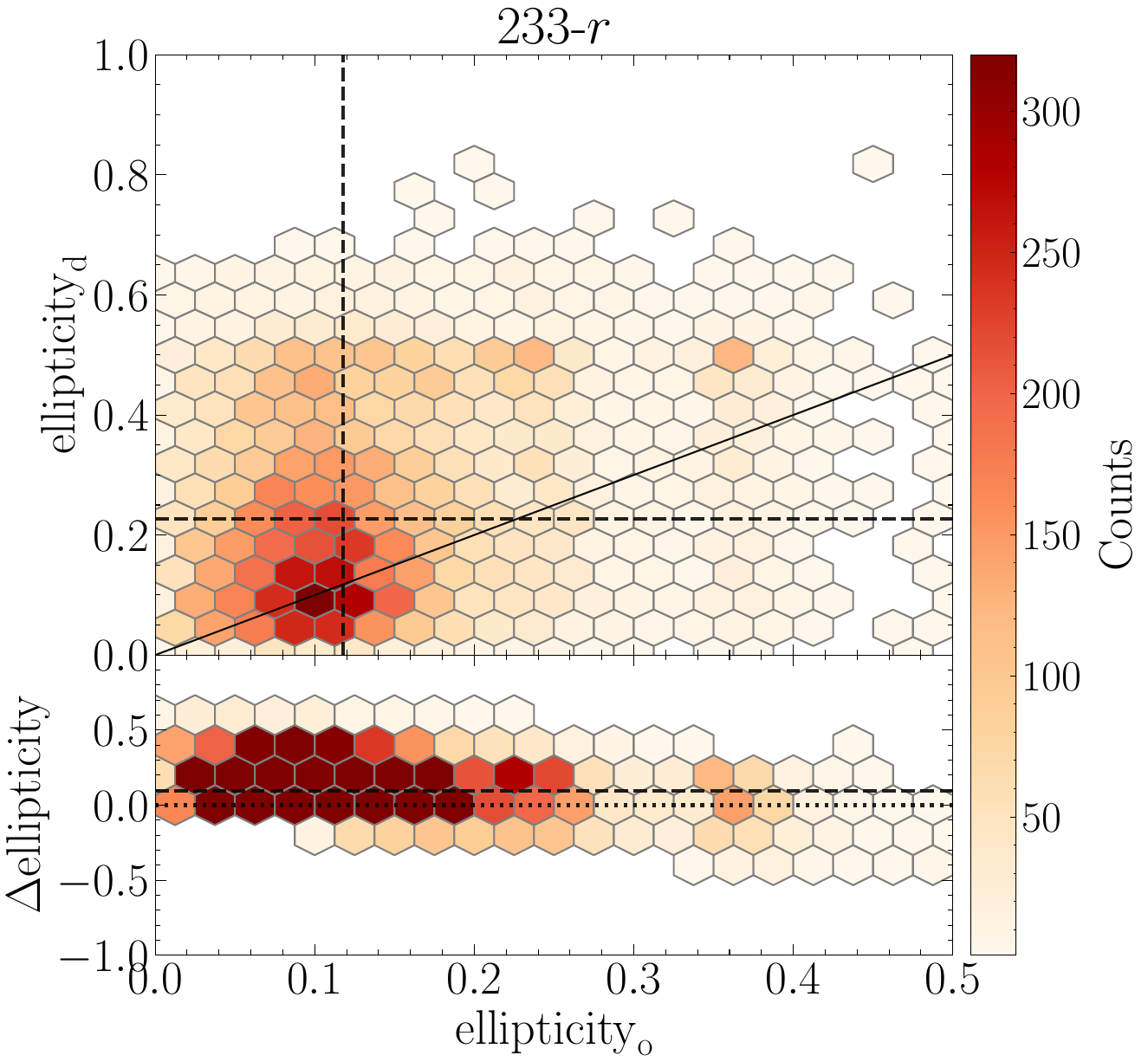}
      \includegraphics[keepaspectratio,width=0.32\linewidth]{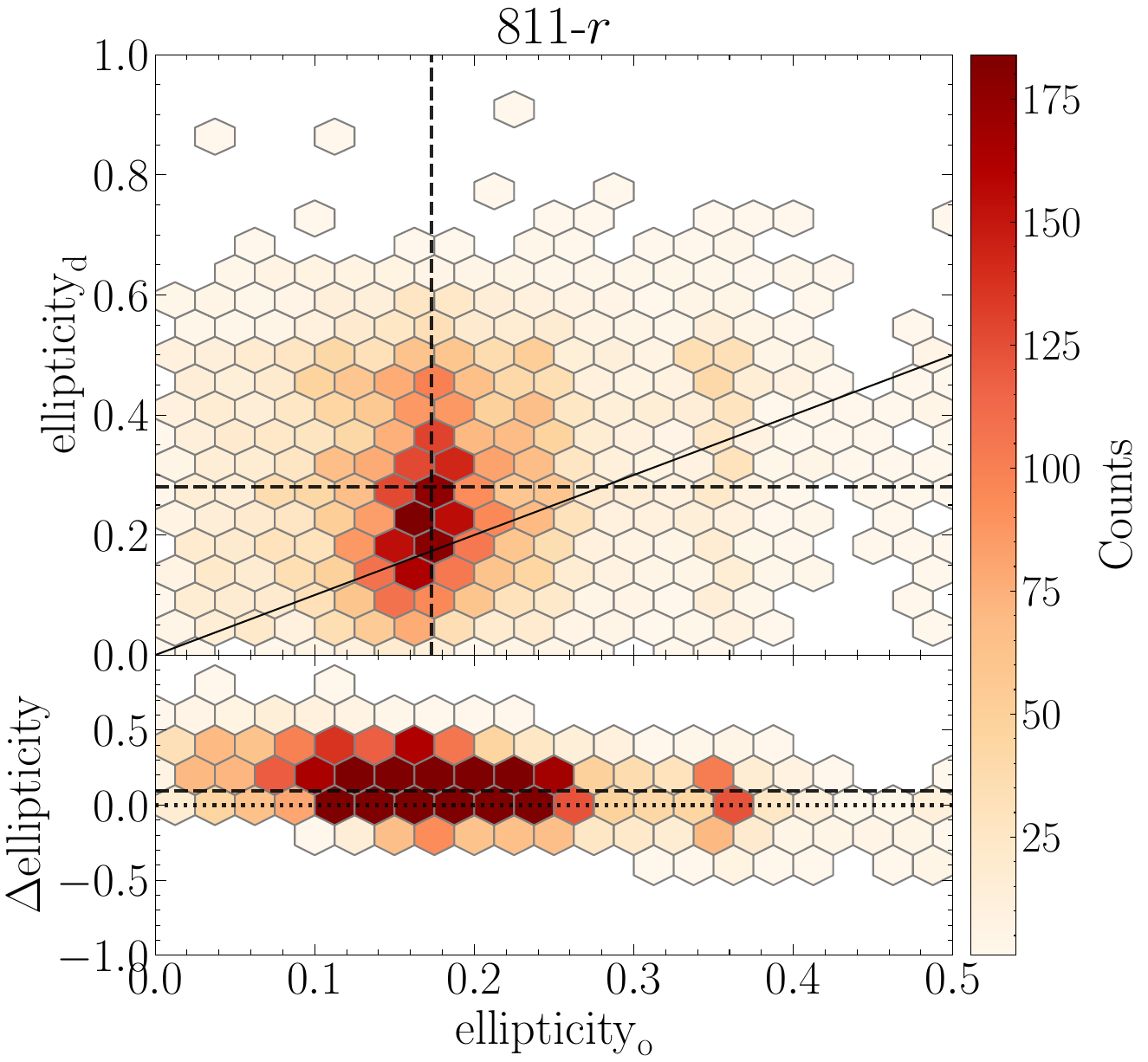}
      \includegraphics[keepaspectratio,width=0.32\linewidth]{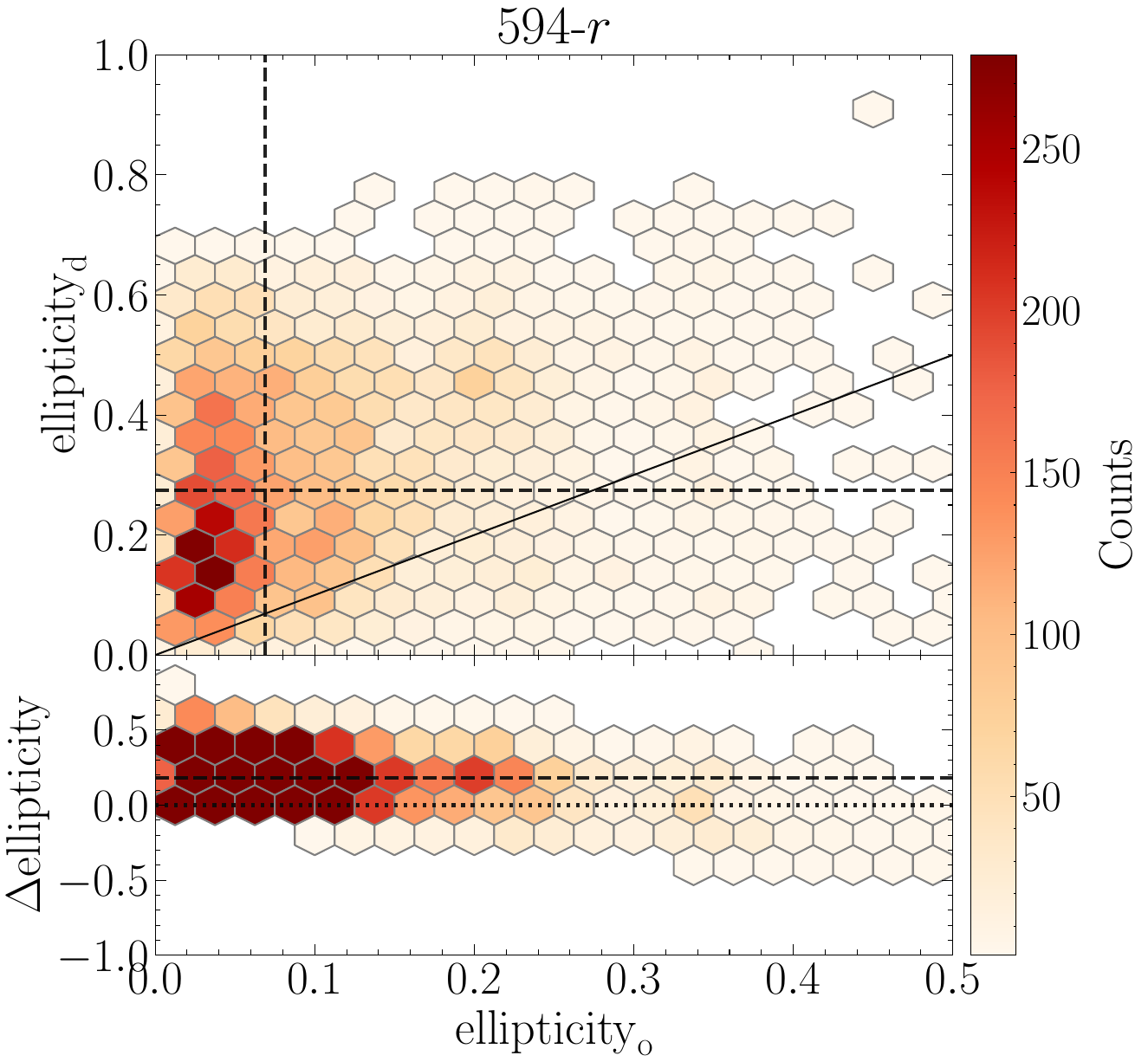}
    \caption{Comparison of ellipticities of the original ($\mathrm{ellipticity}_{\mathrm{o}}$) and the one-to-one matched deconvolved ($\mathrm{ellipticity}_{\mathrm{d}}$) sources for all the $r$ filter images considered in this study. The bottom subpanels of each panel show the difference in the ellipticities, $\Delta \mathrm{ellipticity} = \mathrm{ellipticity}_{\mathrm{d}} - \mathrm{ellipticity}_{\mathrm{o}}$. The solid line in the upper panel and the dotted horizontal line in the bottom subpanels denote perfect agreement $\Delta \mathrm{ellipticity} = 0$. The vertical and horizontal dashed lines in the upper panel denote the median ellipticities. The horizontal dashed line in the lower panel denotes the median $\Delta \mathrm{ellipticity}$. The title of each panel denotes the ZTF field ID. The deconvolved ellipticities are slightly enlarged compared to the original ellipticities by different amounts for the different images, but generally from $\sim$0.1 in the original to $\sim$0.2-0.3 in the deconvolved, except the 626 and 251 ID fields where the increase was smaller. While the typical scatter for original sources is generally small (0.03-0.05 pix), it is enlarged for deconvolved sources (0.1-0.14 pix) except for 626 and 251 ID fields where it is smaller (0.06-0.07 pix). One reason that might contribute to the slightly enlarged ellipticities in the deconvolved is that, as detailed in Sect.~\ref{subsec:experimental-details}, we do not use filtering before detecting deconvolved sources, and since ellipticity is calculated post this smoothing in SExtractor, our calculated deconvolved ellipticities may be slightly overestimated \citep{Holwerda2005}. This corroborates the visual inspection in Sect.~\ref{fig:visual}, where the deconvolved sources were found to be compact but not necessarily circular in shape.} \label{fig:ellipticity-one-to-one-comparison}
\end{figure*}

\begin{figure*}
    \centering
      \includegraphics[keepaspectratio,width=0.32\linewidth]{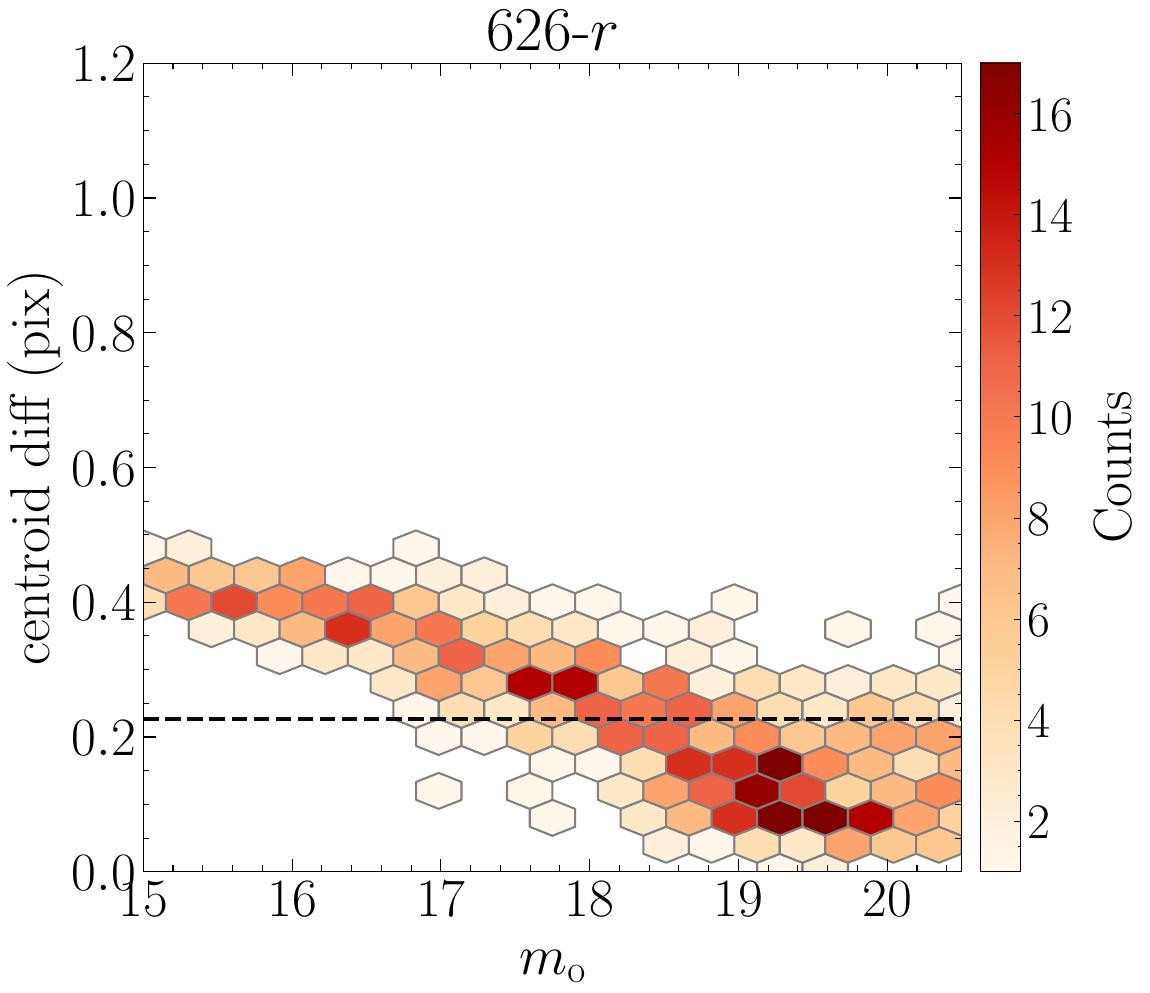}
      \includegraphics[keepaspectratio,width=0.32\linewidth]{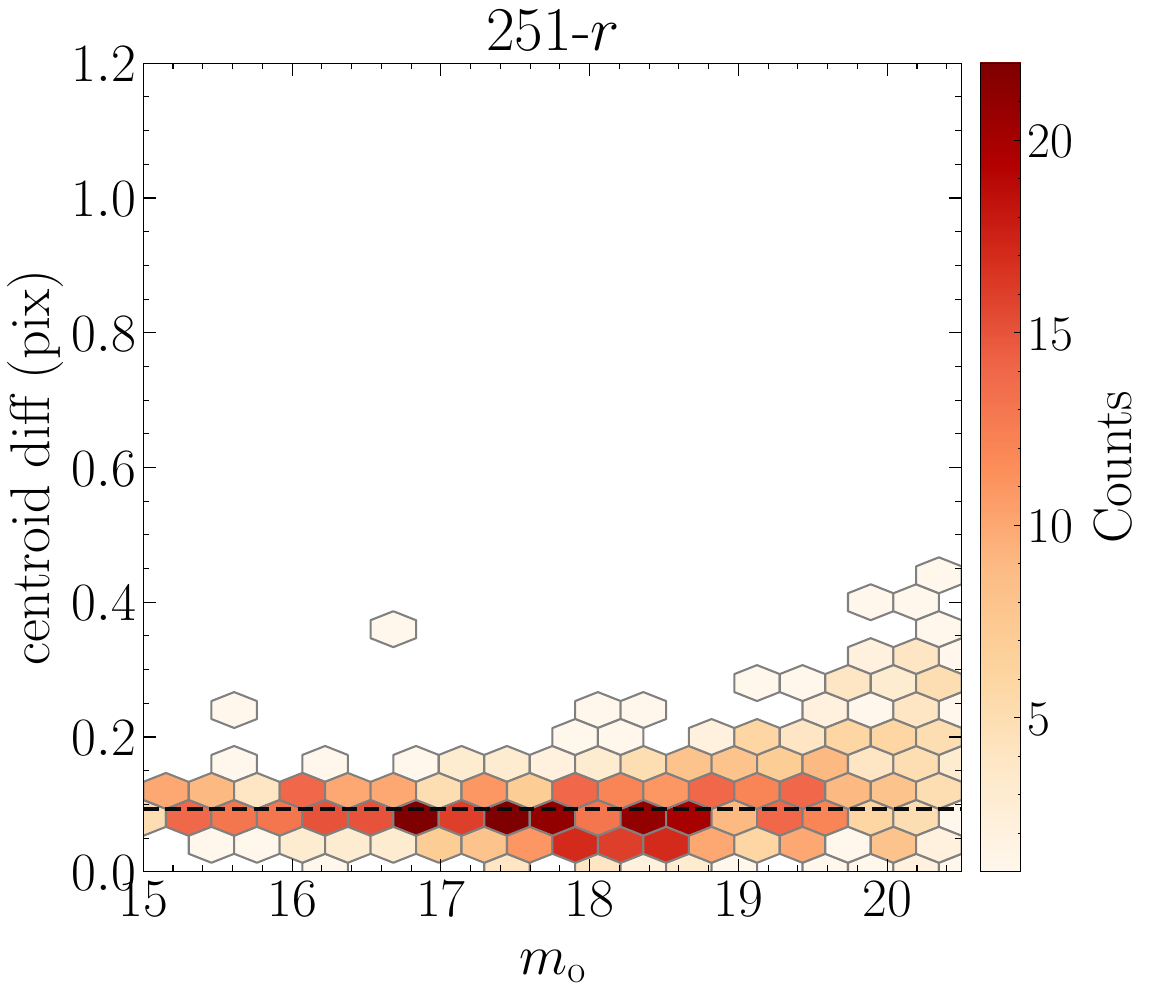}
      \includegraphics[keepaspectratio,width=0.32\linewidth]{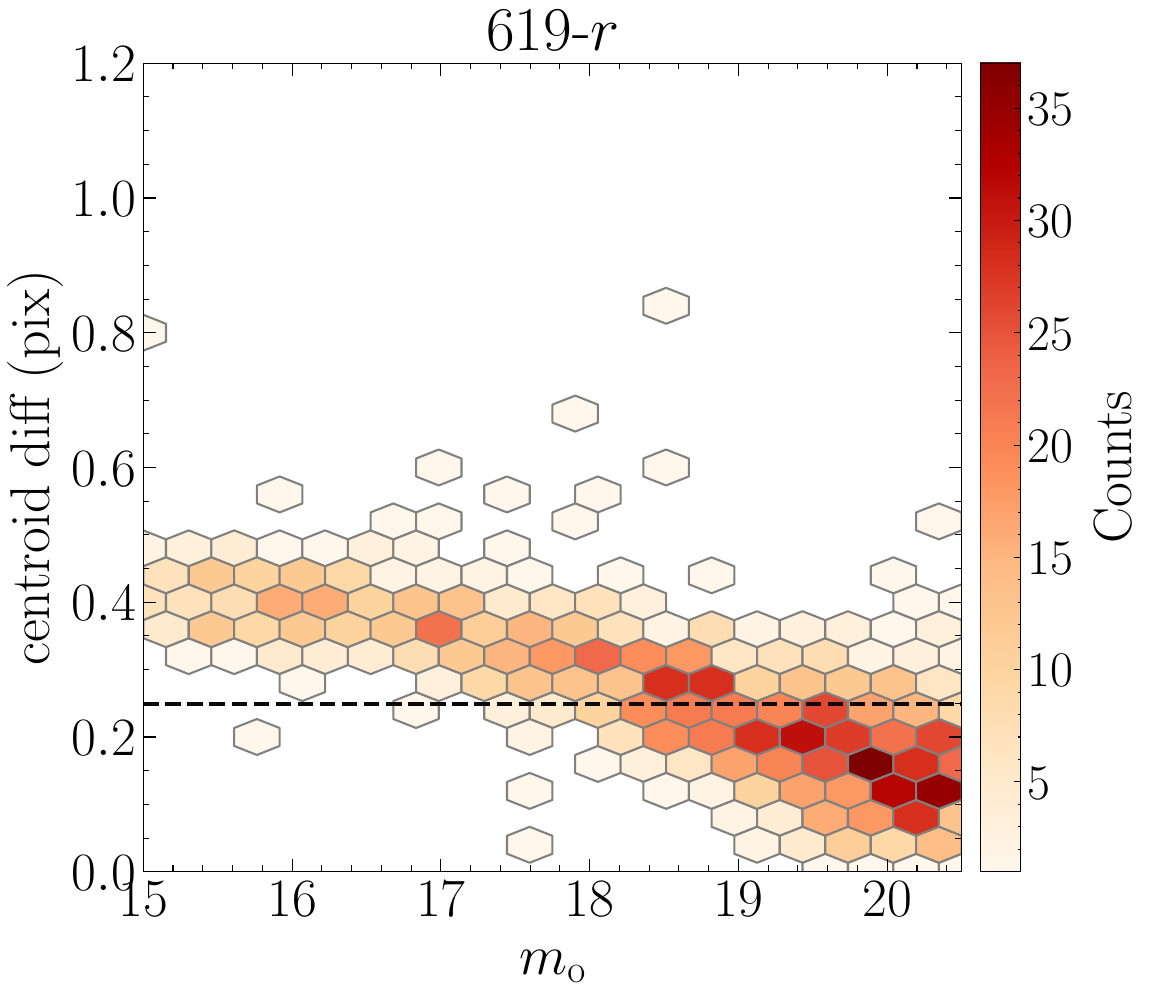}
      \includegraphics[keepaspectratio,width=0.32\linewidth]{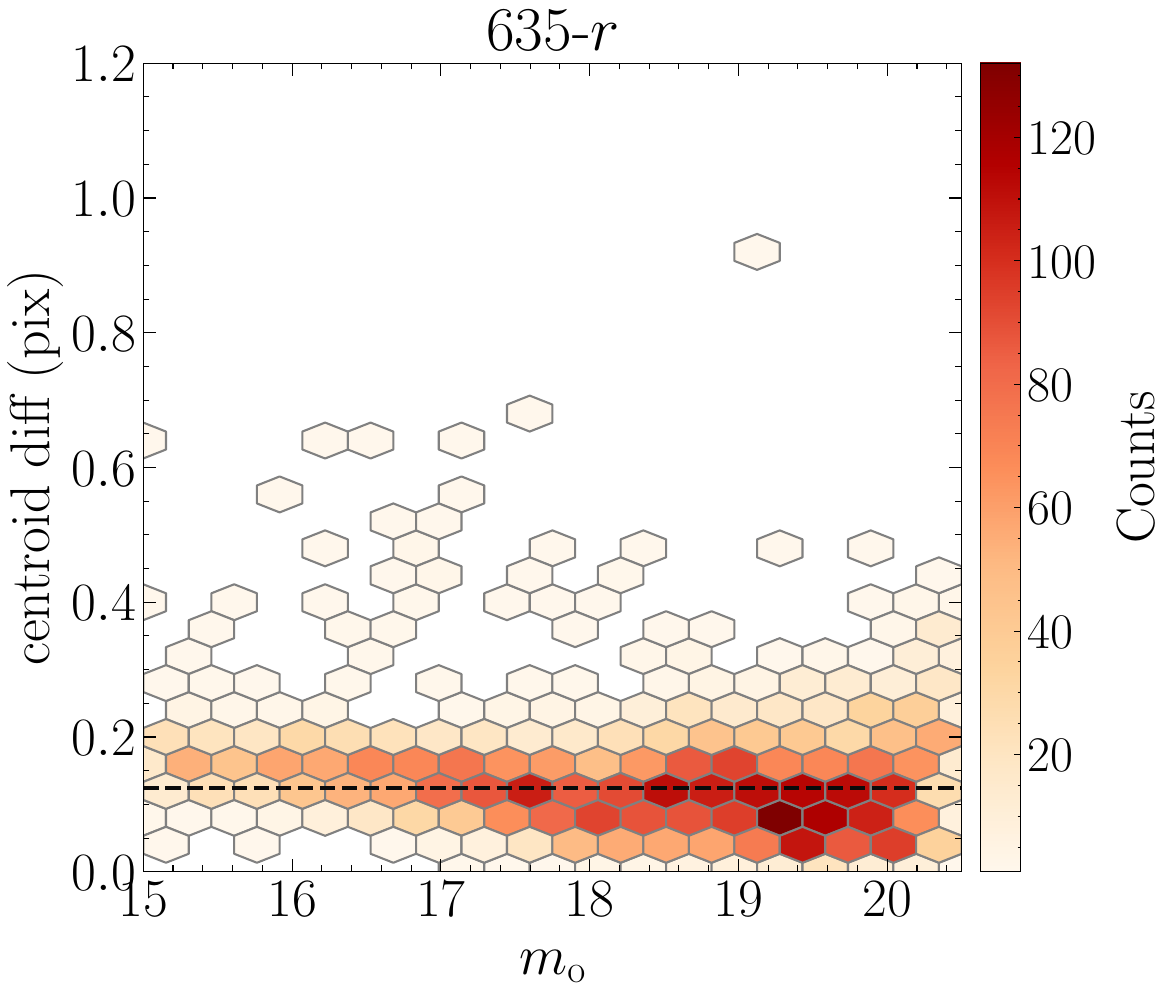}
      \includegraphics[keepaspectratio,width=0.32\linewidth]{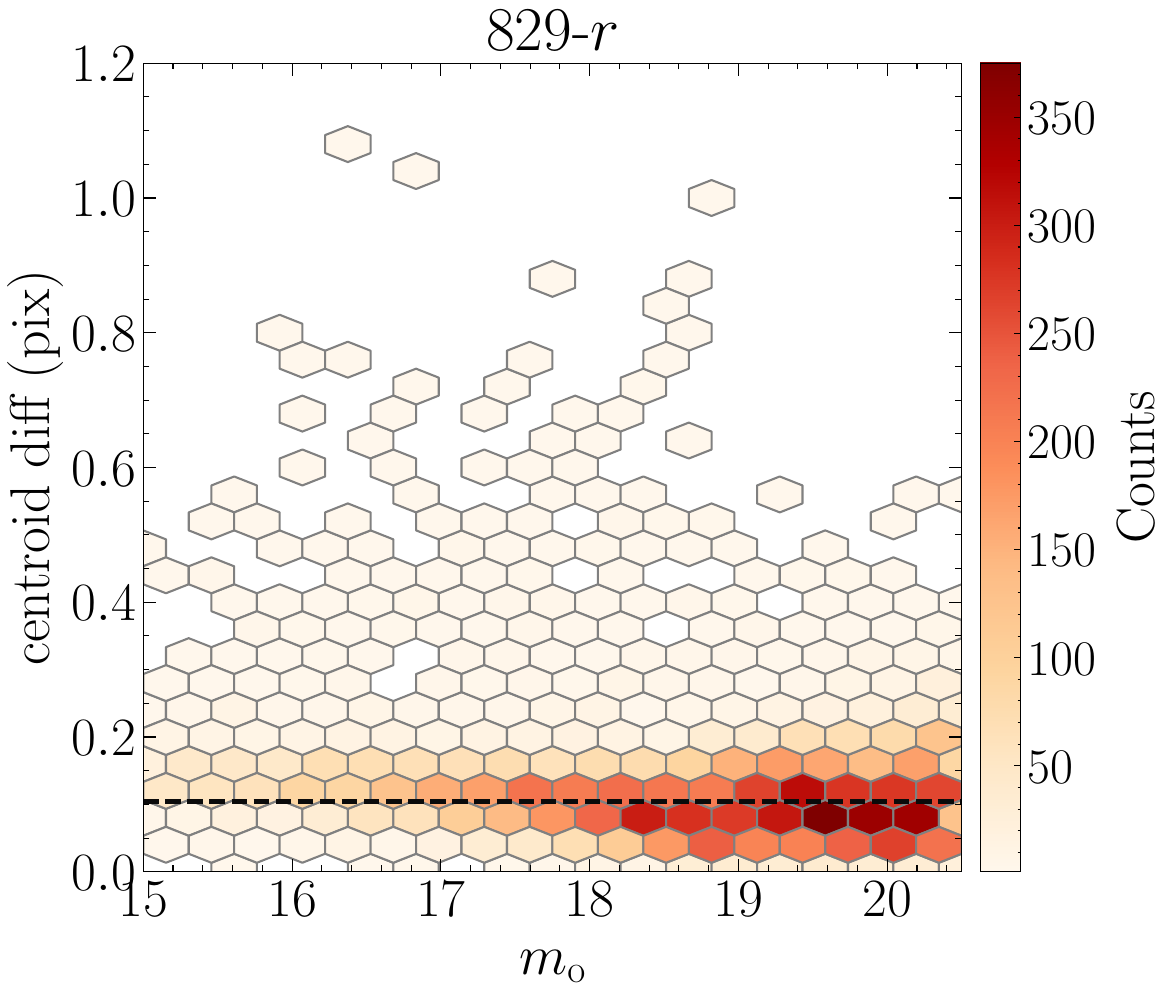}
      \includegraphics[keepaspectratio,width=0.32\linewidth]{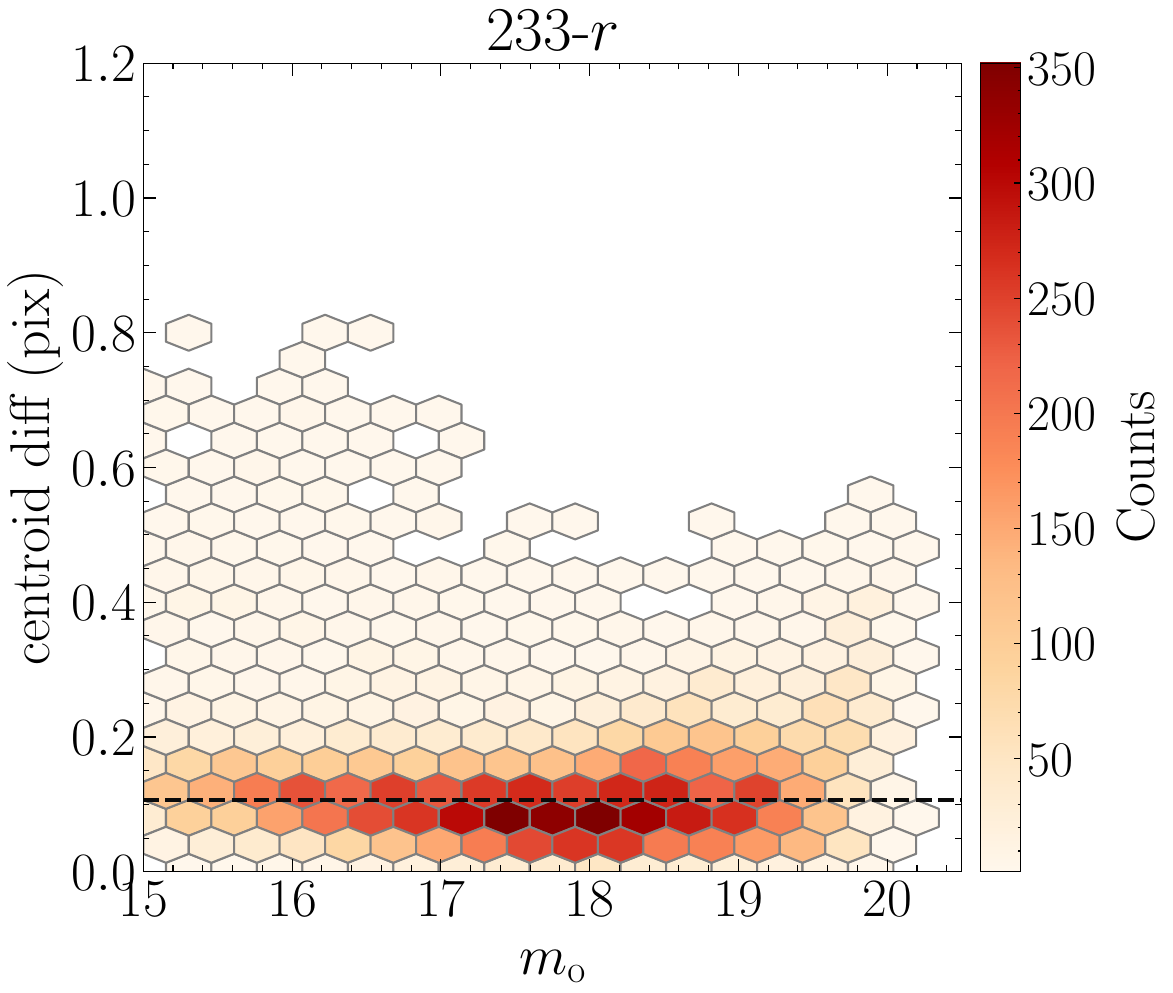}
      \includegraphics[keepaspectratio,width=0.32\linewidth]{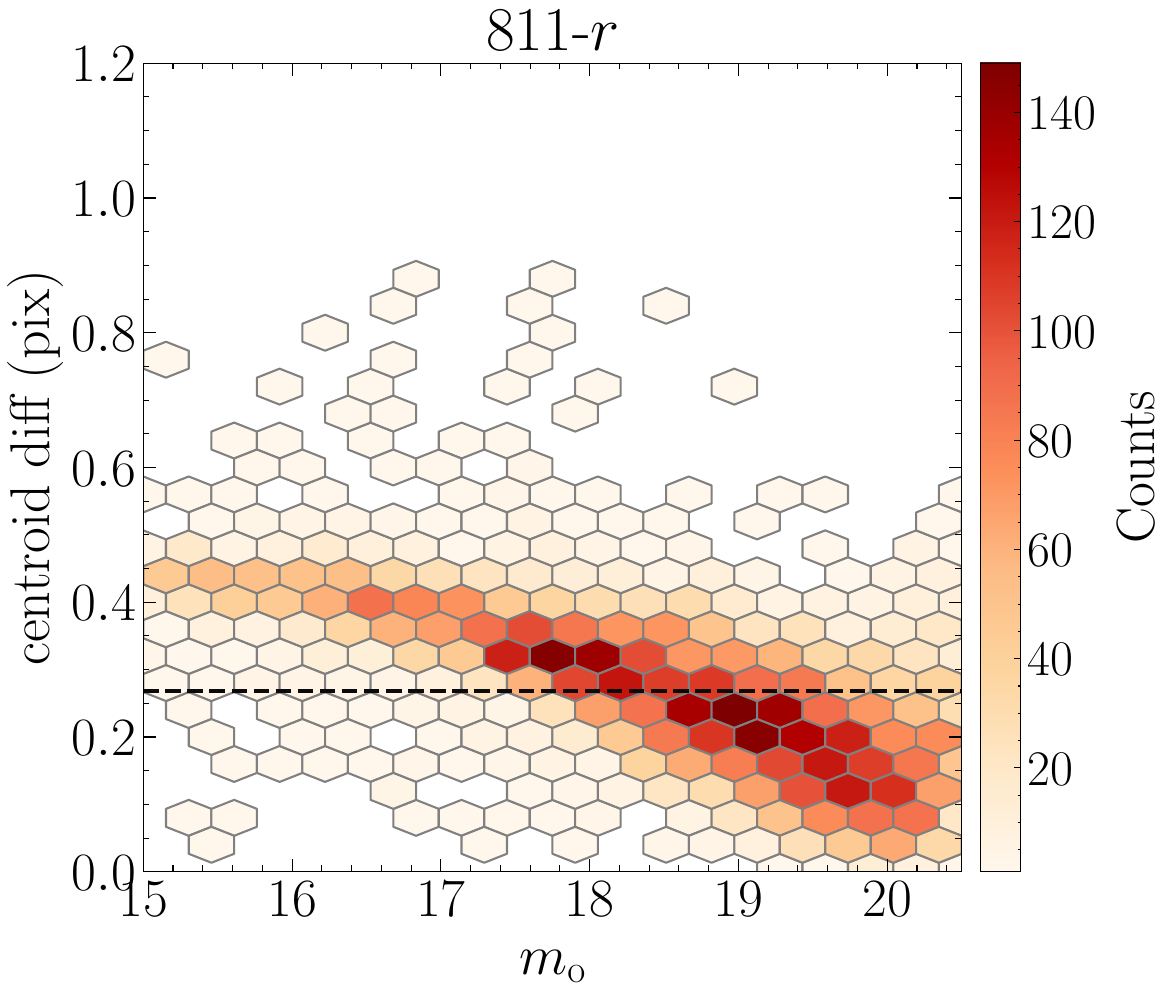}
      \includegraphics[keepaspectratio,width=0.32\linewidth]{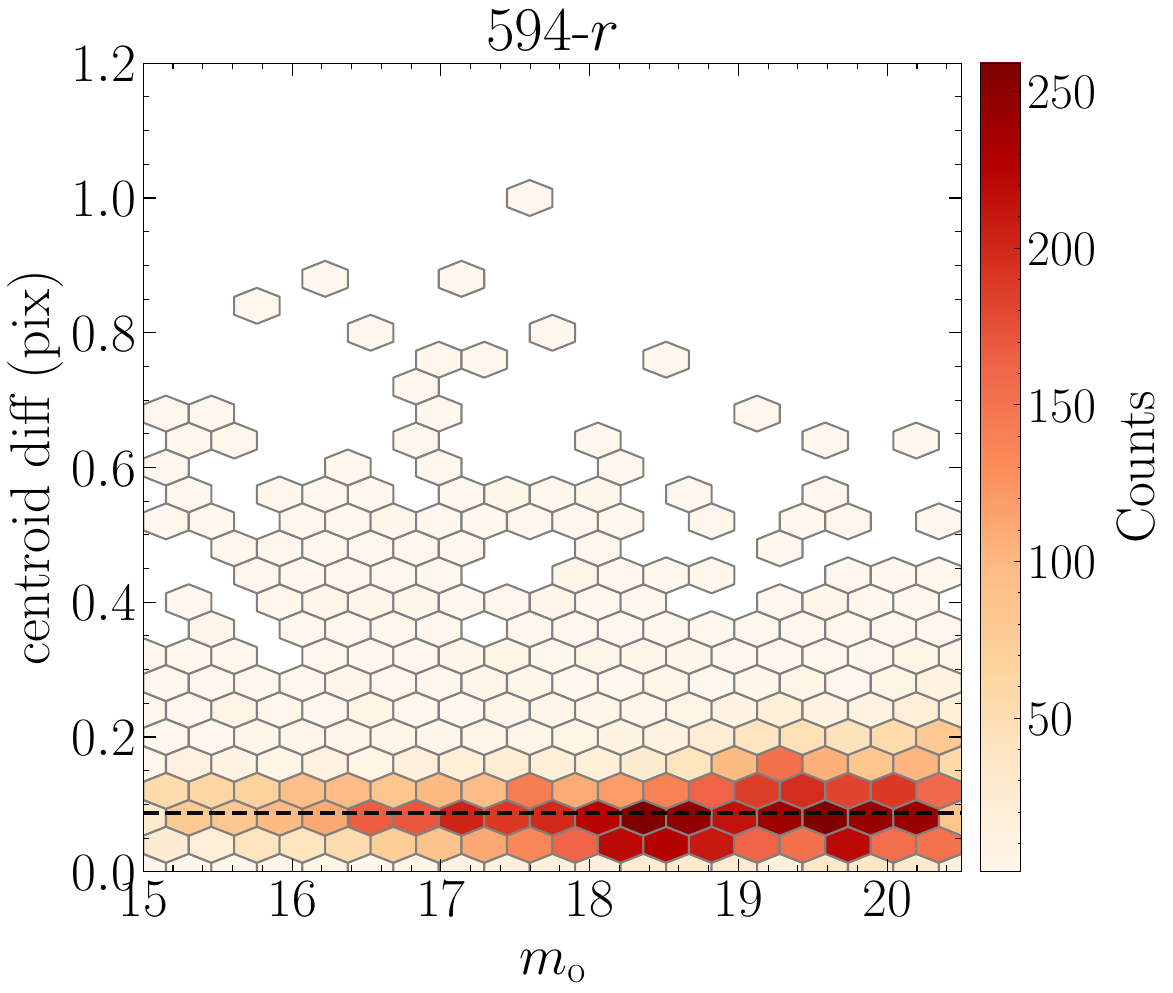}
    \caption{Distance between the centroids of the original and the one-to-one matched deconvolved sources (y-axis) for all the $r$ filter images considered in this study vs. the magnitude of the original source (x-axis). The primary point of interest is in the typical values of the centroid differences, and the original magnitude is only plotted for supplementary analyses (see the specific subsections in Sect.~\ref{sec:field-specific-results}). `Centroid' here means the barycenters of the detected sources, which are the flux-weighted mean coordinates. The distance is the Euclidean distance between the pixel coordinates: centroid diff$\,\, = \sqrt{(x_{\mathrm{o}} - x_{\mathrm{d}})^2 + (y_{\mathrm{o}} - y_{\mathrm{d}})^2}$, where $(x_{\mathrm{o}}, y_{\mathrm{o}})$ and $(x_{\mathrm{d}}, y_{\mathrm{d}})$ are the locations of the original and deconvolved sources, respectively. The title of each panel denotes the ZTF field ID. The horizontal dashed line denotes the median distance. The deconvolved sources are located close to the original sources, with mostly $<$1 pix shifts compared to the original source position across all ranges of magnitudes. This suggests that the deconvolution is able to accurately locate the location of the pixel with the maximum flux, even for the faintest original sources, and the flux distribution of the deconvolved sources is generally arranged in a way that does not affect the centroid location. It is also worthwhile to note that PSFs of all these images are not highly elliptical ($\mathrm{ellipticity} \lesssim 0.1-0.2$, as found by our internal calculations) or distorted in shape, which may partly contribute to the small centroid differences. The fact that the ellipticity of deconvolved sources slightly increases (see Fig.~\ref{fig:ellipticity-one-to-one-comparison}) but their centroids are not severely shifted from the original coordinates indicates that the flux distribution of such deconvolved sources is {\it not} severely distorted compared to the corresponding original sources. Thus, it is still possible to accurately locate the sources in the deconvolved image despite the increased ellipticities.} \label{fig:centroidDiff-one-to-one-comparison}
\end{figure*}

In this section, we present the deconvolution results on the images described in Table~\ref{tab:data-description}. All subsequent subsections discuss results only in the $r$ filter; the other filters, $g$ and $i$ are considered in Sect.~\ref{subsec:diff-filters}. Figs.~\ref{fig:mag-one-to-one-comparison}, \ref{fig:fwhm-one-to-one-comparison}, \ref{fig:ellipticity-one-to-one-comparison}, and \ref{fig:centroidDiff-one-to-one-comparison} show the comparison of the magnitude, FWHM, ellipticity, and Euclidean distance between the original and the corresponding deconvolved source coordinates for the one-to-one matches. Fig.~\ref{fig:unmatched-deconvolved-combined} shows the distribution of the FWHM, ellipticity, and magnitude of the unmatched deconvolved sources. Figs.~\ref{fig:ngc1569-zoomed} and \ref{fig:ngc7006-zoomed} show visualizations of the deconvolution of a dwarf galaxy and a globular cluster, respectively. Appendix~\ref{appn:many-to-one} shows visualizations of the many-to-one matches for possible deblending scenarios different from those in Sect.~\ref{sec:deblending-examples}. Table~\ref{tab:summary-deconv-results} summarizes the deconvolution results for all images.

\begin{table*}
\renewcommand{\arraystretch}{1.1}
\setlength{\tabcolsep}{8pt}
\centering
\caption{Summary of the deconvolution results for all images considered in this study. For the one-to-one matches, the following metrics are shown: median $\Delta m$ (difference in original and deconvolved magnitude), median FWHM and ellipticity of the original and deconvolved sources and the corresponding scatter across all sources using the median absolute deviation (these are the variations in the point estimates across all sources rather than the averaged quoted uncertainties obtained from SExtractor), and the centroid differences of the one-to-one matched sources. The values inside the parentheses in the column denoting $\Delta m$ show the percentage flux preserved in the deconvolved image compared to the original image corresponding to $\Delta m$. `O' denotes original and `D' denotes deconvolved. The median magnitude, FWHM, and ellipticity and the corresponding scatter across all unmatched deconvolved sources that are likely astrophysical (obtained by applying the additional filtering criteria described in Sect.~\ref{subsec:crossmatch-results} to the unmatched deconvolved statistics in Table~\ref{tab:crossmatching-results}), are also shown. See Figs.~\ref{fig:mag-one-to-one-comparison}--\ref{fig:centroidDiff-one-to-one-comparison}, Fig.~\ref{fig:unmatched-deconvolved-combined}, Fig.~\ref{fig:g-and-i-bands}, and Fig.~\ref{fig:unmatched-deconvolved-combined-g-and-i} for the plots.}
\label{tab:summary-deconv-results}
\begin{tabular*}{\textwidth}{p{0.15in} p{0.1in} p{0.8in} p{0.48in} p{0.48in} p{0.48in} p{0.48in} p{0.48in} p{0.5in} p{0.45in} p{0.45in}}
\hline\hline
ZTF Field ID         & Filter & \multicolumn{6}{c}{One-to-one}   & \multicolumn{2}{c}{Unmatched deconvolved} \\ \midrule
\multicolumn{1}{l}{} &        & $\Delta m$ & FWHM (pix) [O] & FWHM (pix) [D] & ellipticity [O] & ellipticity [D] & centroid diff (pix) & $m_{\mathrm{d}}$ & FWHM (pix) & ellipticity                \\ \cmidrule(lr){3-8} \cmidrule(lr){9-11}
626                  & $r$      & -0.121 (112\%)    & 3.09~$\pm$~0.14 & 1.23~$\pm$~0.25     & 0.13~$\pm$~0.04 & 0.2~$\pm$~0.07    & 0.2              & 20.89~$\pm$~0.38 & 1.31~$\pm$~0.37 & 0.4~$\pm$~0.11                   \\
\multirow{2}{*}{}    & $g$      & -0.079 (108\%)    & 3.13~$\pm$~0.12 & 1.1~$\pm$~0.16     & 0.08~$\pm$~0.03 & 0.12~$\pm$~0.06   & 0.1          & 21.06~$\pm$~0.28 & 1.23~$\pm$~0.25 & 0.35~$\pm$~0.12                   \\
                     & $i$      & -0.148 (115\%)    & 3.31~$\pm$~0.14 & 1.27~$\pm$~0.25     & 0.07~$\pm$~0.04 & 0.12~$\pm$~0.06    & 0.1            & 20.8~$\pm$~0.43 & 1.4~$\pm$~0.41 & 0.35~$\pm$~0.12                   \\
251                  & $r$      & -0.061 (106\%)    & 2.93~$\pm$~0.14 & 1.28~$\pm$~0.26     & 0.07~$\pm$~0.04 & 0.11~$\pm$~0.06    & 0.1              & 20.66~$\pm$~0.39 & 1.15~$\pm$~0.23 & 0.34~$\pm$~0.13                  \\ 
619                  & $r$      & -0.132 (113\%)    & 2.34~$\pm$~0.2 & 0.99~$\pm$~0.12     & 0.13~$\pm$~0.03 & 0.24~$\pm$~0.1    & 0.2              & 21.06~$\pm$~0.25 & 1.06~$\pm$~0.15 & 0.38~$\pm$~0.1                  \\
635                  & $r$      & -0.041 (104\%)   & 1.74~$\pm$~0.14 & 0.97~$\pm$~0.1     & 0.11~$\pm$~0.04 & 0.31~$\pm$~0.13    & 0.1              & 20.86~$\pm$~0.32 & 0.96~$\pm$~0.06 & 0.45~$\pm$~0.05                   \\
829                  & $r$      & -0.082 (108\%)    & 1.9~$\pm$~0.14 & 0.94~$\pm$~0.06     & 0.11~$\pm$~0.04 & 0.25~$\pm$~0.13    & 0.1              & 20.94~$\pm$~0.3 & 0.96~$\pm$~0.06 & 0.39~$\pm$~0.1                   \\
\multirow{2}{*}{}    & $g$      & -0.096 (109\%)    & 1.96~$\pm$~0.11 & 0.94~$\pm$~0.06     & 0.07~$\pm$~0.04 & 0.24~$\pm$~0.12    & 0.1              & 21.03~$\pm$~0.25 & 0.96~$\pm$~0.07 & 0.38~$\pm$~0.11                   \\
                     & $i$      & -0.087 (108\%)   & 1.87~$\pm$~0.12 & 0.96~$\pm$~0.07     & 0.09~$\pm$~0.05 & 0.29~$\pm$~0.14    & 0.1            & 20.42~$\pm$~0.45 & 0.95~$\pm$~0.04 & 0.44~$\pm$~0.05                    \\
233                  & $r$      & -0.03 (103\%)   & 2.67~$\pm$~0.13 & 0.97~$\pm$~0.07     & 0.12~$\pm$0.05 & 0.23~$\pm$~0.12     & 0.1              & 19.48~$\pm$~1.15 & 0.96~$\pm$~0.05 & 0.4~$\pm$~0.1                   \\
811                  & $r$      & -0.093 (109\%)    & 2.32~$\pm$~0.16 & 0.98~$\pm$~0.1     & 0.17~$\pm$~0.04 & 0.28~$\pm$~0.12    & 0.3              & 20.75~$\pm$~0.36 & 0.96~$\pm$~0.07 & 0.43~$\pm$~0.06                   \\
594                  & $r$      & -0.112 (111\%)   & 1.87~$\pm$~0.1 & 0.95~$\pm$~0.09     & 0.07~$\pm$~0.04 & 0.28~$\pm$~0.12     & 0.1            & 20.82~$\pm$~0.35 & 0.96~$\pm$~0.07 & 0.41~$\pm$~0.09                   \\

\bottomrule
\end{tabular*}%
\end{table*}


\subsubsection{Detailed examination of two high galactic latitude fields}\label{subsec:high-gal-lat}
Here, we consider $r$-band images with IDs 626 and 251 (called as `A' and `B' henceforth)\footnote{In the discussion of all metrics, the uncertainties in the measurements are excluded. The uncertainties in the properties of the original and deconvolved sources are on vastly different scales: the deconvolved sources have nearly negligible error bars due to the removal of the background level, which is otherwise an origin of uncertainty in brightness estimations.}.

{\bf One-to-one matches}: These matches are most apparent for fields with low crowding, as is the case with high galactic latitude fields. For A and B, the median values of $\Delta m$ are -0.121 and -0.061, respectively, as indicated by the dashed lines in the lower panel of Fig.~\ref{fig:mag-one-to-one-comparison}. This means that the cumulative flux of deconvolved sources is $\approx$112\% of the total flux of original sources for A and $\approx$106\% for B. The vertical dotted lines in the lower panels show that, on average, the original sources with magnitudes up to $m \approx 17.8$ mag for A and $m \approx 20.1$ mag for B are deconvolved with $\Delta m < 0.1$. However, the vertical dashed lines in the lower panels show that only the brightest original sources (with $m \lesssim 15.4$ mag) for A could be deconvolved with $\Delta m \lesssim 0.02$ ($\gtrsim98$\% flux conservation). This number goes as faint as $m \sim 19.6$ for B. This level of agreement in source-by-source magnitudes is noteworthy considering that SGP only ensures the combined flux of all sources is explicitly preserved rather than that of individual sources.

For A, a systematic trend in $\Delta m$ is evident, increasing towards fainter original sources, where the corresponding deconvolved sources tend to be brighter. This bias towards fainter sources may occur because the deconvolution process may prioritize bright sources over fainter sources. Another speculation is that we only select sources 2.5$\sigma$ level above the background to calculate the total flux of sources, which is used as a constraint in SGP. However, potential sources below that threshold still exist in the image whose deconvolution may have taken place. As a result, the flux used in the SGP constraint will generally be underestimated, and this inconsistency may contribute to the photometric bias. However, for B, this bias is less pronounced. This may be due to the presence of nine more bright ($m < 16$) sources in this field whose brightness has been overestimated by the deconvolution compared to A. As a result, this overestimation can be compensated for by a larger number of fainter deconvolved sources for the faint original sources in B compared to A\@.

The scatter in the residuals visibly increases for sources fainter than $m \sim 18-18.5$ mag. This indicates that preserving flux for these dimmer sources was challenging during deconvolution. This outcome is expected, as brighter sources tend to have better flux preservation due to their greater influence on the flux constraint used during deconvolution, which leads them to be prioritized.

The typical FWHM of the original sources for A and B, shown by the dashed vertical line in the upper panel of Fig.~\ref{fig:fwhm-one-to-one-comparison}, is approximately three pixels, which aligns with the seeing values mentioned in Table~\ref{tab:data-description}. The deconvolved FWHM values are smaller than the original FWHM values, as shown by the mainly negative values of $\Delta \mathrm{FWHM}$ in the lower panels. The median reduction in FWHM from the original to the deconvolved images is about two pixels, as shown by the horizontal dashed lines in the lower panels. In terms of ellipticity, Fig.~\ref{fig:ellipticity-one-to-one-comparison} shows that the deconvolved sources show a slight increase in median ellipticity, rising from 0.13 in the original image to 0.2 in the deconvolved image for A, and from 0.07 to 0.11 for B.

We also show the differences in the centroids detected from the original and corresponding deconvolved sources in Fig.~\ref{fig:centroidDiff-one-to-one-comparison}. The deconvolved sources are located close to the corresponding original source locations, with a median Euclidean distance of 0.2 pixels for A and 0.1 pixels for B, as indicated by the horizontal dashed lines. There are no sources with shifts greater than 1 pixel compared to the original source positions. For A, we observe a trend of decreasing centroid differences as the sources become fainter. This pattern may result from differences in how the flux is distributed across the pixels of the deconvolved sources for brighter and fainter sources. However, we skip a detailed investigation of this relation since the centroid differences are already very small and thus inconsequential to our study.

{\bf Unmatched (original)}: We now discuss the unmatched sources from the original image. As shown in Table~\ref{tab:crossmatching-results}, there are three unmatched sources in A and two unmatched sources in B. To analyze why these could not be matched with any deconvolved sources, we first check for the presence of detected deconvolved sources within a rectangular region of $\pm$2 pixels. The high value of this threshold is justified since these fields are not crowded with nearby sources, making it likely that corresponding deconvolved sources are being matched. We search for deconvolved sources using the unfiltered catalog instead of the filtered catalog.

Two out of the three unmatched sources in A, but none out of the two unmatched sources in B, could be associated with a detected deconvolved source using this modified crossmatching criterion; we visually confirmed that the corresponding deconvolved source was indeed being matched. The remaining unmatched source in A, which could not be crossmatched even after increasing the positional threshold, was visually identifiable but could not be detected by SExtractor because it was located close to a saturated source. As discussed in Sect.~\ref{fig:visual}, deconvolution can generate artifacts near extremely bright sources, which led SExtractor to exclude it from detection even though it was visually observed. A different set of detection parameters might be able to detect it.

Out of the two unmatched sources in B, one was also visually detectable but went undetected by SExtractor because it was close to a saturated source with a blooming artifact. The other source remaining in B had the following properties: $m \approx 20.18$, $\mathrm{FWHM} \approx 1.91$ pix, and $\mathrm{ellipticity} \approx 0.35$. It was not too close to the image boundary but was still visually undetectable in the deconvolved version of B. We found that the subdivision in which this source was located had much fewer sources detected than those visually perceivable. Furthermore, the subdivision contained an artifact due to an extremely bright source in the adjacent subdivision (blooming), which probably contributed to this under-detection in the original image\footnote{Such under-detection was not observed for other subdivisions where no such artifacts were visually present, suggesting that the blooming streak was likely responsible for the under-detection.}. Since the deconvolution is total-flux-constrained and we only consider the flux from sources that are detected by SExtractor to impose this constraint, the flux used in the constraint may be underestimated. However, the image still includes the sources whose flux did not contribute to the total. This phenomenon may have led to the suppression of many sources during the projection step of the SGP algorithm. Because the two original sources were faint, they may have been among those suppressed. This discussion illustrates an example where inconsistencies between the sources contributing to the total flux and those used to devise the constraint can cause the current implementation of deconvolution to fail.

\begin{figure*}[hbt!]
    \centering
      \includegraphics[keepaspectratio,width=0.32\linewidth]{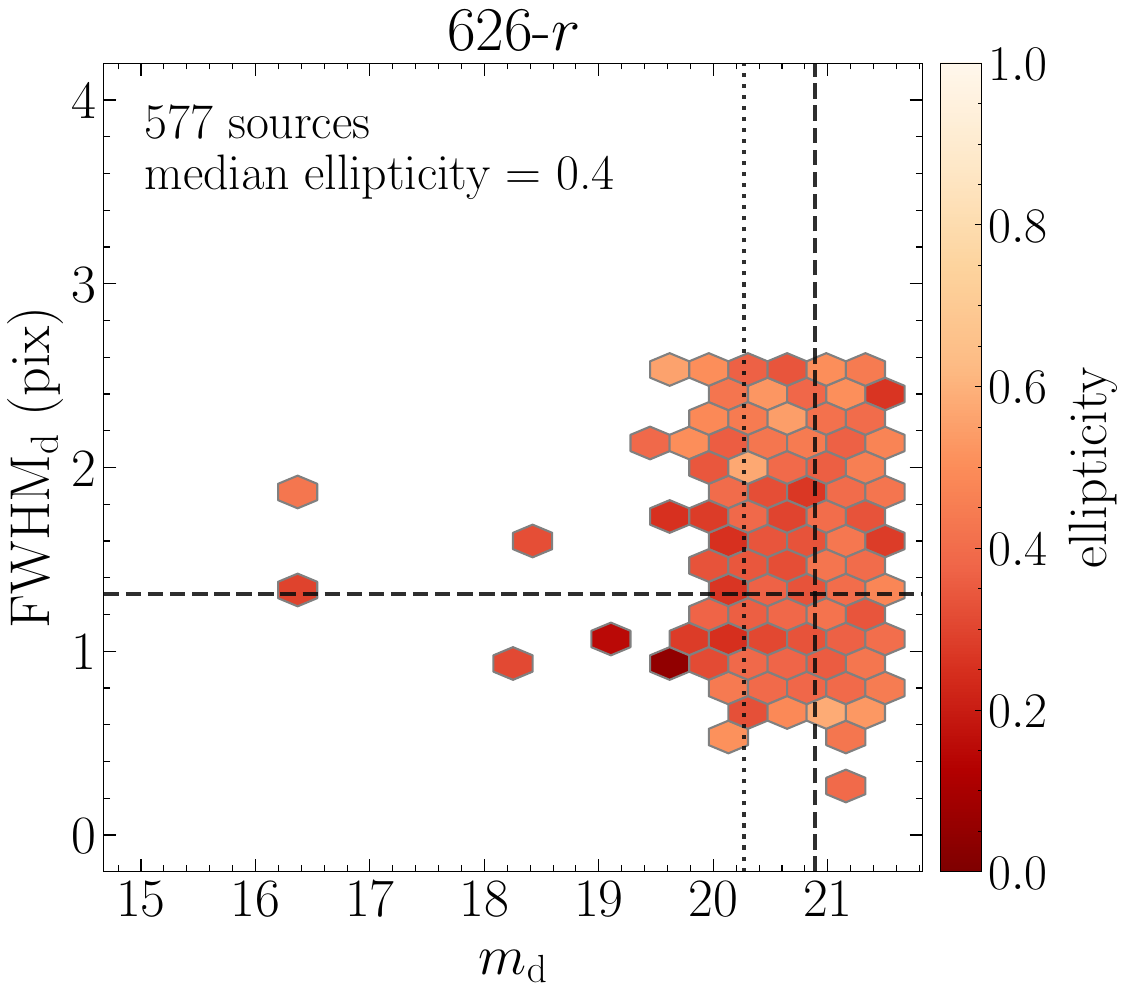}
      \includegraphics[keepaspectratio,width=0.32\linewidth]{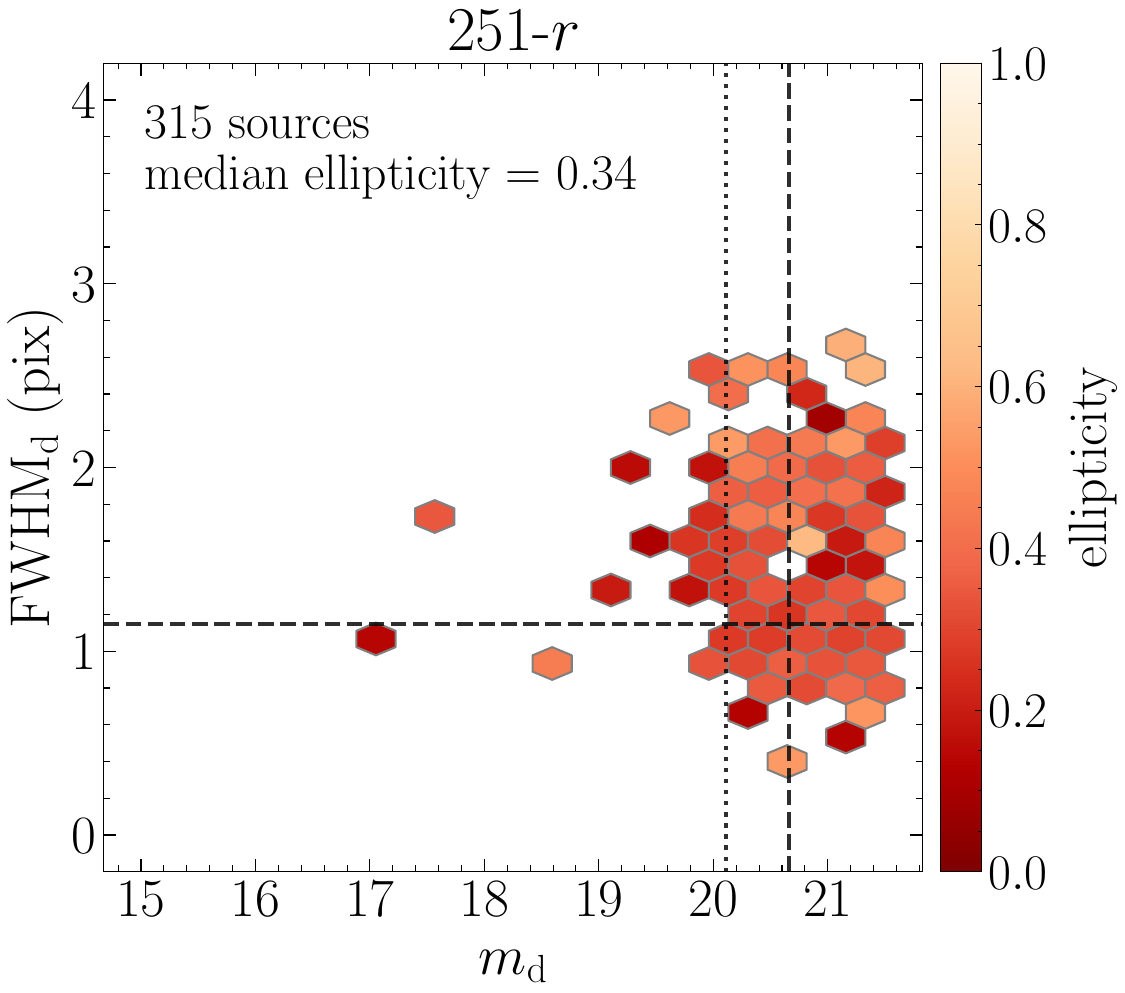}
      \includegraphics[keepaspectratio,width=0.32\linewidth]{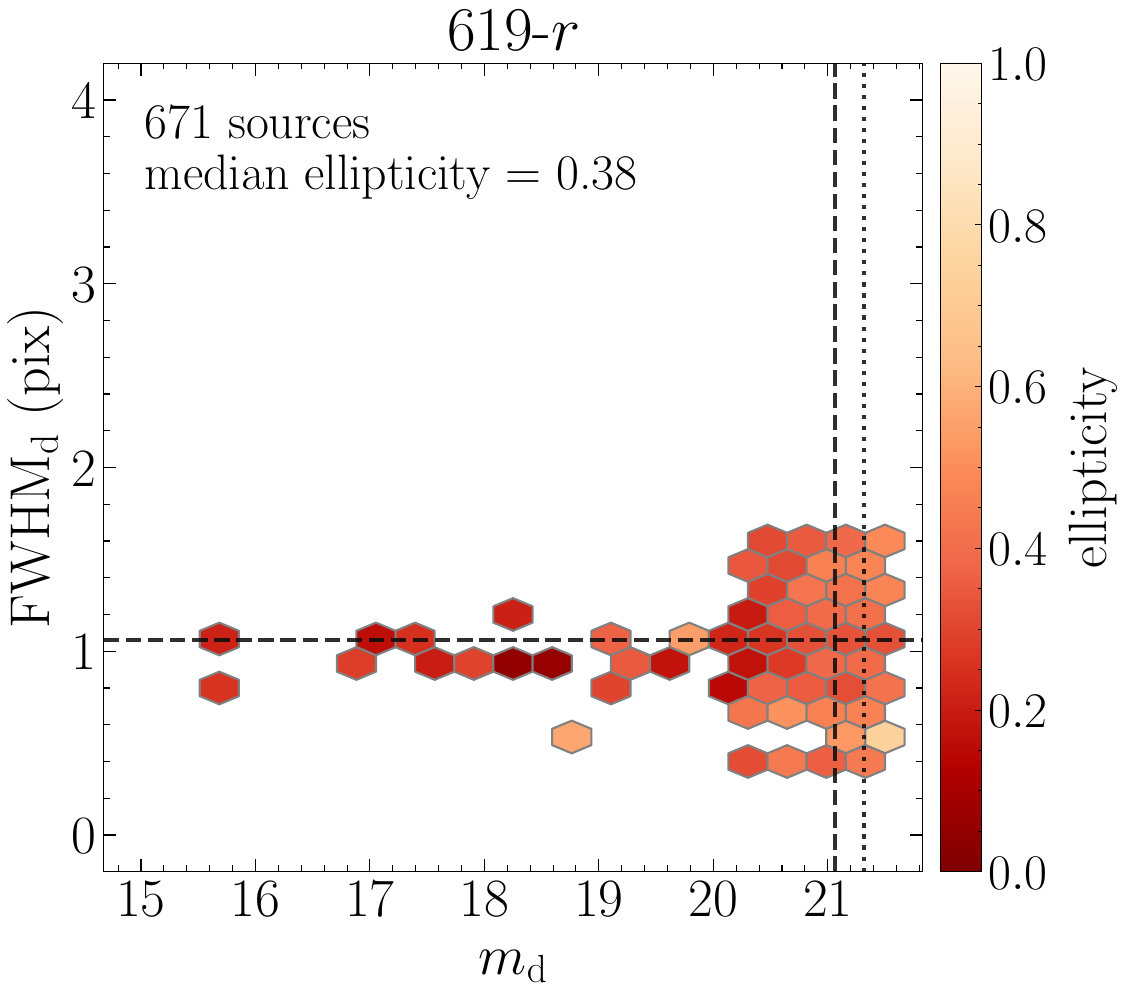}
      \includegraphics[keepaspectratio,width=0.32\linewidth]{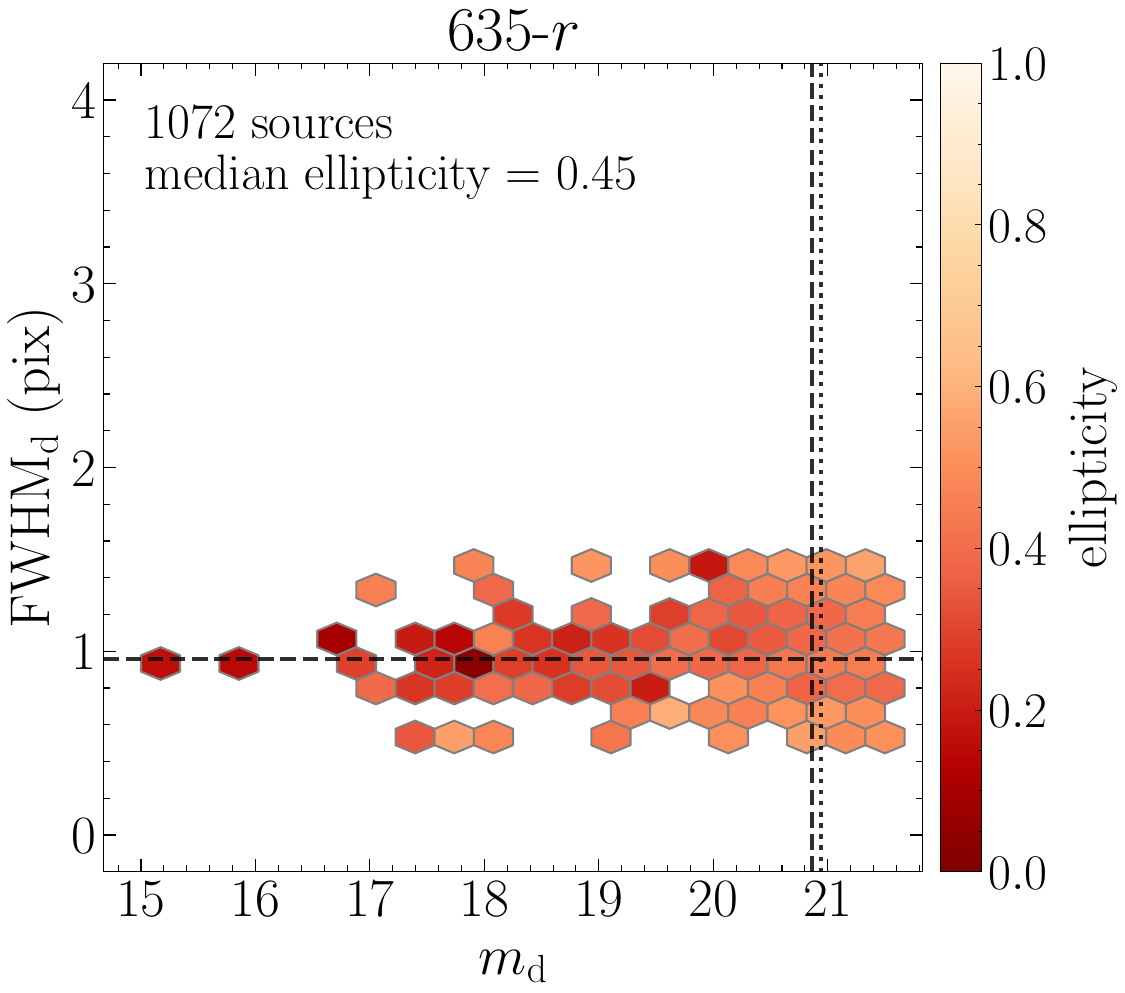}
      \includegraphics[keepaspectratio,width=0.32\linewidth]{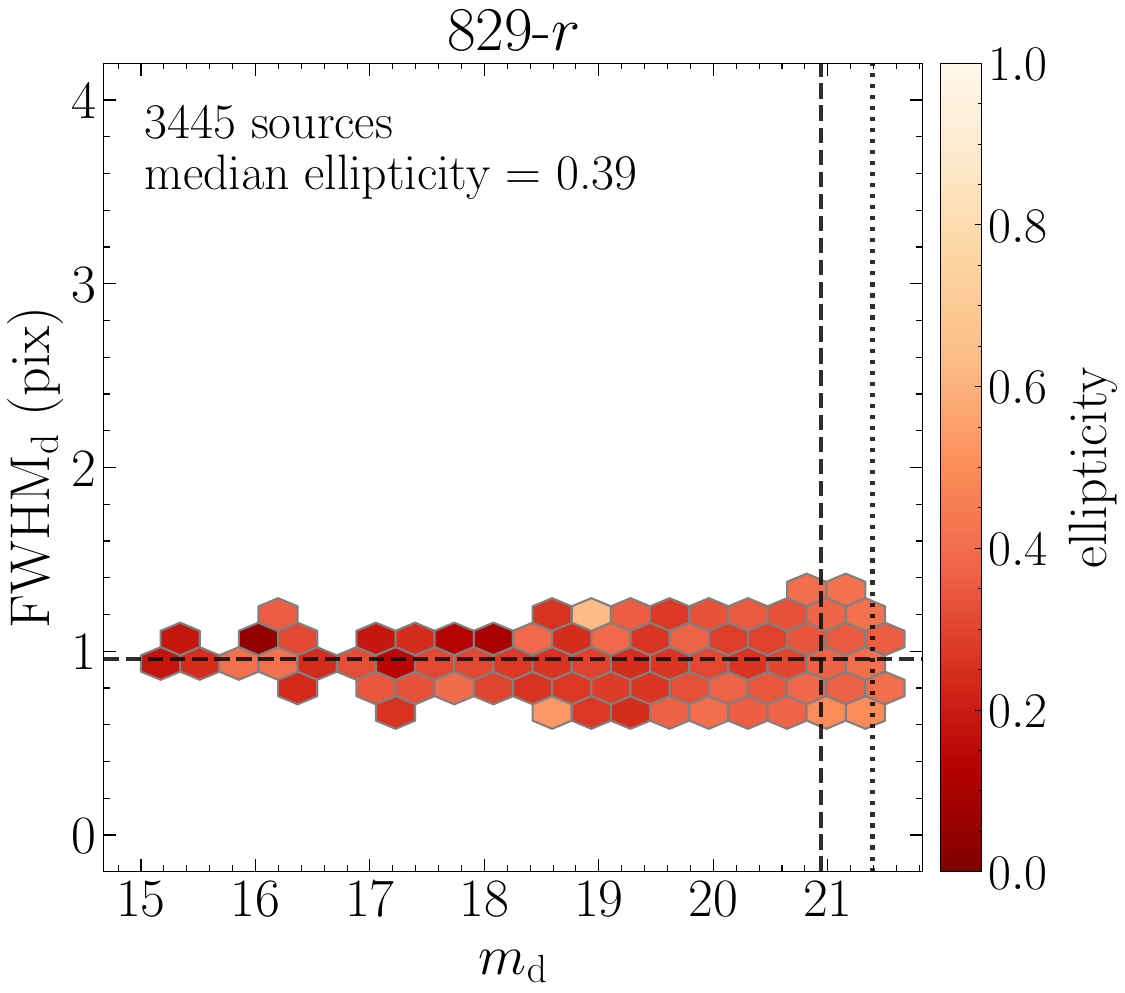}
      \includegraphics[keepaspectratio,width=0.32\linewidth]{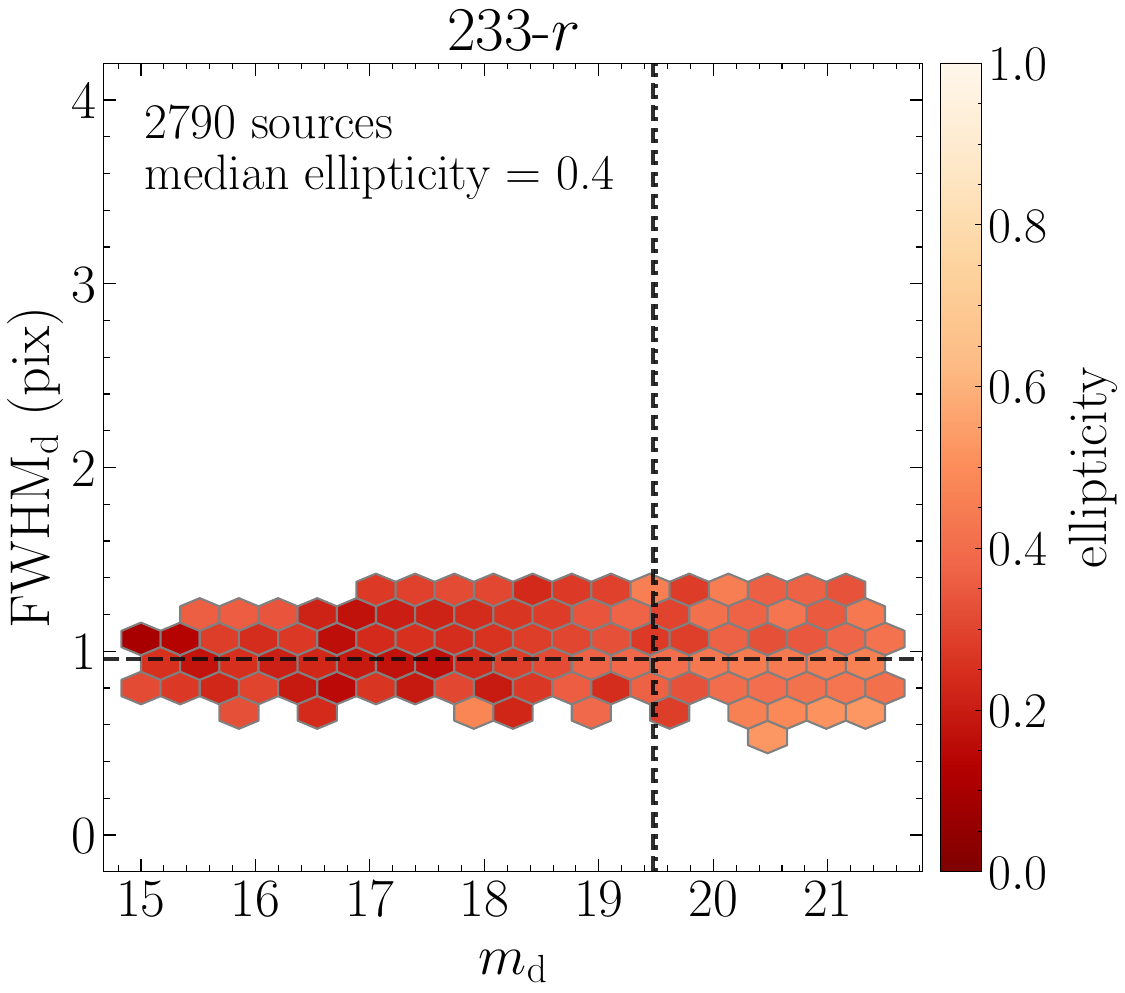}
      \includegraphics[keepaspectratio,width=0.32\linewidth]{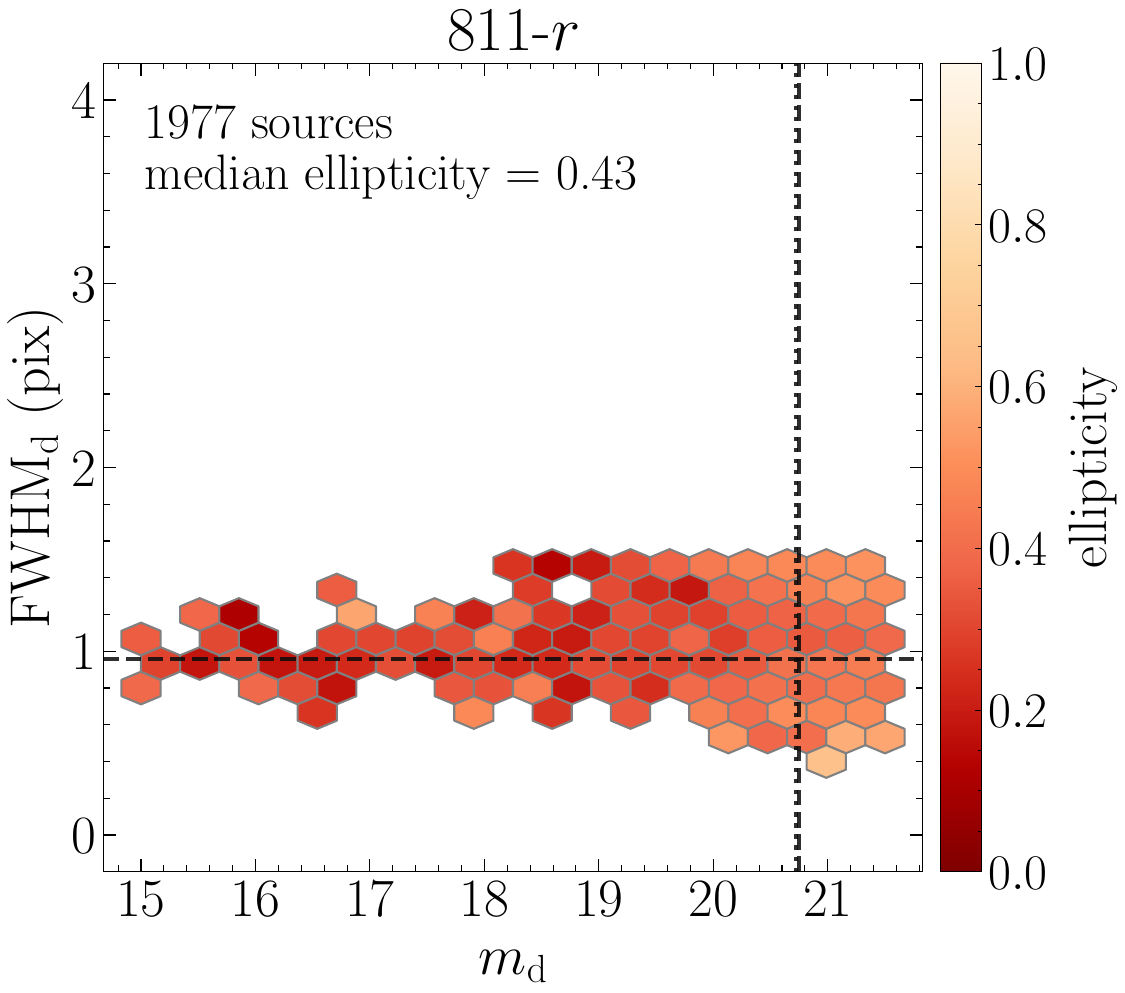}
      \includegraphics[keepaspectratio,width=0.32\linewidth]{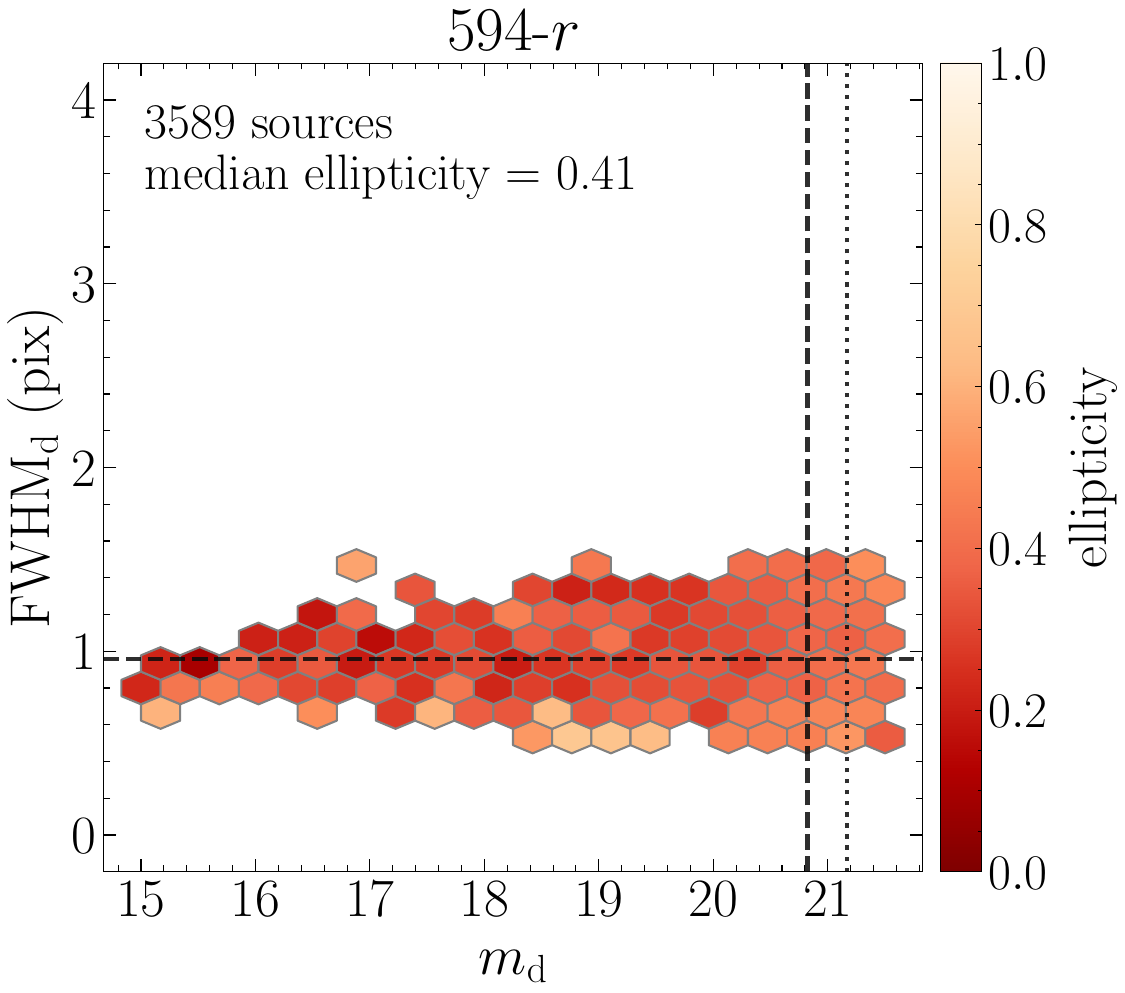}
    \caption{Distribution of the magnitude, FWHM, and ellipticity of the unmatched deconvolved sources (i.e., detected in the deconvolved image but not in the corresponding original image) that are likely astrophysical, for all the $r$ filter images considered in this study. The number of such sources is noted on the plots, and note that these values are less than the unmatched deconvolved crossmatch statistics from Table~\ref{tab:crossmatching-results} because further filtering, \reviewFour{based on additional selection criteria detailed in Sect.~\ref{subsec:crossmatch-results}}, was performed to reduce the chances of including spurious sources in our analysis. The magnitude is shown on the x-axis, the FWHM on the y-axis, and the color bar denotes the ellipticity. Each hexagon bin denotes the average ellipticity of all sources in that bin. The vertical and horizontal dashed lines denote the median magnitude and FWHM, respectively. The text on the plot notes the median ellipticity and is around 0.4 for all cases, which is slightly higher than the typical ellipticities of deconvolved sources from the one-to-one matches. There are weak trends suggesting that fainter deconvolved sources have higher ellipticities than those that are brighter. The vertical dotted line denotes the limiting magnitude of the corresponding science image and is plotted for reference. The title of each panel denotes the ZTF field ID. The median FWHM of these new sources is $\sim$1 pix, and the median magnitude ranges from $\sim$19.5 to 21, depending on the field, which is towards the faint end. Newly detected deconvolved sources fainter than the limiting magnitude of the science image are present, but there are also sources brighter than it. Of the latter category of sources, for each field, some (but fewer than half) are a result of deblending by deconvolution. We have also checked that only a handful to a few tens of additional unmatched deconvolved sources were successfully crossmatched with a source from the unfiltered original catalog (without imposing the selection criteria from Sect.~\ref{subsec:experimental-details}). It is possible that the remaining unmatched deconvolved sources with no match in the original might be because the systematics of the detection procedure may have made the original source undetectable. We recall that the ellipticities may be slightly overestimated due to skipping the filtering step during source detection in deconvolved images, which otherwise could make sources slightly rounder. We have identified that a few of these new deconvolved sources are a result of the deblending of original sources that could not be deblended by SExtractor (see Sect.~\ref{sec:deblending-examples}).} \label{fig:unmatched-deconvolved-combined}
\end{figure*}

{\bf Unmatched (deconvolved)}: Out of the 744 and 362 unmatched deconvolved sources for A and B, respectively, 577 sources for A and 315 sources for B meet the additional astrophysical cuts outlined in Sect.~\ref{subsec:crossmatch-results}. These sources for A and B pass several selection criteria related to magnitude, FWHM, ellipticity, and detection flags, making them likely astrophysical sources rather than dubious detections. Fig.~\ref{fig:unmatched-deconvolved-combined} shows the distribution of the magnitudes, FWHM, and ellipticities of these newly identified deconvolved sources. These sources are faint, with a median magnitude of 20.89 for A and 20.66 for B, are compact, with a median FWHM of 1.31 pixels for A and 1.15 pixels for B, and are moderately elliptical, with median ellipticities of 0.4 for A and 0.34 for B.

Approximately 83\% of these deconvolved sources in A and 88\% in B have magnitudes within the range of $m_{\mathrm{lim}} < m \leq 21.5$. Here, $m_{\mathrm{lim}}$ represents the limiting magnitude of the corresponding original image (see Table~\ref{tab:data-description}), and 21.5 is the faintest magnitude possible for a deconvolved source allowed by our selection criteria outlined in Sect.~\ref{subsec:experimental-details}). This indicates that deconvolution is able to recover many faint sources that were undetectable in the original image.

{\bf Many-to-one/One-to-many matches}: For both fields, we obtain no cases of many-to-one and one-to-many crossmatches.

\subsubsection{Intermediate galactic latitude field}\label{subsec:intermediate-gal-lat}

Here, we consider $r$ band images with IDs 619 and 635 (called `C' and `D' henceforth).

{\bf One-to-one matches}: For C and D, the median values of $\Delta m$ are -0.132 and -0.041, respectively, as indicated by the dashed lines in the lower panel of Fig.~\ref{fig:mag-one-to-one-comparison}. This means that the cumulative flux of deconvolved sources is $\approx$113\% of the total flux of original sources for C and $\approx$104\% for D. The vertical dotted lines in the lower panels show that, on average, the original sources with magnitudes up to $m \approx 18.4$ for C and $m \approx 20.4$ for D are deconvolved with $\Delta m < 0.1$. However, the vertical dashed lines in the lower panels show that only the brightest original sources (with $m \lesssim 16.1$) for C could be deconvolved with $\Delta m \lesssim 0.02$ ($\gtrsim98$\% flux conservation). This number goes as faint as $m \sim 19.6$ for D. The level of agreement in the source-by-source magnitudes of C and D is about a few tenths of magnitude better than the previously discussed high-galactic latitude fields A and B, respectively.

As discussed for A in Sect.~\ref{subsec:high-gal-lat}, a systematic trend in $\Delta m$ can also be observed for C, which increases toward fainter original sources and enlarges rapidly after $m \approx 18$, where the corresponding deconvolved sources tend to be brighter. Similar to the reasons deduced for A, this may be because of how deconvolution prioritizes sources of different magnitudes and inconsistencies in flux values used as constraints in the deconvolution algorithm. However, for D, this bias is much lower, similar to that seen in B. This could also be attributed to similar reasons derived for B, where the overestimation of the brightness of many bright sources is balanced by a larger number of fainter deconvolved sources in D compared to C\@.

The scatter in the residuals increases for sources fainter than $m \sim 18.5-19$ for C and D. This trend is qualitatively similar to that observed in A and B in Sect.~\ref{subsec:high-gal-lat} but occurs at slightly fainter magnitudes than A and B. This indicates that the flux preservation remains stable up to fainter sources in C and D. 

The typical FWHM of the original sources for C and D, shown by the dashed vertical line in the upper panel of Fig.~\ref{fig:fwhm-one-to-one-comparison}, are approximately 2.3 pixels and 1.7 pixels, respectively. This aligns with the seeing values mentioned in Table~\ref{tab:data-description}. The median reduction in FWHM for the deconvolved sources, compared to the original, is approximately 1.3 pixels for C and 0.75 pixels for D, as shown by the horizontal dashed lines in the lower panels. The reductions in FWHM for C and D are smaller than those observed for A and B in Sect.~\ref{subsec:high-gal-lat}. This is likely because the typical FWHM (or seeing) for sources in C and D is lower than that for A and B. Also, the median ellipticities of the sources have increased from 0.13 in the original image to 0.24 in the deconvolved image for C and from 0.11 to 0.31 for D. This increase in ellipticity is greater than what was obtained in the previous section (Sect.~\ref{subsec:high-gal-lat}).

The deconvolved sources are located close to the corresponding original source locations, with a median Euclidean distance of 0.2 pixels for C and 0.1 pixels for D, as shown by the horizontal dashed lines. There are no sources exhibiting shifts greater than one pixel from their original positions. A trend similar to that in A is observed in C, where the differences in centroids decrease as the sources become fainter. However, as discussed in Sect.~\ref{subsec:high-gal-lat}, these sub-pixel shifts are less relevant to our study.

{\bf Unmatched (original)}: There are no unmatched sources for C and seven unmatched sources for D, as shown in Table~\ref{tab:crossmatching-results}. We follow the same procedure used in Sect.~\ref{subsec:high-gal-lat} to analyze these unmatched sources. All seven original sources in D were deblended, as verified by investigating the flags. The deconvolved sources were visually detectable at the same locations in the deconvolved image, but SExtractor could not detect them. When we explicitly searched for detected deconvolved sources in a region of $\pm$7 pixels around the original source locations in the deconvolved image, overplotting their elliptical contours indicated that blended versions of the original sources were detected in the deconvolved image. We hypothesize that this phenomenon results from the number of deblending thresholds used during the deblending process (the number of deblending sub-thresholds, the \texttt{DEBLEND\_NTHRESH} parameter, is set to 4; see Table~\ref{tab:sextractorParams}). As noted in Sect.~\ref{sec:visualization}, the deconvolved sources typically exhibit a higher dynamic range; the peak-flux pixel value is higher, and the profile declines more steeply to zero than the original sources. As a result, SExtractor could not deblend the deconvolved sources, even though it successfully deblended the original sources. We have internally confirmed that four of the seven visible but undetected deconvolved sources could now be deblended by increasing \texttt{DEBLEND\_NTHRESH} to 32 for the deconvolved image, and they were also successfully crossmatched with the corresponding original sources. However, three of them still remain undeblended, even after using more stringent deblending parameters (\texttt{DEBLEND\_NTHRESH} = 64 and \texttt{DEBLEND\_MINCONT} = 0.0003).

{\bf Unmatched (deconvolved)}: Out of the 1277 and 1508 unmatched deconvolved sources for C and D, respectively, 671 sources for C and 1072 sources for D meet the additional astrophysical cuts outlined in Sect.~\ref{subsec:crossmatch-results}. As mentioned previously in Sect.~\ref{subsec:high-gal-lat}, these sources are likely astrophysical sources rather than dubious detections as they pass our several selection criteria. These sources are faint, with a median magnitude of 21.06 for C and 20.86 for D, are compact, with a median FWHM of 1.06 pixels for C and 0.96 pixels for D, and are moderately elliptical, with median ellipticities of 0.38 for C and 0.45 for D. 

Only about 20\% of these deconvolved sources in A and 44\% in B have magnitudes within the range of $m_{\mathrm{lim}} < m \leq 21.5$. This fraction is much smaller than observed in A and B in Sect.~\ref{subsec:high-gal-lat}. The fewer new faint deconvolved sources may partly be because C and D have a fainter limiting magnitude. The visual trends observed in the properties of these deconvolved sources are different from those found for A and B in Sect.~\ref{subsec:high-gal-lat}. Specifically, the deconvolved sources in C and D show smaller variations in FWHM around the median and greater variation in magnitude than A and B (see Fig.~\ref{fig:unmatched-deconvolved-combined}).

{\bf Many-to-one/One-to-many matches}: We select a subset of the many-to-one matches for C and D to identify potential (though unconfirmed) deblending cases, which are detailed in Appendix~\ref{appn:many-to-one}. These examples are distinct from those discussed in Sect.~\ref{sec:deblending-examples}. We found only one such case for each of C and D. No one-to-many crossmatches were found.

\subsubsection{Low galactic latitude field}\label{subsubsec:low-gal}

Here, we consider $r$ band images with IDs 829 and 233 (called `E' and `F' henceforth).

{\bf One-to-one matches}: For E and F, the median values of $\Delta m$ are -0.082 and -0.03, respectively, as shown by the dashed lines in the lower panel of Fig.~\ref{fig:mag-one-to-one-comparison}. This means that the cumulative flux of deconvolved sources is $\approx$108\% of the total flux of original sources for E and $\approx$103\% for F. The vertical dotted lines in the lower panels show that, on average, the original sources with magnitudes up to $m \approx 19.6$ for E and $m \approx 19$ for F are deconvolved with $\Delta m < 0.1$. However, the vertical dashed lines in the lower panels show that only the brightest original sources (with $m \lesssim 16.4$) for E could be deconvolved with $\Delta m \lesssim 0.02$ ($\gtrsim98$\% flux conservation). This number goes up to $m \sim 18.4$ for F. The level of agreement in the source-by-source magnitudes of E is about a few tenths of a magnitude better than that of C. However, the agreement for F is about a magnitude worse than D.

A systematic trend is observed for E, similar to the negative bias observed for A and C in the previous sections, indicating that as the original sources become fainter, the corresponding deconvolved sources tend to be brighter. However, this bias for E is slightly smaller than for A and C. However, for F, the bias towards fainter original sources is positive, as seen for B and D, but this bias is larger than in all previous cases (A--E).

The scatter in the residuals for E remains relatively constant up to $m \approx 19.6$, after which it begins to increase for fainter sources. However, for F, the scatter increases drastically after $m \approx 18$. Therefore, the flux preservation for E is more stable than in previous cases (A--D), while this stability for F is lower than that observed in cases A--E.

The typical FWHM of the original sources for E and F, shown by the dashed vertical line in the upper panel of Fig.~\ref{fig:fwhm-one-to-one-comparison}, are approximately 1.9 pixels and 2.7 pixels, respectively. This is consistent with the seeing values presented in Table~\ref{tab:data-description}. The median reduction in FWHM for the deconvolved sources compared to the original sources is approximately 1 pixel for E and 1.7 pixels for F, as shown by the horizontal dashed lines in the lower panels. The reductions in FWHM for E and F are smaller than those observed for A and B but slightly larger than those for C and D. A consistent pattern has emerged across all cases thus far: as the typical seeing of the image decreases, the reduction in the median FWHM due to deconvolution also decreases. This trend further reassures that the deconvolved FWHM values are similar across all images, irrespective of the original FWHM. The median ellipticities of the sources have increased from 0.11 in the original image to 0.25 in the deconvolved image for E and 0.12 to 0.23 for F. This increase is somewhat similar to the results observed for C and D.

The deconvolved sources are located close to the corresponding original source locations, with a median Euclidean distance of 0.1 pixels for E and F, as shown by the horizontal dashed lines. For E, two of the deconvolved sources are shifted by more than one pixel from the positions of the corresponding original sources. We have verified that the location of the pixel with the peak flux is the same for both the original and deconvolved sources in these cases, and all have $\mathrm{FLAGS} = 0$. This suggests that this large shift is mainly because deconvolution redistributes the flux among the surrounding pixels in a way that alters the overall centroid position. However, for F, none of the deconvolved sources exhibit shifts greater than one pixel.

{\bf Unmatched (original)}: There are 16 unmatched sources for E and six unmatched sources for F, as shown in Table~\ref{tab:crossmatching-results}. Out of the 16 sources in E, 12 were located near a saturated source exhibiting a blooming artifact. Similar to one of the unmatched sources in A (see Sect.~\ref{subsec:high-gal-lat}), all corresponding deconvolved sources for these 12 sources were visually detectable but could not be detected by SExtractor. Three of the remaining four unmatched sources were successfully deblended by SExtractor, while one had a $\mathrm{FLAGS} = 0$, indicating that it was a good detection and not deblended. Following the procedure used in Sect.~\ref{subsec:intermediate-gal-lat}, we found that setting \texttt{DEBLEND\_NTHRESH} to 32 for the deconvolved image helped deblend all four deconvolved sources corresponding to these four original sources. This also enabled successful crossmatching with the corresponding original sources. Also, the unblended source with $\mathrm{FLAGS} = 0$ could be detected in the deconvolved due to the refined deblending parameters. The original image contained an extended, fuzzy emission around this source, so the deblending on the deconvolved image may have separated the fuzzy structure and the deconvolved source. Out of the six unmatched original sources in F, one source was not located in a close vicinity but was still near a saturated source. By using more stringent deblending parameters (\texttt{DEBLEND\_NTHRESH} = 64 and \texttt{DEBLEND\_MINCONT} = 0.0003) for the deconvolved image, we were able to detect and crossmatch the corresponding deconvolved source with this original source. The other five sources were deblended by SExtractor. Four of these five corresponding deconvolved sources, which were visible but undetected previously, were now detectable using \texttt{DEBLEND\_NTHRESH} = 32 and were successfully crossmatched. However, the remaining source was visually observed to be deblended but remained un-deblended by SExtractor even after using the more stringent deblending parameters mentioned above.

{\bf Unmatched (deconvolved)}: Out of the 5466 and 3605 unmatched deconvolved sources for E and F, respectively, 3445 sources for E and 2790 sources for F meet the additional astrophysical cuts outlined in Sect.~\ref{subsec:crossmatch-results}. These sources are likely astrophysical sources rather than dubious detections, as they pass our various selection criteria. These sources are faint, with a median magnitude of 20.94 for E and 19.48 for F, are compact, with a median FWHM of 0.96 pixels for both E and F, and are moderately elliptical, with median ellipticities of 0.39 for E and 0.4 for F.

Only about 9\% of these deconvolved sources in E and 50\% in F have magnitudes within the range of $m_{\mathrm{lim}} < m \leq 21.5$. The small fraction in E is due to its faint limiting magnitude, which is close to 21.5. The visual trends observed in the properties of these deconvolved sources differ from those found in cases A--D. Specifically, the deconvolved sources in E and F show smaller variations in FWHM around the median than C and D (and also A and B) and greater variation in magnitude than C and D (as well as A and B). In particular, in E and F, we observe more new deconvolved sources at the brightest end than in C and D, whereas A and B had hardly any sources at the brightest end (see Fig.~\ref{fig:unmatched-deconvolved-combined}).

{\bf Many-to-one/One-to-many matches}: We select a subset of the many-to-one matches for E and F to identify potential (though unconfirmed) deblending cases, which are detailed in Appendix~\ref{appn:many-to-one}. These examples are distinct from those discussed in Sect.~\ref{sec:deblending-examples}. We found seven such cases in E and three in F. No one-to-many crossmatches were found.

\subsubsection{Field with a dwarf galaxy}\label{subsec:dwarf-galaxy}

Here, we consider the $r$ band low-galactic latitude image with ID 811 (called `G' henceforth). This field contains NGC1569, a dwarf irregular galaxy. We conduct a brief analysis of the deconvolution of this galaxy after summarizing the match and unmatch statistics of all detected sources, as done in the previous sections.

{\bf One-to-one matches}: The median value of $\Delta m$ is -0.093, as shown by the dashed lines in the lower panel of Fig.~\ref{fig:mag-one-to-one-comparison}. This means that the cumulative flux of the deconvolved sources is $\approx$109\% of the total flux of the original sources. The vertical dotted line in the lower panel shows that, on average, original sources with magnitudes up to $m \approx 18.9$ are deconvolved with $\Delta m < 0.1$. However, the vertical dashed line indicates that only the brightest original sources (with $m \lesssim 15.8$) could be deconvolved with $\Delta m \lesssim 0.02$ ($\gtrsim98$\% flux conservation). A negative systematic trend is observed for fainter original sources, similar to trends observed in A, C, and E, where the corresponding deconvolved sources tend to be brighter. The scatter in the residuals remains mostly constant up to $m \approx 19$, after which it increases for fainter sources.

The typical FWHM of the original sources is approximately 2.3 pixels, consistent with the seeing values mentioned in Table~\ref{tab:data-description}. The median reduction in FWHM for the deconvolved sources compared to the original sources is approximately 1.3 pixels, as shown by the horizontal dashed lines in the lower panels. The median ellipticities of the sources have increased from 0.17 in the original image to 0.28 in the deconvolved image. 

The deconvolved sources are located close to the corresponding original source locations, with a median Euclidean distance of 0.3 pixels, as shown by the horizontal dashed lines. There are no sources with shifts greater than 1 pixel compared to the original source positions.

{\bf Unmatched (original)}: There are eight original sources that were unmatched. Seven of these were located near a saturated source that displayed blooming artifacts. The remaining original source had $\mathrm{FLAGS} = 0$ and was present near the far outskirts of NGC1569. A deconvolved source was visibly present in the corresponding location of the deconvolved image, but SExtractor could not detect it. The reason for this absence remains unclear.

{\bf Unmatched (deconvolved)}: Out of the 2755 unmatched deconvolved sources, 1977 sources meet the additional astrophysical cuts outlined in Sect.~\ref{subsec:crossmatch-results}. This means that these sources are likely astrophysical rather than dubious detections as they pass our several selection criteria. These sources are faint, with a median magnitude of 20.75, are compact, with a median FWHM of 0.96 pixels, and are moderately elliptical, with a median ellipticity of 0.43.

Approximately 52\% of these deconvolved sources have magnitudes within the range of $m_{\mathrm{lim}} < m \leq 21.5$. The spread in the FWHM around the median value and the magnitude of these deconvolved sources are visually similar to case D from Sect.~\ref{subsec:intermediate-gal-lat}.

{\bf Many-to-one/One-to-many matches}: We select a subset of the many-to-one matches for this image to identify potential (though unconfirmed) deblending cases, which are detailed in Appendix~\ref{appn:many-to-one}. These examples are distinct from those discussed in Sect.~\ref{sec:deblending-examples}. Seven such cases were found. No one-to-many crossmatches were found.

\begin{figure*}[hbt!]
    \centering
    \includegraphics[keepaspectratio,width=0.8\linewidth]{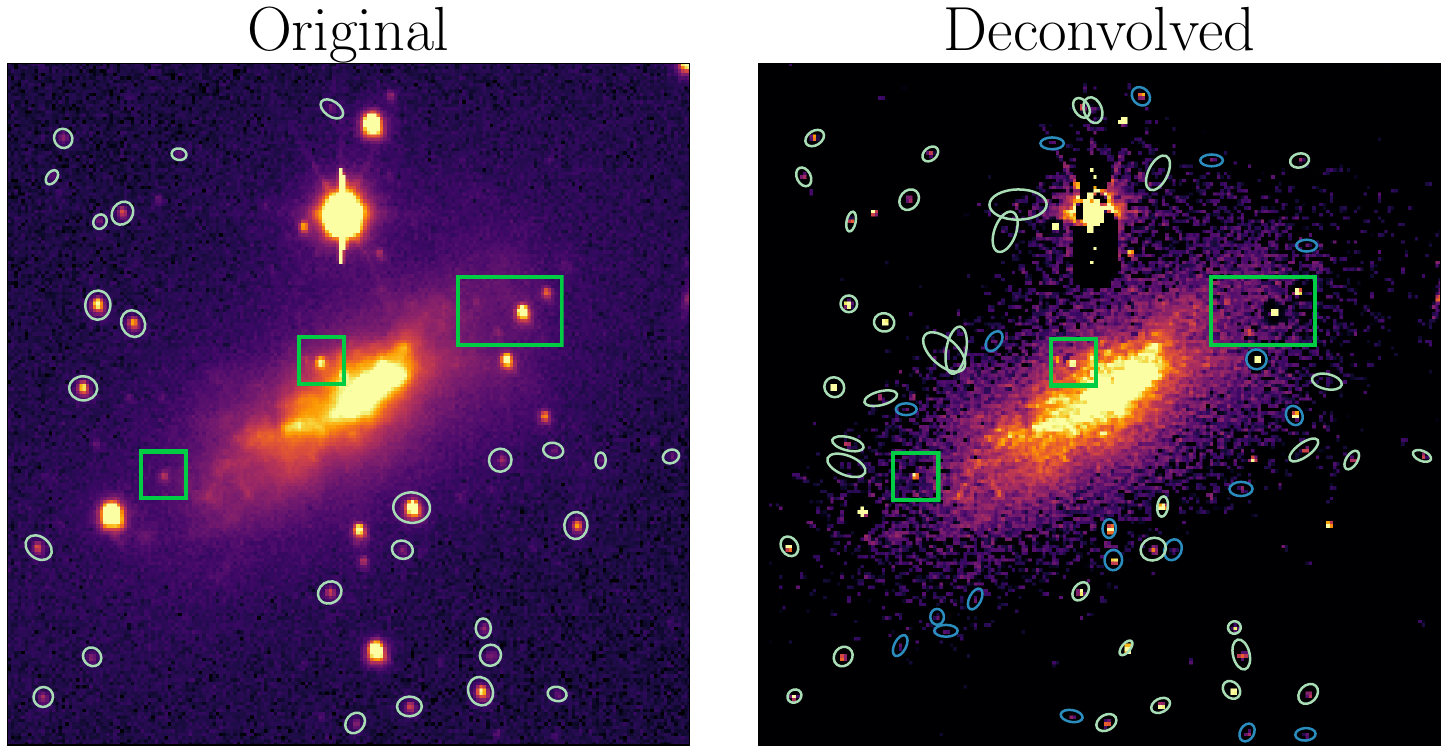}
    \caption{Zoomed cutout of size $200 \times 200$ pixels centered on the galaxy NGC 1569 with coordinates $(l, b) \approx (143.68, 11.24)$ deg extracted from the field ID 811. The whitish ellipses denote detected sources matching the selection criteria described in Sect.~\ref{subsec:experimental-details}, whereas the blue ellipses in the deconvolved image denote unmatched deconvolved sources (i.e., present in the deconvolved image but not the original image). Only deconvolved sources with ellipticity $\leq 0.6$ are marked. All original sources are detected in the deconvolved image except some whose deconvolved counterparts are still visible; such deconvolved sources are either not detected, detected but removed by the selection criteria, or detected but have ellipticity $> 0.6$. The deconvolution highlights a few other sources undetected in the original image, as shown by the blue ellipses: many of these visually seem to be true detections, but some can be visually understood as noise. The green squares show a few examples where the original source was made visually more apparent in the deconvolved, but neither set of sources was detected.} \label{fig:ngc1569-zoomed}
\end{figure*}

{\bf Visualization of deconvolution}: Although the deconvolution of galaxies is not the focus of this study, we present the deconvolution results for NGC1569 located in this field, along with nearby sources, to demonstrate the broad applicability of the deconvolution method. Fig.~\ref{fig:ngc1569-zoomed} shows a zoomed-in cutout centered on the galaxy, with detected sources marked by ellipses. Deconvolution reveals several faint sources at the outskirts of the galaxy. It also makes some sources visually more apparent (indicated by green squares), although it is possible that these could also be detected in the original image with different detection parameters. It is important to note that the SGP deconvolution method used in this paper estimates a non-negative minimizer of the Kullback-Leibler divergence (see Sect.~\ref{sec:deconv} for more mathematical details and also Sect.~\ref{subsec:sgp}). As a result, the solutions are sparse objects consisting of bright spots against black backgrounds. This could explain why the regions surrounding the deconvolved galaxy nucleus are `spotty' in structure.

\subsubsection{Globular cluster}\label{subsubsec:gc}

Here, we consider the $r$ band low-to-intermediate galactic latitude image with ID 594. This field contains NGC7006, a globular cluster located in the outer regions of the Milky Way. We conduct a brief analysis of the deconvolution of this globular cluster after summarizing the match and unmatch statistics of all detected sources, as done in the previous sections.

{\bf One-to-one matches}: The median value of $\Delta m$ is -0.112, as shown by the dashed lines in the lower panel of Fig.~\ref{fig:mag-one-to-one-comparison}. This means that the cumulative flux of the deconvolved sources is $\approx$111\% of the total flux of the original sources. The vertical dotted lines in the lower panels show that, on average, original sources with magnitudes up to $m \approx 18.5$ are deconvolved with $\Delta m < 0.1$. However, the vertical dashed line indicates that only the brightest original sources (with $m \lesssim 15.8$) could be deconvolved with $\Delta m \lesssim 0.02$ ($\gtrsim98$\% flux conservation). A negative systematic trend is observed for the fainter original sources, similar to trends observed in A, C, E, and G, where the corresponding deconvolved sources tend to be brighter. The scatter in the residuals remains mostly constant up to $m \approx 19.2$, after which it increases for fainter sources.

The typical FWHM of original sources is approximately 1.87 pixels, consistent with the seeing values mentioned in Table~\ref{tab:data-description}. The median reduction in FWHM for the deconvolved sources compared to the original sources is approximately 0.9 pixels, as shown by the horizontal dashed lines in the lower panels. The median ellipticities of the sources have increased from 0.07 in the original image to 0.28 in the deconvolved image. 

The deconvolved sources are located close to the corresponding original source locations, with a median Euclidean distance of 0.1 pixels, as shown by the horizontal dashed lines. One deconvolved source was shifted by just over one pixel, but similar to case E in Sect.~\ref{subsubsec:low-gal}, we verified that the peak flux pixel is the same for the original and deconvolved source, and both have $\mathrm{FLAGS} = 0$. Therefore, this shift is likely because of how deconvolution redistributes the flux among the surrounding few pixels.

{\bf Unmatched (original)}: There are 18 unmatched original sources, six of which were located near a saturated (or nearly saturated) source exhibiting blooming artifacts. For these 18 sources, we visually identified a deconvolved source in the corresponding location in the deconvolved image, but SExtractor could not detect it. Among the remaining 12 original sources, we found that using \texttt{DEBLEND\_NTHRESH} = 32 for the deconvolved image helped deblend eight corresponding deconvolved sources. These sources were also successfully crossmatched. However, the remaining four sources could not be deblended even when using stricter deblending parameters (\texttt{DEBLEND\_NTHRESH} = 64 and \texttt{DEBLEND\_MINCONT} = 0.0003).

{\bf Unmatched (deconvolved)}: Out of the 5503 unmatched deconvolved sources, 3589 sources meet the additional astrophysical cuts outlined in Sect.~\ref{subsec:crossmatch-results}. This means that these sources are likely astrophysical rather than dubious detections as they pass our several selection criteria. These sources are faint, with a median magnitude of 20.82, are compact, with a median FWHM of 0.96 pixels, and are moderately elliptical, with a median ellipticity of 0.41.

Only approximately 23\% of these deconvolved sources have magnitudes within the range of $m_{\mathrm{lim}} < m \leq 21.5$. The spread in the FWHM around the median value and the magnitude of these deconvolved sources are visually similar to case G from Sect.~\ref{subsec:dwarf-galaxy}.

{\bf Many-to-one/One-to-many matches}: We select a subset of the many-to-one matches for this field to identify potential (though unconfirmed) deblending cases, which are detailed in Appendix~\ref{appn:many-to-one}. These examples are distinct from those discussed in Sect.~\ref{sec:deblending-examples}. Nine such cases were found. No one-to-many crossmatches were found.

\begin{figure*}[hbt!]
    \centering
    \includegraphics[keepaspectratio,width=0.9\linewidth]{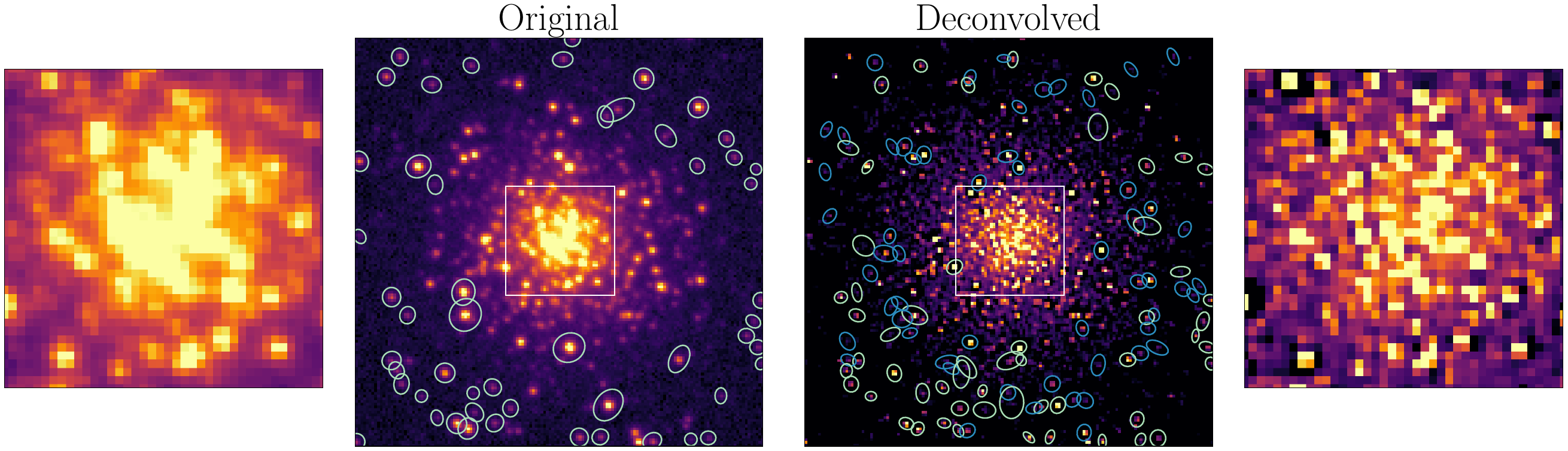}
    \caption{Cutout of size $150 \times 150$ pixels centered on the globular cluster NGC 7006 with $(l, b) \approx (63.77, -19.4)$ deg, extracted from the field ID 594. The whitish ellipses denote detected sources matching the selection criteria described in Sect.~\ref{subsec:experimental-details}, whereas the blue ellipses in the deconvolved image denote unmatched deconvolved sources (i.e., present in the deconvolved image but not the original image). Only deconvolved sources with ellipticity $\leq 0.5$ are marked. All original sources are detected in the deconvolved image except some whose deconvolved counterparts are still visible; such deconvolved sources are either not detected, detected but removed by the selection criteria, or detected but have ellipticity $> 0.5$. The deconvolution highlights a few other sources undetected in the original image, as shown by the blue ellipses: many of these visually seem to be true detections, but a minor fraction can be visually understood as noise. The leftmost and the rightmost insets show the zoomed regions of the core. Although no sources in this region are detected by SExtractor in either the original or the deconvolved (partly because our detection parameters may not be suited for extremely crowded fields), the deconvolved core reveals its granular structure, which was otherwise mostly a continuum emission in the original image. It is possible that deconvolved sources in the core may be detectable using detection parameters more suited for dense crowding in globular clusters.} \label{fig:ngc7006-zoomed}
\end{figure*}

{\bf Visualization of deconvolution}: We visualize the deconvolution result of the globular cluster NGC7006 located in this field. Fig.~\ref{fig:ngc7006-zoomed} shows a zoomed-in cutout centered on the globular cluster, with detected sources marked by ellipses. The deconvolution reveals several sources at the outskirts of the cluster and some near the outskirts of its core, highlighted by blue ellipses. Some sources are more visually apparent and might be identified using different detection parameters than those used for the original image. Additionally, the core of the globular cluster is much better resolved in the deconvolved image than in the original image.

\subsection{Deblending examples}\label{sec:deblending-examples}

In all the $r$-band fields discussed in Sect.~\ref{sec:field-specific-results}, we identified several examples of deblending where the original source was split into multiple deconvolved sources. We selected deblending cases from unmatched deconvolved catalogs in which a deconvolved source was found without a corresponding original source. For this section, we do not use the additional selection criteria for unmatched deconvolved sources described in Sect.~\ref{subsec:crossmatch-results}. Instead, we used simpler selection criteria: we chose only deconvolved sources with an ellipticity $< 0.6$ and a FWHM that is not too close to zero. For simplicity, we only consider cases where the original source was split into exactly two deconvolved sources, as these cases are more common and easier to analyze. Additionally, we excluded original blended sources that had $\mathrm{FLAGS} > 7$, those with the deblending flag set, and those with an ellipticity $< 0.1$. This exclusion is performed because it is unlikely for two deblended sources to produce a small-ellipticity source unless they are extremely close to each other.

To identify blended original sources that have been deblended in the deconvolved image, we first require that an original source (from the unfiltered catalog, rather than the filtered one, to also allow for sources with $\mathrm{FWHM} \gtrsim 4$ pixels) be located within a rectangular region of $\pm$5 pixels around a given unmatched deconvolved source. Additionally, we need another deconvolved source (from the unfiltered catalog) be present in the same region. If this condition is satisfied, we then apply the blending identification criterion outlined by \citet{Dawson2016}. We define $d_{ij}$ as the separation between the deconvolved source $i$ and the original source $j$. $\sigma_i$ and $\sigma_j$ represent the sizes of the deconvolved source convolved with a Gaussian reflecting the seeing conditions of the original image and the size of the original source, respectively. For the purposes of our analysis, we approximate $\sigma_i \approx \sigma_j$ since the convolution of the deconvolved source should ideally resemble the original source. We also choose $k$, the normalizing scale factor, to be one, as used in Dawson et al. If $d_{\mathrm{eff}_{ij}} = \dfrac{d_{ij}}{k (\sigma_i + \sigma_j)} < 1$, then source $j$ may be classified as {\it ambiguously} blended. According to \citet{Dawson2016}, {\it ambiguous} blends refer to situations where two original sources are blended to such an extent that they are detected as a single source. The authors also define {\it conspicuous} blends as cases where two original sources significantly overlap but are detected as individual sources. It is important to note that since we use SExtractor's deblending during the source detection process for both the original and deconvolved images, we are essentially searching for {\it ambiguous} blends that have been converted to {\it conspicuous} blends due to deconvolution. Depending on whether the deblending flag is set for the deconvolved source, one can determine whether the `conspicuousness' is attributable to the deconvolution or to SExtractor's deblending.

\begin{figure*}
    \centering
      \includegraphics[keepaspectratio,width=0.32\linewidth]{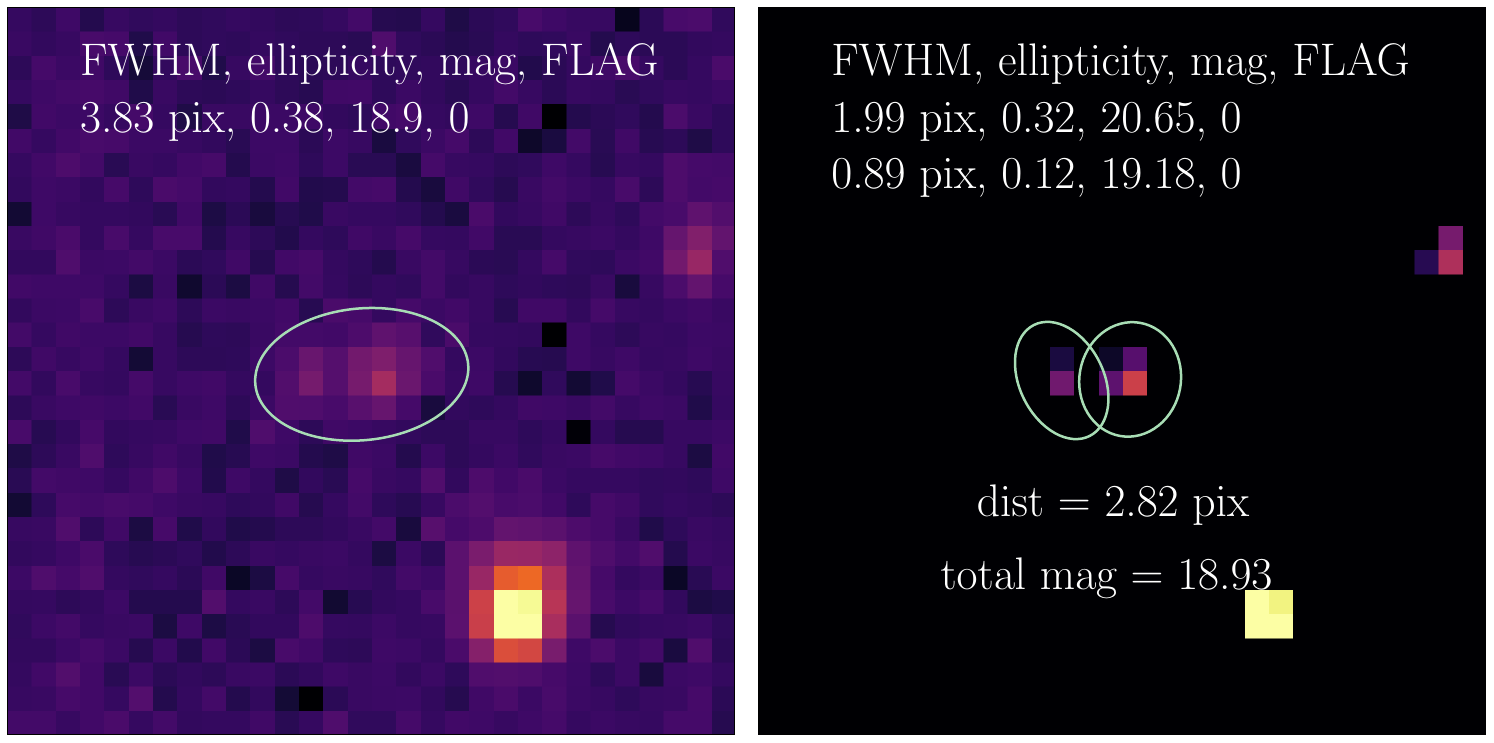}
      \includegraphics[keepaspectratio,width=0.32\linewidth]{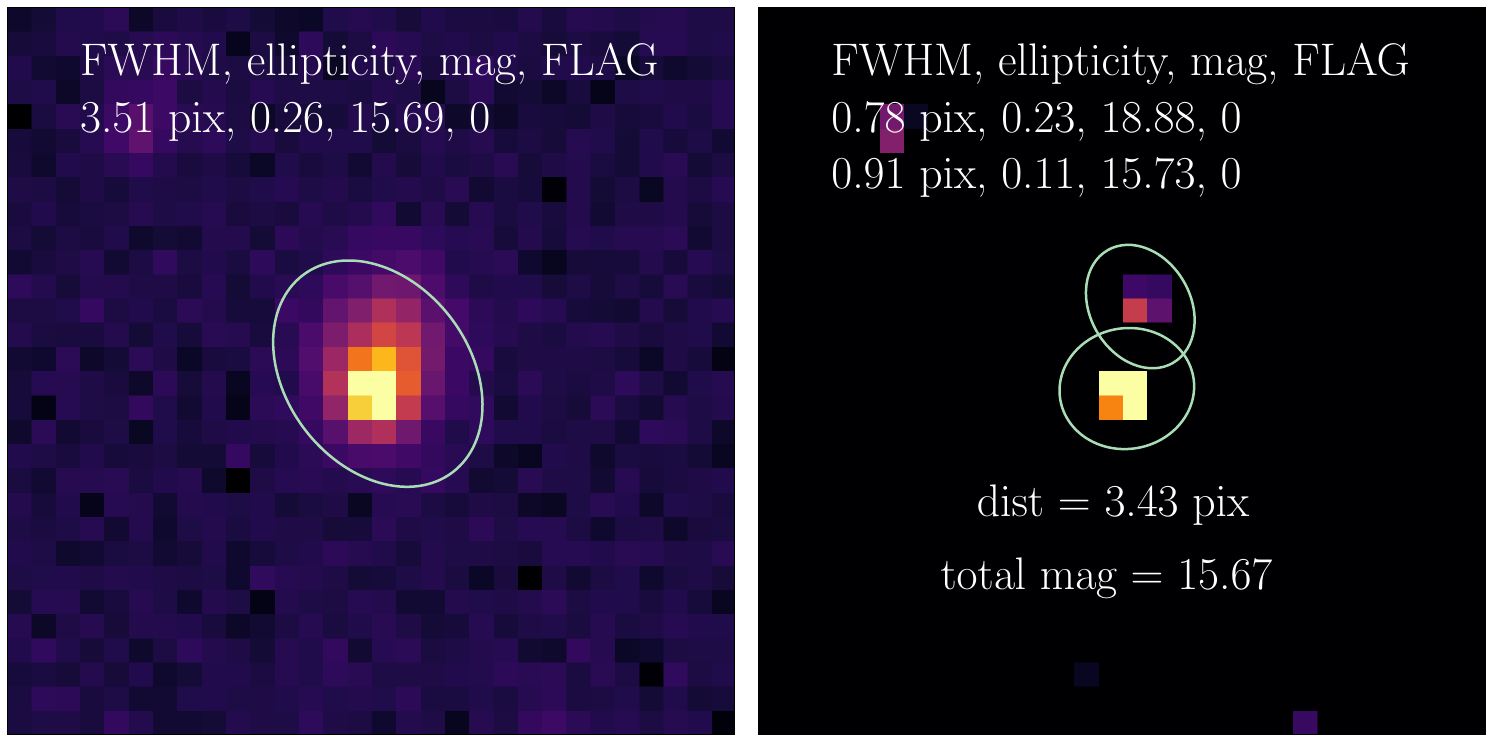}
      \includegraphics[keepaspectratio,width=0.32\linewidth]{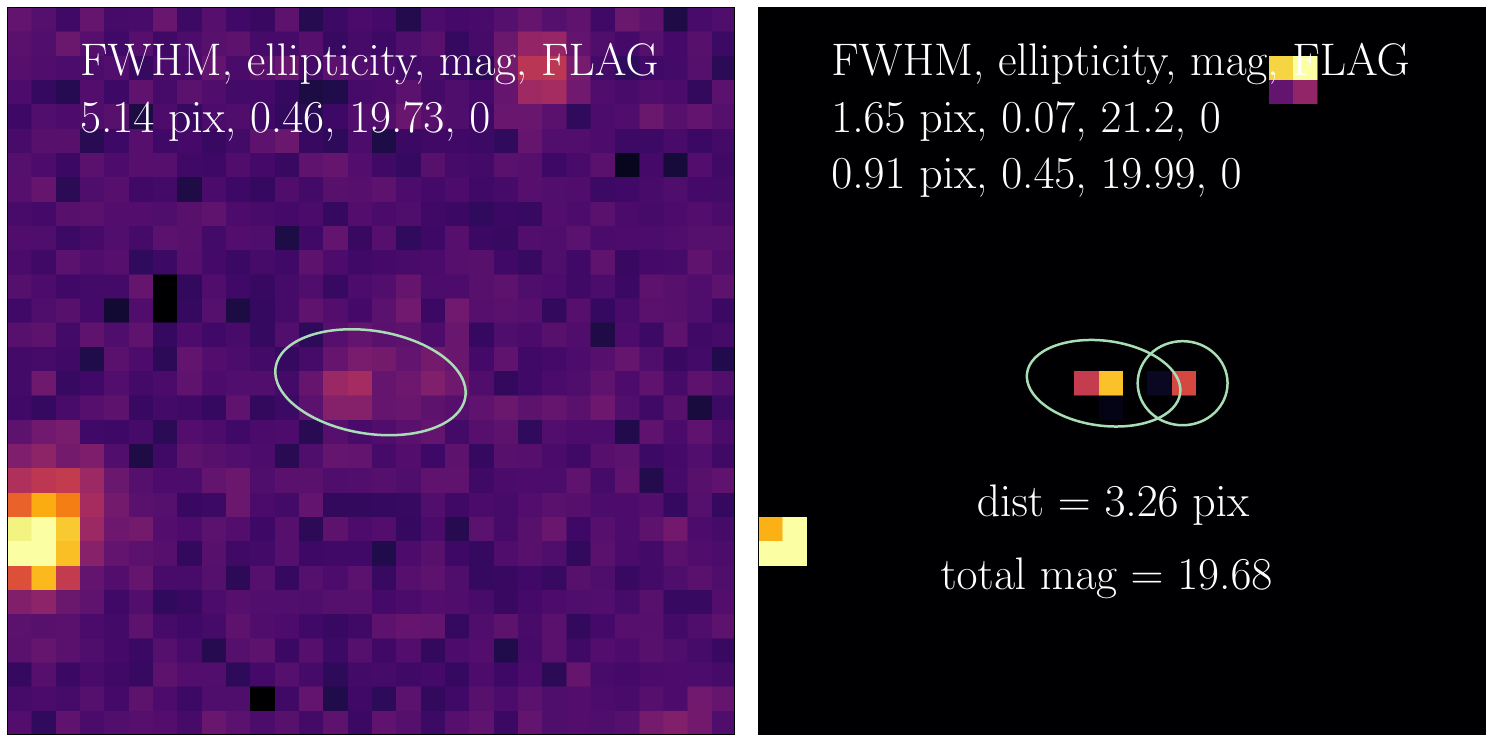}\hfill
      \includegraphics[keepaspectratio,width=0.32\linewidth]{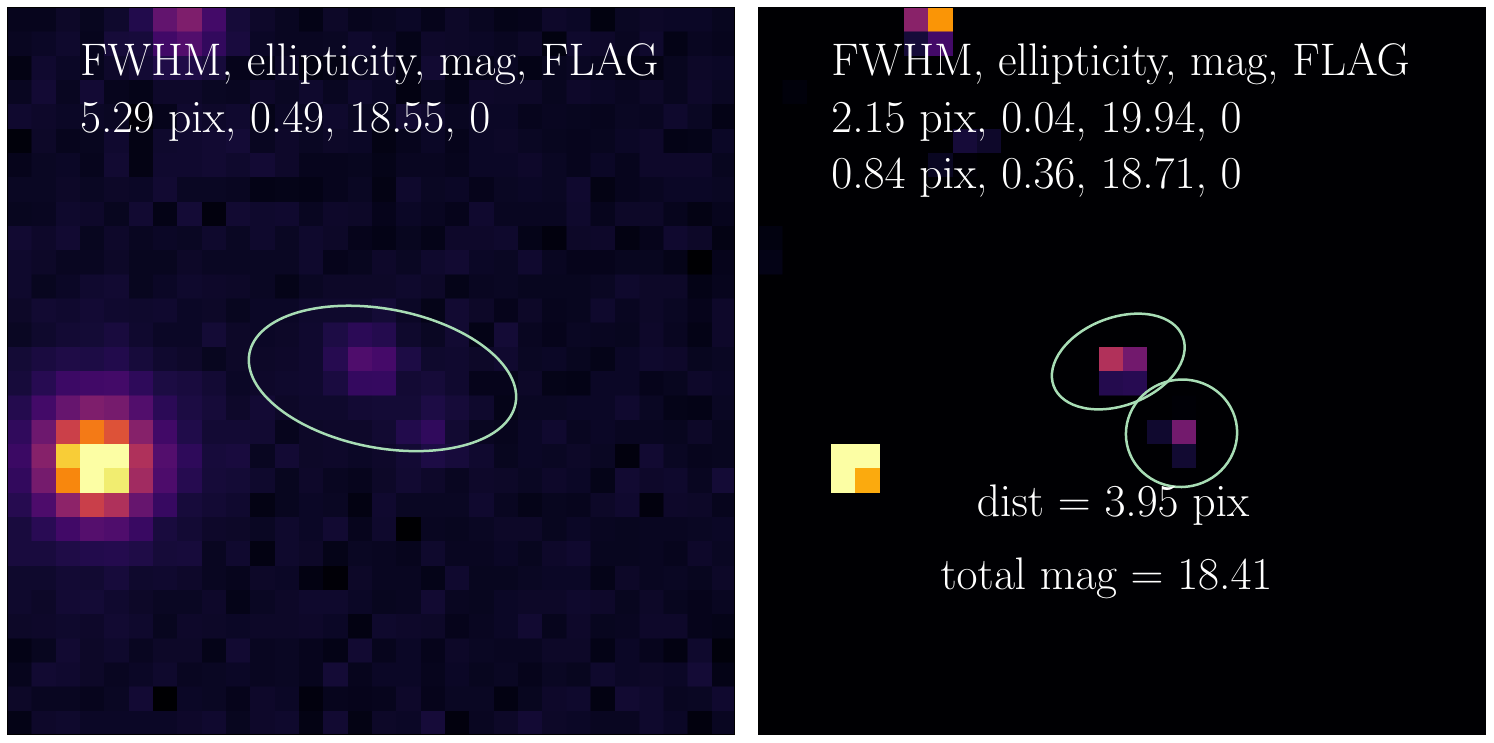}
      \includegraphics[keepaspectratio,width=0.32\linewidth]{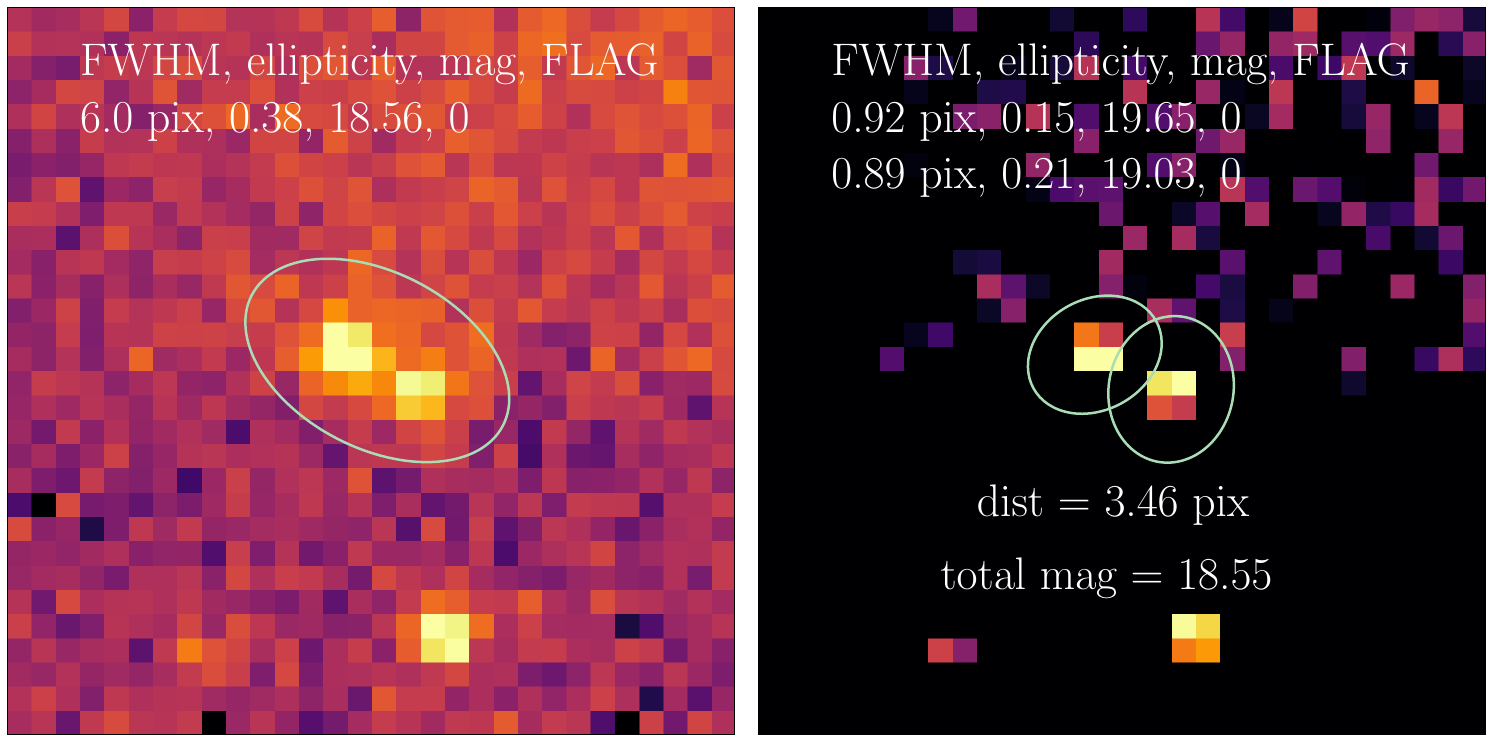}
      \includegraphics[keepaspectratio,width=0.32\linewidth]{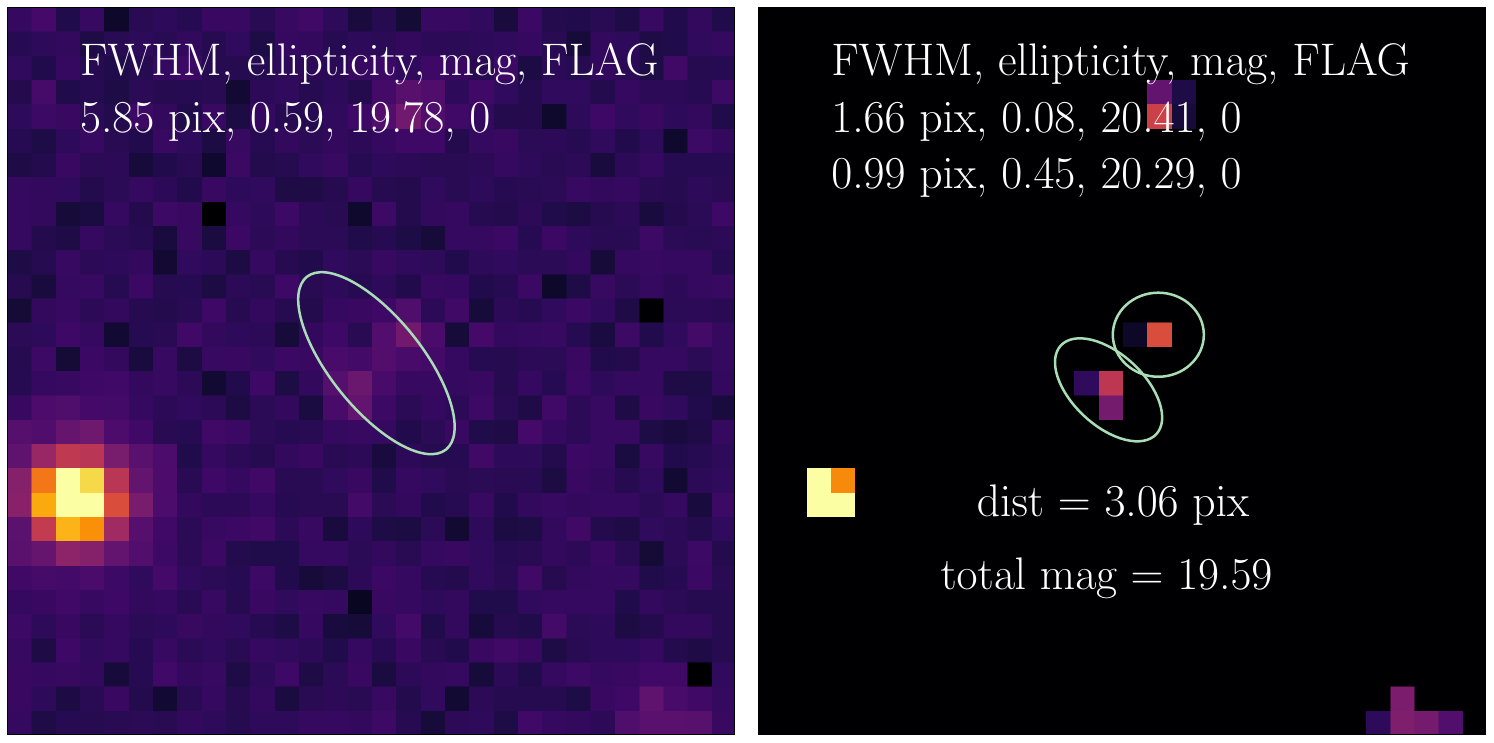}\hfill
    \caption{Few examples of identified {\it ambiguous} blends (two original sources overlapping to an extent being detected as a single source) converted to {\it conspicuous} blends (two sources overlapping but detected as two distinct sources) by the deconvolution, as identified by the employed selection criterion. These examples are across all $r$-band images used in this study. In all examples, the original source is on the left, and the two deconvolved sources are shown on the right, marked by ellipses. The FWHM, ellipticity, magnitude, and SExtractor flags of all sources are shown on top of each image. The FLAGS of all original sources are zero, indicating they are good detections. The distance between the two deconvolved sources and their combined magnitude is also shown on the deconvolved image. Images are shown using a combination of square root stretching and clipping pixel values beyond the central 99.5 percentile. All deconvolved sources have $\mathrm{FLAGS} = 0$ (except in one case where $\mathrm{FLAGS} = 3$, which means the deblending flag is set), suggesting that the deconvolution deblends those sources rather than SExtractor. The SExtractor parameters used are described in Table~\ref{tab:sextractorParams}. Several cases can be seen: deblending into similar brightness sources; deblending into sources with $\sim$16 times flux ratio ($\Delta m \approx 3$); deblending into two smaller ellipticity sources from a larger ellipticity source; deblending into two sources where one of them having higher ellipticity than the original blended source. The distance between the deblended deconvolved sources shown is the Euclidean distance and typically ranges from slightly under three pixels to slightly above five pixels (the pixel size is 1$\arcsec$.012). More examples are shown in Appendix~\ref{appn:more-deblending-examples}. Possible deblends separated by $\approx$1-1.5 pixels that pass our selection criteria are discussed in Appendix~\ref{appn:many-to-one}.} \label{fig:deblend-examples}
\end{figure*}

Using these deblend selection criteria, we found from fewer than five to a few hundred original blended sources depending on the specific field. However, we note that our strict selection criteria are intended to increase the purity of the deblend sample rather than its completeness. Fig.~\ref{fig:deblend-examples} presents several examples of deblending scenarios gathered from all the images considered in this study, with extended visualizations shown in Appendix~\ref{appn:more-deblending-examples}. Typically, the distance between the deconvolved sources is about 3-5 pixels. However, we have not encountered any cases where the deconvolved sources are separated by fewer than two pixels or so that also meet our deblend selection criteria using the unmatched deconvolved source catalog. Potential deblends that are separated by less than two pixels are found using the many-to-one matches and discussed in Appendix~\ref{appn:many-to-one}.

These examples demonstrate that deconvolution is able to deblend an original source into deconvolved sources with similar brightness. It is also capable of deblending when the deconvolved sources differ in brightness by $\sim$6-16 times ($\Delta m =$ 2-3). The combined magnitudes of the two deblended deconvolved sources agree well with the original magnitude, with magnitude differences being $\lesssim$0.1. The FWHM of the deblended deconvolved sources is generally $\lesssim$1-2 pixels, which is much smaller than that of the original blended source. Although it is more common for the deblended deconvolved sources to exhibit ellipticity less than the blended original source, in about $1/3^{\mathrm{rd}}$ of those shown in Fig.~\ref{fig:deblend-examples}, one of the deblended deconvolved sources has a greater ellipticity than the original source. This occurrence is rare because the deblended deconvolved sources are not too close to each other to lead to a lower ellipticity original counterpart. Nonetheless, we include these cases because the ellipticities of the deconvolved sources might be slightly overestimated and because the ellipticities for these sources are not tightly constrained (see Fig.~\ref{fig:ellipticity-one-to-one-comparison}).

These findings suggest that deconvolution, by reversing the effects of the PSF, is naturally suitable for deblending. However, we recognize that some of these original sources deblended by deconvolution could also be deblended by SExtractor if more sensitive deblending parameters are used.

\begin{figure*}
    \centering
      \includegraphics[keepaspectratio,width=0.32\linewidth]{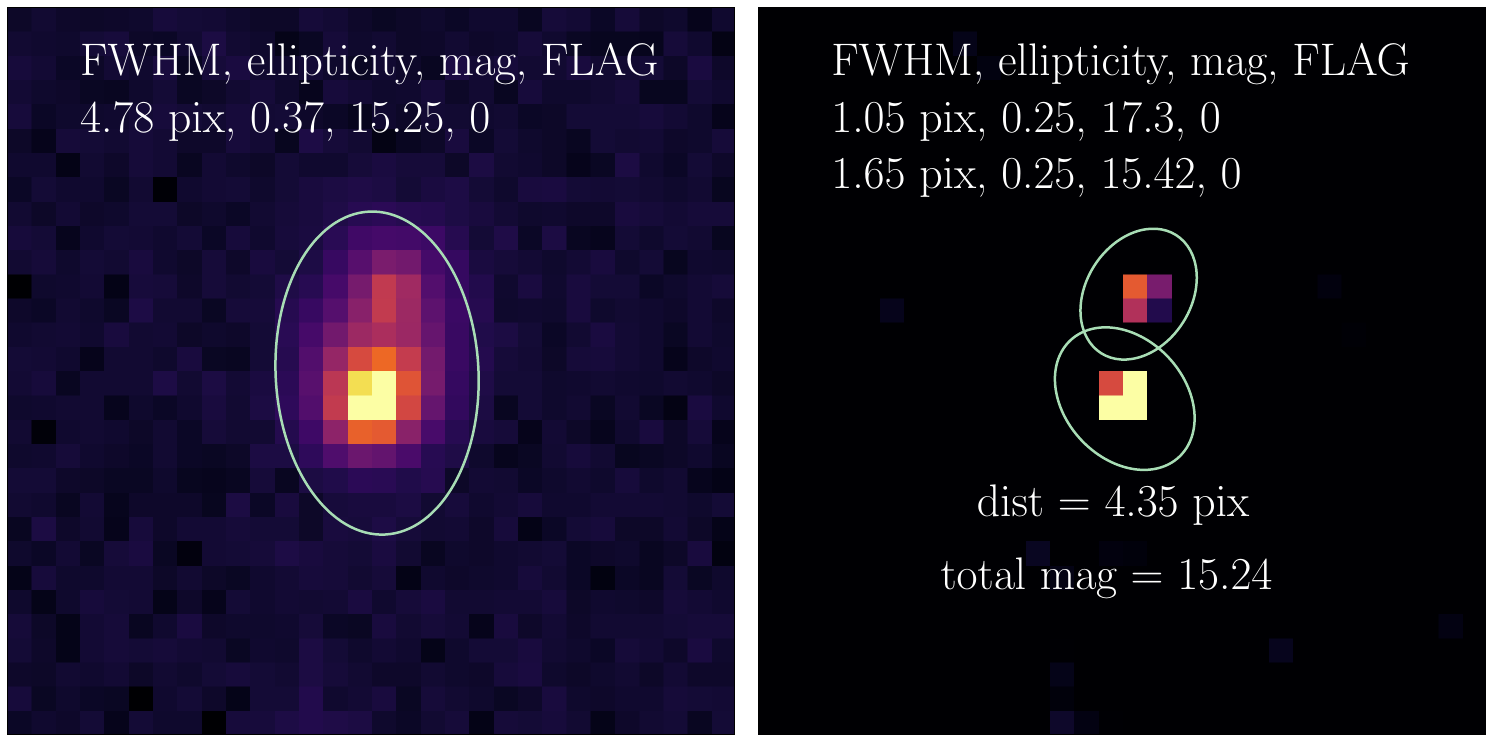}
      \includegraphics[keepaspectratio,width=0.32\linewidth]{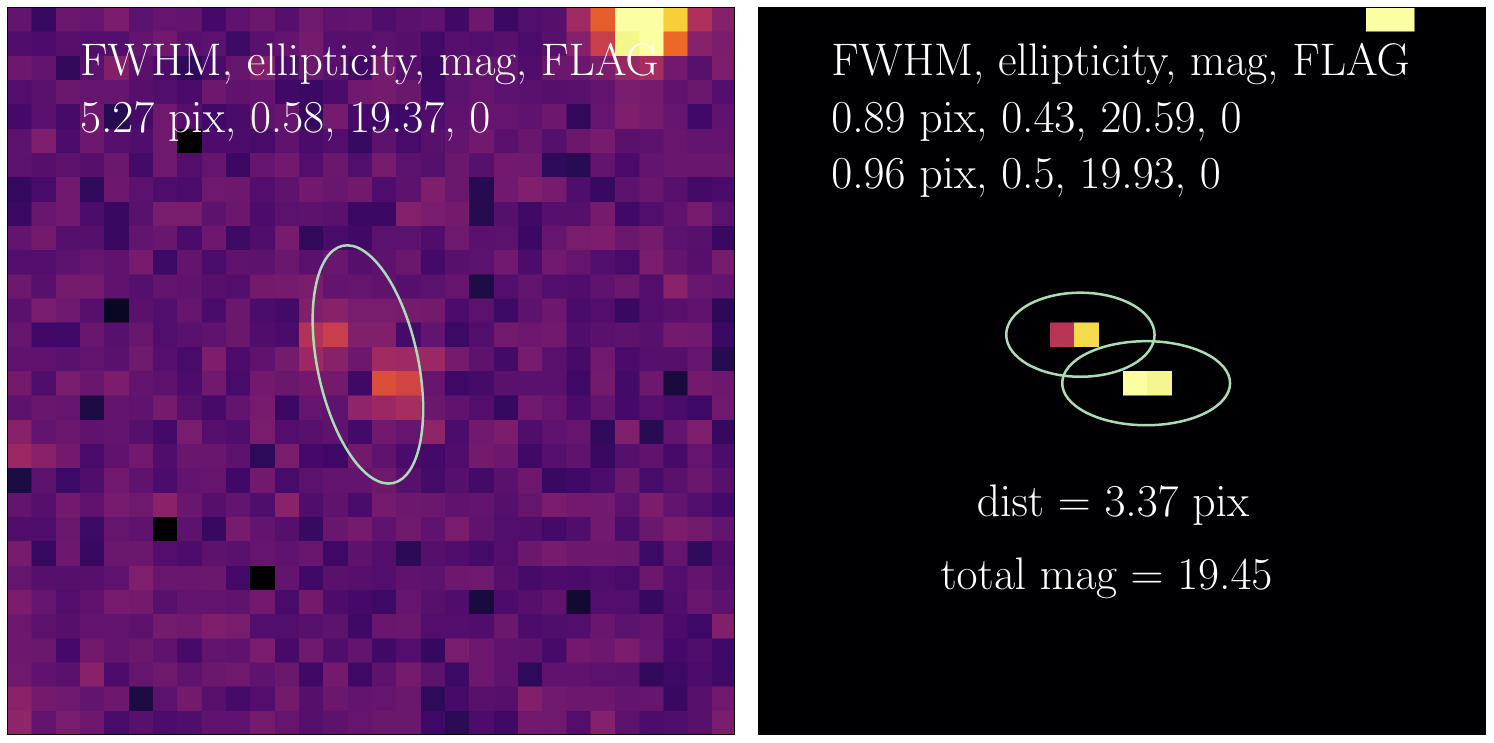}
      \includegraphics[keepaspectratio,width=0.32\linewidth]{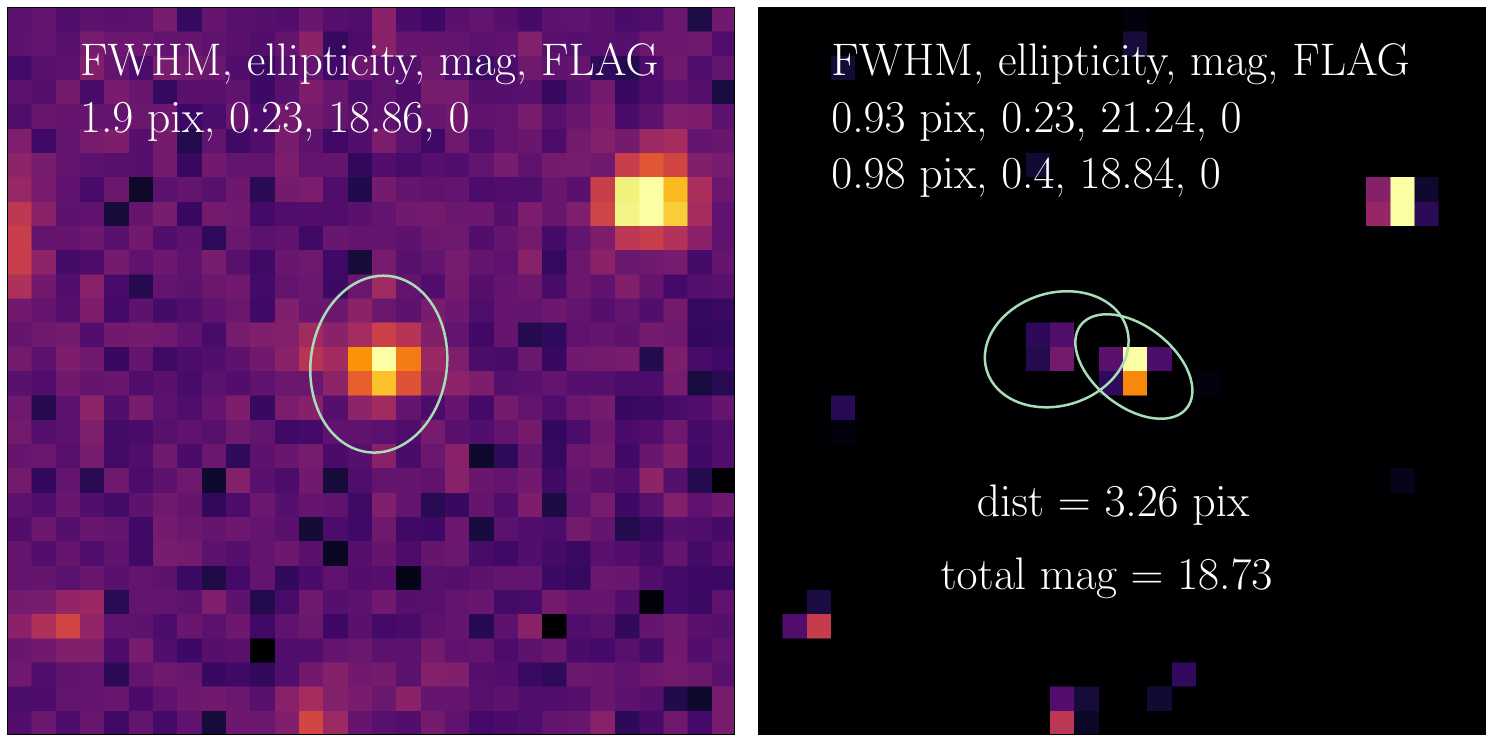}\hfill
      \includegraphics[keepaspectratio,width=0.32\linewidth]{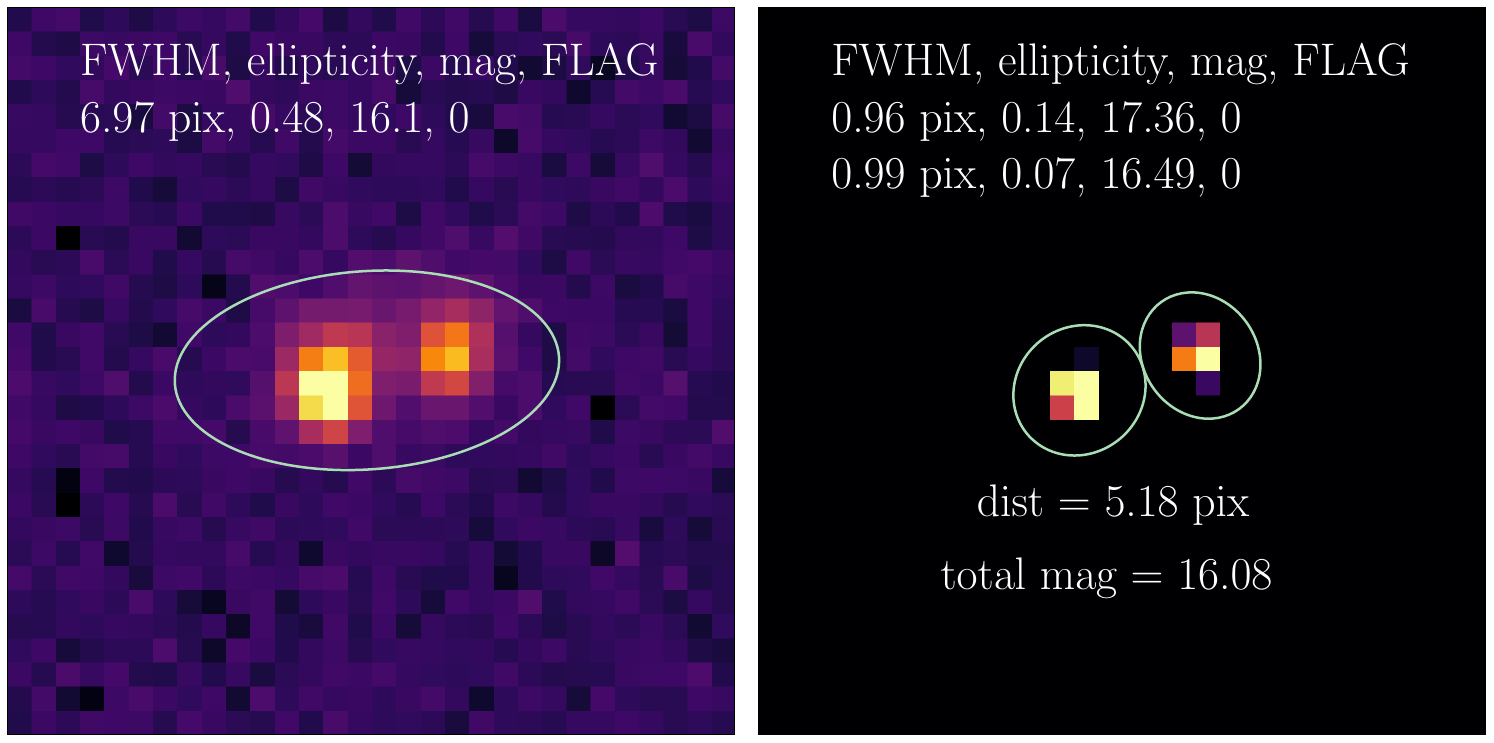}
      \includegraphics[keepaspectratio,width=0.32\linewidth]{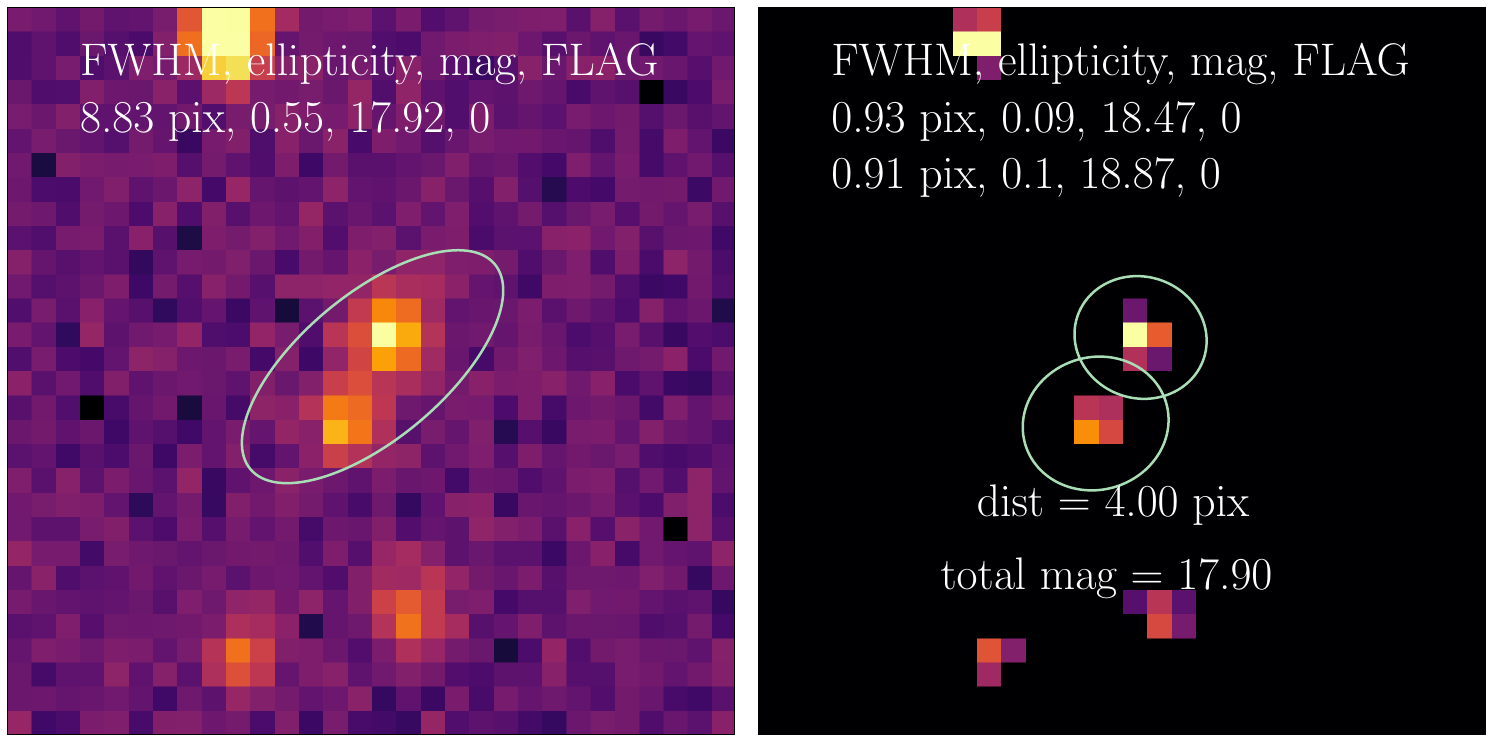}
      \includegraphics[keepaspectratio,width=0.32\linewidth]{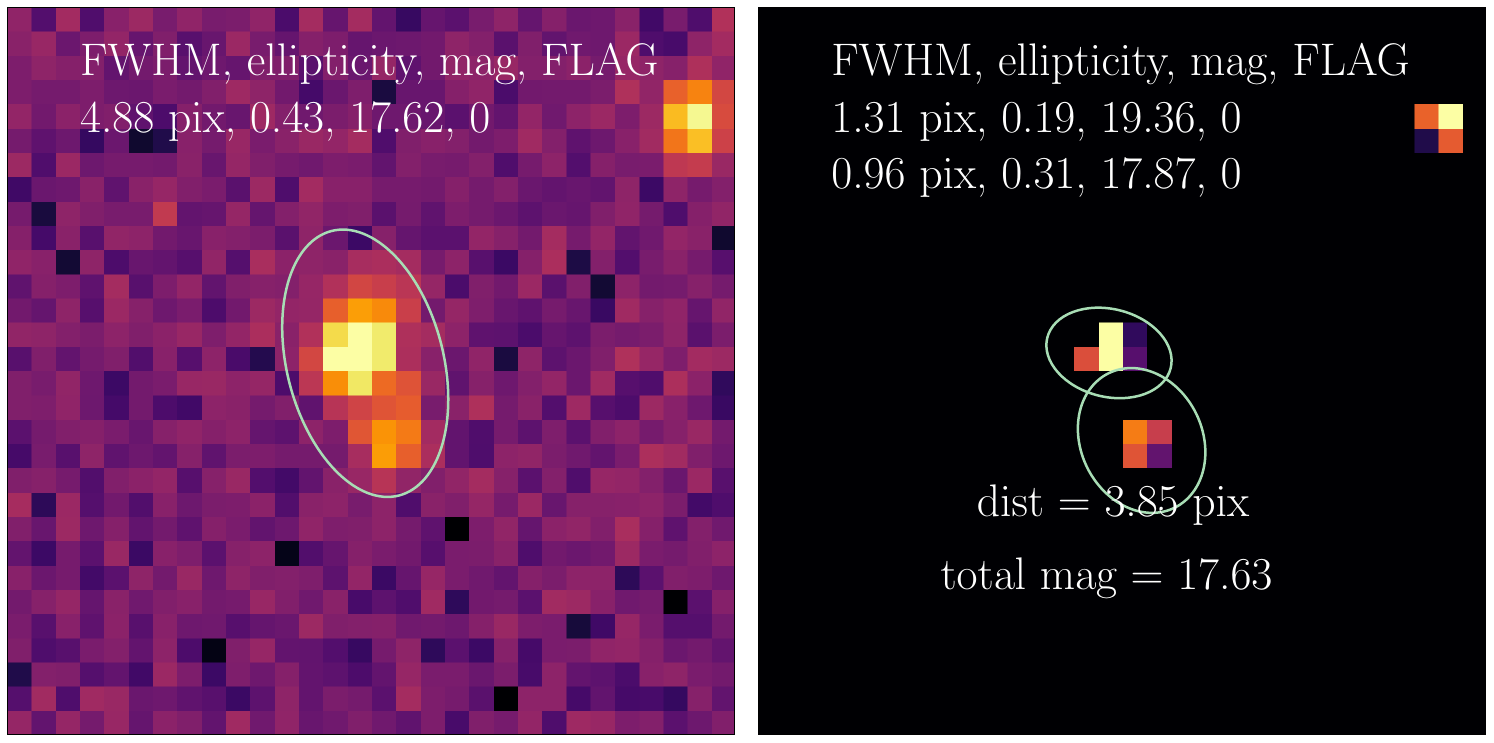}\hfill
      \includegraphics[keepaspectratio,width=0.32\linewidth]{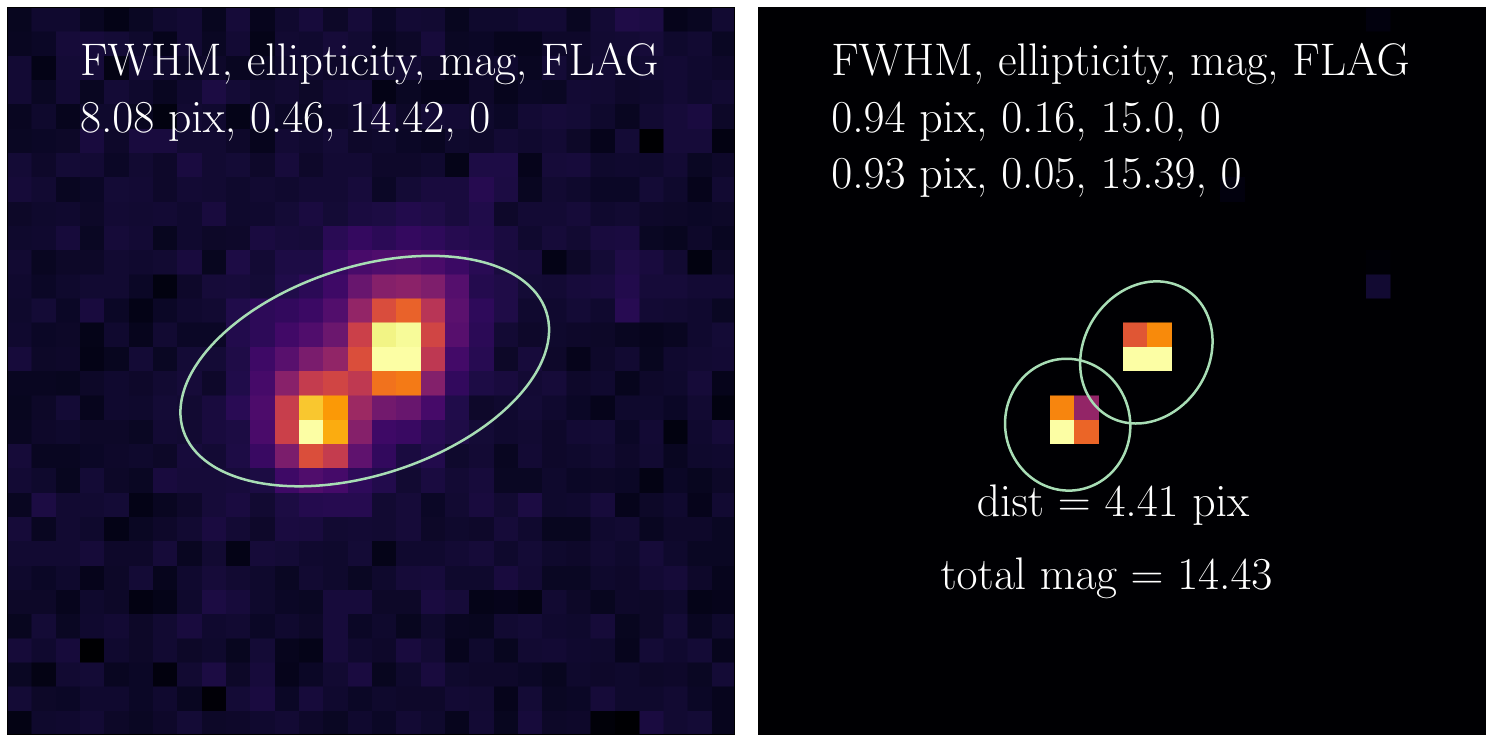}
      \includegraphics[keepaspectratio,width=0.32\linewidth]{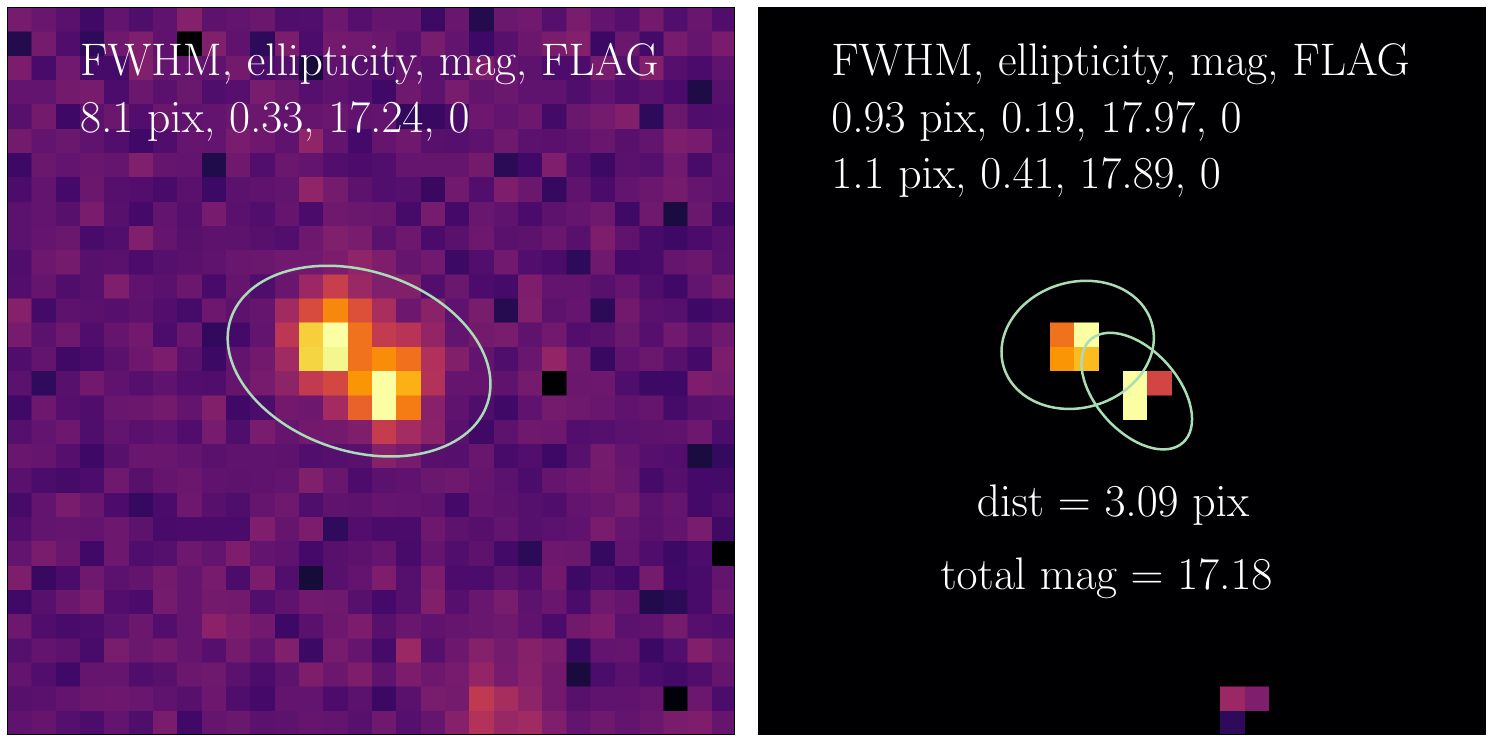}
      \includegraphics[keepaspectratio,width=0.32\linewidth]{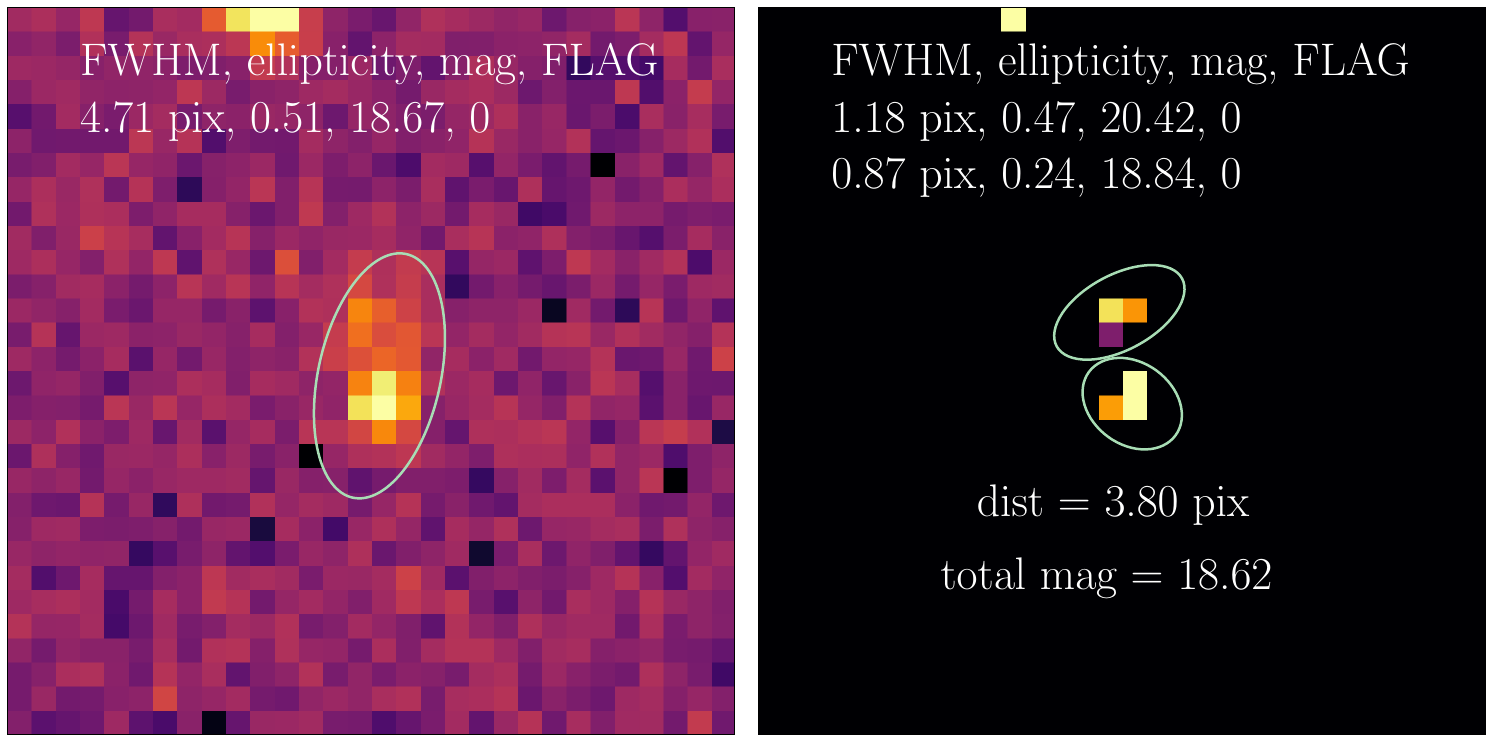}\hfill
      \includegraphics[keepaspectratio,width=0.32\linewidth]{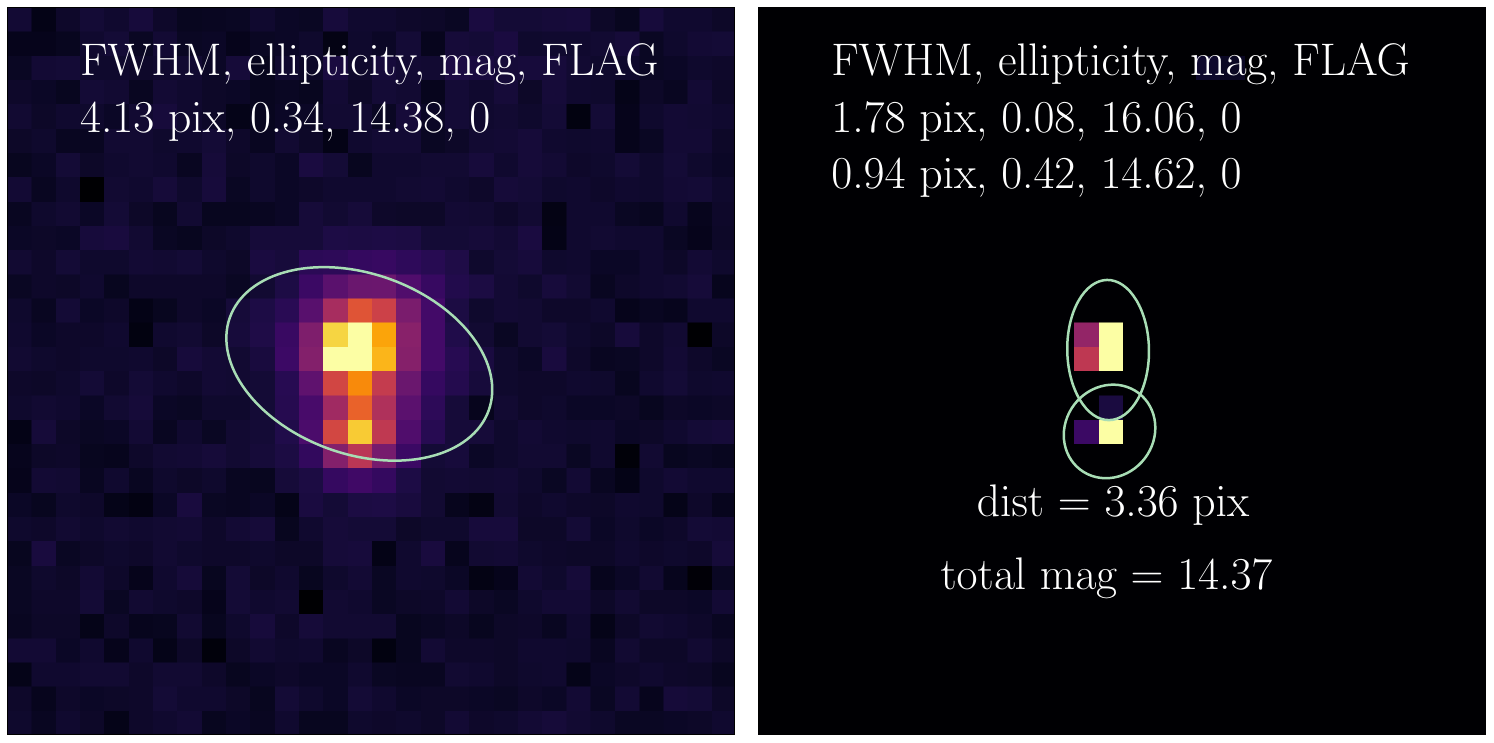}
      \includegraphics[keepaspectratio,width=0.32\linewidth]{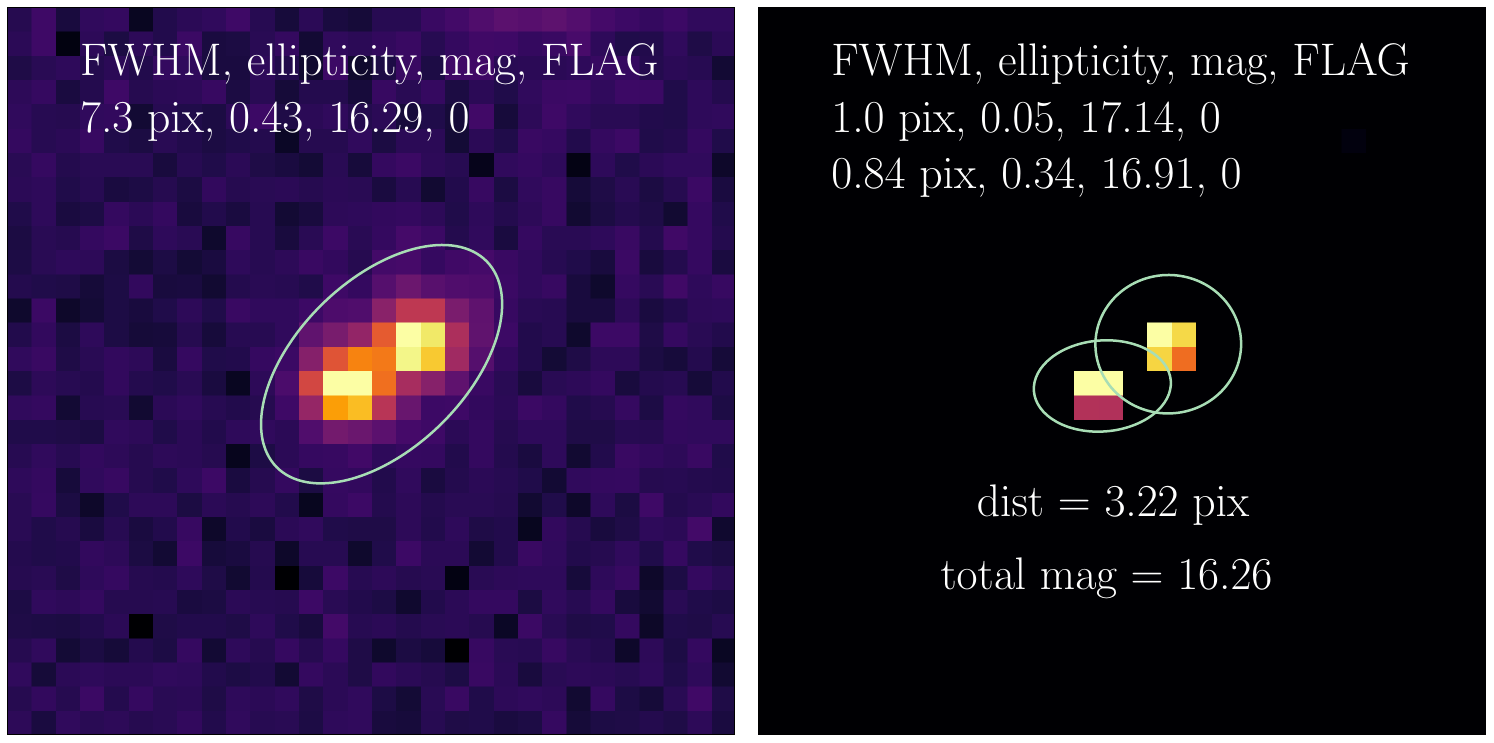}
      \includegraphics[keepaspectratio,width=0.32\linewidth]{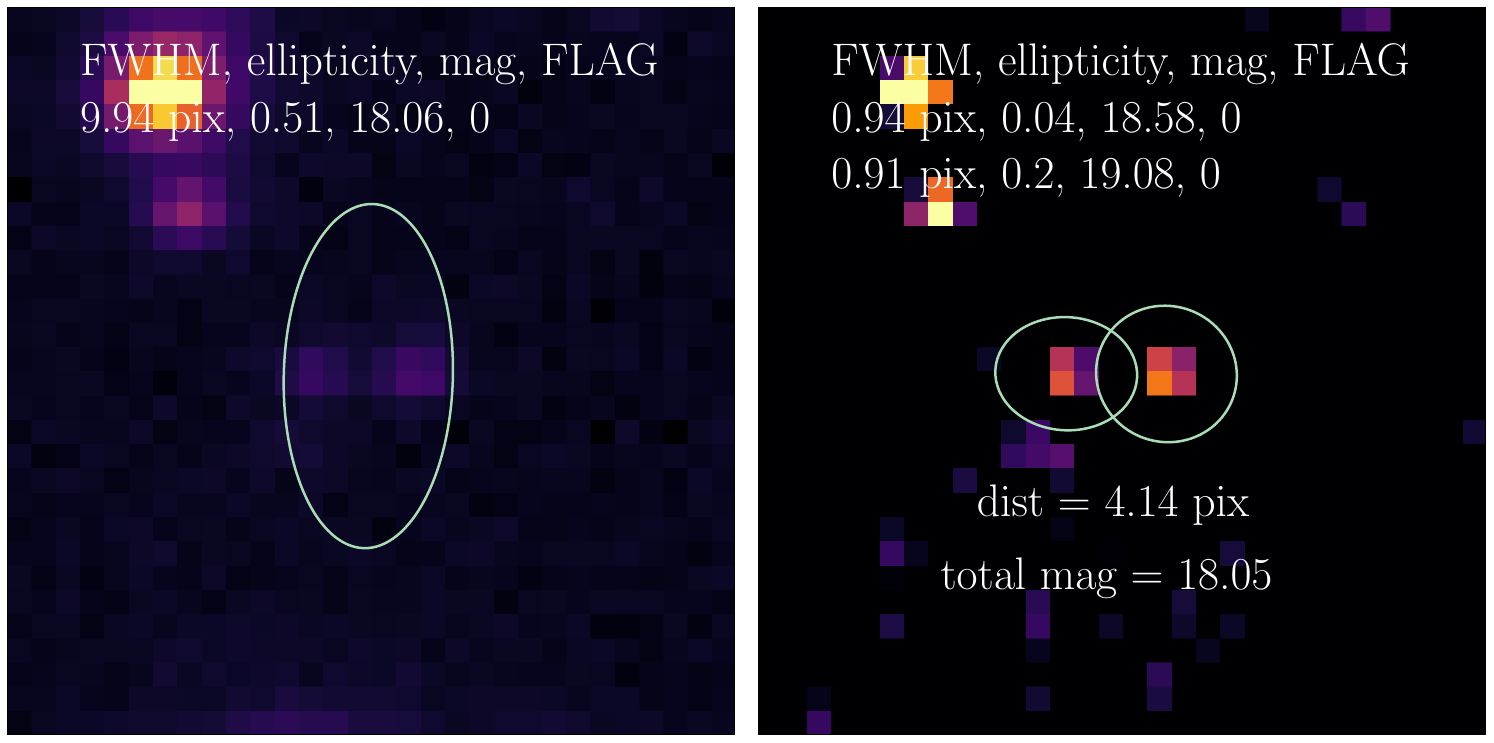}\hfill
      \includegraphics[keepaspectratio,width=0.32\linewidth]{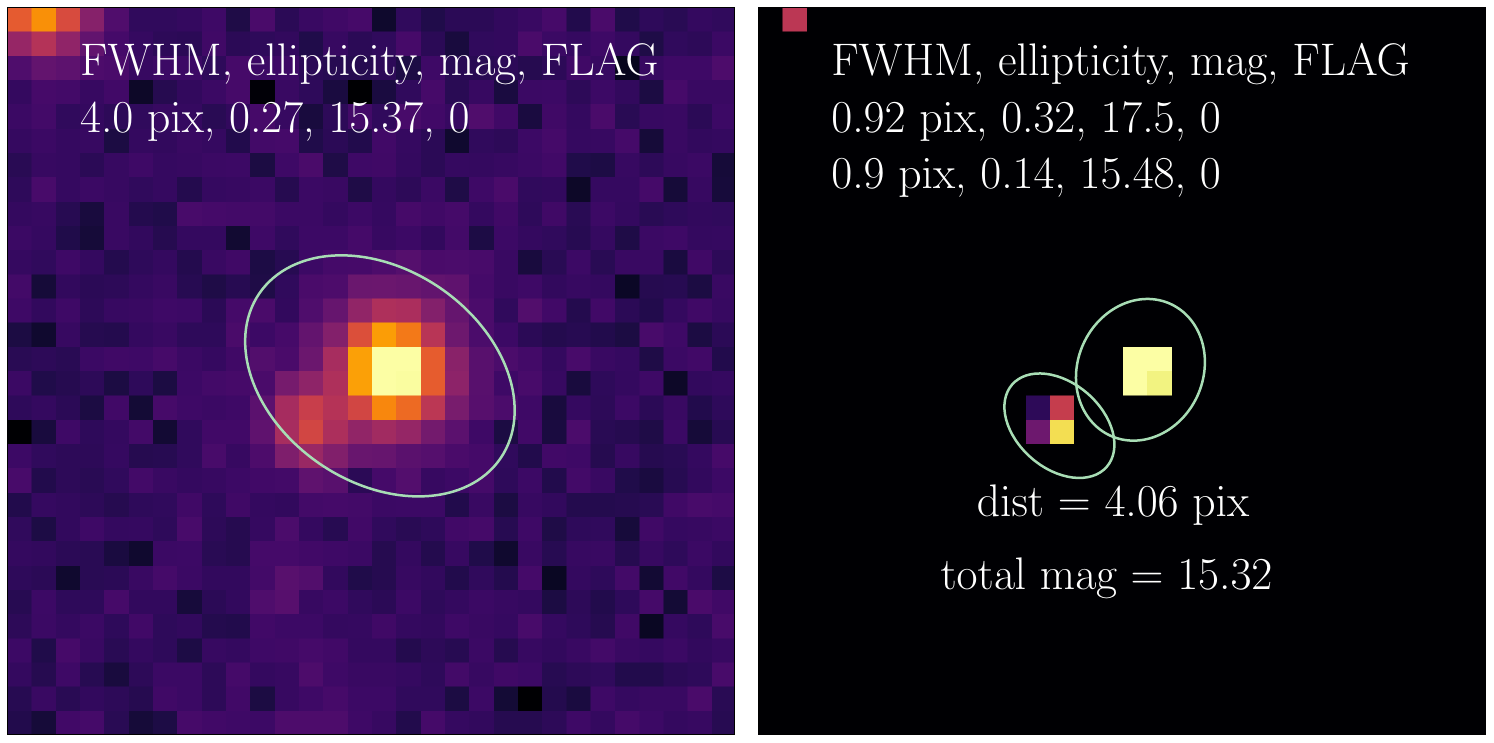}
      \includegraphics[keepaspectratio,width=0.32\linewidth]{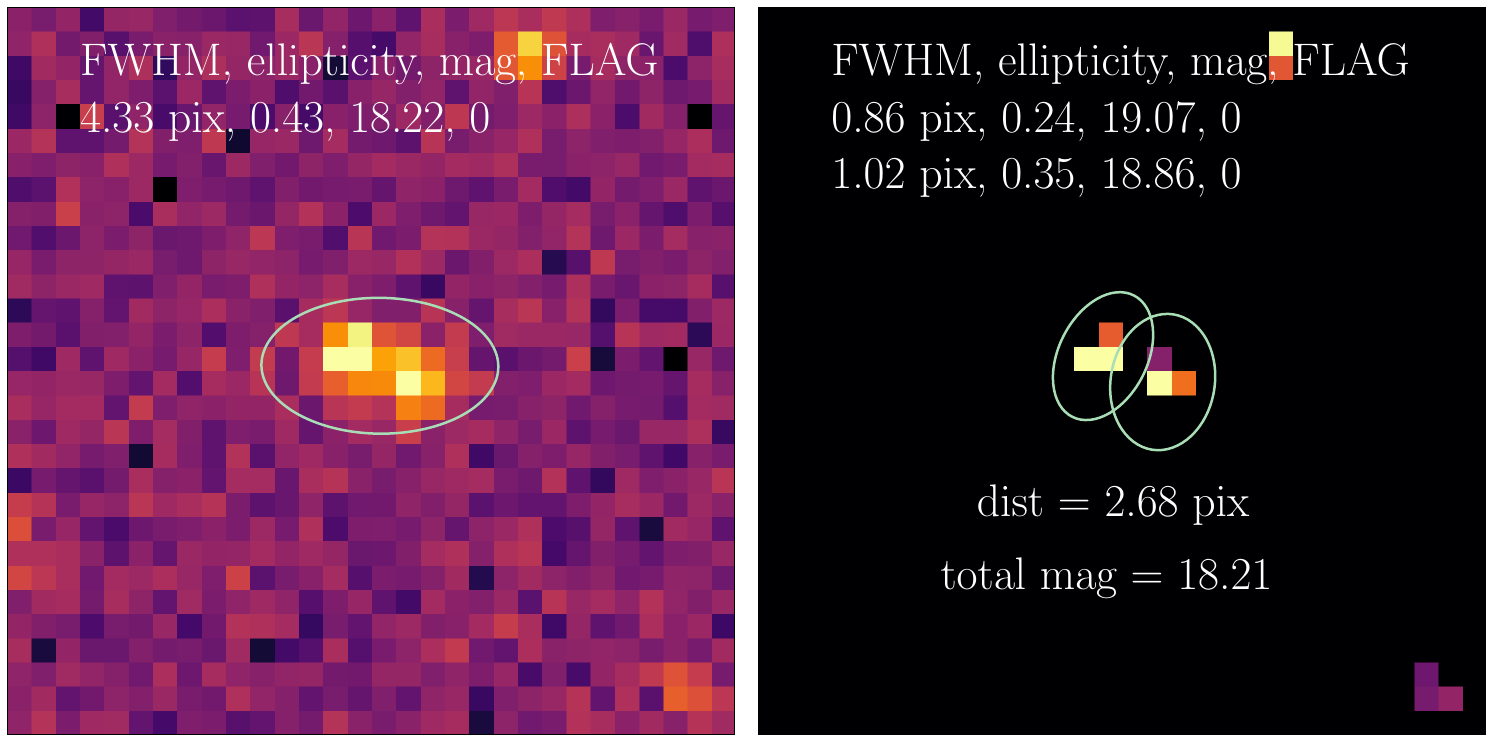}
      \includegraphics[keepaspectratio,width=0.32\linewidth]{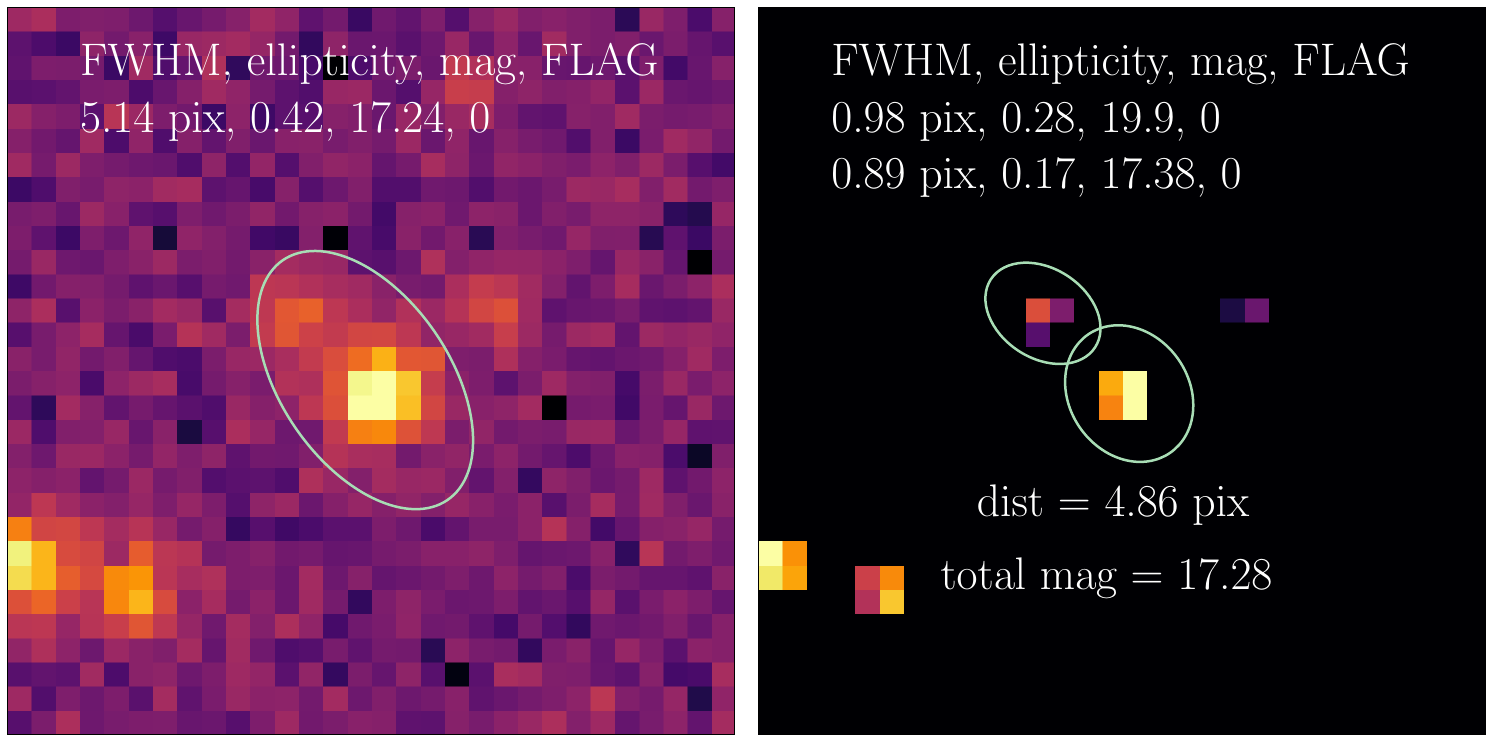}\hfill
      \includegraphics[keepaspectratio,width=0.32\linewidth]{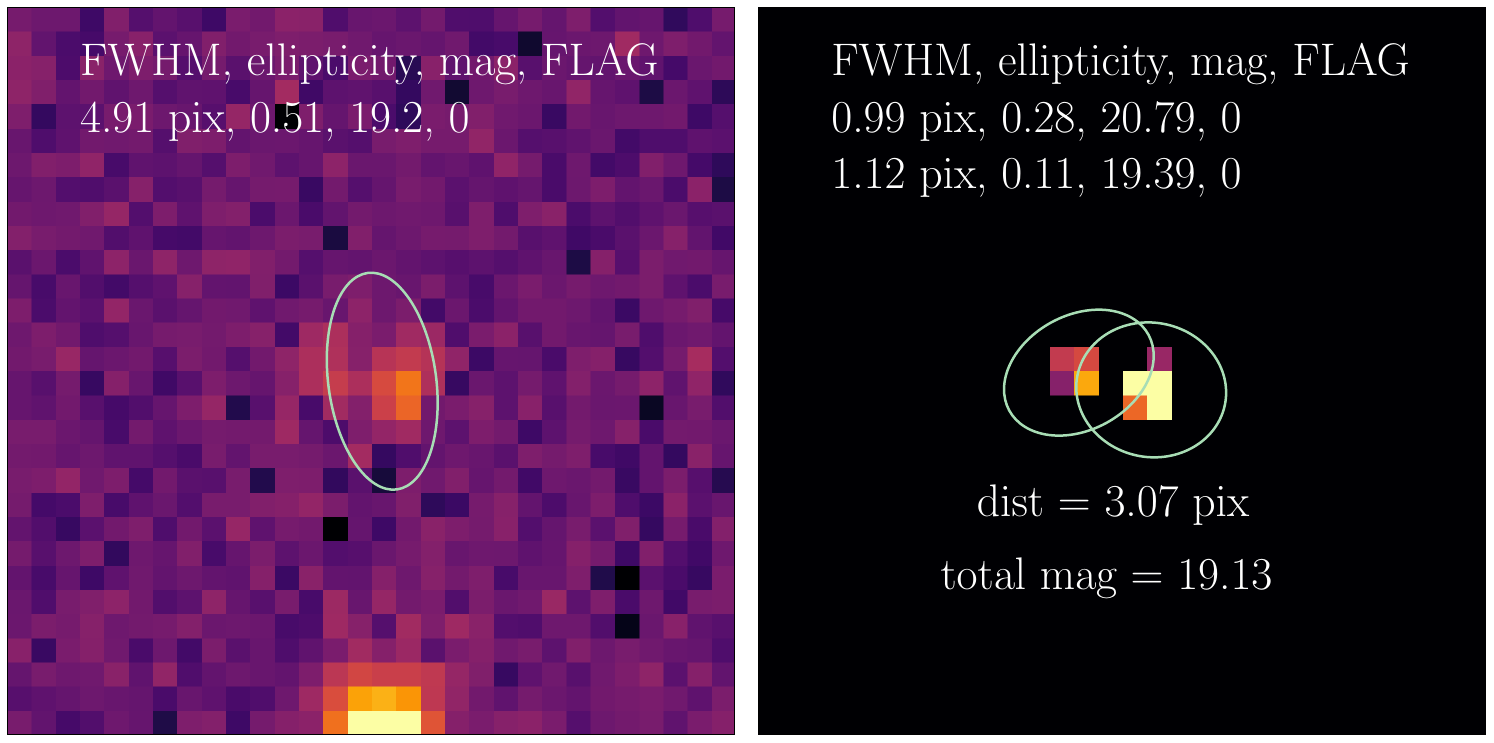}
      \includegraphics[keepaspectratio,width=0.32\linewidth]{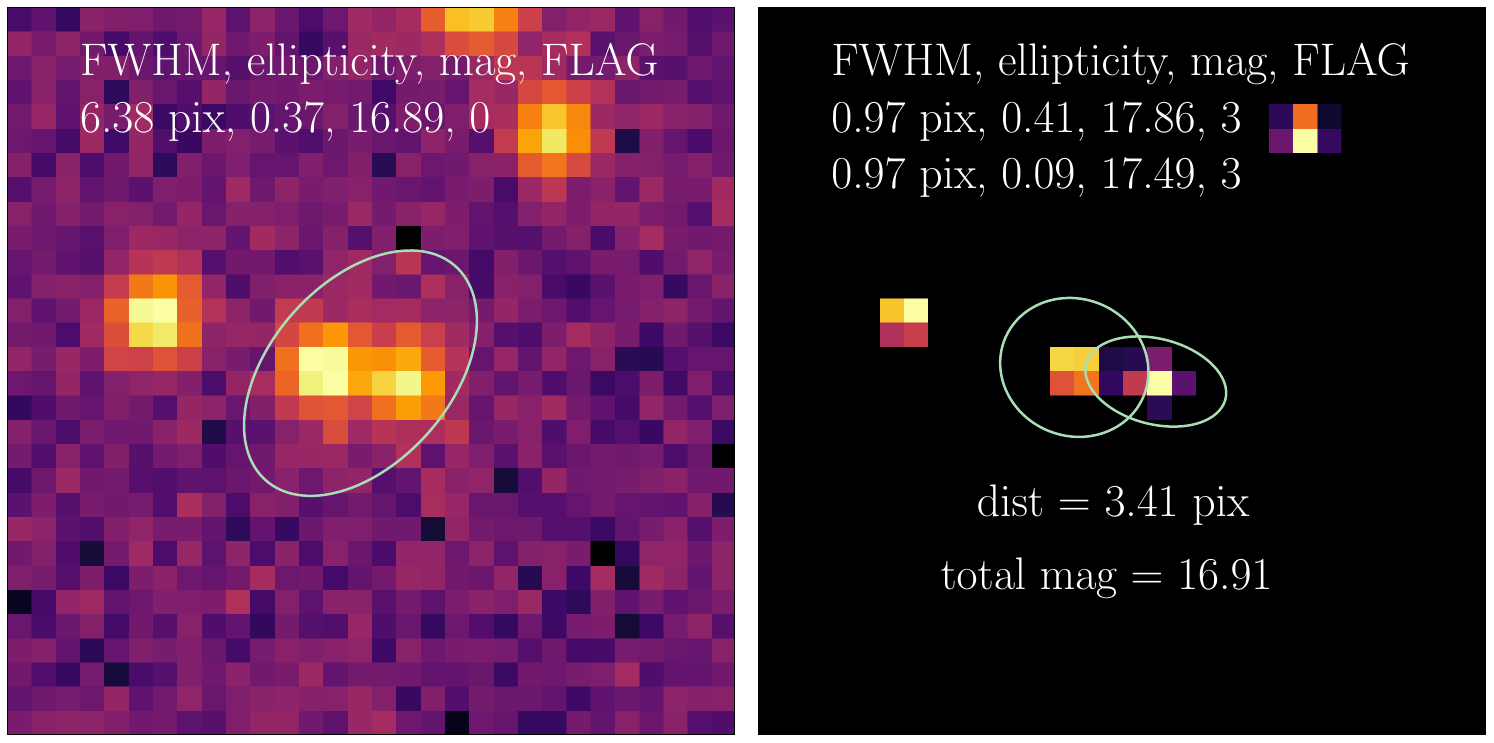}
      \includegraphics[keepaspectratio,width=0.32\linewidth]{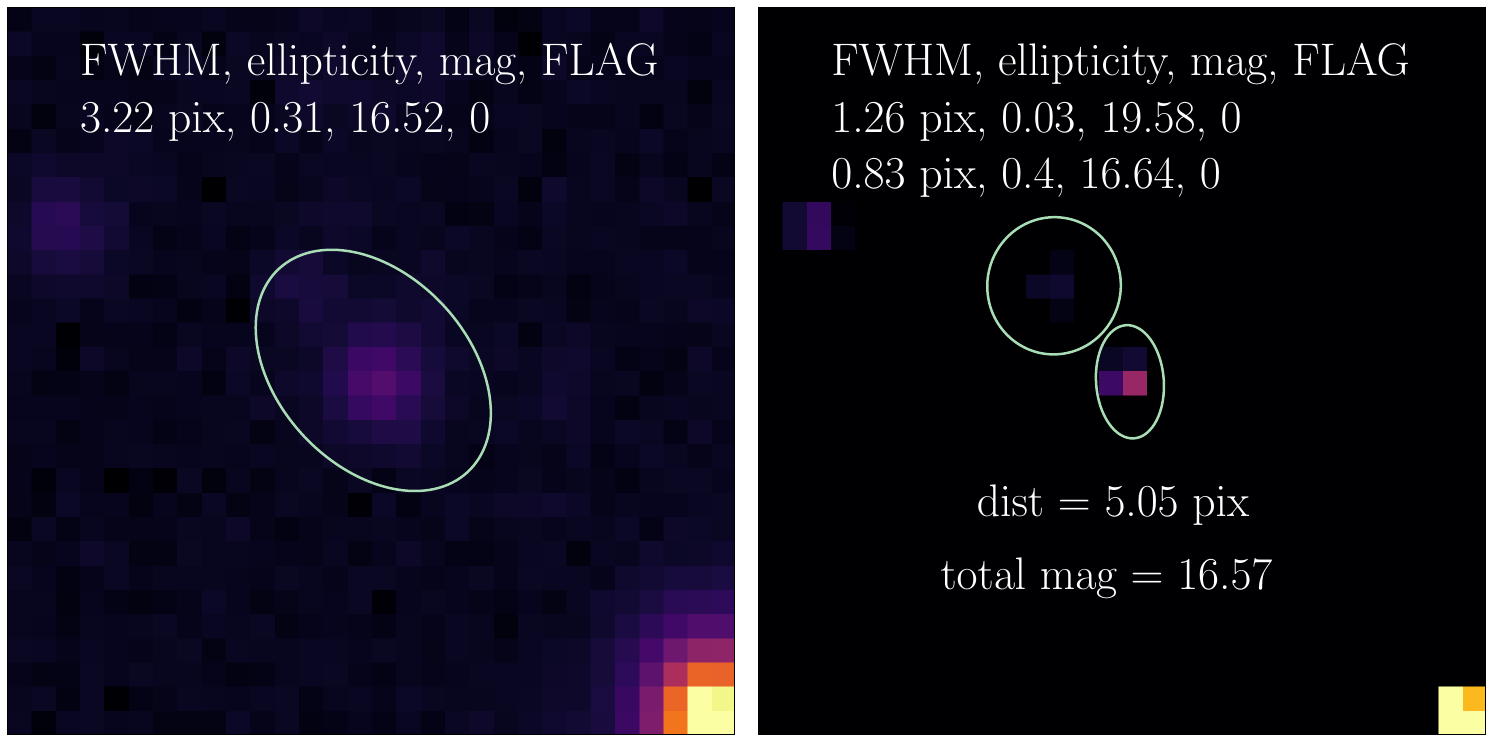}\hfill
      \includegraphics[keepaspectratio,width=0.32\linewidth]{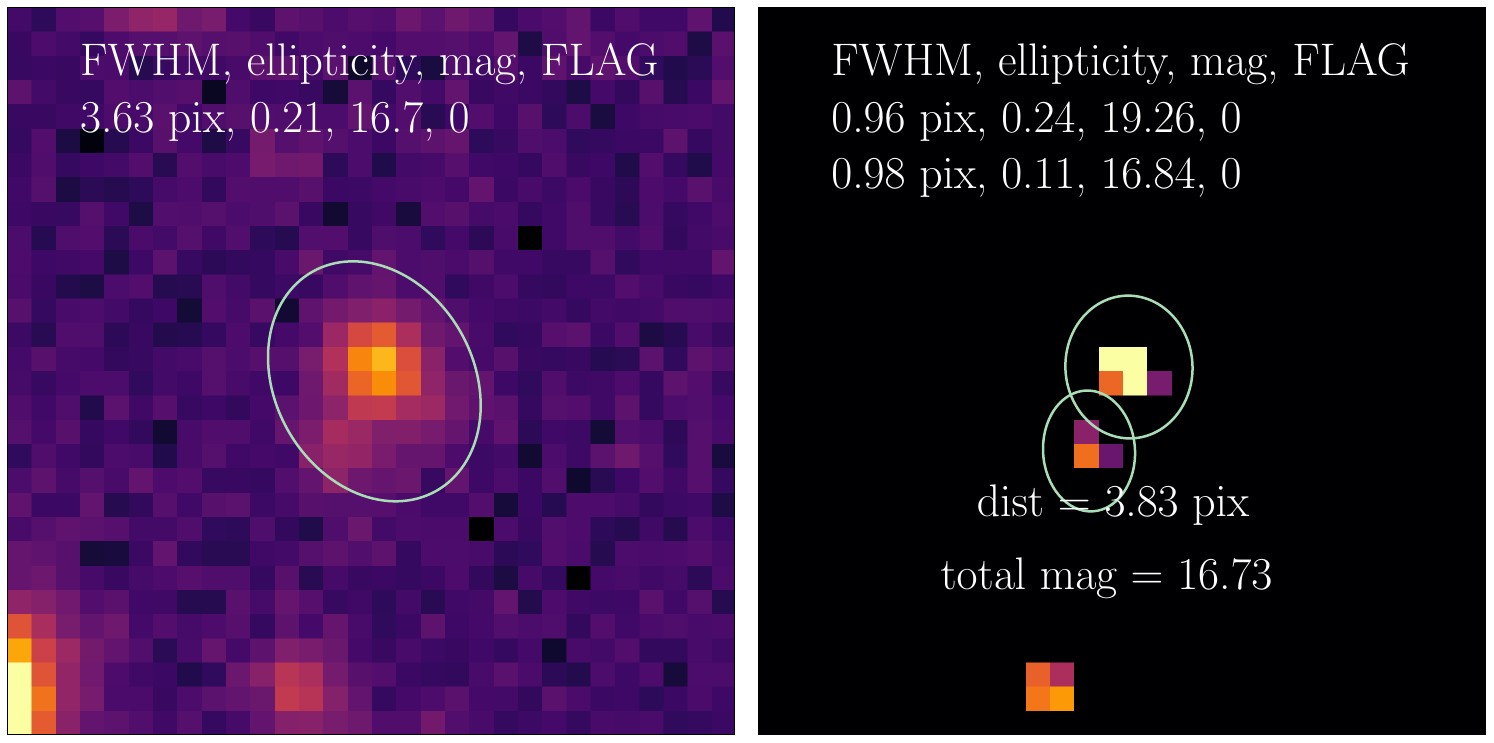}
      \includegraphics[keepaspectratio,width=0.32\linewidth]{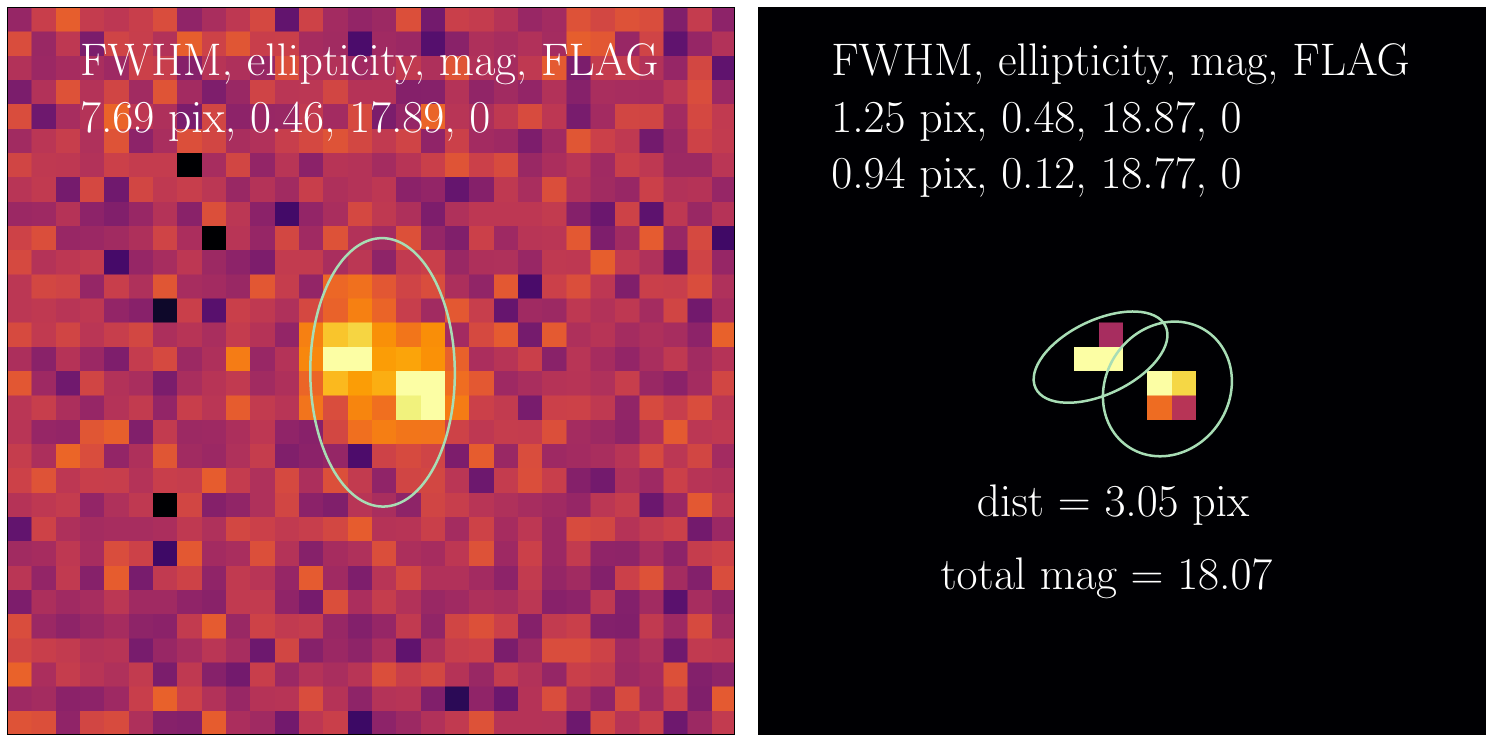}
      \includegraphics[keepaspectratio,width=0.32\linewidth]{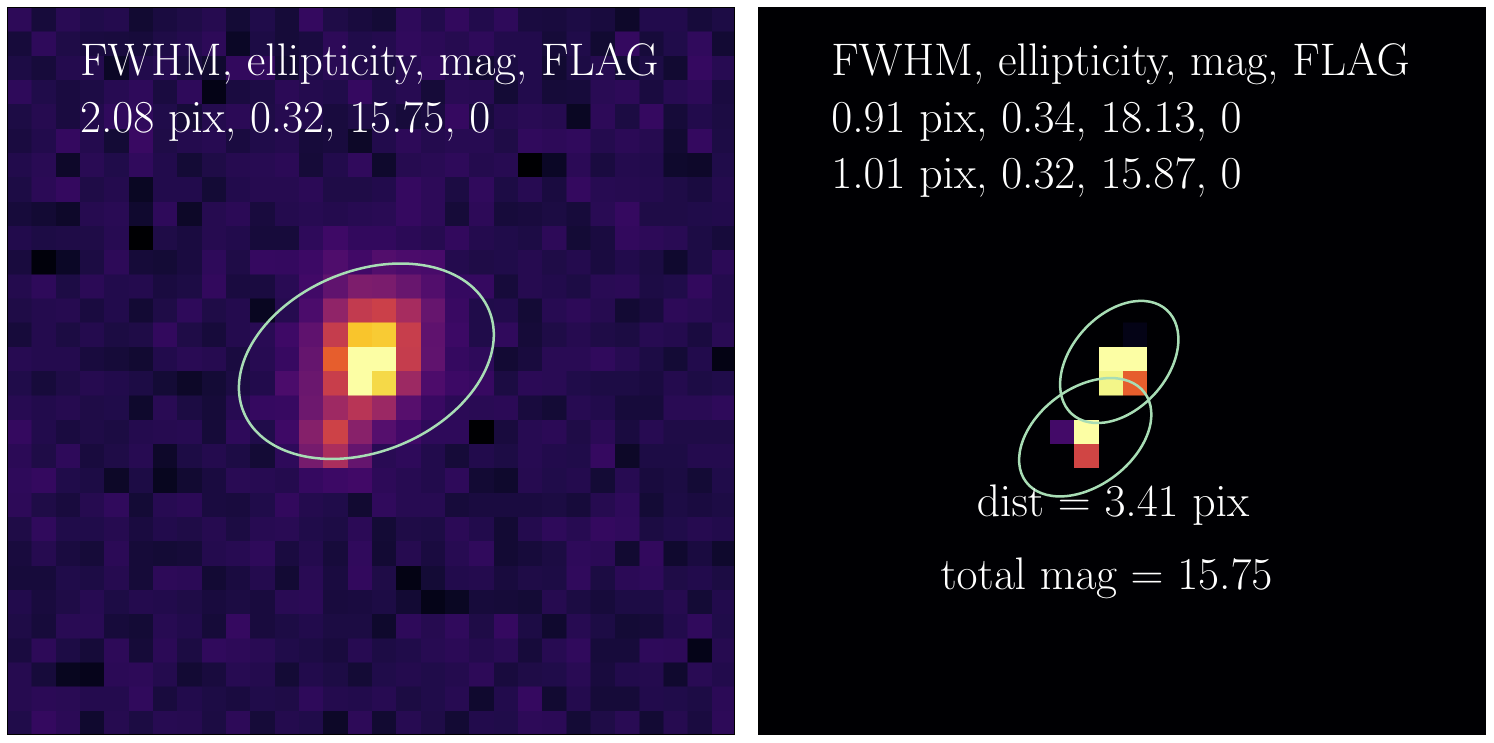}\hfill
\end{figure*}

\subsection{Comparison across different filters}\label{subsec:diff-filters}

\begin{figure*}
      \includegraphics[keepaspectratio,width=0.24\linewidth]{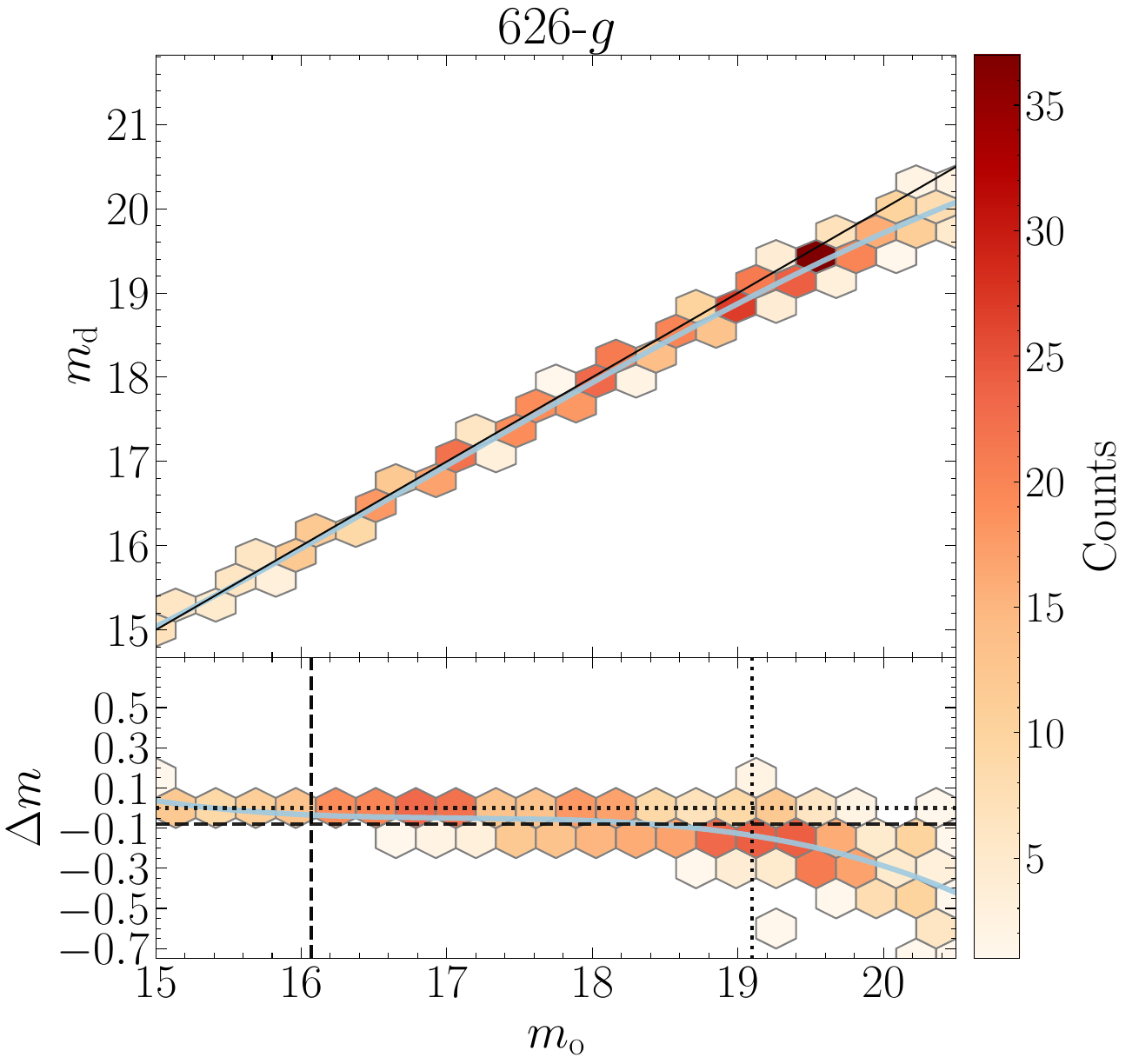}
      \includegraphics[keepaspectratio,width=0.24\linewidth]{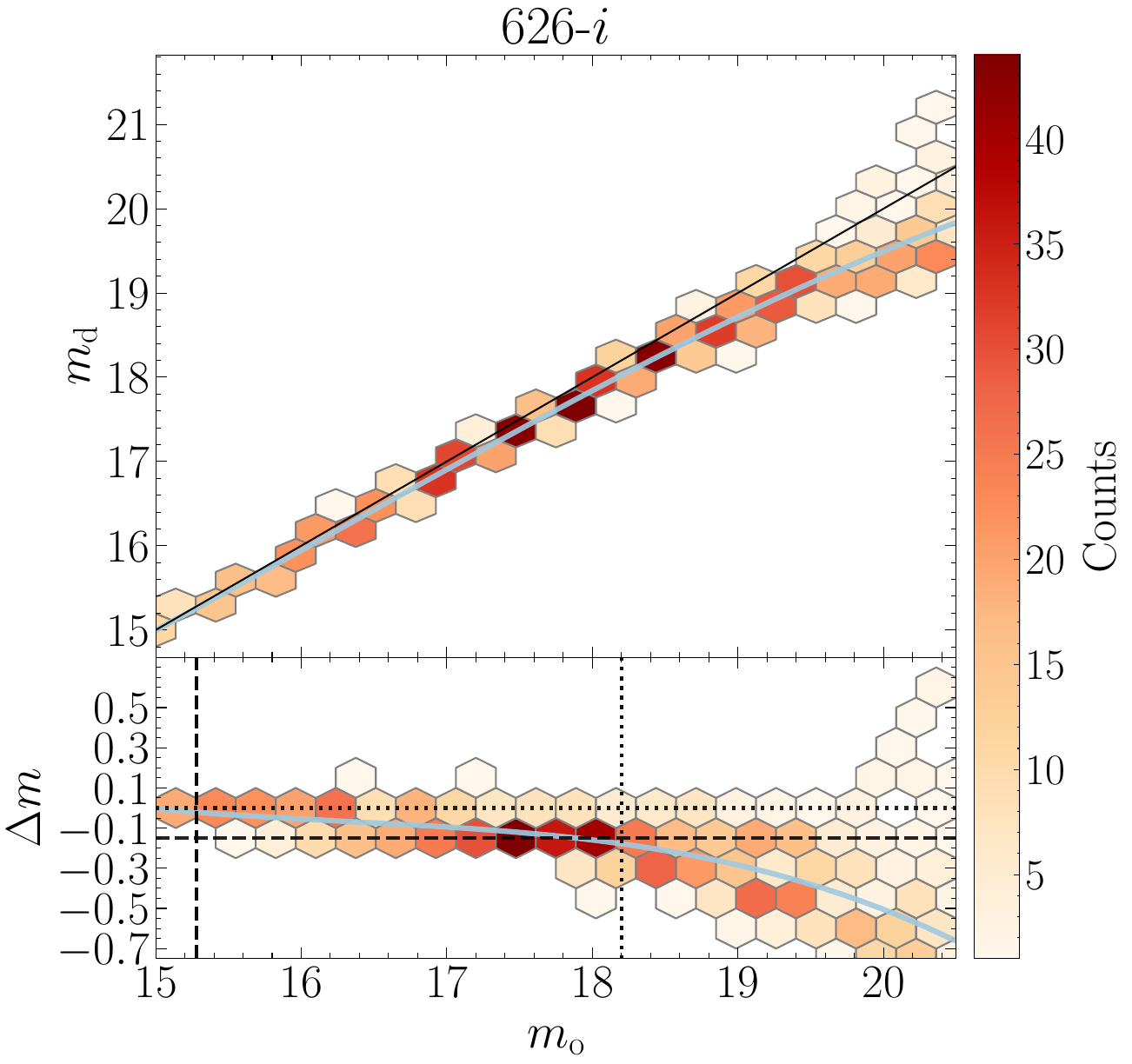}
      \includegraphics[keepaspectratio,width=0.24\linewidth]{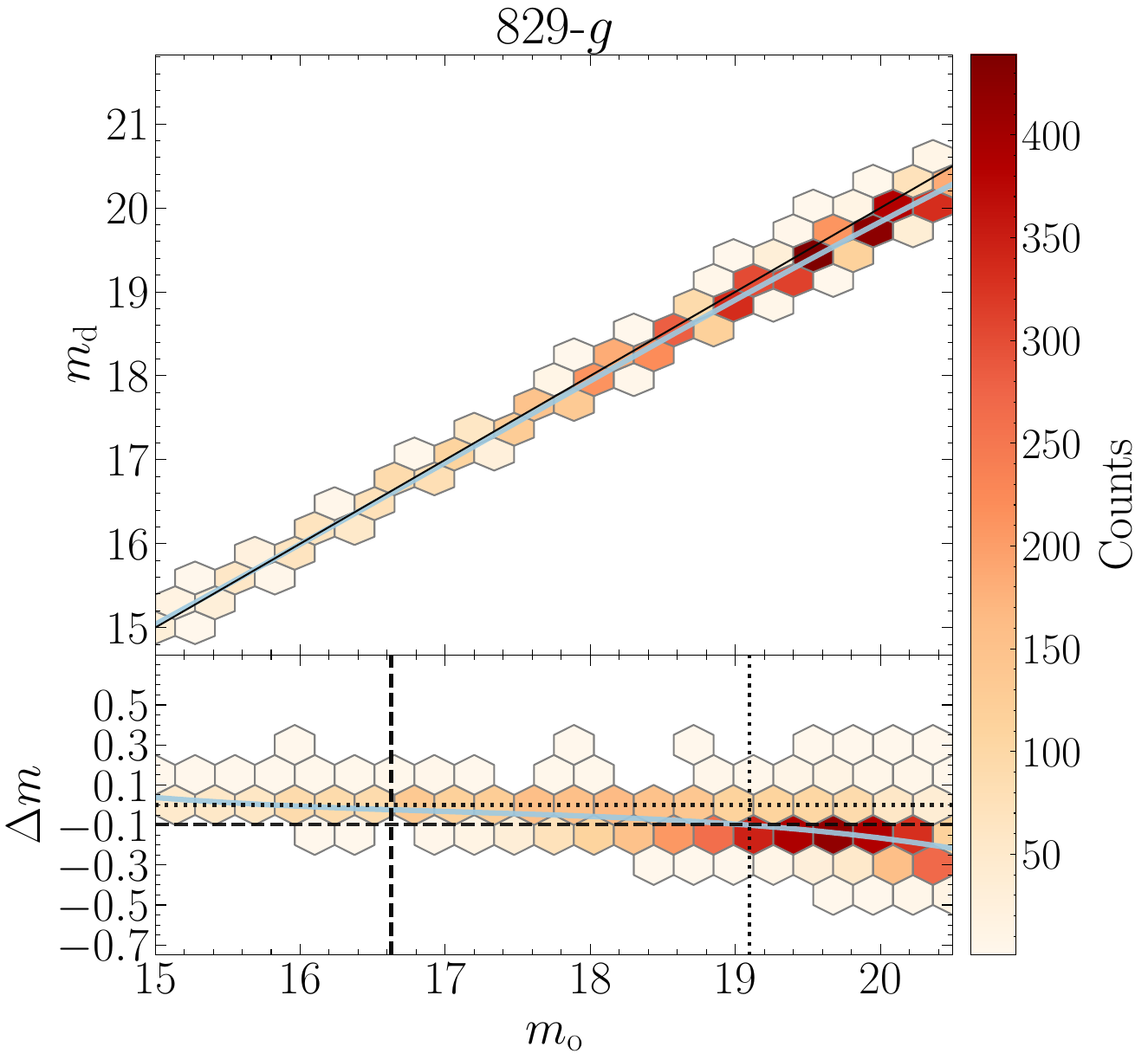}
      \includegraphics[keepaspectratio,width=0.24\linewidth]{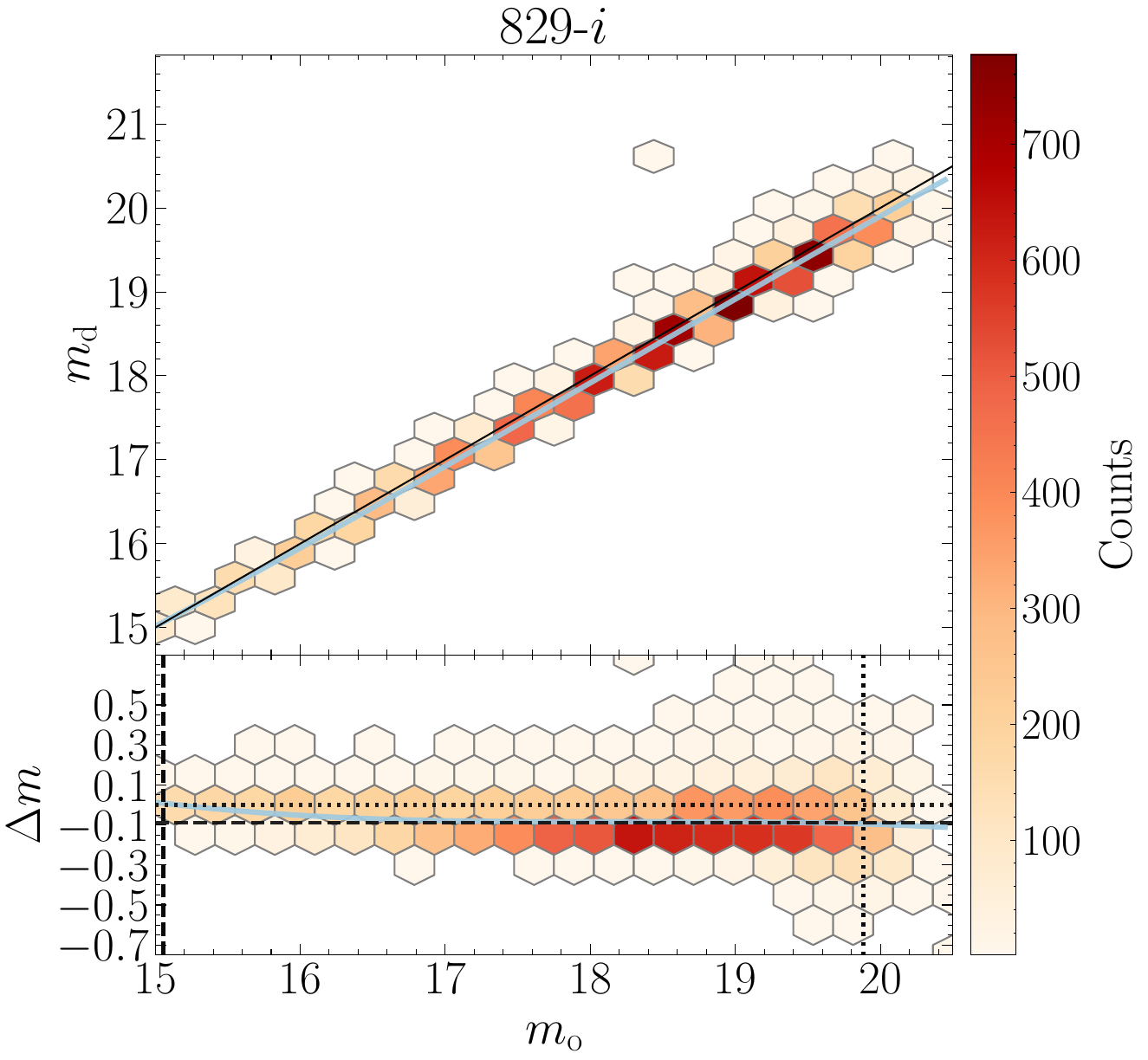}
      \includegraphics[keepaspectratio,width=0.24\linewidth]{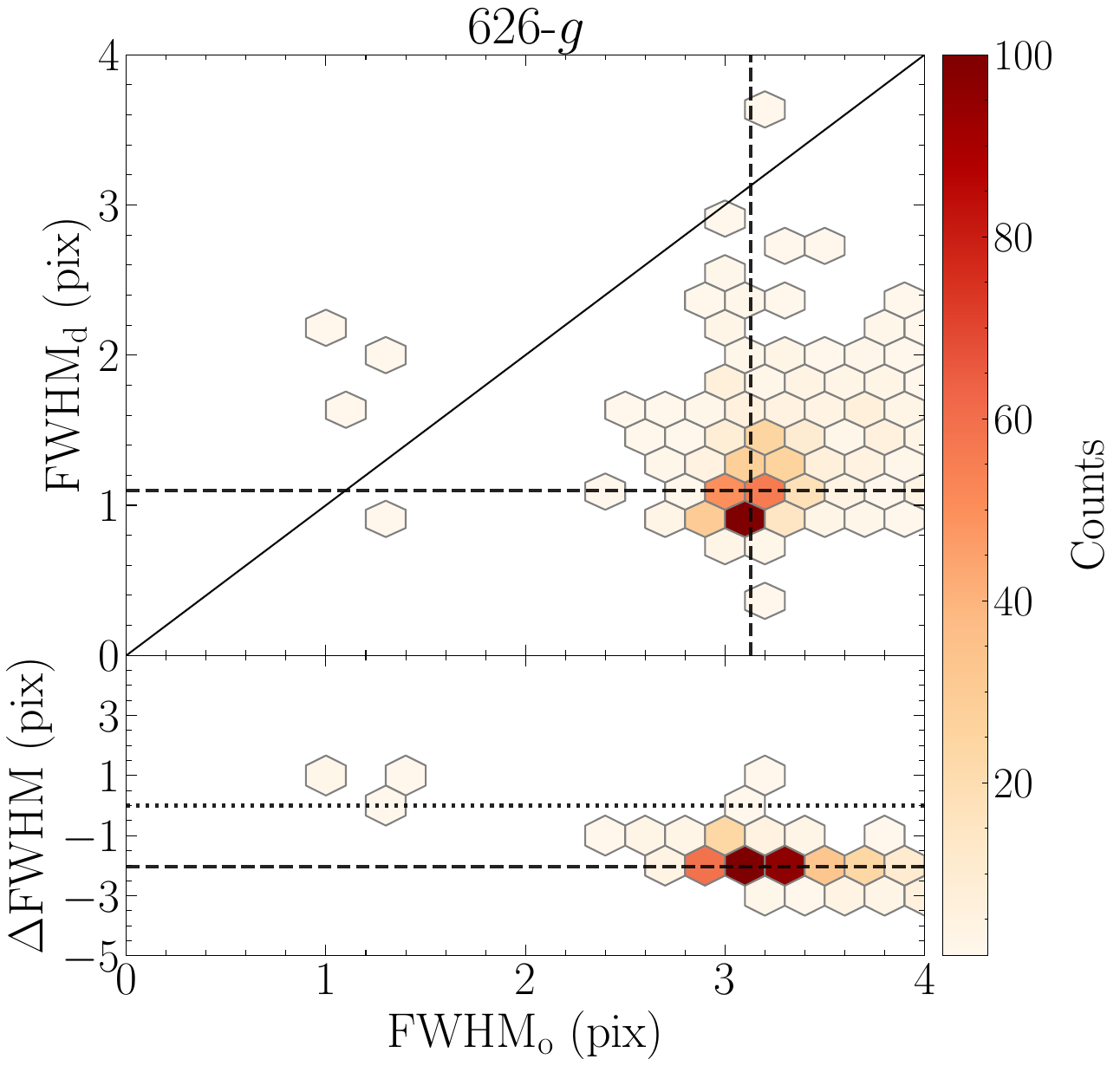}
      \includegraphics[keepaspectratio,width=0.24\linewidth]{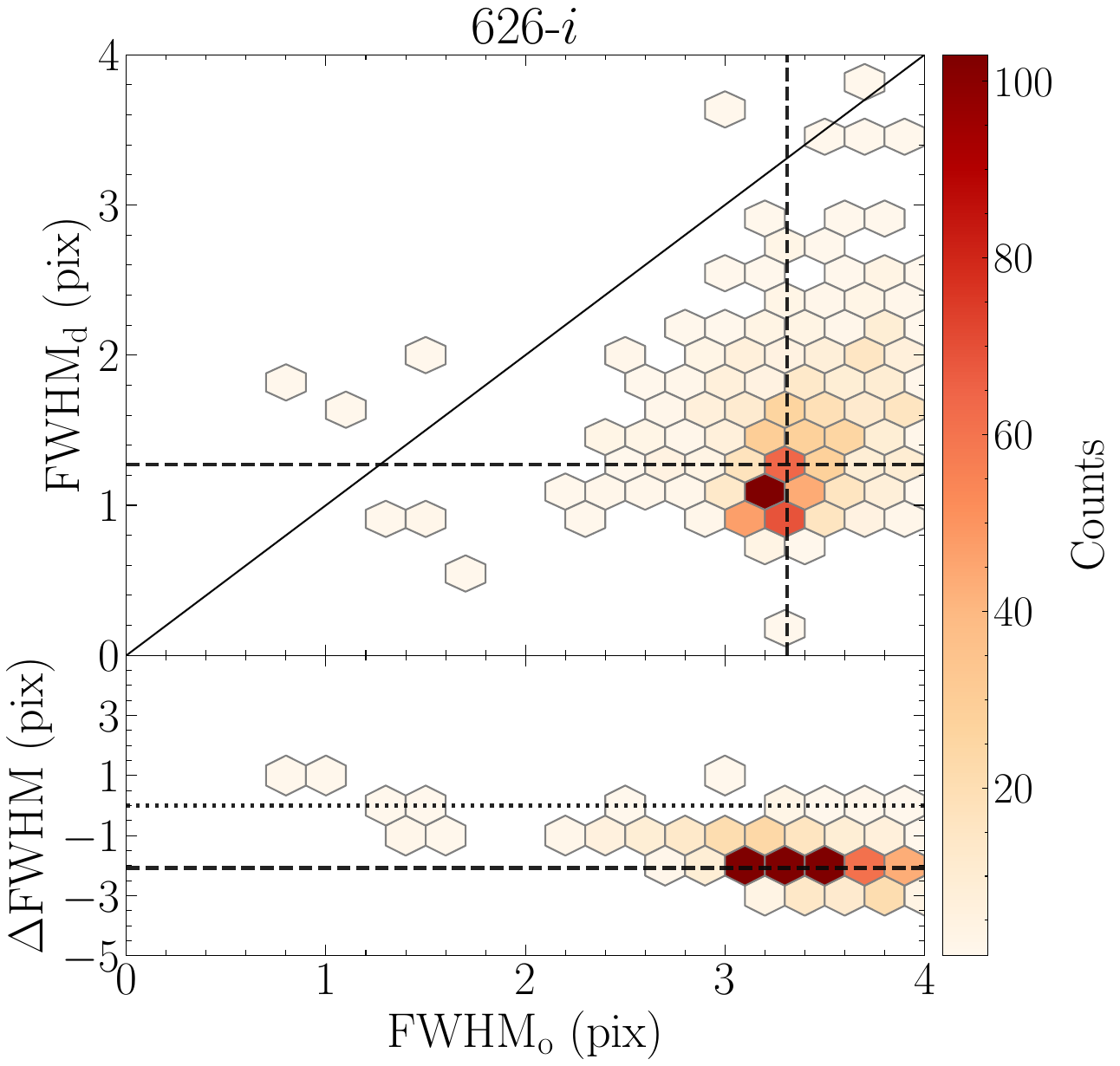}
      \includegraphics[keepaspectratio,width=0.24\linewidth]{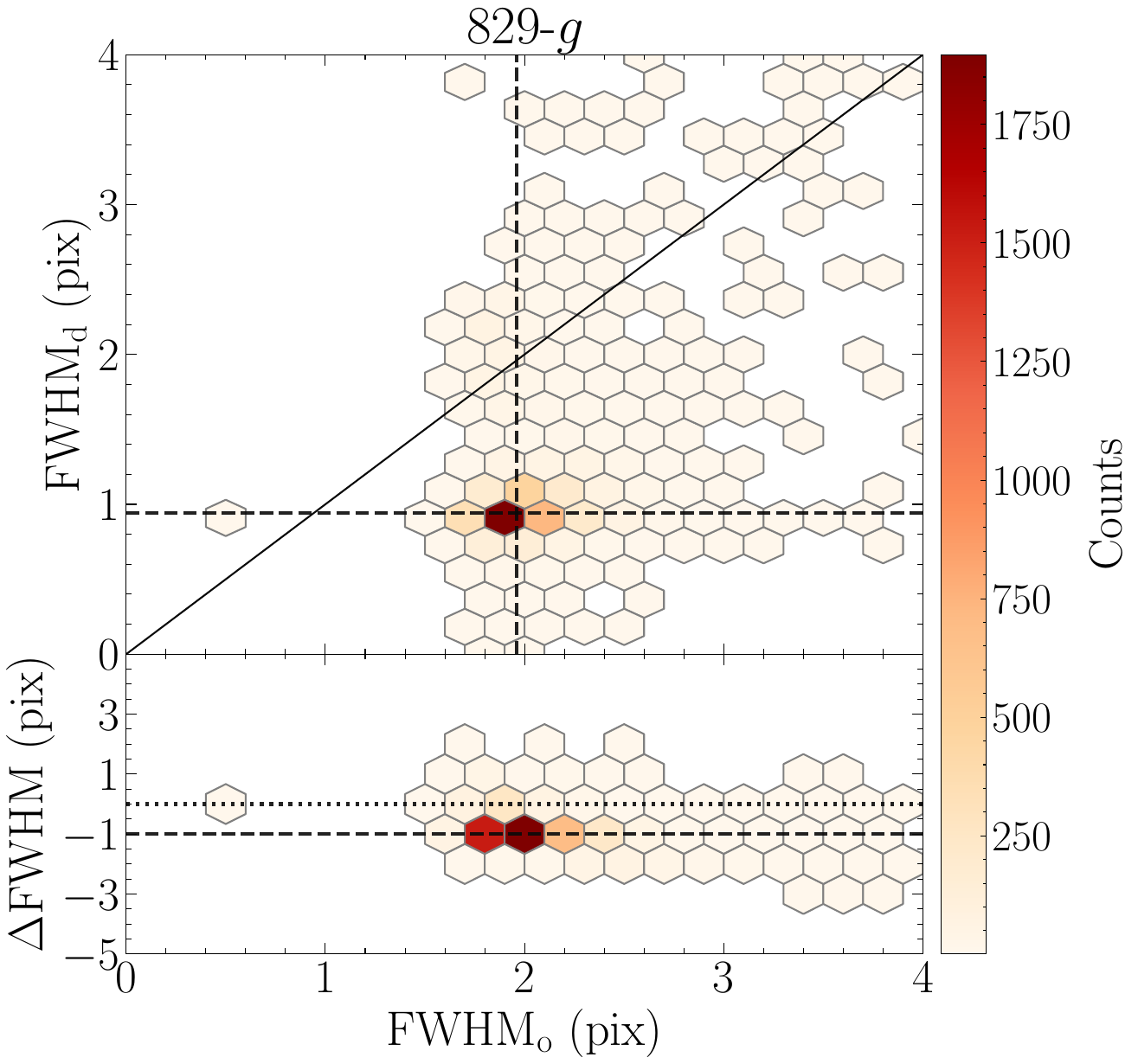}
      \includegraphics[keepaspectratio,width=0.24\linewidth]{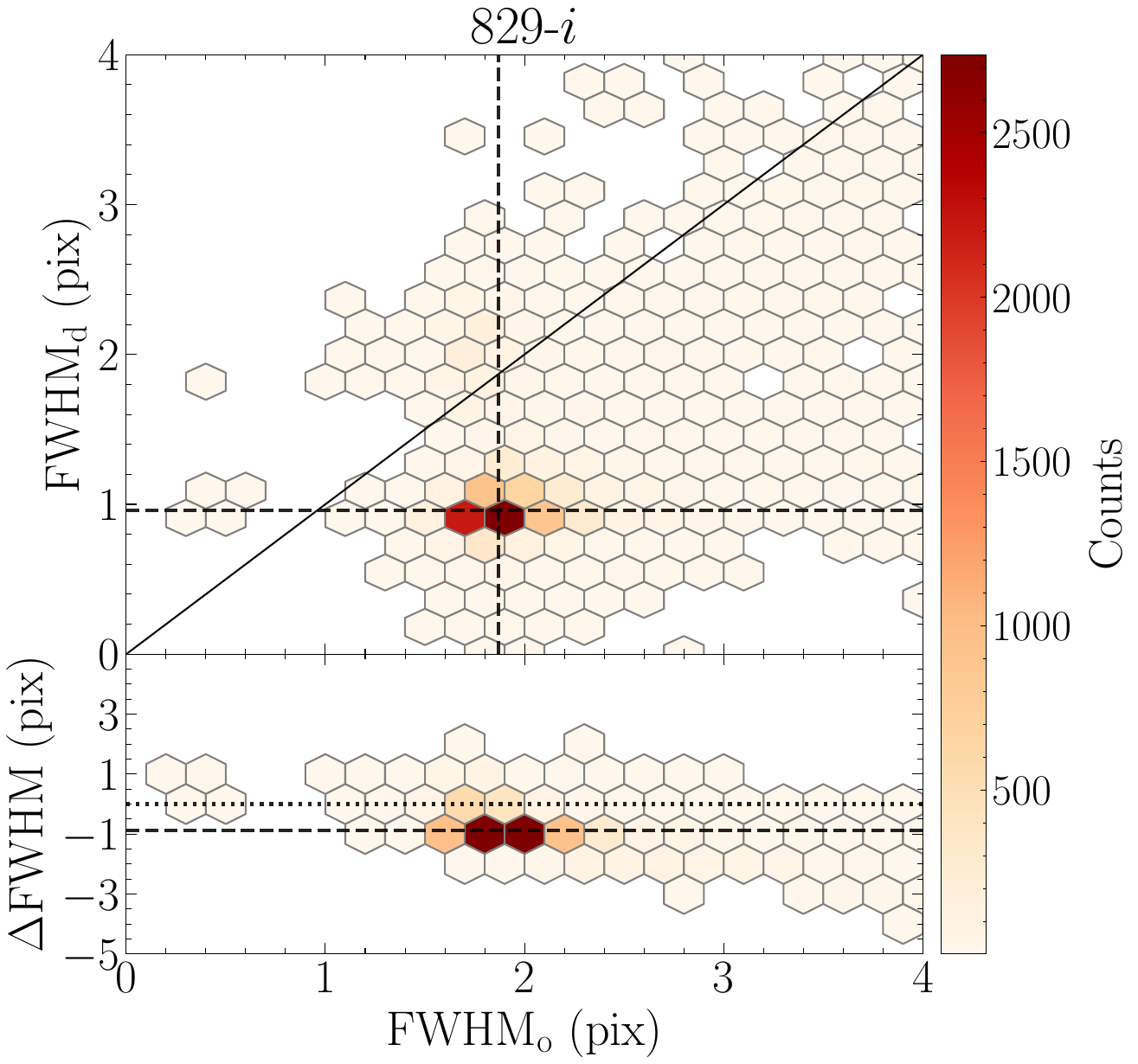}
      \includegraphics[keepaspectratio,width=0.24\linewidth]{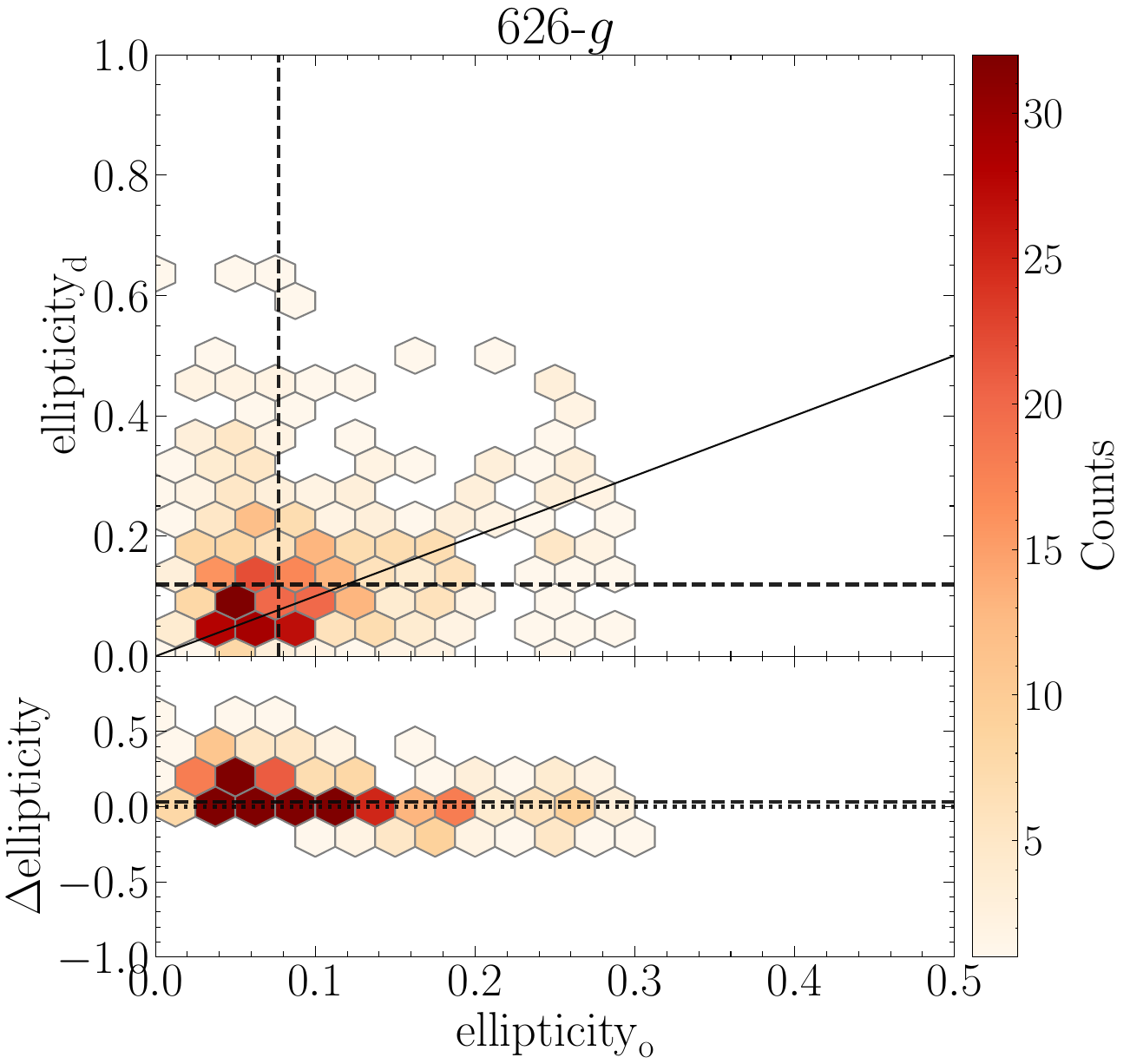}
      \includegraphics[keepaspectratio,width=0.24\linewidth]{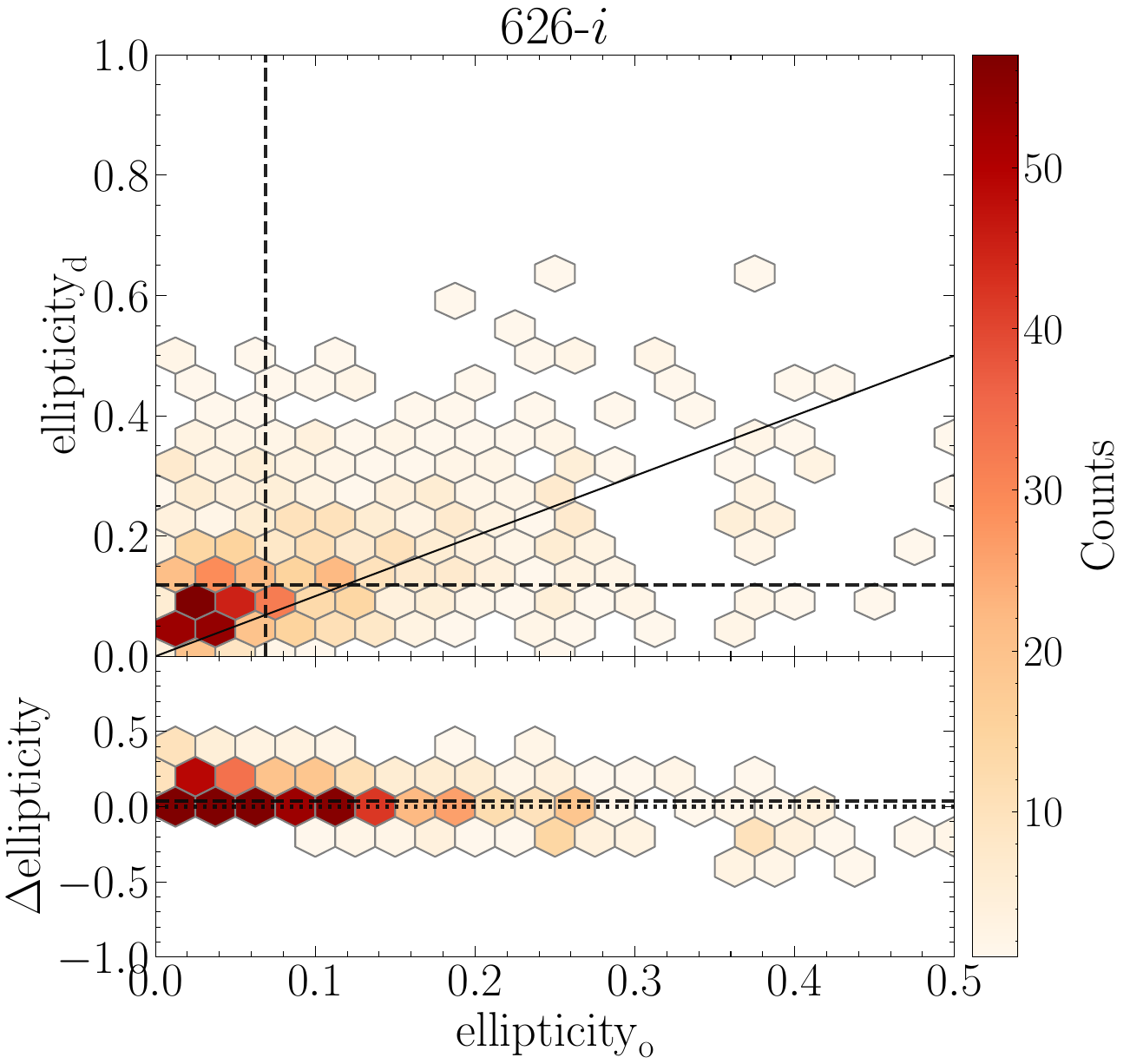}
      \includegraphics[keepaspectratio,width=0.24\linewidth]{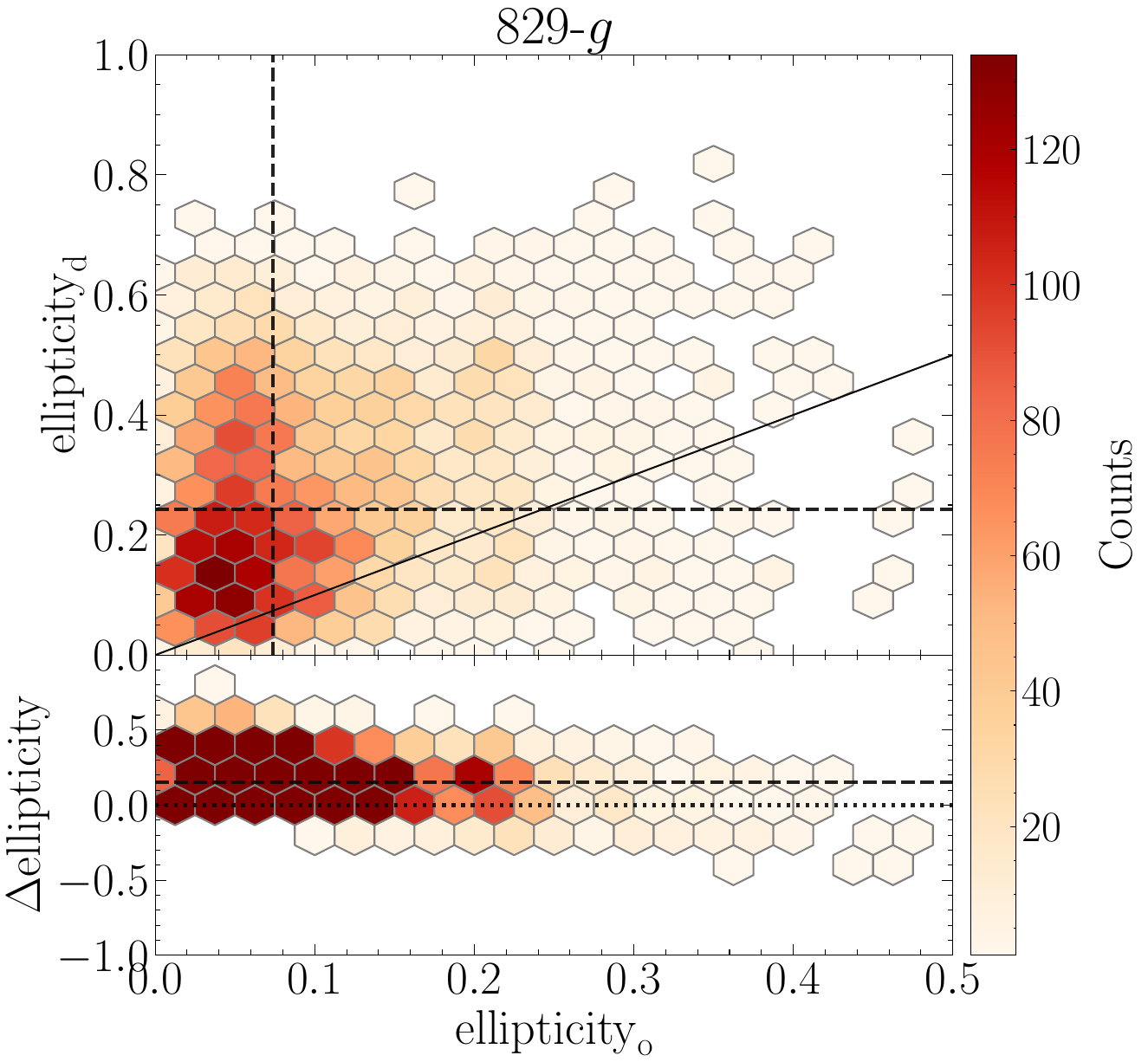}
      \includegraphics[keepaspectratio,width=0.24\linewidth]{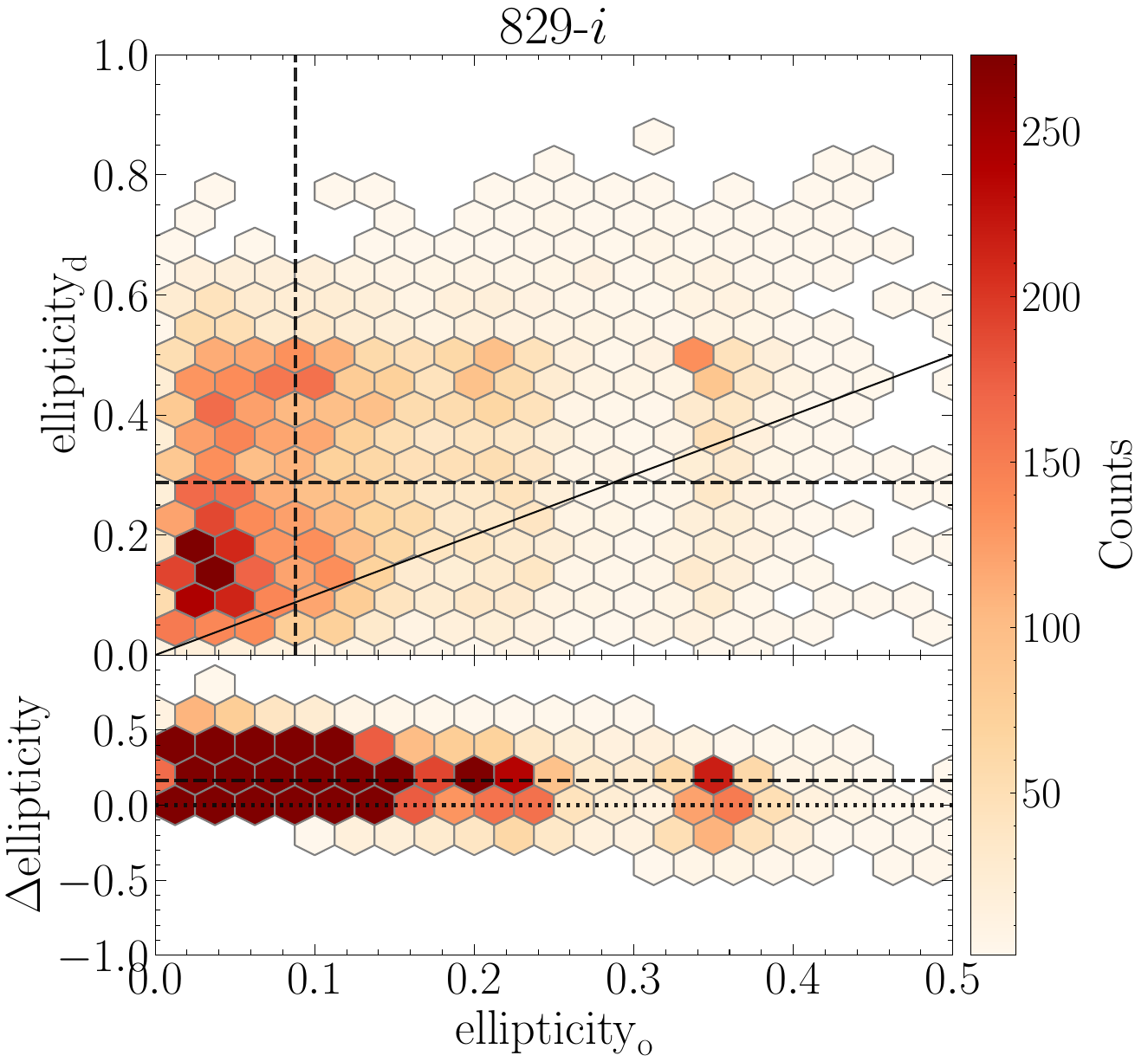}
      \includegraphics[keepaspectratio,width=0.24\linewidth]{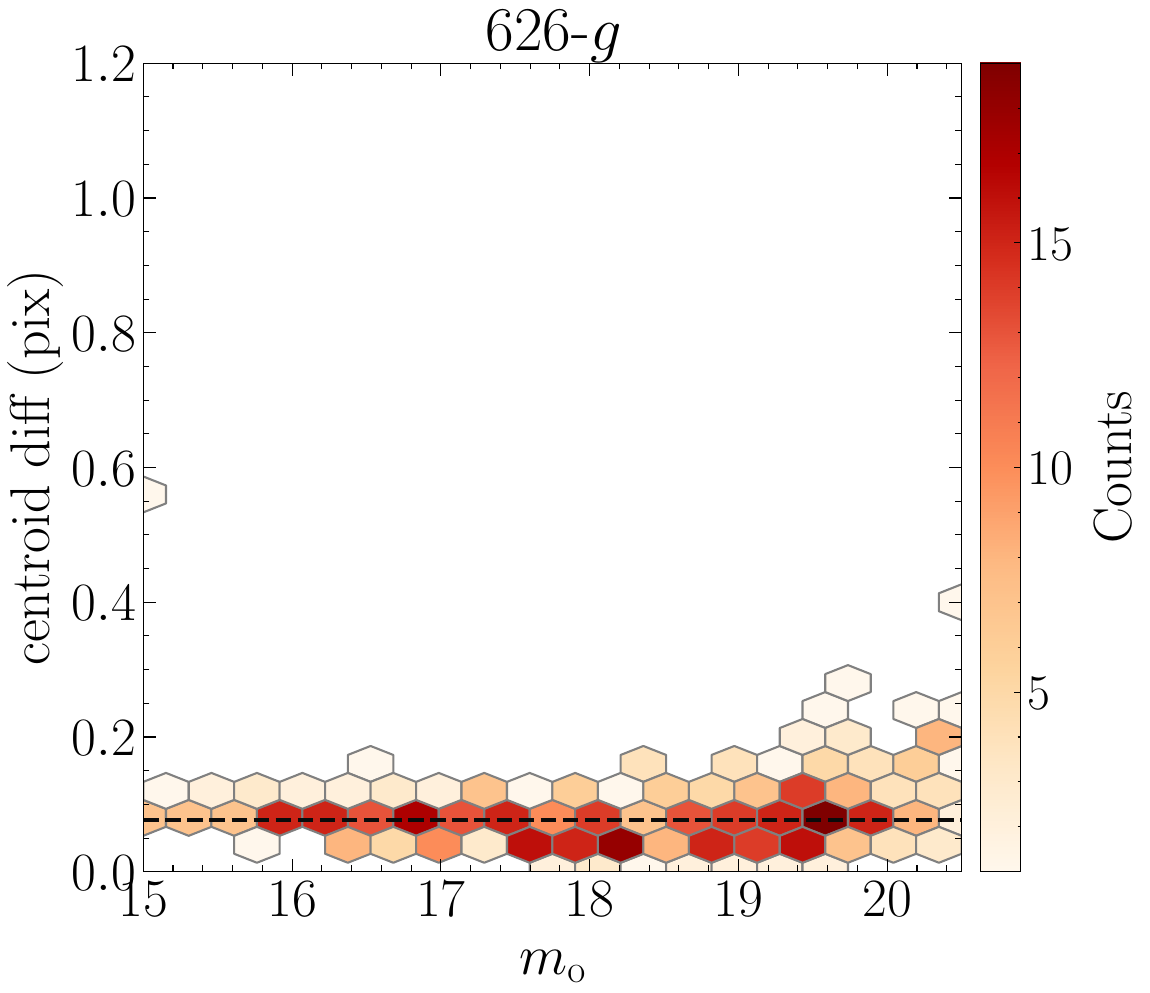}
      \includegraphics[keepaspectratio,width=0.24\linewidth]{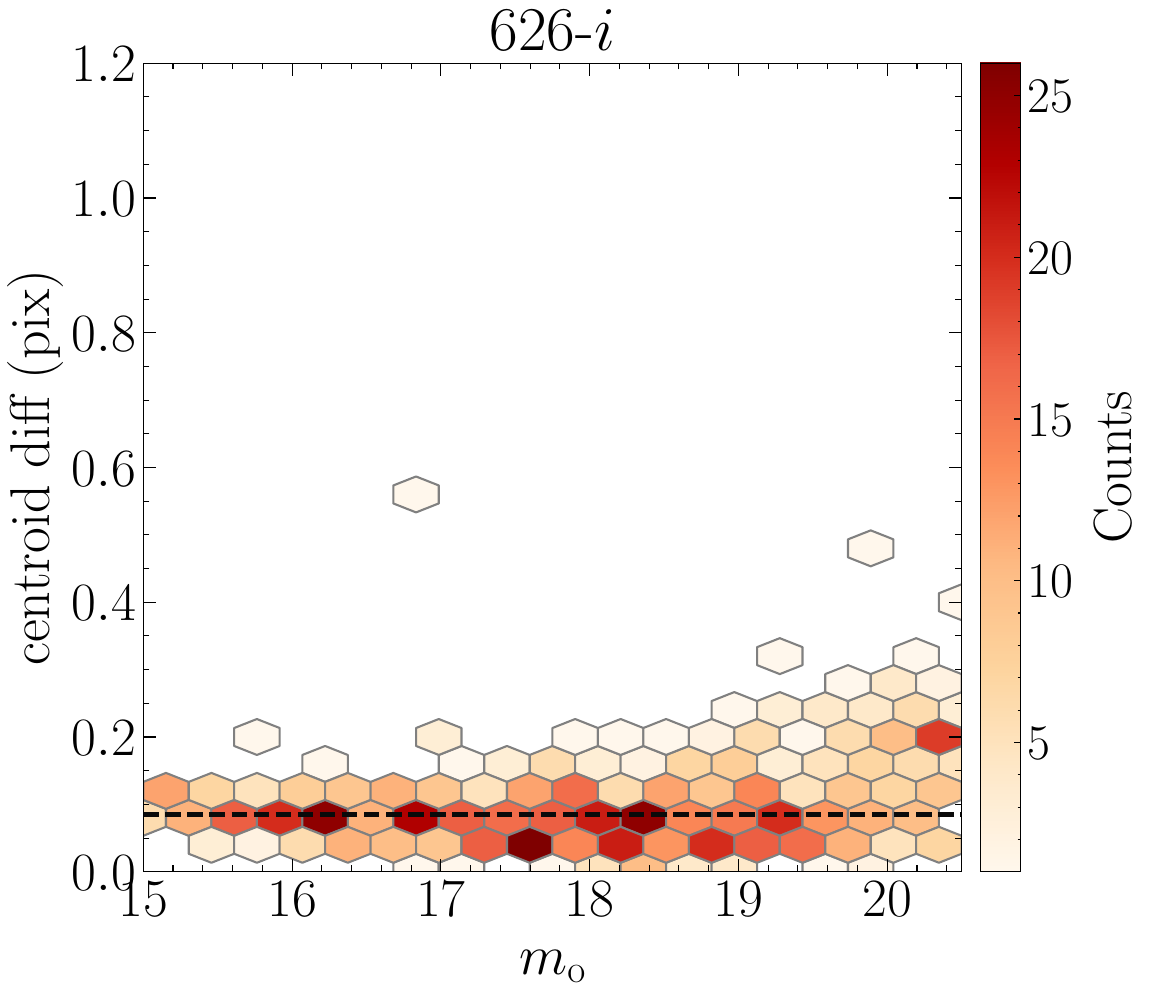}
      \includegraphics[keepaspectratio,width=0.24\linewidth]{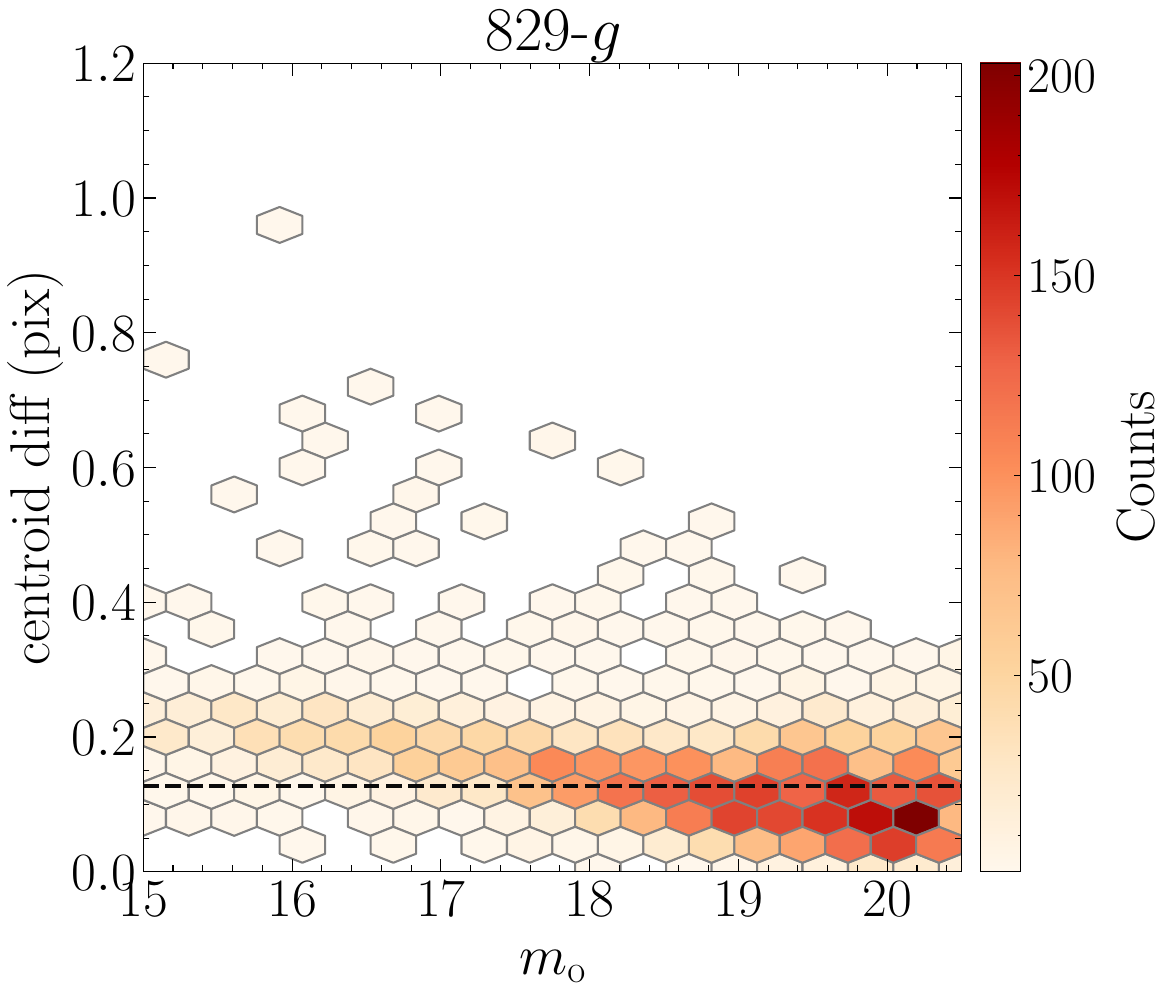}
      \includegraphics[keepaspectratio,width=0.24\linewidth]{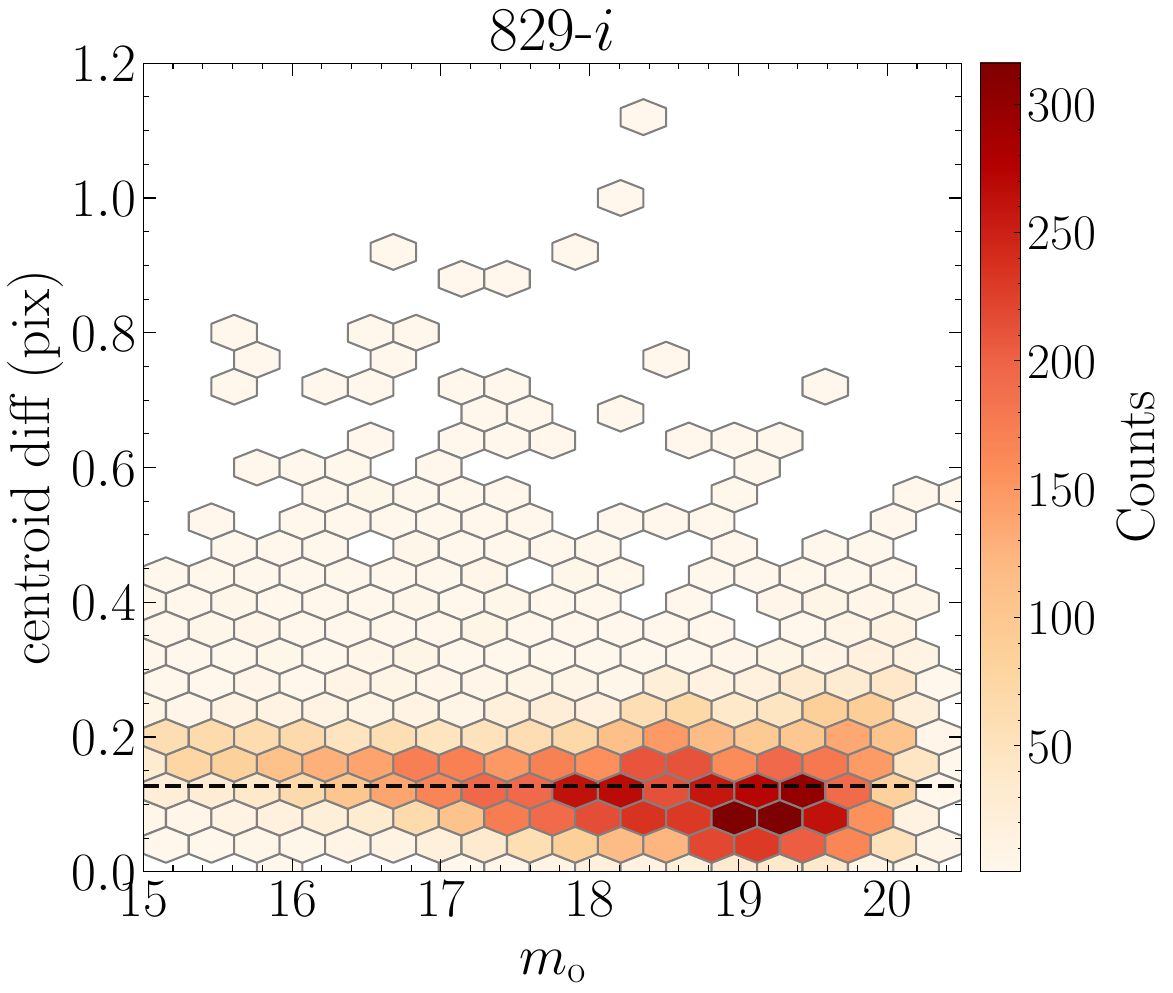}
    \caption{Comparison of the magnitude (top row), FWHM (second row), ellipticity (third row), and the centroid differences (bottom row) of the one-to-one matched original and deconvolved sources for the $g$ and $i$ filter images of fields with IDs 626 and 829. See Figs.~\ref{fig:mag-one-to-one-comparison}--\ref{fig:centroidDiff-one-to-one-comparison} for a description of the panels.} \label{fig:g-and-i-bands}
\end{figure*}

Our current implementation of deconvolution considers only a single band at once, so here we compare the deconvolution results of the $g$ and $i$ bands of the low- and high-galactic latitude fields with IDs 626 and 829. The $r$-band images for these fields were discussed in Sect.~\ref{sec:field-specific-results}. Fig.~\ref{fig:g-and-i-bands} presents various metrics--magnitude, FWHM, ellipticity, and centroid difference--for the one-to-one matches.

{\it Field ID 626}: The flux conservation in the $g$ band is better than that in the $i$ band, which itself is better than that in the $r$ band, as shown by the vertical dotted lines in the lower panels of the magnitude comparison plot. The scatter observed for the faintest original sources ($m \gtrsim 19.5$) is the largest in the $i$ band and lowest in the $g$ band. For all three bands, deconvolved sources corresponding to faint original sources are slightly brighter. The performance in terms of the FWHM is similar across all three bands. The median ellipticities of the original sources in $g$ and $i$ bands are 0.08 and 0.07, respectively, which are smaller than 0.12 in the $r$ band. The median deconvolved ellipticities in $g$ and $i$ are 0.12, which is also less than 0.2 in the $r$ band. The differences in centroids of the original and deconvolved sources for $g$ and $i$ bands do not show the negative trends between centroid differences and the original source magnitude observed for the $r$ band in Fig.~\ref{fig:centroidDiff-one-to-one-comparison}. The median centroid differences are 0.1 pixels for $g$ and $i$ bands compared to the 0.2 pixel difference in the $r$ band.

{\it Field ID 829}: The flux conservation in the $i$ band is better than that in the $r$ band, which itself is better than in the $g$ band, as shown by the vertical dotted lines in the lower panels of the magnitude comparison plot. The scatter observed for the faintest original sources is largest in the $i$ band and lowest in the $g$ band. For all three bands, deconvolved sources corresponding to faint original sources are slightly brighter. The performance in terms of the FWHM is similar across all three bands. The median ellipticities of the original sources in $g$ and $i$ bands are 0.07 and 0.09, respectively, which are smaller than 0.11 in the $r$ band. The median deconvolved ellipticities in $g$ and $i$ are 0.24 and 0.29, but unlike the case in the field with ID 626, these ellipticities are not significantly smaller than 0.25 in the $r$ band. The median centroid differences are 0.1 pixels for all three bands.

Thus, we find that how the FWHM of the sources changes in the deconvolved compared to the original remains most stable across different bands and across both fields. The magnitude agreement is also similar across different bands and fields, but specific patterns may differ across different bands, especially towards fainter original sources. The ellipticity across different bands may depend not only on the band but also on whether it is a low- or high-galactic latitude field. The centroid differences also visually show different patterns, but we consider them mostly inconsequential since the range of values is similar.

\begin{figure*}[hbt!]
    \centering
      \includegraphics[keepaspectratio,width=0.24\linewidth]{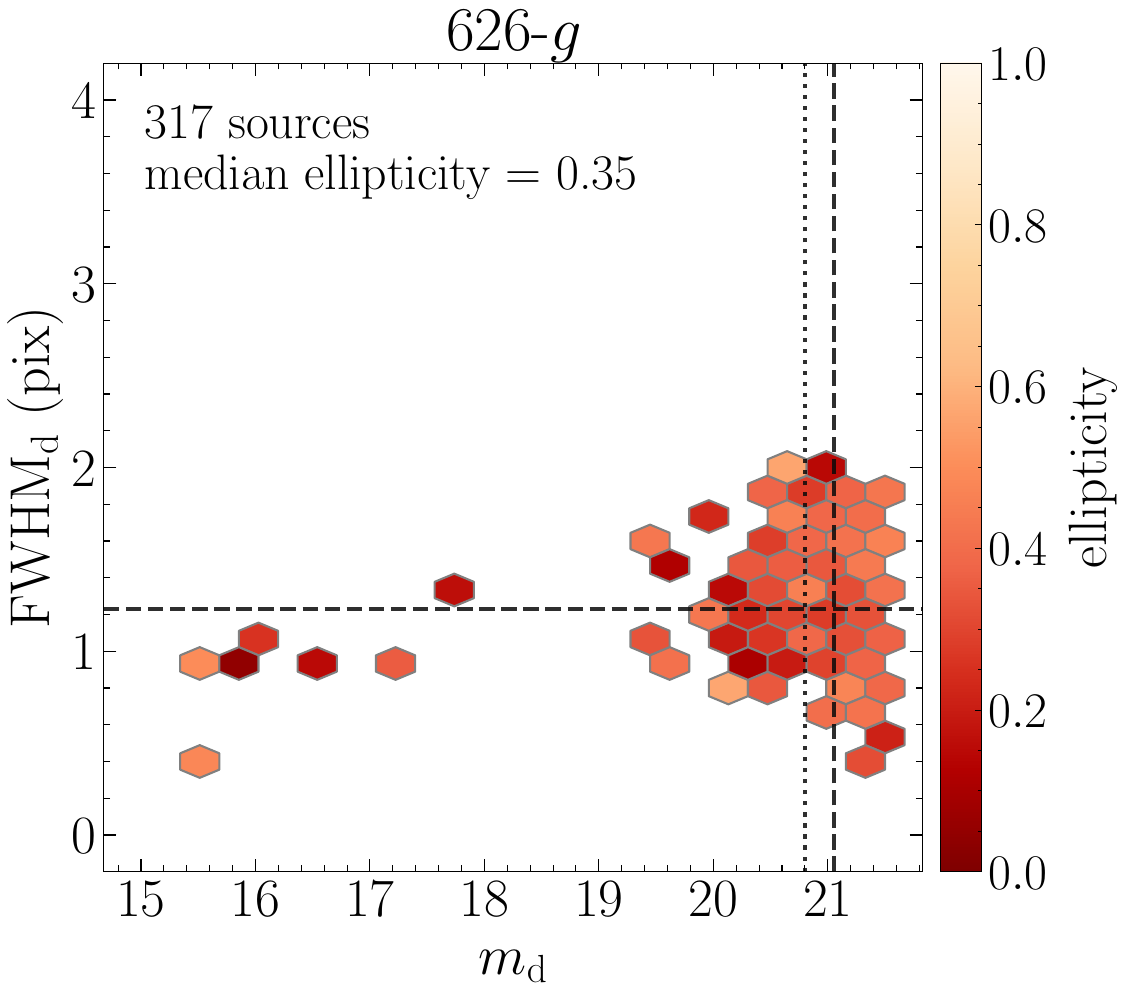}
      \includegraphics[keepaspectratio,width=0.24\linewidth]{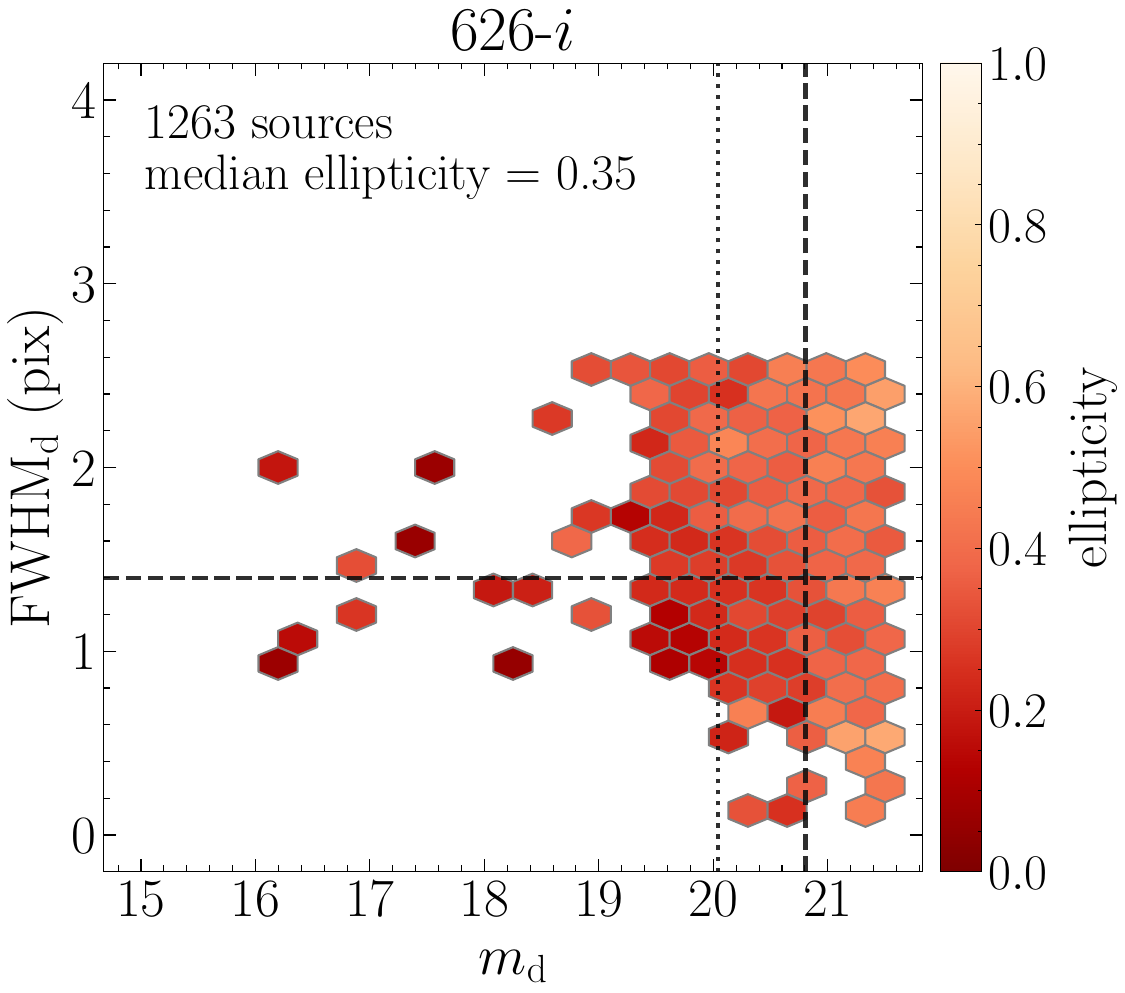}
      \includegraphics[keepaspectratio,width=0.24\linewidth]{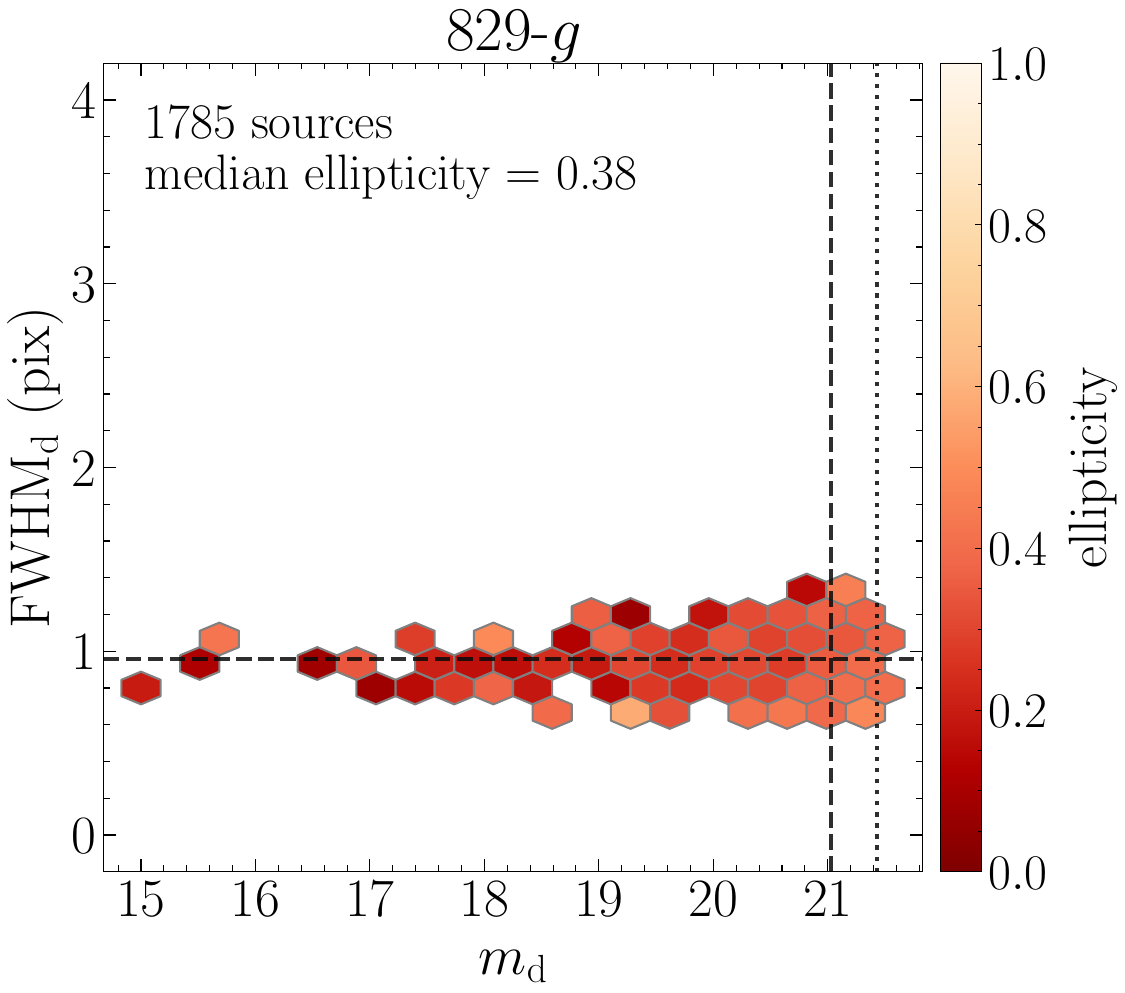}
      \includegraphics[keepaspectratio,width=0.24\linewidth]{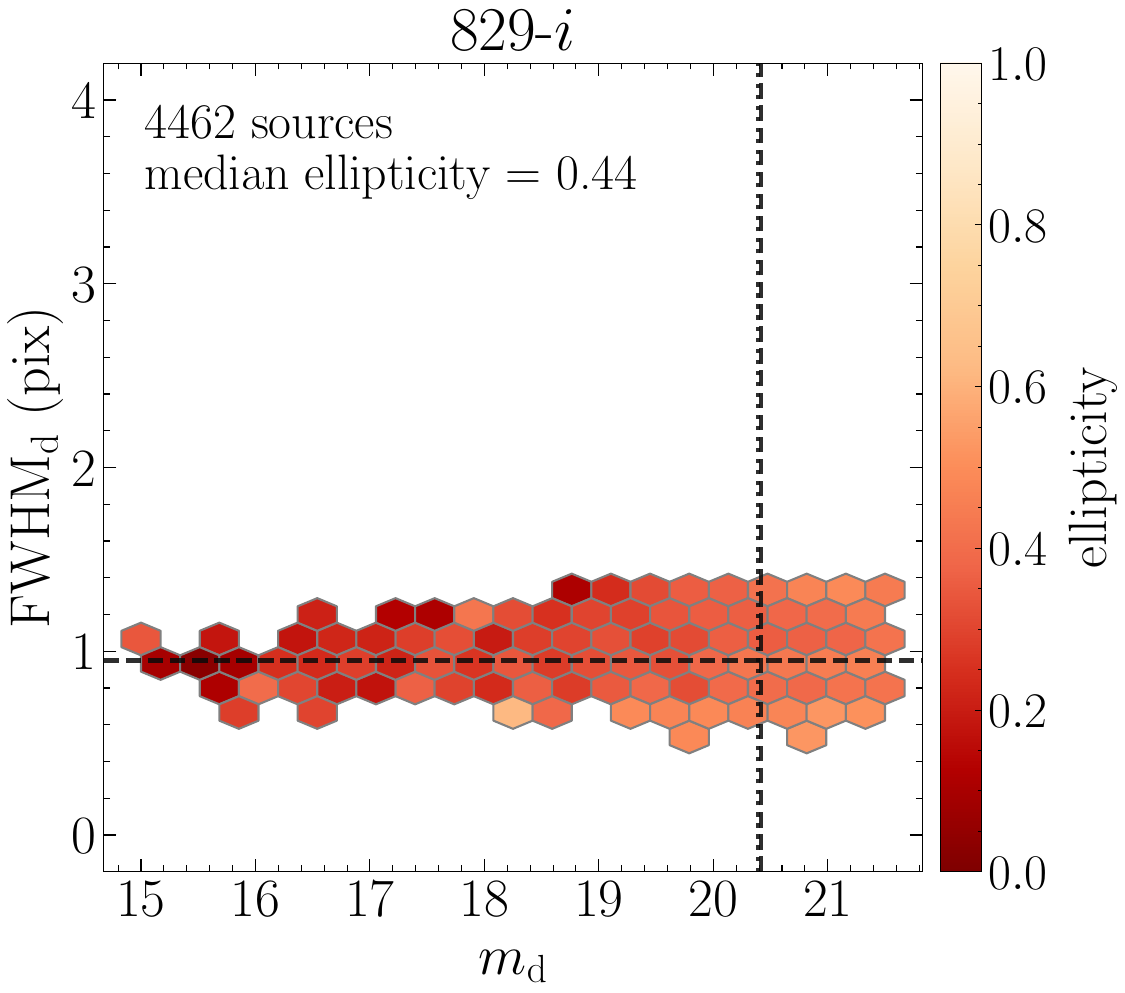}
    \caption{Distribution of the magnitude, FWHM, and ellipticity of the unmatched deconvolved sources for the $g$ and $i$ filter images considered in this study. See Fig.~\ref{fig:unmatched-deconvolved-combined} for a description of panels and related information. The median magnitude, FWHM, and ellipticity of these newly detected deconvolved sources are in a similar range to the corresponding $r$-band examples in Fig.~\ref{fig:unmatched-deconvolved-combined}. Like for the cases shown in Sect.~\ref{sec:deblending-examples} for the $r$-band, we find that some of these new deconvolved sources are a result of the deblending of original sources that SExtractor could not deblend.} \label{fig:unmatched-deconvolved-combined-g-and-i}
\end{figure*}

We now briefly discuss the unmatched deconvolved sources. For both fields, we find that the largest number of newly detected deconvolved sources are found in the $i$ band, followed by the $r$ band, and the least in the $g$ band. Out of the 501, 3012, 1566, and 6277 unmatched deconvolved sources for 626-$g$, 829-$g$, 626-$i$, and 829-$i$, 317, 1785, 1263, and 4462 sources, respectively, meet the additional astrophysical cuts outlined in Sect.~\ref{subsec:crossmatch-results}, making them likely astrophysical. Fig.~\ref{fig:unmatched-deconvolved-combined-g-and-i} shows the distribution of the magnitudes, FWHM, and ellipticities of these newly identified deconvolved sources in the $g$ and $i$ bands. The median FWHM and ellipticities for the $g$ and $i$ bands of field ID 626 are similar to those in the $r$ band, and that of the $g$ and $i$ bands of field ID 829 are also similar to those in the $r$ band.

\subsection{\reviewTwo{Validation of unmatched deconvolved sources with external data}}\label{sec:deepref-experiment}

\reviewTwo{In this section, we perform a series of checks using the DESI Legacy Imaging Surveys \citep{Dey2019} DR10 data\footnote{\url{https://www.legacysurvey.org/}} (LS; going a few magnitudes deeper) and ZTF forced photometry \citep{Masci2023} with the primary objective of determining the validity (real vs. fake) of deconvolved sources not matched to any source in the corresponding science image. As a proof of concept, we apply our procedure to three $r$ filter images of ZTF fields with IDs 619, 626, and 635.}

\reviewTwo{We first retrieve the LS DR10 Tractor catalogs lying in the area of sky spanned by our ZTF fields using a simple cone search protocol in Topcat and crossmatch all deconvolved sources with the retrieved LS catalogs using a crossmatching threshold of 1$\arcsec$.8 (slightly higher than used in previous sections to allow for seeing differences across surveys). To ensure crossmatches are of good quality, we perform the following selection cuts \reviewThree{to select real astrophysical LS sources:} (a) type = `PSF', (b) $\mathrm{fracmasked}_{i} < 0.6$, (c) $\mathrm{nobs}_i > 0$, (d) $\mathrm{fracflux}_i < 0.01$, and (e) $\mathrm{flux\_ivar}_i > 0$, where $i = g$, $r$, and $z$, with specific values taken from the DESI target selection pipeline \citep{Myers2023}. Further, since the LS sources are located in a cone on the sky, we remove excess sources that lie within this region but outside the ZTF fields. We consider deconvolved sources that are successfully crossmatched with LS to be real.}

\reviewTwo{Table~\ref{tab:crossmatching-results-OrigDeconDESI} shows the crossmatching statistics in various cases depending on whether a source was detected in one, two, or all three catalogs. Our primary focus is on sources in columns 3, 6, and 8. The former set of sources is present in the deconvolved image and LS but not in the original image, whereas the latter set of sources is present in the original image and LS but not in the deconvolved image. The total number of deconvolved sources unmatched in the original (i.e., $\bar{O}D$) is mentioned in Fig.~\ref{fig:unmatched-deconvolved-combined}.}

\begin{table*}
\renewcommand{\arraystretch}{1.25}
\centering
\caption{\reviewTwo{Crossmatching results between the original ZTF (``O''), the deconvolved ZTF (``D''), and the LS DR10 catalogs (``L'') for ZTF fields with ID 619, 626, and 635 in the $r$ filter. Columns: (1) the ZTF field ID, (2) one-to-one matches across the three fields, (3) deconvolved sources unmatched in the original image but matched with LS (only one-to-one matches exist between such deconvolved sources and LS, and there are no cases of many-to-one or one-to-many matches): values outside square brackets show one-to-one matches considering only LS sources passing the selection cuts described in the main text, whereas values inside show all one-to-one matches without any cuts on LS sources (for our analysis ahead, we only consider the ones outside the square brackets), (4) deconvolved sources unmatched in LS but with a match in the original image (all are one-to-one matches), (5) LS sources that are unmatched in the deconvolved image and also unmatched in the original image, (6) original sources unmatched in the deconvolved image but matched with LS (only one-to-one matches exist between original and LS), (7) original sources unmatched in the deconvolved image and also unmatched in LS, (8) deconvolved sources unmatched in original image and also unmatched with LS. We use the original and deconvolved ZTF catalogs after applying the basic selection criteria from Sect.~\ref{subsec:experimental-details}. For deconvolved sources for which a match in the original image was not found (columns 3 and 8), we only select deconvolved sources that pass the additional astrophysical selection criteria from Sect.~\ref{subsec:crossmatch-results}. For column 5, we only include LS sources with magnitude $< 21.5$ in the $r$-band as that is the faintest magnitude we consider for ZTF. For column 2, note that there exists a tiny number of many-to-one matches between the deconvolved image and LS and also separately between the deconvolved and the original image (i.e., more than one deconvolved source matched to the same LS or the original source), but we do not include them in these statistics. Notes: (1) One would expect the sum of the no. of sources inside the square brackets (not the ones outside because those are only a subset of those that pass our criteria on LS sources, not all LS sources) in column 3 and in column 8, i.e., $\bar{O}DL$ + $\bar{O}D\bar{L}$ to give exactly those in $\bar{O}D$ and thus match the no. of sources depicted in Fig.~\ref{fig:unmatched-deconvolved-combined}. However, their sum is slightly smaller than the values shown in that figure because there exist cases where a deconvolved source was matched to an LS source lying just outside the sky region spanned by our ZTF field, but that match was removed by our hard cuts on (ra, dec) of LS source mentioned in the main text. (2) When checking for a match/non-match between original and deconvolved catalogs, we stick to the 1$\arcsec$.4 threshold used throughout our discussion of previous results, whereas we use 1$\arcsec$.8 for checking a match/non-match between original and LS or deconvolved and LS catalogs.}}
\label{tab:crossmatching-results-OrigDeconDESI}
\begin{tabular}{@{}crrrrrrr@{}}
\hline\hline
ZTF Field ID & $ODL$ & $\bar{O}DL$ & $OD\bar{L}$ & $\bar{O}\bar{D}L$ & $O\bar{D}L$ & $O\bar{D}\bar{L}$ & $\bar{O}D\bar{L}$\\ 
(1) & (2) & (3) & (4) & (5) & (6) & (7) & (8)\\ \midrule
619 & 1027 & 223 [614] & 26 & 498 & 0 & 0 & 51  \\
626 & 581  & 145 [441] & 10 & 868 & 1 & 0 & 128 \\
635 & 3928 & 517 [1033] & 18 & 3045 & 0 & 0 & 32
\end{tabular}
\end{table*}

\reviewTwo{The table shows that 223 out of 671, 145 out of 577, and 517 out of 1072 deconvolved sources in $\bar{O}D$, for field IDs 619, 626, and 635, respectively, are matched with LS (column 3), and are thus definitely real (corresponding to 33.2\%, 25.3\%, 48.2\% respectively). Here, 671, 577, and 1072 are the total number of likely astrophysical unmatched deconvolved sources discussed in Fig.~\ref{fig:unmatched-deconvolved-combined}. However, the actual fraction of real deconvolved sources is likely greater because the fraction increases significantly when no selection criteria on LS sources are used (614 out of 671, 441 out of 577, and 1033 out of 1072 corresponding to 91.5\%, 76.4\%, and 96.3\%, respectively)\reviewFive{\footnote{We have confirmed internally that the LS sources meeting our selection criteria show similar magnitude and FWHM distributions to those that do not. Therefore, a large fraction of the excluded LS sources is expected to be reliable. This supports the claim that the actual fraction of real deconvolved sources is likely greater than the number of LS sources that meet the selection criteria.}}. Thus, the actual fraction of newly found real deconvolved sources lies somewhere in between, depending on the strictness of the selection criteria applied to select LS sources. Most matches have less than an arcsecond separation, with a handful of cases with a separation of up to roughly 1.3 arcsec.}

\begin{figure*}[hbt!]
    \centering
      \includegraphics[keepaspectratio,width=0.32\linewidth]{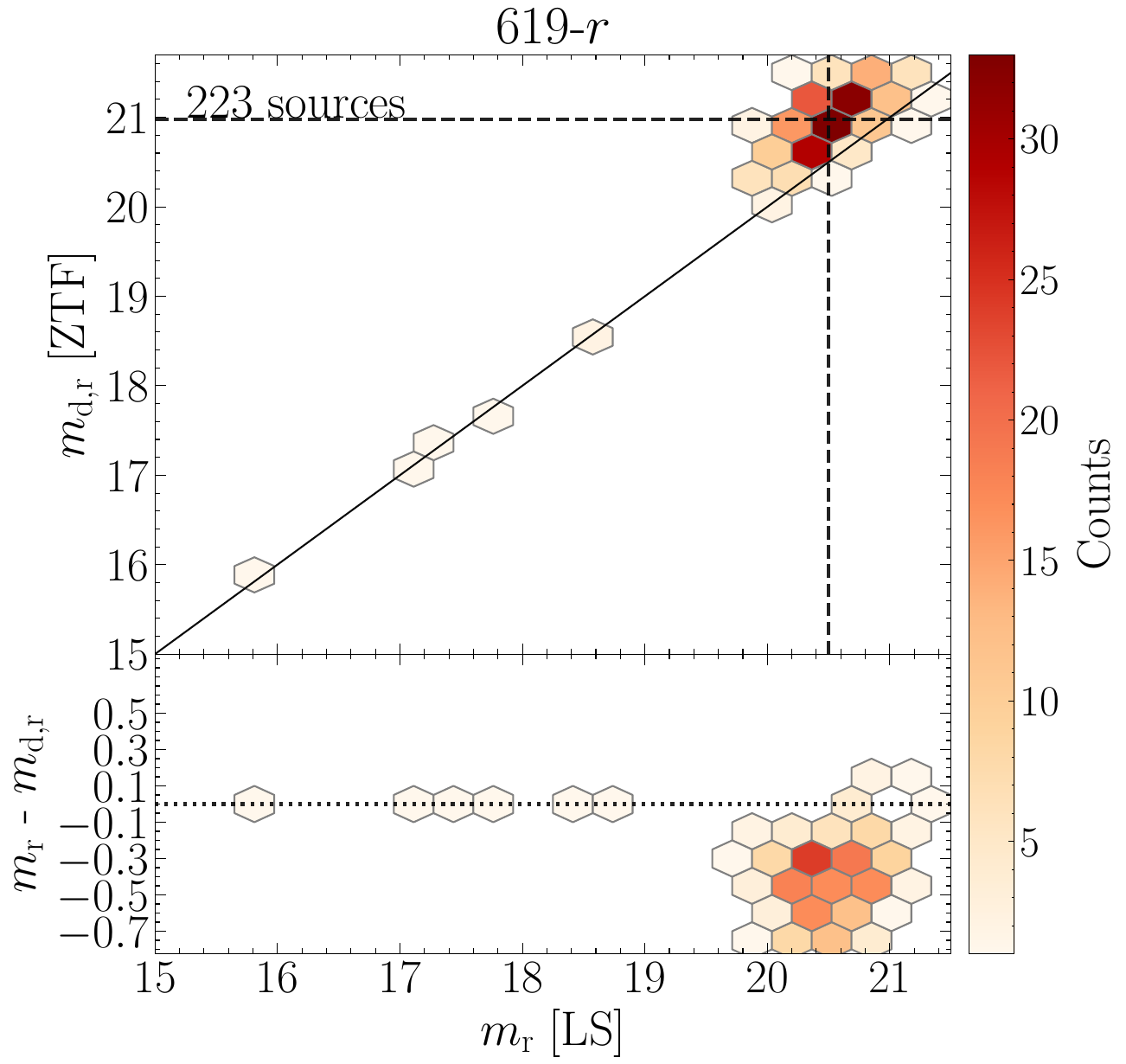}
      \includegraphics[keepaspectratio,width=0.32\linewidth]{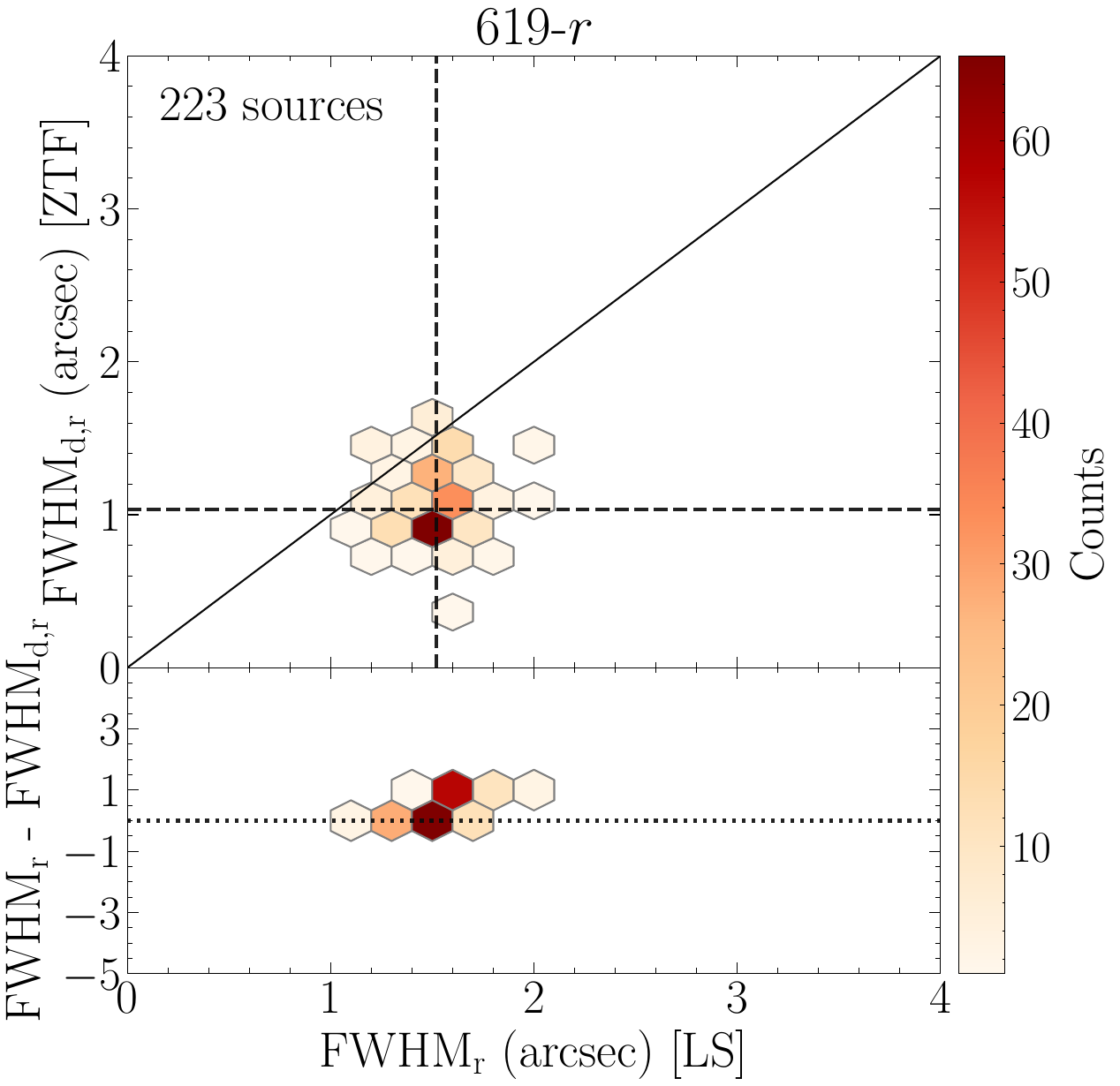}
      \includegraphics[keepaspectratio,width=0.32\linewidth]{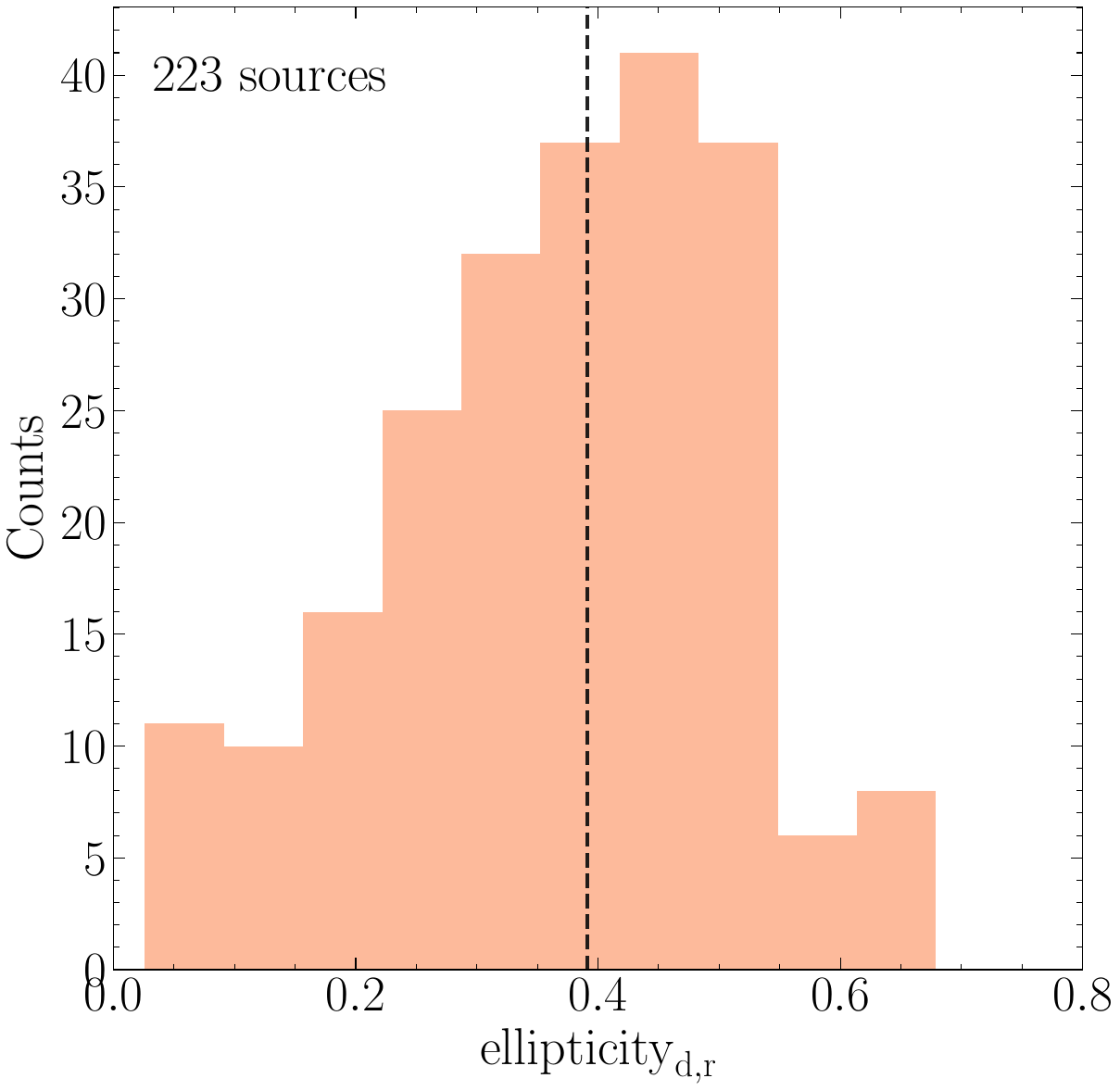}
      \includegraphics[keepaspectratio,width=0.32\linewidth]{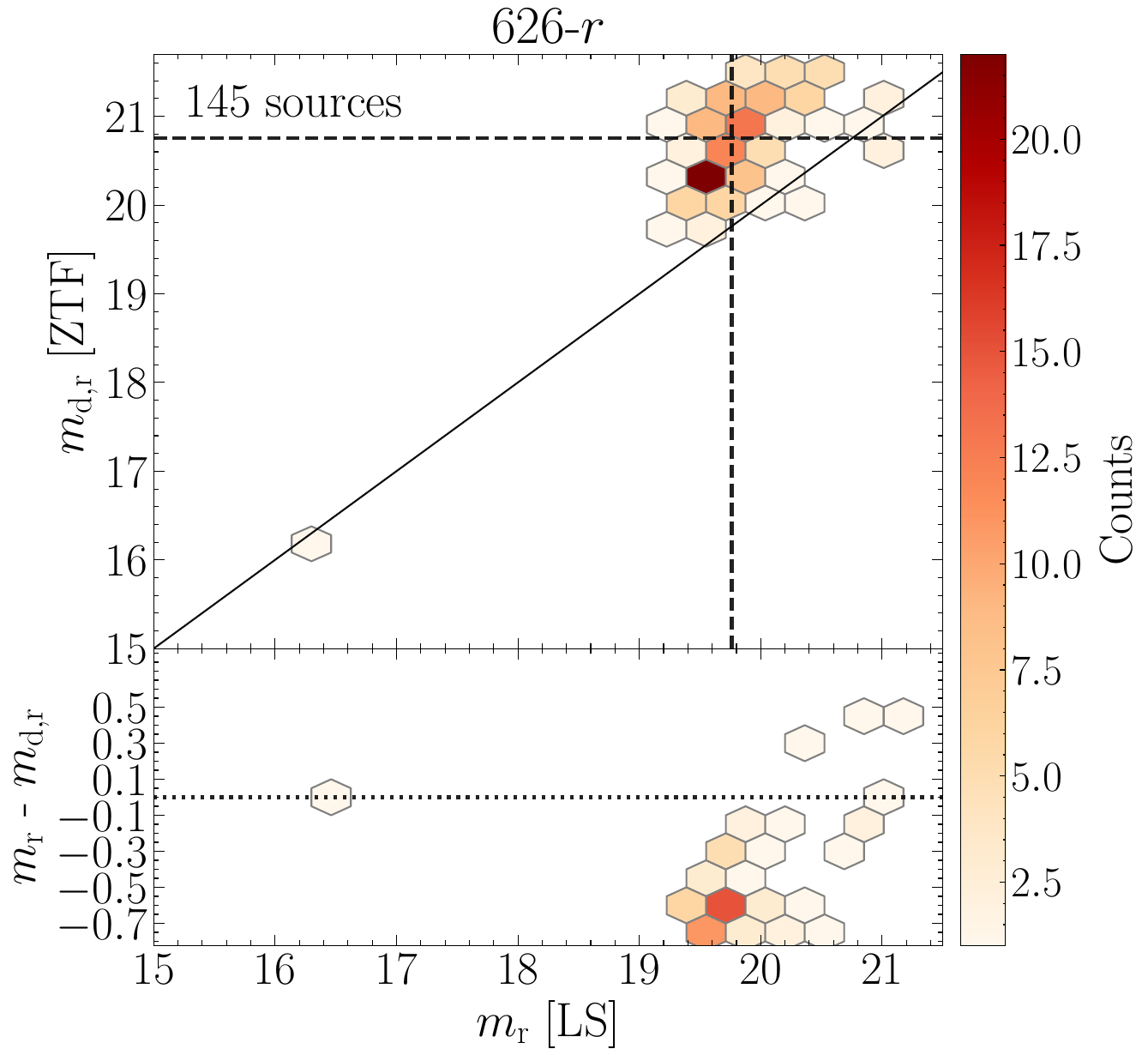}
      \includegraphics[keepaspectratio,width=0.32\linewidth]{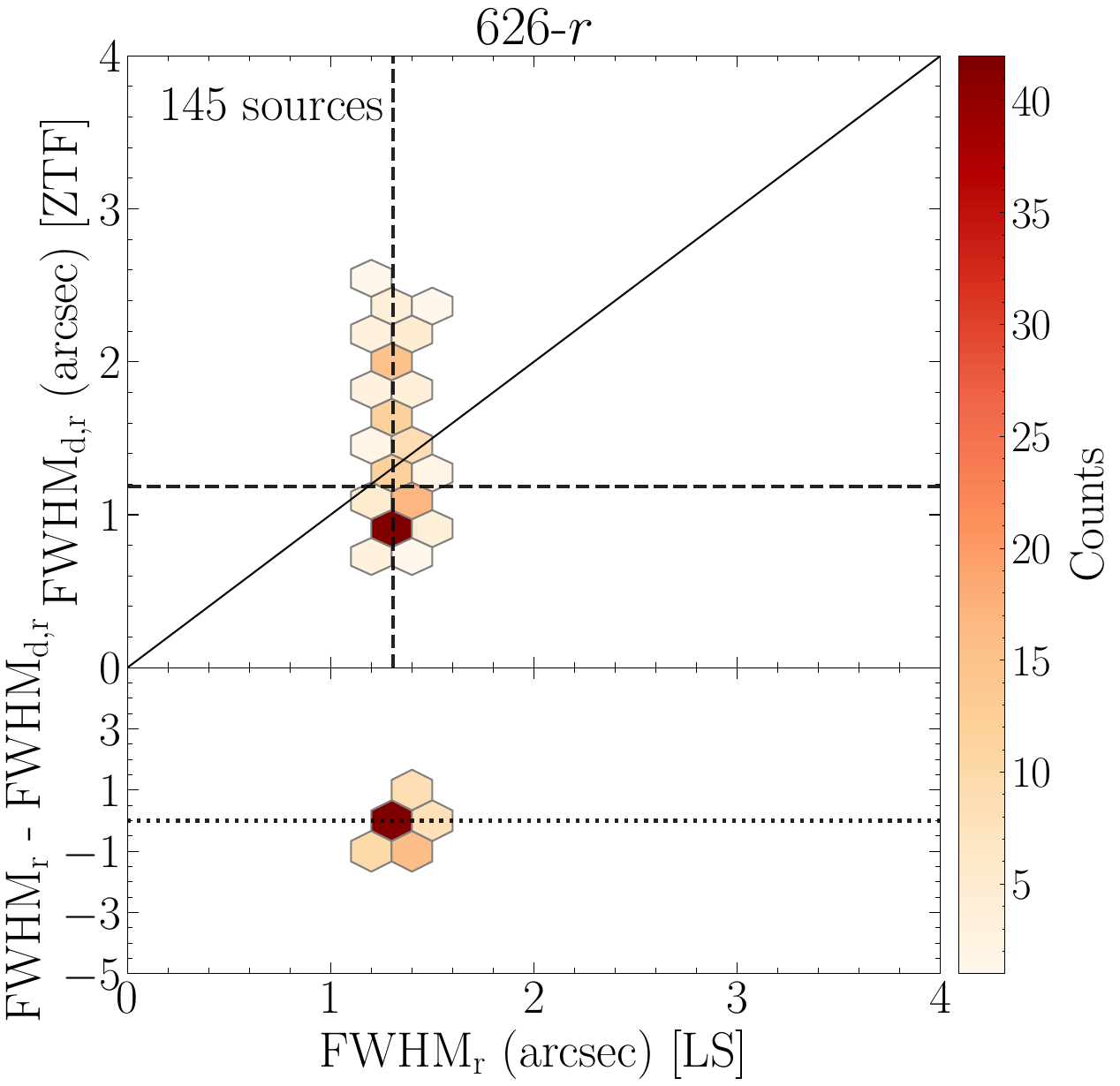}
      \includegraphics[keepaspectratio,width=0.32\linewidth]{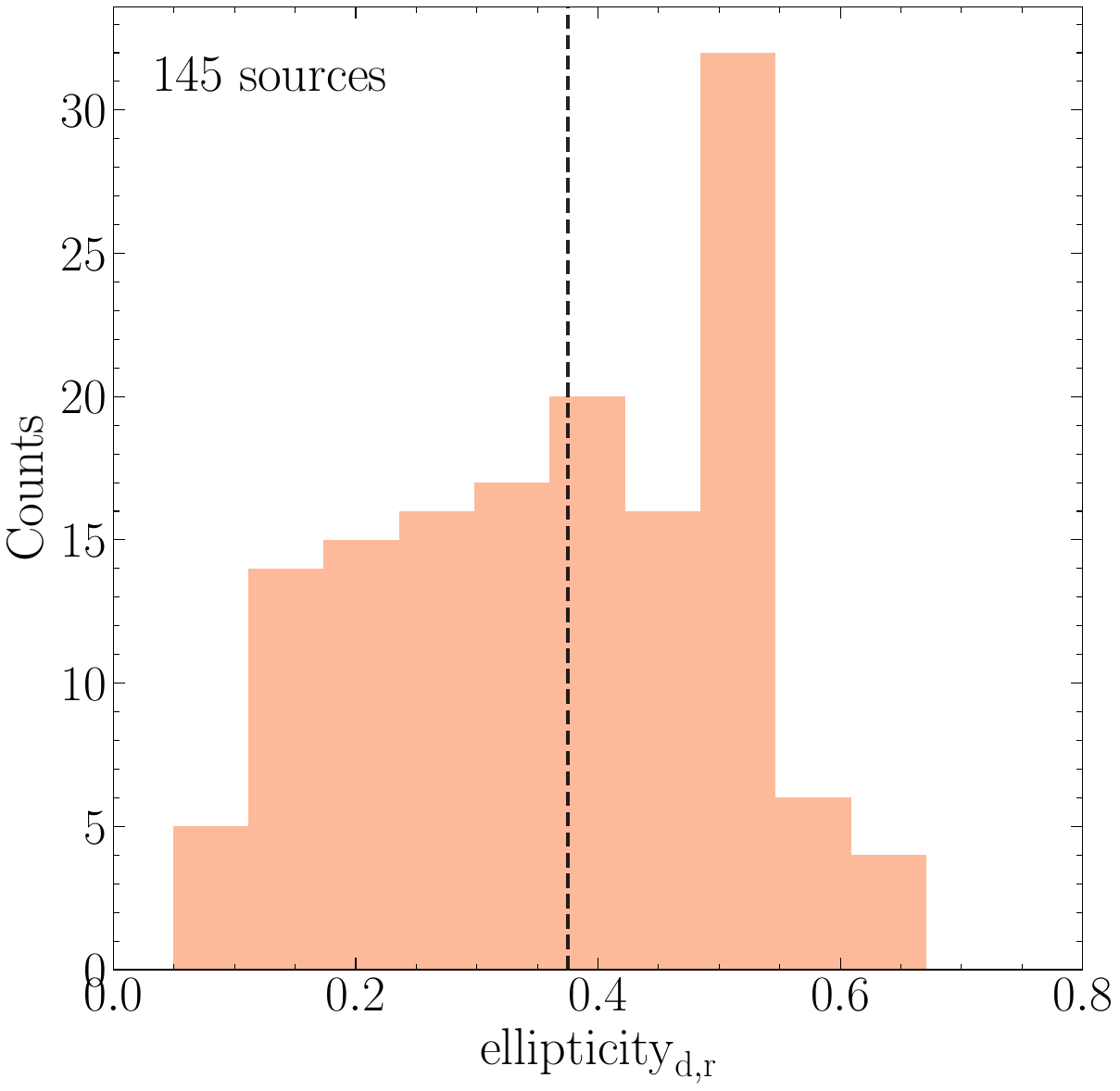}
      \includegraphics[keepaspectratio,width=0.32\linewidth]{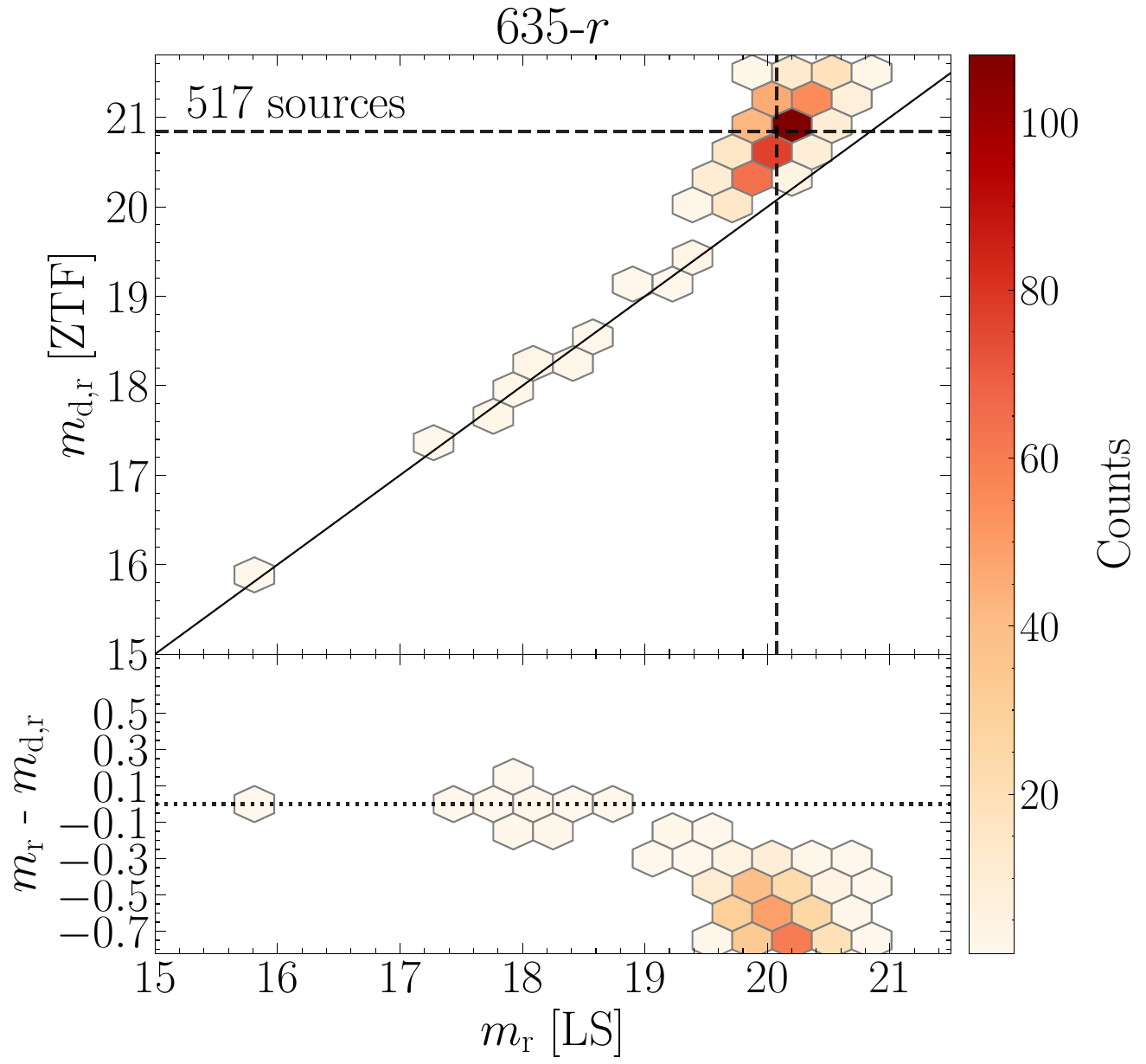}
      \includegraphics[keepaspectratio,width=0.32\linewidth]{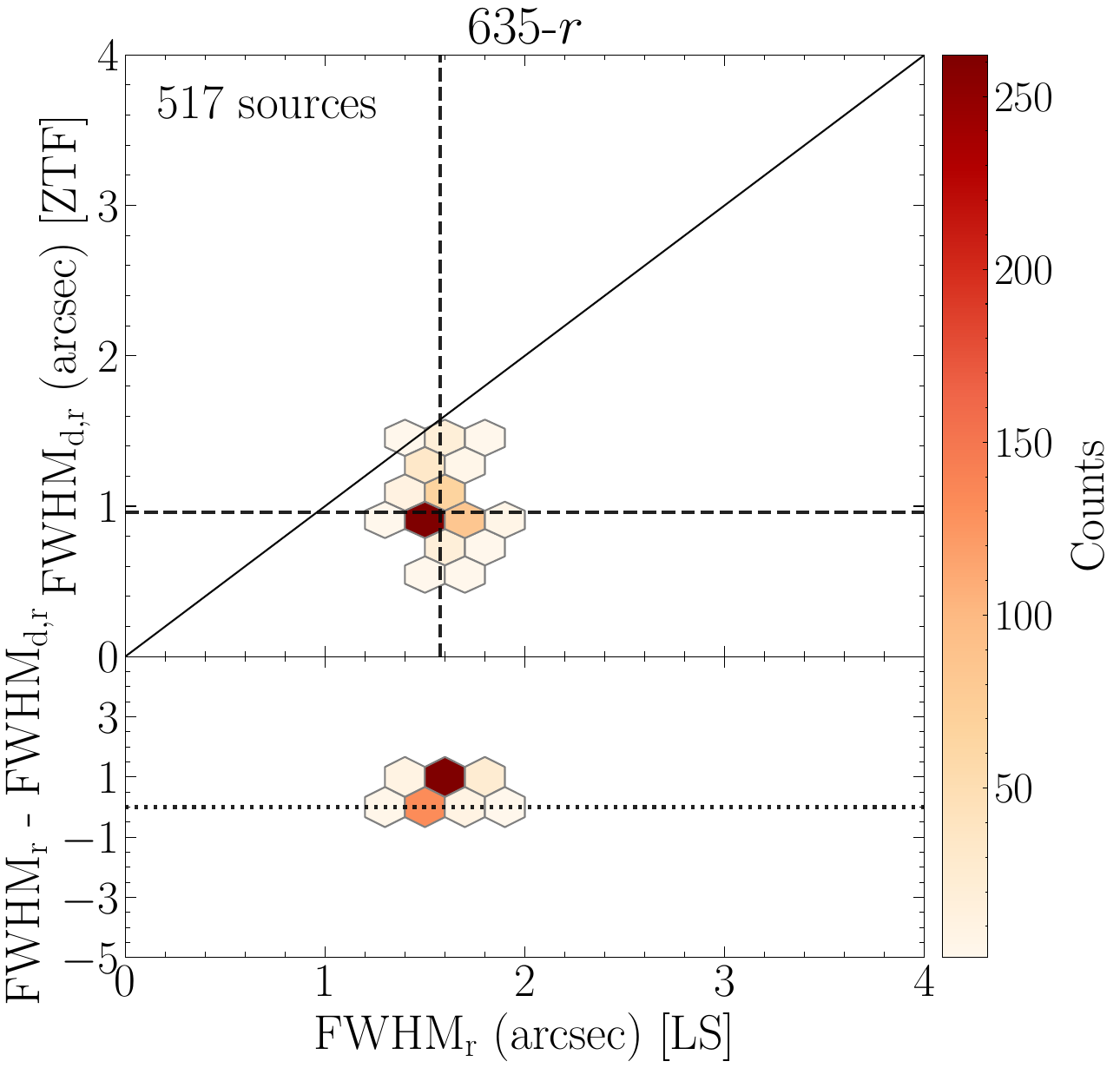}
      \includegraphics[keepaspectratio,width=0.32\linewidth]{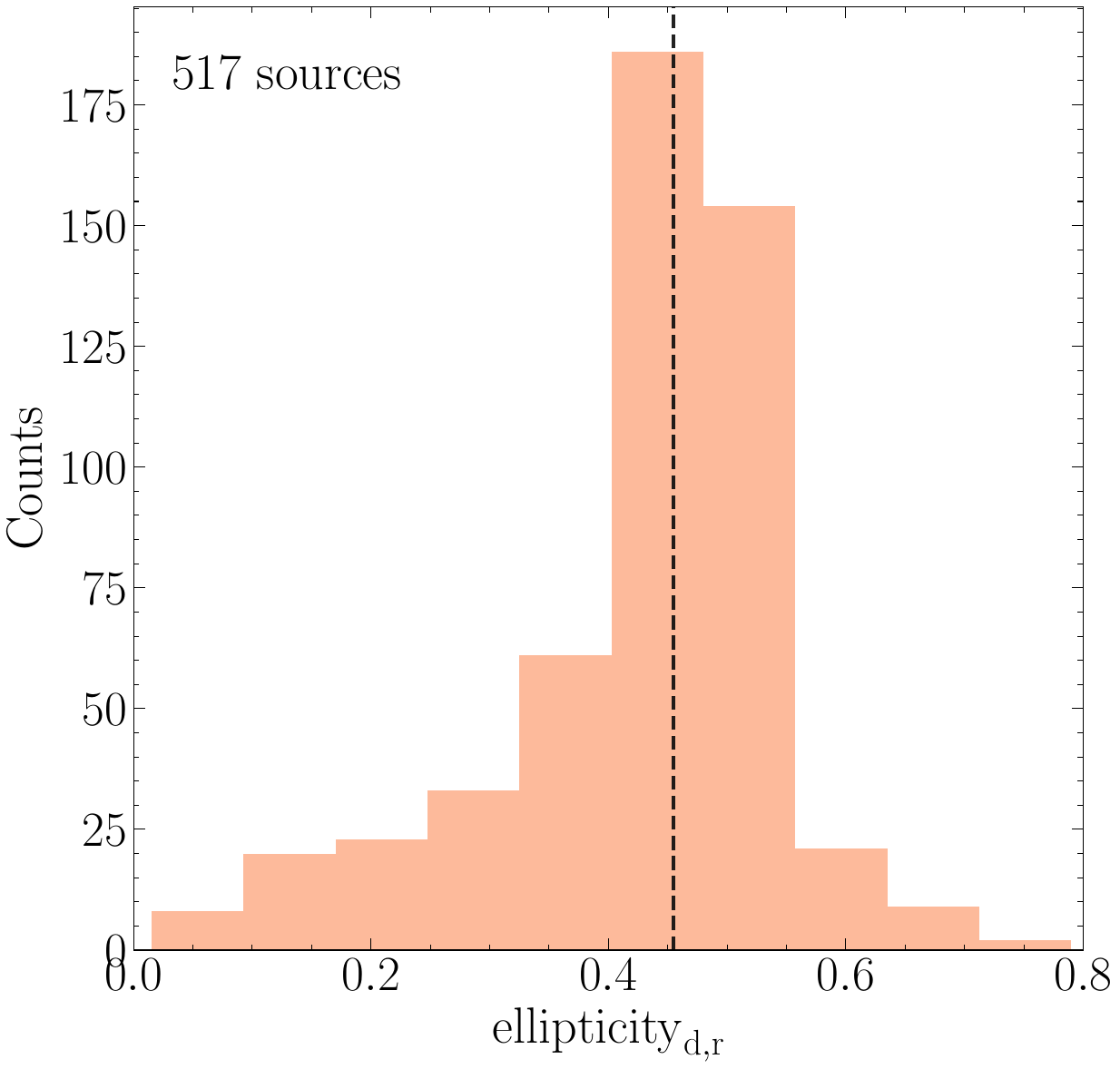}
    \caption{\reviewTwo{{\it First column}: Comparison of magnitudes of the one-to-one matched LS ($m_{\mathrm{r}}$) and deconvolved ($m_{\mathrm{d, r}}$) sources. The vertical dashed lines denote the median magnitude, and the horizontal dotted lines in the lower subpanels denote $m_{\mathrm{r}} = m_{\mathrm{d, r}}$. Ideally, a proper comparison would include converting both to AB magnitudes for photometric consistency, but that is beyond the scope of this work. {\it Second column}: Similar to column 1, but comparing the FWHM of the LS and deconvolved sources. {\it Third column}: Histogram of ellipticities of deconvolved sources matched one-to-one with LS sources. LS catalogs do not include ellipticities of stellar sources, so a one-to-one comparison is omitted.}} \label{fig:crossmatching-results-OrigDeconDESI-plots}
\end{figure*}

\reviewTwo{The properties of the deconvolved sources in column 3 (outside the square brackets) and comparison with corresponding LS sources are shown in Fig.~\ref{fig:crossmatching-results-OrigDeconDESI-plots}. Except for a few at the bright end (discussed below), such sources are concentrated toward the faint end ($m > 19$), as shown in the magnitude comparison plots. The magnitude of the faintest deconvolved sources reaches $m \approx 21.5$; although it should be noted that the deconvolved magnitudes in this faint regime could be offset by as much as $\lvert \Delta m \rvert \approx 0.7$, as seen in the lower subpanels in Fig.~\ref{fig:mag-one-to-one-comparison}, this still indicates that deconvolution is able to identify sources closer to and fainter than the limiting magnitude of the original images. The typical FWHM of deconvolved sources is around an arcsecond, which is slightly smaller than the corresponding LS sources. However, we do see the FWHM of some deconvolved sources as being broader than the corresponding LS sources for field ID 626. The deconvolved sources are more elliptical (median ellipticity $\sim$ 0.4) than those matched to a source in the original images in Fig.~\ref{fig:ellipticity-one-to-one-comparison}.}

\reviewTwo{As indicated in the caption of Fig.~\ref{fig:unmatched-deconvolved-combined}, the presence of relatively bright deconvolved sources (with $m < 19$) that do not match any sources in the original image may be because deconvolution could have deblended the original source in such a manner that it no longer corresponds to the original detection, or because there was a match that was filtered out by our selection criteria, or because SExtractor did not detect the original source or flagged it. In our analysis, we have found three such bright deconvolved sources in field ID 619 that arise from likely instances of deblending. However, the remaining ones for field ID 619 and all bright deconvolved sources for field IDs 626 and 635 that did not match any original source can be attributed to the original sources being undetected, even in the unfiltered detection catalogs.}

\reviewTwo{The statistics in column 6 show that there is only one example of a source present in the original image and LS but not in the deconvolved image. The location of the ZTF source that was undetected after deconvolution is (192.642, 27.135) deg, and it was located close to a saturated source, although we observe that the deconvolved source is visually perceivable at the location. As discussed in Sect.~\ref{sec:field-specific-results}, SExtractor tends to miss deconvolved sources near saturated sources. This analysis confirms that finding.}

\begin{figure*}[hbt!]
    \centering
      \includegraphics[keepaspectratio,width=0.32\linewidth]{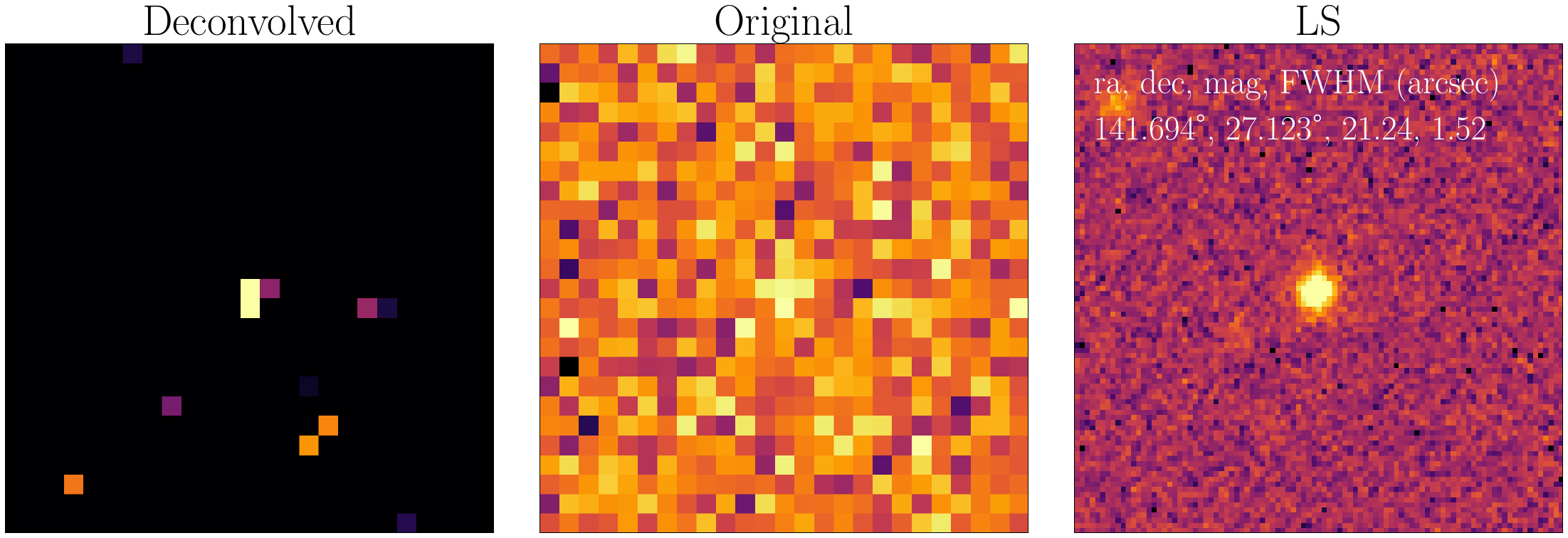}
      \includegraphics[keepaspectratio,width=0.32\linewidth]{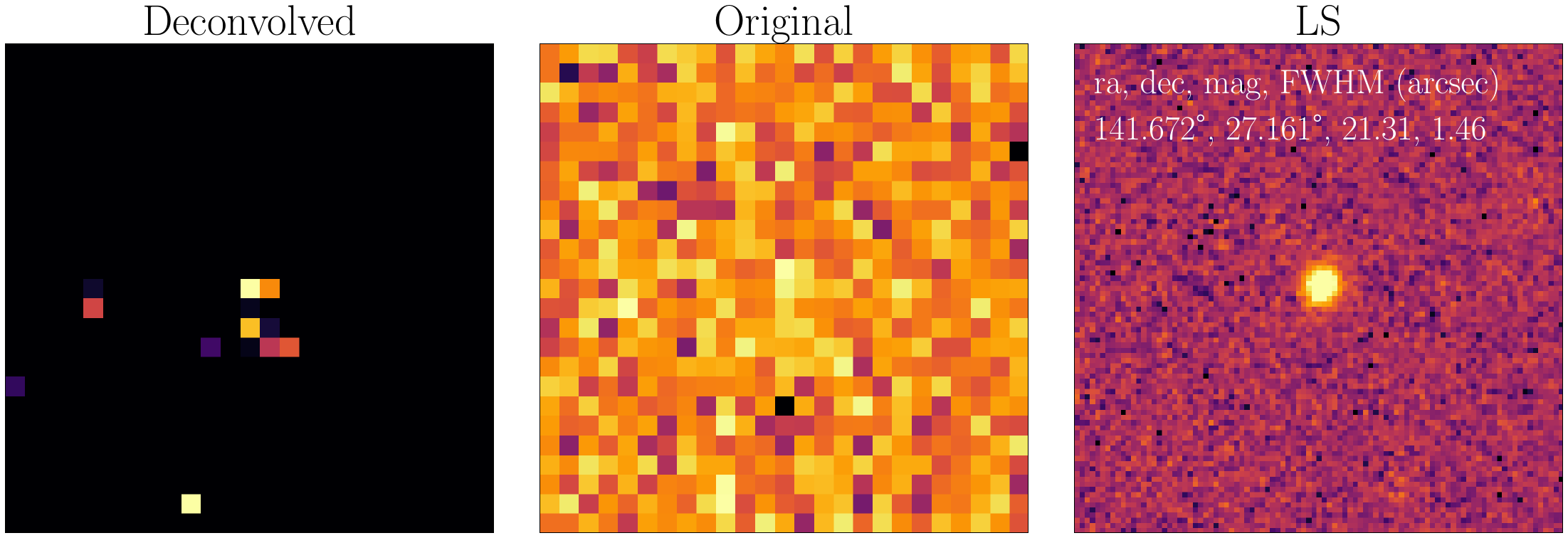}
      \includegraphics[keepaspectratio,width=0.32\linewidth]{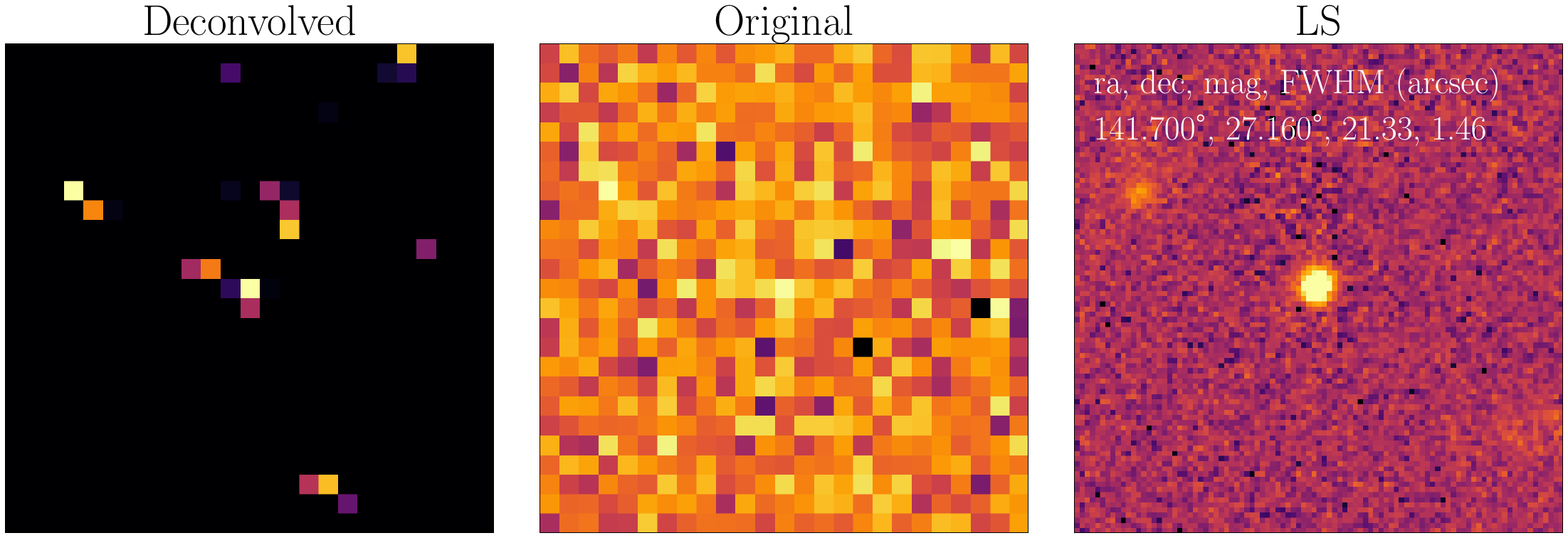}
      \includegraphics[keepaspectratio,width=0.32\linewidth]{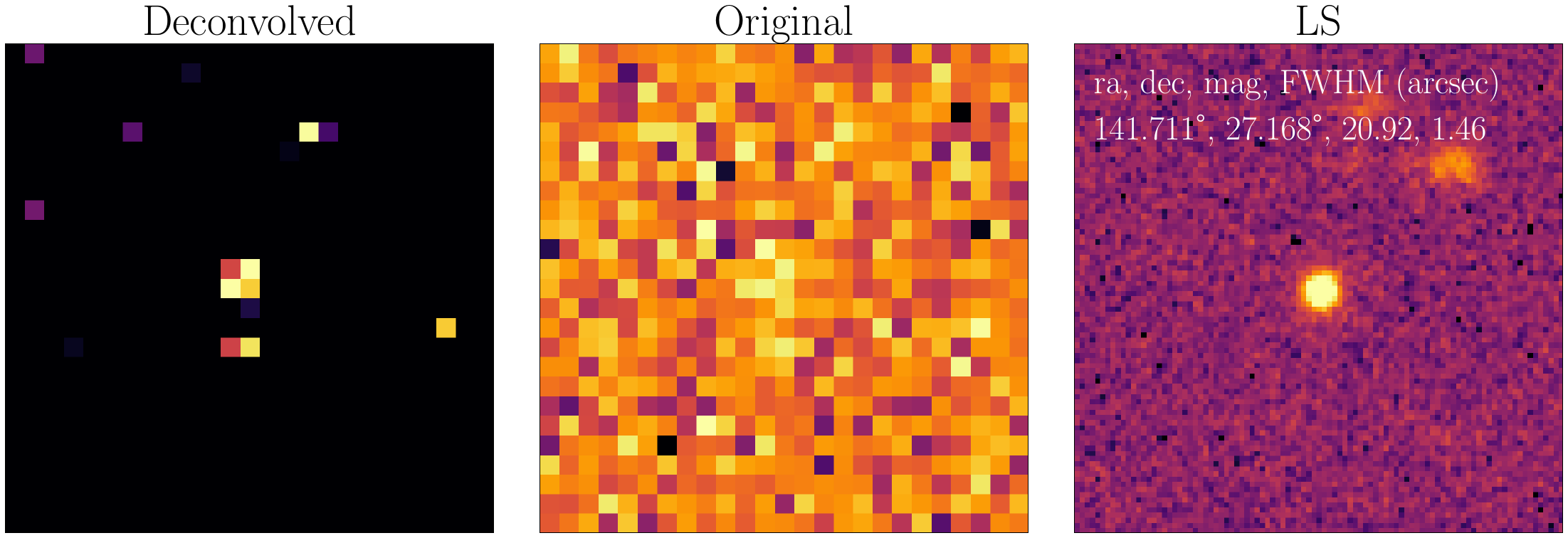}
      \includegraphics[keepaspectratio,width=0.32\linewidth]{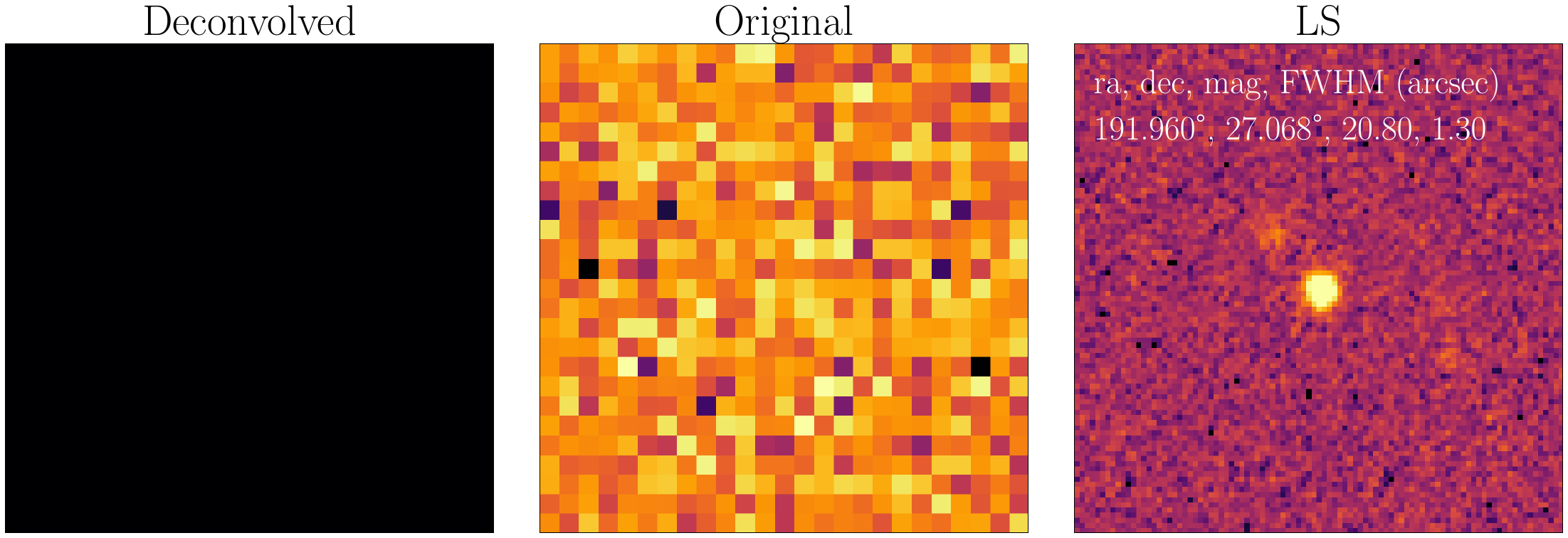}
      \includegraphics[keepaspectratio,width=0.32\linewidth]{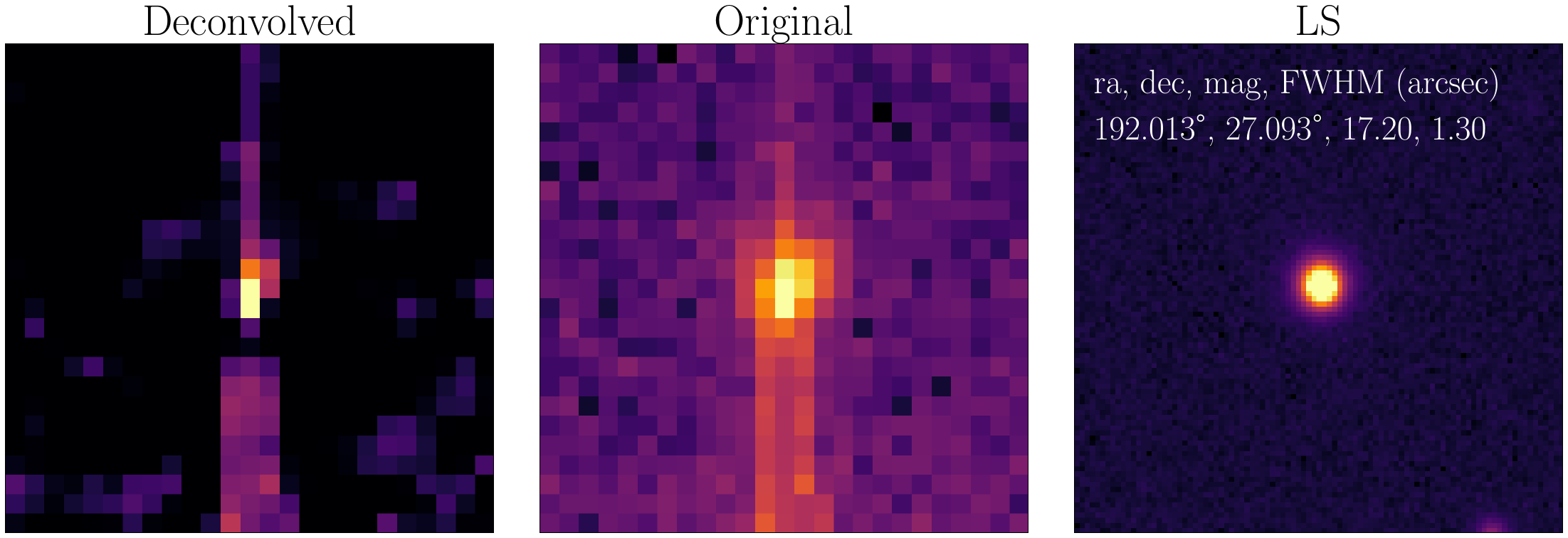}
      \includegraphics[keepaspectratio,width=0.32\linewidth]{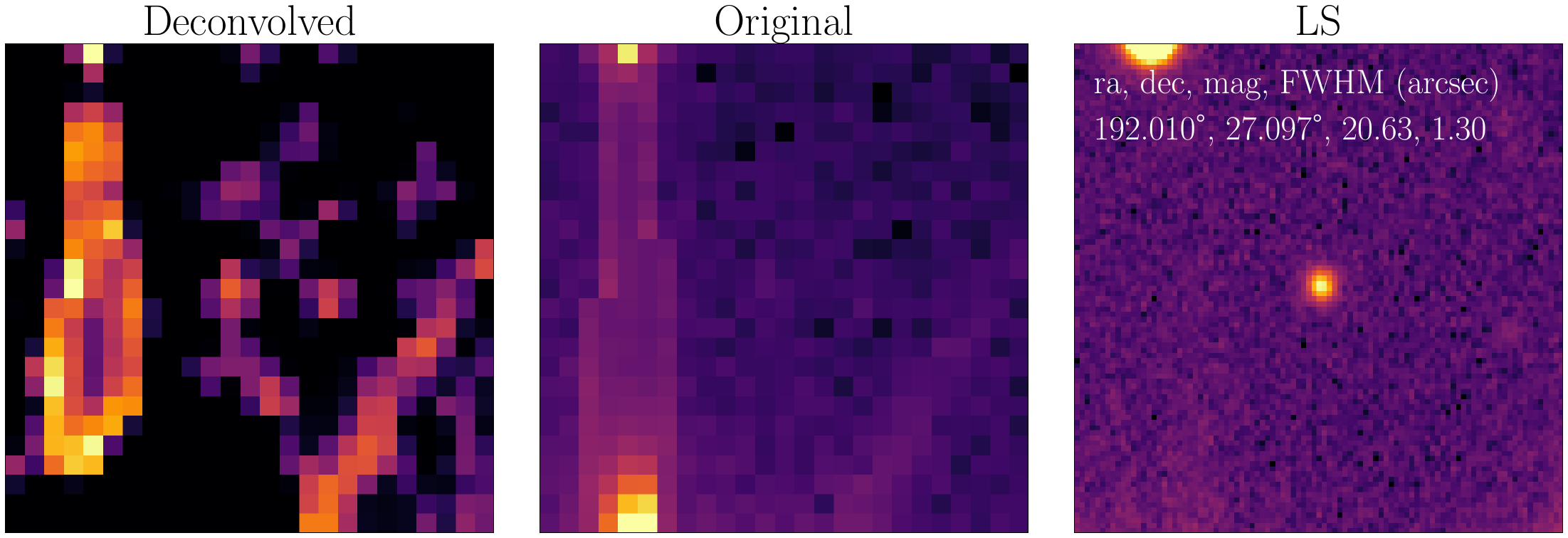}
      \includegraphics[keepaspectratio,width=0.32\linewidth]{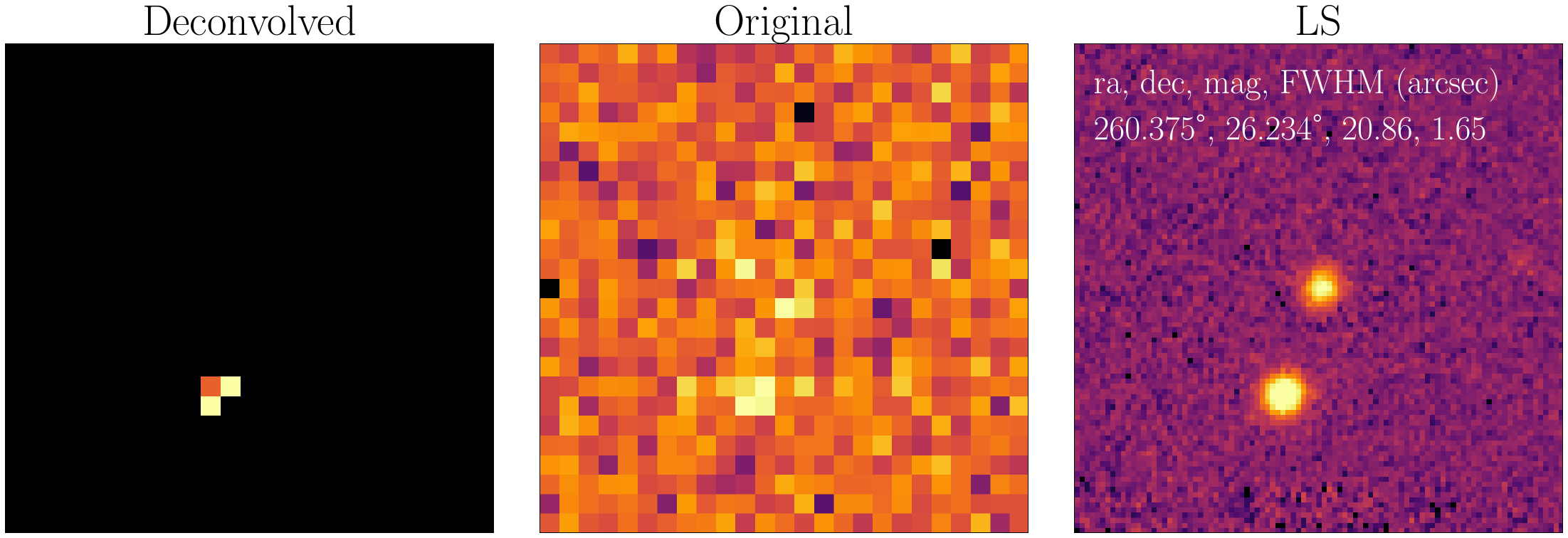}
      \includegraphics[keepaspectratio,width=0.32\linewidth]{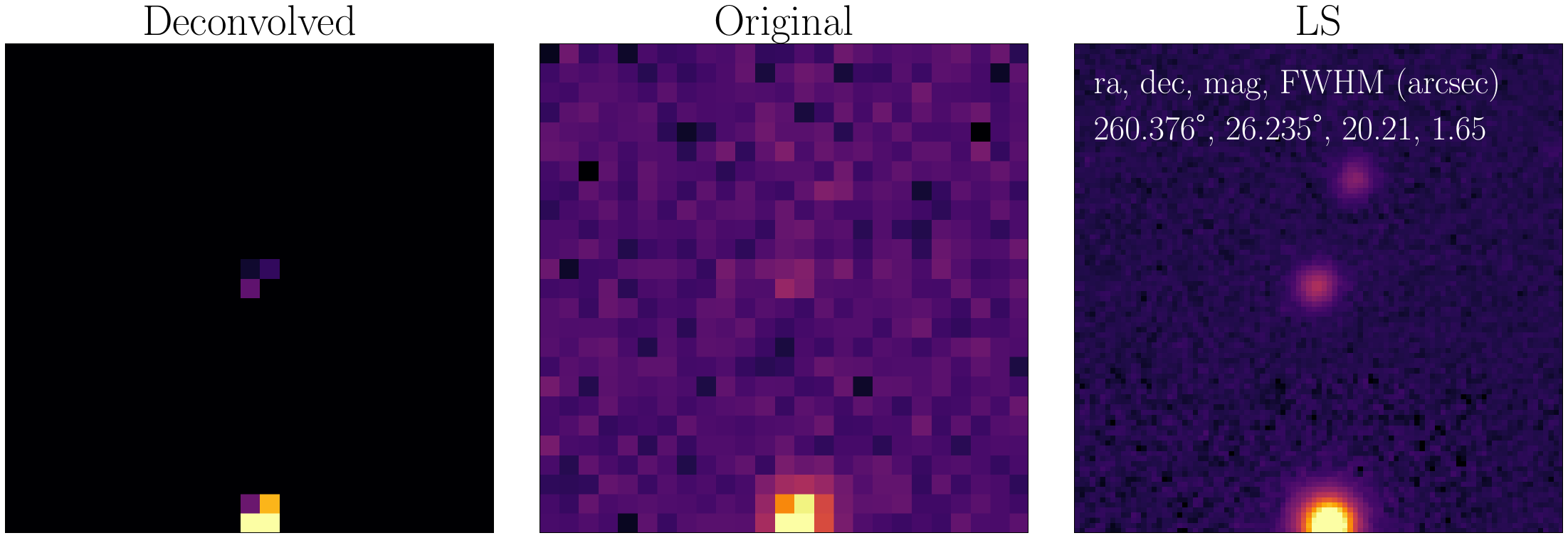}
    \caption{\reviewTwo{Example cutouts of size $25 \times 25$ arcseconds from ZTF deconvolved (left), ZTF original (middle), and LS (right) images centered at the locations of LS sources not detected in either ZTF original or deconvolved images. Each row contains three examples. The (ra, dec) locations of the LS sources and their magnitude and FWHM in the $r$ filter are shown on top of each example. The pixel scale of LS DR10 images is 0$\arcsec$.262 per pixel. Images are shown using a combination of square root stretching and clipping pixel values beyond the central 99.5 percentile.}} \label{fig:crossmatching-results-OrigDeconDESI-plots-cutouts}
\end{figure*}

\reviewTwo{The statistics in column 5 show the number of sources detected only in LS and in neither the original nor the deconvolved ZTF images. The median magnitude of these LS sources is 21.11, 20.77, and 20.86 for field IDs 619, 626, and 635, and most of these LS sources have $m > 19.5$. The FWHM of these LS sources is tightly distributed around a median value of roughly 1$\arcsec$.5. We show some example cutouts at the locations of the detected LS sources across all three fields in Fig.~\ref{fig:crossmatching-results-OrigDeconDESI-plots-cutouts}. The original images have visual marginal detection for which deconvolution could not reveal the source. There were also cases where the LS sources were located at the edges of the ZTF field images with partial detection within the field of view, so they are expected to be unmatched.}

\reviewTwo{As mentioned in the caption of Table~\ref{tab:crossmatching-results-OrigDeconDESI}, we find a few examples of many-to-one matches between the deconvolved and LS sources; however, such LS sources are not elongated, so we do not find sufficient visual evidence that deconvolution deblends the LS sources.}

\begin{figure*}[hbt!]
    \centering
      \includegraphics[keepaspectratio,width=0.32\linewidth]{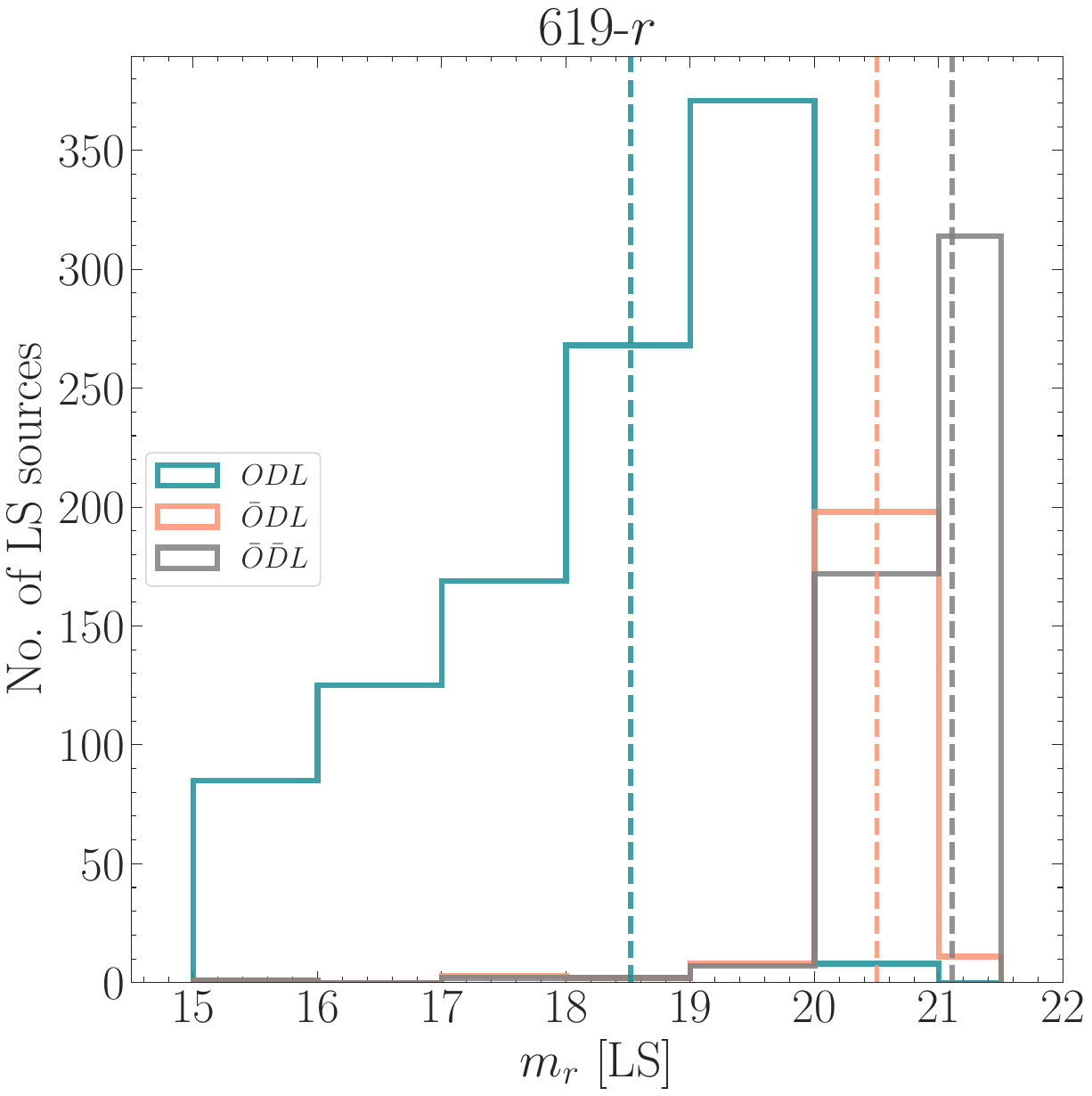}
      \includegraphics[keepaspectratio,width=0.32\linewidth]{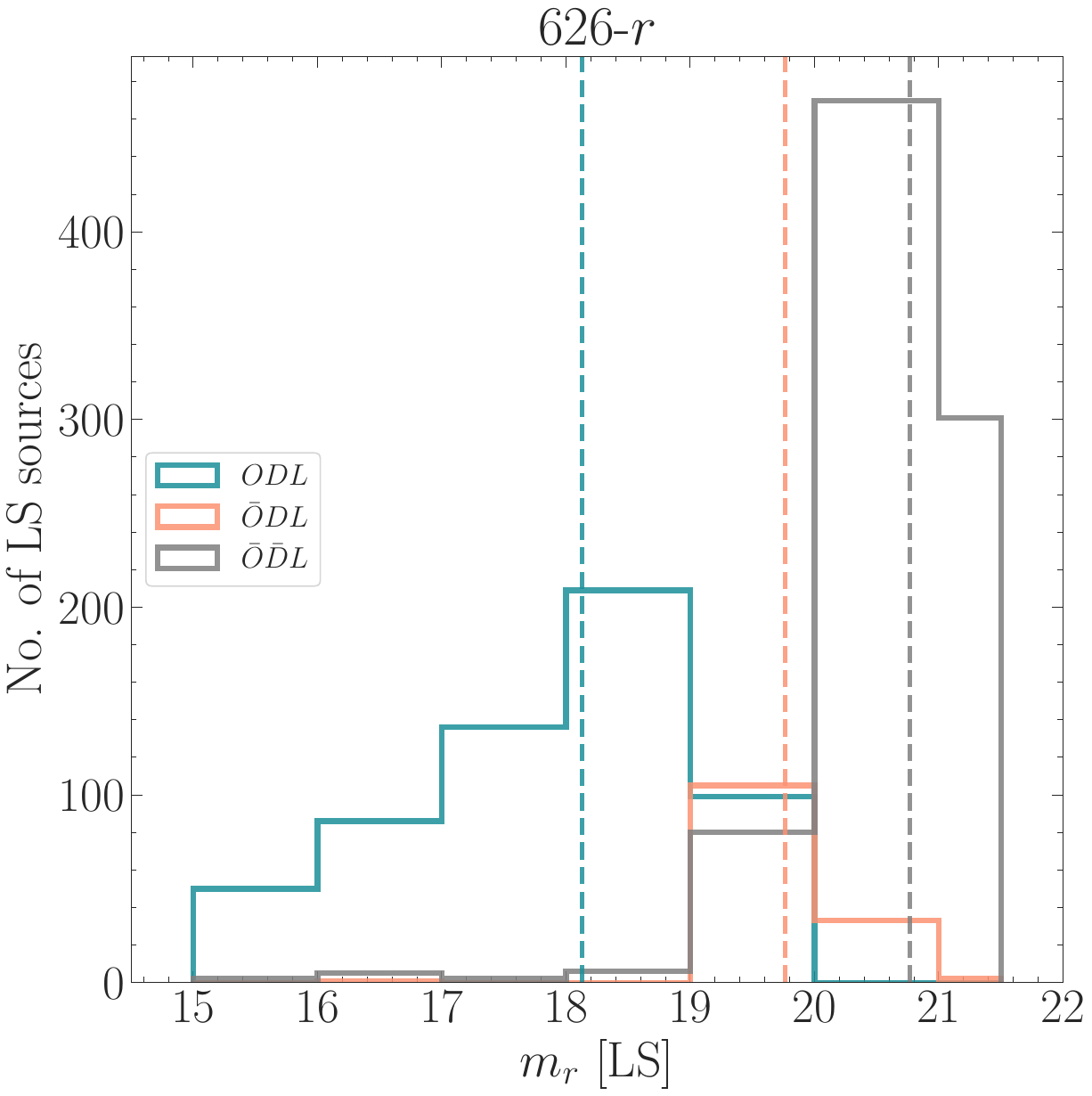}
      \includegraphics[keepaspectratio,width=0.32\linewidth]{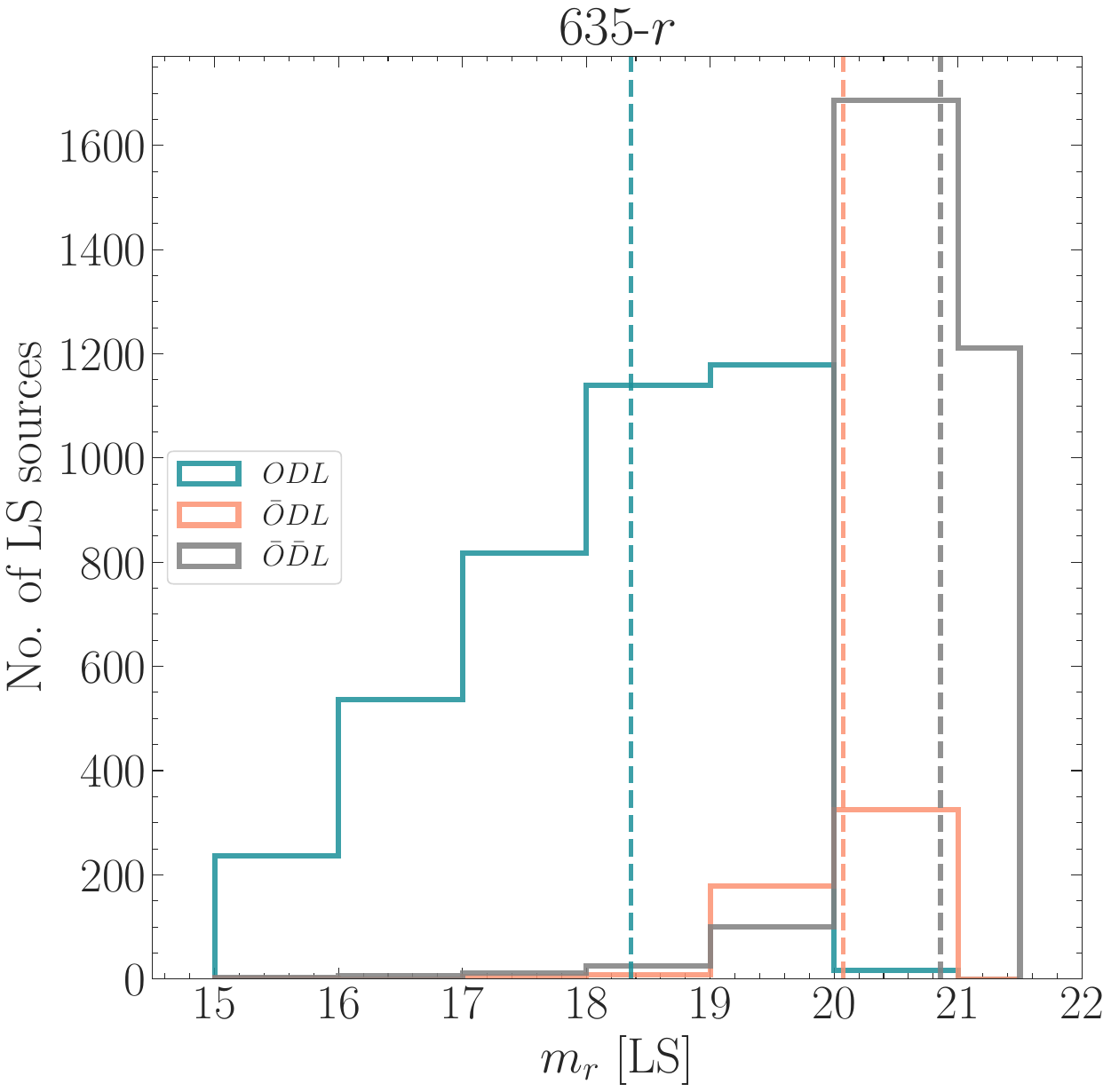}
    \caption{\reviewTwo{Histograms of the magnitude of LS sources in the $r$ filter for three cases: $ODL$ (sources present in all three fields), $\bar{O}DL$ (sources present in the deconvolved image and LS, but not in the original), $\bar{O}\bar{D}L$ (sources present neither in the original nor the deconvolved image, but present in LS). The vertical dashed lines denote the median magnitude.}} \label{fig:mag-hists-odl-obardbarl-obardl}
\end{figure*}

\reviewTwo{In Fig.~\ref{fig:mag-hists-odl-obardbarl-obardl}, we plot the distribution of the magnitudes of the LS sources from columns 2, 3, and 5. For each of the three fields, the median magnitude shifts progressively fainter from roughly 18.5 to 20.5 to 21.1, in that order, for field ID 619. For ID 626, it shifts from 18.2 to 19.8 to 20.8, and for ID 635, it shifts from 18.4 to 20 to 20.8. Although not shown in the figure, the median FWHM of the LS sources across $ODL$, $\bar{O}DL$, and $\bar{O}\bar{D}L$ for all three fields are similar (roughly 1$\arcsec$.5). These population-level trends suggest that deconvolution discovers several sources with $21 > m_r > 20$ (and even a few with $21.5 > m_r > 21$ for field ID 619) that could not be found in the original image. There are several sources with $21.5 > m_r > 20$ that are present only in LS, which indicates that deconvolution could not identify them. As these sources were not detected in the original image either, the impact is less critical.}

\reviewTwo{We perform forced photometry of ZTF sources at the positions of deconvolved sources that remain unmatched with both original and LS (column 8 in the table). These are ZTF sources polled in the $r$- and $g$ filter from the start of the survey that match a source in the Pan-STARRS1 database within a radius of 1$\arcsec$.8. We require the resulting light curves to have at least two distinct detections across both filters, and photometry flags $< 32768$, and ``infobits'' $< 33554432$ corresponding to usable photometry. If these conditions are satisfied, we consider the deconvolved source to be a real candidate, given its detection in Pan-STARRS1, which probes deeper than ZTF.}

\begin{figure*}[hbt!]
    \centering
      \includegraphics[keepaspectratio,width=0.48\linewidth]{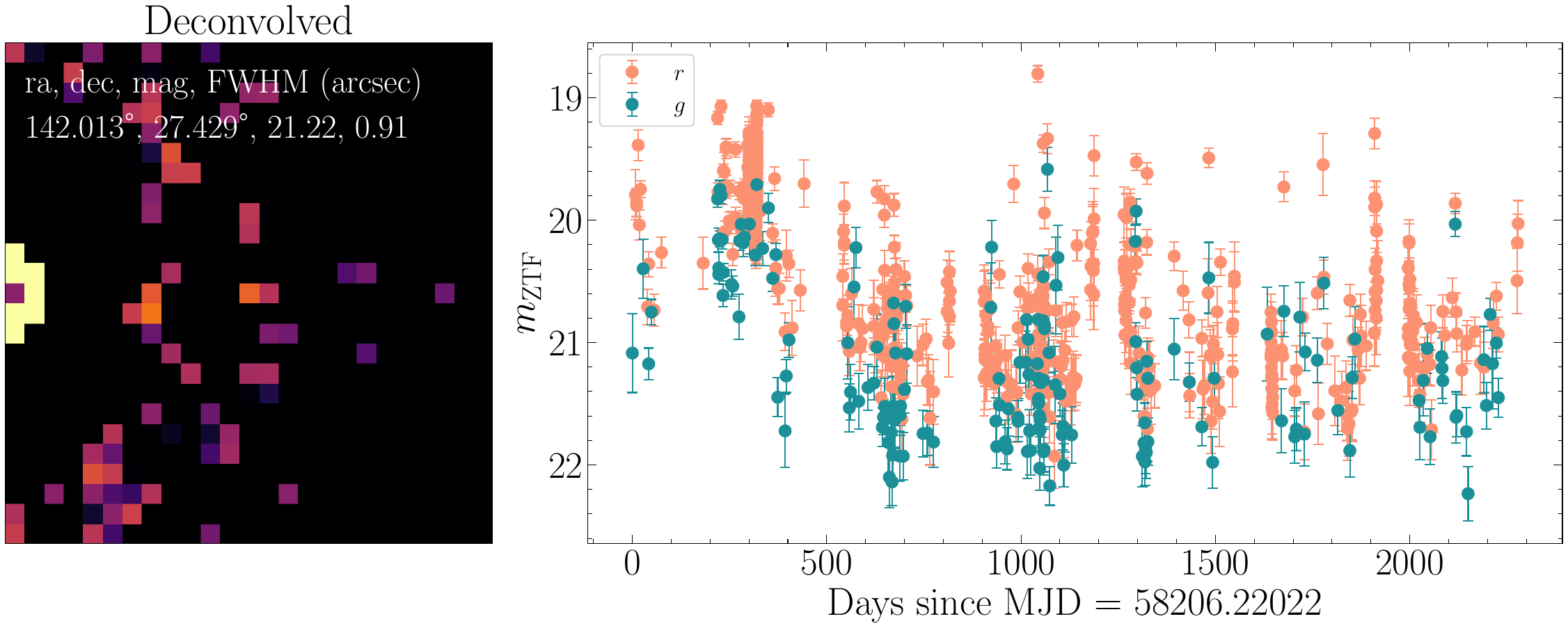}
      \includegraphics[keepaspectratio,width=0.48\linewidth]{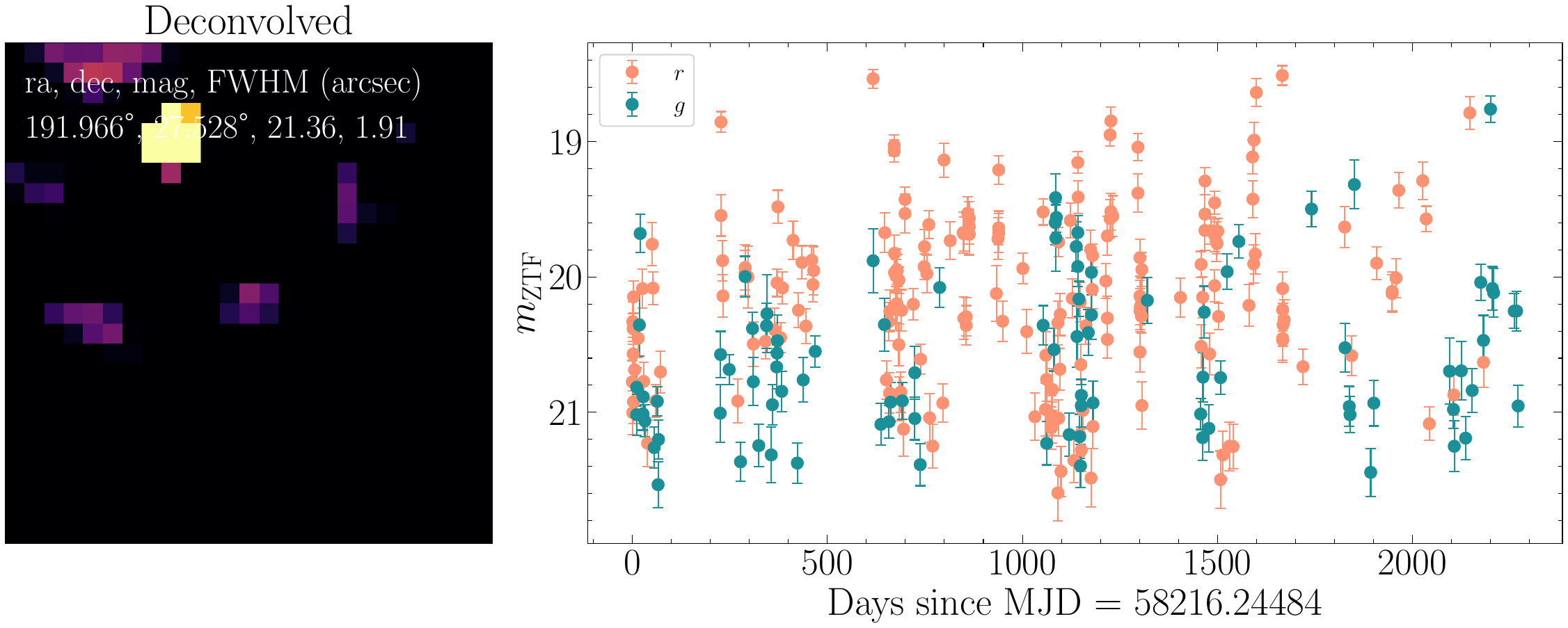}
      \includegraphics[keepaspectratio,width=0.48\linewidth]{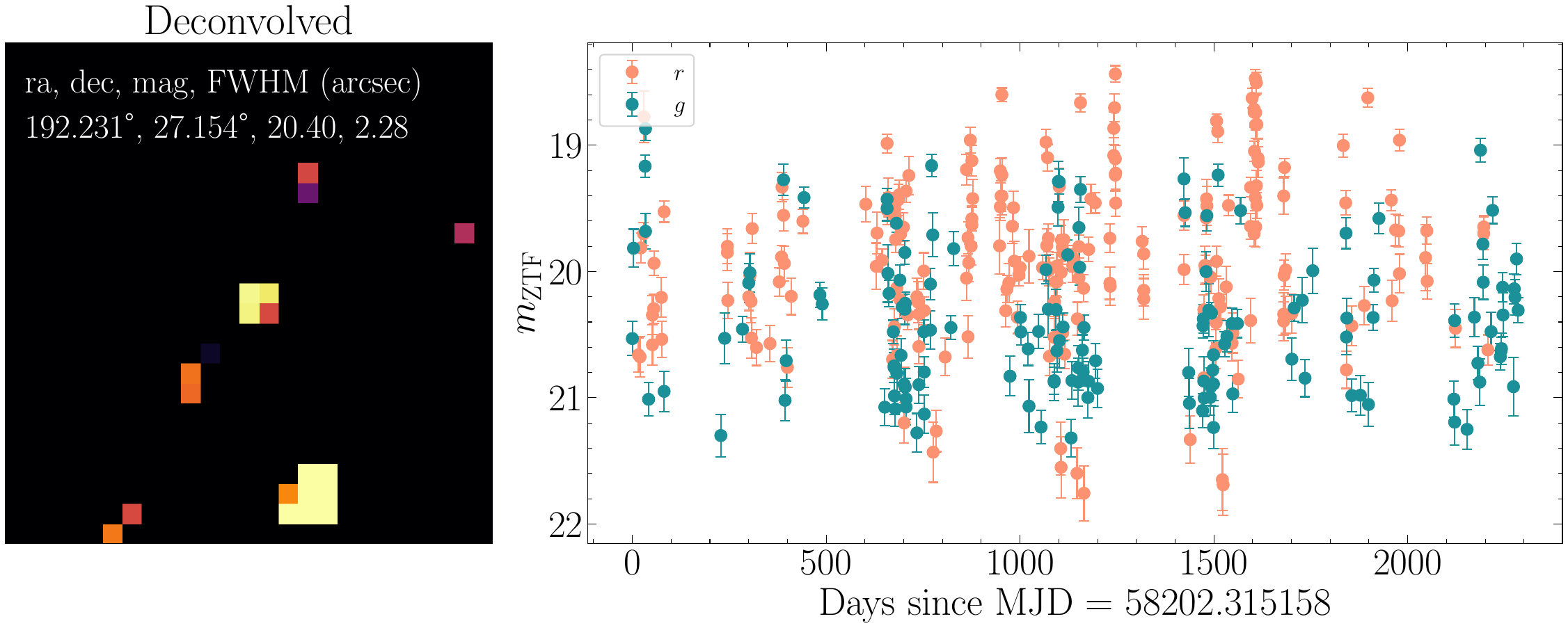}
      \includegraphics[keepaspectratio,width=0.48\linewidth]{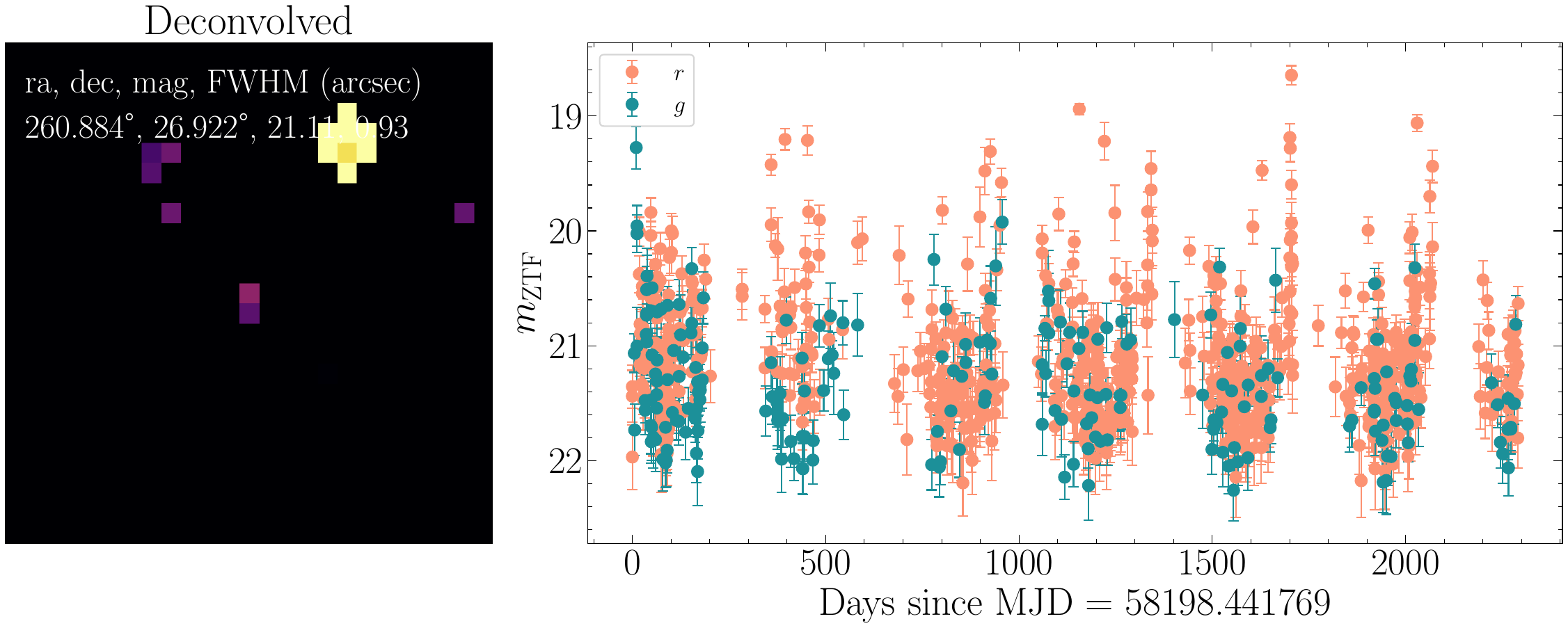}
      \includegraphics[keepaspectratio,width=0.48\linewidth]{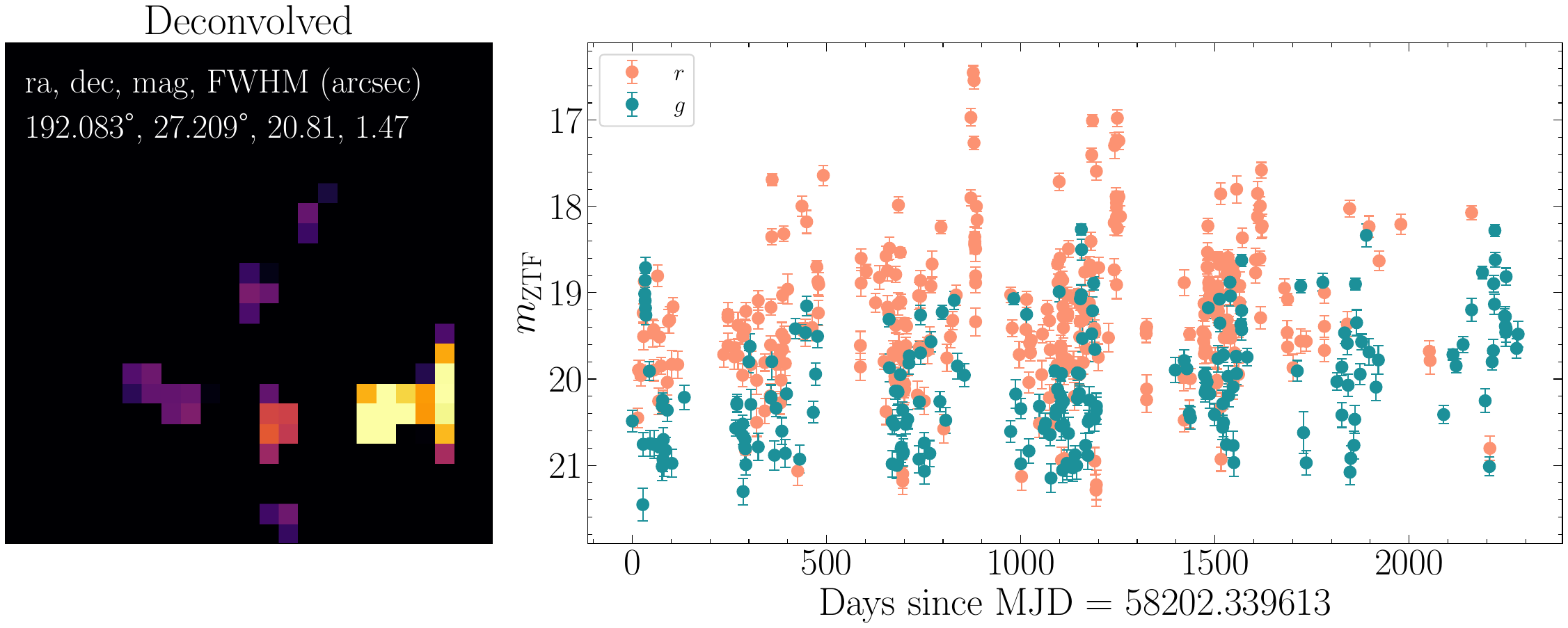}
      \includegraphics[keepaspectratio,width=0.48\linewidth]{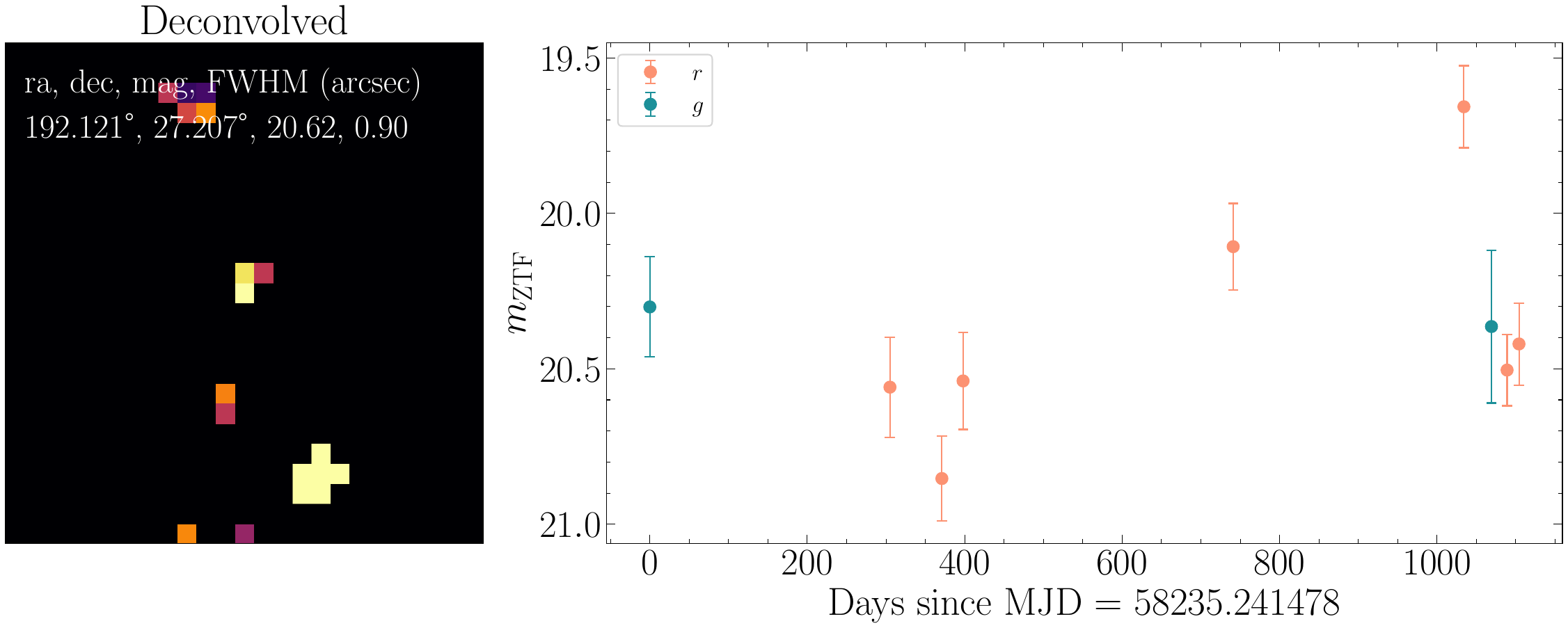}
      \includegraphics[keepaspectratio,width=0.48\linewidth]{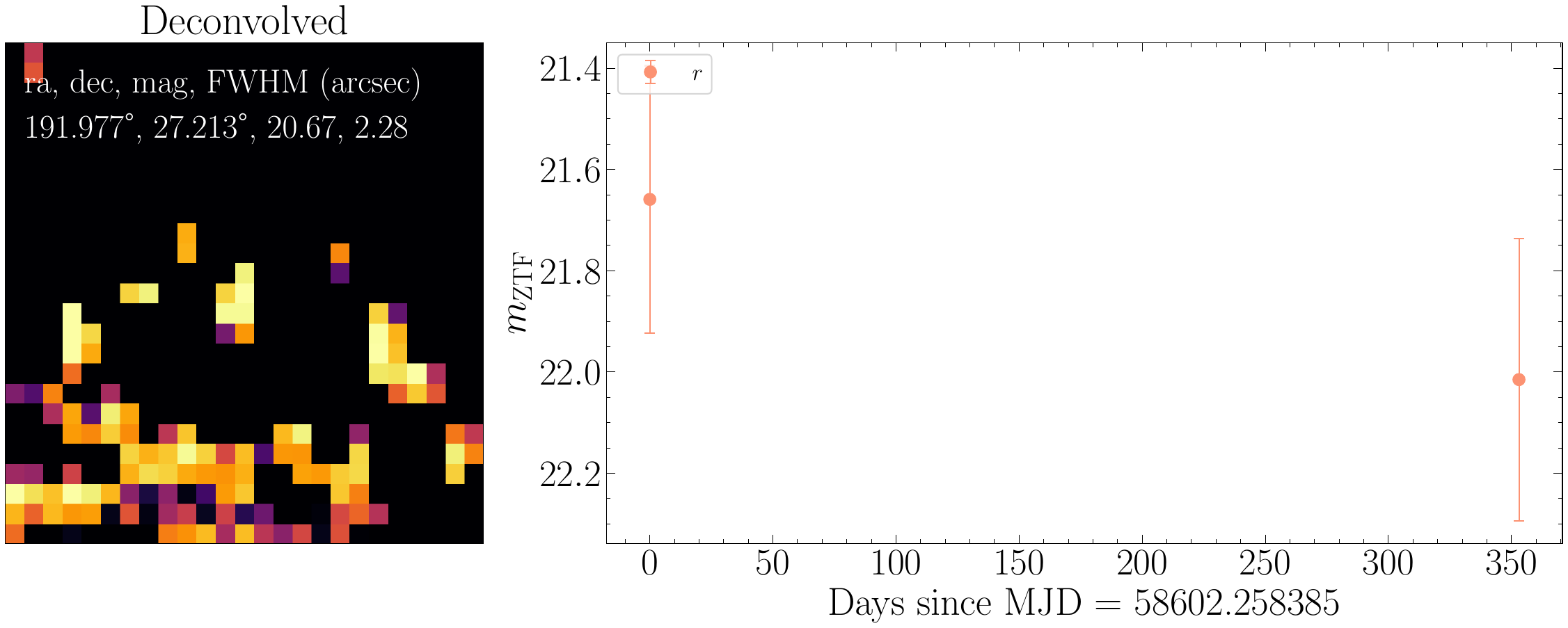}\hfill
      \includegraphics[keepaspectratio,width=0.48\linewidth]{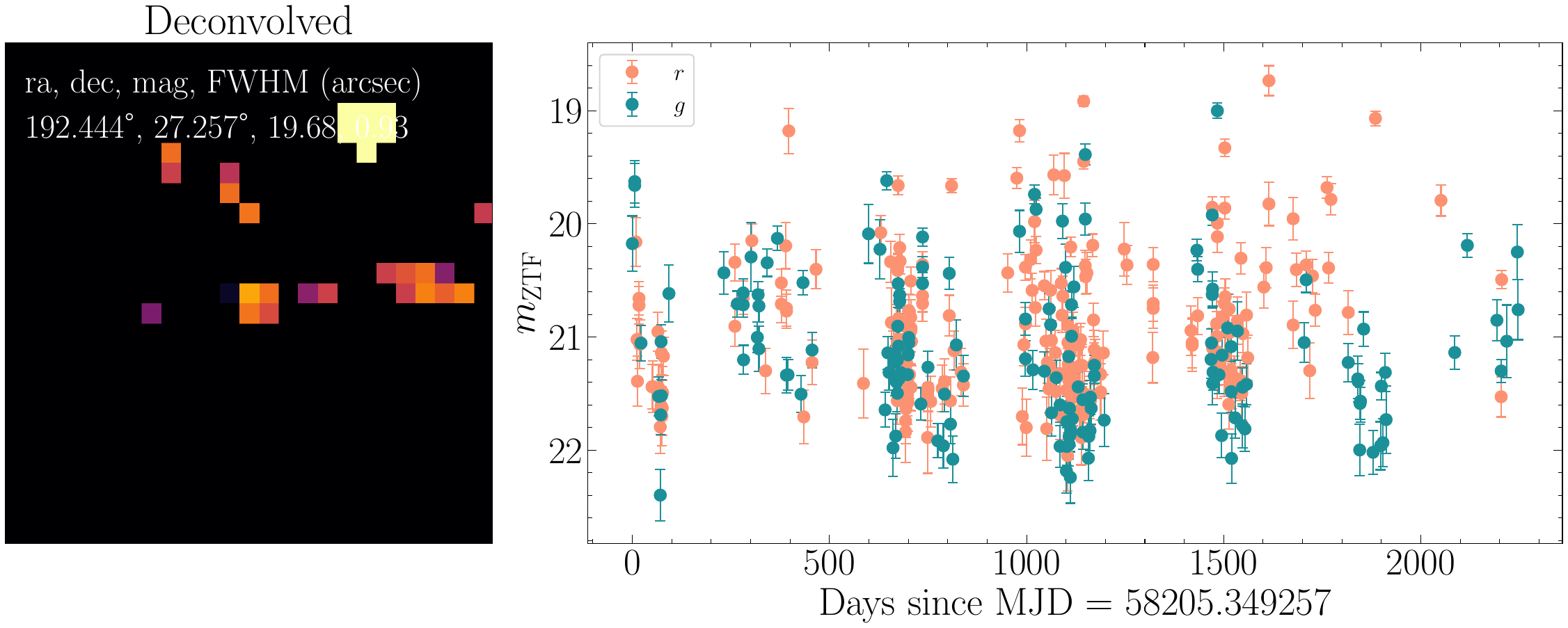}
    \caption{\reviewTwo{Example cutouts of size $25 \times 25$ arcseconds from ZTF deconvolved images centered at the locations of deconvolved sources not detected in either ZTF original or LS images but matched to a Pan-STARRS1 source. The ZTF light curves (magnitude with error bars vs. modified Julian date) of detections matching a Pan-STARRS1 source are shown beside every example. The (ra, dec) locations of the deconvolved sources and their magnitude and FWHM in the $r$ filter are shown on top of each example. Images are shown using a combination of log stretching and clipping pixel values beyond the central 98 percentile to better highlight the faint detections.}} \label{fig:crossmatching-results-OrigDeconDESI-plots-cutouts-onlyInDecon}
\end{figure*}

\reviewTwo{Using these selection criteria, 6 out of 51, 20 out of 128, and 2 out of 32 such deconvolved sources are real. In Fig.~\ref{fig:crossmatching-results-OrigDeconDESI-plots-cutouts-onlyInDecon}, we show a visualization of these deconvolved sources and the forced photometry light curves, with extended visualizations in Appendix~\ref{appn:crossmatching-results-OrigDeconDESI-plots-cutouts-onlyInDecon-extended}. Most of these light curves contain detections over several years. Most of the deconvolved source magnitudes lie towards the fainter end of the light curves, whereas there is one at the brighter end (last row, second column in Fig.~\ref{fig:crossmatching-results-OrigDeconDESI-plots-cutouts-onlyInDecon}). However, as noted earlier in this section, the deconvolved magnitudes in the faint regime could be off by several tenths of a magnitude. One of them only had two detections, which makes this case unreliable (last row, first column in Fig.~\ref{fig:crossmatching-results-OrigDeconDESI-plots-cutouts-onlyInDecon}).}

\reviewTwo{The remaining 45, 108, and 30 spurious deconvolved sources (unmatched in neither of the original image, LS, or Pan-STARRS1) are concentrated toward the faint end, similar to the confirmed real astrophysical deconvolved sources shown in Fig.~\ref{fig:crossmatching-results-OrigDeconDESI-plots}, with median magnitudes of 21.15, 21.1, and 21.01, respectively. We find no clear evidence that their FWHM or ellipticity distributions differ from those of confirmed sources. However, due to the relatively small sample sizes, definitive conclusions remain limited.}

\reviewTwo{Thus, through this analysis, we conclude that 229 (223 + 6) sources out of 671 (34\%), 165 (145 + 20) sources out of 577 (29\%), and 519 (517 + 2) sources out of 1072 (48\%) likely astrophysical sources for field IDs 619, 626, and 635, respectively are confirmed (definitely real) astrophysical deconvolved sources not present in the corresponding original images found through crossmatching with LS and matching with Pan-STARRS1. The actual fraction of real deconvolved sources is likely larger depending on the strictness of quality filtering applied to LS sources, with \reviewFive{our calculated upper bounds to be} 92\% (620/671), 80\% (461/577), and 97\% (1035/1072) for the three fields, respectively. The remaining 8\%, 20\%, or 3\% deconvolved sources are likely spurious but could include faint variables, asteroids, and transients, or even galaxies presenting non-stellar profiles. Here, we have made no attempt to use additional astrophysical observables, such as colors or proper motions, to get exact numbers, but users wishing to use the methodology on their own data are encouraged to use such additional data as befits their requirements.}

\subsection{Comparison with Richardson-Lucy deconvolution}\label{subsec:rl-sgp-compare-sec}

Here, we compare the deconvolution performance of SGP with the RL algorithm on a high-galactic latitude field with ID 626 in the $r$-band as an example.

In our internal experiments, we found that the traditional RL algorithm produces unwanted speckling artifacts, as discussed in Sect.~\ref{sec:deconv}. To address these issues, we experimented with two regularized versions of the RL algorithm: the damped RL method proposed by \citet{White1994} and the spatial regularization introduced by \citet{Bratsolis2001}. Although both methods did not produce drastically different restoration results visually, we found that the spatial regularization approach was better in terms of the number of crossmatches with sources in the original image. Therefore, we chose it for comparison. The spatial regularization approach adds an additional term to the traditional RL iterations (see Eqn.~\ref{eqn:traditional-rl}) due to which the pixel values are made dependent on those of their nearest neighbors as follows:
\begin{equation}
	\mathbf{f}^n = (1 - \lambda) \mathbf{f}^{n-1} \odot A^\intercal \frac{\mathbf{g}}{A\mathbf{f}^{n-1} + \mathbf{b}} + \lambda R \ast \mathbf{f}^{n-1}
\end{equation}
where $\lambda$ is a hyperparameter of the method. Setting $\lambda = 0$ yields the normal RL iteration. We choose $\lambda = 0.05$ and the matrix $R = \begin{bmatrix} 0 & 1/4 & 0\\ 1/4 & 0 & 1/4 \\ 0 & 1/4 & 0\end{bmatrix}$, which is taken from \citet{Bratsolis2001}. Note that, as in the normal RL algorithm introduced in Sect.~\ref{sec:deconv}, the inclusion of $\mathbf{b}$ means that flux conservation would not be achieved despite this modification. We have confirmed that this simple change significantly reduces artifacts in the deconvolved result. For a fair comparison, we apply the same termination criterion and initialization of the deconvolved image used for SGP.

For simplicity, we only focus on comparing the properties of the one-to-one matches and their execution times. A total of 365 one-to-one crossmatches were obtained using RL deconvolution after applying the selection criteria discussed in Sect.~\ref{subsec:experimental-details}. For SGP, there were 722 one-to-one crossmatches (see Table~\ref{tab:crossmatching-results}). Out of the 365 RL crossmatches, 351 original sources were also present in the crossmatches obtained with deconvolution using SGP. As a result, for a clearer comparison, we will only consider these 351 crossmatches for both algorithms and compare the properties of the respective sets of deconvolved sources.

\begin{figure*}[hbt!]
    \centering
      \includegraphics[keepaspectratio,width=0.245\linewidth]{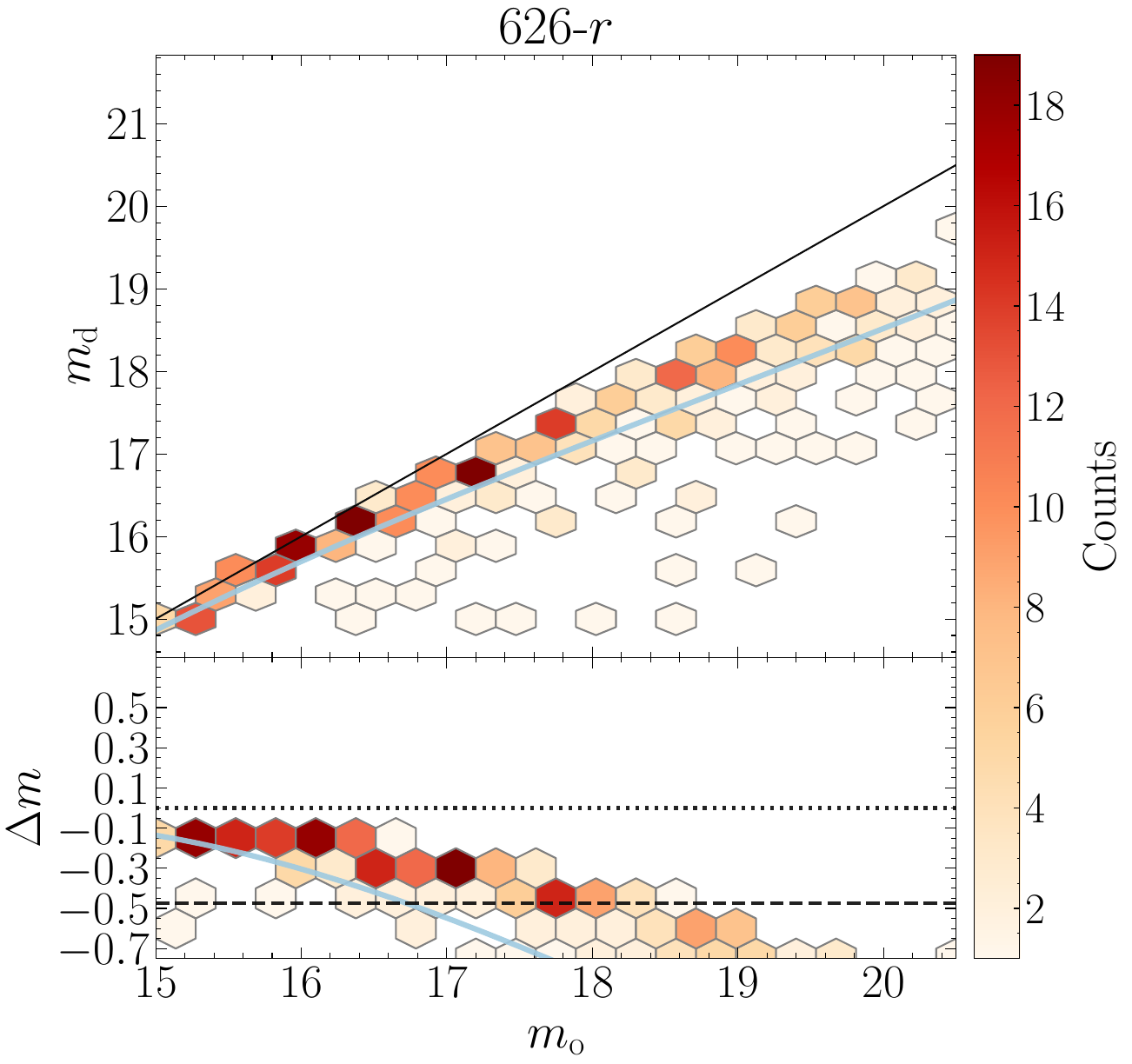}
      \includegraphics[keepaspectratio,width=0.245\linewidth]{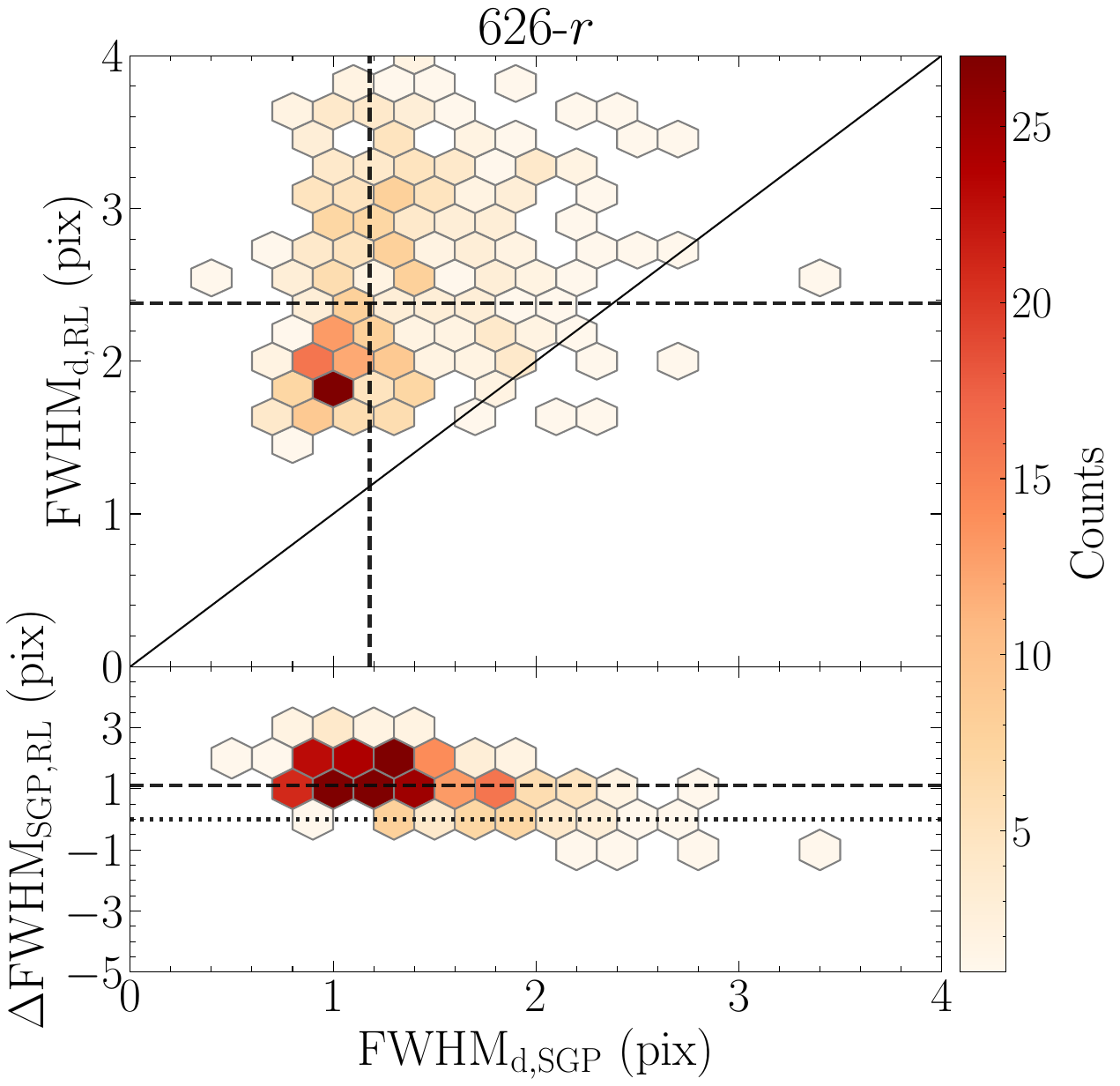}
      \includegraphics[keepaspectratio,width=0.245\linewidth]{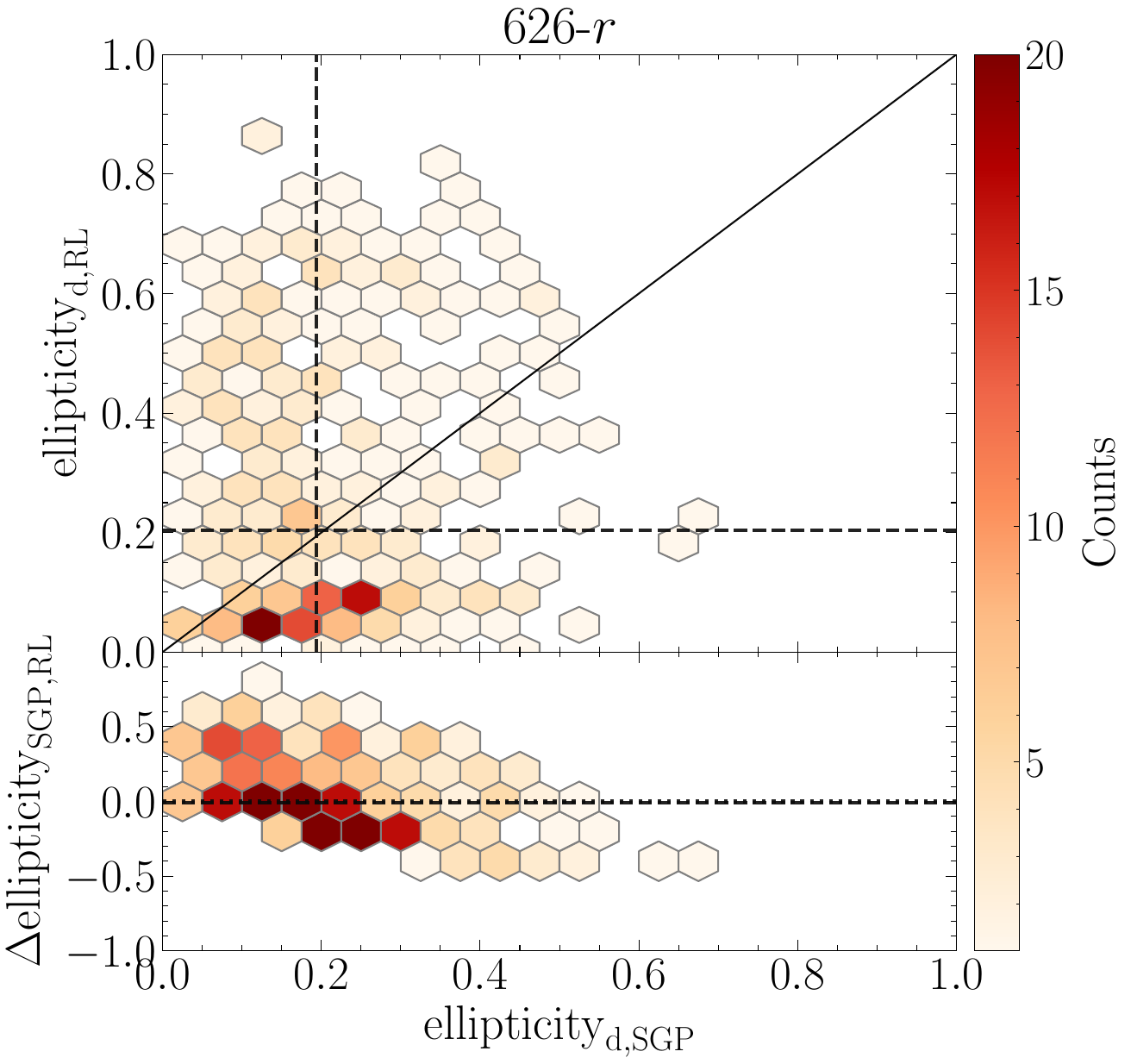}
      \includegraphics[keepaspectratio,width=0.245\linewidth]{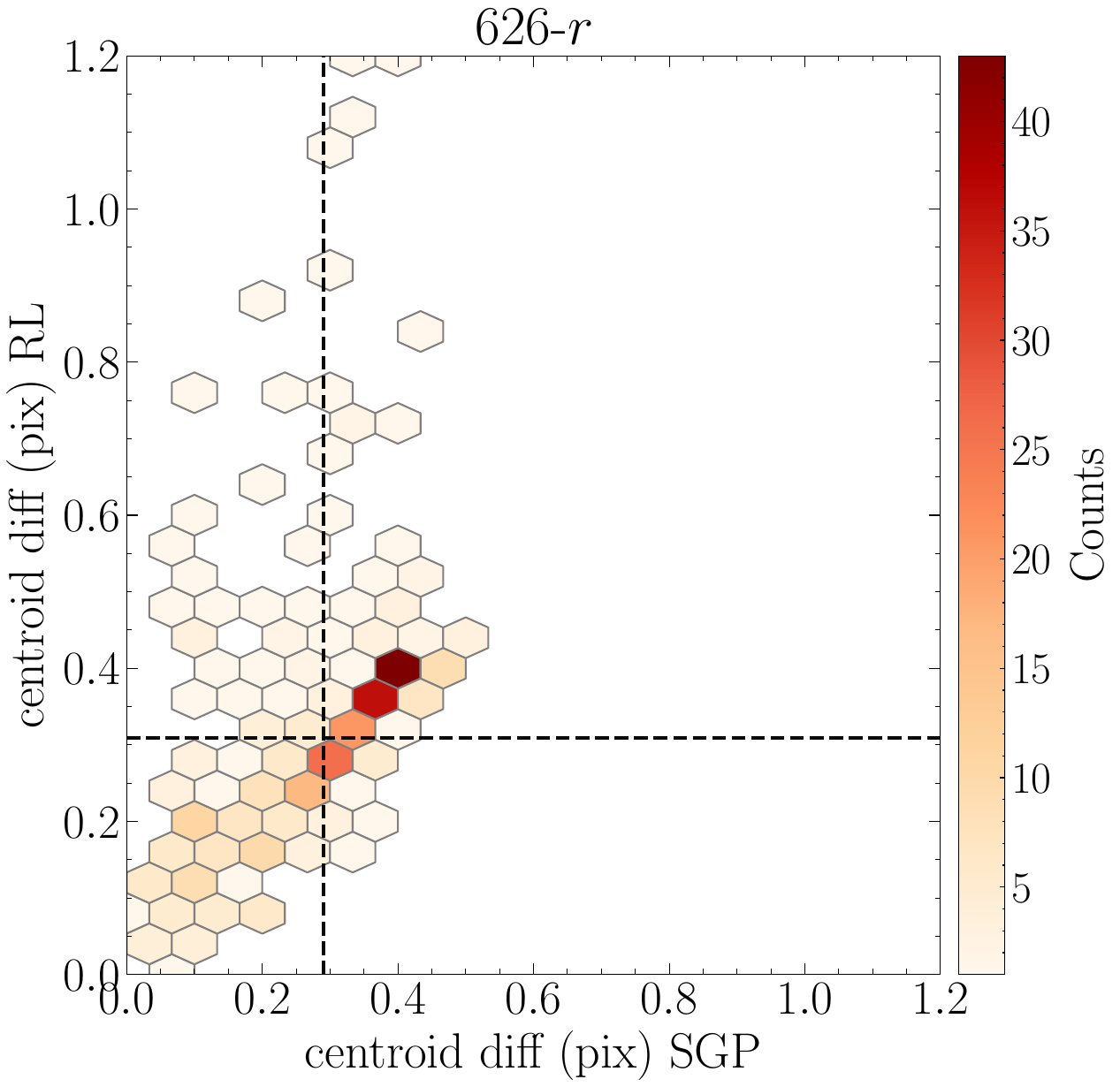}
        \caption{Comparison of magnitudes of the original ($m_{\mathrm{o}}$) and the one-to-one matched deconvolved ($m_{\mathrm{d}}$) sources for the RL algorithm (leftmost panel), followed by the comparison of SGP and RL algorithms in terms of FWHM and ellipticity of deconvolved sources, and difference in centroids of original and deconvolved sources for the field with ID 626 in the $r$-band. For RL, there are 365 one-to-one matches, whereas 722 matches for SGP. We have found that 351 original sources from the RL one-to-one crossmatches are also present in the SGP one-to-one crossmatches. The plots shown here compare the properties of these common 351 deconvolved sources produced by the two algorithms. The lower panels in the SGP-RL comparison show the difference between the properties of deconvolved sources from RL and SGP.} \label{fig:RL-SGP-one-to-one-comparison}
\end{figure*}

Fig.~\ref{fig:RL-SGP-one-to-one-comparison} shows this comparison. Due to the inclusion of background, denoted as $\mathbf{b}$, in the RL iterations, it is evident that the magnitudes of the deconvolved sources do not agree with the magnitudes of the original sources. Even for the brightest sources, the magnitude differences are around -0.15, and the flux is overestimated by more than a magnitude for the faintest sources. On the other hand, SGP explicitly imposes flux preservation through the special projection step. The FWHM of the RL-deconvolved sources is generally larger than that of the SGP-deconvolved sources. The median value for RL is 2.4 pixels compared to 1.2 pixels for SGP. The median ellipticities of both sets of deconvolved sources are approximately 0.2. Differences in centroids of the original and corresponding deconvolved sources are similar for RL than for SGP. However, RL produces several cases where centroid shifts are greater than one pixel, which is not the case with SGP. Overall, our implementation of SGP outperforms the specific variant of the RL algorithm used in this analysis.

It is known that RL is slow to converge. Our implementation of RL requires $\sim$150-300 iterations for a single subdivision, which is roughly ten times more than required for SGP (see Sect.~\ref{sec:exec-time}). The total execution time for RL to restore the entire field is 40 minutes, approximately eight times slower than SGP. This slowdown is primarily due to the large number of iterations required by RL, even though the per-iteration costs of RL and SGP are relatively similar.

\section{Discussion}\label{sec:discussion}

\subsection{\reviewFour{Summary and Outlook}}

The deconvolution of photometric images from ongoing and future astronomical surveys is expected to have several spin-off applications that may allow these surveys to reveal new scientific information from available observed images. We have demonstrated the Scaled Gradient Projection (SGP) algorithm, an efficient (faster and better convergence properties) version of the famous Richardson-Lucy algorithm, that performs maximum likelihood estimation in the case of data with Poisson noise, which is a good approximation for astronomical images. \reviewThree{Apart from this common assumption regarding the noise model, our deconvolution pipeline assumes a known PSF model but is otherwise survey‑agnostic; \reviewFour{see Sect.~\ref{subsec:survey-specific-tuning} for some survey-specific considerations when applying the method}.} We deconvolved a dozen field images observed by ZTF across different filter bands (mostly $r$, but also $g$ and $i$), seeing FWHM ($\sim$$1\arcsec.7 - 3\arcsec.3$), and the crowdedness of the field (low to high galactic latitude). Due to the lack of any ground truth, we crossmatched sources detected in the observed (which we call `original') and the corresponding deconvolved images and performed an extensive comparison of the star-like matched and unmatched sources in terms of photometric accuracy, astrometric shifts, FWHM, and ellipticity.

Our results start with a visual inspection, where we confirmed that deconvolution does not generate unnecessary artifacts such as ringing around deconvolved sources or noise amplification. The deconvolution method does not vanish sources detected in the original images, even faint ones near bright sources. These characteristic features of the algorithm can help address some common skepticism about whether deconvolution can reliably be applied to observed data.

We conducted detailed tests using several evaluation metrics, which we now discuss in the following. Our tests suggest that sources are deconvolved with $\gtrsim$90\% flux conservation up to $m \sim 17.8$ - $20.4$ depending on the field, and generally only the brightest sources ($m \lesssim 16$) are conserved with $\gtrsim$98\% flux conservation. However, we find three fields (with $b \approx -60, 30, 0$ deg; high, intermediate, and a low galactic latitude field, respectively), where original sources up to $m \sim 18.4$ - $19.6$ were conserved with $\gtrsim$98\% flux conservation. The scatter in the deconvolved magnitudes typically increases and becomes unfeasible for sources fainter than $m \sim 18$ - $19$, depending on the field. \reviewTwo{There is a slight photometric bias in the magnitudes of the deconvolved sources towards fainter original sources having $m \gtrsim 19$ where, depending on the field, the deconvolved sources were slightly brighter or dimmer than expected}. These levels of agreement are not excellent, which may be a consequence of the deconvolution algorithm being total-flux constrained rather than constraining individual source fluxes and because the total-flux constraint tends to be dominated by the relatively brighter sources. However, a particular advantage of deconvolution is that since the deconvolution removes the background emission, the uncertainties in the deconvolved magnitudes are reduced even for the faintest sources (see Appendix~\ref{appn:photo-error-compare}).

The deconvolved source ellipticities are typically enlarged by 0.05 - 0.2, depending on the field. The increase in ellipticity is because deconvolution removes relatively fainter pixels in the wings of the stellar sources and incorporates the emission in its core, but it does so in a way that leads to a non-symmetrical shape (see, e.g., Fig.~\ref{fig:visual}). Another possible reason is a methodological consequence since we skip the filtering step used before detection inside SExtractor for deconvolved sources, which otherwise smooths and leads to slightly rounder ellipticities. The spread in the ellipticities of deconvolved sources was also found to be higher than that of the original sources. A possible reason for this is that SExtractor defines ellipticities based on second-order moments of the detected sources, which may be particularly sensitive to flux distribution across pixels because we do not use the filtering step in SExtractor.

When crossmatching, we typically found only a fraction of a percentage (less than a few tenths of a percentage) of sources that could not be detected in the deconvolved. A detailed investigation of such sources suggested that they typically tend to be close (within several tens of pixels) to a saturated source with a blooming streak. Although our selection criteria aim to pick only relatively good detections, it may be possible that some of these unmatched original sources are actually artifacts, and thus, there may be little consequence. However, it is important to clarify that we currently do not have a robust mechanism to handle saturated sources during deconvolution. Therefore, plausible sources around saturated sources may not be reliably deconvolved. Handling this is prioritized in future work where bad pixels can be simply excluded in SGP's optimization routine \citep[see, e.g.,][for an application in a different algorithm]{Hirsch2011}. There have been some cases where the unmatched original source resulted from the deblend routine inside SExtractor, and rather, the blended version of that source was detected in the deconvolved even though we allow deblending in the deconvolved image. We have found this is mainly due to the \texttt{DEBLEND\_NTHRESH} parameter in SExtractor that controls the number of subthresholds: we hypothesized that, since the deconvolved images have a much higher dynamic range in their source profiles (characterized by a small FWHM and profiles that abruptly fall to zero), some deconvolved sources may require a larger number of thresholds. Many of these `un-deblended' deconvolved sources were deblended with higher thresholds, confirming our hypothesis. Nevertheless, we visually identified deconvolved sources at locations corresponding to the unmatched original sources in almost all cases, but they were not detected by SExtractor.

\reviewTwo{Several sources undetected in the original image but newly detected by deconvolution pass the additional selection criteria we devised to improve the astrophysical purity of the new detections.} In general, deconvolution finds faint sources, not only those that are isolated but also those that exist in the vicinity of relatively brighter sources. These sources have FWHM values similar to the typical FWHM of deconvolved sources that were matched with a source in the original images. However, the ellipticities of these newly detected deconvolved sources are slightly higher than those obtained from the one-to-one matches. This anomaly may happen because this catalog of newly detected deconvolved sources has magnitudes towards the fainter end than the one-to-one matches, and we have observed weak trends suggesting that ellipticities of fainter deconvolved sources tend to be higher (see Figs.~\ref{fig:unmatched-deconvolved-combined} and \ref{fig:unmatched-deconvolved-combined-g-and-i}). Furthermore, we note that deconvolution also identified several sources brighter than the limiting magnitude of the original image. This fact may be counterintuitive at first, since those sources should have been detected in the original image. However, upon closer inspection, it appears that deconvolution may have made previously undetectable sources detectable, as whether a source can be detected depends on the specifics of the detection procedure and the choices of parameters used in SExtractor. We have further confirmed that a few tens of percent of such sources resulted from serendipitous deblending performed by deconvolution, which separated the deblended deconvolved sources further away from an original source beyond the crossmatching threshold, and thus could not be crossmatched. Additionally, a small fraction of these unmatched deconvolved sources could be crossmatched using the original catalog before applying our selection criteria.

A careful investigation of the newly detected sources in the deconvolved image revealed that many of these sources are deblended by deconvolution without being deblended by SExtractor. We show several visualizations of deblending scenarios in Figs.~\ref{fig:deblend-examples} and \ref{fig:more-deblend-examples}. We find that the deblended deconvolved sources are separated by $\sim$3-5 pixels and generally find good agreement between the combined magnitudes of the deconvolved and those of the original. These visualizations have highlighted that deconvolution can separate similarly bright sources, but also when one of them is $\approx$16 times fainter than the other in brightness ($\Delta m \approx 3$). Many of these cases are such that a larger ellipticity (original) source is deblended into two smaller ellipticity sources, but there are a few cases where a smaller ellipticity source is deblended such that one of the components has a larger ellipticity. The latter types of cases are generally less common, but we caution that these may be plausible because of the tendency to overestimate the ellipticity of the deconvolved sources (as discussed above). Finally, we also found potential deblending candidates highlighted by the deconvolution in Appendix~\ref{appn:many-to-one} that are separated only slightly greater than a pixel, at least some of which seem to visually be true deblends -- in this case, we found examples where the deblended sources differ in magnitude by $\Delta m \approx 1-1.5$. This suggests that deconvolution also has the potential to deblend very close overlapping sources. \reviewTwo{It might be possible to ascertain the deblending examples we have obtained by cross-correlating with surveys having higher spatial resolution, such as the DESI Legacy used above or space-based telescope data (e.g., from JWST).}

\reviewTwo{Because the newly detected deconvolved sources that pass our astrophysical cuts were still not confirmed to be real, we crossmatched them with source catalogs from the deeper DESI Legacy DR10 (or `LS' for short) and analyzed the crossmatching results. For demonstration, we applied our procedure to three ZTF fields at high- and intermediate galactic latitudes. We applied cuts on LS catalogs based on photometric quality and the number of observations in each band, and flagged those deconvolved sources that successfully matched a source in LS as real detections. Depending on the field, at least 30-50\% of sources that were newly found by deconvolution (i.e., not present in the original image) and previously considered likely astrophysical were confirmed to be real through LS crossmatching. However, the true fraction of real astrophysical deconvolved sources is expected to be larger if the photometric quality filtering on LS sources is made less strict. In the case where no such cuts are applied, we find that the fraction can reach 80 to 97\%, depending on the field. For the remaining deconvolved sources (i.e., those only detected in the deconvolved image and not in the original or LS images), we ran forced photometry of ZTF sources that matched a source from Pan-STARRS1 at the deconvolved source locations. The deconvolved sources at whose locations the light curves passed our set of basic photometric quality criteria were considered to be real candidates. Only about 10-15\% of the remaining deconvolved sources were flagged as real, while the rest are potentially anomalous detections. The confirmed real deconvolved sources not present in the corresponding original images found using the above procedure were concentrated toward the faint end ($m > 19$), and we also identified deconvolved sources around and fainter than the limiting magnitude of the original image}.

The results presented above were obtained using the $r$-band. Comparing these results with the $g$ and $i$ bands, we find that the FWHM of the deconvolved sources remains generally similar across the different bands. The magnitude agreement of individual sources does not show significant differences across bands, but different trends and patterns emerge as the observed sources become fainter. The ellipticities of the deconvolved sources vary the most across the three bands. The SGP deconvolution can be extended to multi-band deconvolution, which may allow for the incorporation of more information through cross-filter correlations and potentially improve the deconvolution results.

We performed a brief visual analysis of the deconvolution results for the NGC 1569 dwarf galaxy and the NGC 7006 globular cluster. For the dwarf galaxy, we obtained a `speckled' deconvolution, which is a characteristic feature of the RL algorithm \citep{Hanisch1994}. The speckled structure for this extended source is not unexpected because, apart from early stopping, no explicit regularization is imposed during the iterations of SGP. Implementing additional regularization may help smooth out the result. Additionally, the deconvolution visually resolved the globular cluster's core, which was previously a continuum in the original image. In both cases, we also detected some new sources.

Our current Python implementation of SGP is not optimized for speed, so it takes roughly 5 minutes to process the entire 3k $\times$ 3k pixel field on the CPU. This execution time is the total time required to deconvolve 49 subdivisions of size $512 \times 512$ pixels with a 10-pixel overlap, which means that, effectively, a field with slightly larger dimensions than the original is being deconvolved. Our current implementation does not use multi-threading or a GPU, so it is envisioned that these tools will be used in the future to provide substantial computational benefits. Several options exist to make our code scalable to large-scale datasets from surveys: using a GPU-based implementation of SGP following \citet{Prato2012}, using third-party libraries such as {\sc JAX} \citep{Jax2018} for quicker gradient calculations, or allowing parallelization using multiple cores to take advantage of the independent deconvolution of different subdivisions of the entire field. However, we note that our method is prescriptive, unlike predictive models such as deep learning-based models, so there is no requirement for training or retraining models on different datasets, which may be desirable.

We compared SGP with a regularized RL algorithm on a test example and found that the number of one-to-one crossmatches is more complete when using SGP than RL, i.e., more original sources crossmatched with deconvolved sources. The RL algorithm does not preserve the flux of sources, which is expected as the background level is non-zero, whereas SGP is significantly better in flux conservation due to the special projection step contained in the algorithm. The general trend is that the FWHM of deconvolved sources are lower when using SGP, whereas their ellipticities are similar for both algorithms. RL produces several cases where the centroid of the deconvolved source is shifted by more than a pixel, while this was not observed for SGP. Finally, RL is found to be computationally prohibitive due to its slower convergence, which requires 40 minutes to deconvolve the entire field, which is eight times slower than SGP. Therefore, it is apparent that SGP is the preferred choice considering both physical plausibility and computational expense.

We have also discussed that our method based on SGP may produce small biases in the photometric magnitudes, and it may be compelling to experiment with a modified flux constraint in which the contribution of bright and faint sources can be appropriately weighed. We note that there exist debiasing procedures that can be used following deconvolution \citep[see, for example,][]{Akhaury2024}. In fact, changes in the deconvolved SGP estimate during iterations can be adapted differently for bright and faint sources, which may be in general beneficial \citep[e.g.,][]{White1994, Lee2017}, and it might also improve photometric accuracy in the deconvolved solution. Also, while our deconvolution accounts for spatial variations of the background across the field of view using subdivisions, other sources of noise, particularly correlated noise, fall outside the assumptions of our imaging model. In this paper, we have not discussed how correlated noise may affect deconvolution; however, as long as the noise has a small correlation length, the model should not be invalidated and the results are expected to remain fairly robust.

The clearest advantage of the deconvolution method presented here might be in locating previously undetected sources (e.g., extremely faint sources) and in deblending closely separated sources. These new detections can be located in the corresponding original image, and photometry can be performed on them. The deconvolved images may be used as astrometric references and may also be directly useful for downstream scientific analyses, but we caution that while individual source flux conservation is exceptional for some fields, it is generally moderately good. As a result, it may currently be less reliable for studies requiring precision flux preservation, such as variable detection through image subtraction or similar approaches.

\subsection{\reviewFour{Survey-specific tuning}}\label{subsec:survey-specific-tuning}

\reviewFour{While the application of the deconvolution algorithm itself is data agnostic, post-processing filtering criteria should be tailored to scientific goals. Our deconvolved source selection criteria in this work were intentionally conservative rather than perfect for magnitude depth: although deconvolution could (and has indeed) found previously uncataloged sources up to the zeropoint magnitude, we apply several cuts, including $m \leq 21.5$, to prioritize purity at the cost of excluding some very faint ($m > 21.5$) but plausible deconvolved sources. Therefore, it is possible that deconvolution finds new astrophysical detections in the extremely faint regime, but expect to have substantial spurious detections there.}

\reviewFour{If deconvolved sample purity is desired, restrictive cuts like ours, but suitably adapted to other surveys, can be a good starting choice for application to new data: for example, the $m = 21.5$ deconvolved limit we used for ZTF, which is one magnitude deeper than the typical limiting magnitude of ZTF, can be replaced with around 1 mag deeper than that survey's limiting magnitude. However, if completeness at the very faint ends (e.g., fainter than the limiting magnitude of the survey) is a priority, the magnitude limit can be set much deeper than the limiting magnitude. This flexibility is analogous to how adjusting classification thresholds in machine learning allows trading precision for recall, or vice versa. For stellar sources, a FWHM criterion like ours, based on stellarity cuts and modified z-scores, can be a natural first choice for other surveys because, using ZTF data, we have shown that deconvolved FWHMs show a tight distribution irrespective of the original source seeing. The criterion would require modifications to include non-stellar sources. Note that our FWHM thresholds are pixel-based, which, for ZTF, is roughly the same as in arcsec units, so other surveys can use the same criterion expressed in arcseconds. While we find deconvolved ellipticities are not constrained as well as FWHM, ellipticity cuts can be tuned to new data if desired, depending on observed PSF ellipticity. Likewise, cuts based on detection flags can be set according to photometric purity requirements.}

\reviewFour{A note on PSF sampling: the FWHM of our (stellar) deconvolved sources for all fields is between 0.95 and 1.3 arcsec, irrespective of the FWHM of the original sources. In an ideal scenario, deconvolution can approach the diffraction limit, which in the $r$ band for the 1.2 m telescope of ZTF is $\approx$0.11$\arcsec$, provided that the data is well sampled. However, our deconvolved FWHM are larger than this theoretical value. This is expected due to several factors, including data noise, imperfect PSF modeling, and imperfect deconvolution due to the regularization used in our algorithm. It is also important to note that, as per the Nyquist-Shannon sampling theorem, a minimum of two pixels are required across the seeing profile to sample a source optimally. The ZTF camera pixel size is 1$\arcsec$.012 per pixel, which means that it can optimally sample sources with $\mathrm{FWHM} \gtrsim 2\arcsec.024$. Five images in our dataset had a seeing FWHM that was slightly smaller than this threshold, suggesting that these images were undersampled. As the difference is not huge, we expect that it did not have a substantial negative effect on our deconvolution. However, deconvolution can help in superresolution of severely undersampled astronomical images using, for instance, multiple dithered exposures \citep{Starck2002}, but the extension of the SGP algorithm used here to multi-epoch deconvolution is left to future work.}

\reviewFour{Nevertheless, when PSFs are not adequately sampled, it is important to adapt the changes to the filtering criteria discussed above accordingly. We have also found that the typical differences in the centroids (flux-weighted average positions) of the original and deconvolved sources are much smaller than a pixel. This negligible difference is expected since deconvolution reveals the core emission, which has a dominant contribution to flux weighting. For severely distorted PSFs, unlike those used in this study, the centroid differences might increase, but the deconvolved centroids should then be viewed as astrometric improvements. Another relevant application is blind deconvolution, where the PSF is estimated along with the deconvolved image, and SGP has been adapted in at least one such study before \citep{Jia2017}.}

\reviewFour{On a side note, the alternative treatment by \citet{Magain1998} can constrain the shapes of the deconvolved point sources to the extent that they become completely known. They do this by deconvolving with a narrower PSF to prevent violating the sampling theorem. Our quick internal check suggested that this straightforward modification does yield more circular deconvolved stellar sources. However, a more detailed investigation of this approach is required, especially because their approach involves solving another deconvolution problem, and small errors in intermediate estimations may affect the final result \citep{Starck2002}. If proven effective, this approach could directly inform the design of robust filtering criteria for deconvolved sources.}

\reviewFour{A systematic approach to design survey-specific selection criteria can be obtained by two independent approaches: injecting simulated sources including magnitudes fainter than the limiting magnitude of the survey, or purely observationally by crossmatching deconvolved detections with deeper catalogs from the literature. In the first case, one can compare detection recovery as a function of original source magnitude and identify what selection cuts on deconvolved sources help identify the faintest sources, and in the second case, one can derive empirical guidance on what selection cuts maximize true crossmatches. In the observational approach, vetted ambiguous deconvolved sources can also be used to train or fine-tune real-bogus classifiers.}

\reviewFour{For the reasons mentioned below, we have adopted the wholly observational approach, but without claiming our deconvolved selection cuts to be optimal. We hope that future applications make informed choices depending on their goals when analyzing deconvolution results.}

\reviewFour{Finally, a note on source injecting experiments: while seemingly straightforward, it brings with it problems related to proper photometric, spatial, and temporal distributions, making things often more difficult, or at times seemingly simpler, than with real data, and it can be a big undertaking on its own. But our method is amenable to such testing and can be undertaken by groups getting ready for specific new datasets, such as from the Rubin observatory, to quantify completeness limits on magnitudes and constrain observable properties of deconvolved sources when the method is applied.}

\section{Conclusion}\label{sec:conclusion}
We have presented extensive applications of the SGP algorithm, an effective framework to solve convex, constrained optimization problems, using it for the deconvolution of several observed images from ZTF. Our conclusions are as follows:

\begin{enumerate}
\item The deconvolved images preserved total photon flux, but we found that the agreement in the flux between individual observed and deconvolved sources is also moderately good, with small biases towards fainter sources.
\item The FWHM of the deconvolved sources was reduced and tightly constrained, being roughly an arcsecond on average, irrespective of the observed seeing.
\item \reviewTwo{The deconvolution identified several new sources at the faint end that were not detected in the observed images and which were flagged as likely astrophysical by our series of selection cuts on photometric quality and properties. We have confirmed the validity of several of these deconvolved sources either by successfully identifying them in the deeper DESI Legacy surveys images or by follow-up confirmation through forced photometry of ZTF sources matching Pan-STARRS1 data.}
\item Deconvolution effectively deblended overlapping sources, even when they differed significantly in brightness. We also extracted some challenging deblending cases where sources are separated by just slightly more than a pixel (which, for ZTF's pixel scale, also turns out to be slightly more than an arcsecond) and found potential deblending candidates, as identified by the deconvolution.
\item The SGP algorithm outperformed the Richardson-Lucy algorithm in terms of execution time, detectability using SExtractor, and photometric properties of deconvolved sources, which makes it more scientifically useful and practical. 
\end{enumerate}

\reviewThree{Overall, this work demonstrates the universal benefits of deconvolving observed images degraded by atmospheric and instrumental anomalies.}

Since our current implementation requires around 5 minutes to process a 3k $\times$ 3k pixel field, it may not yet be scalable for real-time application since we do not use GPUs or multi-threading. In the future, we plan to release a software package incorporating recipes to address the current downsides of our method, such as improved handling of bad pixels/saturated sources, more efficient computation, incorporation of robust statistics, or more effective regularization within SGP for improved results, and intrinsic handling of multi-band imaging for direct application to sky survey data. 

The results presented in this paper are more transparent than those obtained through machine learning, as most test sets in machine learning do not include all possible real-world situations, thus generally providing an incomplete understanding of their generalization to new data and posing risks for deployment. \reviewThree{In contrast, our approach can be applied to any astronomical image without these issues, as long as a reasonable PSF estimate is obtained beforehand. Testing this method on different astronomical surveys will help assess its astrophysical utility, particularly in preparation for data from next-generation observatories such as the Rubin Observatory.} 

\begin{acknowledgments}
A.M. acknowledges support from the NSF (1545949, 1640818, AST-1440341, AST-1815034) and from from IUSSTF (JC-001/2017). A.M., A.K.K., and M.J.G. also acknowledge Sajeeth Philip, Kaushal Sharma, Shubhranshu Singh, Andrew Drake, and Dmitry Duev for the early discussions on deconvolution using other methods, and Theophile du Laz and Avery Wold for technical help. M.S. would like to acknowledge the hospitality of the Centre for Cosmology and Science Popularization (CCSP), SGT University, Gurugram, India.
\end{acknowledgments}

\facilities{ZTF \citep{Bellm2019,Masci2019}} 

\software{
    SExtractor \citep{Bertin1996}; \textsc{STILTS} \citep[][; version 3.4-7]{Taylor2006}; Astropy \citep{AstropyCollaboration2013,AstropyCollaboration2018}; reproject \citep{Robitaille2024}; scikit-learn \citep{scikit-learn}. Our code implementation is available at this GitHub repository: 
    \url{https://github.com/Yash-10/deconv_ztf}. 
    We also release all SExtractor catalogs, including source properties and detection flags, for users to apply custom selection criteria based on their specific aims.
}

\newpage
\appendix

\section{Comparison of photometric magnitude uncertainties}\label{appn:photo-error-compare}

\begin{figure*}[hbt!]
      \centering
      \includegraphics[keepaspectratio,width=0.24\linewidth]{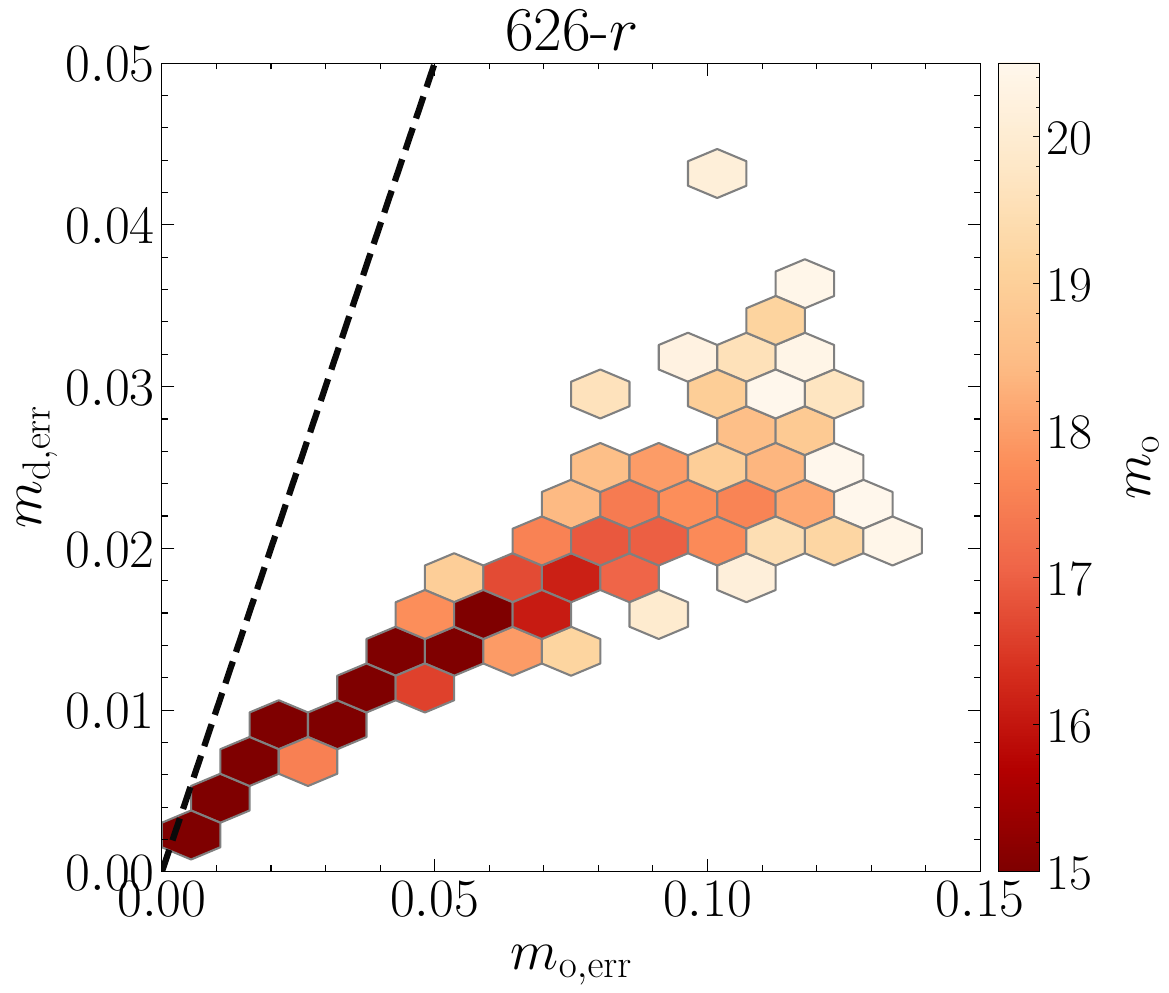}
      \includegraphics[keepaspectratio,width=0.24\linewidth]{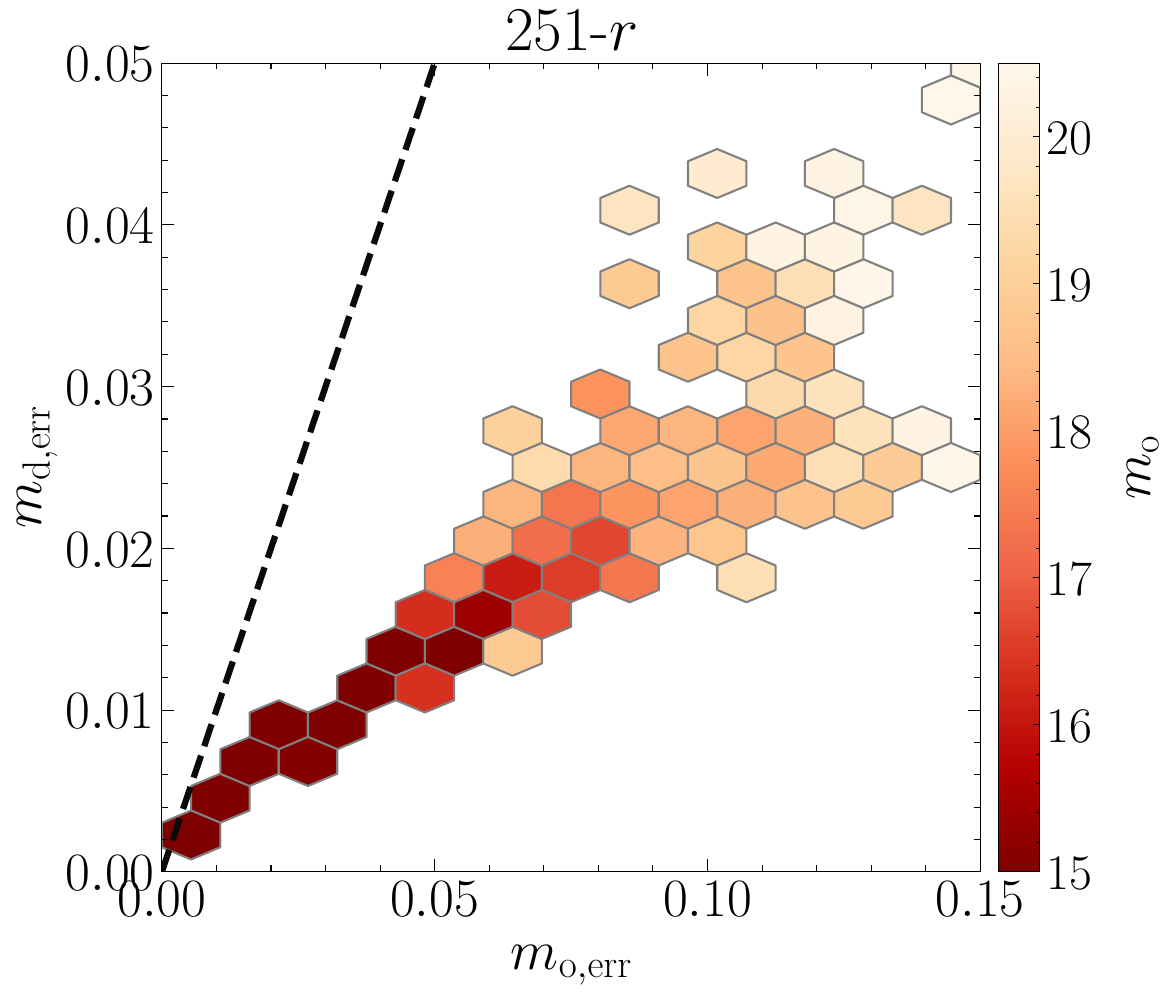}
      \includegraphics[keepaspectratio,width=0.24\linewidth]{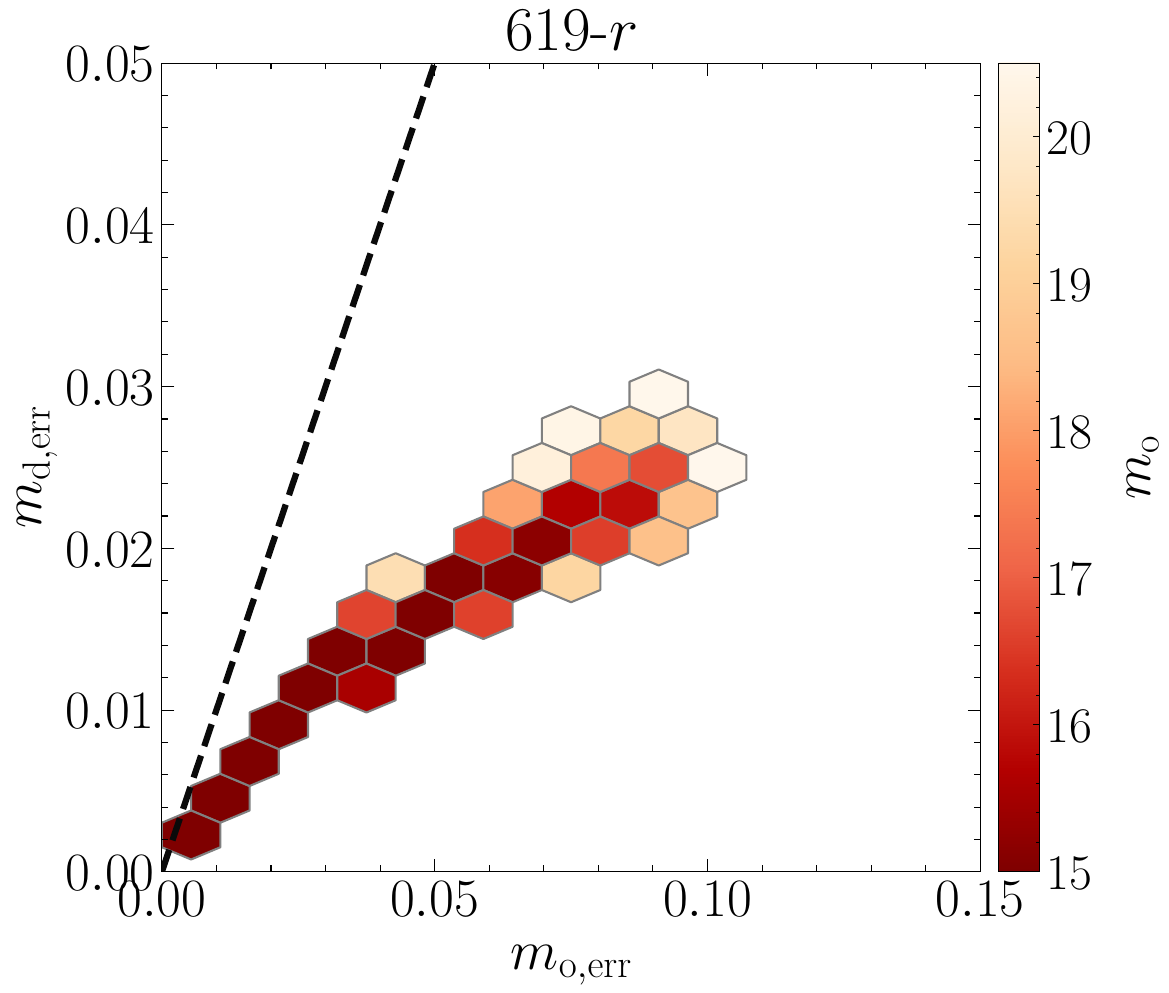}
      \includegraphics[keepaspectratio,width=0.24\linewidth]{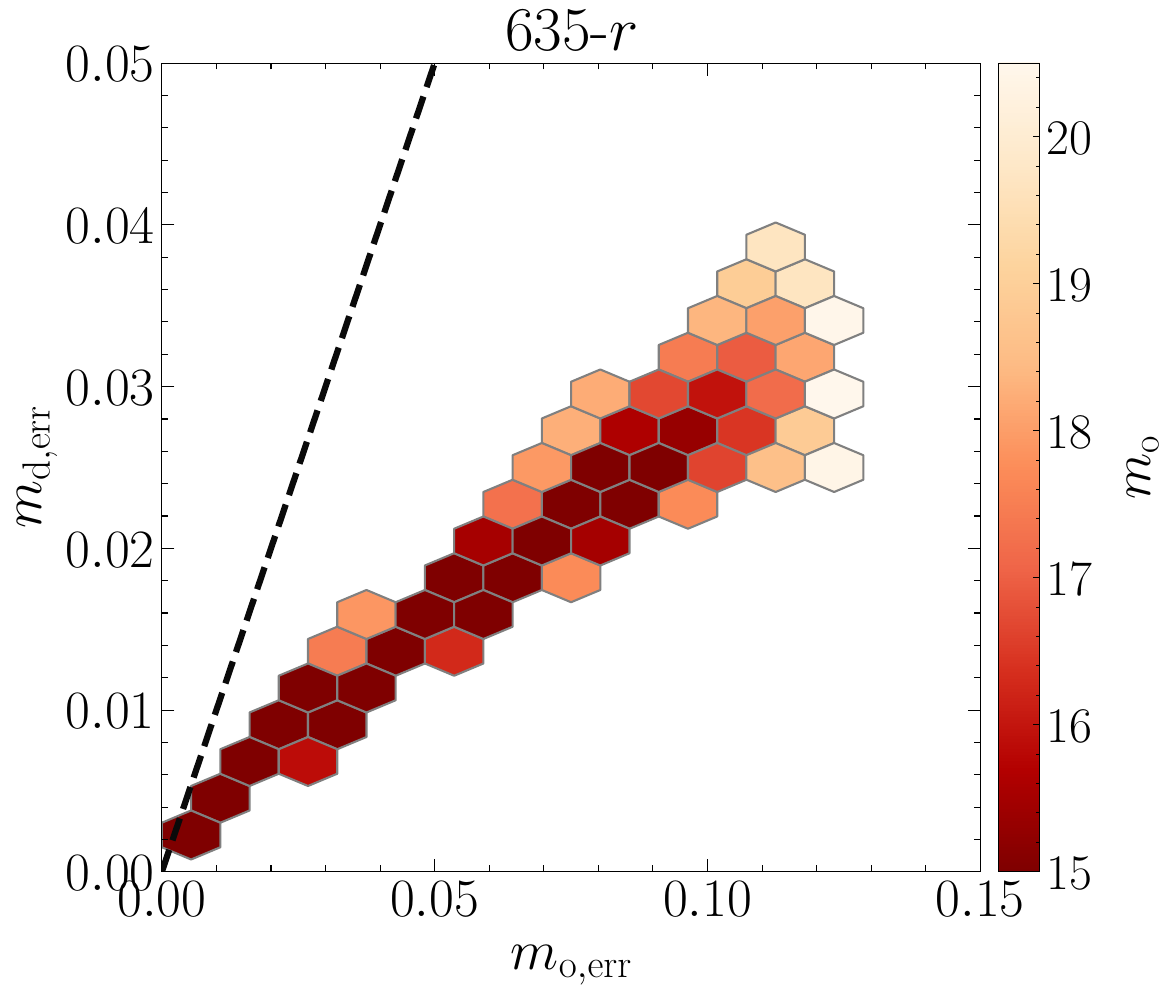}\hfill
      \includegraphics[keepaspectratio,width=0.24\linewidth]{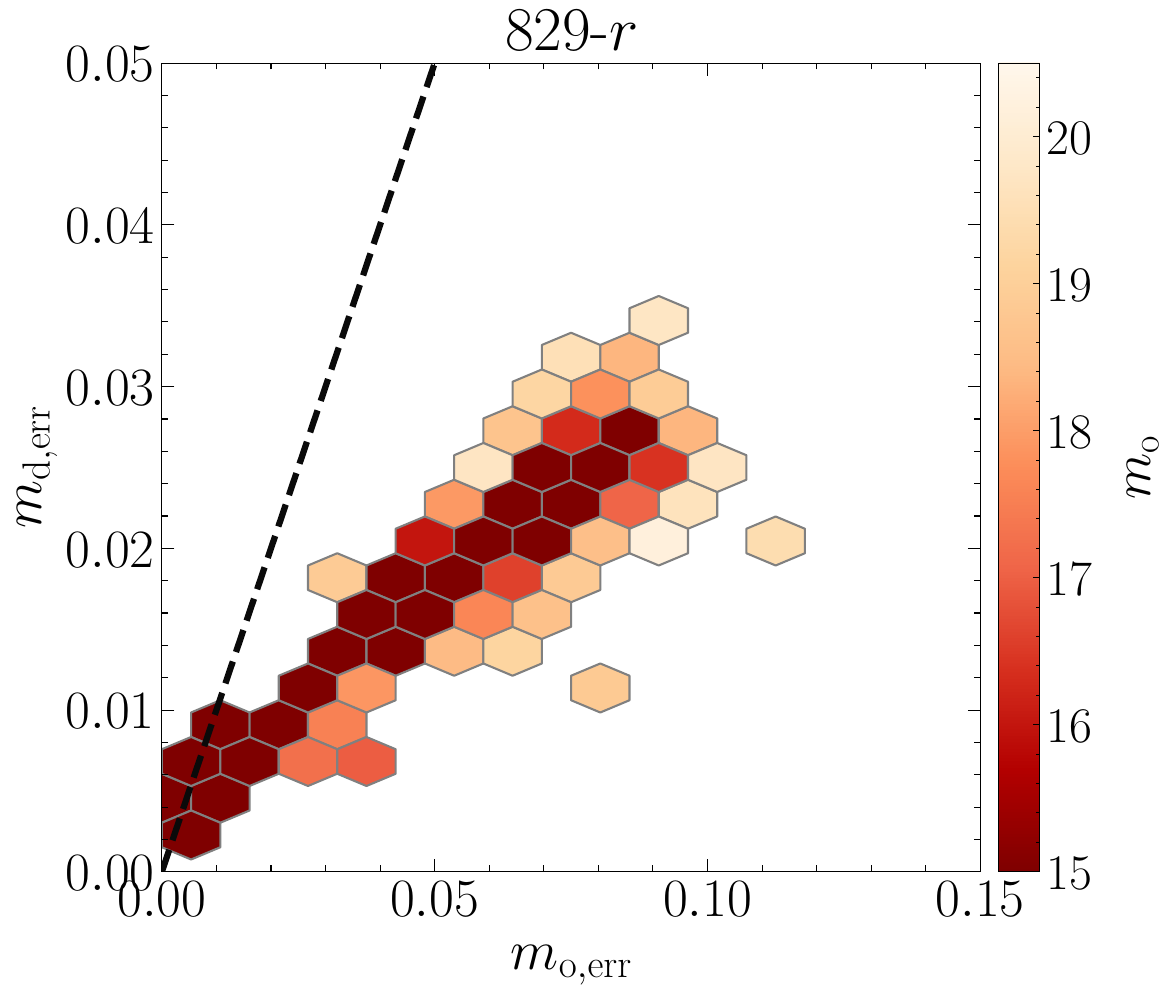}
      \includegraphics[keepaspectratio,width=0.24\linewidth]{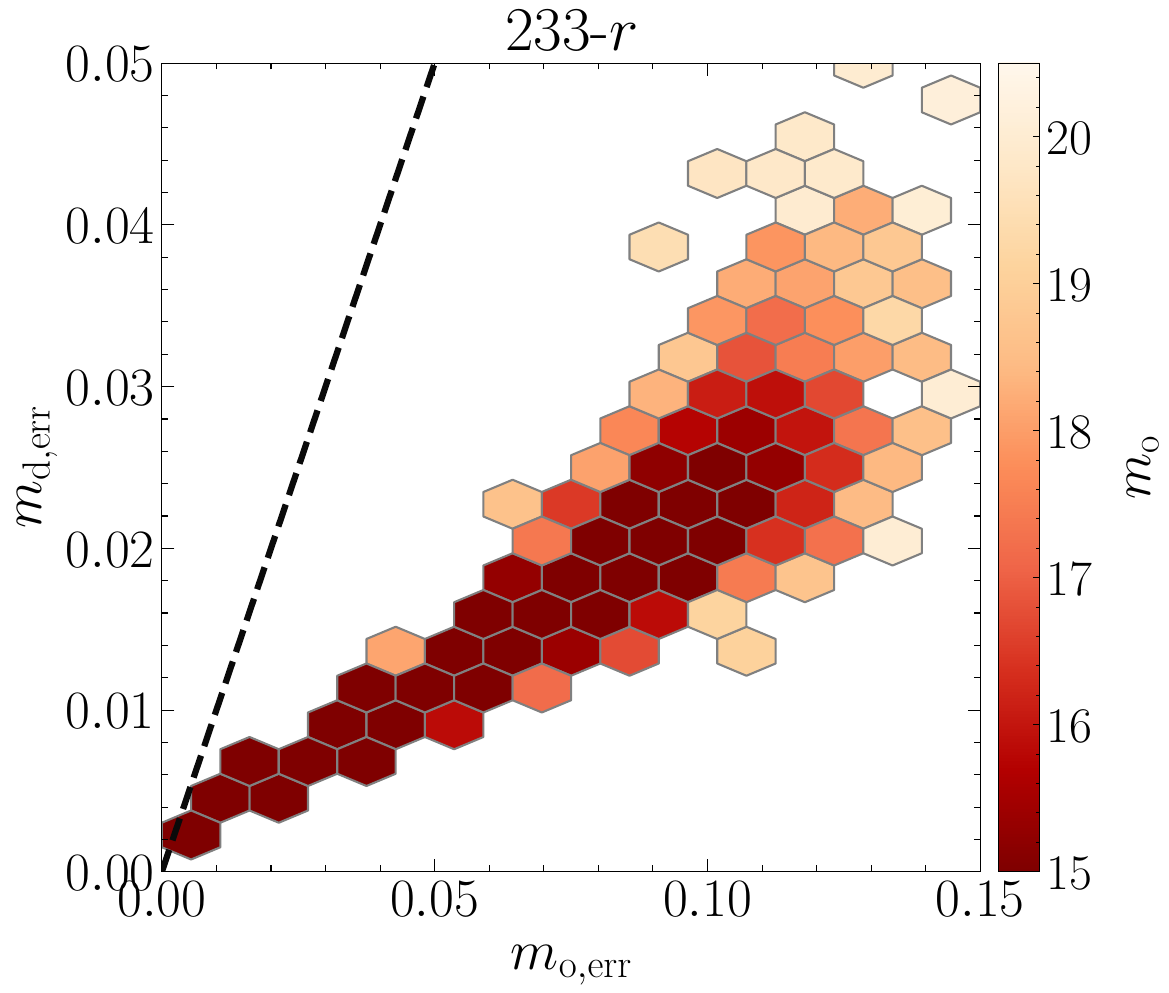}
      \includegraphics[keepaspectratio,width=0.24\linewidth]{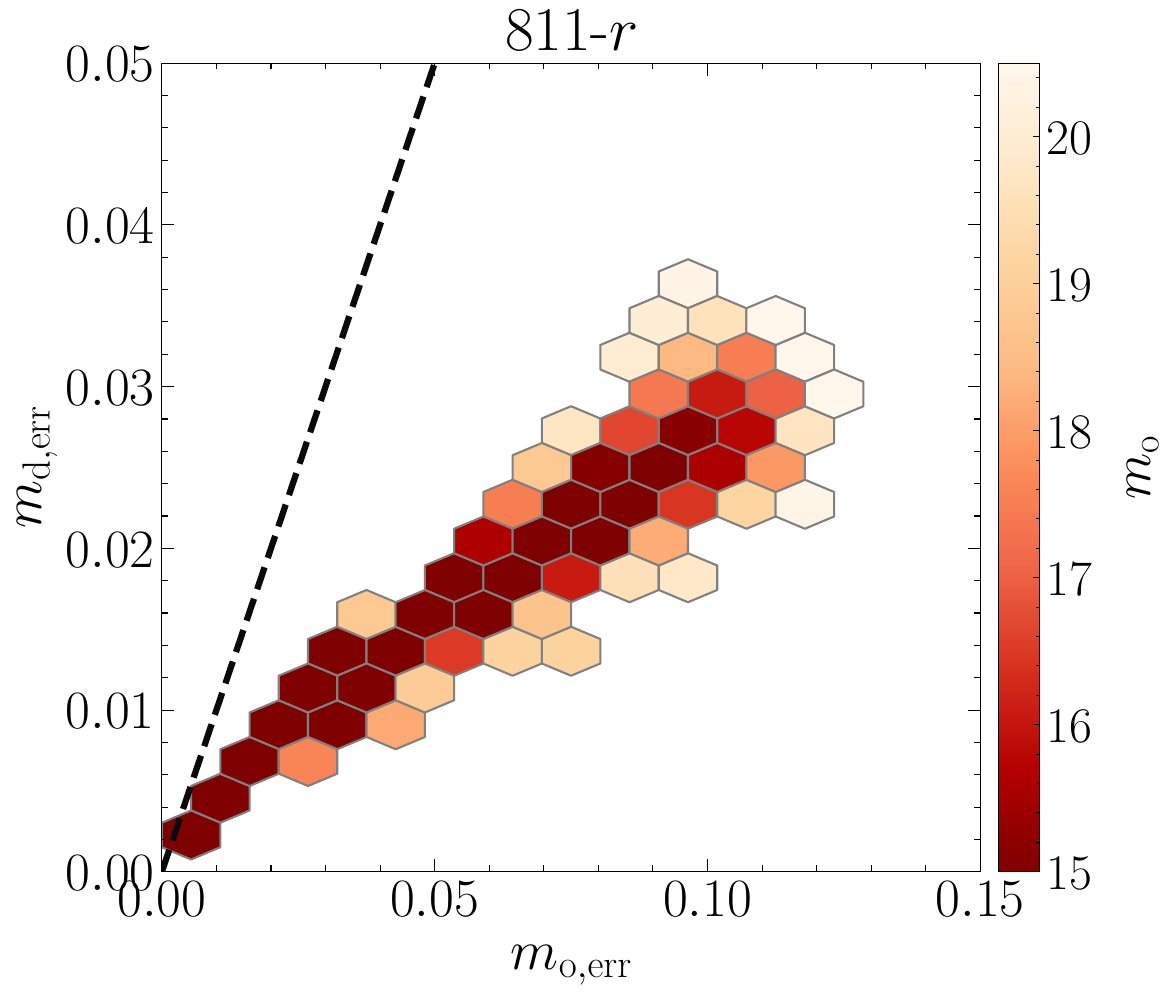}
      \includegraphics[keepaspectratio,width=0.24\linewidth]{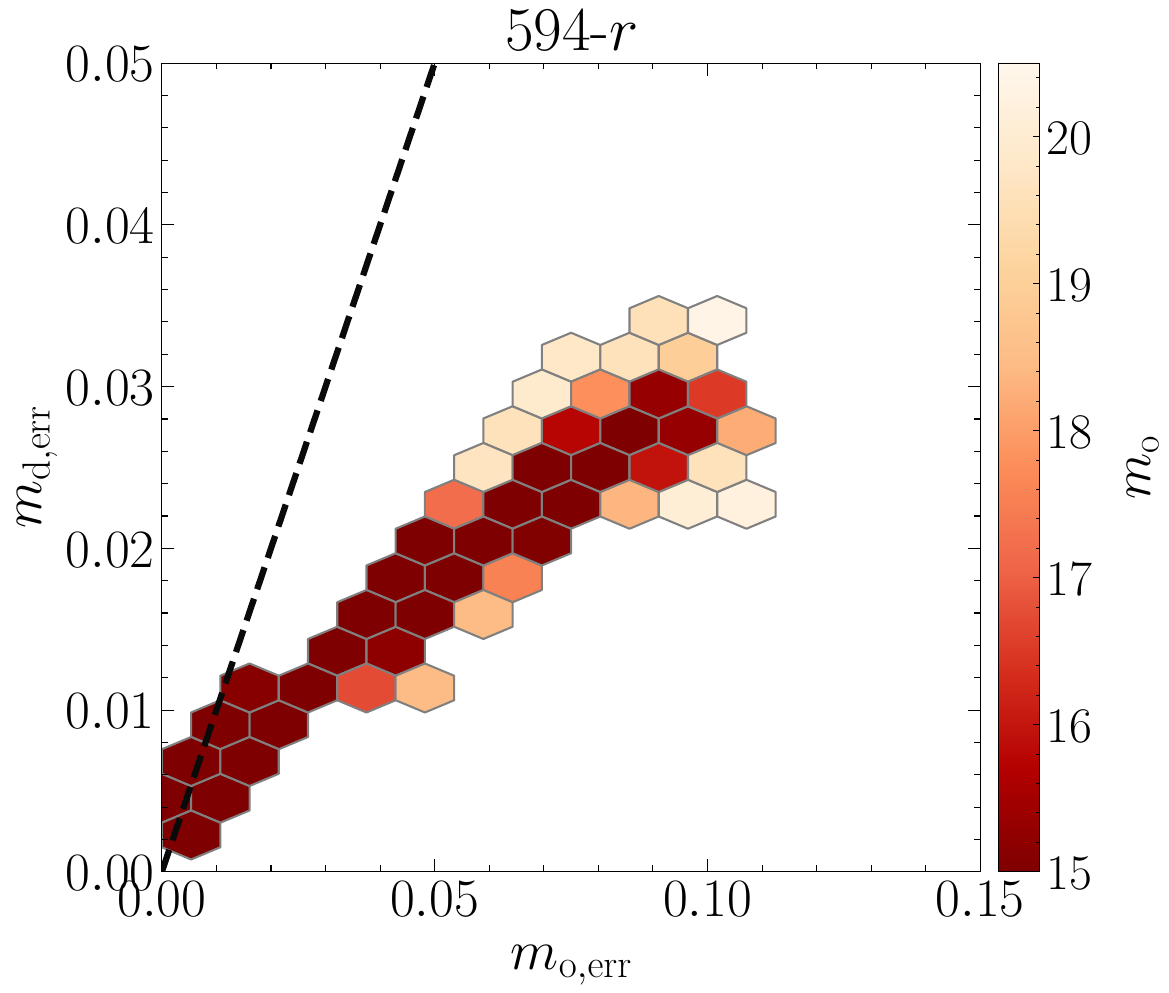}\hfill
      \includegraphics[keepaspectratio,width=0.24\linewidth]{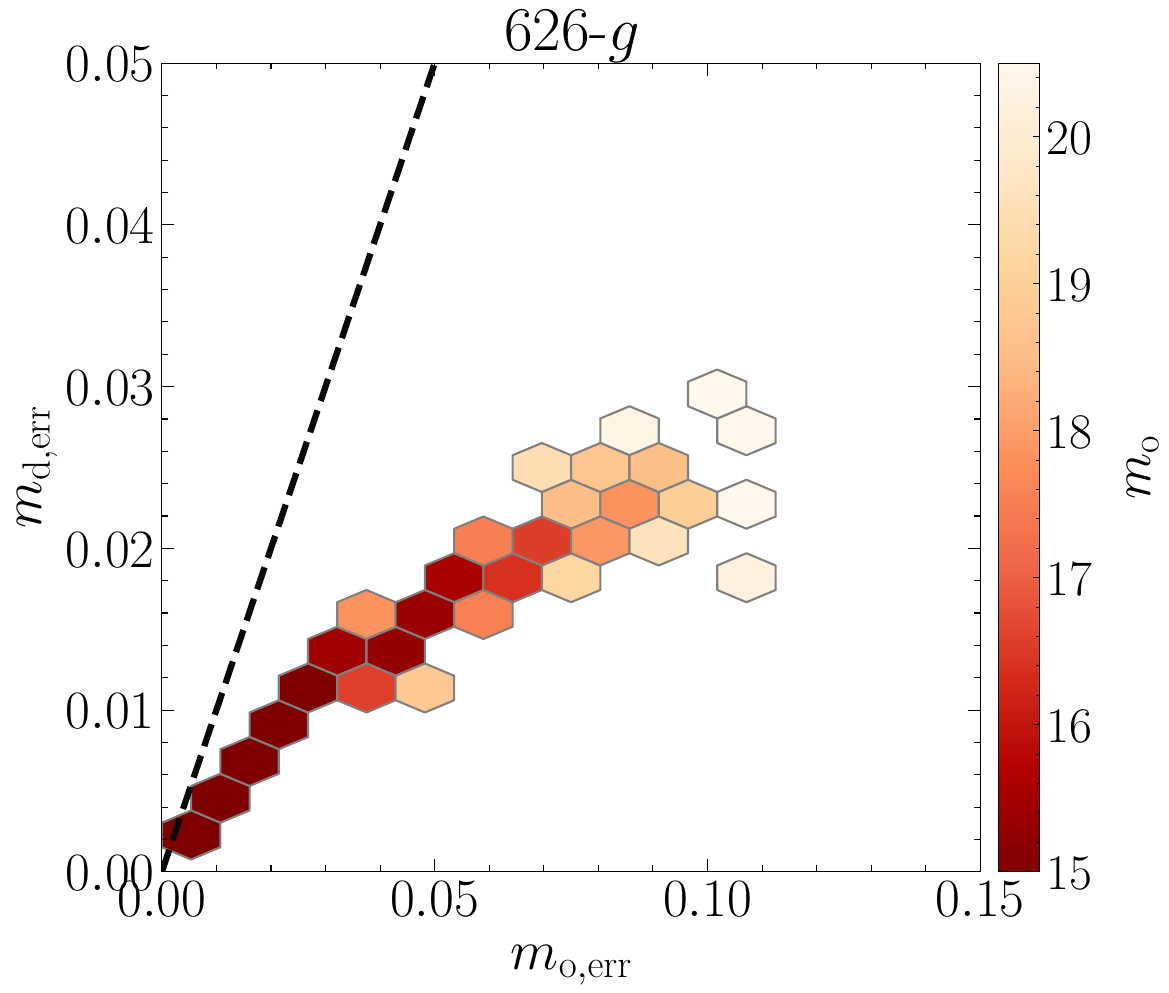}
      \includegraphics[keepaspectratio,width=0.24\linewidth]{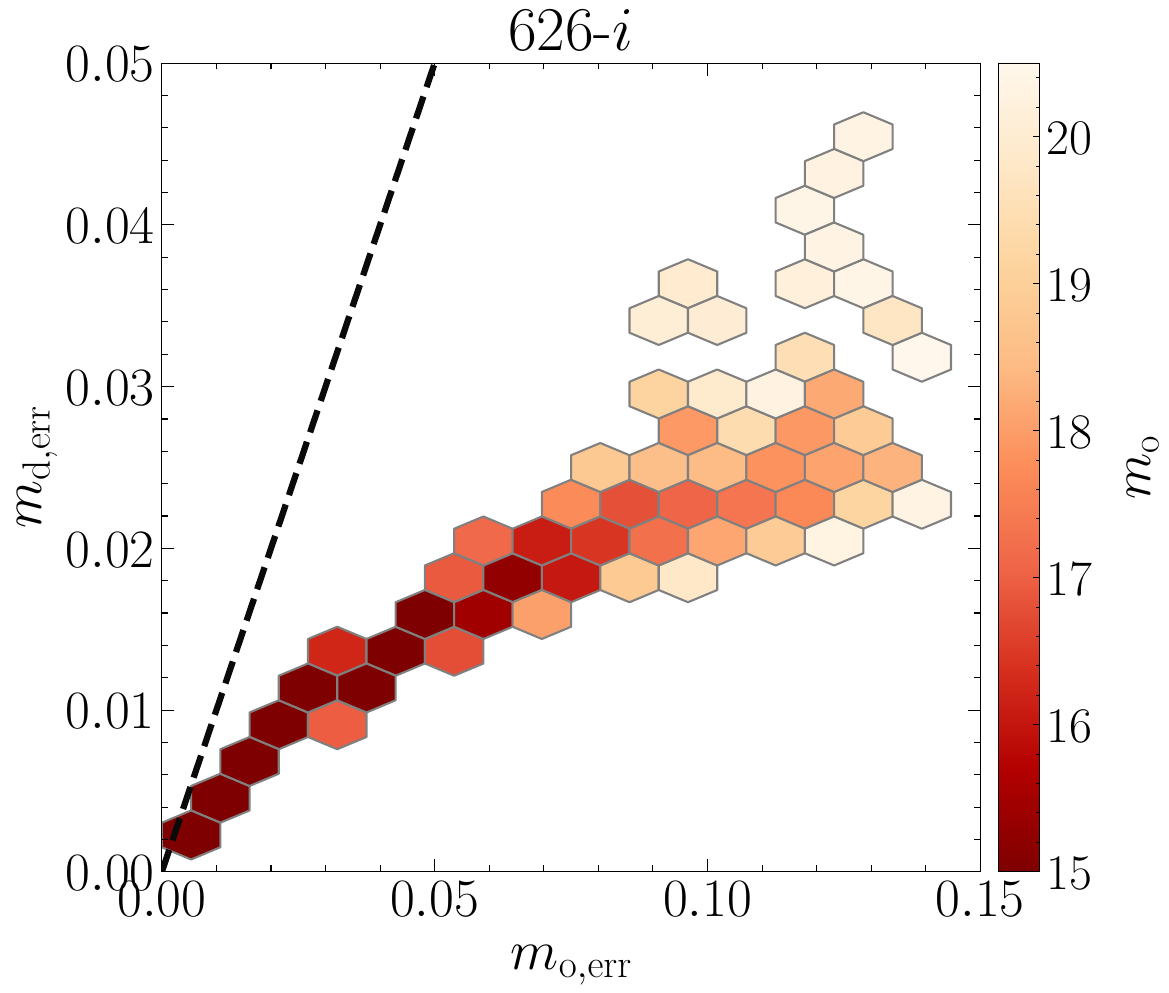}
      \includegraphics[keepaspectratio,width=0.24\linewidth]{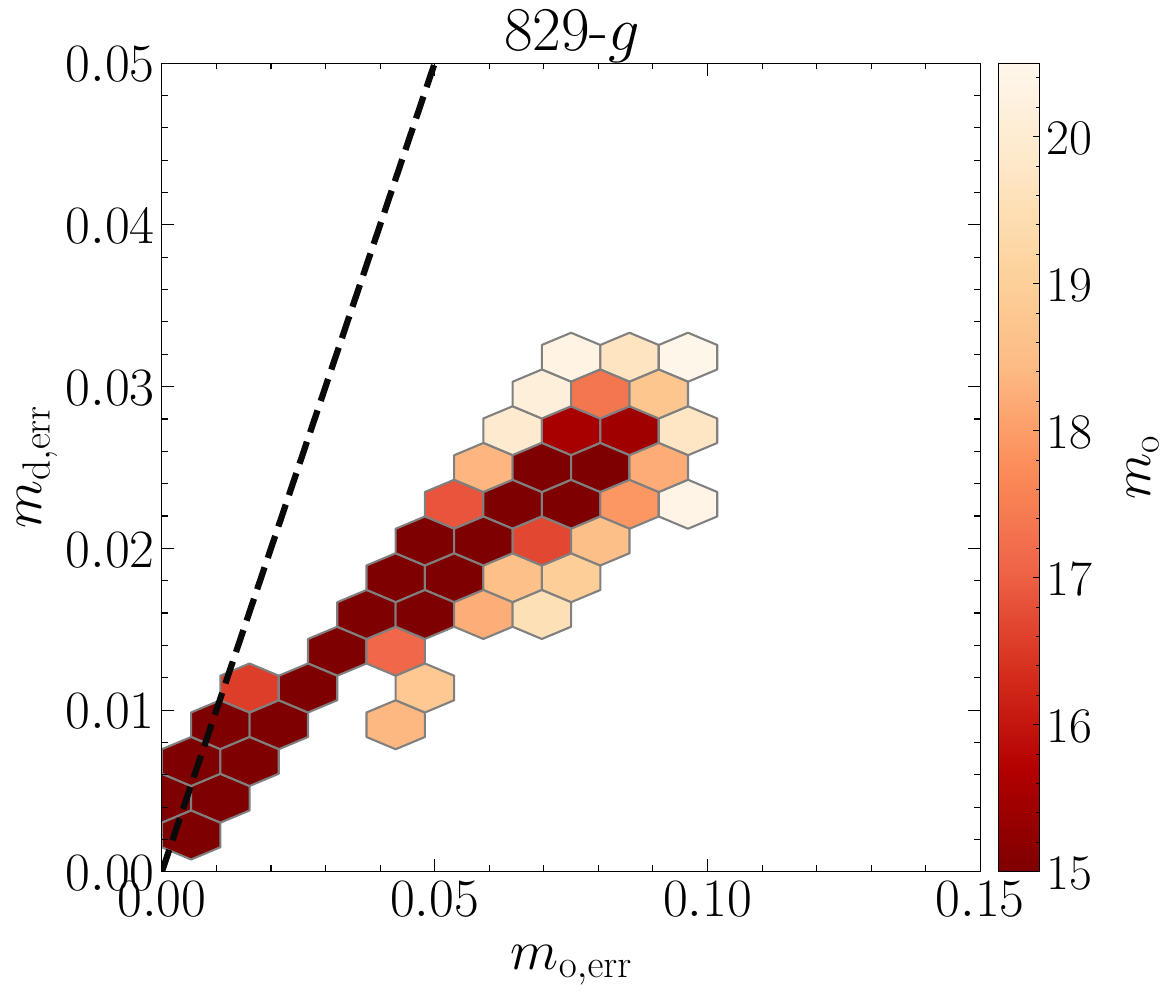}
      \includegraphics[keepaspectratio,width=0.24\linewidth]{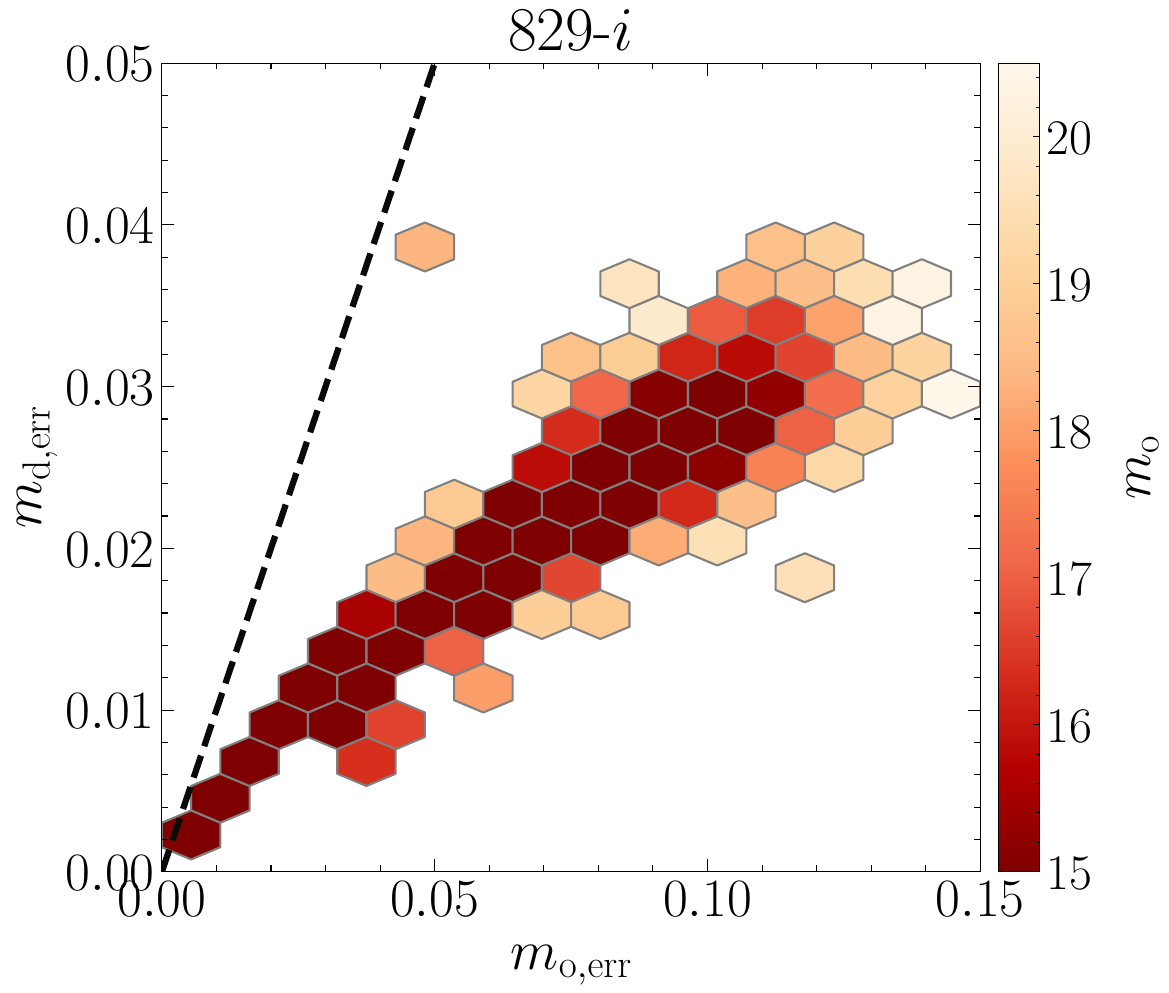}\hfill
    \caption{Comparison of the magnitude uncertainties of the original sources (x-axis) and that of the corresponding one-to-one matched deconvolved sources (y-axis) for all twelve images considered in this study. These include $r$-band images in the top two rows and $g$ and $i$ band images in the bottom row. The title of each panel denotes the ZTF field ID along with the band. The color bar denotes the magnitude of the original source. Each hexagon bin denotes the combined magnitude of all sources in that bin (obtained by finding the magnitude corresponding to the summed flux). The dotted line denotes the $y = x$ line, i.e., the same magnitude errors for the original and deconvolved sources. It can be seen that as the original source gets fainter, the uncertainties in the magnitude increase for both original and deconvolved sources. However, the uncertainties for the deconvolved sources are smaller and increase much more slowly than those for the original sources. As a result, for the faintest original sources, the deconvolved magnitude uncertainties are $\approx$3-4 times smaller than the corresponding original uncertainties.} \label{fig:photo-err-compare}
\end{figure*}

Fig.~\ref{fig:photo-err-compare} shows a comparison of the magnitude uncertainties for the original and deconvolved sources from the one-to-one matches. We note that these uncertainties are those reported by SExtractor and are not the difference between the original and deconvolved magnitudes. The deconvolved sources have much lower magnitude uncertainties and scale better as the sources become fainter. This behavior is expected because deconvolution removes the background and thus removes one of the contributions to the uncertainty term. Thus, the contribution from Poisson noise is mainly left.

\section{Visualizations of many-to-one matches}\label{appn:many-to-one}

We recall that in this study, a successful many-to-one match has a source from the original source catalog matched to multiple distinct sources in the deconvolved source catalog within a distance of 1.383 pixels. Although, as discussed in the caption of Table~\ref{tab:crossmatching-results}, almost all cases are two-source matches. Our primary point of interest is to search for possible deblending scenarios in which the deconvolved sources are separated by distances ranging from slightly smaller than $1\arcsec$ to up to $2\arcsec$, which are the most challenging cases. In this regard, we present a visualization of the many-to-one matches as possible deblended candidates through deconvolution.

We have used the deblending criterion described in Sect.~\ref{sec:deblending-examples} to select potential deblending situations. To obtain more reliability, we excluded original sources that have $\mathrm{FLAGS} > 7$ and those that have the deblending flag set. We also remove many-to-one matches in which at least one of the deconvolved sources has FWHM close to zero and where the distance between the two deconvolved sources is $< 0.5$ pix. We do not remove cases where the original sources have small ellipticities because of the possibility that blends with close separation may give rise to small ellipticities.

\begin{figure*}
      \centering
      \includegraphics[keepaspectratio,width=0.32\linewidth]{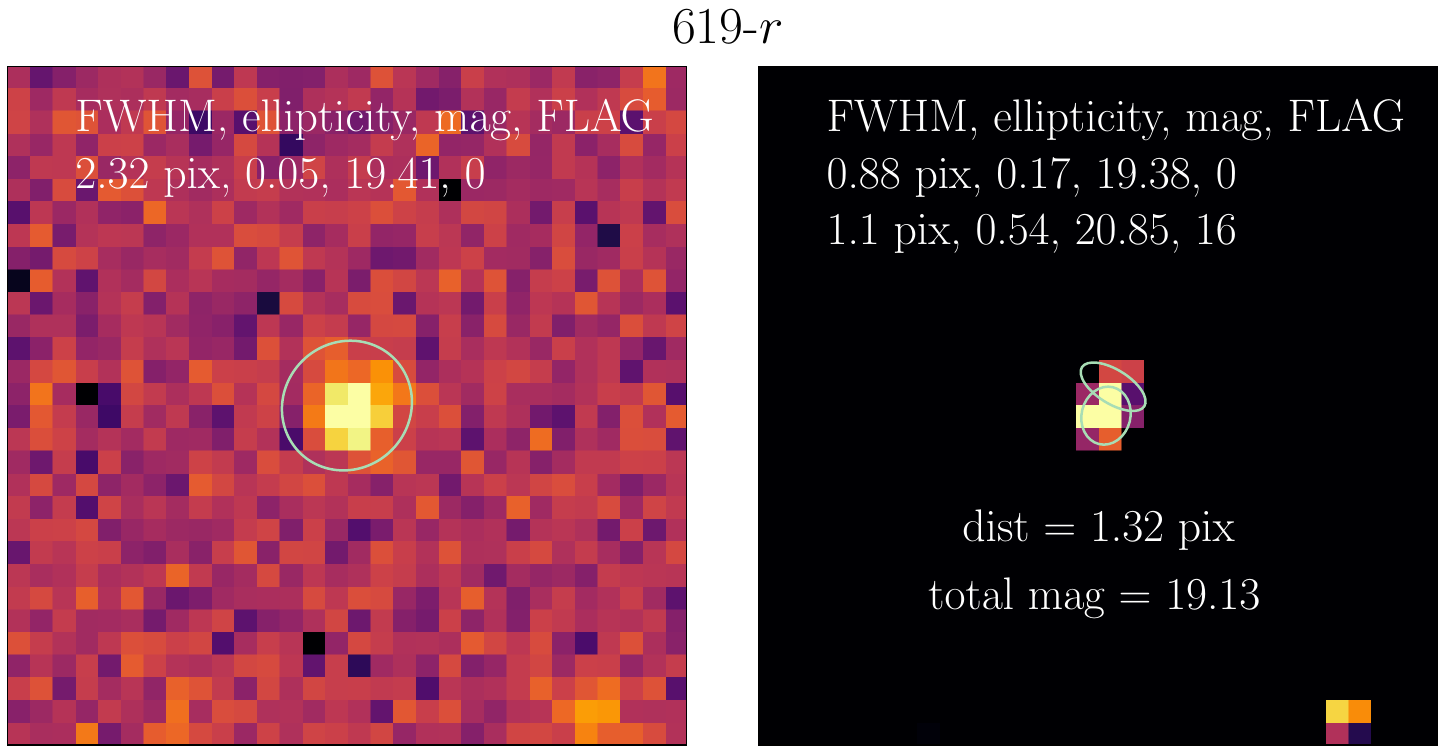}
      \includegraphics[keepaspectratio,width=0.32\linewidth]{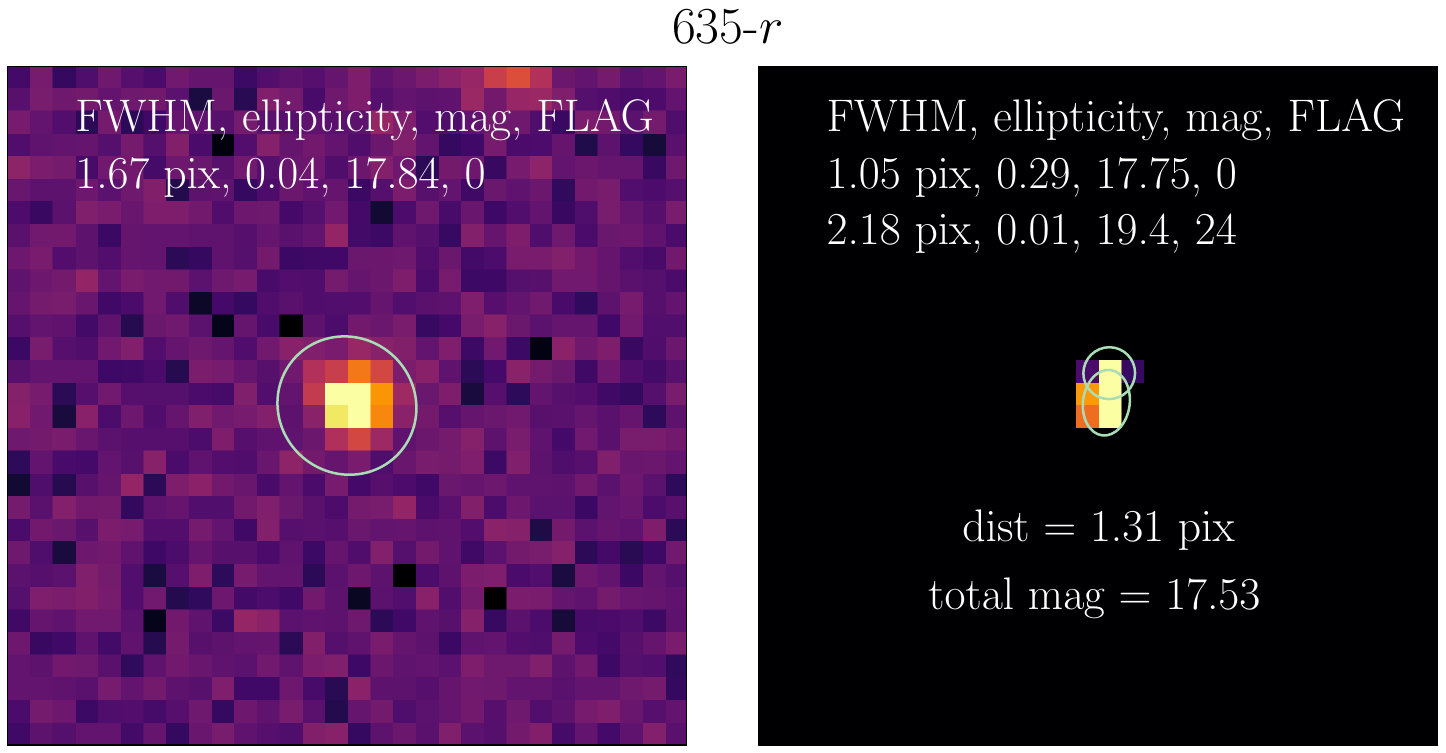}
      \includegraphics[keepaspectratio,width=0.32\linewidth]{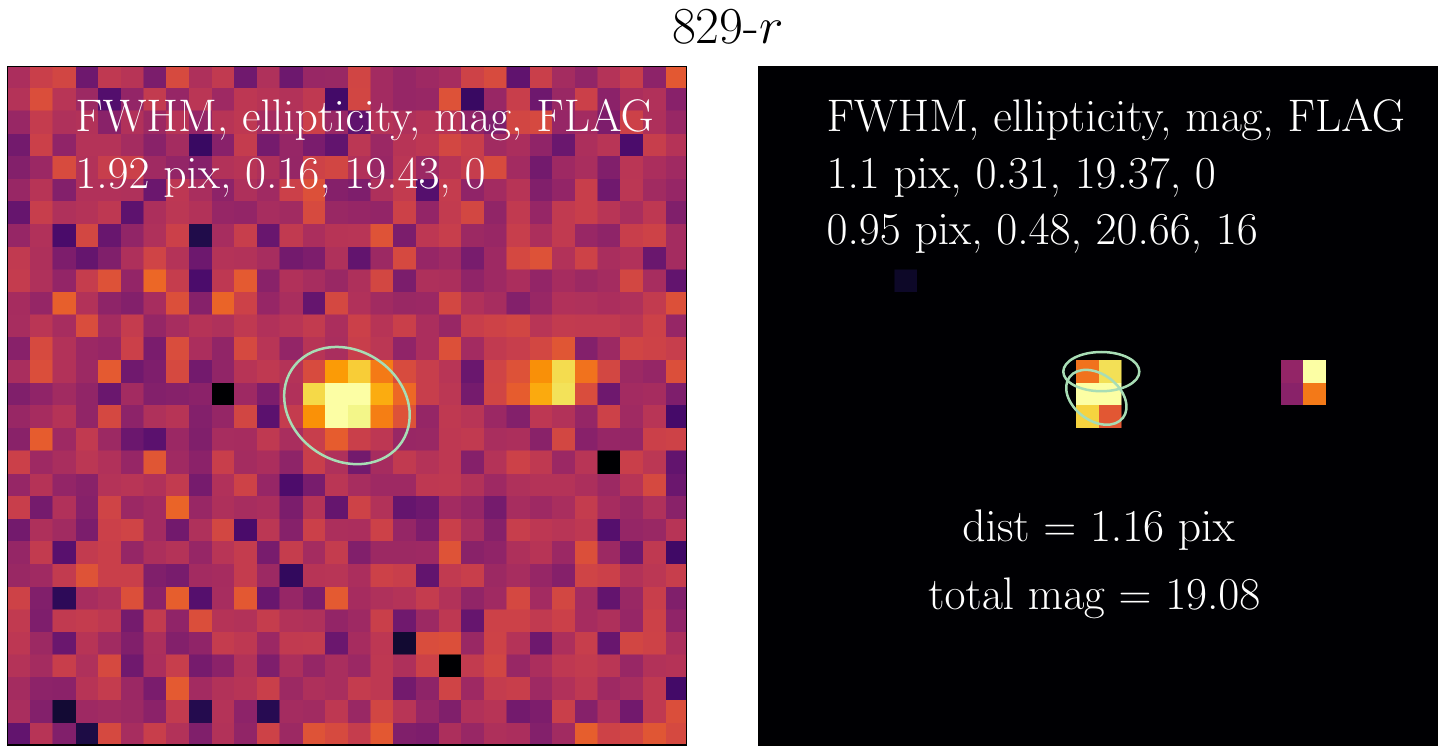}\hfill
      \includegraphics[keepaspectratio,width=0.32\linewidth]{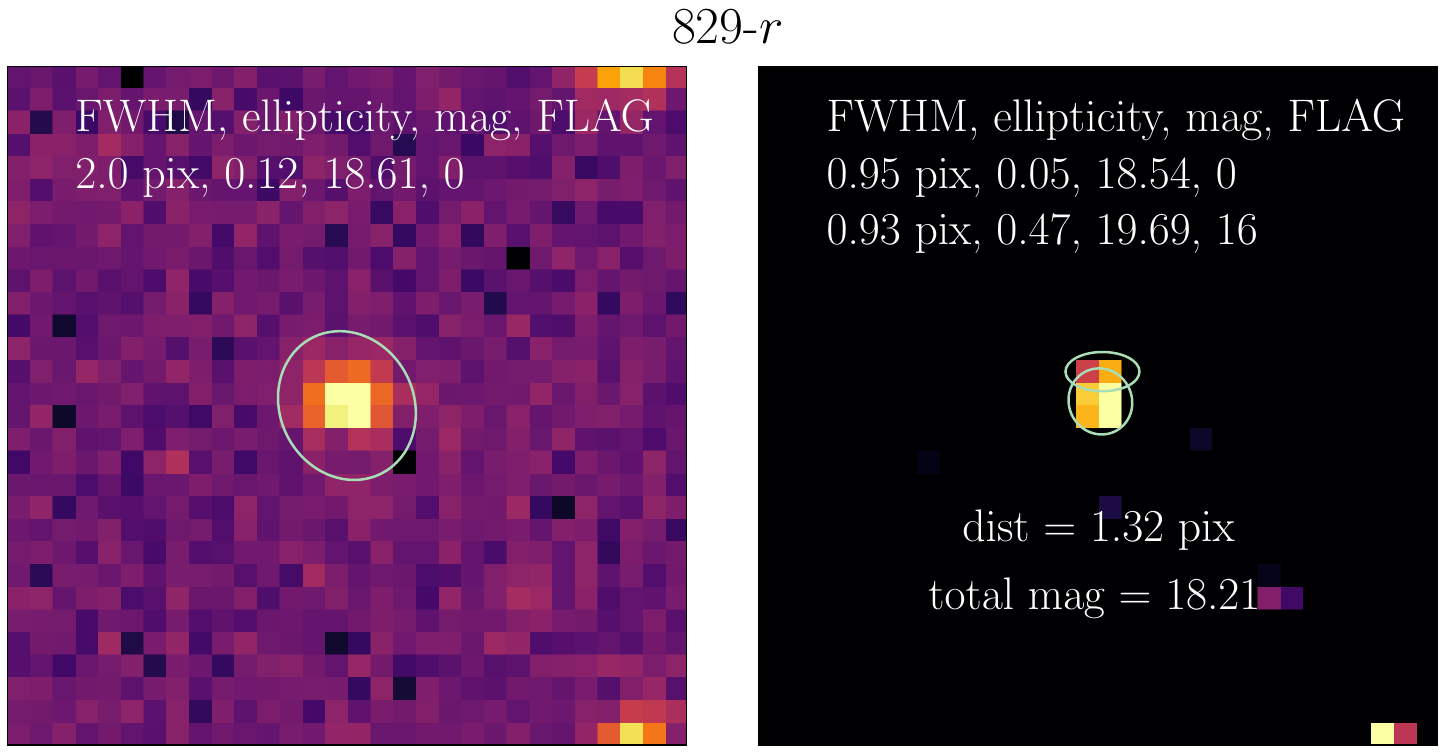}
      \includegraphics[keepaspectratio,width=0.32\linewidth]{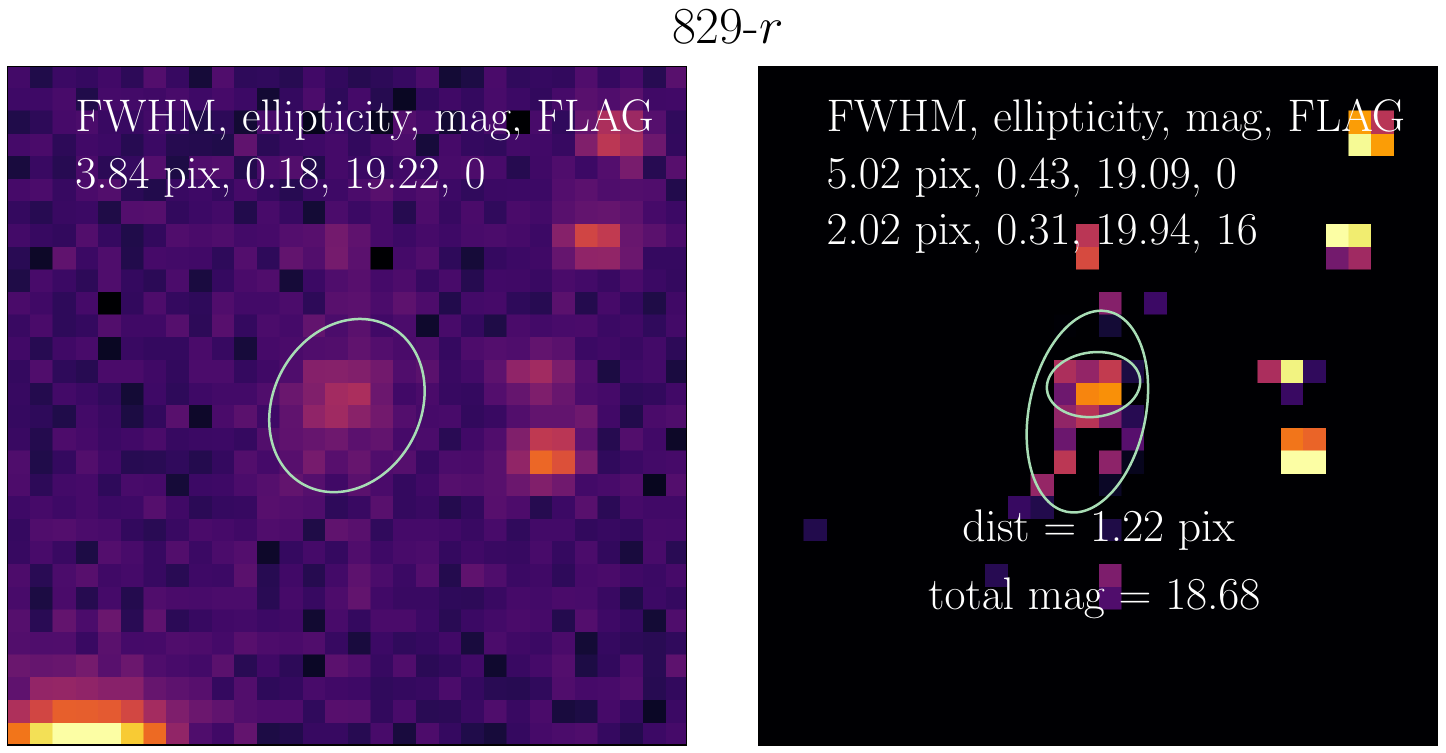}
      \includegraphics[keepaspectratio,width=0.32\linewidth]{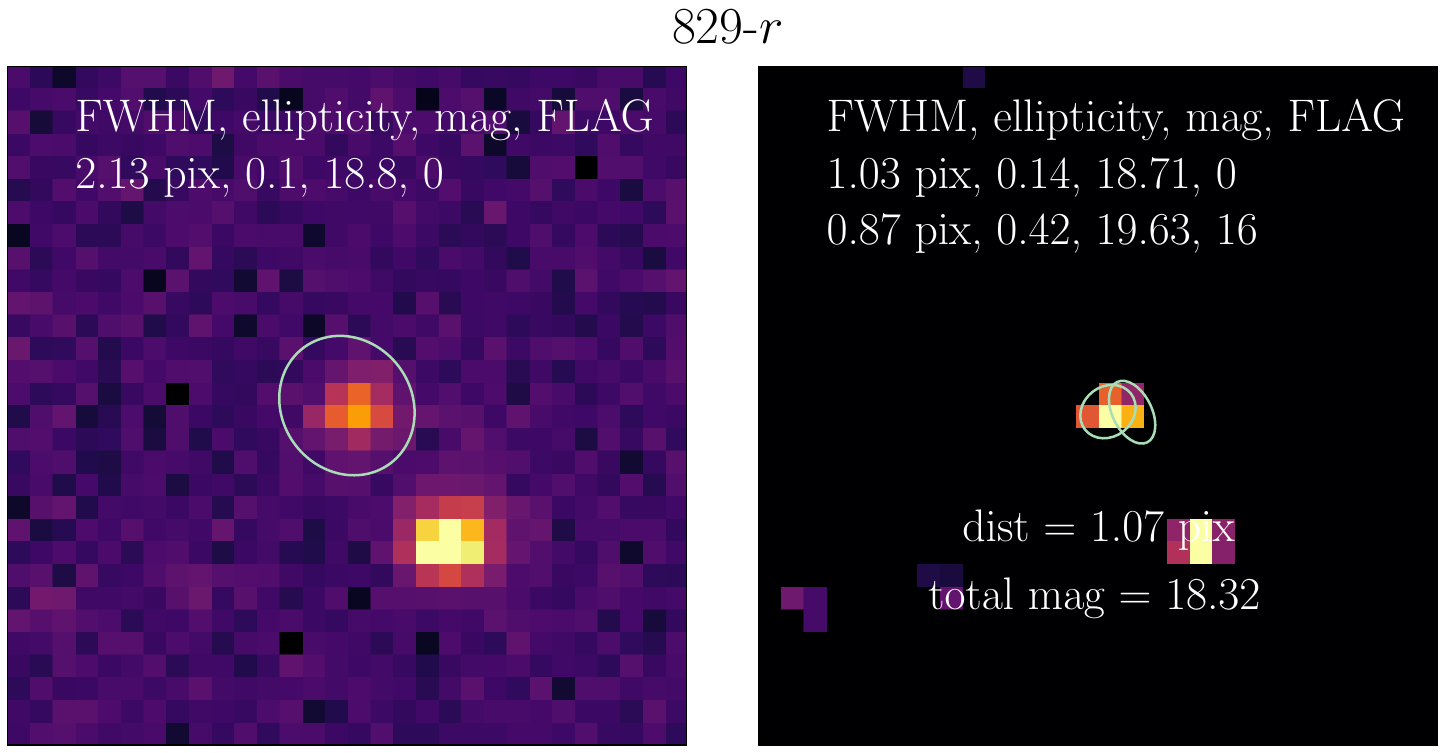}\hfill
      \includegraphics[keepaspectratio,width=0.32\linewidth]{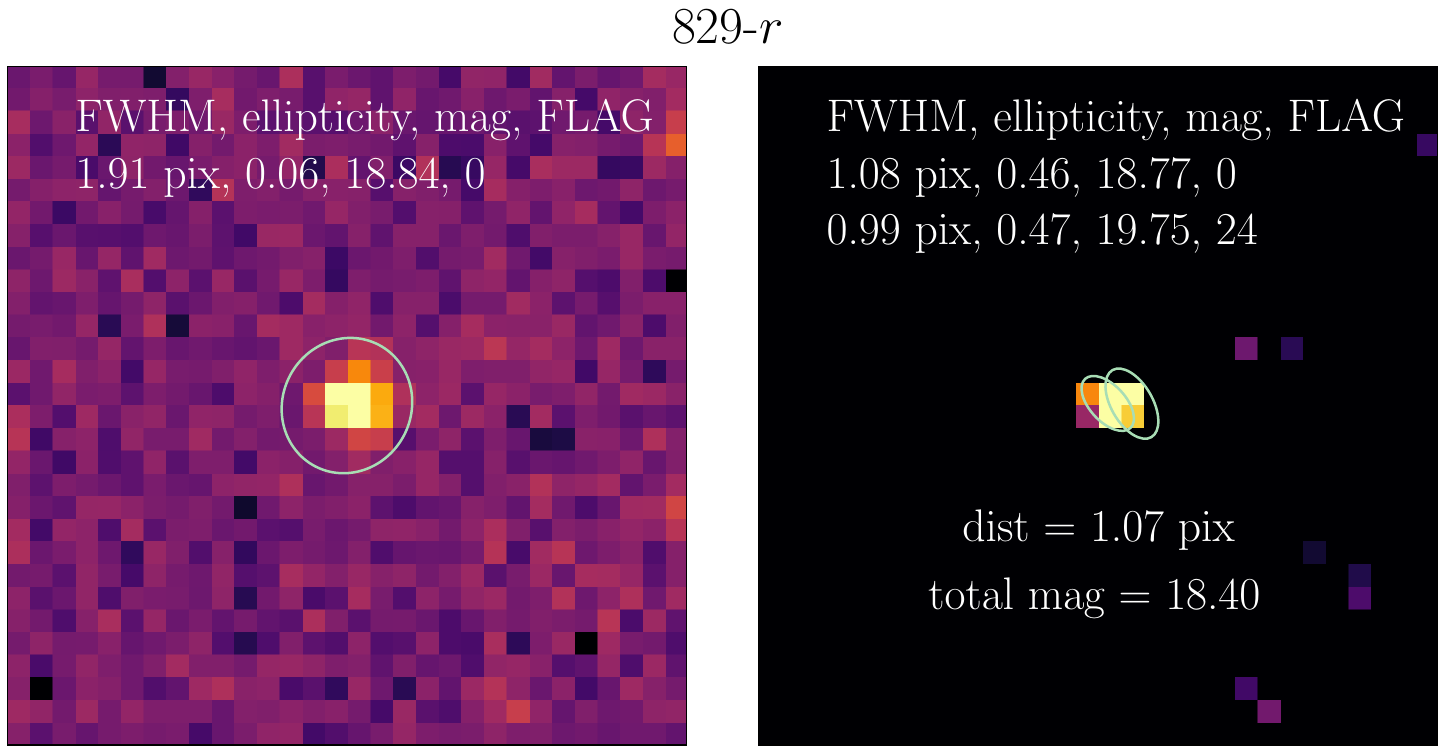}
      \includegraphics[keepaspectratio,width=0.32\linewidth]{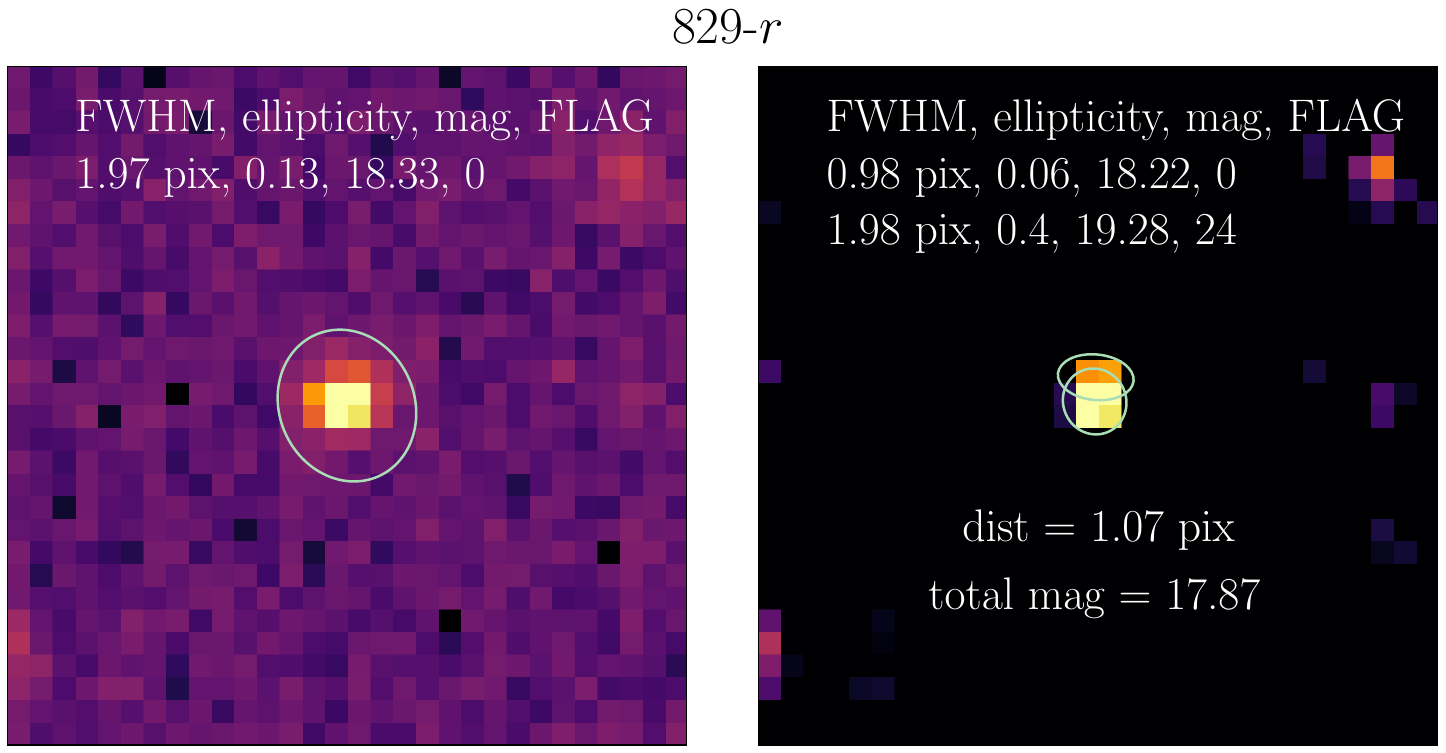}
      \includegraphics[keepaspectratio,width=0.32\linewidth]{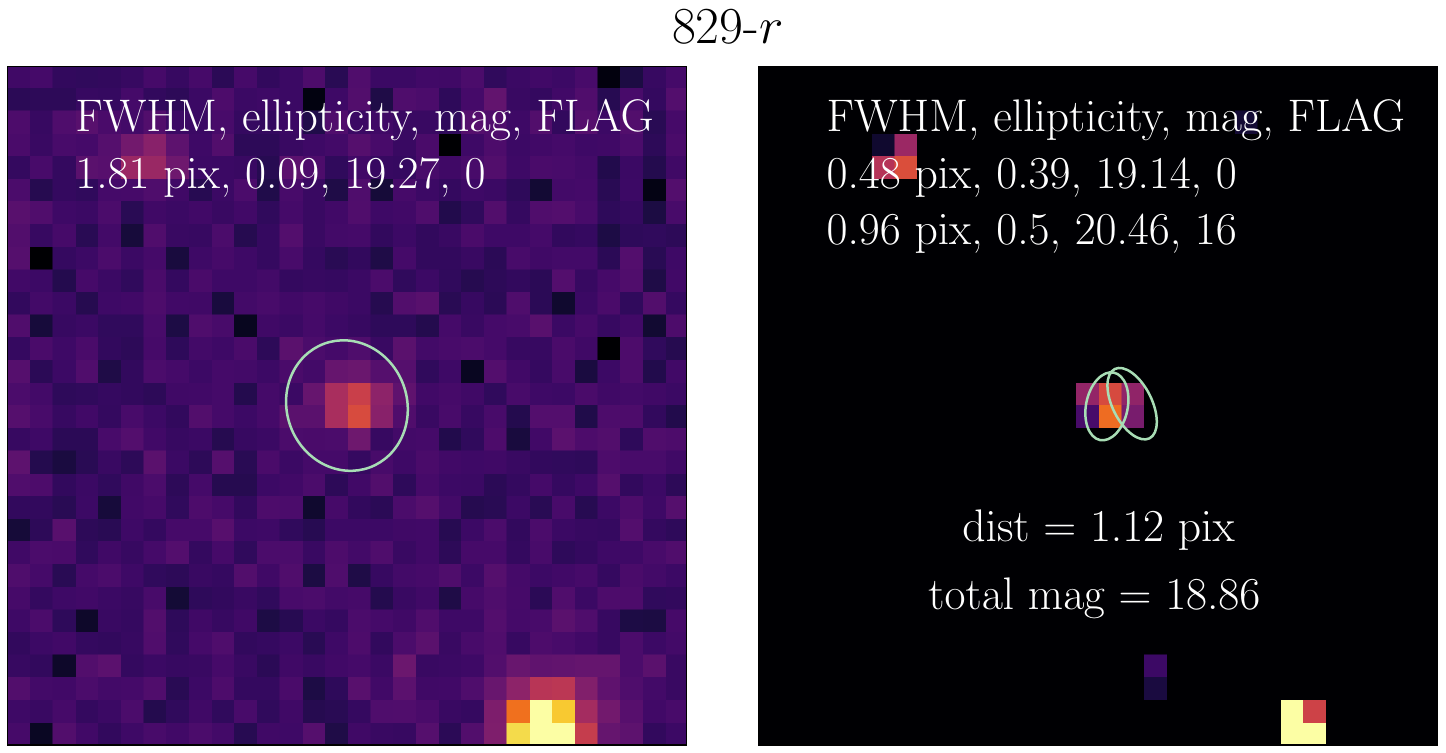}\hfill
      \includegraphics[keepaspectratio,width=0.32\linewidth]{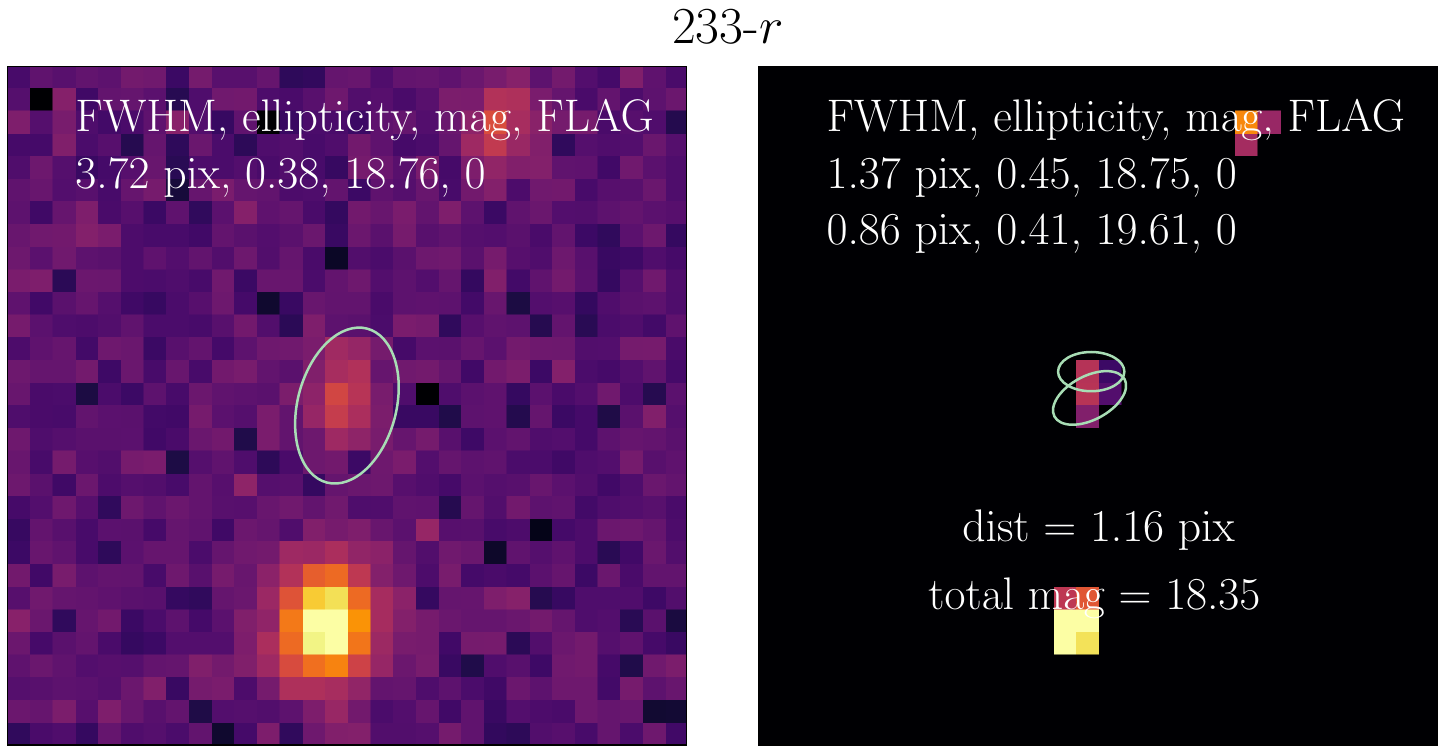}
      \includegraphics[keepaspectratio,width=0.32\linewidth]{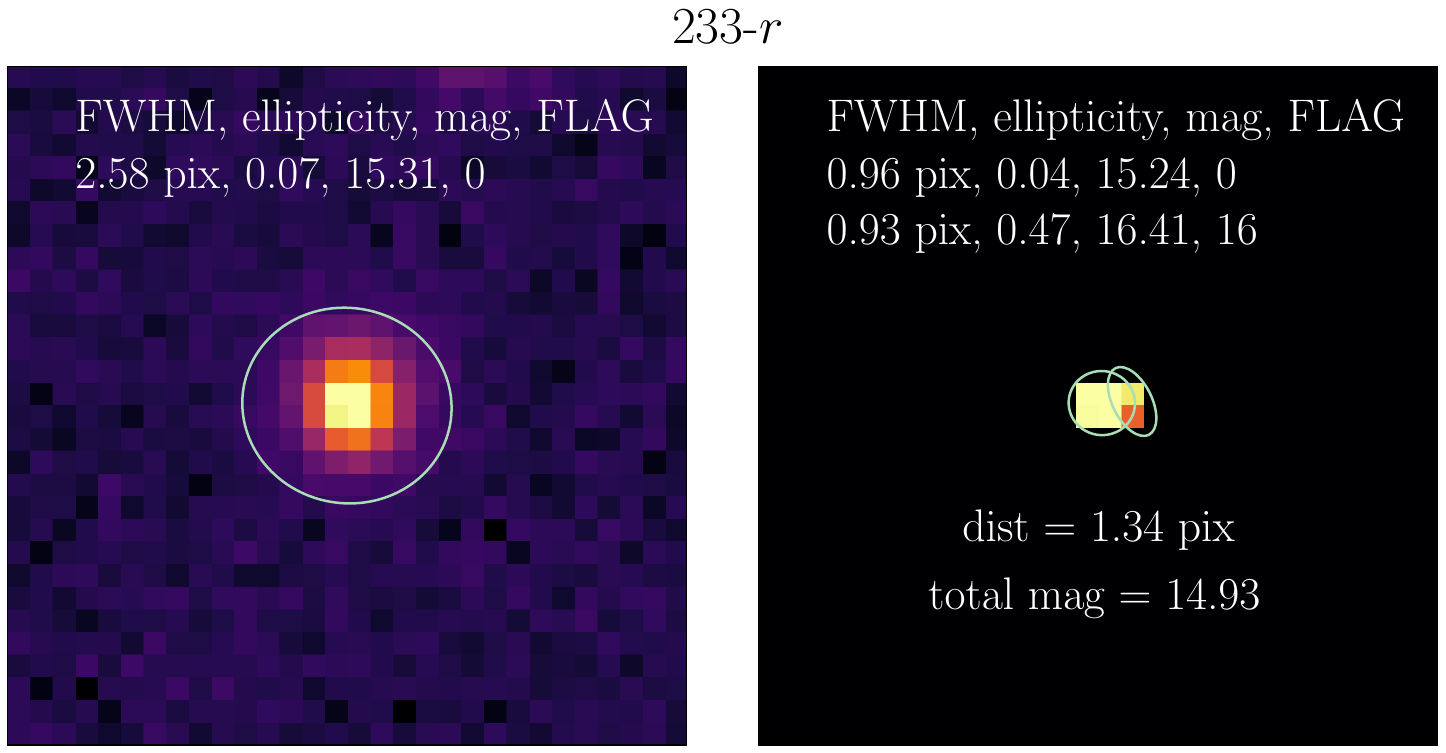}
      \includegraphics[keepaspectratio,width=0.32\linewidth]{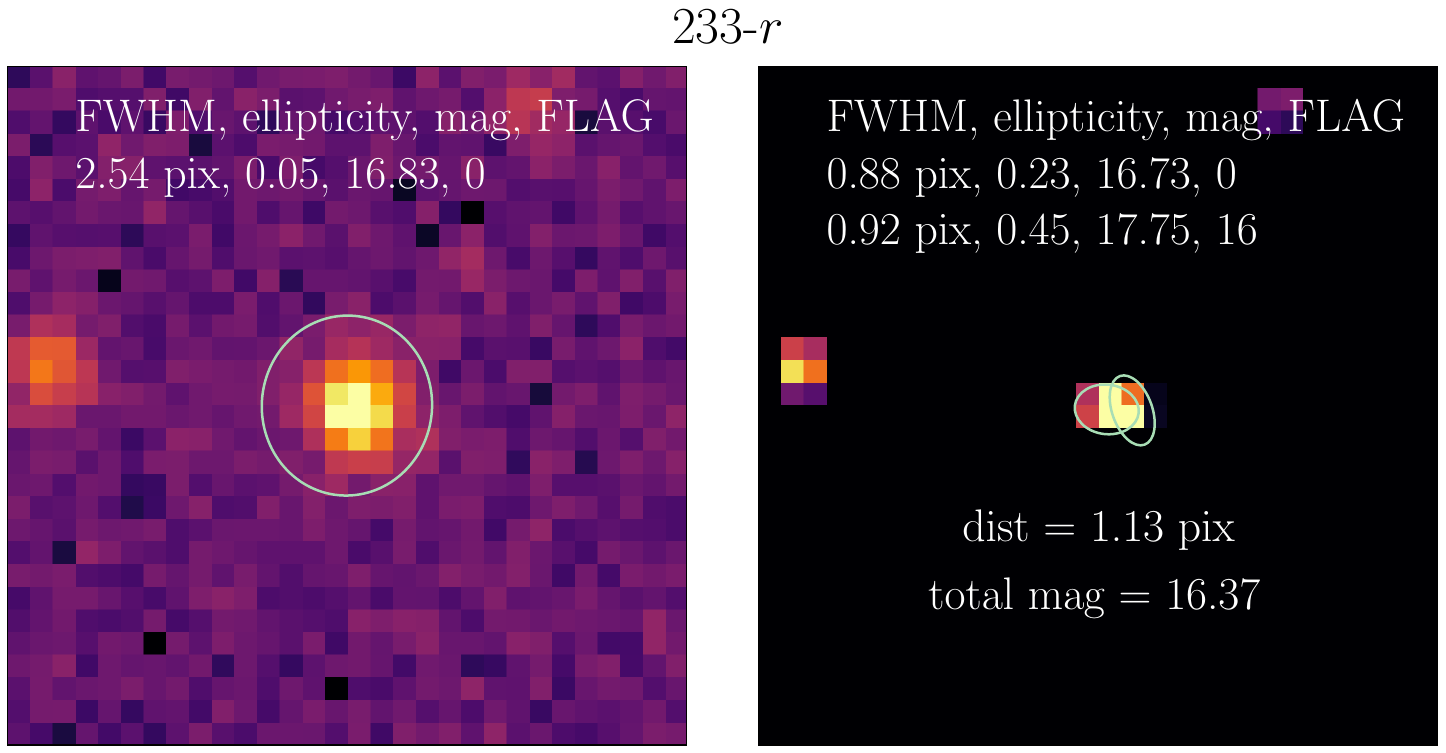}\hfill
      \includegraphics[keepaspectratio,width=0.32\linewidth]{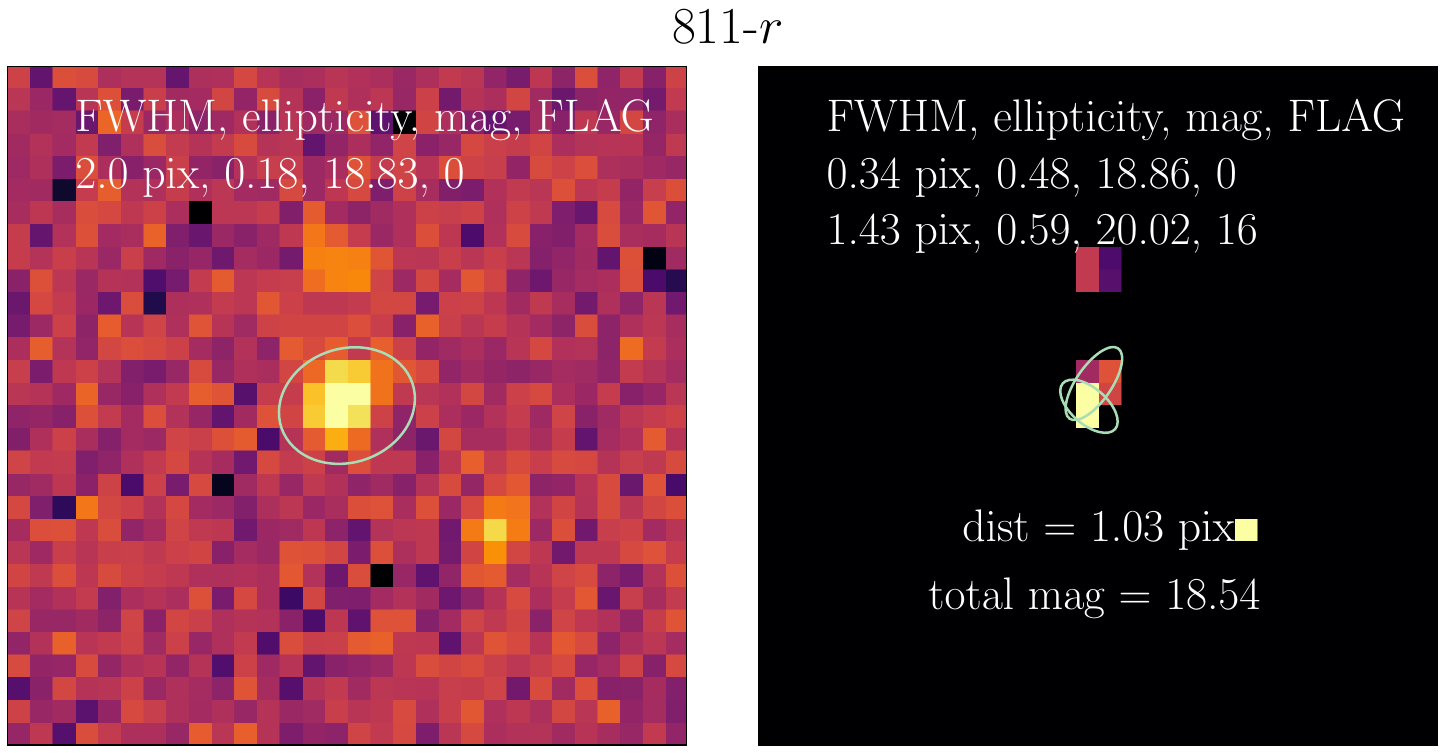}
      \includegraphics[keepaspectratio,width=0.32\linewidth]{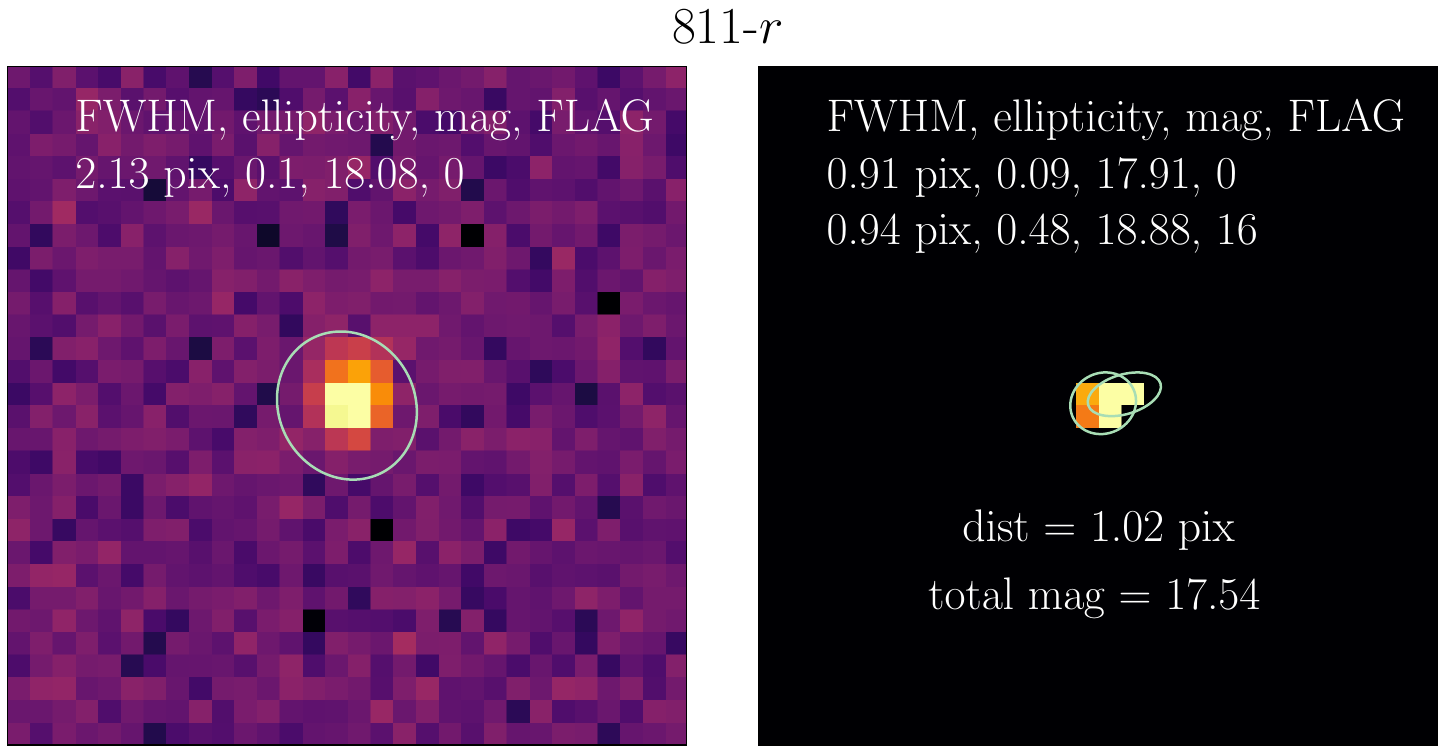}
      \includegraphics[keepaspectratio,width=0.32\linewidth]{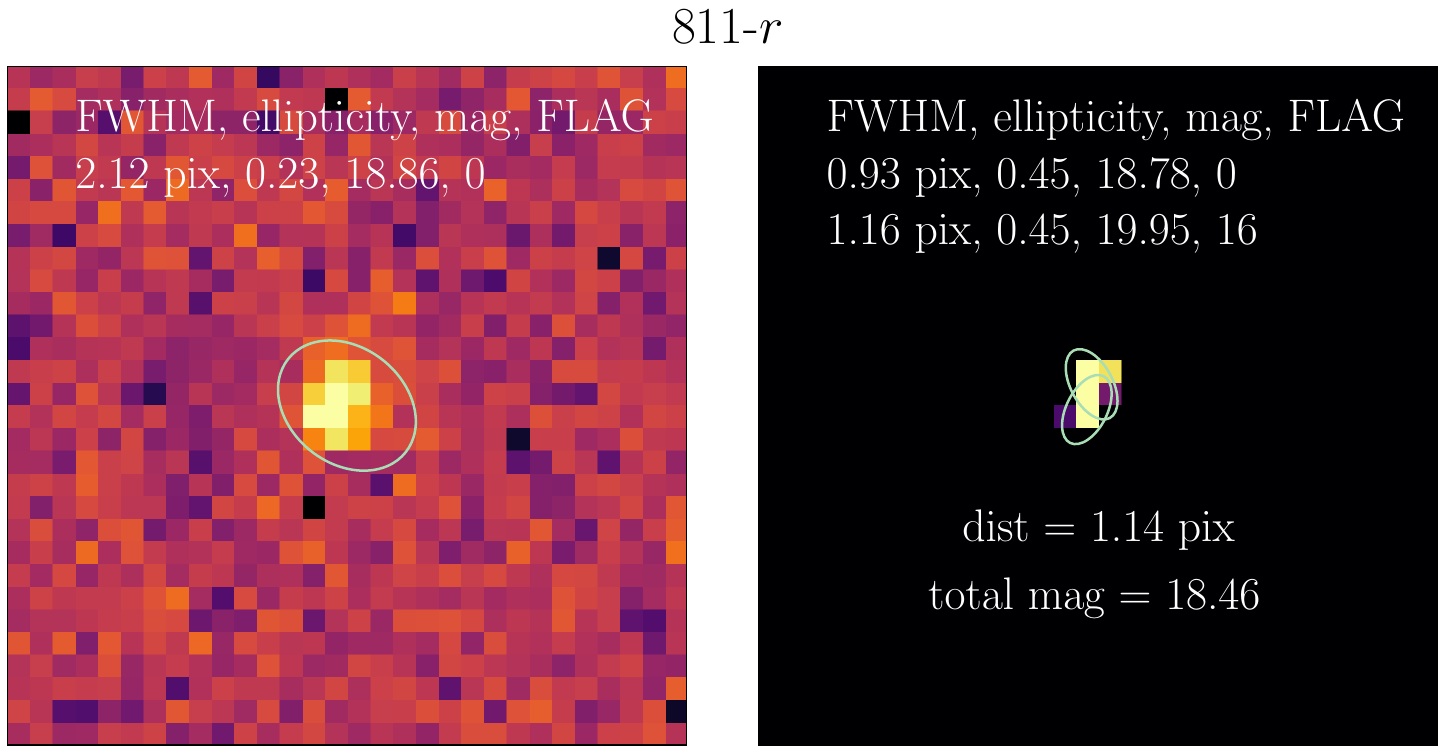}\hfill
      \includegraphics[keepaspectratio,width=0.32\linewidth]{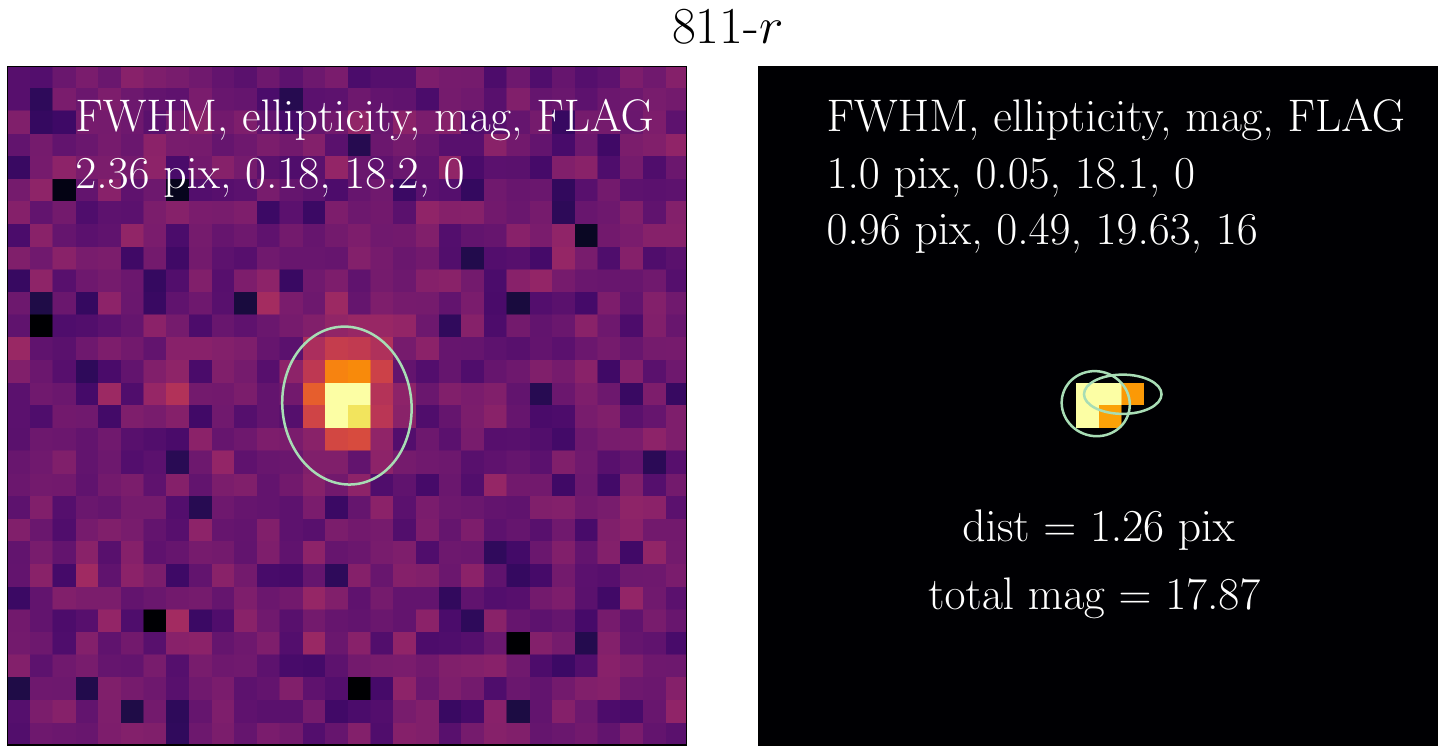}
      \includegraphics[keepaspectratio,width=0.32\linewidth]{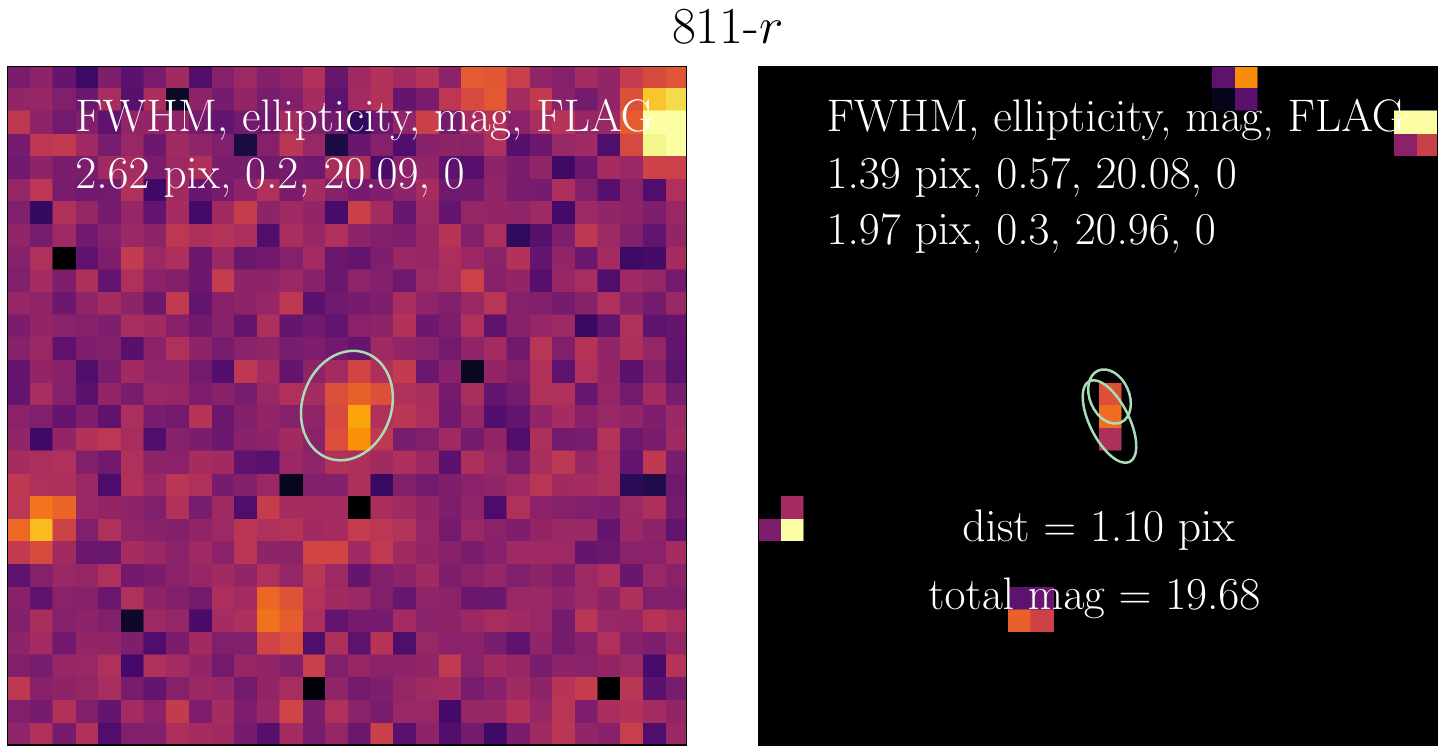}
      \includegraphics[keepaspectratio,width=0.32\linewidth]{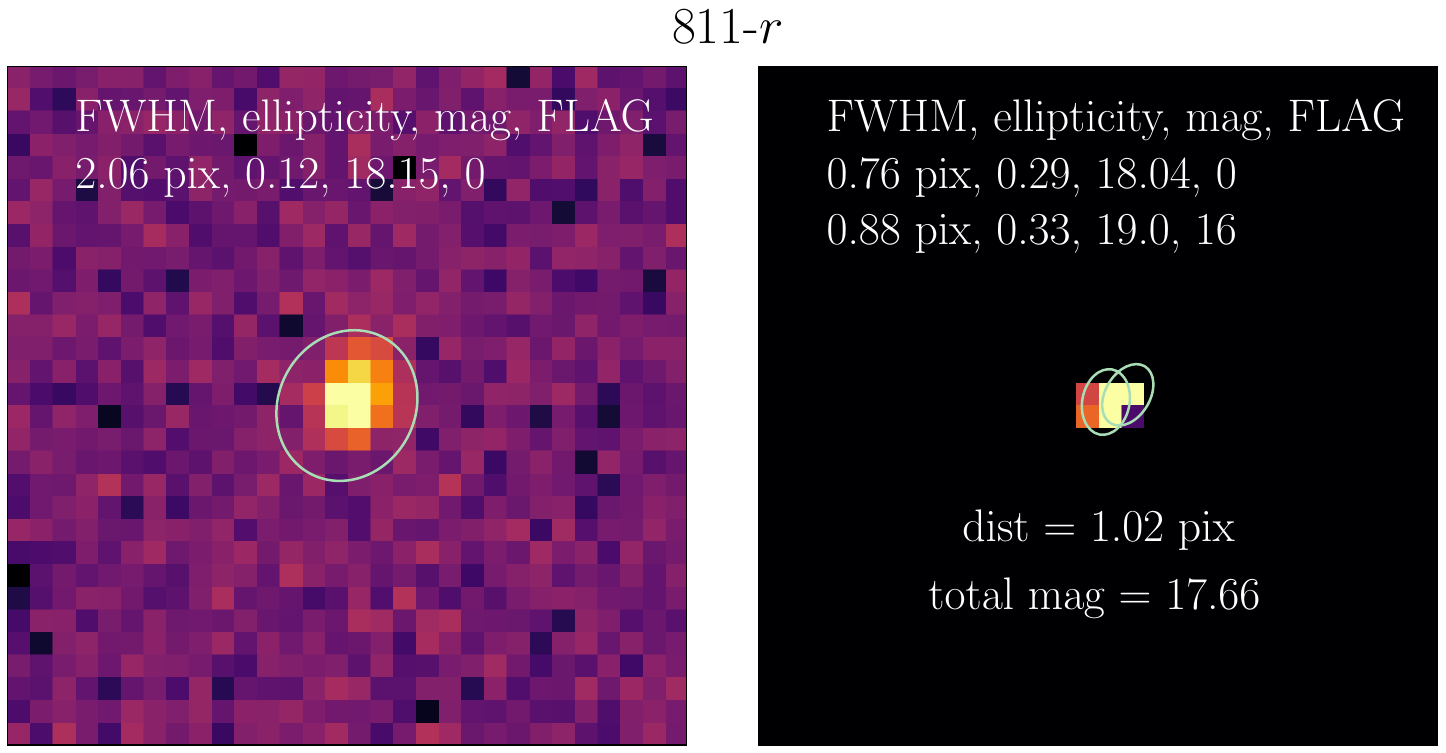}\hfill
      \includegraphics[keepaspectratio,width=0.32\linewidth]{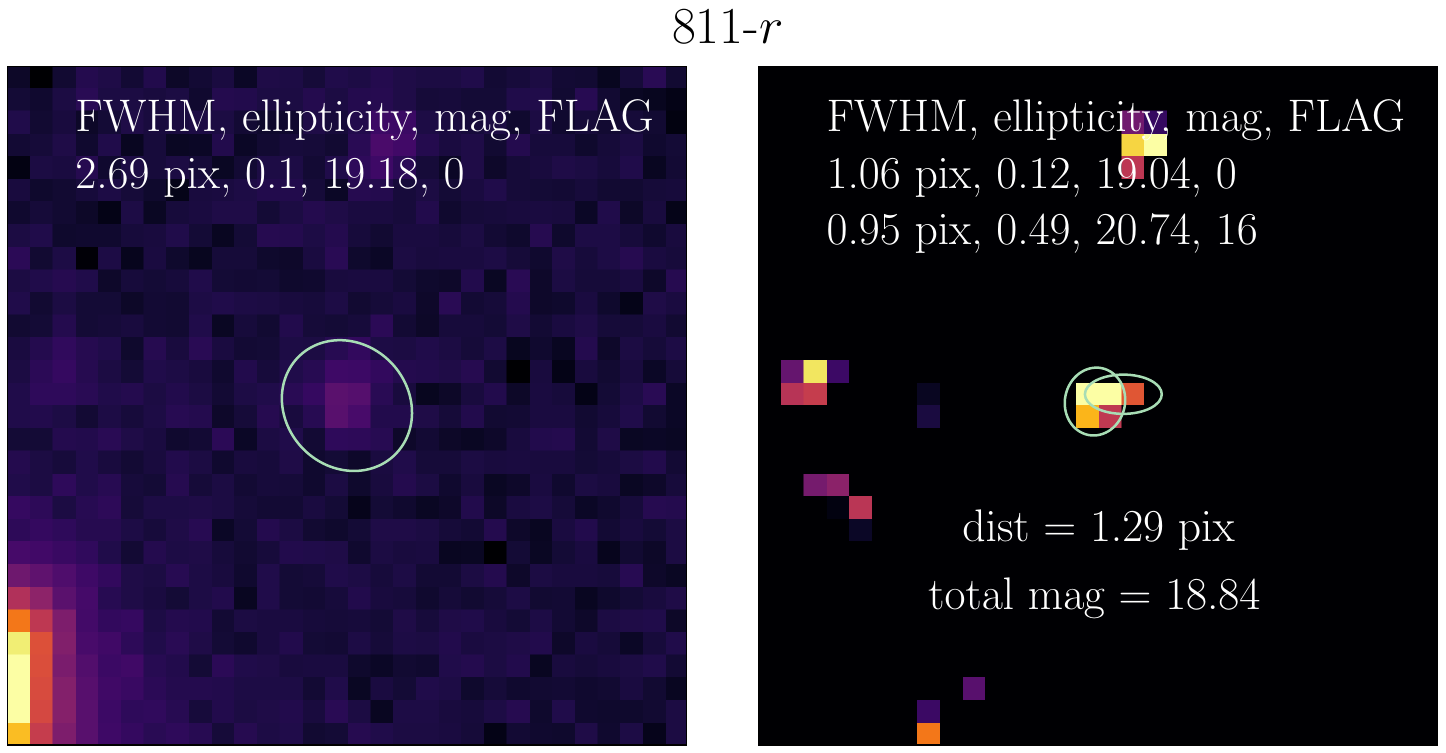}
      \includegraphics[keepaspectratio,width=0.32\linewidth]{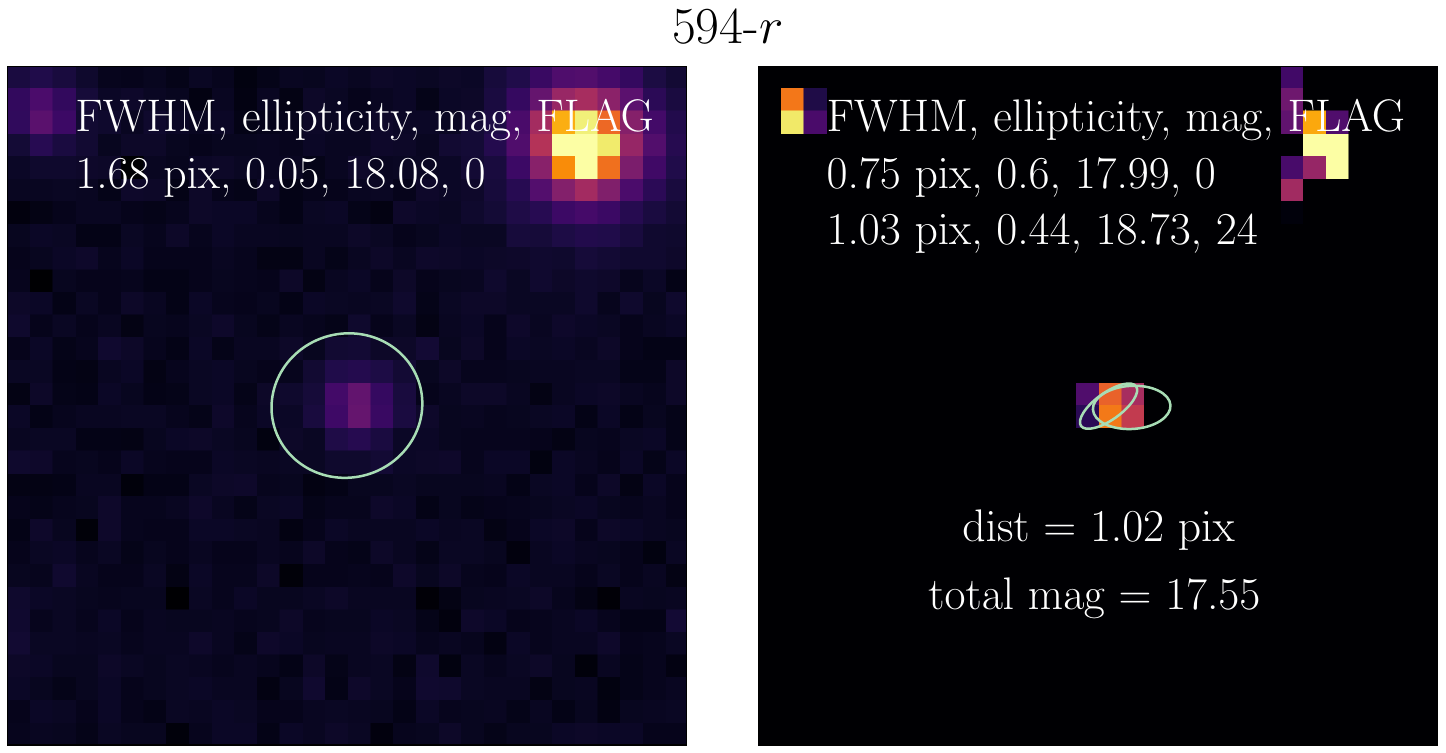}
      \includegraphics[keepaspectratio,width=0.32\linewidth]{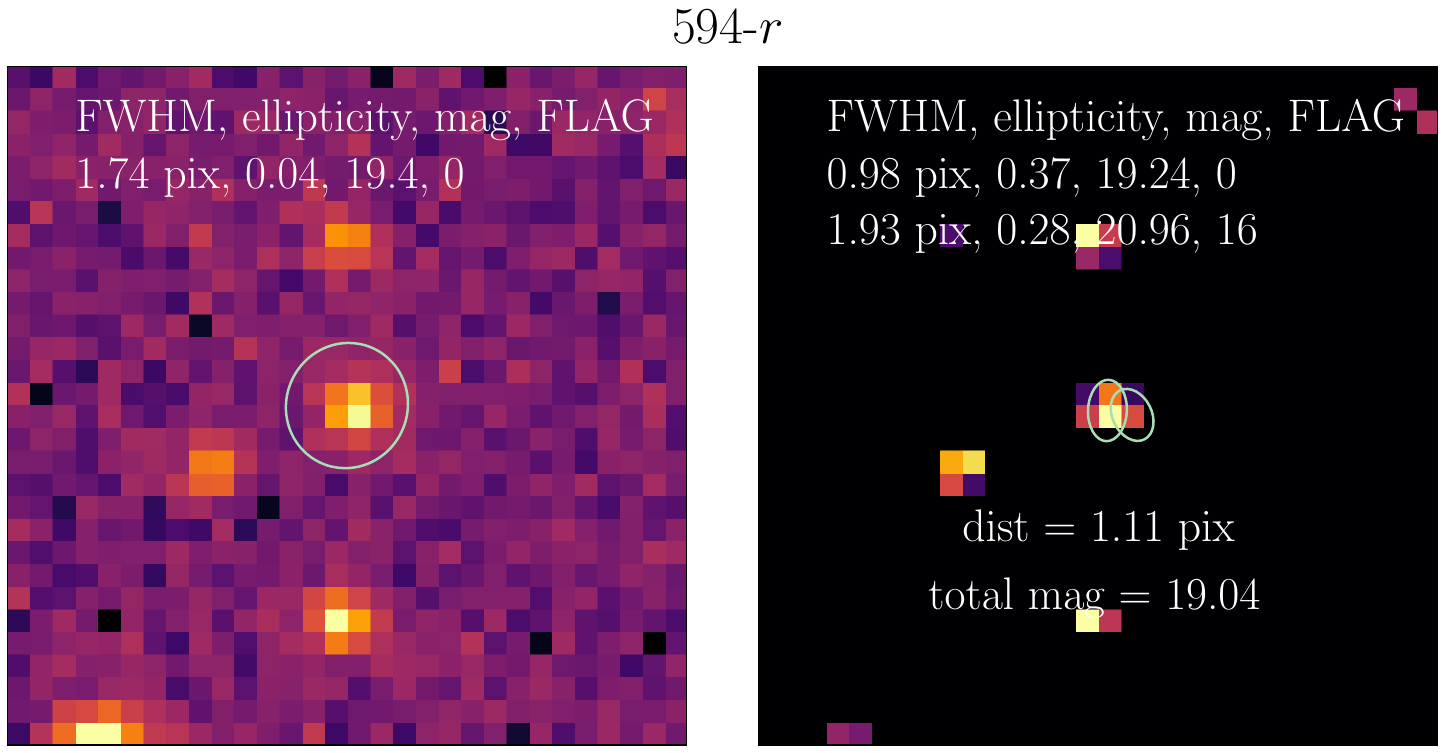}\hfill
\end{figure*}
\begin{figure*}
    \centering
      \includegraphics[keepaspectratio,width=0.32\linewidth]{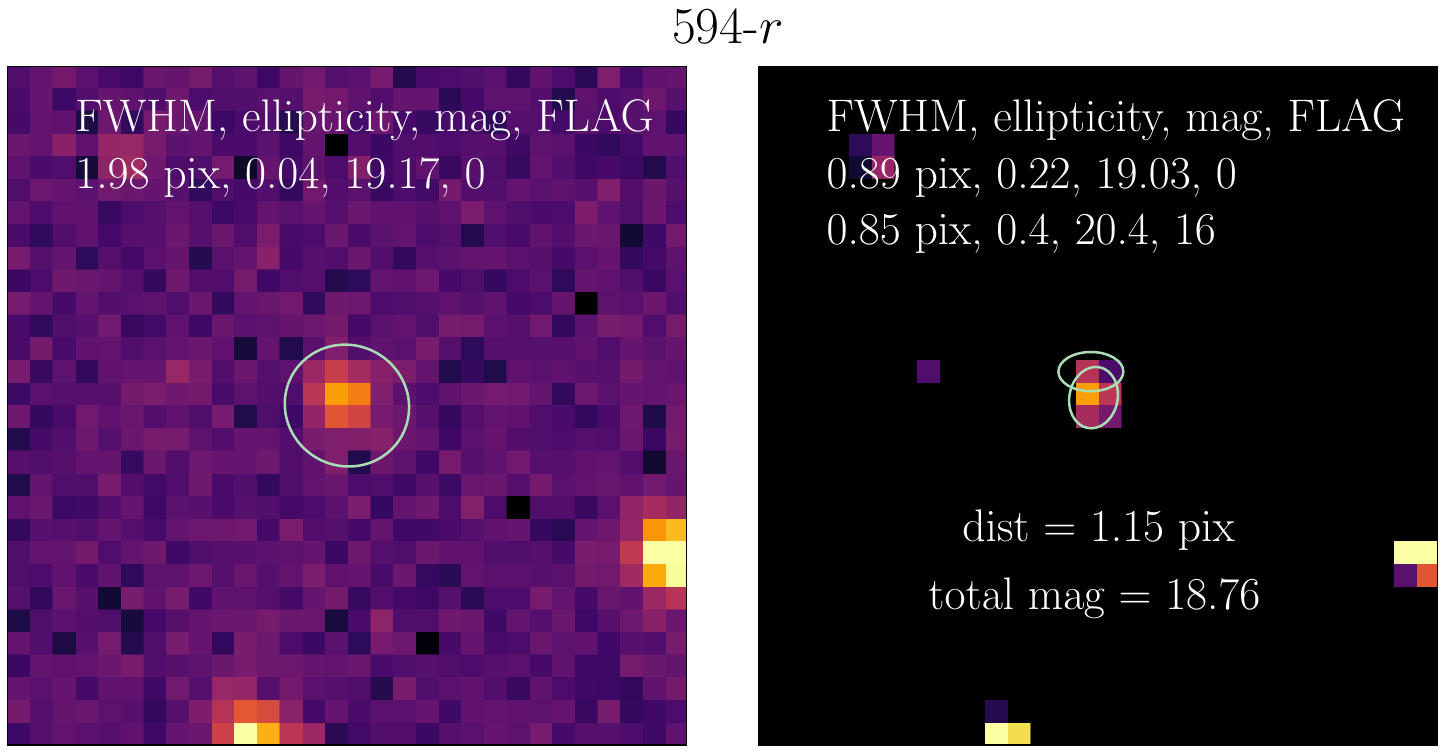}
      \includegraphics[keepaspectratio,width=0.32\linewidth]{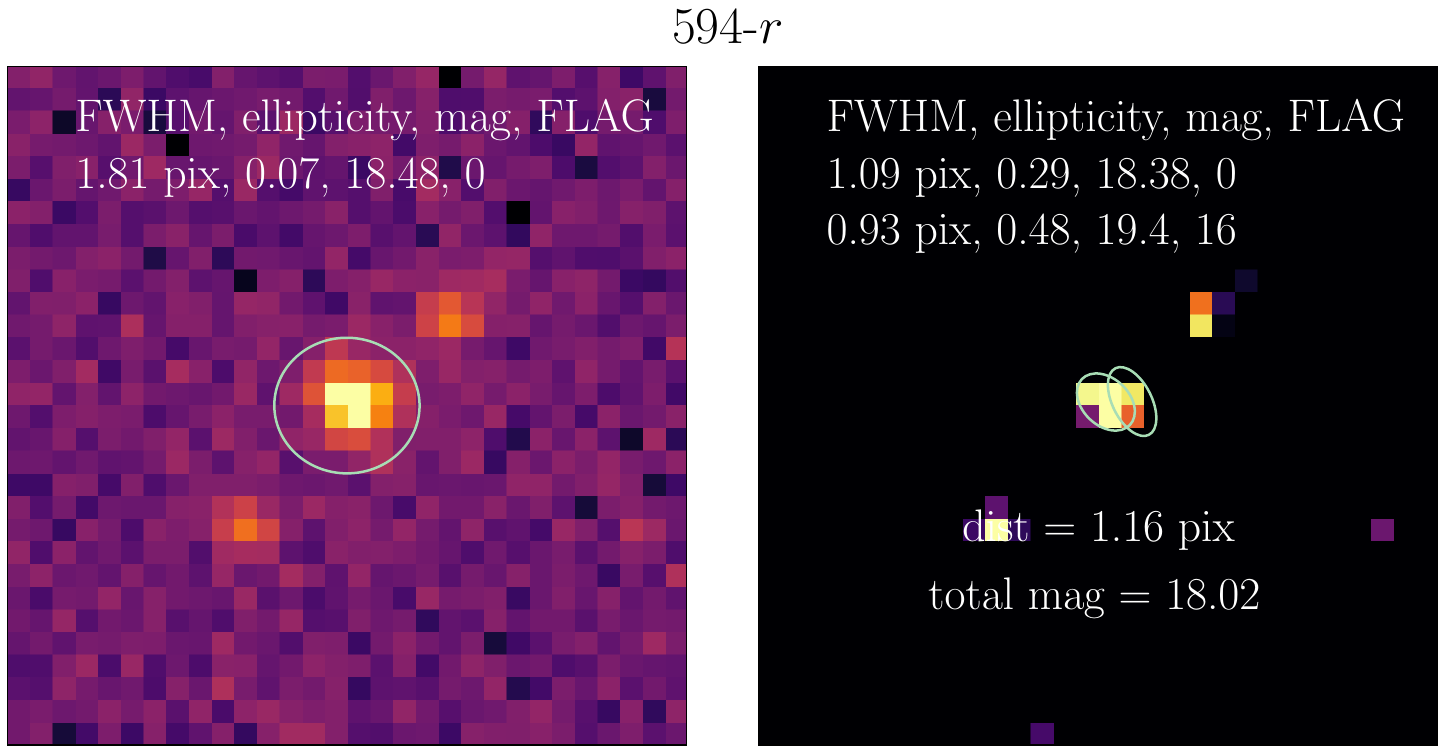}\hfill
      \includegraphics[keepaspectratio,width=0.32\linewidth]{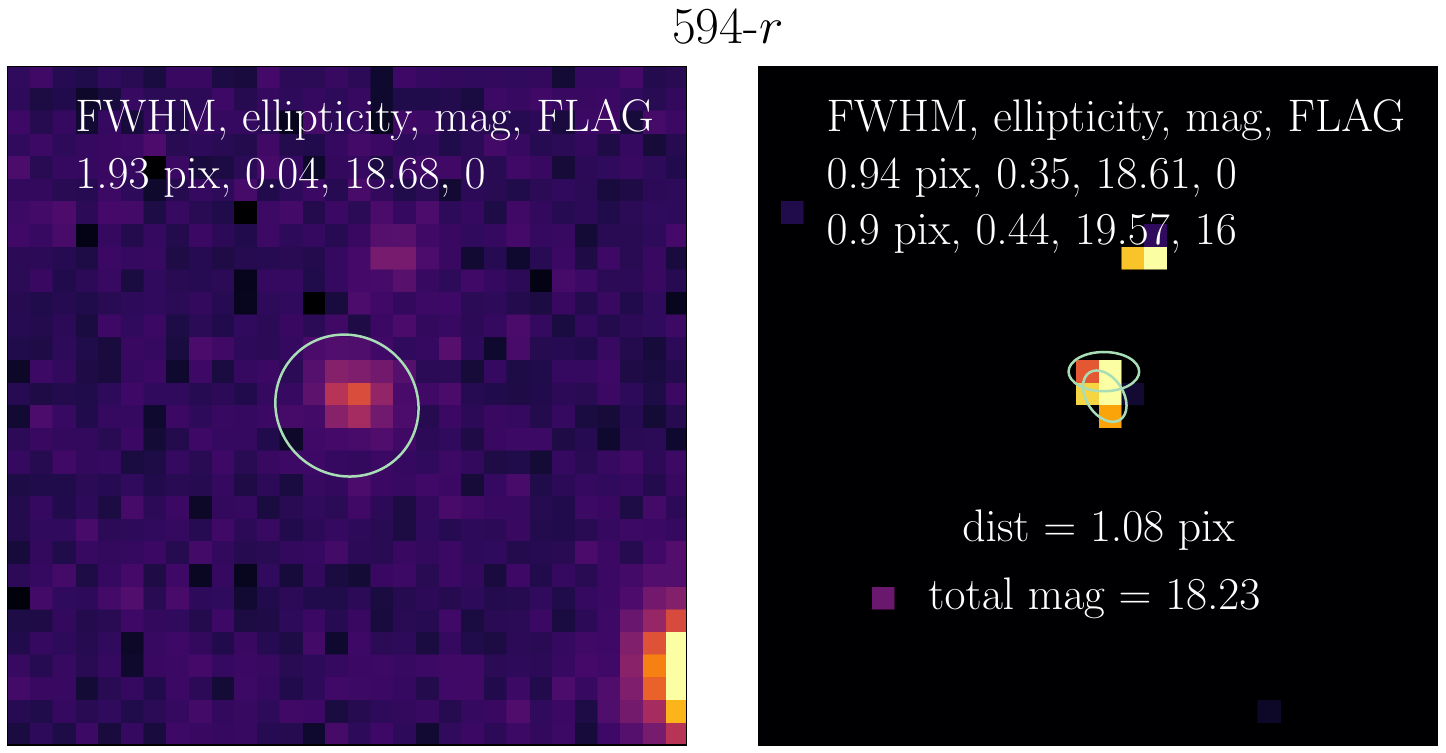}
      \includegraphics[keepaspectratio,width=0.32\linewidth]{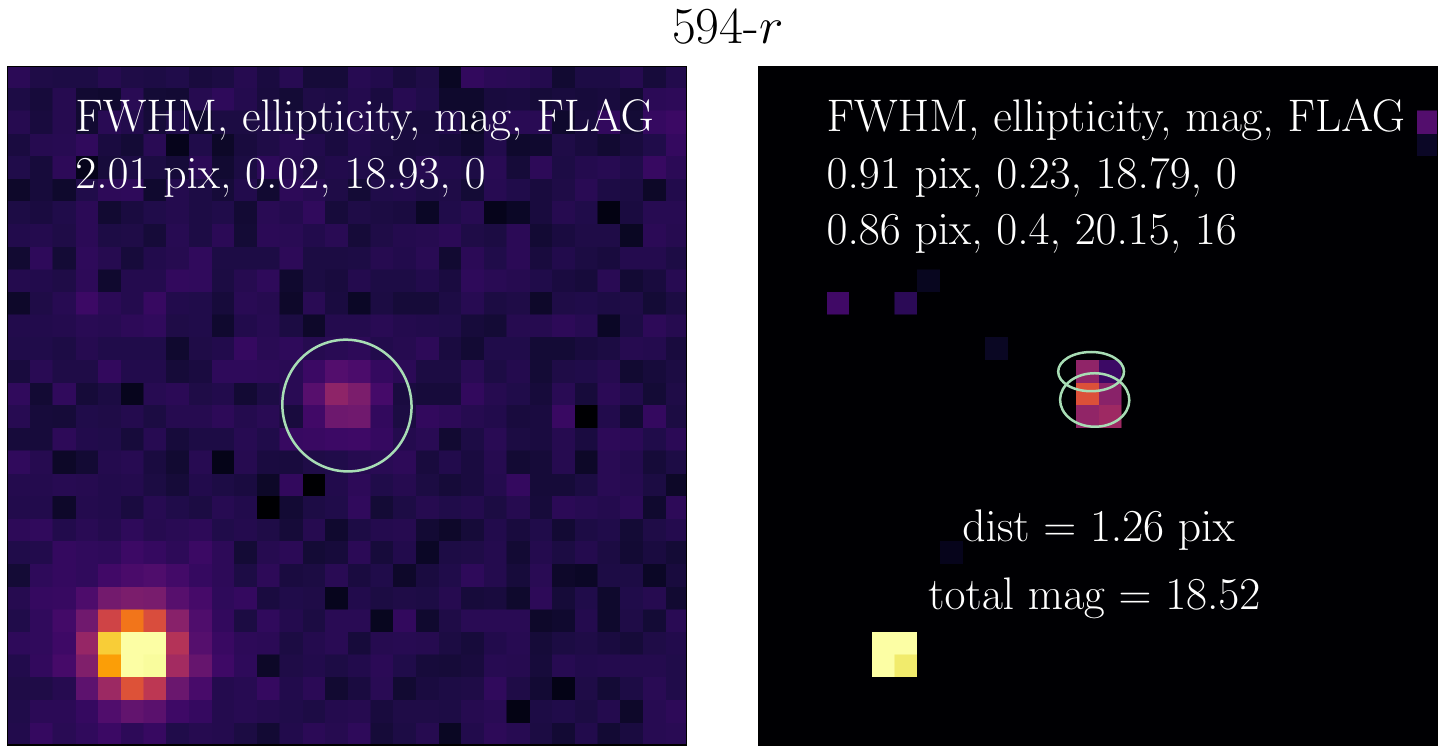}
      \includegraphics[keepaspectratio,width=0.32\linewidth]{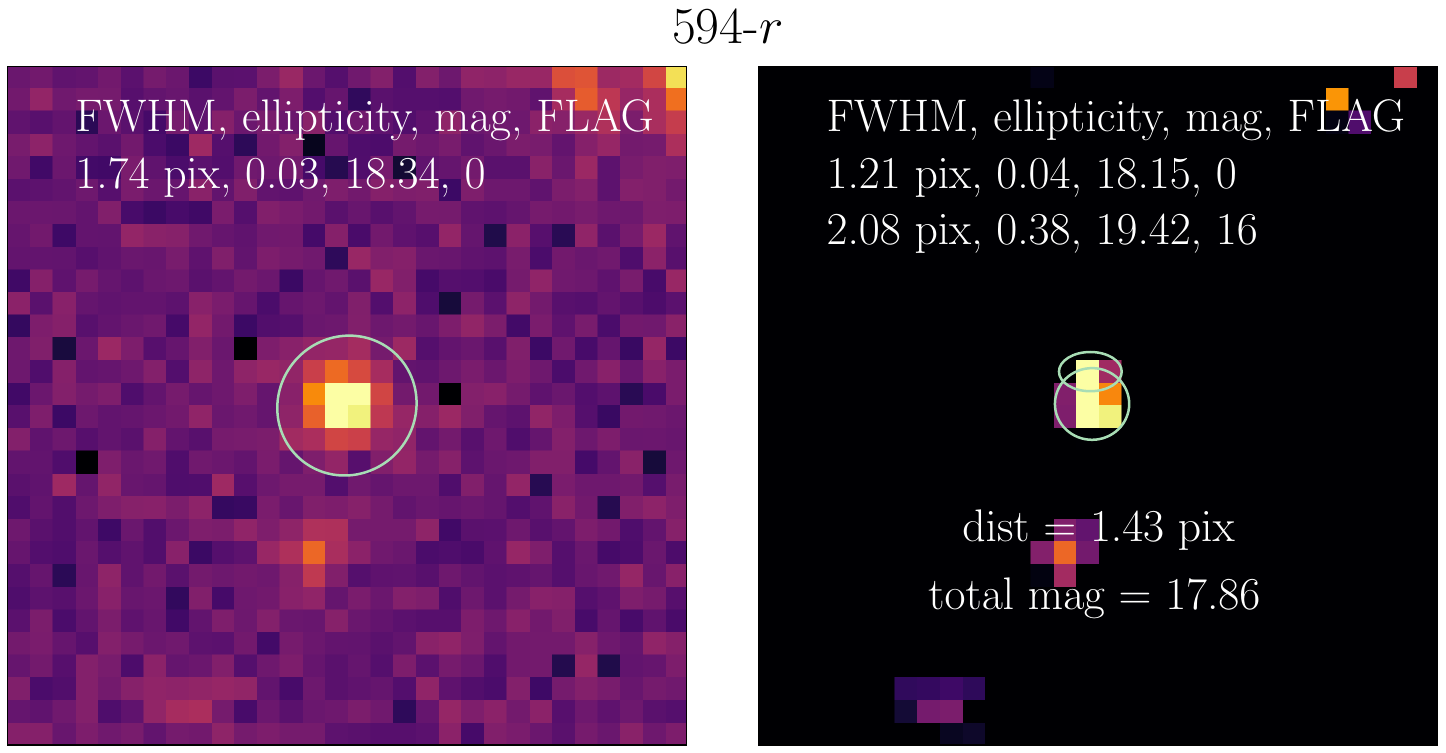}\hfill
      \includegraphics[keepaspectratio,width=0.32\linewidth]{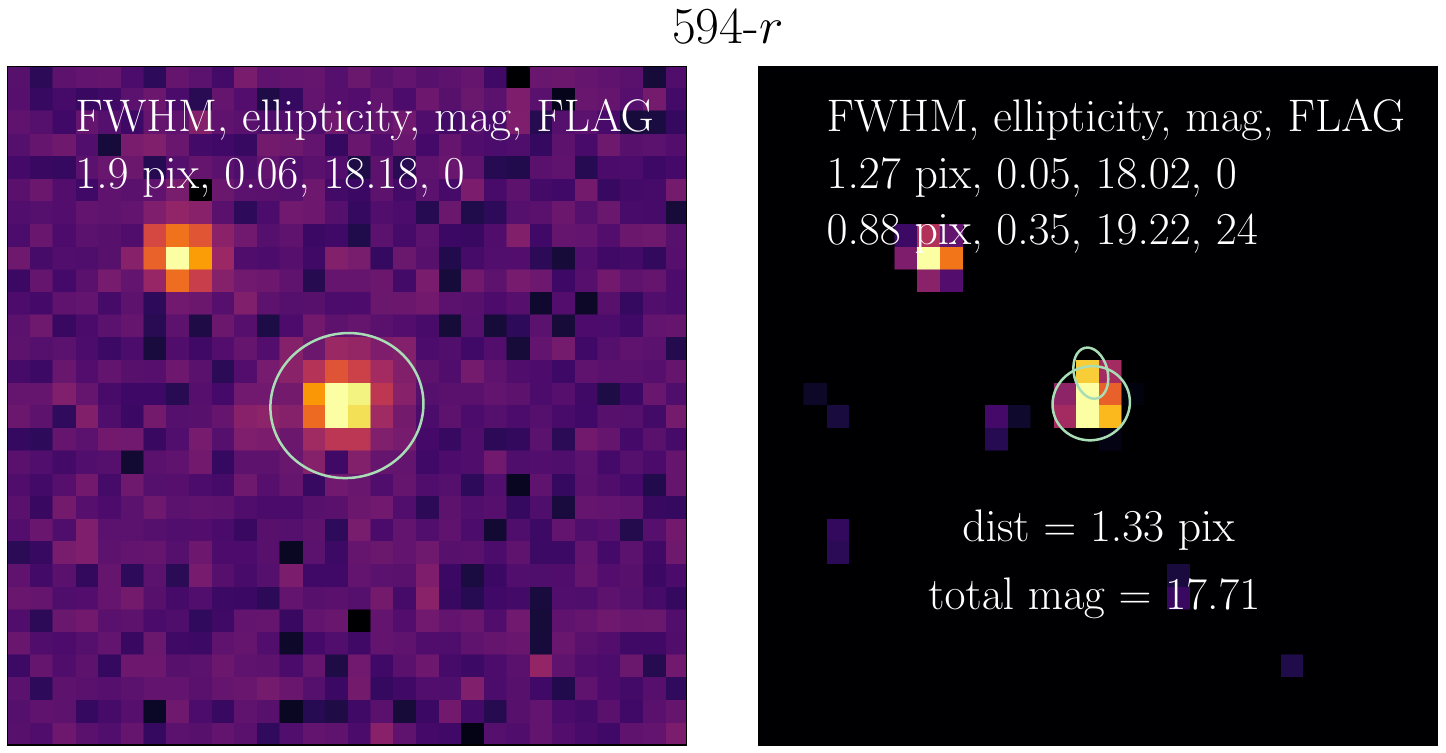}
    \caption{Examples of many-to-one matches that are possible {\it ambiguous} blends (two original sources overlapping to an extent being detected as a single source) converted to {\it conspicuous} blends (two sources overlapping but detected as two distinct sources) by the deconvolution (see Sect.~\ref{appn:many-to-one} for details on how these cases were selected). These examples are across all $r$-band images used in this study. In all examples, the original source is on the left, and the two deconvolved sources are shown on the right, marked by ellipses. The FWHM, ellipticity, magnitude, and SExtractor flags of all sources are shown on top of each image. The FLAGs of all original sources are zero, indicating they are good detections. The distance between the two deconvolved sources and their combined magnitude is also shown on the deconvolved image. Images are shown using a combination of square root stretching and clipping pixel values beyond the central 99.5 percentile. One of the deconvolved sources tends to have $\mathrm{FLAGS} = 0$, and the other a non-zero flag, but none of these have a deblending flag set, which means the deconvolution deblends these sources rather than SExtractor. The SExtractor parameters used are described in Table~\ref{tab:sextractorParams}. All cases show deblending into sources with $\approx$1-1.5 orders of magnitude difference, and the ellipticities of the deconvolved sources are generally enlarged compared to the original. The combined magnitudes are brighter than the corresponding original magnitudes, which may be due to the significant overlap in the source profiles. The distance between the deblended deconvolved sources shown is the Euclidean distance and typically ranges from $\approx$1-1.3 pixels (the pixel size is 1$\arcsec$.012).} \label{fig:appn-many-to-one}
\end{figure*}

Fig.~\ref{fig:appn-many-to-one} shows the visualization of some selected many-to-one matches obtained across all the $r$-band images considered in this study. It is challenging to visually confirm whether the cases shown are true deblends, but a few cases do look visually plausible. The separation between these sources is only slightly greater than a pixel, which may hint at the fact that the deconvolution has the potential to deblend significantly overlapping sources. However, a definitive conclusion is difficult to obtain unless supplementary data is used to ascertain these blends and deblends.

\section{Additional deblending visualizations}\label{appn:more-deblending-examples}

Fig.~\ref{fig:more-deblend-examples} shows additional deblending visualizations to provide a better idea of the deblends found in this study across all $r$-band images considered in this study.

\begin{figure*}
      \centering
      \includegraphics[keepaspectratio,width=0.32\linewidth]{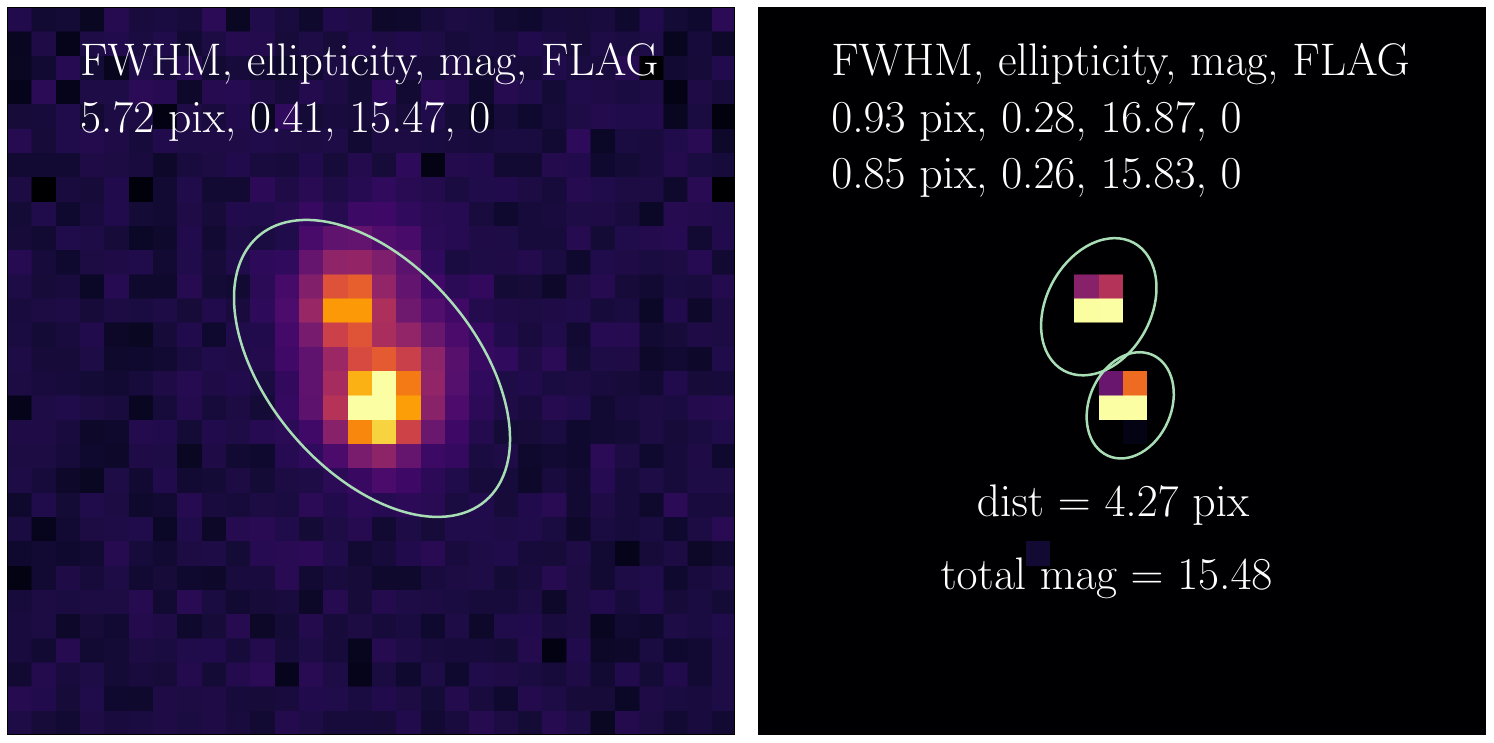}
      \includegraphics[keepaspectratio,width=0.32\linewidth]{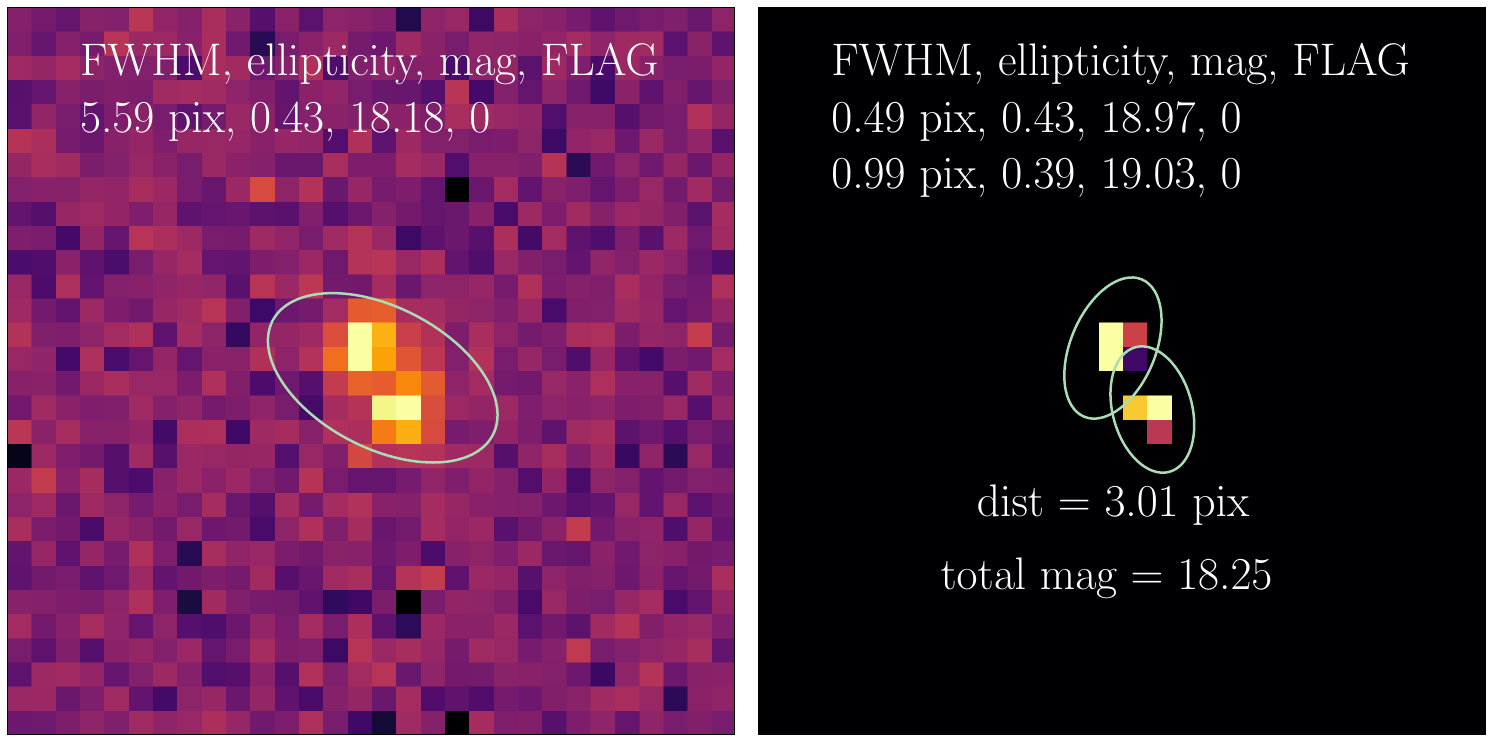}
      \includegraphics[keepaspectratio,width=0.32\linewidth]{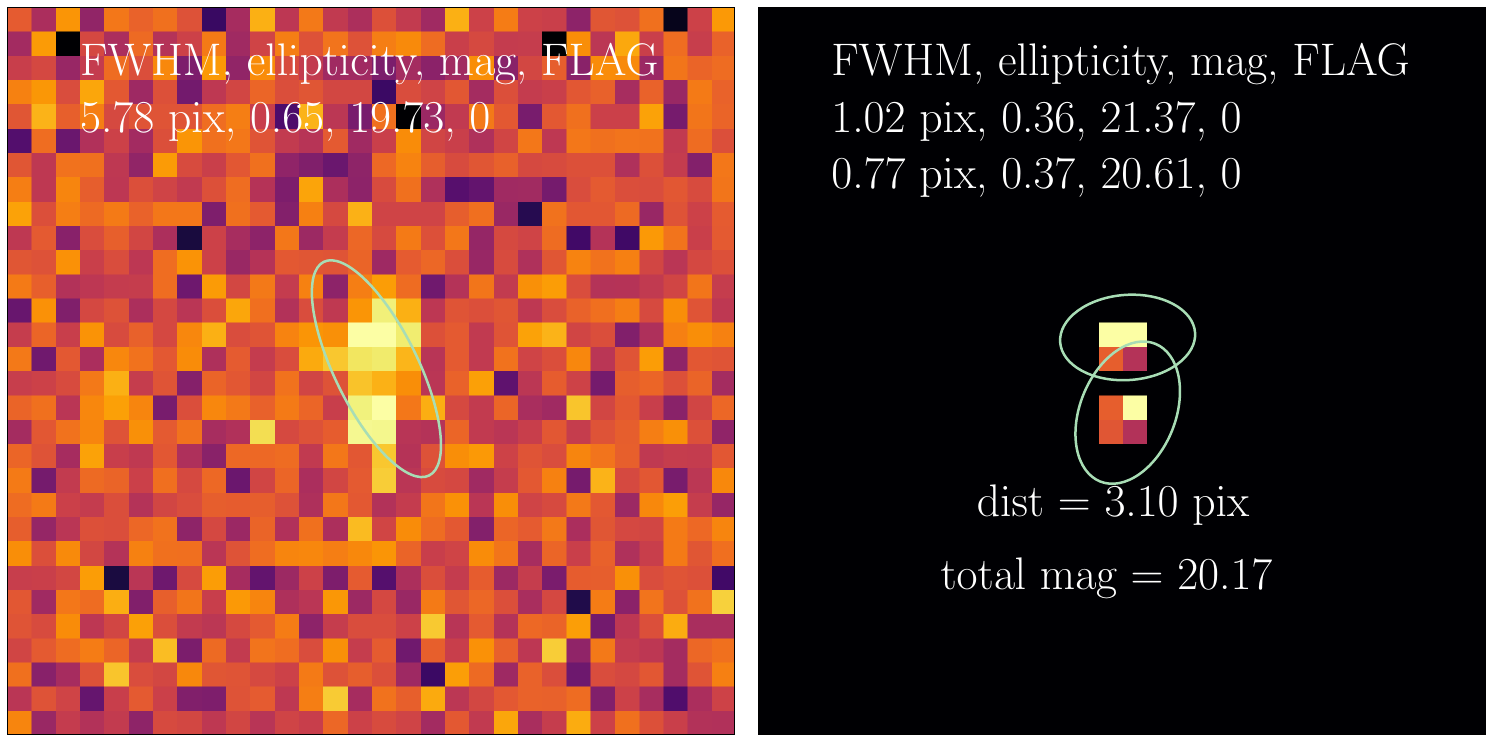}\hfill
      \includegraphics[keepaspectratio,width=0.32\linewidth]{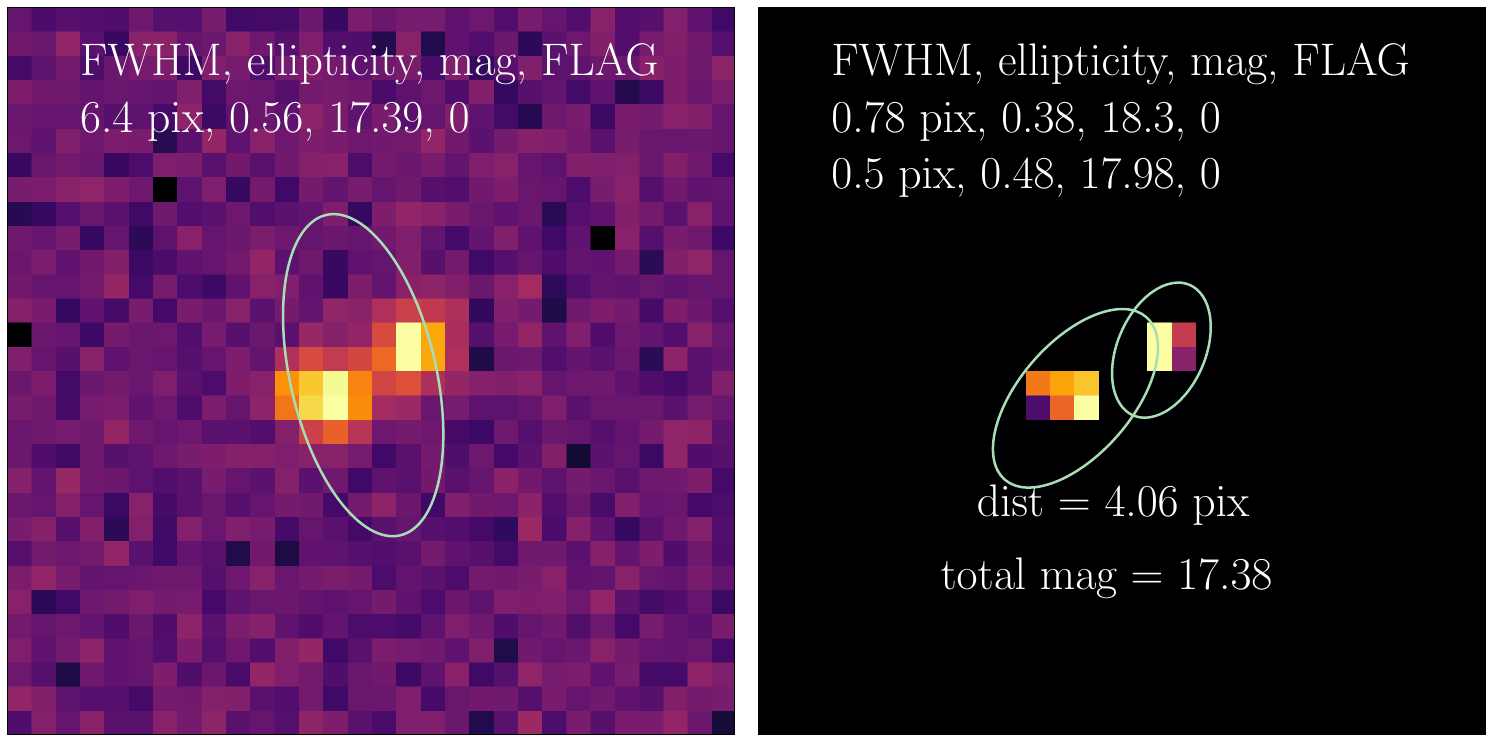}
      \includegraphics[keepaspectratio,width=0.32\linewidth]{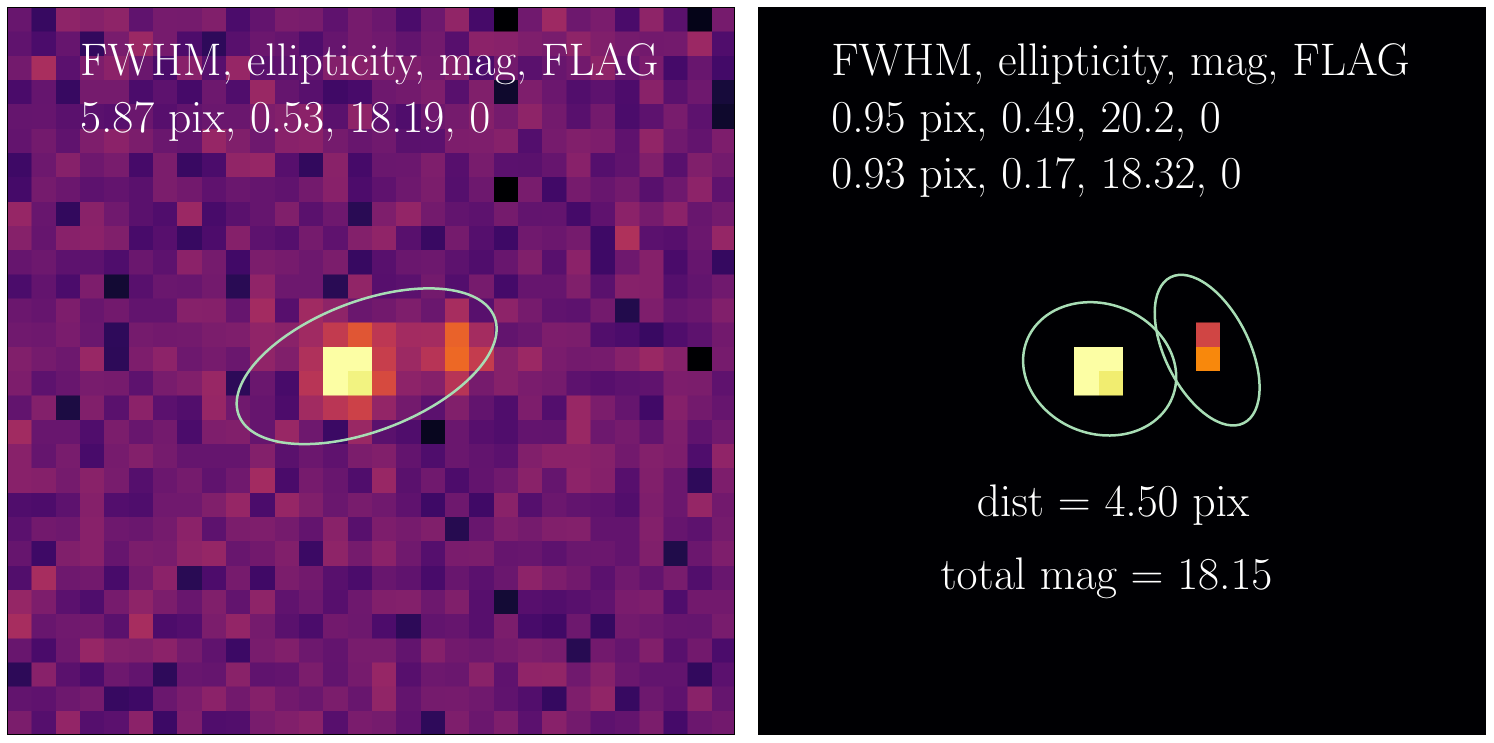}
      \includegraphics[keepaspectratio,width=0.32\linewidth]{Paper/figures/deblend_example_233_33.pdf}\hfill
      \includegraphics[keepaspectratio,width=0.32\linewidth]{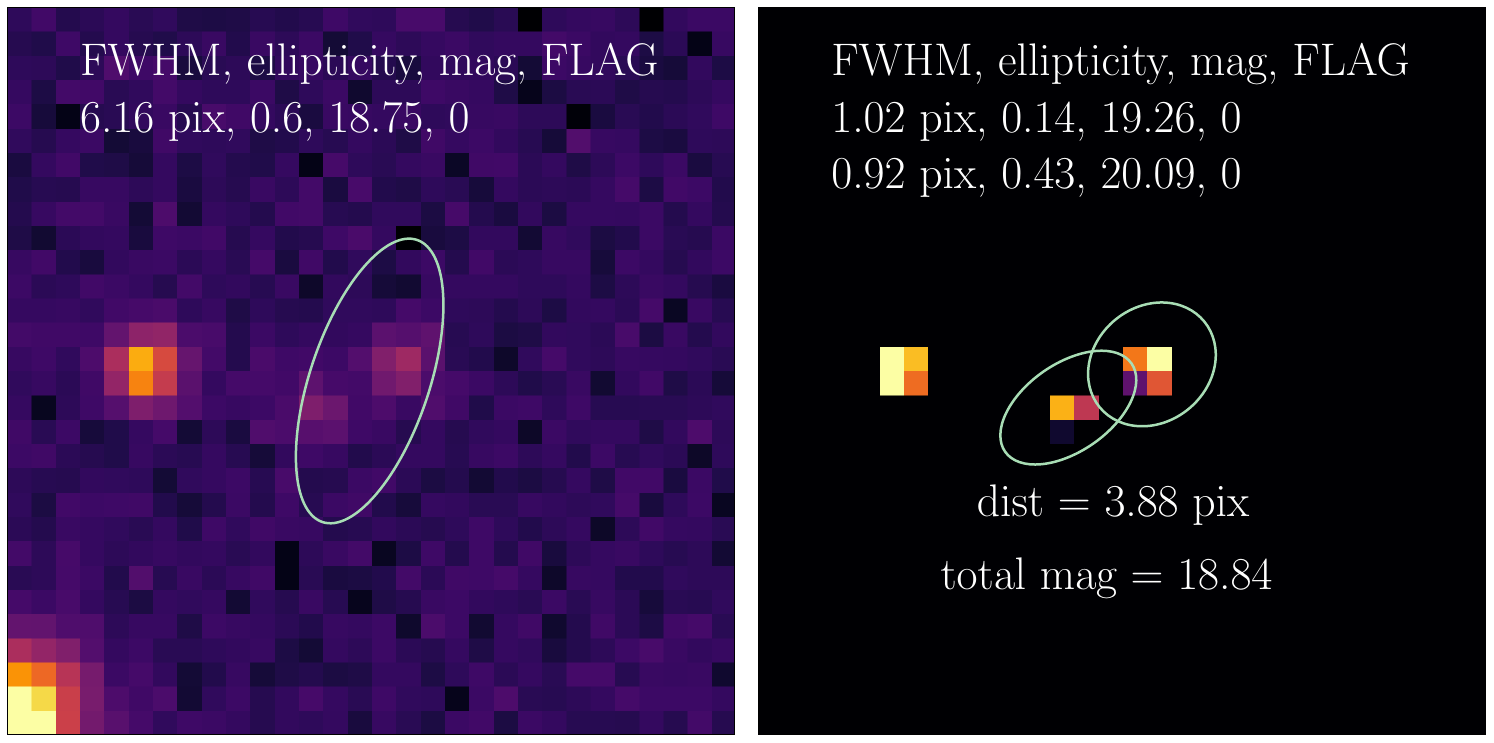}
      \includegraphics[keepaspectratio,width=0.32\linewidth]{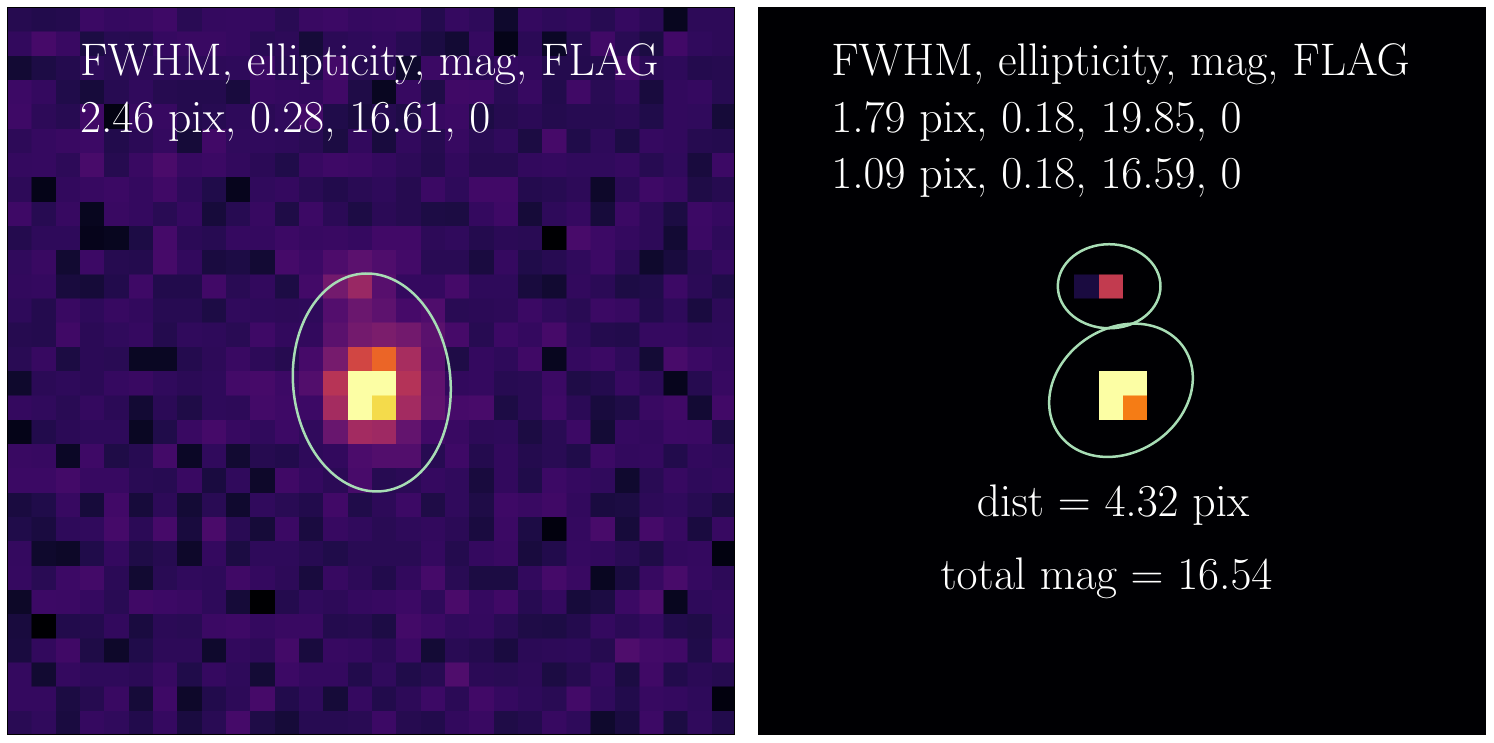}
      \includegraphics[keepaspectratio,width=0.32\linewidth]{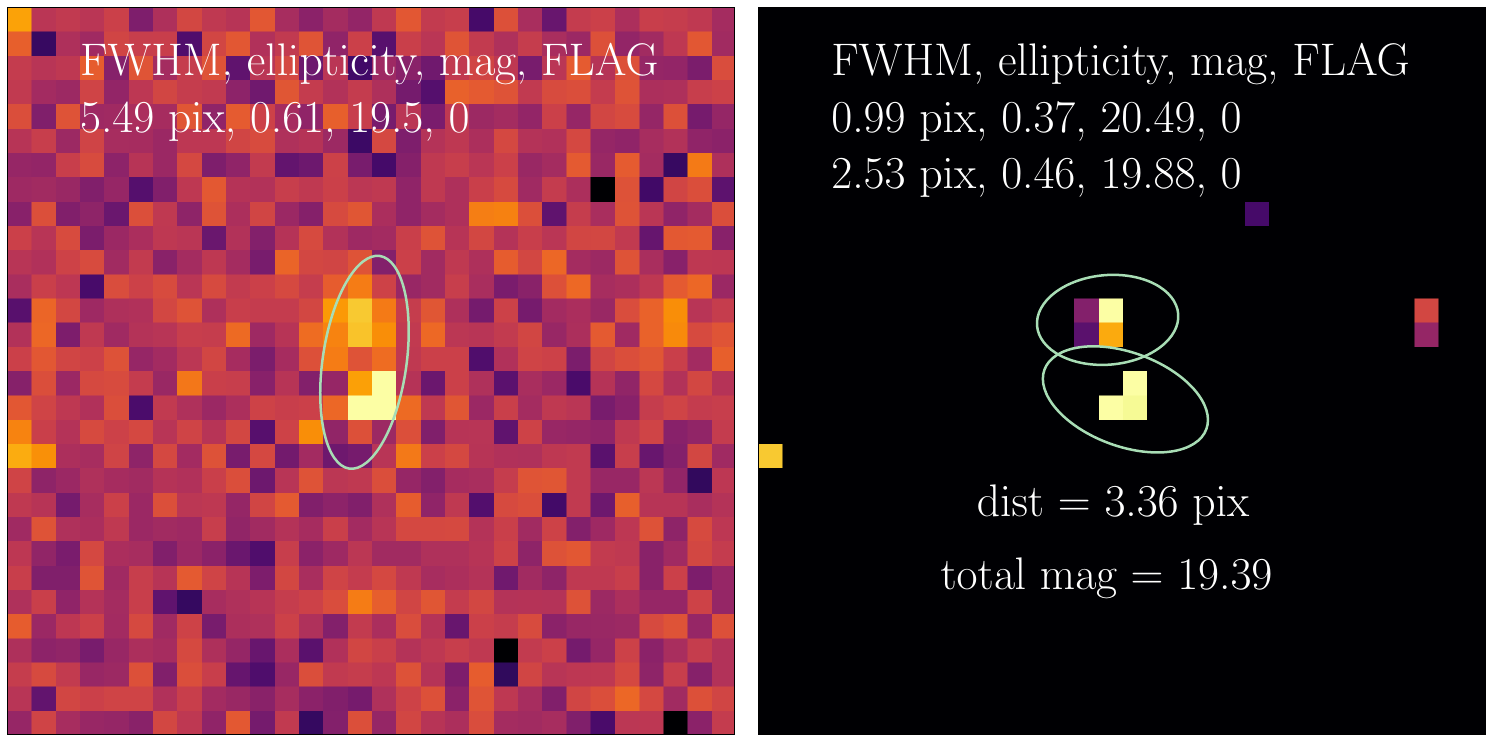}\hfill
      \includegraphics[keepaspectratio,width=0.32\linewidth]{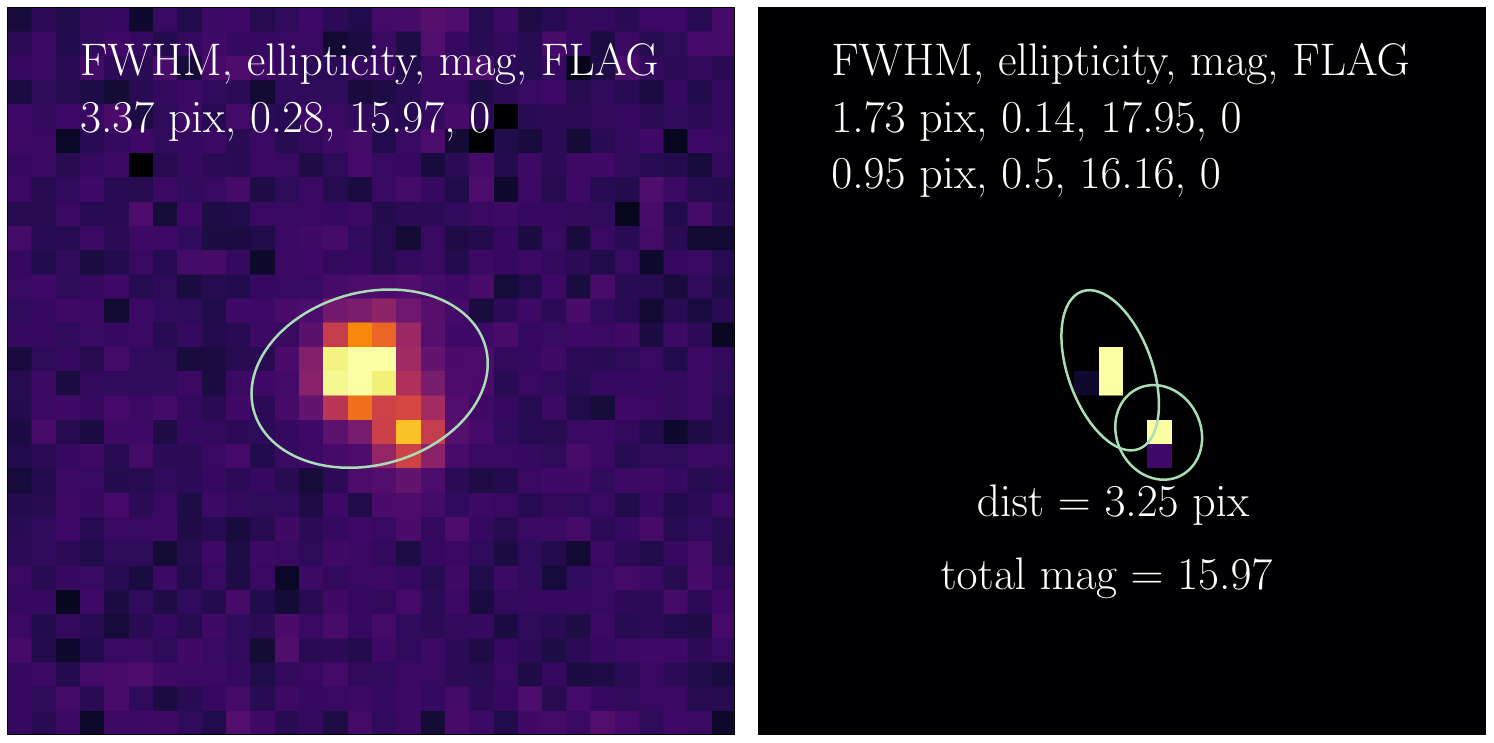}
      \includegraphics[keepaspectratio,width=0.32\linewidth]{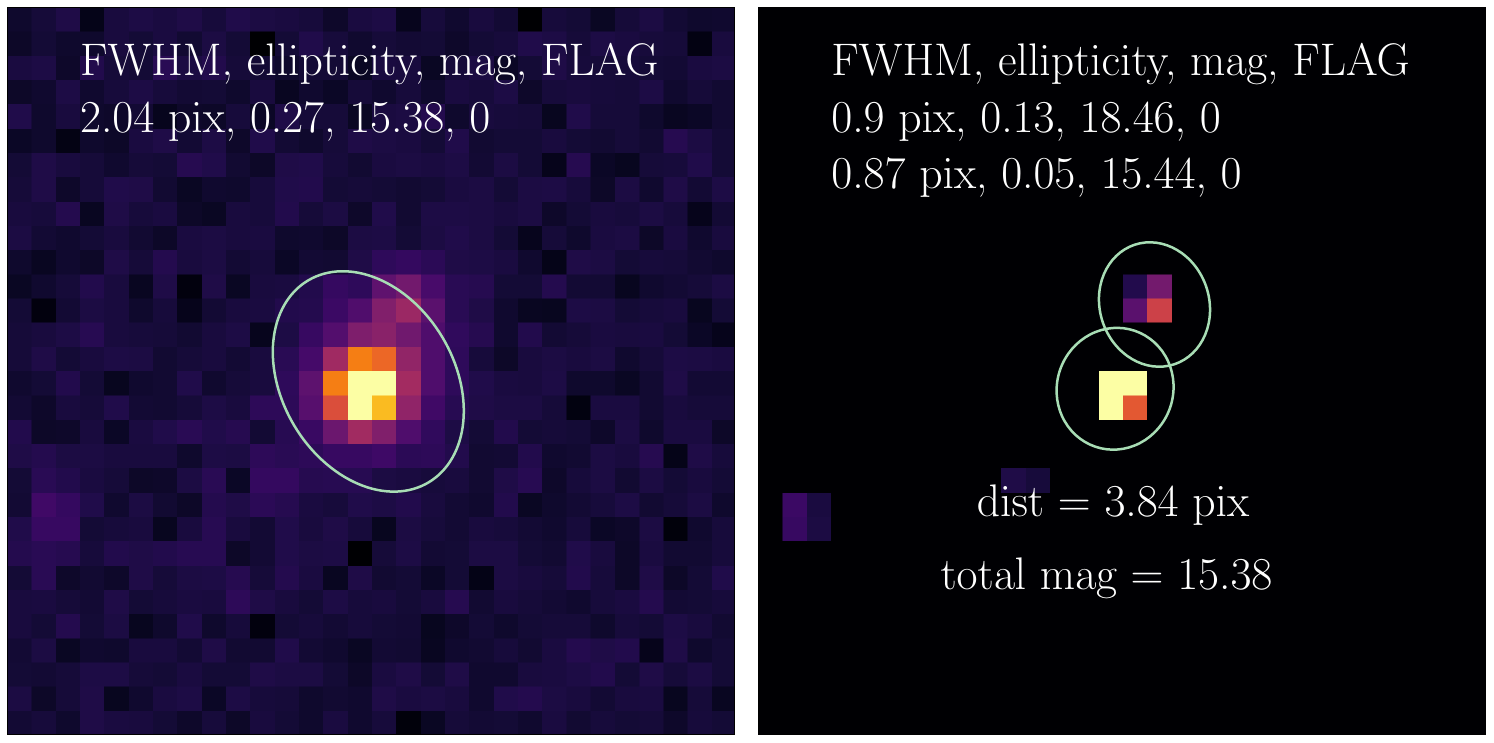}
      \includegraphics[keepaspectratio,width=0.32\linewidth]{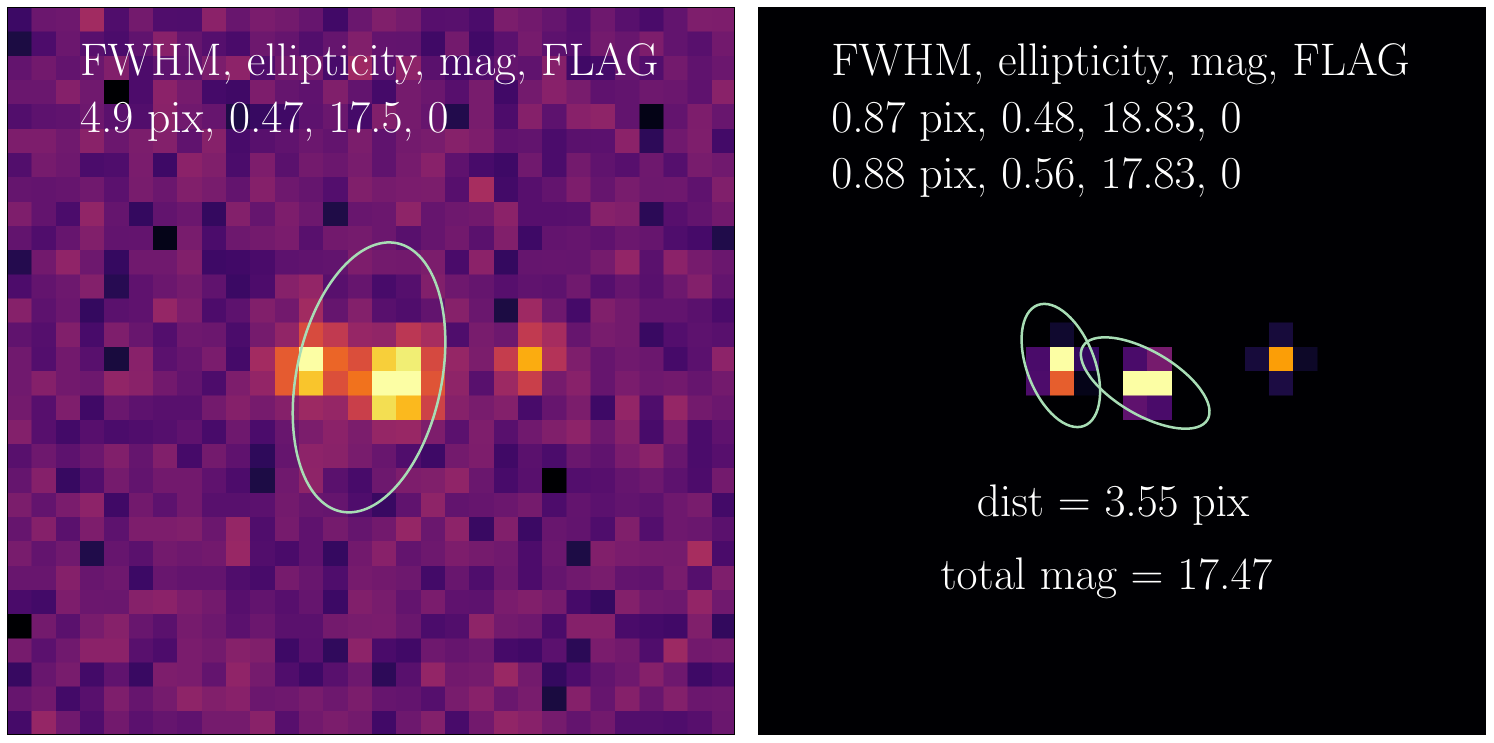}\hfill
      \includegraphics[keepaspectratio,width=0.32\linewidth]{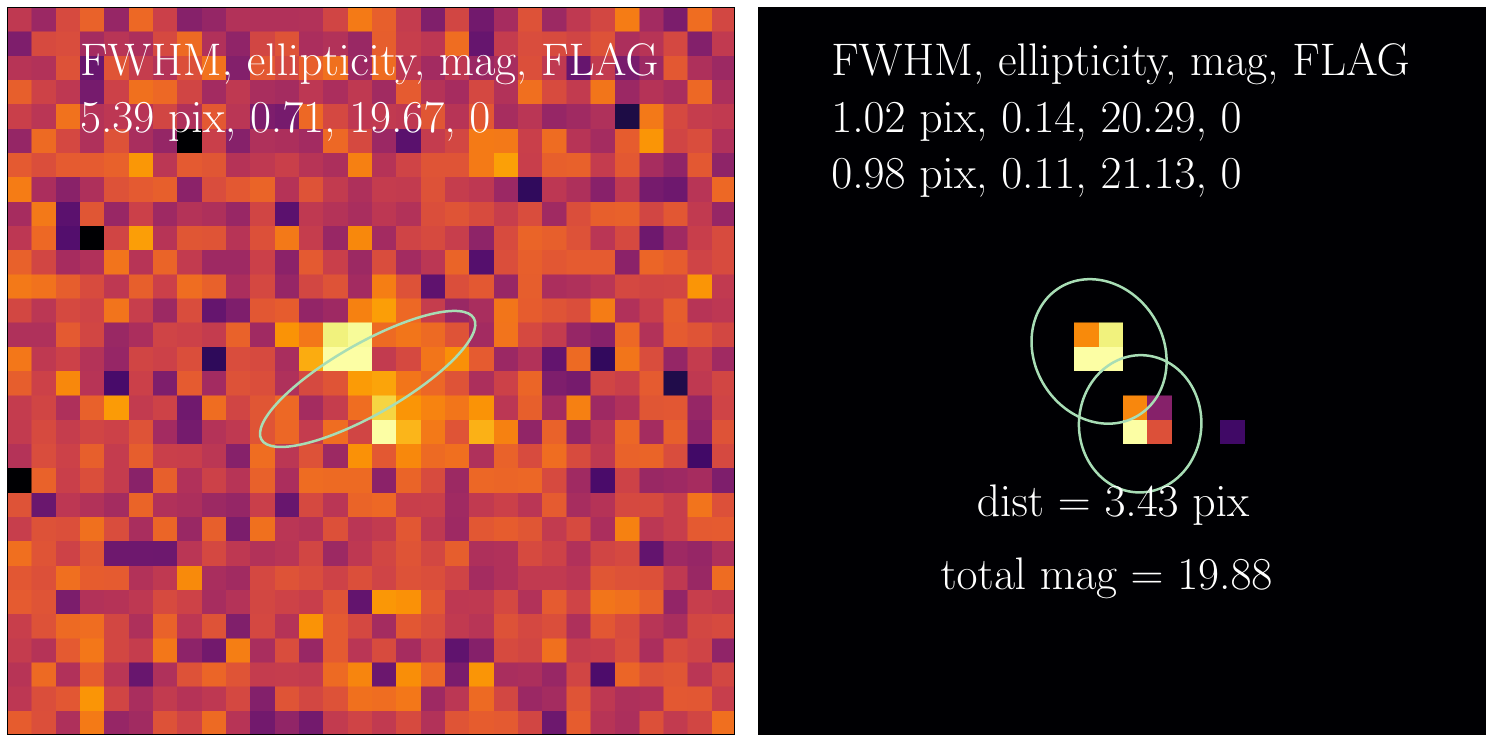}
      \includegraphics[keepaspectratio,width=0.32\linewidth]{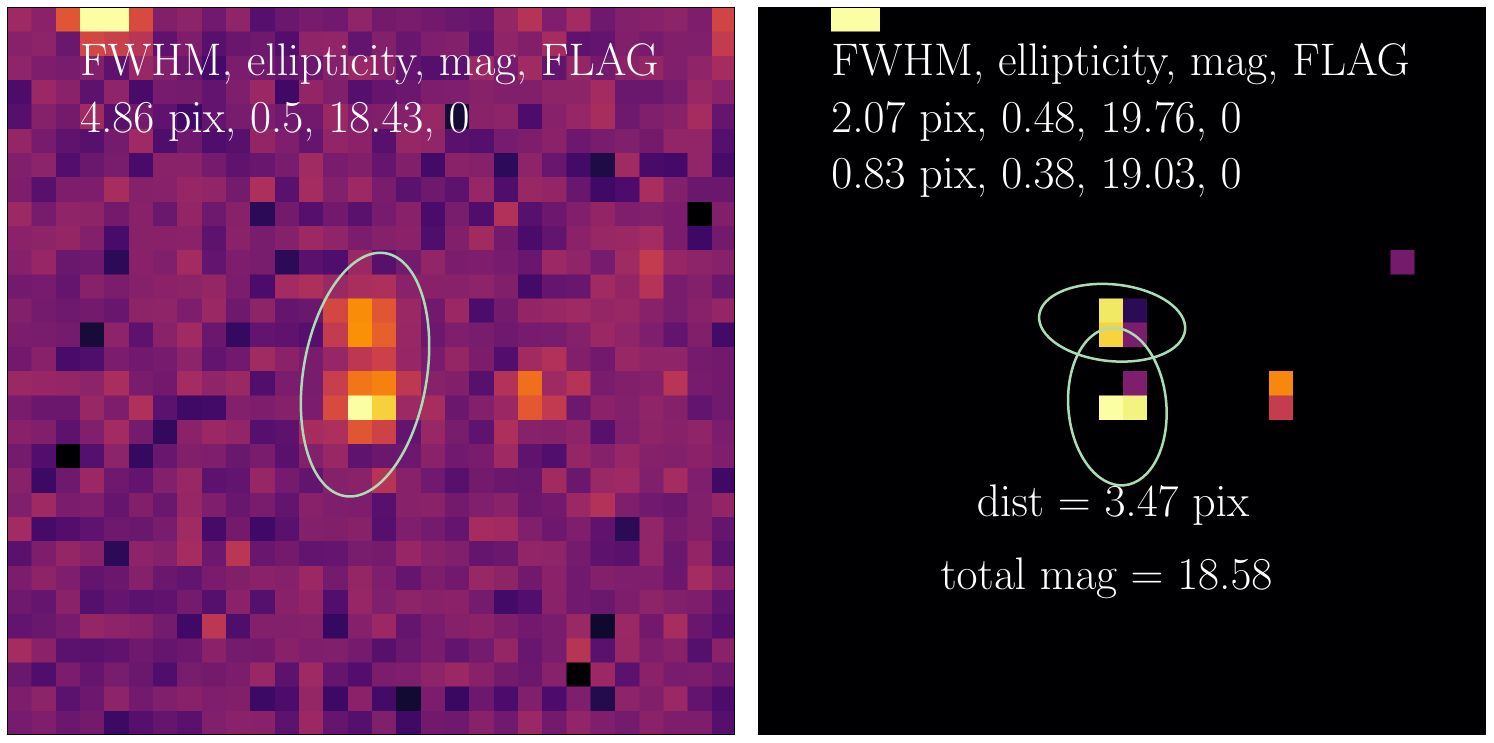}
      \includegraphics[keepaspectratio,width=0.32\linewidth]{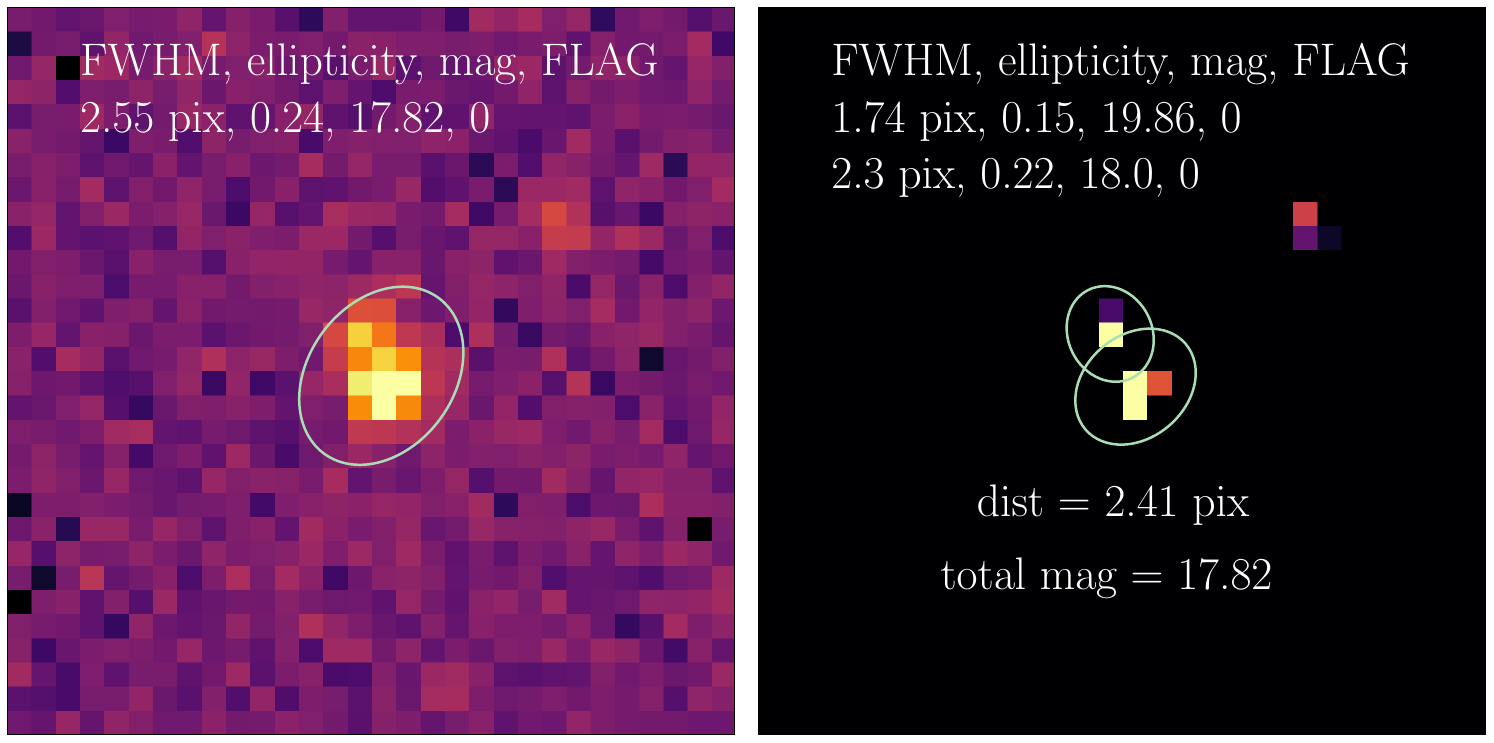}\hfill
      \includegraphics[keepaspectratio,width=0.32\linewidth]{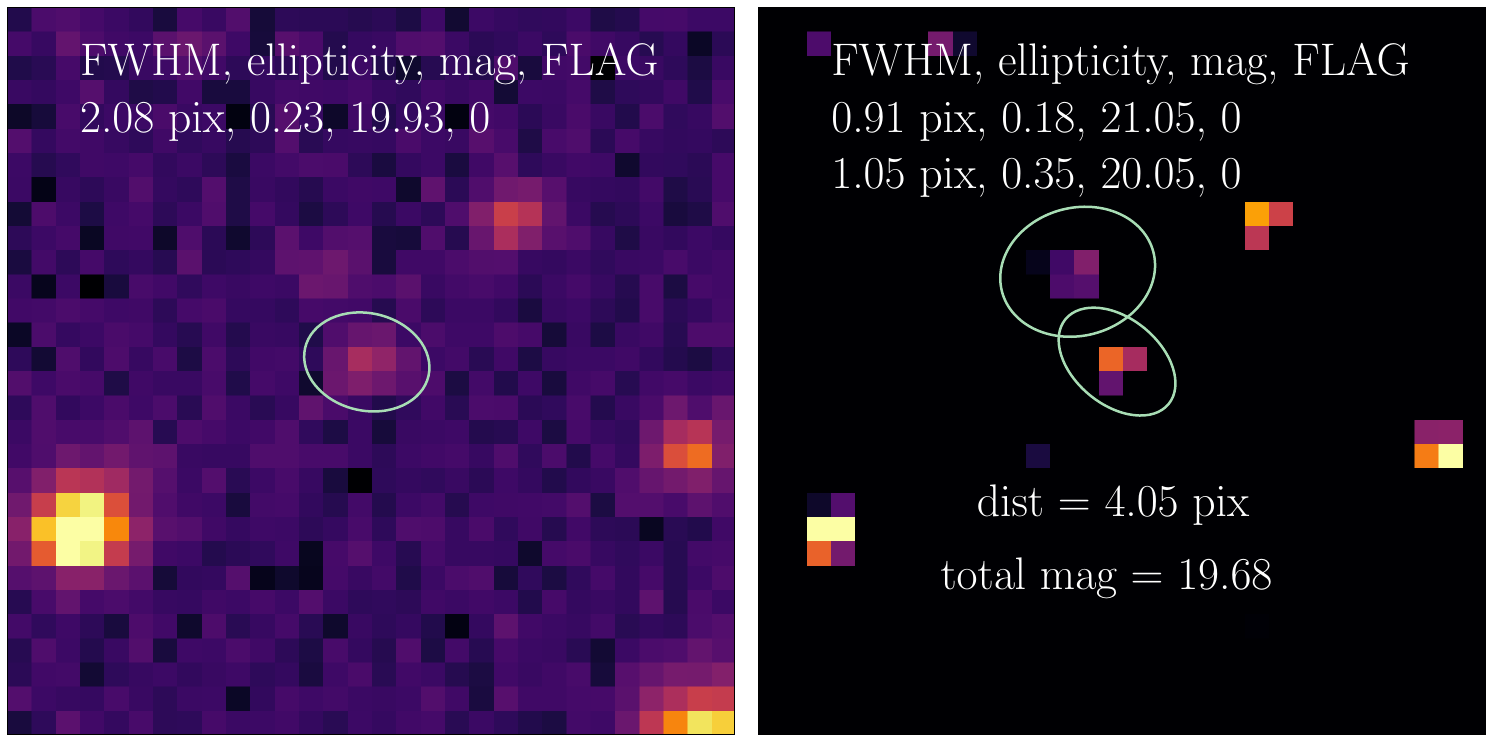}
      \includegraphics[keepaspectratio,width=0.32\linewidth]{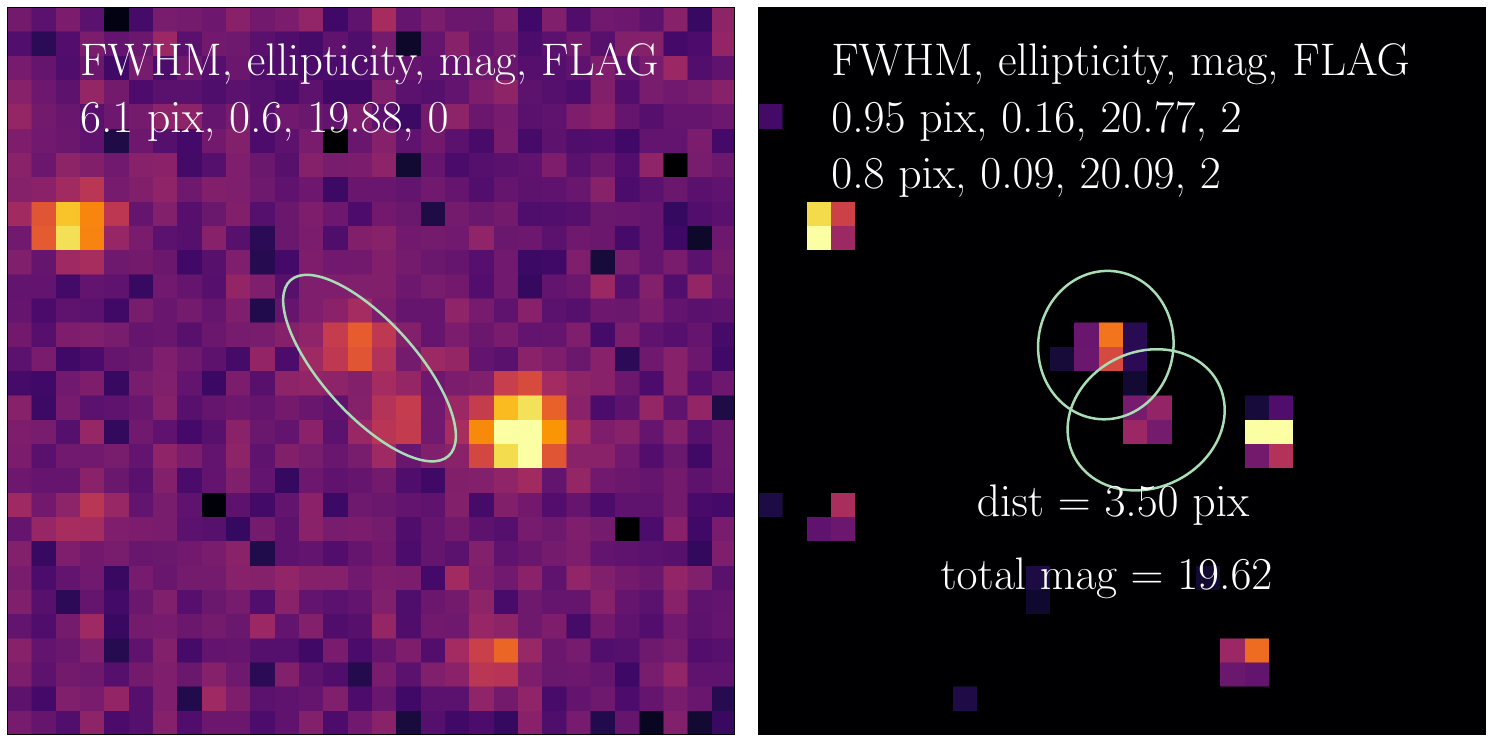}
      \includegraphics[keepaspectratio,width=0.32\linewidth]{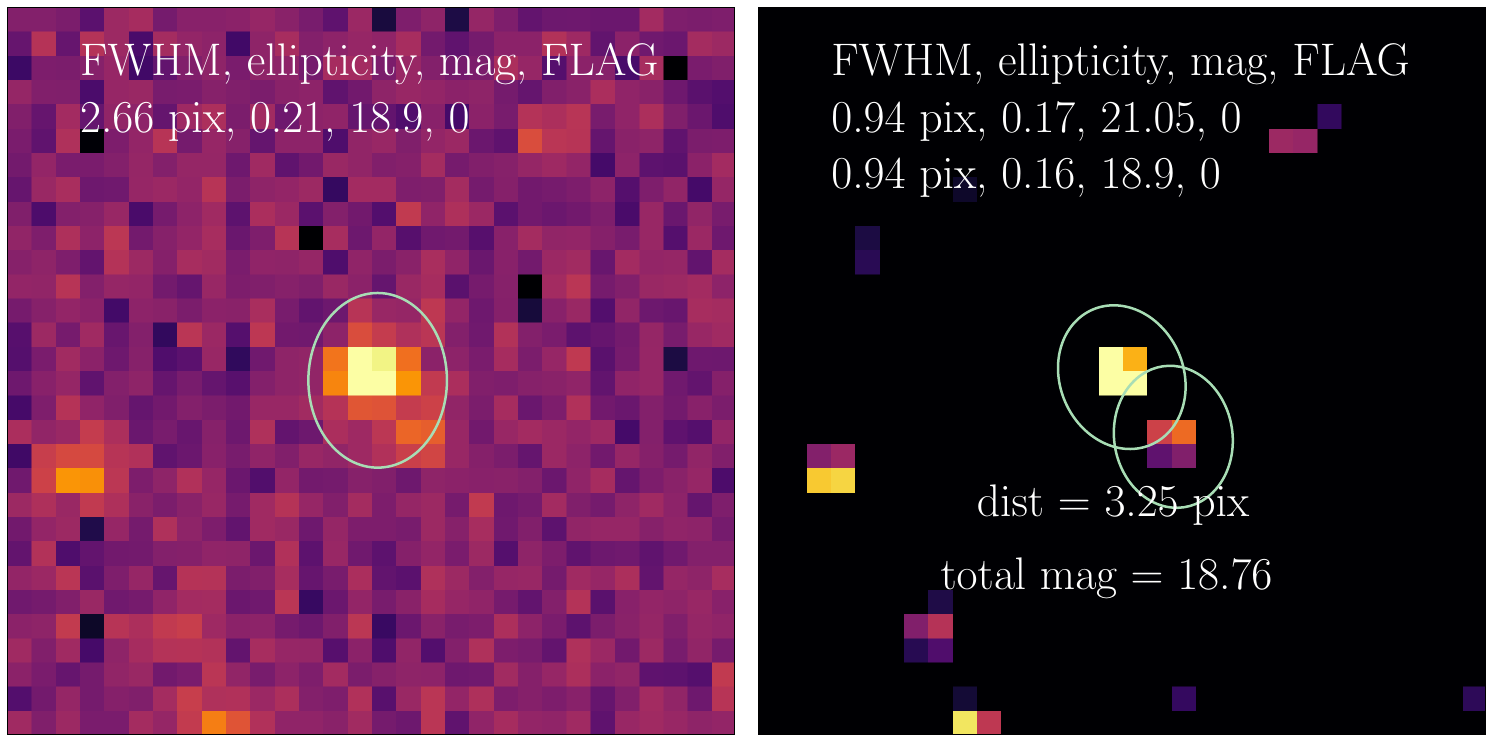}\hfill
      \includegraphics[keepaspectratio,width=0.32\linewidth]{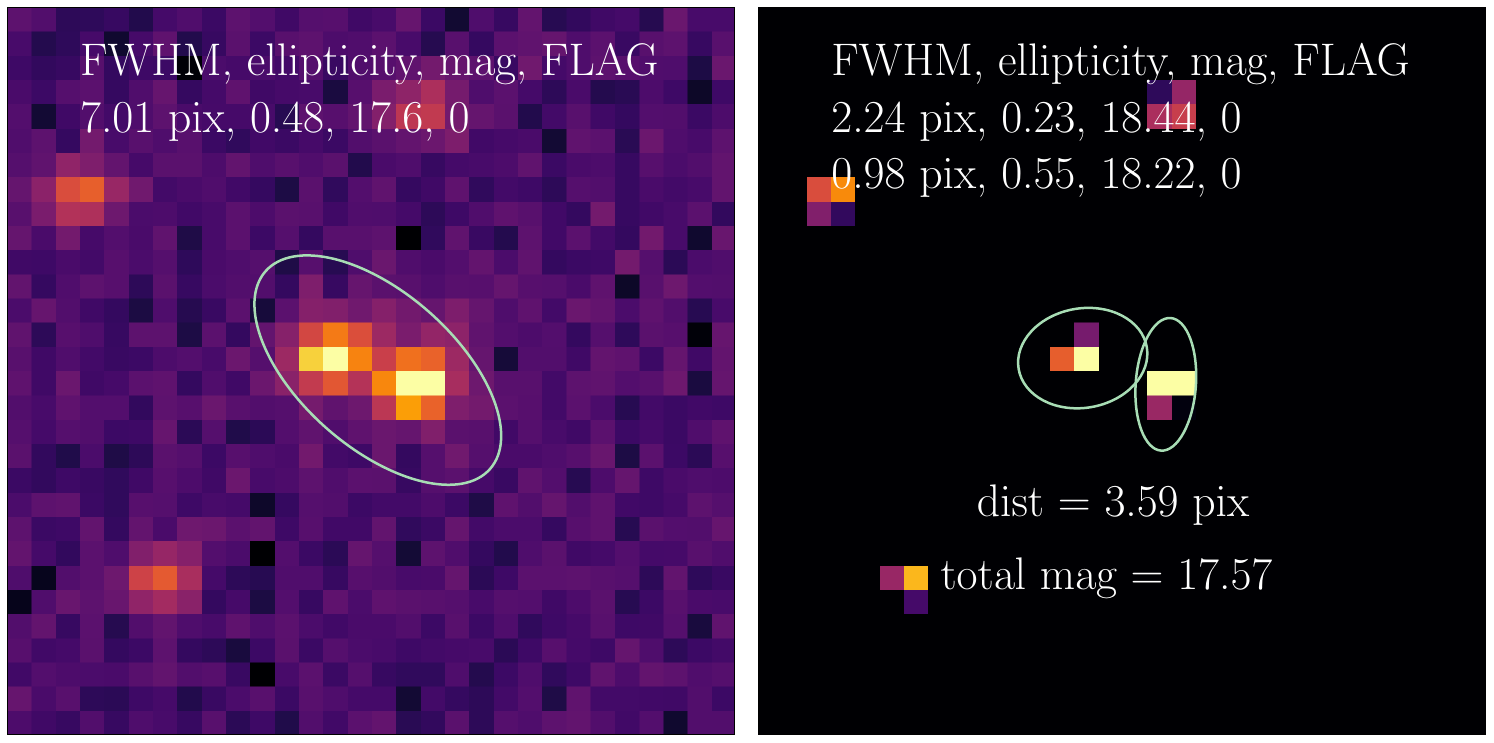}
      \includegraphics[keepaspectratio,width=0.32\linewidth]{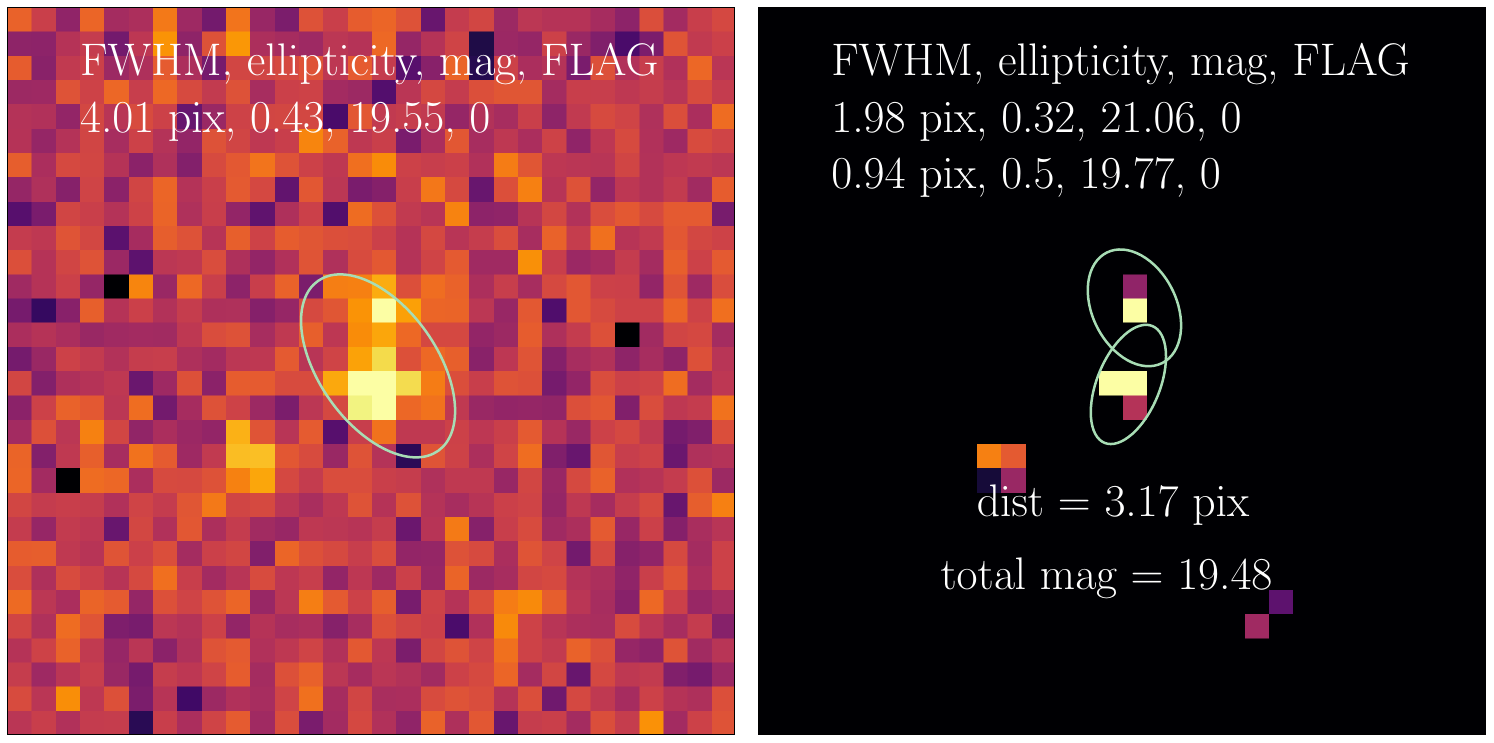}
      \includegraphics[keepaspectratio,width=0.32\linewidth]{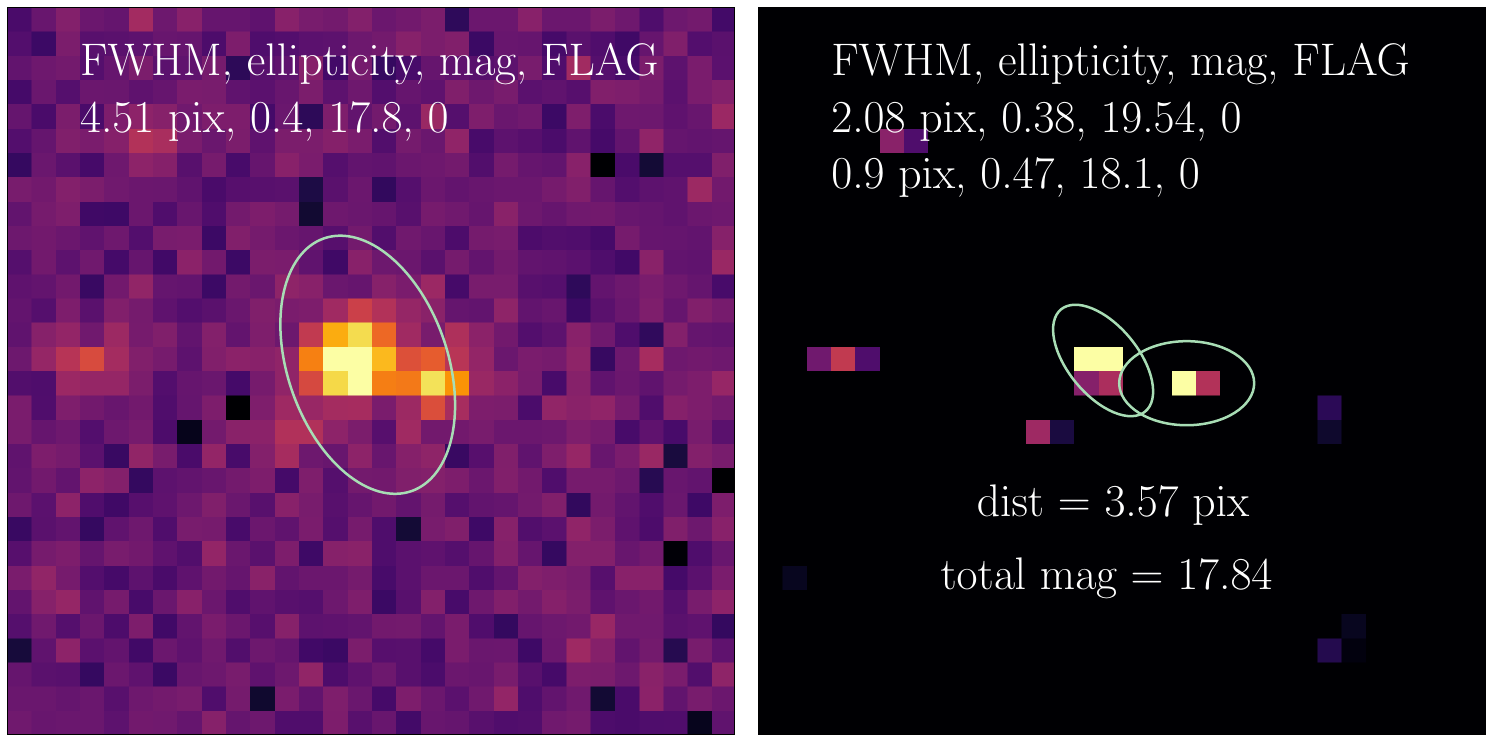}\hfill
      \includegraphics[keepaspectratio,width=0.32\linewidth]{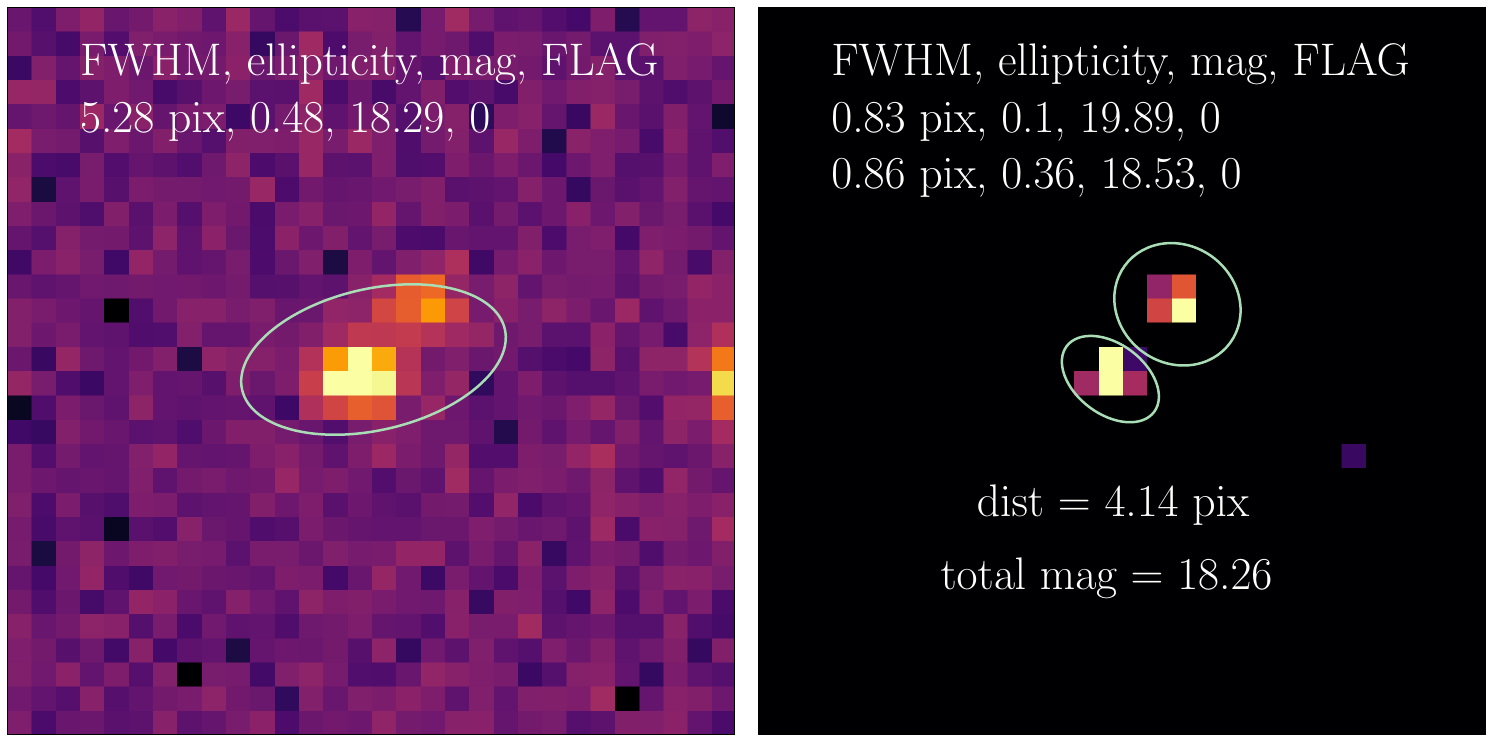}
      \includegraphics[keepaspectratio,width=0.32\linewidth]{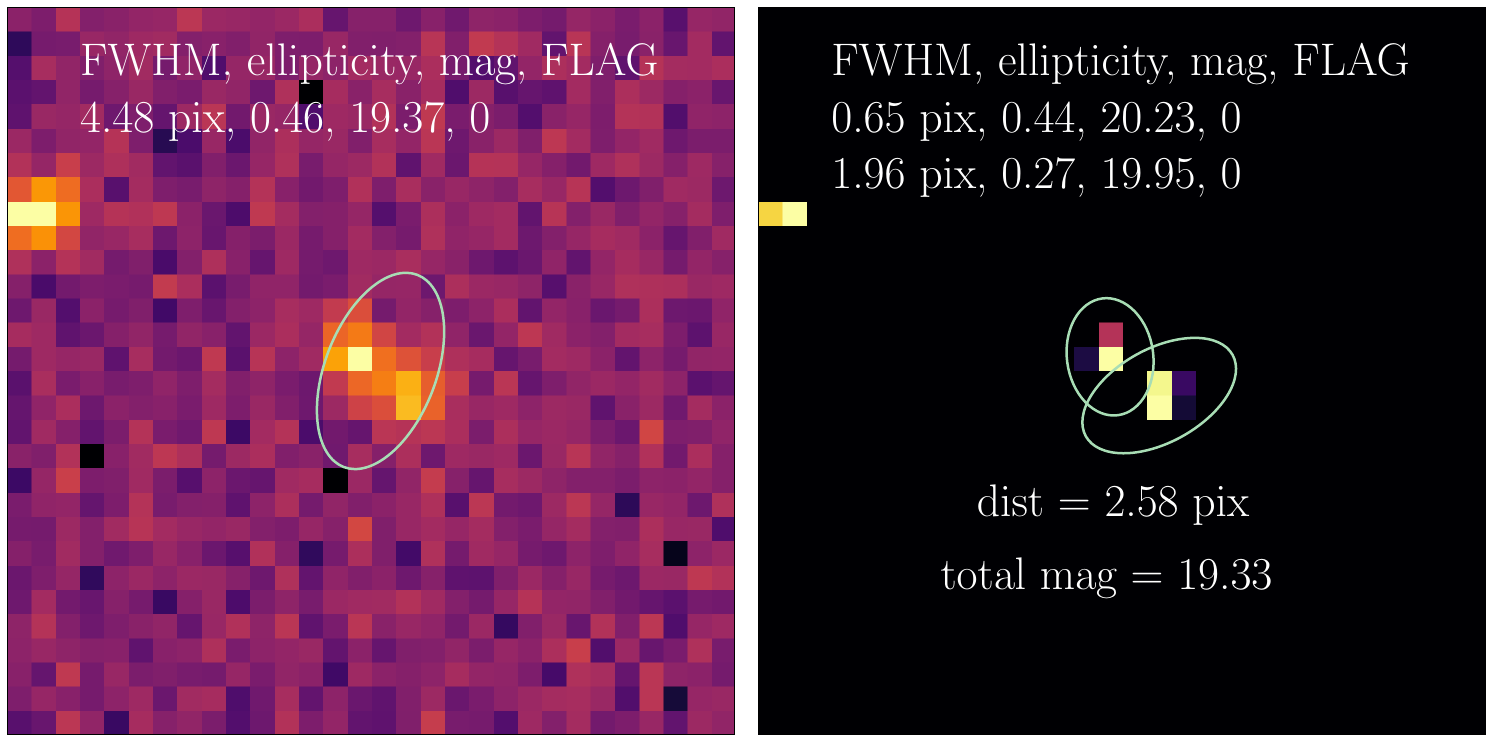}
      \includegraphics[keepaspectratio,width=0.32\linewidth]{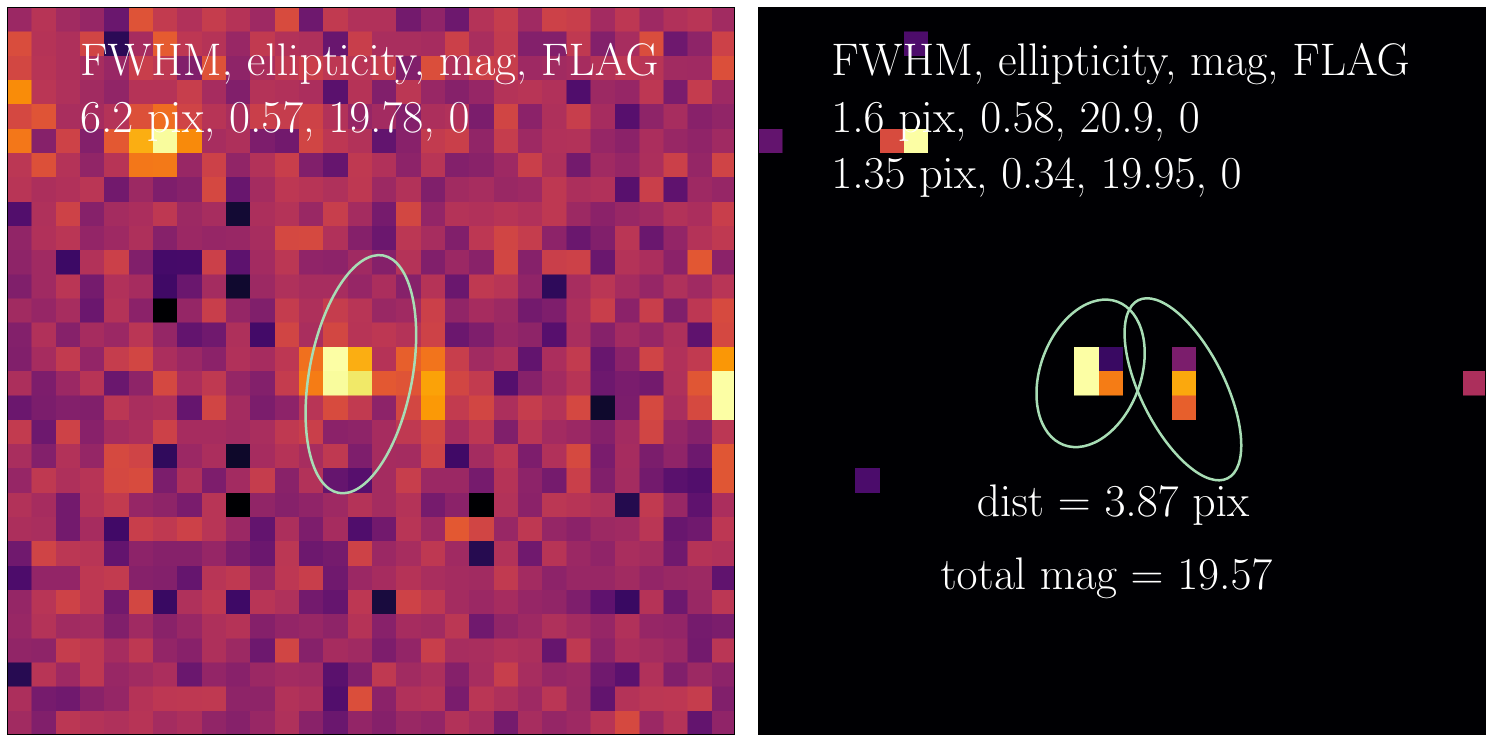}\hfill
\end{figure*}
\begin{figure*}
      \centering
      \includegraphics[keepaspectratio,width=0.32\linewidth]{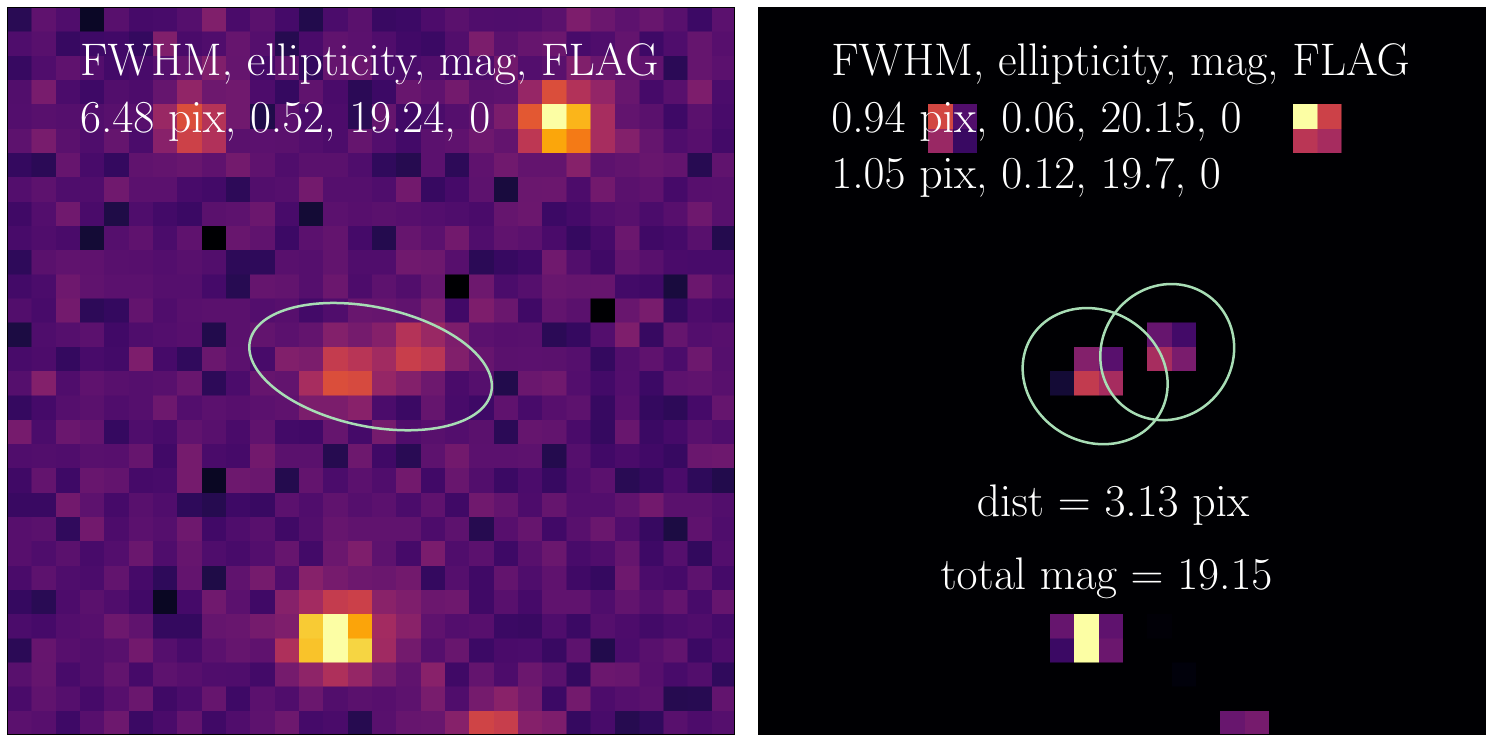}
      \includegraphics[keepaspectratio,width=0.32\linewidth]{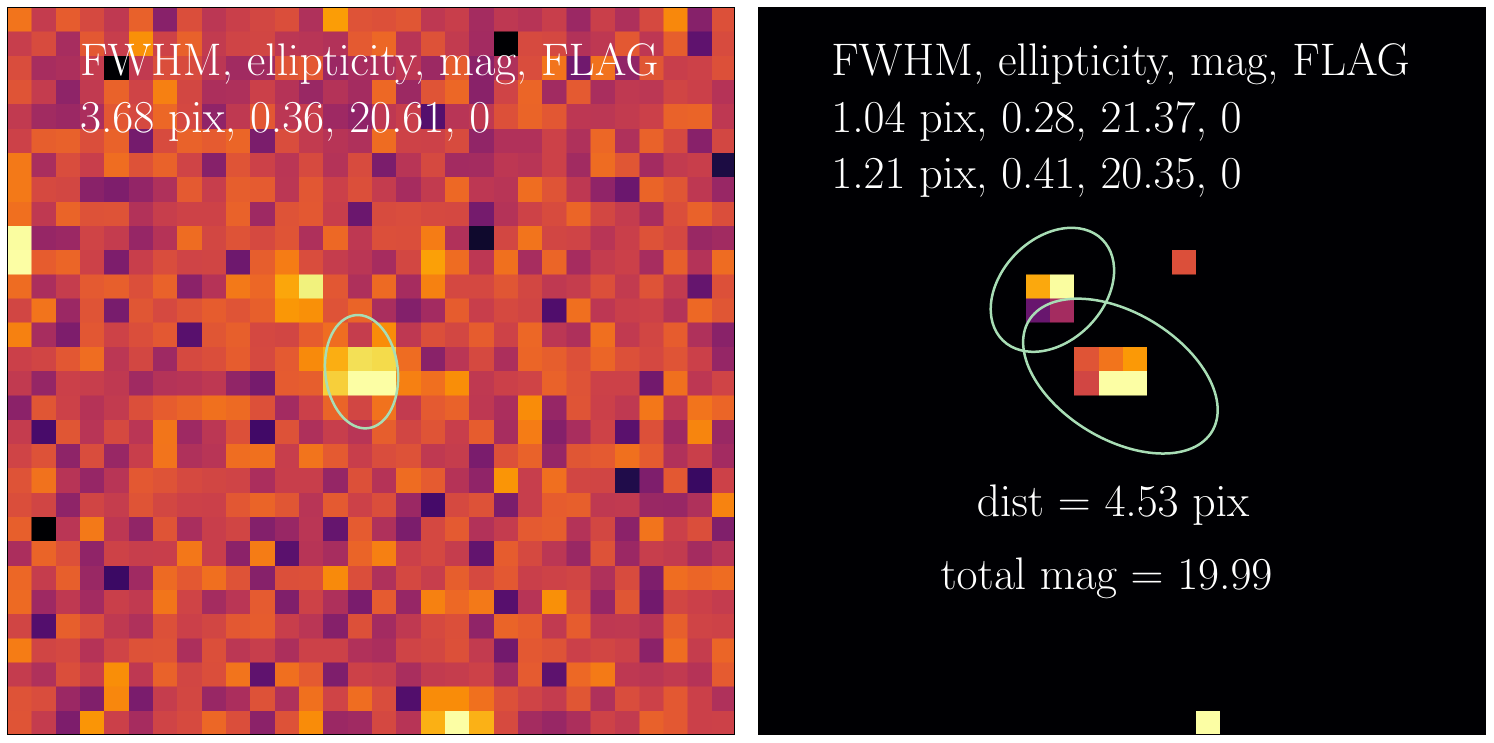}
      \includegraphics[keepaspectratio,width=0.32\linewidth]{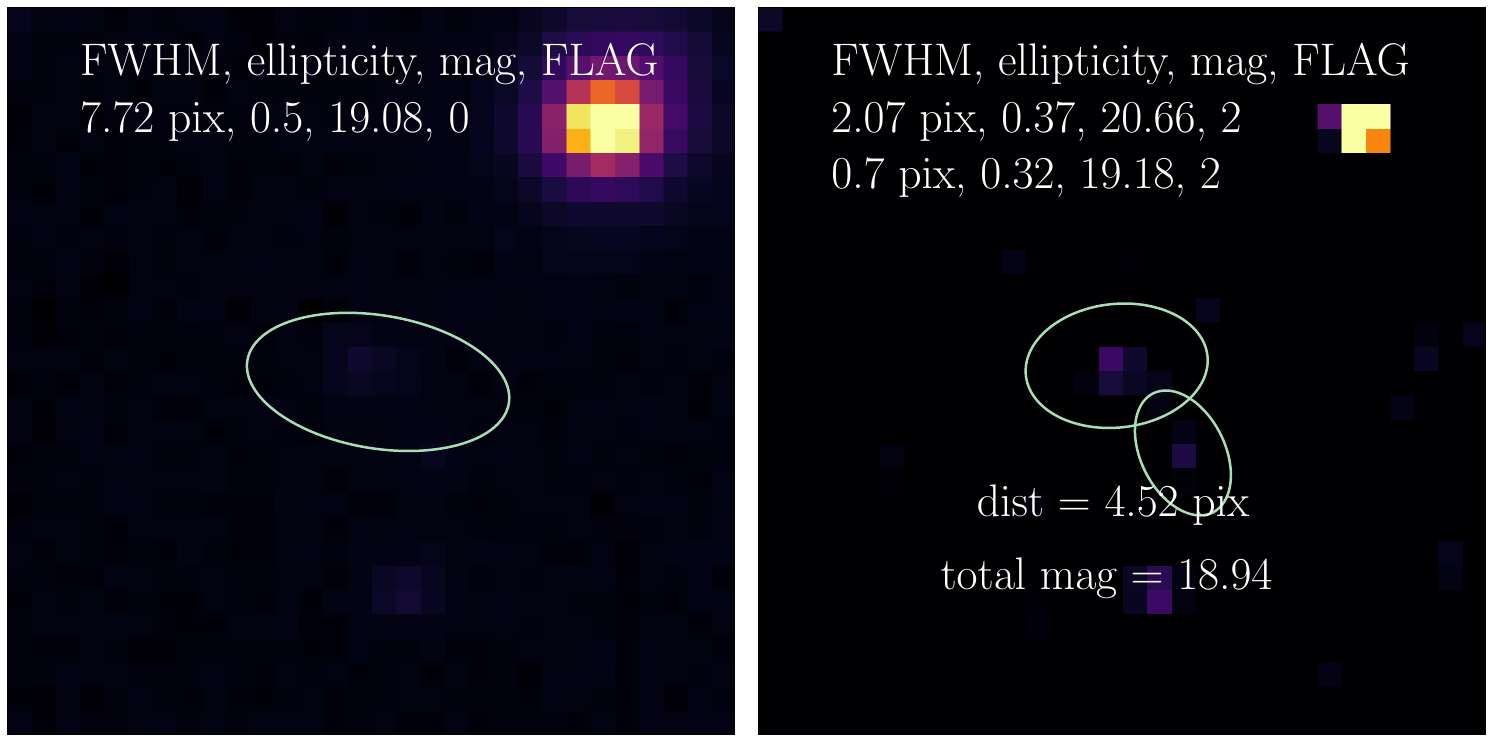}\hfill
      \includegraphics[keepaspectratio,width=0.32\linewidth]{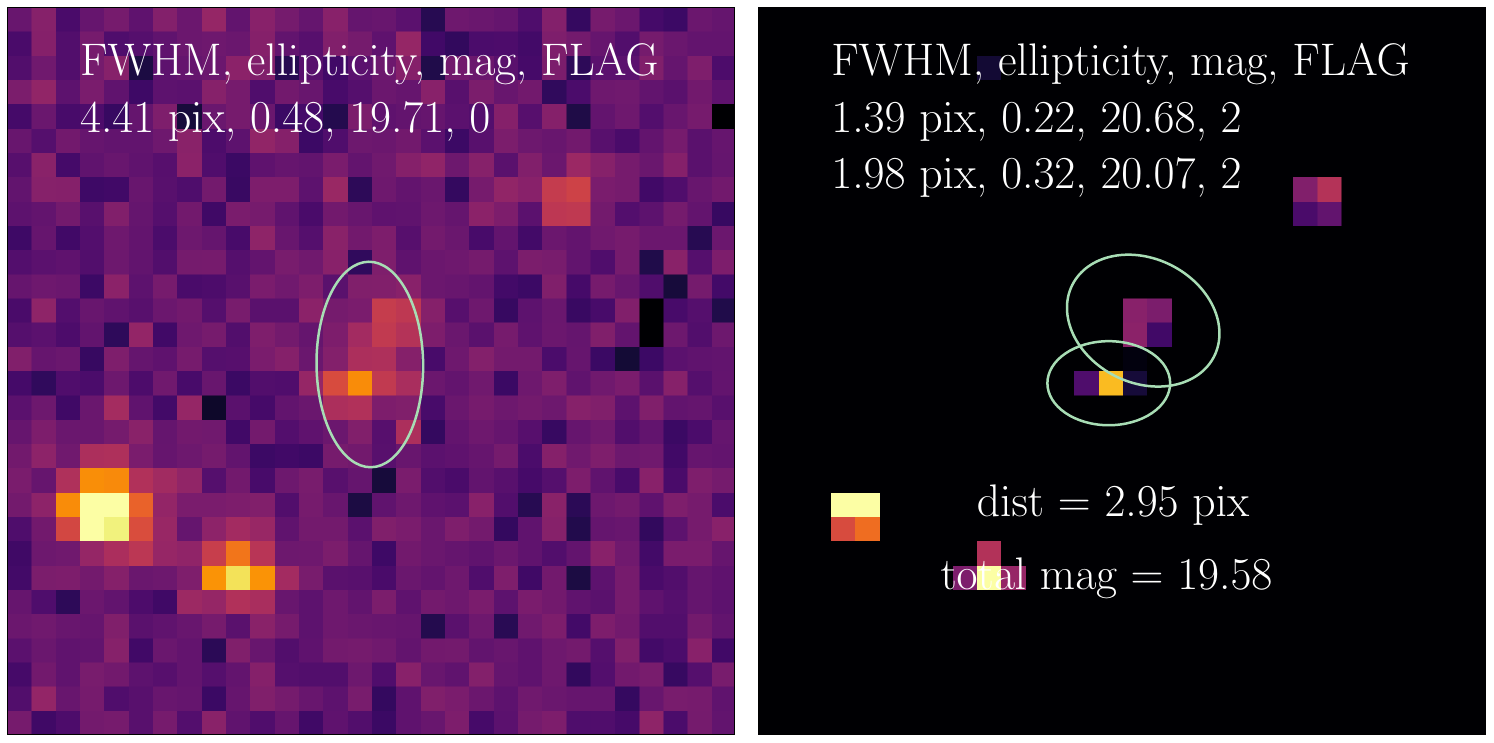}
      \includegraphics[keepaspectratio,width=0.32\linewidth]{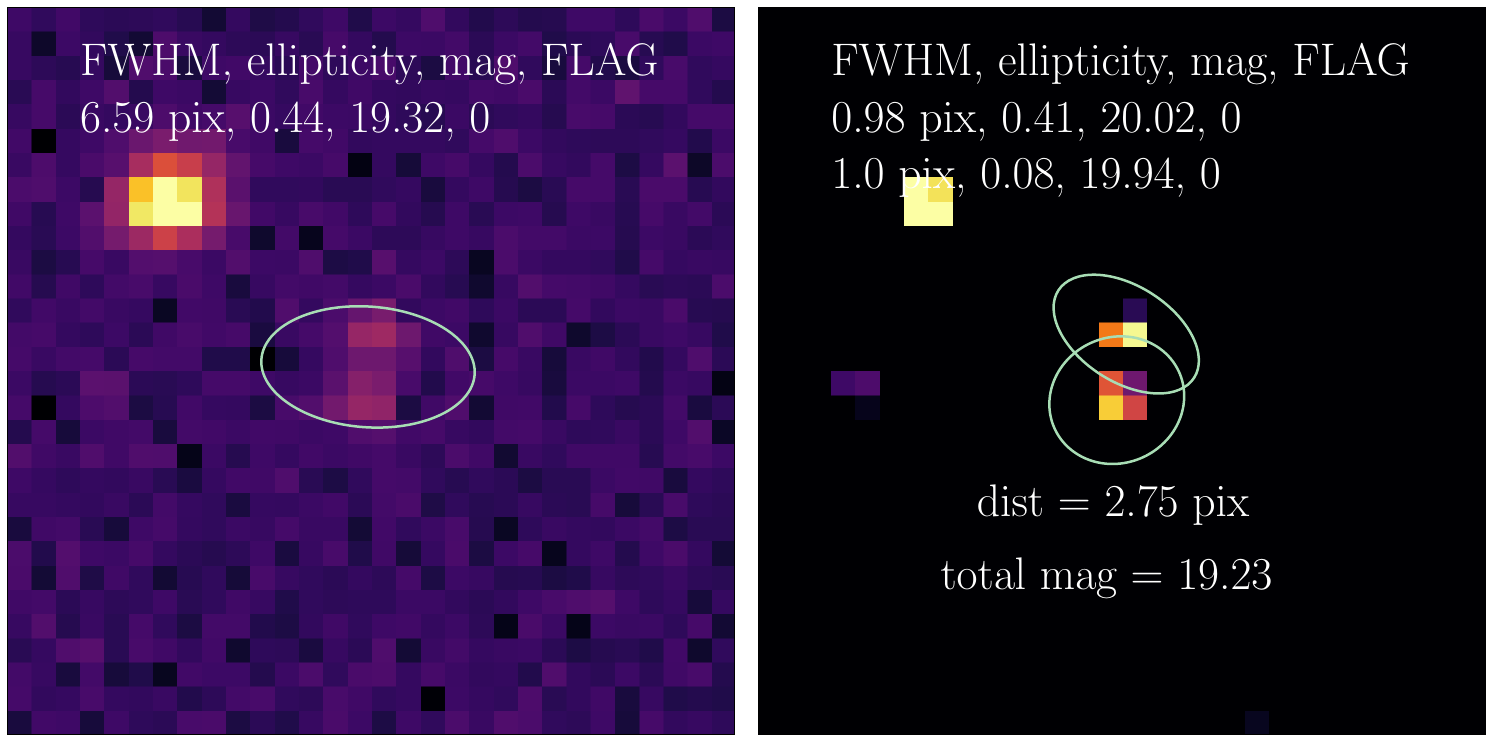}
      \includegraphics[keepaspectratio,width=0.32\linewidth]{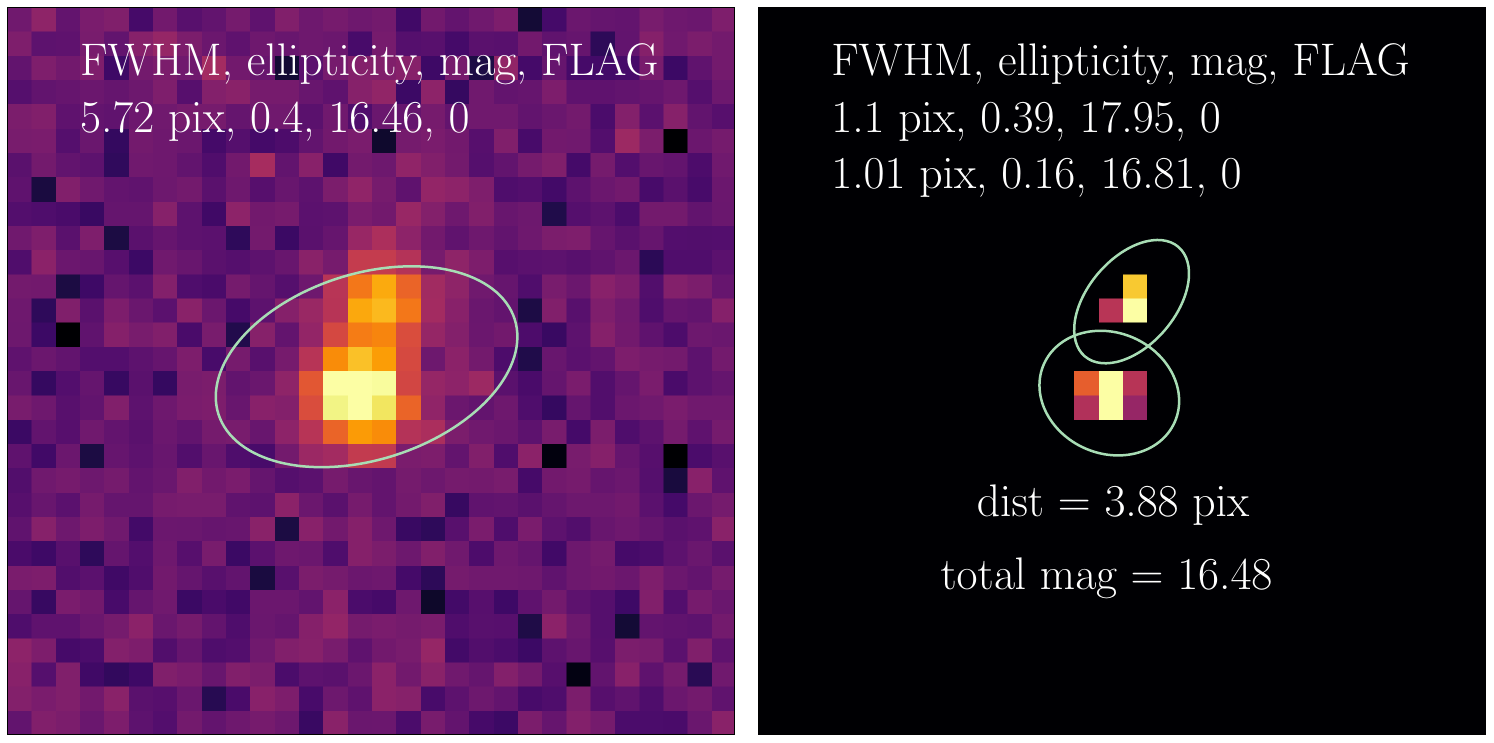}\hfill
      \includegraphics[keepaspectratio,width=0.32\linewidth]{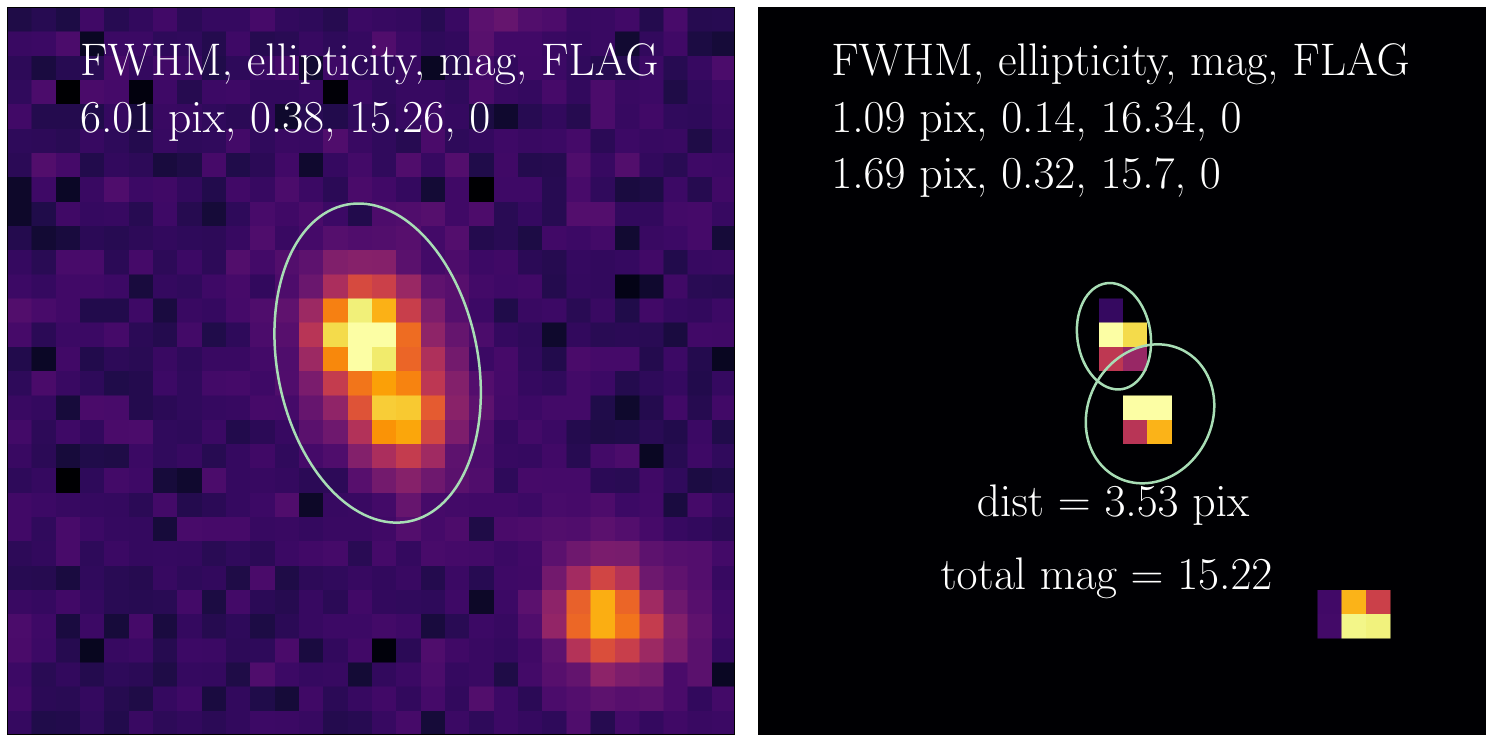}
      \includegraphics[keepaspectratio,width=0.32\linewidth]{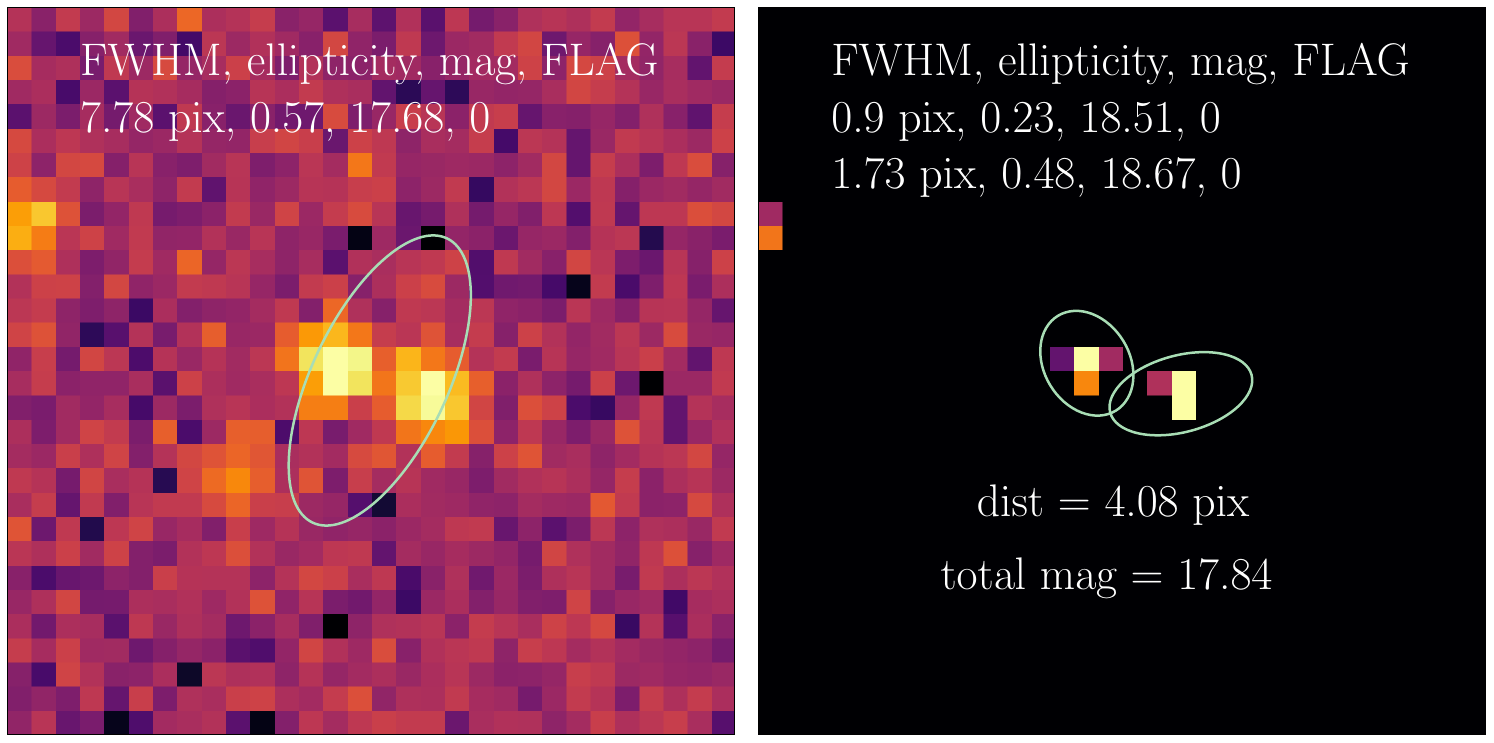}
      \includegraphics[keepaspectratio,width=0.32\linewidth]{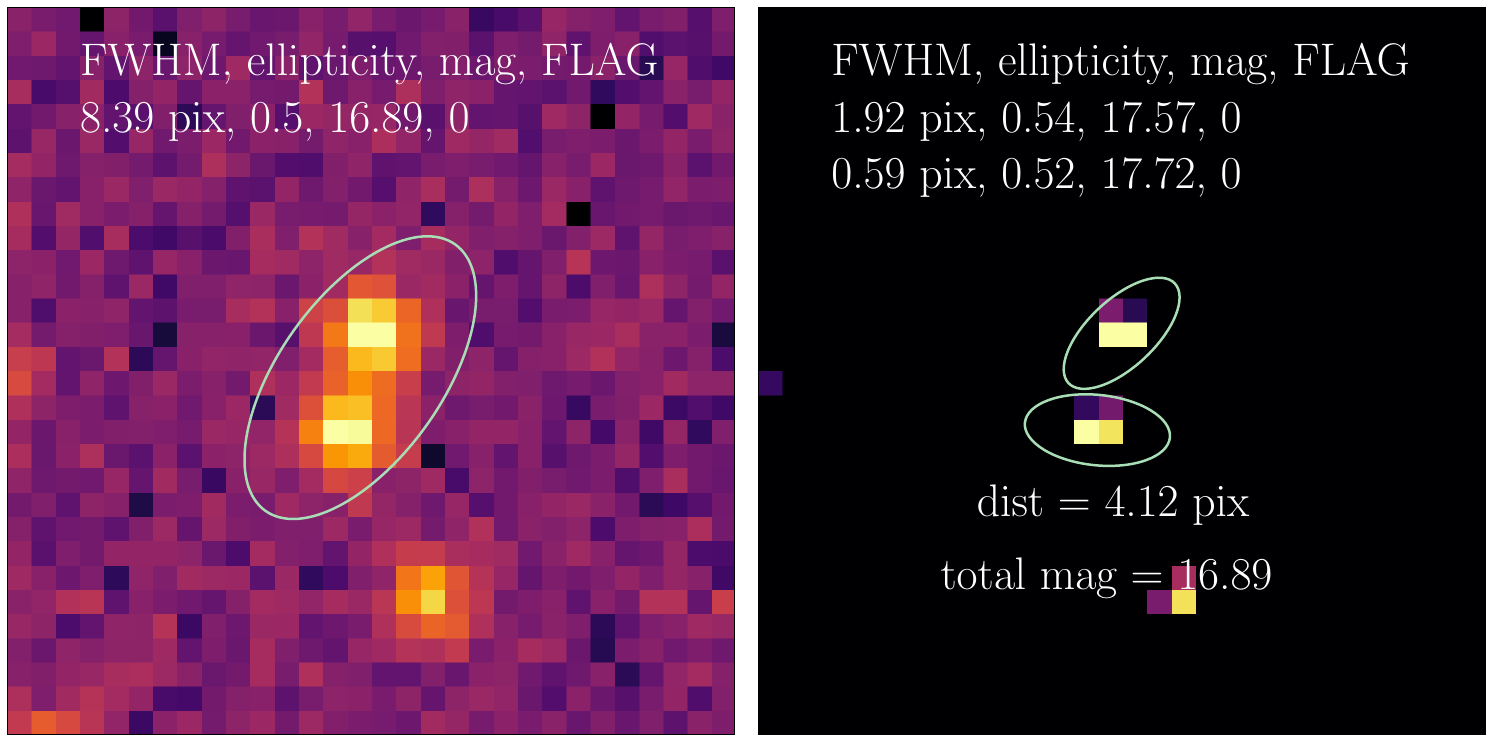}\hfill
      \centering
      \includegraphics[keepaspectratio,width=0.32\linewidth]{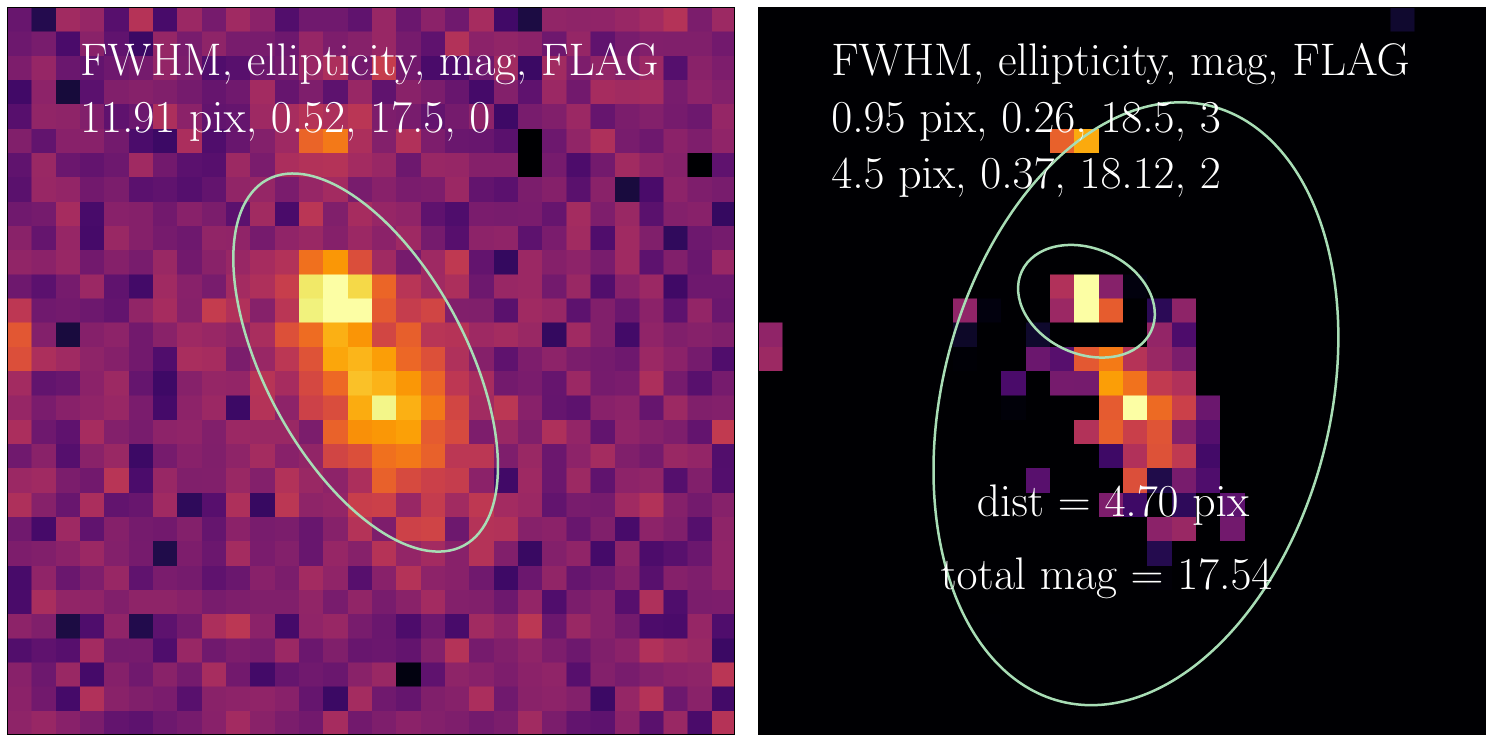}
    \caption{Continuation of deblending visualizations from Fig.~\ref{fig:deblend-examples}} \label{fig:more-deblend-examples}
\end{figure*}

\section{\reviewTwo{Extended visualizations of likely real unmatched deconvolved sources}}\label{appn:crossmatching-results-OrigDeconDESI-plots-cutouts-onlyInDecon-extended}

Fig.~\ref{fig:crossmatching-results-OrigDeconDESI-plots-cutouts-onlyInDecon-extended-fig} shows additional visualizations like in Fig.~\ref{fig:crossmatching-results-OrigDeconDESI-plots-cutouts-onlyInDecon}.

\begin{figure*}[hbt!]
    \centering
      \includegraphics[keepaspectratio,width=0.48\linewidth]{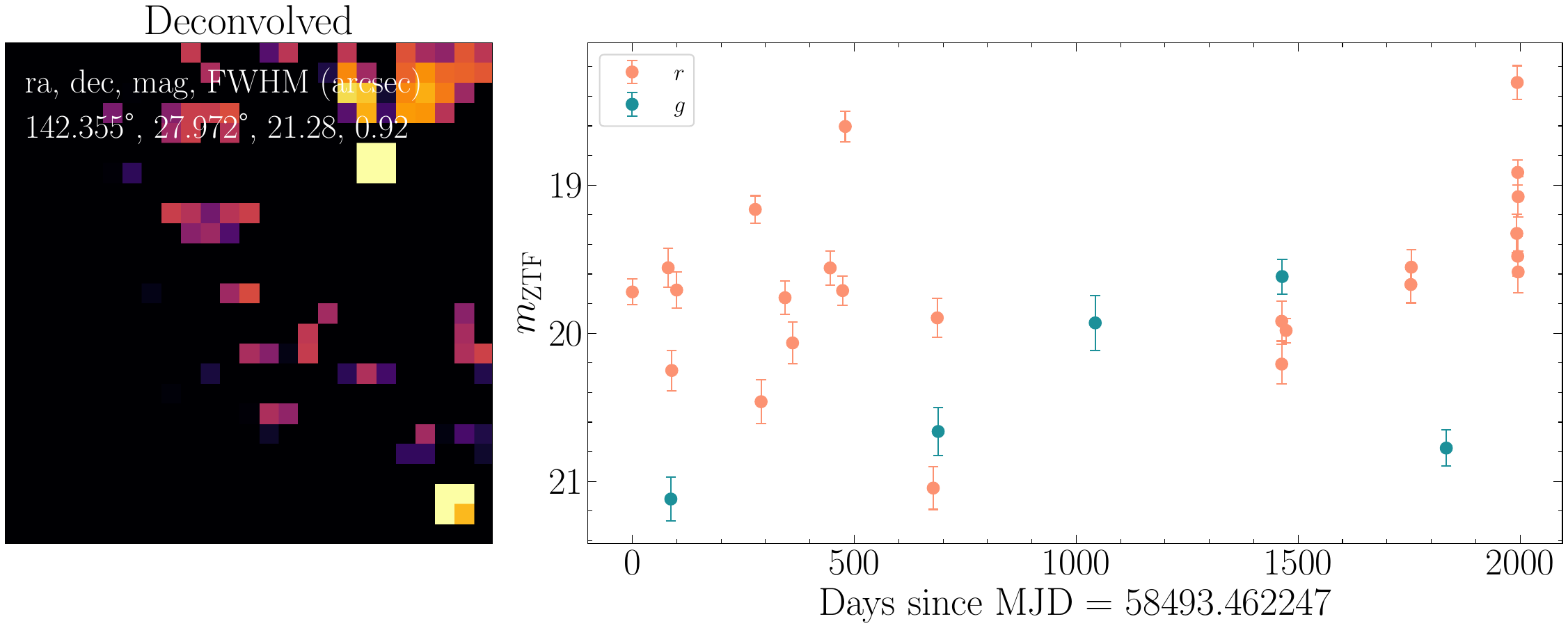}
      \includegraphics[keepaspectratio,width=0.48\linewidth]{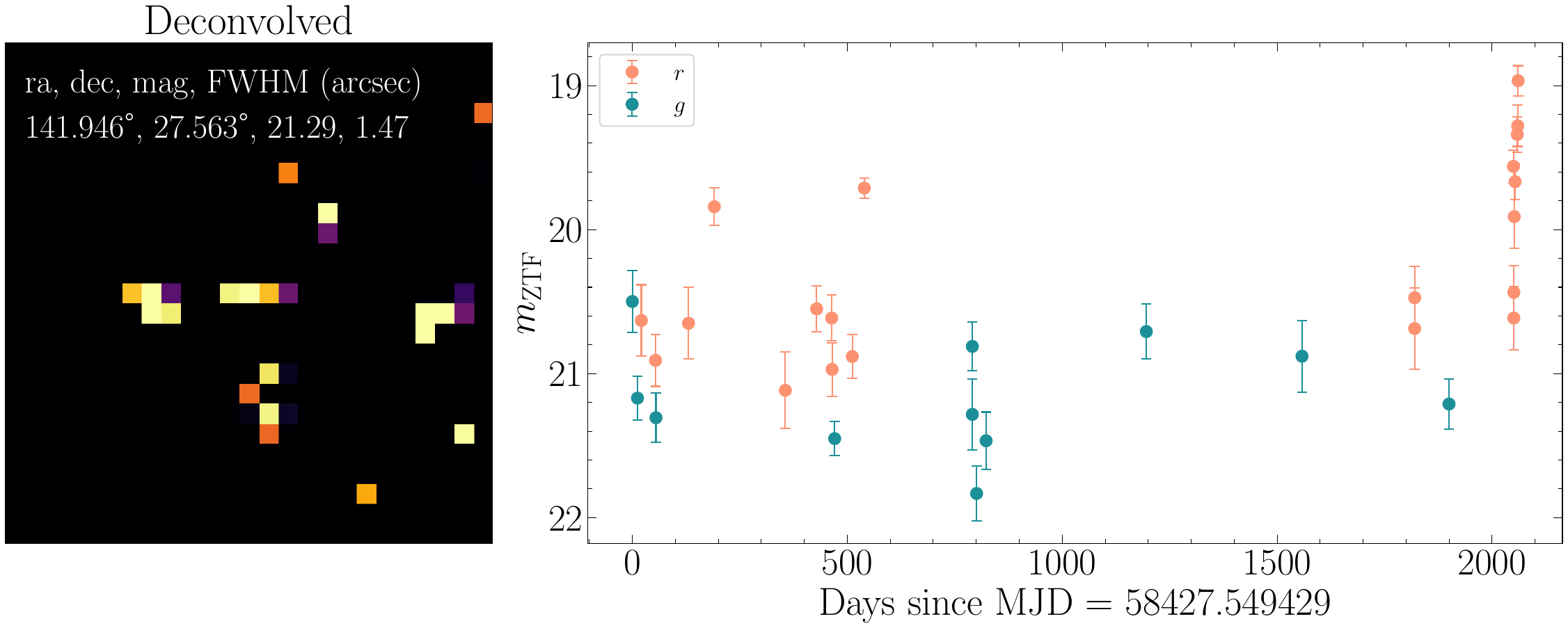}
      \includegraphics[keepaspectratio,width=0.48\linewidth]{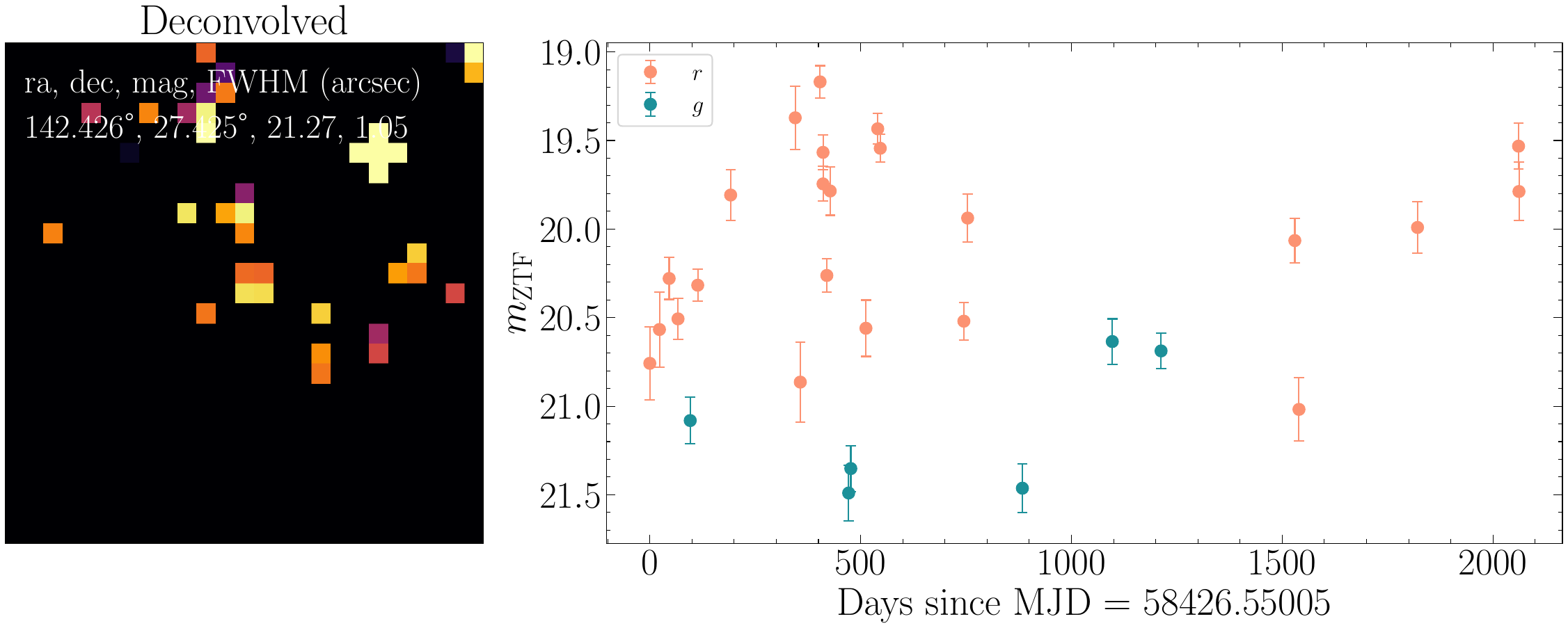}
      \includegraphics[keepaspectratio,width=0.48\linewidth]{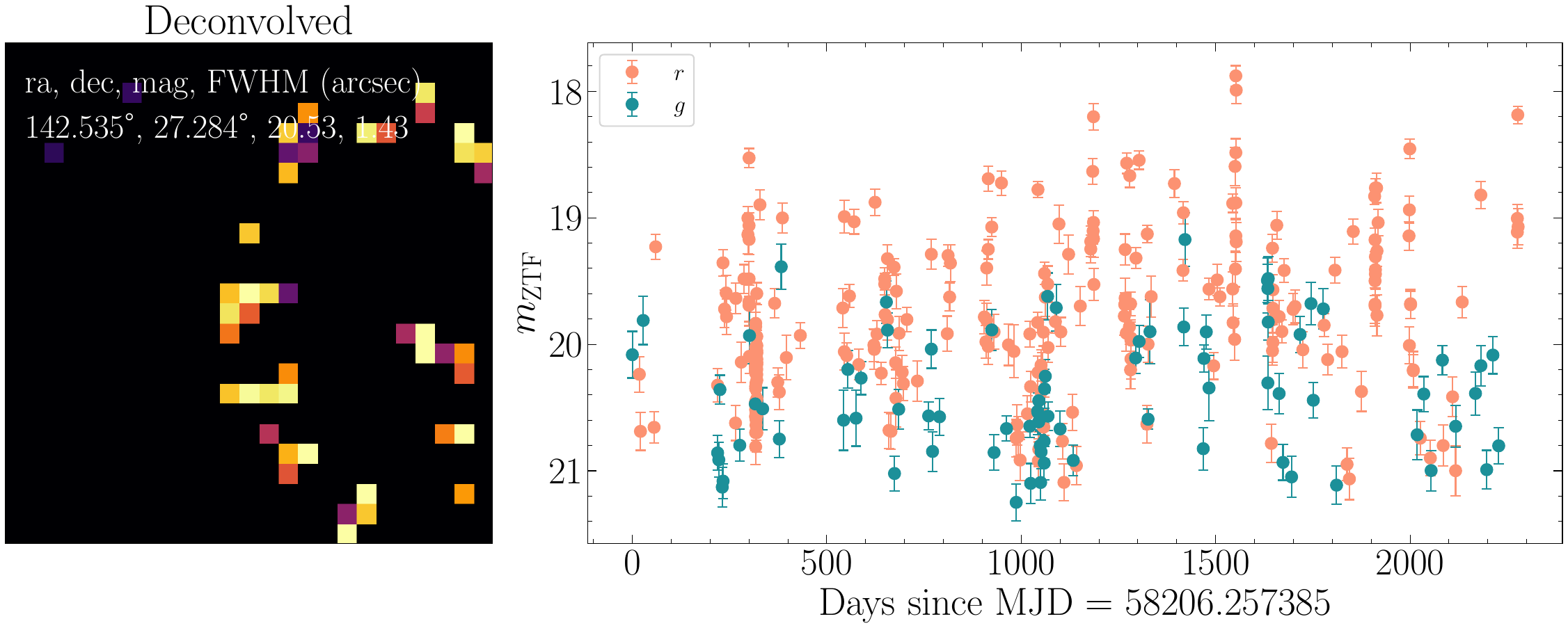}
      \includegraphics[keepaspectratio,width=0.48\linewidth]{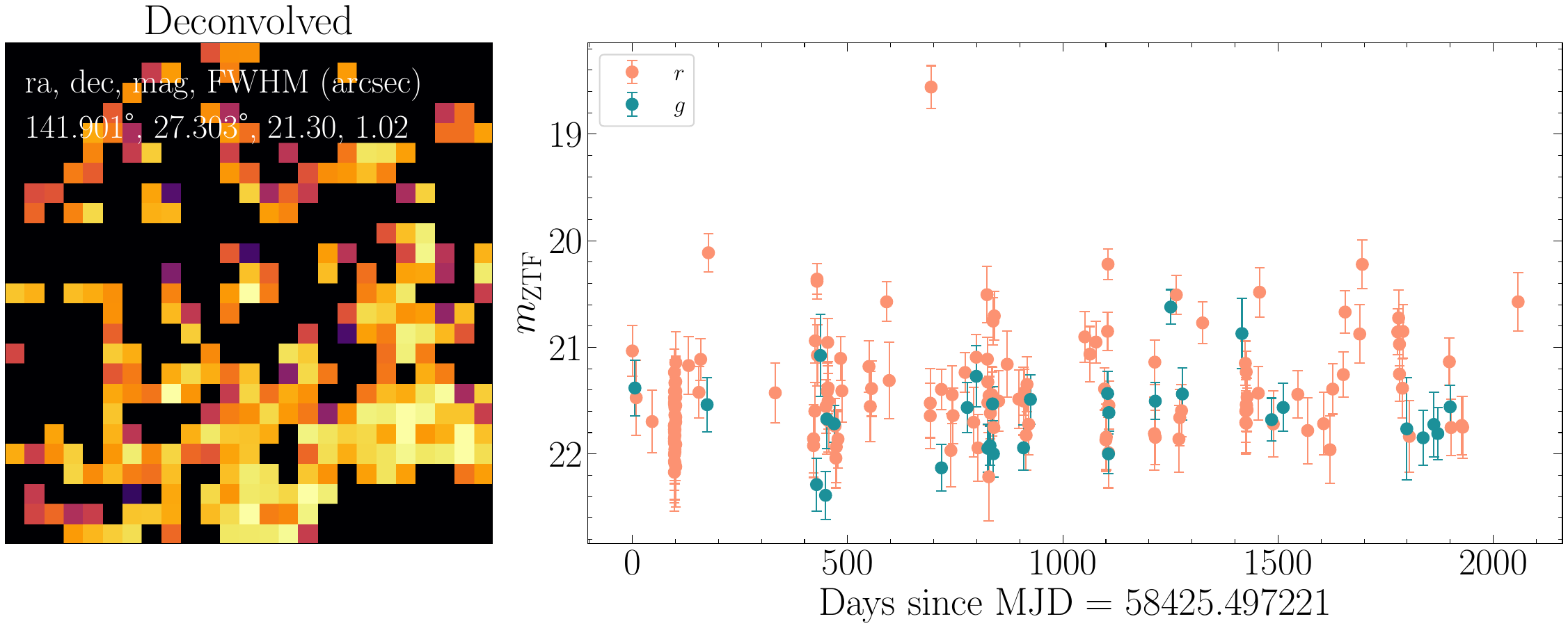}
      \includegraphics[keepaspectratio,width=0.48\linewidth]{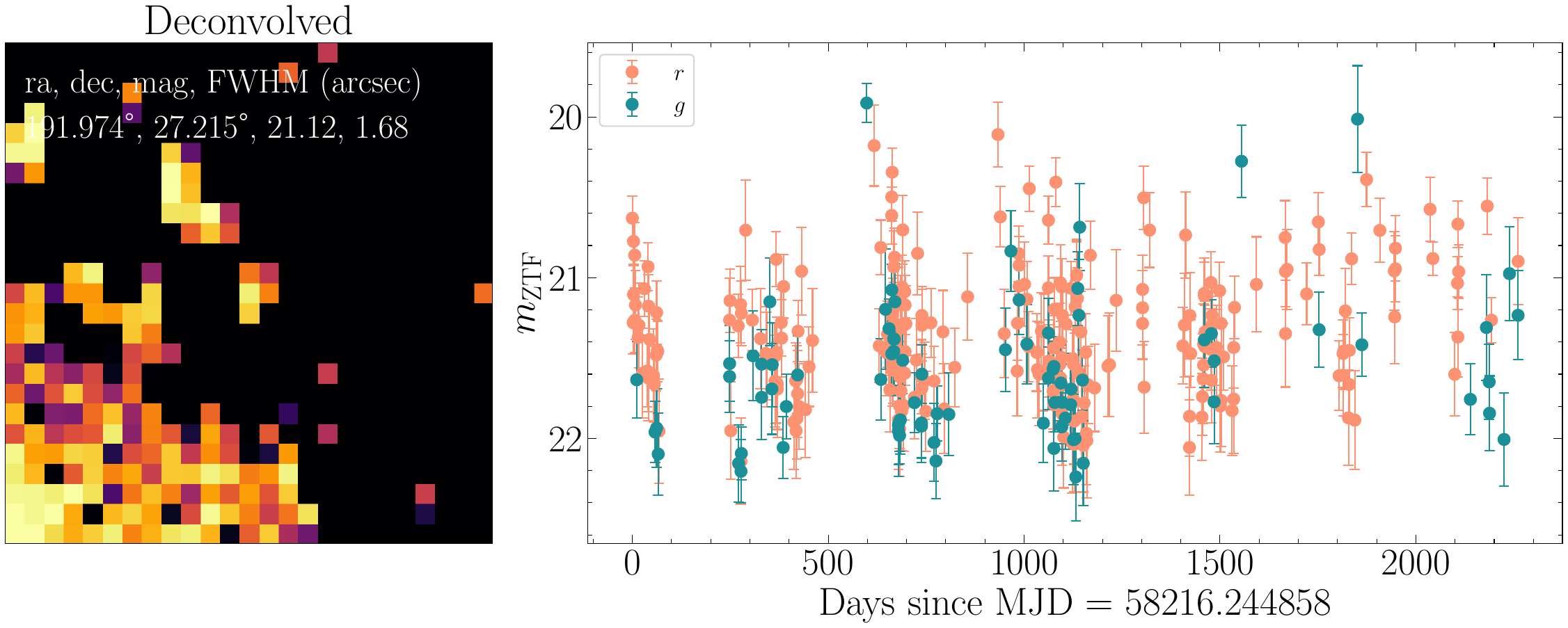}
      \includegraphics[keepaspectratio,width=0.48\linewidth]{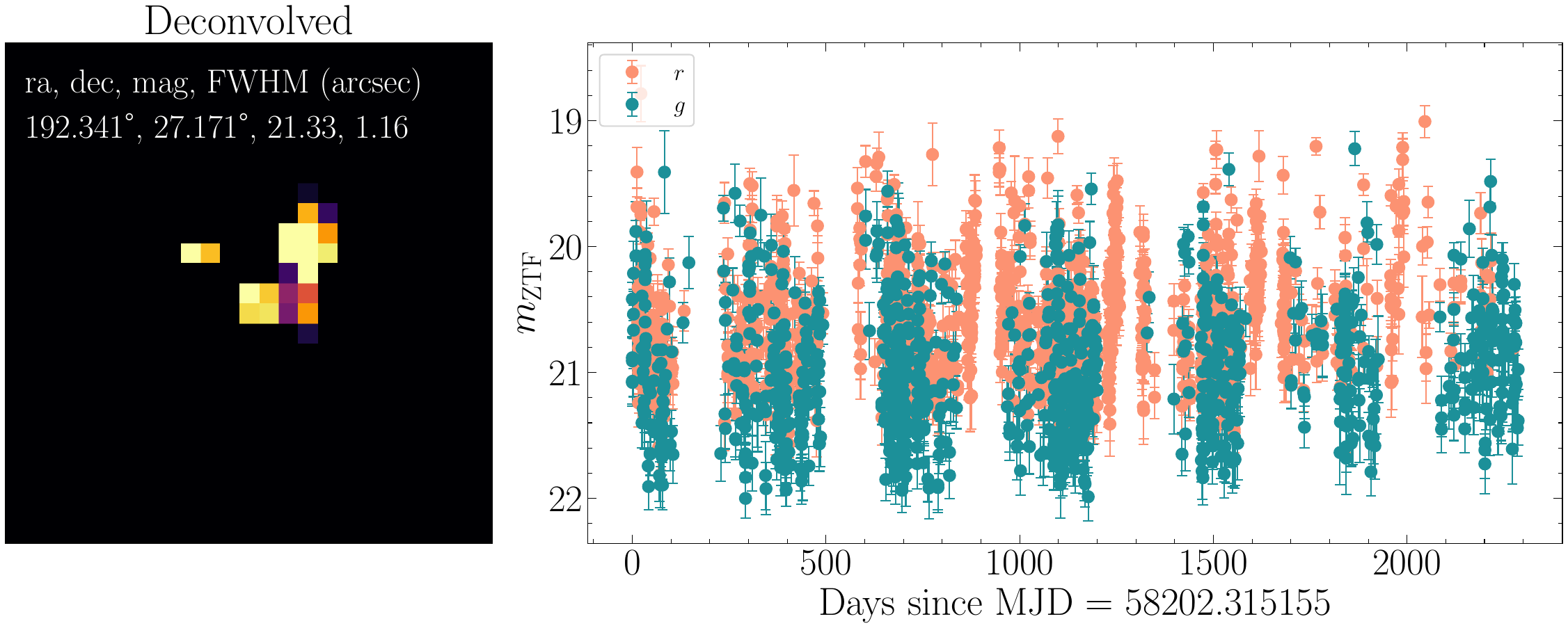}
      \includegraphics[keepaspectratio,width=0.48\linewidth]{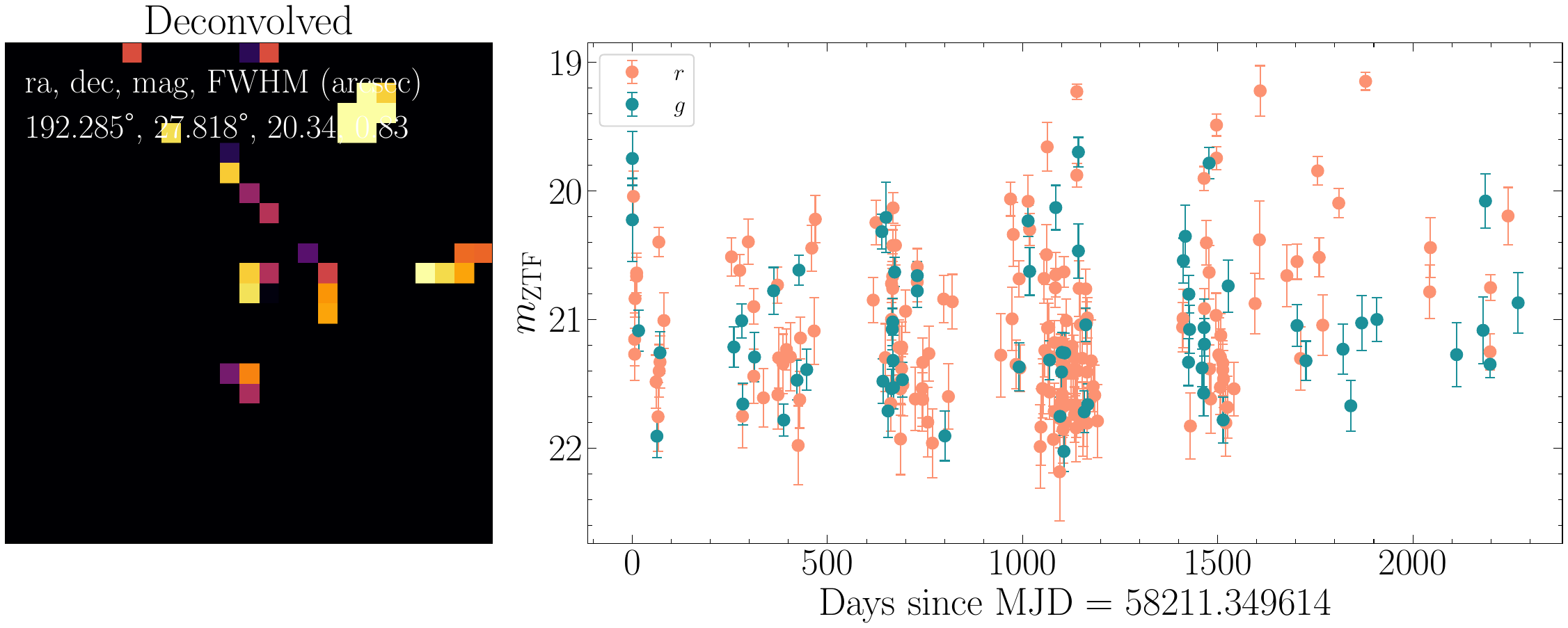}
      \includegraphics[keepaspectratio,width=0.48\linewidth]{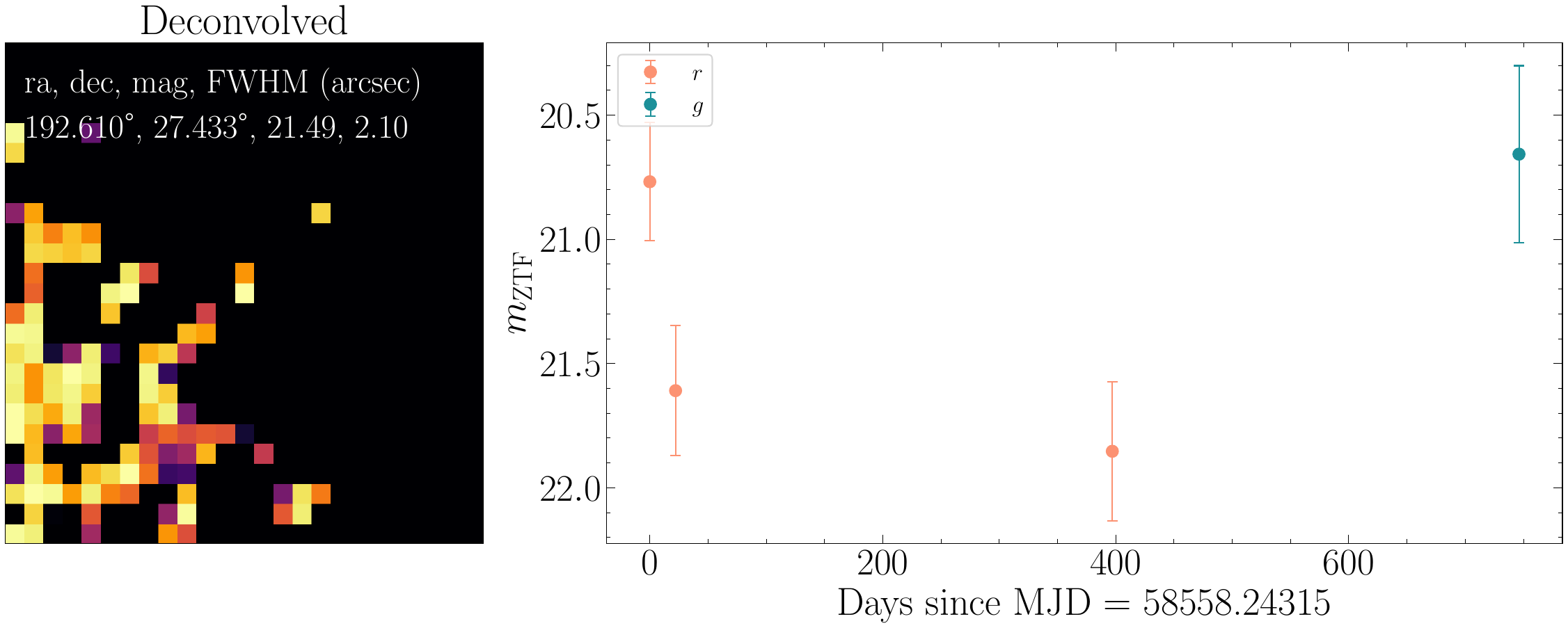}
      \includegraphics[keepaspectratio,width=0.48\linewidth]{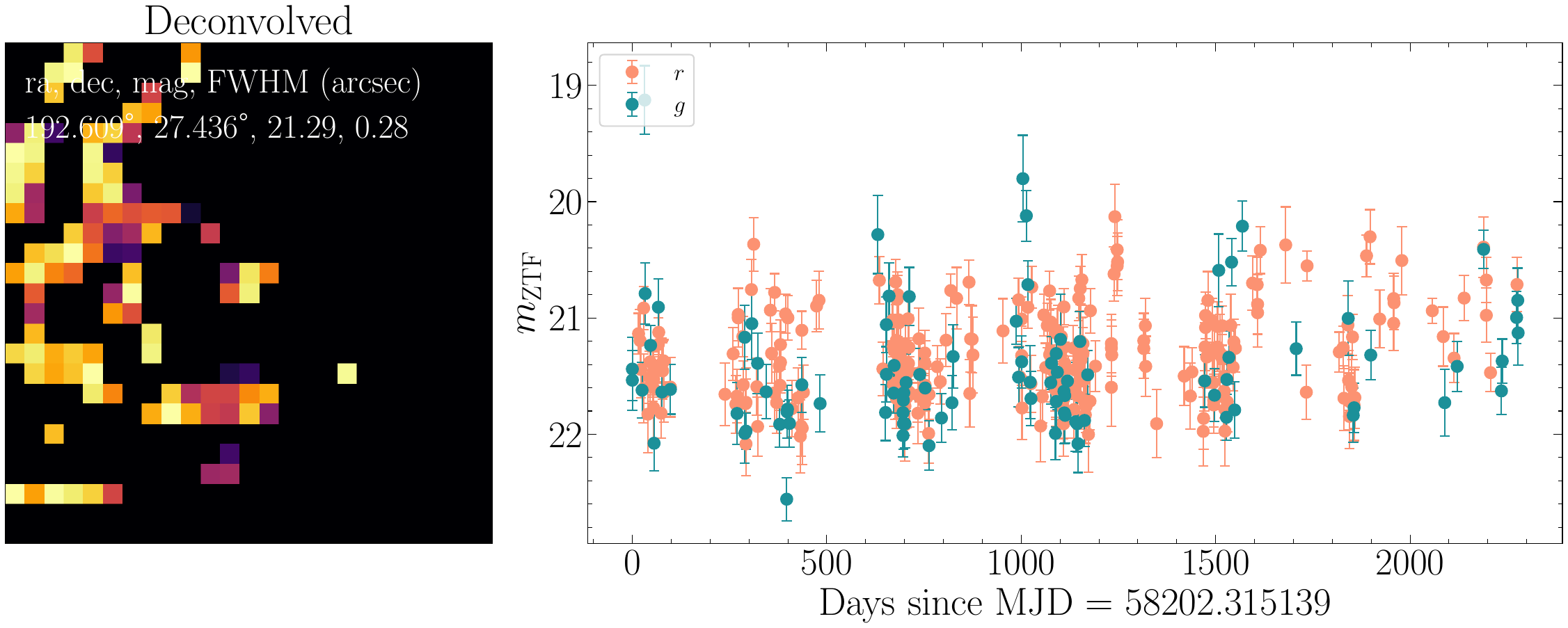}
      \includegraphics[keepaspectratio,width=0.48\linewidth]{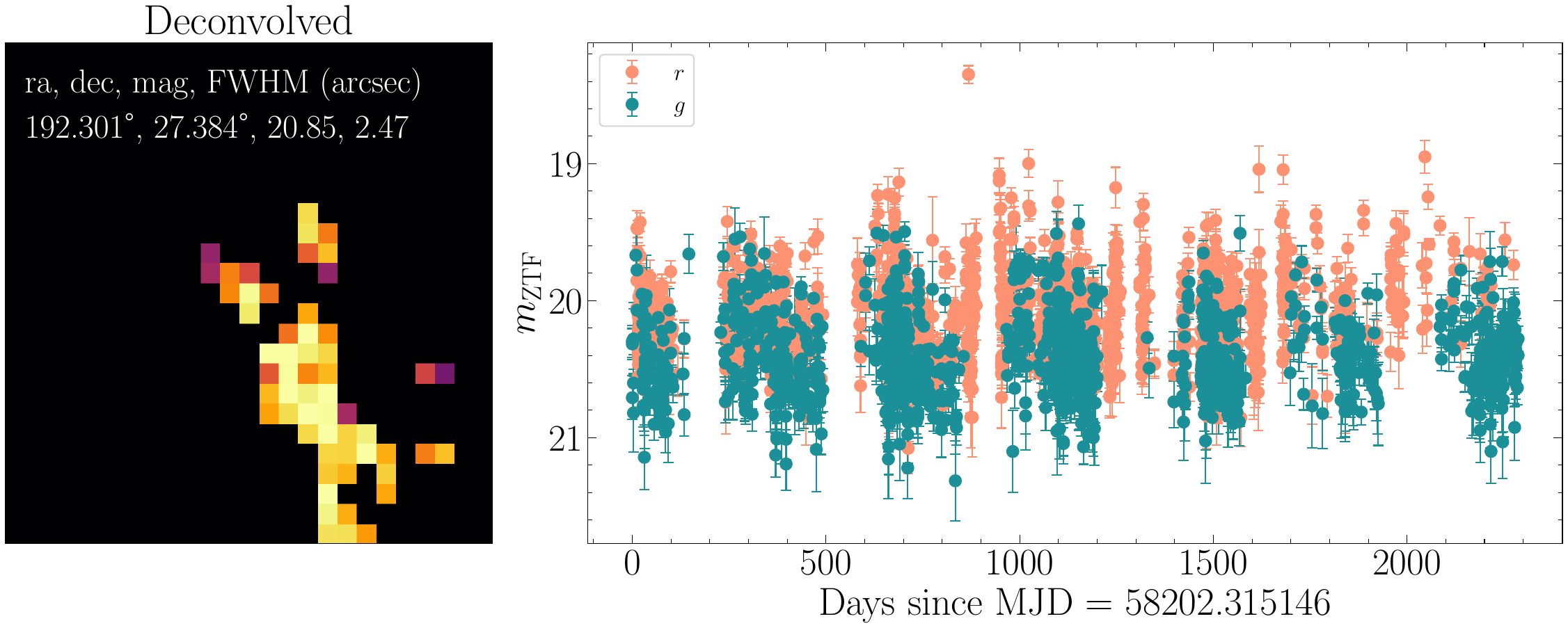}
      \includegraphics[keepaspectratio,width=0.48\linewidth]{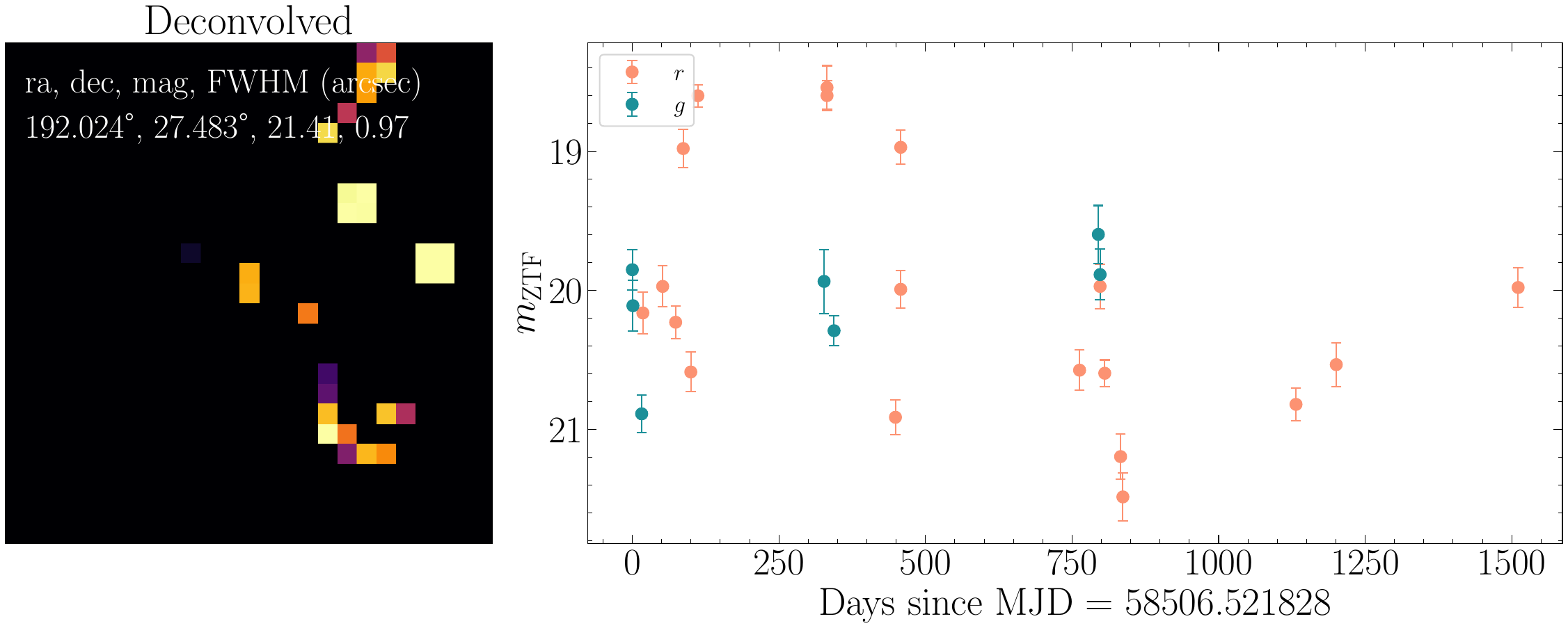}
      \includegraphics[keepaspectratio,width=0.48\linewidth]{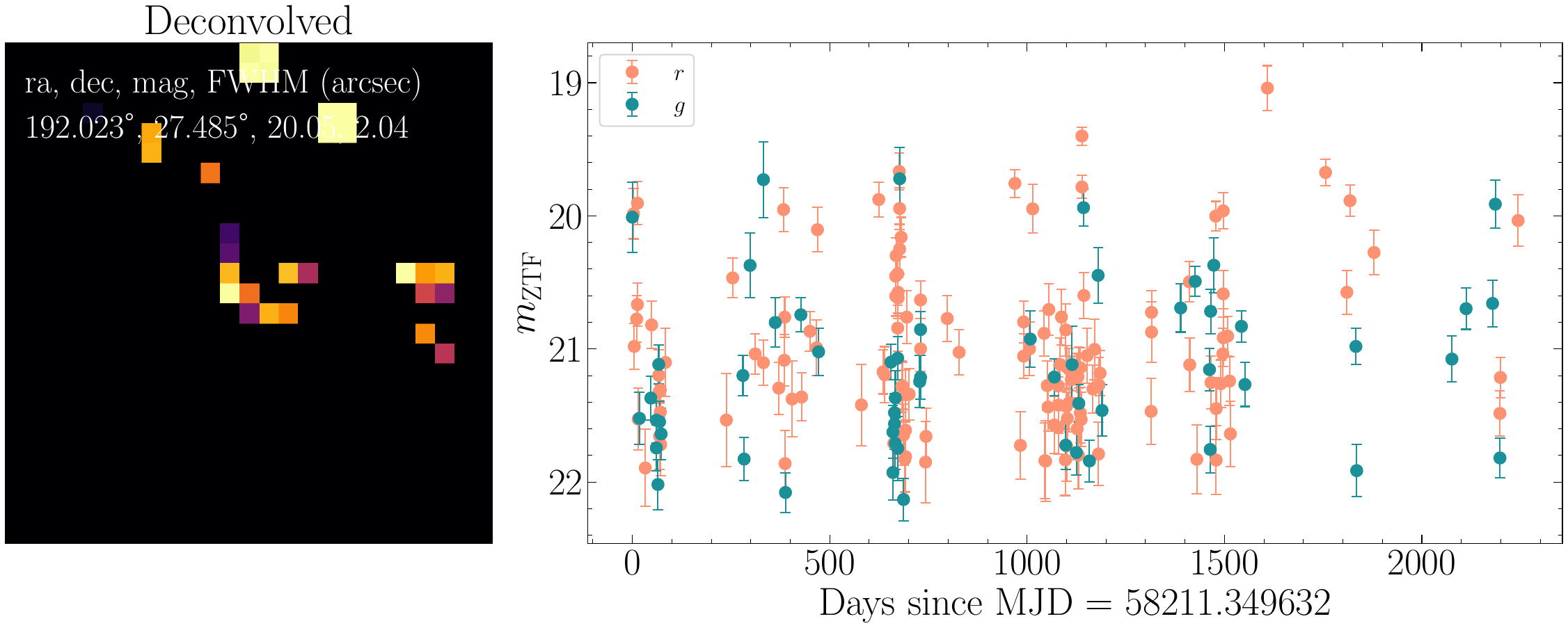}
      \includegraphics[keepaspectratio,width=0.48\linewidth]{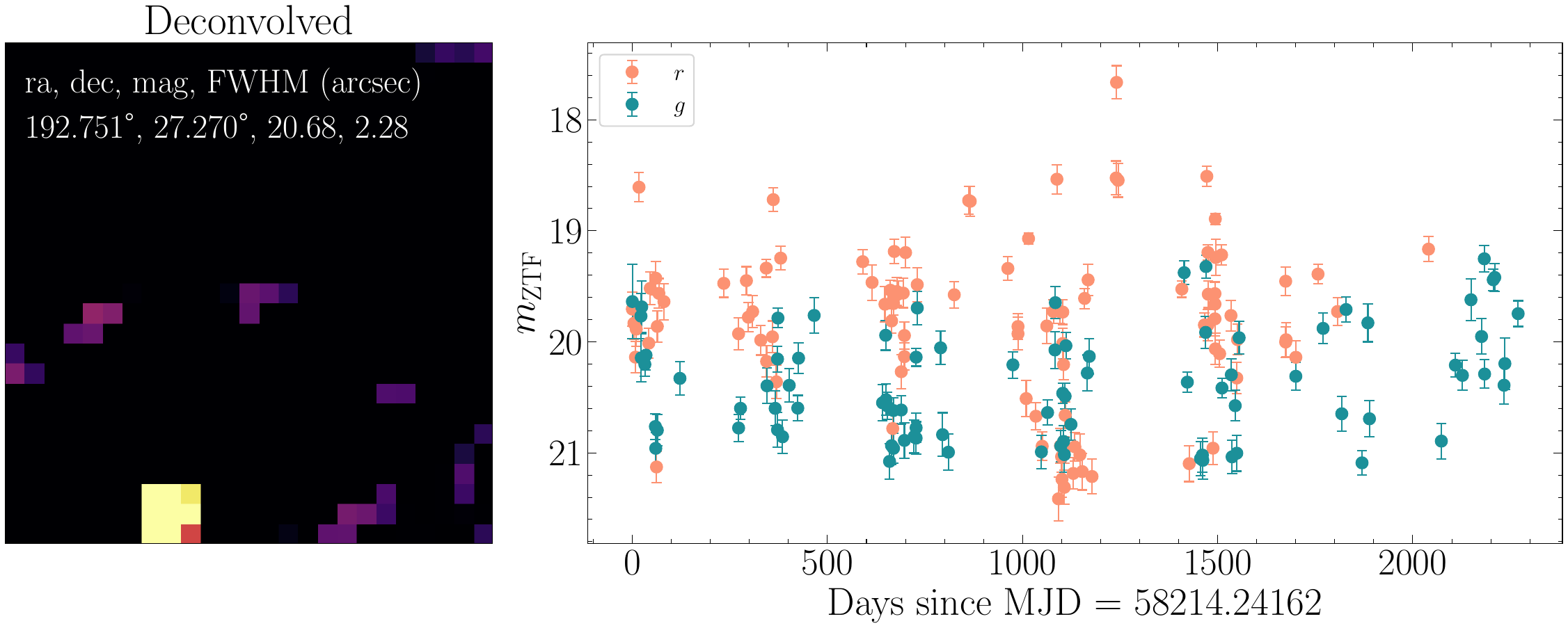}
\end{figure*}
\begin{figure*}[hbt!]
      \includegraphics[keepaspectratio,width=0.48\linewidth]{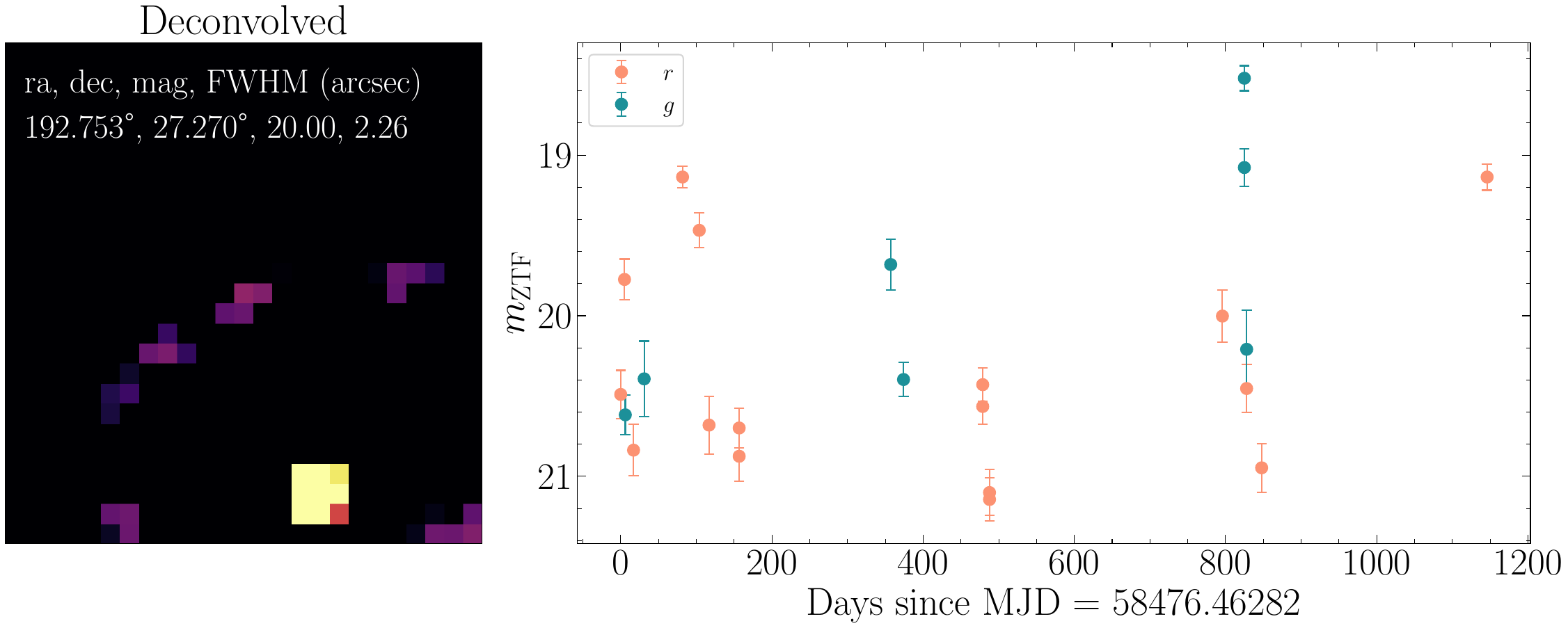}
      \includegraphics[keepaspectratio,width=0.48\linewidth]{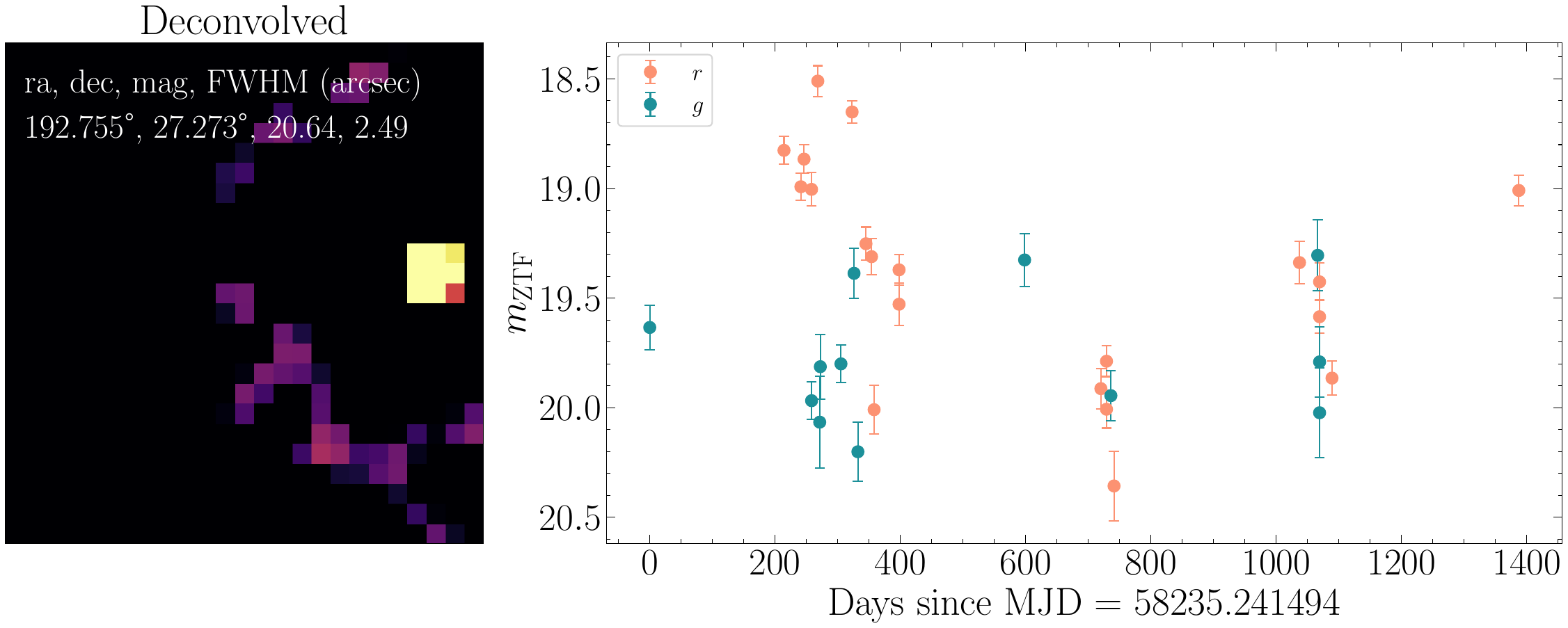}
      \includegraphics[keepaspectratio,width=0.48\linewidth]{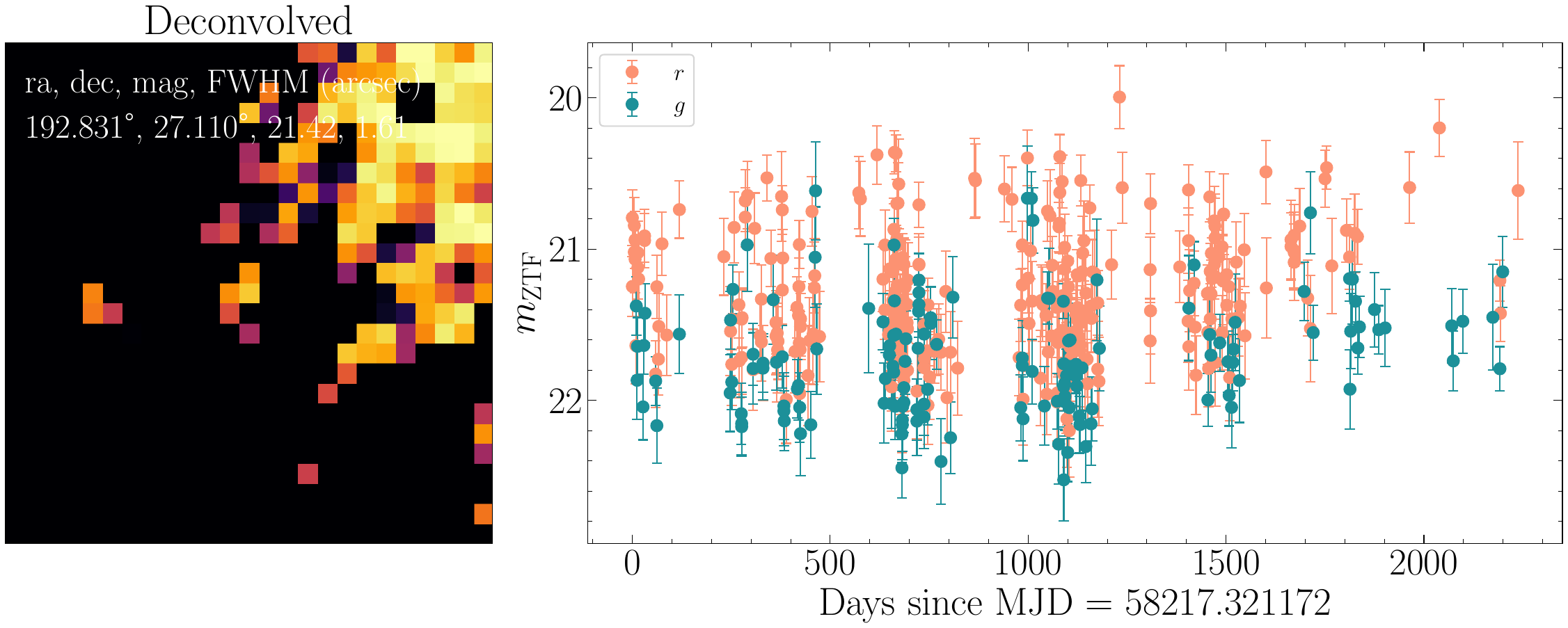}
      \includegraphics[keepaspectratio,width=0.48\linewidth]{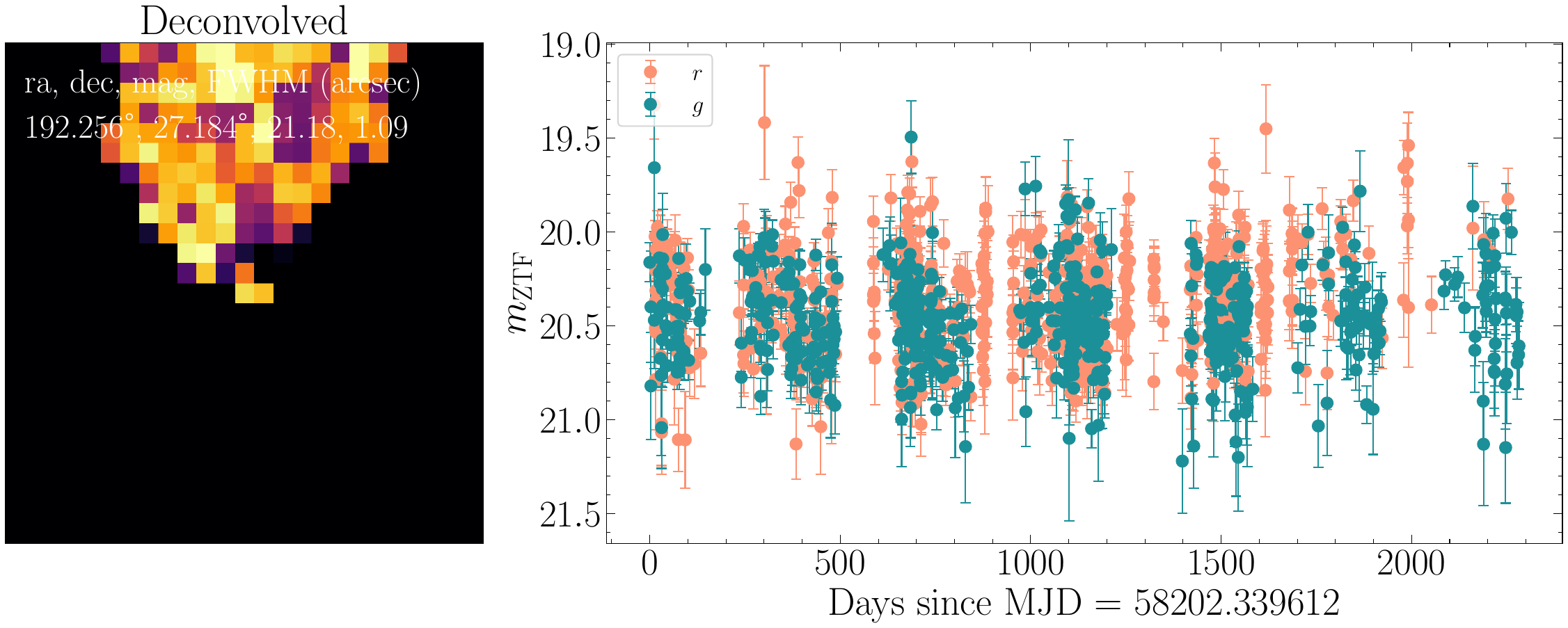}
      \includegraphics[keepaspectratio,width=0.48\linewidth]{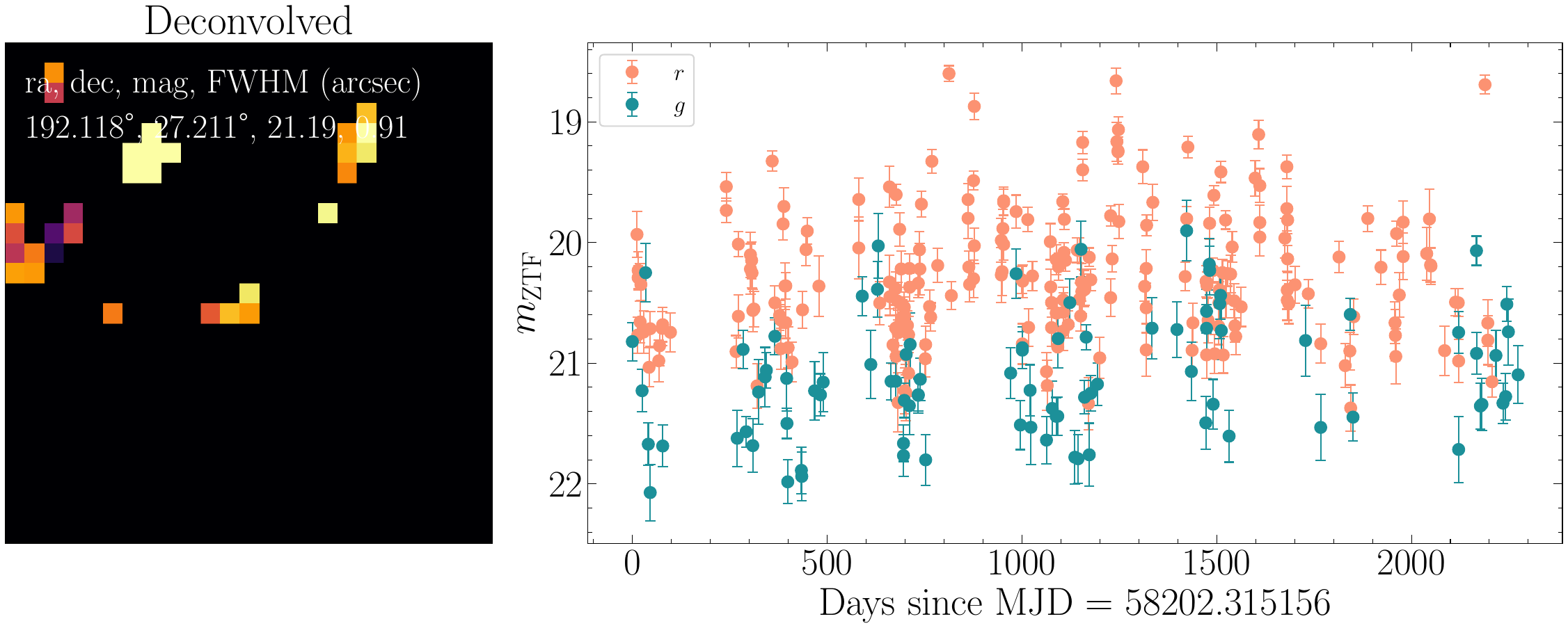} 
      \includegraphics[keepaspectratio,width=0.48\linewidth]{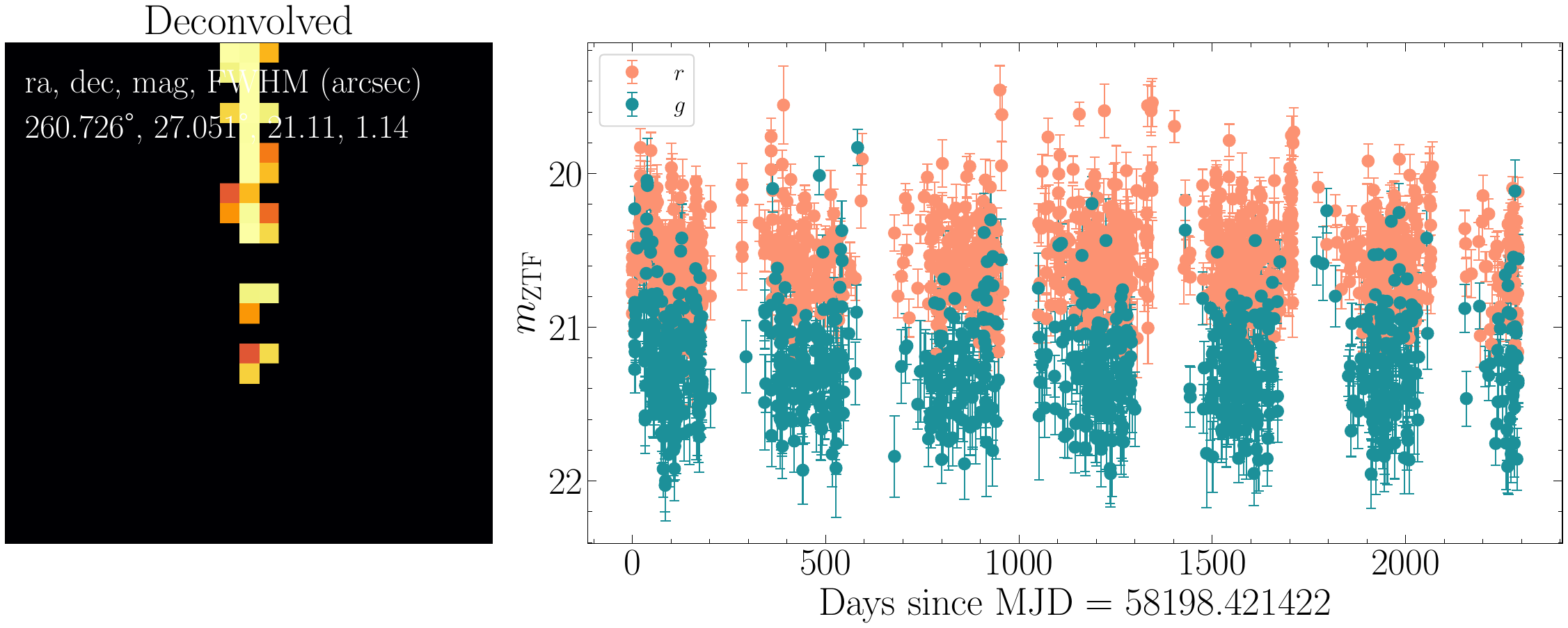}
    \caption{Continuation of Fig.~\ref{fig:crossmatching-results-OrigDeconDESI-plots-cutouts-onlyInDecon}. See the figure there for a description of the panels.} \label{fig:crossmatching-results-OrigDeconDESI-plots-cutouts-onlyInDecon-extended-fig}
\end{figure*}

\bibliography{main}{}
\bibliographystyle{aasjournal}

\end{document}